%% file: manuscript_with_supplement_for_arXiv.tex
\documentclass[11pt,english]{article}

\usepackage{pdfpages}
\usepackage[T1]{fontenc}
\usepackage[latin9]{inputenc}
\usepackage[letterpaper]{geometry}
\usepackage{color}
\usepackage{babel}
\usepackage{array}
\usepackage{booktabs}
\usepackage{units}
\usepackage{mathrsfs}
\usepackage{mathtools}
\usepackage{amsmath}
\usepackage{amsthm}
\usepackage{amssymb}
\usepackage{setspace}
\usepackage{wasysym}
\usepackage[authoryear]{natbib}
\PassOptionsToPackage{normalem}{ulem}
\usepackage{ulem}
\usepackage[unicode=true,pdfusetitle,
 bookmarks=true,bookmarksnumbered=false,bookmarksopen=false,
 breaklinks=true,pdfborder={0 0 0},pdfborderstyle={},backref=false,colorlinks=true]
 {hyperref}
\hypersetup{
 linkcolor=blue, citecolor=blue}
\usepackage{breakurl}

\makeatletter

\providecommand{\tabularnewline}{\\}

\numberwithin{equation}{section}

\theoremstyle{plain}
\newtheorem{assumption}{\protect\assumptionname}
\theoremstyle{plain}
\newtheorem{thm}{\protect\theoremname}[section]
\theoremstyle{plain}
\newtheorem{cor}{\protect\corollaryname}[section]
\theoremstyle{plain}
\newtheorem{lem}{\protect\lemmaname}[section]

\usepackage{titletoc}
\dottedcontents{section}[5.5em]{}{3.2em}{1pc}
\dottedcontents{subsection}[11em]{}{3.2em}{2pc}

\usepackage{bbm}

\usepackage{scalerel,stackengine}
\stackMath
\newcommand\reallywidecheck[1]{%
\savestack{\tmpbox}{\stretchto{%
  \scaleto{%
    \scalerel*[\widthof{\ensuremath{#1}}]{\kern-.6pt\bigwedge\kern-.6pt}%
    {\rule[-\textheight/2]{1ex}{\textheight}}
  }{\textheight}%
}{0.5ex}}%
\stackon[1pt]{#1}{\scalebox{-1}{\tmpbox}}%
}
\date{}

\newtheoremstyle{remboldstyle}
  {}{}{}{}{\bfseries}{.}{.5em}{{\thmname{#1 }}{\thmnumber{#2}}{\thmnote{ (#3)}}}
\theoremstyle{remboldstyle}
\newtheorem{rembold}{Remark}[section]

\makeatletter
\renewenvironment{proof}[1][\proofname]{%
  \par\pushQED{\qed}\normalfont%
  \topsep6\p@\@plus6\p@\relax
  \trivlist\item[\hskip\labelsep\bfseries#1\@addpunct{.}]%
  \ignorespaces
}{%
  \popQED\endtrivlist\@endpefalse
}
\makeatother

\newtheoremstyle{ntnboldstyle}
  {}{}{}{}{\bfseries}{.}{.5em}{{\thmname{#1}}{\thmnote{(#2)}}}
\theoremstyle{ntnboldstyle}
\newtheorem{ntnbold}{Notation}

\makeatother

\providecommand{\assumptionname}{Assumption}
\providecommand{\corollaryname}{Corollary}
\providecommand{\lemmaname}{Lemma}
\providecommand{\theoremname}{Theorem}

\usepackage{appendix}

\makeatletter
\@addtoreset{section}{part}
\makeatother

\begin{document}

\title{\textbf{Inference for First-Price Auctions with Guerre, Perrigne,
and Vuong's Estimator}\footnote{We thank the editor, Han Hong, the associate editor and two anonymous
referees, whose comments have greatly improved the paper. We also
thank Emmanuel Guerre, Jinyong Hahn, Nianqing Liu, Ryo Okui, Joris
Pinkse, Christoph Rothe, and Quang Vuong for their helpful comments.
Jun Ma's research is supported by the Fundamental Research Funds for
the Central Universities, and the Research Funds of Renmin University
of China, \#15XNF014 and fund for building world-class universities
(disciplines) of Renmin University of China. Vadim Marmer gratefully
acknowledges the financial support of the Social Sciences and Humanities
Research Council of Canada under grants 435-2013-0331 and 435-2017-0329. }
\let\thefootnote\relax\footnotetext{
\textcopyright 2019. This manuscript version is made available under the Creative Commons CC-BY-NC-ND 4.0 license \url{http://creativecommons.org/licenses/by-nc-nd/4.0/}. First version: March 3, 2016, This version: \today.}}

\author{Jun Ma\thanks{School of Economics, Renmin University of China, 59 Zhongguancun Street,
Haidian District, Beijing, China. Email: jun.ma@ruc.edu.cn. Tel.:
8610 62511102.}\and Vadim Marmer\thanks{Corresponding author. Vancouver School of Economics, University of
British Columbia, 6000 Iona Drive, Vancouver, BC, V6T 1L4, Canada.
Email: vadim.marmer@ubc.ca. Tel.: 1 (604) 822 8217.} \and Artyom Shneyerov\thanks{Department of Economics, Concordia University, 1455 de Maisonneuve
Blvd. West, Montreal, Quebec, H3G 1M8, Canada. Deceased 24 October
2017.}}
\maketitle
\begin{abstract}
We consider inference on the probability density of valuations in
the first-price sealed-bid auctions model within the independent private
value paradigm. We show the asymptotic normality of the two-step nonparametric
estimator of \citet*{Guerre_Perrigne_Vuong_Auction_Econometrica_2000}
(GPV), and propose an easily implementable and consistent estimator
of the asymptotic variance. We prove the validity of the pointwise
percentile bootstrap confidence intervals based on the GPV estimator.
Lastly, we use the intermediate Gaussian approximation approach to
construct bootstrap-based asymptotically valid uniform confidence
bands for the density of the valuations. \\


\end{abstract}

\tableofcontents
\newpage

\section{Introduction}

The structural estimation of auctions is an important and rapidly
growing subfield at the junction of econometrics and industrial organization.
Since the seminal work of \citet*[GPV hereafter]{Guerre_Perrigne_Vuong_Auction_Econometrica_2000},
much of theoretical and applied work has focused on nonparametric
estimation of first-price, sealed-bid auctions.\footnote{\citet*{hendricks2007empirical} survey the empirical auction literature,
while \citet*{athey2007nonparametric} survey the nonparametric identification
approaches. \citet{hickman2012structural} provide a recent review.} The object of interest is the probability density function (PDF)
of latent valuations, which can then be used for a variety of policy
counterfactuals such as the optimal reserve price (\citealp{paarsch1997deriving}
and \citealp{li2003semiparametric}).\footnote{In auctions, the valuation of a bidder is simply his or her willingness
to pay for the object.}

The focus on nonparametric estimation is due to several reasons. First,
in empirical applications, large auction datasets are often available.\footnote{E.g., \citet{kawai2014detecting} utilize a dataset of $40,000$ auctions
in their study of collusion in Japan, while \citet{augenblick2015sunk}
employs a dataset $160,000$ penny auctions.} Second, nonparametric methods are flexible since no functional form
assumptions are needed. Third, in auctions, nonparametric estimators
are built directly from the identification arguments, and are often
easy to implement.\footnote{See \citet{athey2002identification} for a number of additional identification
results.} 

GPV proposed a two-step nonparametric estimator of the PDF of valuations
in first-price auctions, and showed that it is uniformly consistent
and attains the minimax optimal uniform convergence rate. However,
it has been an open question whether this estimator also converges
in a distributional sense, which would allow empirical researchers
to perform inferences, thereby increasing the scope of applications.

Recently, \citet{Marmer_Shneyerov_Quantile_Auctions} developed an
alternative quantile-based estimator of the PDF of valuations and
showed its asymptotic normality.\footnote{More recently, quantile methods have been used in the context of auctions
in \citet{gimenes2016quantile}, \citet{gimenes2017econometrics},
\citet{liu2017nonparametric}, and \citet{luo2017integrated}.} However, the GPV estimator is well established in the literature
and is used in all the empirical applications we are aware of. Moreover,
our results imply that the GPV estimator has a smaller asymptotic
variance than that of the quantile-based estimator, as long as the
two estimators use the same second-order kernel.

Inference and the closely related problem of nonparametric testing
in structural auction models are important and have been receiving
increasing attention in the literature. Beginning with the fundamental
\citet{haile2003nonparametric}'s test for common values, recent contributions
include testing for the monotonicity of bidding strategies (\citealp{liuvuong2013}),
endogenous entry (\citealp{li2009entry} and \citealp{marmer2013model}),
common versus private values (\citealp{hill2013there}), the affiliation
of bidder valuations (\citealp{jun2010consistent}, \citealp{li2010testing}
and \citealp{de2010testing}), and inferences on bidder risk attitudes
(\citealp{fang2014inference}). In the absence of the asymptotic distribution
framework for the GPV estimator, these papers have adopted problem-specific
approaches in each case.

The first main result we show in this paper is that the GPV estimator
is asymptotically normal. The key difficulty is the presence of the
nonparametric first step, which provides nonparametric estimates of
the valuations in each auction. In the second step, the kernel density
estimator is applied to those estimates, rather than the true valuations.
This creates a unique challenge, to our knowledge not previously addressed
in the econometrics literature. Our main insight is that the leading
term in an asymptotic expansion of the estimator can be viewed as
a V-statistic with a kernel that depends on the bandwidth. A projection
argument shows that the distribution of this V-statistic is asymptotically
normal. Using maximal inequalities for empirical processes and U-processes
developed in recent literature, we show that the remainder term is
uniformly negligible. The proof is rather long due to an intricate
nature of the estimator, and involves some delicate steps.

Note that a working paper version of GPV \citep{GPV1995} also has
an asymptotic normality result for the GPV estimator. However, the
result therein is of a limited nature as it relies on a particular
choice of tuning parameters, which insures that only the second stage
of the estimating procedure contributes to the asymptotic variance.
Thus, in their approach, the uncertainty due to the estimation of
valuations in the first stage can be ignored asymptotically, which
is achieved by applying different rates of smoothing of \emph{auction-specific
covariates} at both stages. The approach is restrictive in two respects:
(i) While GPV's smoothing strategy makes first-stage estimation errors
negligible asymptotically, in finite samples their contribution to
the variance may still be significant. Our approach takes into account
the contribution of both stages and, as a result, is more accurate
in finite samples. (ii) Equally importantly, their approach cannot
be applied in cases with no auction-specific covariates or when covariates
are modeled semi-parametrically as in \citet{haile2003nonparametric}.
Note that treating auction-specific characteristics semi-parametrically
is particularly appealing to practitioners. 

One unusual feature of our asymptotic normality result concerns the
form of the asymptotic variance of the GPV estimator. Typically, the
asymptotic variances of kernel density estimators depend on the integral
of the squared kernel, which is a known constant that can be easily
computed analytically or numerically.\footnote{See, e.g., \citet{li2007net}.}
However, in the case of the GPV estimator, this constant is replaced
by a convoluted integral transformation that involves the kernel function,
its derivative, and the derivatives of the bidding strategy. This
is a consequence of the two-step nature of the GPV estimator and happens
due to the impact of the estimation errors from the first stage of
the procedure on the distribution of the estimator. Since the bidding
strategy is unknown, this fact complicates estimation of the asymptotic
variance of the GPV estimator. 

Our second contribution is to propose a consistent estimator of the
asymptotic variance that avoids estimation of the bidding strategy
and its derivative. Its uniform rate of convergence is established
using the maximal inequalities. Our third contribution is to show
the validity of the percentile bootstrap for the GPV estimator, which
allows constructing confidence intervals without estimation of the
asymptotic variance.

Our pointwise asymptotic normality results can be used for inference
on the optimal reserve price, as the latter is determined by a nonlinear
equation in the PDF of valuations \citep[see][]{haile2003iim}. In
our fourth contribution, however, we extend the pointwise results
and develop valid uniform confidence bands for the PDF. The uniform
confidence bands can be used, e.g., for specification of valuations'
density. The extension utilizes the uniform rates of convergence of
the remainder terms in our V-statistic approximation of the GPV estimator
and its Hoeffding decomposition; it also relies on Gaussian anti-concentration
inequalities and Gaussian coupling theorems developed in recent literature.
This approach, referred to as the Intermediate Gaussian Approximation
(IGA, hereafter) in the literature, is based on the seminal work of
\citet{chernozhukov2014anti,chernozhukov2014gaussian,chernozhukov2016empirical}.
\citet{chernozhukov2014gaussian} showed that although a random function
based on nonparametric estimation errors does not typically weakly
converge to any tight Gaussian random element, under certain conditions
the supremum of its studentized version can be often approximated
by the supremum of a tight Gaussian random element, the distribution
of which changes with the sample size. \citet{chernozhukov2016empirical,chernozhukov2014anti}
showed that under certain conditions the distribution of the Gaussian
supremum can be approximated by bootstrapping, and the bootstrap consistency
can be shown by applying the coupling theorems and the Gaussian anti-concentration
inequality developed in these papers.\footnote{See, e.g., \citet{Kato_Sasaki_2,Kato_Sasaki_1} for recent applications
of these theorems for constructing confidence bands for different
nonparametric curves. } Our paper is one of the first applications of these results. Our
Monte Carlo simulation results show that the IGA approach produces
confidence bands with excellent finite-sample coverage properties.

Our paper is also related to the recent literature on nonparametrically
generated regressors in nonparametric regression. See, e.g., \citet{rilstone1996nonparametric},
\citet{pinkse2001nonparametric}, and \citet{mammen2012nonparametric}.
Note, however, that while that literature is concerned with nonparametrically
estimated exogenous covariates, we deal with kernel estimation of
the density of a nonparametrically generated ``dependent'' variable,
potentially in presence of observable conditioning variables.

The rest of the paper proceeds as follows. Section \ref{sec:Data-Generating-Process-(DGP)}
introduces the data-generating process (DGP) and describes the GPV
estimator in detail. Due to complexity of the estimator, in Section
\ref{sec:Asymptotic Normality} we show the asymptotic normality of
the GPV estimator in a simplified model that has a constant number
of bidders across auctions and no auction-specific heterogeneity.
Such a simplification allows us to present the main ideas in a more
transparent fashion. In Section \ref{sec:Pointwise-Confidence-Intervals},
we derive an estimator for the asymptotic variance and establish its
uniform rate of convergence. We also show consistency of the percentile
bootstrap confidence intervals. Section \ref{sec:Uniform-Confidence-Bands}
provides results on constructing valid confidence bands within the
same simplified framework. Proofs of the results in Sections \ref{sec:Asymptotic Normality}\textendash \ref{sec:Uniform-Confidence-Bands}
are given in the Appendix. Section \ref{sec:Auction-Specific-Heterogeneity}
provides corresponding theorems in the general model with a random
number of bidders and auction-specific heterogeneity. The proofs of
these results can be found in the Supplement (included). Section \ref{sec:Binding-Reserve-Price}
discusses how our approach can be extended to auctions with binding
reserve prices. We report the results from our Monte Carlo study in
Section \ref{sec:Monte-Carlo-Simulations}. Section \ref{sec:Conclusions}
concludes.

\begin{ntnbold} ``$a\coloneqq b$'' is understood as ``$a$ is
defined by $b$''. ``$a\eqqcolon b$'' is understood as ``$b$
is defined by $a$''. $\mathbbm{1}\left(\cdot\right)$ denotes the
indicator function, and we also denote $\mathbbm{1}_{A}\coloneqq\mathbbm{1}\left(\cdot\in A\right)$.
Let $\mathbb{H}\left(\boldsymbol{c},r\right)$ denote a closed hypercube
centered at some real vector $\boldsymbol{c}$ with edge $r$. Let
$\boldsymbol{c}^{\mathrm{T}}$ denote the transpose of $\boldsymbol{c}$.
``$\overset{d}{=}$'' means ``equal in distribution''. Let $\ell^{\infty}\left(A\right)$
be the class of bounded functions defined on $A$. For any $f\in\ell^{\infty}\left(A\right)$,
let $\left\Vert f\right\Vert _{A}\coloneqq\underset{x\in A}{\mathrm{sup}}\left|f\left(x\right)\right|$
be the sup-norm. \end{ntnbold}

\section{Data-Generating Process (DGP) and the GPV Estimator\label{sec:Data-Generating-Process-(DGP)}}

The econometrician observes data from $L$ auctions. Let $\boldsymbol{X}_{l}$
denote the $d$-dimensional relevant characteristics for the object
in the $l$-th auction. Let $N_{l}$ denote the number of bidders
in the $l$-th auction. Let $B_{il}$ denote the bid submitted by
the $i$-th bidder in the $l$-th auction. The data observed by the
econometrician is given by $\left\{ \left(B_{il},\boldsymbol{X}_{l},N_{l}\right):i=1,...,N_{l},\,l=1,...,L\right\} .$
Unobserved bidders' valuations of the $l$-th auctioned object are
denoted by $\left\{ V_{il}:i=1,...,N_{l},\,l=1,...,L\right\} .$ The
following assumption describes the DGP.\footnote{Assumption \ref{assu:DGP} is similar to Assumptions A1 and A2 of
GPV and \citet[Assumption 1]{Marmer_Shneyerov_Quantile_Auctions}.
Part (f) imposes the condition that the valuations and the random
number of bidders are independent conditionally on the characteristics.
See Footnote 14 of GPV.} 
\begin{assumption}[\textbf{DGP}]
\textup{\label{assu:DGP}(a) $\left\{ \left(\boldsymbol{X}_{l},N_{l}\right):l=1,...,L\right\} $
are i.i.d.}

\textup{(b) The marginal PDF of $\boldsymbol{X}_{1}$, denoted by
$\varphi\left(\cdot\right)$, is strictly positive and continuous
on its support $\mathcal{X}\coloneqq\left[\underline{x},\overline{x}\right]^{d}$
for some $\underline{x}<\overline{x}$ assumed to be known and admits
up to $R+1$ ($R\geq2$) continuous partial derivatives.}

\textup{(c) The conditional probability mass function of $N_{1}$
given $\boldsymbol{X}_{1}=\boldsymbol{x}$, denoted by $\pi\left(\cdot|\boldsymbol{x}\right)$,
has a known support $\mathcal{N}\coloneqq\left\{ \underline{n},...,\overline{n}\right\} $
for all $\boldsymbol{x}\in\mathcal{X}$, $\underline{n}\geq2$.}

\textup{(d) For all $n\in\mathcal{N}$, $\pi\left(n|\cdot\right)$
is strictly positive and admits up to $R+1$ continuous partial derivatives.}

\textup{(e) $\left(n,\boldsymbol{x}\right)\mapsto\pi\left(n|\boldsymbol{x}\right)\varphi\left(\boldsymbol{x}\right)$
is bounded above and away from zero on its support $\mathcal{N}\times\mathcal{X}$. }

\textup{(f) For each $l=1,...,L$, given $\boldsymbol{X}_{l}=\boldsymbol{x}$
and $N_{l}=n$, $\left\{ V_{il}:i=1,...,n\right\} $ are i.i.d. with
conditional PDF $f\left(\cdot|\boldsymbol{x}\right)$ and conditional
cumulative distributional function (CDF) $F\left(\cdot|\boldsymbol{x}\right)$.}

\textup{(g) For each $n\in\mathcal{N}$, the support of $\left(V_{11},\boldsymbol{X}_{1}\right)$
is $\mathcal{S}_{V,\boldsymbol{X}}\coloneqq\left\{ \left(v,\boldsymbol{x}\right):\boldsymbol{x}\in\mathcal{X},\,v\in\left[\underline{v}\left(\boldsymbol{x}\right),\overline{v}\left(\boldsymbol{x}\right)\right]\right\} ,$
with some positive boundary functions $\underline{v}\left(\cdot\right)$
and $\overline{v}\left(\cdot\right)$.}

\textup{(h) $f\left(\cdot|\cdot\right)$ is strictly positive and
bounded away from zero and admits up to $R$ continuous partial derivatives
on $\mathcal{S}_{V,\boldsymbol{X}}$.}
\end{assumption}
Bidders' valuations are not directly observable. Following GPV, we
assume that $B_{il}$ is the equilibrium bid of risk-neutral bidder
$i$ submitted in the $l$-th auction. Therefore the valuations are
linked to the observed bids through the Bayesian Nash equilibrium
(BNE) bidding strategy:
\begin{equation}
B_{il}=s\left(V_{il},\boldsymbol{X}_{l},N_{l}\right)\coloneqq V_{il}-\frac{1}{\left(F\left(V_{il}|\boldsymbol{X}_{l}\right)\right)^{N_{l}-1}}\int_{\underline{v}\left(\boldsymbol{X}_{l}\right)}^{V_{il}}\left(F\left(u|\boldsymbol{X}_{l}\right)\right)^{N_{l}-1}\mathrm{d}u.\footnotemark\label{eq:BNE strategy}
\end{equation}
\footnotetext{See Equations (1) and (8) of GPV.}Under Assumptions
\ref{assu:DGP}(a) and \ref{assu:DGP}(f), $\left\{ B_{il}:i=1,...,N_{l}\right\} $
are conditionally i.i.d. draws given $\boldsymbol{X}_{l}$ and $N_{l}$.
Let $\overline{b}\left(\boldsymbol{x},n\right)\coloneqq s\left(\overline{v}\left(\boldsymbol{x}\right),\boldsymbol{x},n\right)$
and $\underline{b}\left(\boldsymbol{x}\right)\coloneqq\underline{v}\left(\boldsymbol{x}\right)$.
Proposition 1(i) of GPV shows that the support of $\left(B_{il},\boldsymbol{X}_{l},N_{l}\right)$
is $\left\{ \left(b,\boldsymbol{x},n\right):n\in\mathcal{N},\,\left(b,\boldsymbol{x}\right)\in\mathcal{S}_{B,\boldsymbol{X}}^{n}\right\} ,$
where $\mathcal{S}_{B,\boldsymbol{X}}^{n}\coloneqq\left\{ \left(b,\boldsymbol{x}\right):\boldsymbol{x}\in\mathcal{X},\,b\in\left[\underline{b}\left(\boldsymbol{x}\right),\overline{b}\left(\boldsymbol{x},n\right)\right]\right\} .$

Let $G\left(\cdot|\boldsymbol{x},n\right)$ denote the conditional
CDF of $B_{il}$ given $\boldsymbol{X}_{l}=\boldsymbol{x}$ and $N_{l}=n$.
Let $g\left(\cdot|\boldsymbol{x},n\right)$ be the corresponding conditional
PDF. GPV established identification of the inverse bidding strategy:
\begin{equation}
V_{il}=\xi\left(B_{il},\boldsymbol{X}_{l},N_{l}\right)\coloneqq B_{il}+\frac{1}{N_{l}-1}\frac{G\left(B_{il}|\boldsymbol{X}_{l},N_{l}\right)}{g\left(B_{il}|\boldsymbol{X}_{l},N_{l}\right)}.\label{eq:inverse bidding strategy}
\end{equation}
By replacing $G\left(\cdot|\cdot,\cdot\right)$ and $g\left(\cdot|\cdot,\cdot\right)$
in (\ref{eq:inverse bidding strategy}) with their nonparametric estimators,
GPV proposed an estimator of $\xi\left(\cdot,\cdot,\cdot\right)$,
denoted by $\widehat{\xi}\left(\cdot,\cdot,\cdot\right)$. The GPV
estimator of $f\left(v|\boldsymbol{x}\right)$ is the kernel density
estimator that, in place of the true valuations, uses the so-called
pseudo valuations $\left\{ \widehat{V}_{il}\coloneqq\widehat{\xi}\left(B_{il},\boldsymbol{X}_{l},N_{l}\right):i=1,...,N_{l},\,l=1,...,L\right\} $. 

Below we provide the details of GPV's estimation procedure. Let $K_{0}$
and $K_{1}$ be univariate kernel functions of different orders satisfying
the following assumption: 
\begin{assumption}[\textbf{Kernel}]
\label{assu:Assumption 2 Kernel}\textup{(a) $K_{0}$ and $K_{1}$
are symmetric, compactly supported on $[-1,1]$ and twice continuously
differentiable on $\mathbb{R}$ with Lipschitz derivatives.}

\textup{(b) $K_{0}$ is of order $R$, and $K_{1}$ is of order $1+R$:
$\int K_{0}(u)\mathrm{d}u=1$ and $\int u^{k}K_{0}(u)\mathrm{d}u=0$
for $k=1,...,R-1$; $\int K_{1}(u)\mathrm{d}u=1$ and $\int u^{k}K_{1}(u)\mathrm{d}u=0$
for $k=1,...,R$.}
\end{assumption}
With $K_{g}\coloneqq K_{1}$ and the multi-dimensional product kernels
\[
K_{f}\left(v,\boldsymbol{x}\right)\coloneqq K_{0}\left(v\right)\cdot\prod_{k=1}^{d}K_{0}\left(x_{k}\right)\textrm{ and }K_{\boldsymbol{X}}\left(\boldsymbol{x}\right)\coloneqq\prod_{k=1}^{d}K_{1}\left(x_{k}\right),\textrm{ for }v\in\mathbb{R},\,\boldsymbol{x}=\left(x_{1},...,x_{d}\right)\in\mathbb{R}^{d},
\]
define the following nonparametric estimators:
\[
\widehat{\varphi}\left(\boldsymbol{x}\right)\coloneqq\frac{1}{L}\sum_{l=1}^{L}\frac{1}{h^{d}}K_{\boldsymbol{X}}\left(\frac{\boldsymbol{X}_{l}-\boldsymbol{x}}{h}\right)\textrm{ and }\widehat{\pi}\left(n|\boldsymbol{x}\right)\coloneqq\frac{1}{\widehat{\varphi}\left(\boldsymbol{x}\right)L}\sum_{l=1}^{L}\mathbbm{1}\left(N_{l}=n\right)\frac{1}{h^{d}}K_{\boldsymbol{X}}\left(\frac{\boldsymbol{X}_{l}-\boldsymbol{x}}{h}\right),
\]
where $\widehat{\varphi}\left(\cdot\right)$ is the kernel density
estimator of $\varphi$ and $\widehat{\pi}\left(\cdot|\cdot\right)$
is the Nadaraya-Watson estimator of the conditional probability mass
function $\pi\left(\cdot|\cdot\right)$. Based on these, we define
below the nonparametric estimators of the conditional CDF and PDF
of the bids: 
\begin{gather*}
\widehat{G}\left(b|\boldsymbol{x},n\right)\coloneqq\frac{1}{\widehat{\pi}\left(n|\boldsymbol{x}\right)\widehat{\varphi}\left(\boldsymbol{x}\right)L}\sum_{l=1}^{L}\mathbbm{1}\left(N_{l}=n\right)\frac{1}{N_{l}}\sum_{i=1}^{N_{l}}\mathbbm{1}\left(B_{il}\leq b\right)\frac{1}{h^{d}}K_{\boldsymbol{X}}\left(\frac{\boldsymbol{X}_{l}-\boldsymbol{x}}{h}\right),\\
\widehat{g}\left(b|\boldsymbol{x},n\right)\coloneqq\frac{1}{\widehat{\pi}\left(n|\boldsymbol{x}\right)\widehat{\varphi}\left(\boldsymbol{x}\right)L}\sum_{l=1}^{L}\mathbbm{1}\left(N_{l}=n\right)\frac{1}{N_{l}}\sum_{i=1}^{N_{l}}\frac{1}{h^{1+d}}K_{g}\left(\frac{B_{il}-b}{h}\right)K_{\boldsymbol{X}}\left(\frac{\boldsymbol{X}_{l}-\boldsymbol{x}}{h}\right).
\end{gather*}

Consider a partition of $\mathbb{R}^{d}$ with generic half-open hypercubes
of side $h_{\partial}>0$:
\[
\Pi_{k_{1},...,k_{d}}\coloneqq\left[k_{1}h_{\partial},\left(k_{1}+1\right)h_{\partial}\right)\times\cdots\times\left[k_{d}h_{\partial},\left(k_{d}+1\right)h_{\partial}\right),
\]
where $\left(k_{1},...,k_{d}\right)$ runs over $\mathbb{Z}^{d}$.
Let $\Pi_{h_{\partial}}\left(\boldsymbol{x}\right)$ denote the hypercube
that contains $\boldsymbol{x}$ in this partition. Define 
\begin{gather}
\widehat{\overline{b}}\left(\boldsymbol{x},n\right)\coloneqq\mathrm{\mathrm{max}}\left\{ B_{pl}:p=1,...,N_{l},\,\boldsymbol{X}_{l}\in\Pi_{h_{\partial}}\left(\boldsymbol{x}\right),\,N_{l}=n,\,l=1,...,L\right\} ,\nonumber \\
\widehat{\underline{b}}\left(\boldsymbol{x}\right)\coloneqq\mathrm{min}\left\{ B_{pl}:p=1,...,N_{l},\,\boldsymbol{X}_{l}\in\Pi_{h_{\partial}}\left(\boldsymbol{x}\right),\,l=1,...,L\right\} \label{eq:bids support estimator}
\end{gather}
to be the estimators of the boundaries of the support. Note that the
estimators of the boundaries are super consistent. Let $\mathcal{\widehat{S}}_{B,\boldsymbol{X}}^{n}\coloneqq\left\{ \left(b,\boldsymbol{x}\right):\boldsymbol{x}\in\mathcal{X},\,b\in\left[\widehat{\underline{b}}\left(\boldsymbol{x}\right),\widehat{\overline{b}}\left(\boldsymbol{x},n\right)\right]\right\} .$
The support of $\left(B_{il},\boldsymbol{X}_{l},N_{l}\right)$ then
can be estimated by $\left\{ \left(b,\boldsymbol{x},n\right):n\in\mathcal{N},\,\left(b,\boldsymbol{x}\right)\in\mathcal{\widehat{S}}_{B,\boldsymbol{X}}^{n}\right\} .$

The kernel density estimator $\widehat{g}\left(b|\boldsymbol{x},n\right)$
is asymptotically biased when $\left(b,\boldsymbol{x}\right)$ is
near the boundaries of the support. GPV suggested that trimming should
be applied to the observations near the estimated boundaries using
the trimming factor $\mathbb{T}_{il}\coloneqq\mathbbm{1}\left(\mathbb{H}\left(\left(B_{il},\boldsymbol{X}_{l}\right),2h\right)\subseteq\mathcal{\widehat{S}}_{B,\boldsymbol{X}}^{N_{l}}\right).$
The two-step nonparametric estimator of $f\left(v|\boldsymbol{x}\right)$
developed by GPV is
\begin{equation}
\widehat{f}_{GPV}\left(v|\boldsymbol{x}\right)\coloneqq\frac{1}{\widehat{\varphi}\left(\boldsymbol{x}\right)L}\sum_{l=1}^{L}\frac{1}{N_{l}}\sum_{i=1}^{N_{l}}\mathbb{T}_{il}\frac{1}{h^{1+d}}K_{f}\left(\frac{\widehat{V}_{il}-v}{h},\frac{\boldsymbol{X}_{l}-\boldsymbol{x}}{h}\right).\label{eq:f_hat_GPV definition}
\end{equation}

For deriving the asymptotic properties of the GPV estimator, we make
the following assumption on the bandwidths $h$ and $h_{\partial}$.\footnote{Assumption \ref{assu: rate of bandwidth}(a) is the same as the assumption
on the rate of bandwidth for \citet{Marmer_Shneyerov_Quantile_Auctions}'s
quantile-based estimator. See Assumption 3 therein. }
\begin{assumption}[\textbf{Bandwidth}]
\label{assu: rate of bandwidth}\textup{ (a) The bandwidth $h$ is
of the form $h=\lambda_{1}L^{-\gamma_{1}}$, for some strictly positive
constants $\lambda_{1}$ and $\gamma_{1}$ satisfying $\nicefrac{1}{\left(2R+3+d\right)}\leq\gamma_{1}<\nicefrac{1}{\left(3+d\right)}$.}

\textup{(b) When $d>0$, the ``boundary'' bandwidth is of the form
$h_{\partial}=\lambda_{\partial}\left(\nicefrac{\mathrm{log}\left(L\right)}{L}\right)^{\nicefrac{1}{\left(1+d\right)}}$,
where $\lambda_{\partial}$ is a strictly positive constant.}
\end{assumption}
GPV showed that the optimal uniform convergence rate of their estimator
is attained when the bandwidth $h$ is of order $O\left(\left(\nicefrac{\mathrm{log}\left(L\right)}{L}\right)^{\nicefrac{1}{\left(2R+3+d\right)}}\right)$.
Note that the bandwidth in Assumption \ref{assu: rate of bandwidth}
is of smaller order. Under-smoothing imposed in Assumption \ref{assu: rate of bandwidth}
is needed to control the asymptotic bias of the GPV estimator, which
is important for the validity of inference.

\section{Asymptotic Normality of the GPV Estimator\label{sec:Asymptotic Normality}}

For clarity of the presentation of the main ideas and results, in
this section we first establish pointwise asymptotic normality of
the GPV estimator in a simplified version of the model that has a
fixed number of bidders and no auction-specific heterogeneity. When
there are covariates capturing auction-specific heterogeneity present,
these results can be used by treating the covariates additively semi-parametrically
as in \citet[Section 6]{haile2003nonparametric}. In that case, there
is no kernel smoothing over the covariates, as the GPV procedure would
be applied to the ``homogenized'' bids, which are constructed as
residuals from the parametric regression of the bids against the covariates.

In the simplified model, the econometrician observes data on bids
in $L$ identical auctions, with a fixed number of bidders $N$ in
each auction: $\left\{ B_{il}:i=1,\ldots,N,\,l=1,\ldots,L\right\} .$
Under Assumption \ref{assu:DGP}, the valuations $\left\{ V_{il}:i=1,\ldots,N,\,l=1,\ldots,L\right\} $
are i.i.d. with a compact support $\left[\underline{v},\overline{v}\right]\subseteq\mathbb{R}_{+}$,
PDF $f$ and CDF $F$. The object of interest is the PDF of the valuation
at interior points of $\left[\underline{v},\overline{v}\right]$.
Suppose that $v_{l}>\underline{v}$, $v_{u}<\overline{v}$ and $I\coloneqq\left[v_{l},v_{u}\right]$
is an inner closed sub-interval of $\left[\underline{v},\overline{v}\right]$.
Fix 
\[
\overline{\delta}\coloneqq\mathrm{min}\left\{ \nicefrac{\left(\overline{v}-v_{u}\right)}{2},\nicefrac{\left(v_{l}-\underline{v}\right)}{2}\right\} .
\]

Under Assumption \ref{assu:DGP}, $f$ is strictly positive and bounded
away from zero on its support and admits at least $R$ continuous
derivatives. Lemma A1 of GPV showed that under Assumption \ref{assu:DGP},
the BNE bidding strategy is strictly increasing and $R+1$ times continuously
differentiable. In this simplified framework, the inverse of the BNE
bidding strategy is
\begin{equation}
\xi\left(b\right)\coloneqq b+\frac{1}{N-1}\frac{G\left(b\right)}{g\left(b\right)},\label{eq: inverse bid strat}
\end{equation}
where $G$ and $g$ are the CDF and PDF of bids respectively. Denote
$\overline{b}\coloneqq s\left(\overline{v}\right)$ and $\underline{b}\coloneqq s\left(\underline{v}\right)$.
Proposition 1(ii) of GPV shows that under Assumption \ref{assu:DGP},
$g$ is also bounded away from zero on its support $\left[\underline{b},\overline{b}\right]$:
\begin{equation}
\underline{C}_{g}\coloneqq\underset{b\in\left[\underline{b},\overline{b}\right]}{\mathrm{inf}}g\left(b\right)>0.\label{eq:density of bids bounded  from 0}
\end{equation}

The inverse bidding strategy (\ref{eq: inverse bid strat}) can be
estimated by
\[
\widehat{\xi}\left(b\right)\coloneqq b+\frac{1}{N-1}\frac{\widehat{G}\left(b\right)}{\widehat{g}\left(b\right)},
\]
where we use the usual nonparametric estimators of $G$ and $g$:
\[
\widehat{G}\left(b\right)\coloneqq\frac{1}{N\cdot L}\sum_{i,l}\mathbbm{1}\left(B_{il}\leq b\right)\textrm{ and }\widehat{g}\left(b\right)\coloneqq\frac{1}{N\cdot L}\sum_{i,l}\frac{1}{h}K_{g}\left(\frac{B_{il}-b}{h}\right),
\]
where $\sum_{i,l}$ is understood as $\sum_{l=1}^{L}\sum_{i=1}^{N}$. 

Let $\widehat{\overline{b}}\coloneqq\mathrm{max}\left\{ B_{il}:i=1,\ldots,N,\,l=1,...,L\right\} $,
and $\widehat{\underline{b}}\coloneqq\mathrm{min}\left\{ B_{il}:i=1,\ldots,N,\,l=1,...,L\right\} $.
The trimming factor is now simply $\mathbb{T}_{il}\coloneqq\mathbbm{1}\left(\widehat{\underline{b}}+h\leq B_{il}\leq\widehat{\overline{b}}-h\right)$.
The GPV estimator of $f\left(v\right)$ is now given by
\[
\widehat{f}_{GPV}\left(v\right)=\frac{1}{N\cdot L}\sum_{i,l}\mathbb{T}_{il}\frac{1}{h}K_{f}\left(\frac{\widehat{V}_{il}-v}{h}\right),
\]
where $K_{f}=K_{0}$ in this simplified framework. 

We derive the following stochastic expansion of $\widehat{f}_{GPV}\left(v\right)$
around $f\left(v\right)$:
\begin{equation}
\widehat{f}_{GPV}\left(v\right)-f\left(v\right)=\frac{1}{\left(N\cdot L\right)^{2}}\sum_{i,l}\widetilde{\mathbb{T}}_{il}\frac{1}{h^{2}}K_{f}'\left(\frac{V_{il}-v}{h}\right)\left(\widehat{V}_{il}-V_{il}\right)+o_{p}\left(\left(Lh^{3}\right)^{-\nicefrac{1}{2}}\right),\label{eq:the first stochastic approximation result}
\end{equation}
where $\widetilde{\mathbb{T}}_{il}\coloneqq\mathbbm{1}\left(\left|V_{il}-v\right|\leq\overline{\delta}\right)$
is an infeasible trimming factor and the remainder term is uniform
in $v\in I$. In the above expression, the derivative $K_{f}'$ of
the kernel function appears due to the linearization of $K_{f}\left(\nicefrac{\left(\widehat{V}_{il}-v\right)}{h}\right)$
around $K_{f}\left(\nicefrac{\left(V_{il}-v\right)}{h}\right)$. The
result in (\ref{eq:the first stochastic approximation result}) shows
that the distribution of the GPV estimator depends not only on the
variation in $V_{il}$'s, but also on the estimation errors of pseudo
valuations. In other words, the errors from estimation of the inverse
bidding strategy affect the asymptotic distribution of the GPV estimator.

Since $\widehat{G}$ has a faster rate of convergence than $\widehat{g}$,
the discrepancy between $V_{il}$ and $\widehat{V}_{il}$ depends
on that between the true PDF $g\left(B_{il}\right)$ and the estimated
PDF $\widehat{g}\left(B_{il}\right)$, which in turn depends on the
averaged discrepancy between $K_{g}\left(\nicefrac{\left(B_{jm}-B_{il}\right)}{h}\right)$
and $g\left(B_{il}\right)$, where the averaging is across $B_{jm}$'s.
Lemma \ref{lem:lemma 2} establishes a further asymptotic expansion
for the GPV estimator:
\begin{equation}
\widehat{f}_{GPV}\left(v\right)-f\left(v\right)=\frac{1}{\left(N-1\right)}\frac{1}{\left(N\cdot L\right)^{2}}\sum_{i,l}\sum_{j,k}\mathcal{M}\left(B_{il},B_{jk};v\right)+o_{p}\left(\left(Lh^{3}\right)^{-\nicefrac{1}{2}}\right),\label{eq:f_GPV - f leading term}
\end{equation}
where the remainder term is uniform in $v\in I$, and
\begin{equation}
\mathcal{M}\left(b,b';v\right)\coloneqq-\frac{1}{h^{2}}K_{f}'\left(\frac{\xi\left(b\right)-v}{h}\right)\frac{G\left(b\right)}{g\left(b\right)^{2}}\left(\frac{1}{h}K_{g}\left(\frac{b'-b}{h}\right)-g\left(b\right)\right).\label{eq:V_stat kernel definition}
\end{equation}
For any fixed $v$, the leading term in (\ref{eq:f_GPV - f leading term})
is a V-statistic with a kernel that depends on the bandwidth $h$.
We now apply Hoeffding decomposition to this leading term. Define
\begin{eqnarray}
\mathcal{M}_{1}\left(b;v\right) & \coloneqq & \int\mathcal{M}\left(b,b';v\right)\mathrm{d}G\left(b'\right)\nonumber \\
 & = & -\frac{1}{h^{2}}K_{f}'\left(\frac{\xi\left(b\right)-v}{h}\right)\frac{G\left(b\right)}{g\left(b\right)^{2}}\left(\mathrm{E}\left[\frac{1}{h}K_{g}\left(\frac{B_{11}-b}{h}\right)\right]-g(b)\right),\label{eq: M1 explained}
\end{eqnarray}
and further,
\[
\mathcal{M}_{2}\left(b;v\right)\coloneqq\int\mathcal{M}\left(b',b;v\right)\mathrm{d}G\left(b'\right),\textrm{ and }\mu_{\mathcal{M}}\left(v\right)\coloneqq\int\int\mathcal{M}\left(b,b';v\right)\mathrm{d}G\left(b\right)\mathrm{d}G\left(b'\right).
\]
Note that $\mu_{\mathcal{M}}\left(v\right)=\mathrm{E}\left[\mathcal{M}_{1}\left(B_{11};v\right)\right]=\mathrm{E}\left[\mathcal{M}_{2}\left(B_{11};v\right)\right]$.
The Hoeffding decomposition yields
\begin{align}
 & \frac{1}{\left(N\cdot L\right)^{2}}\sum_{i,l}\sum_{j,k}\mathcal{M}\left(B_{il},B_{jk};v\right)\nonumber \\
= & \mu_{\mathcal{M}}\left(v\right)+\left\{ \frac{1}{N\cdot L}\sum_{i,l}\mathcal{M}_{1}\left(B_{il};v\right)-\mu_{\mathcal{M}}\left(v\right)\right\} +\left\{ \frac{1}{N\cdot L}\sum_{i,l}\mathcal{M}_{2}\left(B_{il};v\right)-\mu_{\mathcal{M}}\left(v\right)\right\} \nonumber \\
 & +\frac{1}{\left(N\cdot L\right)\left(N\cdot L-1\right)}\sum_{\left(i,l\right)\neq\left(j,k\right)}\left\{ \mathcal{M}\left(B_{il},B_{jk};v\right)-\mathcal{M}_{1}\left(B_{il};v\right)-\mathcal{M}_{2}\left(B_{jk};v\right)+\mu_{\mathcal{M}}\left(v\right)\right\} \nonumber \\
 & +\frac{1}{\left(N\cdot L\right)^{2}}\sum_{i,l}\mathcal{M}\left(B_{il},B_{il};v\right)-\frac{1}{\left(N\cdot L\right)^{2}\left(N\cdot L-1\right)}\sum_{\left(i,l\right)\neq\left(j,k\right)}\mathcal{M}\left(B_{il},B_{jk};v\right).\label{eq:M Hoeffding decomposition}
\end{align}

In the proof of Theorem \ref{thm:Asymptotic Normality GPV} below,
we use results for empirical processes and U-processes to show that
the terms in the third and fourth lines of (\ref{eq:M Hoeffding decomposition})
and $\left(N\cdot L\right)^{-1}\sum_{i,l}\mathcal{M}_{1}\left(B_{il};v\right)$
are asymptotically negligible uniformly in $v\in I$. As is apparent
from the definition of $\mathcal{M}_{1}$ in (\ref{eq: M1 explained}),
the contribution of the $\mathcal{M}_{1}\left(b;v\right)$ terms is
negligible because they depend on the difference between the expectation
$\mathrm{E}\left[h^{-1}K_{g}\left(\nicefrac{\left(B_{il}-b\right)}{h}\right)\right]$
and $g\left(b\right)$, i.e., the bias of the kernel density estimator,
which is of order $O\left(h^{1+R}\right)$. Thus, the asymptotic distribution
of the GPV estimator is driven solely by
\begin{equation}
\frac{1}{(N-1)}\frac{1}{N\cdot L}\sum_{i,l}\left(\mathcal{M}_{2}\left(B_{il};v\right)-\mu_{\mathcal{M}}\left(v\right)\right),\label{eq:leading term of the stochastic expansion identical auctions}
\end{equation}
where $\mathcal{M}_{2}\left(B_{il};v\right)-\mu_{\mathcal{M}}\left(v\right)$,
$i=1,...,N$, $l=1,...,L$ are independent, zero-mean and depend on
the bandwidth. 

We also show in the proof of Theorem \ref{thm:Asymptotic Normality GPV}
that the rescaled variance of (\ref{eq:leading term of the stochastic expansion identical auctions})
satisfies
\begin{align}
 & \mathrm{E}\left[\frac{Lh^{3}}{\left(N-1\right)^{2}}\left(\frac{1}{N\cdot L}\sum_{i,l}\left(\mathcal{M}_{2}\left(B_{il};v\right)-\mu_{\mathcal{M}}\left(v\right)\right)\right)^{2}\right]\nonumber \\
= & \frac{1}{N\left(N-1\right)^{2}h^{3}}\int\left\{ \int K_{f}'\left(\frac{\xi\left(b'\right)-v}{h}\right)\frac{G\left(b'\right)}{g\left(b'\right)^{2}}K_{g}\left(\frac{b-b'}{h}\right)\mathrm{d}G\left(b'\right)\right\} ^{2}\mathrm{d}G\left(b\right)+O\left(h^{3}\right)\nonumber \\
\eqqcolon & \mathrm{V}_{\mathcal{M}}\left(v\right)+O\left(h^{3}\right),\label{eq:E_m2^2 expansion}
\end{align}
where the remainder term is uniform in $v\in I$. Thus, the asymptotic
variance of the GPV estimator is the limit of the leading term in
(\ref{eq:E_m2^2 expansion}) as $h\downarrow0$. Note that
\begin{multline}
\lim_{h\downarrow0}\frac{1}{h^{3}}\int\left\{ \int K_{f}'\left(\frac{\xi\left(b'\right)-v}{h}\right)\frac{G\left(b'\right)}{g\left(b'\right)^{2}}K_{g}\left(\frac{b-b'}{h}\right)\mathrm{d}G\left(b'\right)\right\} ^{2}\mathrm{d}G\left(b\right)\\
=\frac{G(s(v))^{2}(s'(v))^{2}}{g(s(v))}\int\left\{ \int K_{f}'\left(u\right)K_{g}\left(w-s'\left(v\right)u\right)\mathrm{d}u\right\} ^{2}\mathrm{d}w.\label{eq:variance limit simple}
\end{multline}
We have the following result.
\begin{thm}[\textbf{Asymptotic Normality}]
\label{thm:Asymptotic Normality GPV}Suppose Assumptions \ref{assu:DGP}
- \ref{assu: rate of bandwidth} hold. Then for any interior point
$v\in\left(\underline{v},\overline{v}\right)$, $\left(Lh^{3}\right)^{\nicefrac{1}{2}}\left(\widehat{f}_{GPV}\left(v\right)-f\left(v\right)\right)\rightarrow_{d}\mathrm{N}\left(0,\mathrm{V}_{GPV}\left(v\right)\right),$
where
\begin{equation}
\mathrm{V}_{GPV}\left(v\right)\coloneqq\frac{1}{N\left(N-1\right)^{2}}\frac{F\left(v\right)^{2}f\left(v\right)^{2}}{g\left(s\left(v\right)\right)^{3}}\int\left\{ \int K_{f}'\left(u\right)K_{g}\left(w-s'\left(v\right)u\right)\mathrm{d}u\right\} ^{2}\mathrm{d}w.\label{eq:asymptotic covariance}
\end{equation}
\end{thm}
\begin{rembold}The asymptotic variance $\mathrm{V}_{GPV}\left(v\right)$
has an interesting and non-standard feature. Typically, the asymptotic
variance of a kernel estimator involves a constant which is a simple
functional of the kernel and does not involve any elements of the
DGP. However, in the case of the GPV estimator, the constant has a
convolution form. Moreover, this constant also involves the derivative
of the unknown bidding function. As can be seen from (\ref{eq:the first stochastic approximation result}),
the convolution form is due to the fact that the variance of the GPV
estimator is determined not only by the variation of $V_{il}$, but
also by the estimation errors $\widehat{V}_{il}-V_{il}$. The $s'(v)$
term appears due to averaging of $\xi(b)$'s in a small neighborhood
of $v$, as one can see from (\ref{eq:E_m2^2 expansion}).

The presence of the $s'(v)$ term inside the integral in (\ref{eq:asymptotic covariance})
can cause additional complications in estimation of the asymptotic
variance $\mathrm{V}_{GPV}(v)$, as the econometrician now also has
to estimate the functional of the kernel function. Potentially, one
could estimate $s'(v)$ and then estimate the convolution by the plug-in
approach. However, we show in the next section that the asymptotic
variance can be estimated directly without separate estimation of
$s'(v)$ and the convolution by using the sample analogue of the expression
in the second line of (\ref{eq:E_m2^2 expansion}). \end{rembold}

\begin{rembold}[Extensions]Theorem \ref{thm:Asymptotic Normality GPV}
also implies asymptotic normality of some modified GPV estimators.
\citet{henderson_2012_JOE} proposed to modify the standard GPV approach
by estimating $\xi$ under a monotonicity constraint implied by the
structural model. However, if the auction model is correctly specified,
the unconstrained estimator of $\xi$ will be monotone with probability
approaching one. Hence, we expect the standard GPV estimator and the
estimator proposed in \citet{henderson_2012_JOE} to be first-order
asymptotically equivalent.

\citet{hickman_hubbard_2014_JAE} proposed another modified GPV estimator
by replacing sample trimming used in the standard GPV procedure with
a boundary correction. Their method uses a boundary-bias-corrected
kernel estimator in estimation of the density of bids. The corrected
estimator is uniformly consistent over the entire support of the distribution
of bids. When estimating the PDF of valuations away from the boundaries,
our proof of (\ref{eq:f_GPV - f leading term}) can be adapted to
show that the same V-statistic approximation holds for the estimator
in \citet{hickman_hubbard_2014_JAE}. Hence, their estimator is first-order
asymptotically equivalent to the standard GPV estimator, when $v$
is chosen away from the boundaries. \end{rembold}

\begin{rembold}[Asymptotic Bias]In the proof of Theorem \ref{thm:Asymptotic Normality GPV},
we incorporate the bias term:

\begin{multline}
\left(Lh^{3}\right)^{\nicefrac{1}{2}}\left(\widehat{f}_{GPV}\left(v\right)-f\left(v\right)-\frac{1}{R!}f^{\left(R\right)}\left(v\right)\left(\int K_{f}\left(u\right)u^{R}\mathrm{d}u\right)h^{R}+o\left(h^{R}\right)\right)\\
\rightarrow_{d}\mathrm{N}\left(0,\mathrm{V}_{GPV}\left(v\right)\right).\label{eq:asymptotic normality with bias term}
\end{multline}
The leading bias term of the GPV estimator is the same as that of
the infeasible estimator constructed using the unobserved true valuations.
This is due to the following feature of the first-price auction model:
at interior points, the bid density has $1+R$ continuous derivatives
rather than $R$. It is natural to incorporate this structural feature
in the estimation procedure by using a higher-order kernel in the
first stage. \end{rembold}

\begin{rembold}[Comparison with the Quantile-Based Estimator]\label{rem:comparison QB}The
quantile-based (QB) estimator of \citet{Marmer_Shneyerov_Quantile_Auctions}
does not require estimation of latent valuations, and instead relies
on a direct representation of the PDF of valuations using the distribution
functions of observable bids. While the two estimators have the same
rate of convergence, the limiting distribution of the GPV estimator
shows that it indeed improves on the QB estimator in the following
sense. One can show that the GPV estimator has a smaller asymptotic
variance than that of the quantile-based estimator, as long as the
two estimators use the same second-order kernel functions. Suppose
now $K_{f}=K_{g}=K$, for some second-order kernel function $K$.
Let $\mathrm{V}_{QB}\left(v\right)$ be the asymptotic variance of
the QB estimator in the simplest setting without auction-specific
heterogeneity.\footnote{See \citet[Theorem 2]{Marmer_Shneyerov_Quantile_Auctions} for the
expression of the asymptotic variance of the quantile-based estimator.} By Jensen's inequality and standard calculus techniques, it can be
easily shown that $\nicefrac{\mathrm{V}_{QB}\left(v\right)}{\mathrm{V}_{GPV}\left(v\right)}\geq1$.
The details of the proof can be found in the Supplement.

Consider the following family of PDF's: $f_{\theta}(v)=\theta v^{\theta-1}\cdot\mathbbm{1}\left(0\leq v\leq1\right)$
for some $\theta>0$. The corresponding BNE bidding strategy is $s(v)=\left(1-\left(\theta(N-1)+1\right)^{-1}\right)v$.
Note that in this case $s'(v)$ is constant, and we can compute the
ratio $\nicefrac{\mathrm{V}_{QB}\left(v\right)}{\mathrm{V}_{GPV}\left(v\right)}$
analytically. In the case of the triweight kernel $K$, ratio is found
to be, e.g., 1.3259 when $\left(\theta,N\right)=\left(1,2\right)$
and 2.3038 when $\left(\theta,N\right)=\left(2,7\right)$. Thus, depending
on the model, the GPV estimator could be substantially more precise
than the QB estimator.\footnote{We believe that this finding is of interest in a more general context
outside of the empirical auctions literature. It illustrates that
two-step nonparametric estimators can outperform more direct estimators
that avoid first-stage estimation of latent variables.}\end{rembold}

\section{Pointwise Confidence Intervals\label{sec:Pointwise-Confidence-Intervals}}

If the asymptotic variance $\mathrm{V}_{GPV}(v)$ can be consistently
estimated by some estimator $\widehat{\mathrm{V}}_{GPV}\left(v\right)$,
one can construct an asymptotically valid pointwise confidence interval
for $f\left(v\right)$ as 
\begin{equation}
CI^{\dagger}\left(v\right)\coloneqq\left[\widehat{f}_{GPV}\left(v\right)-z_{1-\nicefrac{\alpha}{2}}\sqrt{\frac{\widehat{\mathrm{V}}_{GPV}\left(v\right)}{Lh^{3}}},\widehat{f}_{GPV}\left(v\right)+z_{1-\nicefrac{\alpha}{2}}\sqrt{\frac{\widehat{\mathrm{V}}_{GPV}\left(v\right)}{Lh^{3}}}\right],\label{eq:CI_n definitions}
\end{equation}
where $z_{1-\nicefrac{\alpha}{2}}$ denotes the $1-\nicefrac{\alpha}{2}$
quantile of the standard normal distribution. 

While the formula for the asymptotic variance in (\ref{eq:asymptotic covariance})
can be used for plug-in estimation of $\mathrm{V}_{GPV}(v)$, such
an estimator would be difficult to implement in practice. Firstly,
it would require estimating the bidding strategy and its derivative.
Secondly, even with an estimate of $s'\left(v\right)$, it is not
always easy to compute analytically the double integral in the definition
of $\mathrm{V}_{GPV}\left(v\right)$. This issue becomes even more
severe when there is auction-specific heterogeneity, as we discuss
in Section \ref{sec:Auction-Specific-Heterogeneity}. In that case,
one would need to evaluate a multidimensional integral. 

To avoid those issues, we propose an alternative approach to estimation
of the asymptotic variance. As we discuss in the previous section,
the asymptotic variance of the GPV estimator is the limit of the expression
in (\ref{eq:E_m2^2 expansion}). The leading term on the right-hand
side of (\ref{eq:E_m2^2 expansion}) can be estimated using a U-type-statistic,
while replacing the unknown $G$, $g$, and $\xi$ with $\widehat{G}$,
$\widehat{g}$, and $\widehat{\xi}$ respectively. The resulting estimator
is given by 
\begin{eqnarray}
\widehat{\mathrm{V}}_{GPV}\left(v\right) & \coloneqq & \frac{1}{N\left(N-1\right)^{2}h^{3}}\frac{1}{\left(N\cdot L\right)\left(N\cdot L-1\right)\left(N\cdot L-2\right)}\nonumber \\
 &  & \times\sum_{i,l}\sum_{\left(j,k\right)\neq\left(i,l\right)}\sum_{\left(j',k'\right)\neq\left(i,l\right),\,\left(j',k'\right)\neq\left(j,k\right)}\eta_{il,jk}(v)\eta_{il,j'k'}(v),\text{ where}\nonumber \\
\eta_{il,jk}(v) & \coloneqq & \mathbb{T}_{jk}K_{f}'\left(\frac{\widehat{V}_{jk}-v}{h}\right)\frac{\widehat{G}\left(B_{jk}\right)}{\widehat{g}\left(B_{jk}\right)^{2}}K_{g}\left(\frac{B_{il}-B_{jk}}{h}\right).\label{eq:definition estimator of asymptotic variance}
\end{eqnarray}
The estimator avoids estimation of the bidding strategy and its derivative
and evaluation of multidimensional integrals. It is very easily implementable
in practice since it depends only on the bids, $\widehat{G}$, $\widehat{g}$,
and the pseudo valuations. The next theorem shows consistency of the
proposed estimator and provides an estimate of its uniform convergence
rate.
\begin{thm}[\textbf{Variance Estimation}]
\label{thm:variance estimator}Suppose Assumptions \ref{assu:DGP}
- \ref{assu: rate of bandwidth} hold. Then,
\[
\underset{v\in I}{\mathrm{sup}}\left|\widehat{\mathrm{V}}_{GPV}\left(v\right)-\mathrm{V}_{\mathcal{M}}\left(v\right)\right|=O_{p}\left(\left(\frac{\mathrm{log}\left(L\right)}{Lh^{3}}\right)^{\nicefrac{1}{2}}+h^{R}\right).
\]
\end{thm}
An alternative to the confidence interval (\ref{eq:CI_n definitions})
is the bootstrap. We show below that the bootstrap approximation to
the distribution of $S\left(v\right)\coloneqq\left(Lh^{3}\right)^{\nicefrac{1}{2}}\left(\widehat{f}_{GPV}\left(v\right)-f\left(v\right)\right)$
is asymptotically valid. Our focus is on the percentile bootstrap
as it does not require estimation of the asymptotic variance, which
makes it fairly popular among practitioners.

Let $\left\{ B_{il}^{*}:i=1,\ldots,N,l=1,\ldots,L\right\} $ denote
the bootstrap sample, i.e., a set of independent random variables
drawn from the distribution $\widehat{G}$ conditionally on the original
sample of bids. Let $\widehat{G}^{*}$ and $\widehat{g}^{*}$ denote
the bootstrap analogues of $\widehat{G}$ and $\widehat{g}$ respectively:
they are constructed by following exactly the same procedure as that
for constructing $\widehat{G}$ and $\widehat{g}$, however using
the (empirical) bootstrap sample instead of the original sample. Let
$\widehat{\xi}^{*}$ be the bootstrap analogue of $\widehat{\xi}$
defined using $\widehat{G}^{*}$ and $\widehat{g}^{*}$ in place of
$\widehat{G}$ and $\widehat{g}$. We generate bootstrap samples of
pseudo values as $\widehat{V}_{il}^{*}\coloneqq\widehat{\xi}^{*}\left(B_{il}^{*}\right)$.
Lastly, we construct a bootstrap analogue of $\widehat{f}_{GPV}\left(v\right)$:
\[
\widehat{f}_{GPV}^{*}\left(v\right)\coloneqq\frac{1}{N\cdot L}\sum_{i,l}\mathbb{T}_{il}^{*}\frac{1}{h}K_{f}\left(\frac{\widehat{V}_{il}^{*}-v}{h}\right),
\]
where $\mathbb{T}_{il}^{*}\coloneqq\mathbbm{1}\left(\widehat{\underline{b}}+h\leq B_{il}^{*}\leq\widehat{\overline{b}}-h\right)$. 

Let $q_{\tau}^{*}(v)$ be the $\tau$-th quantile of the conditional
distribution of $\widehat{f}_{GPV}^{*}\left(v\right)$ given the original
sample. The percentile bootstrap confidence interval is
\[
CI^{*}\left(v\right)\coloneqq\left[q_{\nicefrac{\alpha}{2}}^{*}(v),\,q_{1-\nicefrac{\alpha}{2}}^{*}(v)\right]=\left[\widehat{f}_{GPV}\left(v\right)+\frac{s_{\nicefrac{\alpha}{2}}^{*}(v)}{\sqrt{Lh^{3}}},\,\widehat{f}_{GPV}\left(v\right)+\frac{s_{\nicefrac{1-\alpha}{2}}^{*}(v)}{\sqrt{Lh^{3}}}\right],
\]
where $s_{\tau}^{*}(v)$ is the $\tau-$th quantile of the conditional
distribution of
\[
S^{*}\left(v\right)\coloneqq\left(Lh^{3}\right)^{\nicefrac{1}{2}}\left(\widehat{f}_{GPV}^{*}\left(v\right)-\widehat{f}_{GPV}\left(v\right)\right)
\]
given the original sample. The conditional distributions of the bootstrap
statistics $\widehat{f}_{GPV}^{*}\left(v\right)$ and $S^{*}\left(v\right)$
given the original sample can be easily approximated by Monte Carlo
methods. We show below that the bootstrap estimator of the finite-sample
distribution of $S\left(v\right)$ is consistent. Let $\mathrm{P}^{*}\left[\cdot\right]$
denote the conditional probability given the original sample of bids. 
\begin{thm}[\textbf{Bootstrap Consistency}]
 \label{thm: bootstrap consistency}Suppose Assumptions \ref{assu:DGP}
- \ref{assu: rate of bandwidth} hold. Then for any interior point
$v\in\left(\underline{v},\overline{v}\right)$, 
\[
\underset{z\in\mathbb{R}}{\mathrm{sup}}\left|\mathrm{P}^{*}\left[S^{*}\left(v\right)\leq z\right]-\mathrm{P}\left[S\left(v\right)\leq z\right]\right|\rightarrow_{p}0,\,\textrm{as \ensuremath{L\uparrow\infty}}.
\]
\end{thm}
\begin{rembold}\label{rem: bootstrap validity pointwise confidence interval}Theorems
\ref{thm:Asymptotic Normality GPV}, \ref{thm: bootstrap consistency}
and P\'{o}lya's theorem yield
\[
\underset{z\in\mathbb{R}}{\mathrm{sup}}\left|\mathrm{P}^{*}\left[S^{*}\left(v\right)\leq z\right]-\mathrm{P}\left[\mathrm{N}\left(0,\mathrm{V}_{GPV}\left(v\right)\right)\leq z\right]\right|\rightarrow_{p}0,\,\textrm{as \ensuremath{L\uparrow\infty}},
\]
for each $v\in\left(\underline{v},\overline{v}\right)$. The above
result and standard arguments (see, e.g., \citealp[Lemma 23.3]{VanDerVaart_Asymptotic_Statistics_Book})
yield the asymptotic validity (consistency) of the percentile bootstrap
confidence interval $CI^{*}\left(v\right)$, i.e., $\mathrm{P}\left[f\left(v\right)\in CI^{*}\left(v\right)\right]\rightarrow1-\alpha$
as $\textrm{\ensuremath{L\uparrow\infty}}.$\end{rembold}

\begin{rembold}One can also studentize $S\left(v\right)$ using our
estimator of the asymptotic variance:
\begin{equation}
Z\left(v\right)\coloneqq\frac{\widehat{f}_{GPV}\left(v\right)-f\left(v\right)}{\left(Lh^{3}\right)^{-\nicefrac{1}{2}}\widehat{\mathrm{V}}_{GPV}\left(v\right)^{\nicefrac{1}{2}}}.\label{eq:Z definition}
\end{equation}
Since $\widehat{\mathrm{V}}_{GPV}\left(v\right)$ is consistent for
the asymptotic variance, $Z\left(v\right)$ is asymptotically distributed
as a standard normal random variable. Let $\widehat{\mathrm{V}}_{GPV}^{*}\left(v\right)$
be the bootstrap analogue of $\widehat{\mathrm{V}}_{GPV}\left(v\right)$.
The bootstrap analogue of $Z\left(v\right)$ is $Z^{**}\left(v\right)\coloneqq\nicefrac{S^{*}\left(v\right)}{\sqrt{\widehat{\mathrm{V}}_{GPV}^{*}\left(v\right)}}$.
Let $z_{\tau}^{**}(v)$ be the $\tau-$th quantile of the conditional
distribution of $Z^{**}(v)$ given the original sample. The ``bootstrap-$t$''
(or studentized bootstrap) confidence intervals can be obtained by
replacing the critical value $z_{1-\nicefrac{\alpha}{2}}$ in $CI^{\dagger}\left(v\right)$
with their bootstrap counterpart $z_{\tau}^{**}(v)$. The asymptotic
validity of this alternative bootstrap confidence interval easily
follows as a corollary to Theorem \ref{thm: bootstrap consistency}.\end{rembold}

\section{Uniform Confidence Bands\label{sec:Uniform-Confidence-Bands}}

Consider the stochastic process
\begin{equation}
\mathit{\Gamma}\left(v\right)\coloneqq\frac{1}{N^{\nicefrac{1}{2}}\left(N-1\right)}\frac{1}{\left(N\cdot L\right)^{\nicefrac{1}{2}}}\sum_{i,l}\frac{\mathcal{M}_{2}\left(B_{il};v\right)-\mu_{\mathcal{M}}\left(v\right)}{\mathrm{Var}\left[N^{-\nicefrac{1}{2}}\left(N-1\right)^{-1}\mathcal{M}_{2}\left(B_{11};v\right)\right]^{\nicefrac{1}{2}}},\,v\in I.\label{eq:GAMMA defnition}
\end{equation}
Note that $\mathrm{E}\left[\mathit{\Gamma}\left(v\right)\right]=0$
and $\mathrm{E}\left[\mathit{\Gamma}\left(v\right)^{2}\right]=1$
for all $v\in I$. 

The following theorem shows that (a version of) the centered Gaussian
process with index set $I$ and covariance function $\mathrm{E}\left[\mathit{\Gamma}\left(v\right)\mathit{\Gamma}\left(v'\right)\right]$,
for $\left(v,v'\right)\in I^{2}$, is a tight random element in $\ell^{\infty}\left(I\right)$.
This Gaussian process, denoted by $\left\{ \mathit{\Gamma}_{G}\left(v\right):v\in I\right\} $,
is the \emph{intermediate} Gaussian process. The tightness of $\varGamma_{G}$
as a random element in $\ell^{\infty}\left(I\right)$ can be established
using standard results (see, e.g., \citealp[Lemma 2.1]{chernozhukov2014gaussian}).
The following theorem also shows that one can approximate the distribution
of the sup-norm $\left\Vert Z\right\Vert _{I}=\underset{v\in I}{\mathrm{sup}}\left|Z\left(v\right)\right|$
with that of $\mathit{\Gamma}_{G}$. The result follows from uniform
approximations of (\ref{eq:leading term of the stochastic expansion identical auctions})
and $\widehat{\mathrm{V}}_{GPV}\left(v\right)$ and uses the coupling
theorem for suprema of empirical processes of \citet{chernozhukov2014gaussian}
and the Gaussian anti-concentration inequality of \citet{chernozhukov2014anti}.
\begin{thm}
\label{thm:Gaussian coupling}Suppose Assumptions \ref{assu:DGP}
- \ref{assu: rate of bandwidth} hold. Then there exists a tight Gaussian
random element $\varGamma_{G}$ in $\ell^{\infty}\left(I\right)$
that has mean zero and the same covariance structure as that of $\mathit{\Gamma}$.
Moreover, 
\[
\underset{z\in\mathbb{R}}{\mathrm{sup}}\left|\mathrm{P}\left[\left\Vert Z\right\Vert _{I}\leq z\right]-\mathrm{P}\left[\left\Vert \mathit{\Gamma}_{G}\right\Vert _{I}\leq z\right]\right|\rightarrow0,\,\textrm{as \ensuremath{L\uparrow\infty}}.
\]
\end{thm}
The next result shows that the distribution of the sup-norm of the
(empirical) bootstrap process 
\begin{equation}
Z^{*}\left(v\right)\coloneqq\frac{\widehat{f}_{GPV}^{*}\left(v\right)-\widehat{f}_{GPV}\left(v\right)}{\left(Lh^{3}\right)^{-\nicefrac{1}{2}}\widehat{\mathrm{V}}_{GPV}\left(v\right)^{\nicefrac{1}{2}}},\,v\in I\label{eq:Z_star process}
\end{equation}
can be similarly approximated by that of $\mathit{\Gamma}_{G}$.
\begin{thm}
\label{thm:Gaussian coupling for empirical bootstrap process}Suppose
Assumptions \ref{assu:DGP} - \ref{assu: rate of bandwidth} hold.
Then,
\[
\underset{z\in\mathbb{R}}{\mathrm{sup}}\left|\mathrm{P}^{*}\left[\left\Vert Z^{*}\right\Vert _{I}\leq z\right]-\mathrm{P}\left[\left\Vert \mathit{\Gamma}_{G}\right\Vert _{I}\leq z\right]\right|\rightarrow_{p}0,\,\textrm{as \ensuremath{L\uparrow\infty}}.
\]
\end{thm}
Since the distributions of the suprema of (the absolute values of)
$\left\{ Z\left(v\right):v\in I\right\} $ and that of $\left\{ Z^{*}\left(v\right):v\in I\right\} $
are both well approximated by that of $\left\{ \varGamma_{G}\left(v\right):v\in I\right\} $,
one can use the bootstrap critical values based on $\left\Vert Z^{*}\right\Vert _{I}$
for construction of uniform confidence bands. Let 
\begin{equation}
\zeta_{L,\alpha}^{*}\coloneqq\mathrm{inf}\left\{ z\in\mathbb{R}:\mathrm{P}^{*}\left[\left\Vert Z^{*}\right\Vert _{I}\leq z\right]\geq1-\alpha\right\} \label{eq:zeta_L^* definition}
\end{equation}
be the $\left(1-\alpha\right)$-quantile of the conditional distribution
of $\left\Vert Z^{*}\right\Vert _{I}$ given the original sample.
The uniform confidence band is given by
\[
CB^{*}\left(v\right)\coloneqq\left[\widehat{f}_{GPV}\left(v\right)-\zeta_{L,\alpha}^{*}\sqrt{\frac{\widehat{\mathrm{V}}_{GPV}\left(v\right)}{Lh^{3}}},\,\widehat{f}_{GPV}\left(v\right)+\zeta_{L,\alpha}^{*}\sqrt{\frac{\widehat{\mathrm{V}}_{GPV}\left(v\right)}{Lh^{3}}}\right],\,\textrm{for \ensuremath{v\in I}}.
\]
The following corollary establishes its asymptotic validity and provides
an estimate of the order of the bootstrap critical value $\zeta_{L,\alpha}^{*}$.\footnote{It also implies that the (supremum) width of the band $CB^{*}$ is
of order $O_{p}\left(\mathrm{log}\left(h^{-1}\right)^{\nicefrac{1}{2}}\left(Lh^{3}\right)^{-\nicefrac{1}{2}}\right)$.}
\begin{cor}[\textbf{Validity of Bootstrap Confidence Band}]
\label{cor:consistency of IGA bootstrap}Suppose Assumptions \ref{assu:DGP}
- \ref{assu: rate of bandwidth} hold. Then, $\mathrm{P}\left[f\left(v\right)\in CB^{*}\left(v\right),\,\textrm{for all \ensuremath{v\in I}}\right]\rightarrow1-\alpha$
as $L\uparrow\infty$. Moreover, $\zeta_{L,\alpha}^{*}=O_{p}\left(\mathrm{log}\left(h^{-1}\right)^{\nicefrac{1}{2}}\right)$. 
\end{cor}
\begin{rembold}[Limiting Distribution of Uniform Error]Since the
seminal work of \citet{bickel1973some}, it has been found that in
many cases a suitable normalization of the supremum of a studentized
absolute difference between a nonparametric curve and its kernel-based
estimator converges in distribution to standard Gumbel distribution.
Asymptotically valid uniform confidence bands can be based on the
Gumbel approximation. We do not pursue such an approach in this paper,
since it is known that the accuracy of such approximation is poor.
See \citet[Section 2.7]{gine2015mathematical} and \citet{chernozhukov2014anti}
for discussion. On the other hand, for the auction model one can show
that (a suitable normalization of) $\left\Vert Z\right\Vert _{I}$
converges in distribution to standard Gumbel distribution when the
true bidding strategy is linear. Derivation of the limiting distribution
for the general case is interesting but beyond the scope of this paper.
See the Supplement for more discussion.\end{rembold}

\begin{rembold}Taking the IGA approach, we establish consistency
of bootstrap uniform confidence bands by showing that the distributions
of both $\left\Vert Z\right\Vert _{I}$ and its empirical bootstrap
counterpart can be approximated by the distribution of $\left\Vert \mathit{\Gamma}_{G}\right\Vert _{I}$.
The proof hinges on using the coupling theorems of \citet{chernozhukov2014gaussian,chernozhukov2016empirical}
and the Gaussian anti-concentration inequality of \citet{chernozhukov2014anti}.
In the literature, the multiplier bootstrap is used when the IGA approach
is taken to construct confidence bands for nonparametric curves. See,
e.g., \citet{chernozhukov2014anti} and \citet{Kato_Sasaki_2,Kato_Sasaki_1}.
Here, we chose the empirical bootstrap since it is practically convenient
given the two-step nature of the GPV estimator.\end{rembold}

\section{Auction-Specific Heterogeneity\label{sec:Auction-Specific-Heterogeneity}}

\subsection{Asymptotic Normality and Estimation of the Asymptotic Variance}

We now turn to the general model with auction-specific heterogeneity
and a random number of bidders. Firstly, we establish the asymptotic
normality of the GPV estimator by following the same approach and
steps as in the case of the simplified model in Section \ref{sec:Asymptotic Normality}.
While handling the general case is complicated by much heavier notations,
all the results provided in this section can be viewed as straightforward
generalizations of the results in Sections \ref{sec:Asymptotic Normality}-\ref{sec:Uniform-Confidence-Bands}.
The proofs of the results for the general case can be found in the
Supplement.

In comparison with the simplified model, one of the main differences
is in the form of the asymptotic variance. Recall that in the simplified
case, the asymptotic variance of the GPV estimator depends on the
derivative of the bidding strategy. As we show below, when there is
auction-specific heterogeneity, the asymptotic variance also involves
the partial derivatives of the bidding strategy with respect to the
auction-specific characteristics. This is in addition to the partial
derivative with respect to the valuation. 

For some fixed $\boldsymbol{x}$ which is an interior point of $\mathcal{X}$,
let $I\left(\boldsymbol{x}\right)\coloneqq\left[v_{l}\left(\boldsymbol{x}\right),v_{u}\left(\boldsymbol{x}\right)\right]$
be an inner closed sub-interval of $\left[\underline{v}\left(\boldsymbol{x}\right),\overline{v}\left(\boldsymbol{x}\right)\right]$.
The fact that the conditional density of the valuations given $\boldsymbol{X}=\boldsymbol{x}$
and $N=n$ is $f\left(\cdot|\boldsymbol{x}\right)$ under Assumption
\ref{assu:DGP} motivates the following two-step estimator of $f\left(v|\boldsymbol{x}\right)$:
\[
\widehat{f}_{GPV}\left(v|\boldsymbol{x},n\right)\coloneqq\frac{1}{\widehat{\pi}\left(n|\boldsymbol{x}\right)\widehat{\varphi}\left(\boldsymbol{x}\right)L}\sum_{l=1}^{L}\mathbbm{1}\left(N_{l}=n\right)\frac{1}{N_{l}}\sum_{i=1}^{N_{l}}\mathbb{T}_{il}\frac{1}{h^{1+d}}K_{f}\left(\frac{\widehat{V}_{il}-v}{h},\frac{\boldsymbol{X}_{l}-\boldsymbol{x}}{h}\right).
\]
Note that the above estimator only uses data from auctions with $N_{l}=n$.
Since the PDF of valuations does not depend on the number of bidders,
an estimator for $f(v|\boldsymbol{x})$ can be constructed as a weighted
average of $\left\{ \widehat{f}_{GPV}\left(v|\boldsymbol{x},n\right):n\in\mathcal{N}\right\} $.
E.g., GPV suggested using estimates of the conditional probabilities
of drawing $N_{l}=n$ as the weights: 
\begin{equation}
\widehat{f}_{GPV}\left(v|\boldsymbol{x}\right)=\sum_{n\in\mathcal{N}}\widehat{\pi}\left(n|\boldsymbol{x}\right)\widehat{f}_{GPV}\left(v|\boldsymbol{x},n\right).\label{eq:f_hat_GPV heterogeneity}
\end{equation}
Note that this gives an expression that is the same as the right hand
side of (\ref{eq:f_hat_GPV definition}).

By repeating the steps from Section \ref{sec:Asymptotic Normality},
one can show that the following analogue of the linearization results
in (\ref{eq:f_GPV - f leading term}) holds for the general model:
\begin{multline}
\widehat{f}_{GPV}\left(v|\boldsymbol{x},n\right)-f\left(v|\boldsymbol{x}\right)\\
=\frac{1}{\widehat{\pi}\left(n|\boldsymbol{x}\right)\widehat{\varphi}\left(\boldsymbol{x}\right)L^{2}}\sum_{l=1}^{L}\sum_{k=1}^{L}\mathcal{M}^{n}\left(\left(\boldsymbol{B}_{\cdot l},\boldsymbol{X}_{l},N_{l}\right),\left(\boldsymbol{B}_{\cdot k},\boldsymbol{X}_{k},N_{k}\right);v\right)+o_{p}\left(\left(Lh^{3+d}\right)^{-\nicefrac{1}{2}}\right),\label{eq:auction heterogeneity V statistic approximation}
\end{multline}
where $\boldsymbol{B}_{\cdot l}\coloneqq\left(B_{1l},...,B_{N_{l}l}\right)$,
and the remainder term is uniform in $v\in I\left(\boldsymbol{x}\right)$.
For $\boldsymbol{b}_{\cdot}\coloneqq(b_{1},\ldots,b_{m})$ , the kernel
function $\mathcal{M}^{n}$ is given by:
\begin{eqnarray}
 &  & \mathcal{M}^{n}\left(\left(\boldsymbol{b}_{\cdot},\boldsymbol{z},m\right),\left(\boldsymbol{b}_{\cdot}',\boldsymbol{z}',m'\right);v\right)\nonumber \\
 & \coloneqq & -\mathbbm{1}\left(m=n\right)\frac{1}{m}\sum_{i=1}^{m}\frac{1}{h^{2+d}}K_{f}'\left(\frac{\xi\left(b_{i},\boldsymbol{z},m\right)-v}{h},\frac{\boldsymbol{z}-\boldsymbol{x}}{h}\right)\frac{G\left(b_{i},\boldsymbol{z},m\right)}{\left(m-1\right)g\left(b_{i},\boldsymbol{z},m\right)^{2}}\nonumber \\
 &  & \times\left(\mathbbm{1}\left(m'=m\right)\frac{1}{m'}\sum_{j=1}^{m'}\frac{1}{h^{1+d}}K_{g}\left(\frac{b_{j}'-b_{i}}{h}\right)K_{\boldsymbol{X}}\left(\frac{\boldsymbol{z}'-\boldsymbol{z}}{h}\right)-g\left(b_{i},\boldsymbol{z},m\right)\right),\label{eq:auction heterogeneity V statistic kernel}
\end{eqnarray}
where $K_{f}'\left(\cdot,\cdot\right)$ denotes the partial derivative
function of $K_{f}$ with respect to its first argument, 
\[
G\left(b,\boldsymbol{z},m\right)\coloneqq G\left(b|\boldsymbol{z},m\right)\pi\left(m|\boldsymbol{z}\right)\varphi\left(\boldsymbol{z}\right)\text{ and }g\left(b,\boldsymbol{z},m\right)\coloneqq g\left(b|\boldsymbol{z},m\right)\pi\left(m|\boldsymbol{z}\right)\varphi\left(\boldsymbol{z}\right).
\]

Note that the leading term on the right-hand side of (\ref{eq:auction heterogeneity V statistic approximation})
involves a V-statistic (with a kernel that depends on the bandwidth)
and, therefore, can be analyzed using the Hoeffding decomposition.
Thus, (\ref{eq:M Hoeffding decomposition}) can be generalized as
\begin{align*}
 & \frac{1}{L^{2}}\sum_{l=1}^{L}\sum_{m=1}^{L}\mathcal{M}^{n}\left(\left(\boldsymbol{B}_{\cdot l},\boldsymbol{X}_{l},N_{l}\right),\left(\boldsymbol{B}_{\cdot m},\boldsymbol{X}_{m},N_{m}\right);v\right)\\
= & \mu_{\mathcal{M}^{n}}\left(v\right)+\left\{ \frac{1}{L}\sum_{l=1}^{L}\left(\mathcal{M}_{1}^{n}\left(\boldsymbol{B}_{\cdot l},\boldsymbol{X}_{l},N_{l};v\right)-\mu_{\mathcal{M}^{n}}\left(v\right)\right)\right\} \\
 & +\left\{ \frac{1}{L}\sum_{l=1}^{L}\left(\mathcal{M}_{2}^{n}\left(\boldsymbol{B}_{\cdot l},\boldsymbol{X}_{l},N_{l};v\right)-\mu_{\mathcal{M}^{n}}\left(v\right)\right)\right\} +o_{p}\left(\left(Lh^{3+d}\right)^{-\nicefrac{1}{2}}\right),
\end{align*}
where the remainder term is uniform in $v\in I\left(\boldsymbol{x}\right)$,
\begin{align*}
\mathcal{M}_{1}^{n}\left(\boldsymbol{b}.,\boldsymbol{z},m;v\right) & \coloneqq\mathrm{E}\left[\mathcal{M}^{n}\left(\left(\boldsymbol{b}.,\boldsymbol{z},m\right),\left(\boldsymbol{B}_{\cdot1},\boldsymbol{X}_{1},N_{1}\right);v\right)\right],\\
\mathcal{M}_{2}^{n}\left(\boldsymbol{b}.,\boldsymbol{z},m;v\right) & \coloneqq\mathrm{E}\left[\mathcal{M}^{n}\left(\left(\boldsymbol{B}_{\cdot1},\boldsymbol{X}_{1},N_{1}\right),\left(\boldsymbol{b}.,\boldsymbol{z},m\right);v\right)\right],
\end{align*}
and $\mu_{\mathcal{M}^{n}}\left(v\right)\coloneqq\mathrm{E}\left[\mathcal{M}^{n}\left(\left(\boldsymbol{B}_{\cdot1},\boldsymbol{X}_{1},N_{1}\right),\left(\boldsymbol{B}_{\cdot2},\boldsymbol{X}_{2},N_{2}\right);v\right)\right]$. 

The projection term $\mathcal{M}_{1}^{n}$ is the expectation of the
kernel $\mathcal{M}^{n}\left(\left(\boldsymbol{b}.,\boldsymbol{z},m\right),\left(\boldsymbol{B}_{\cdot1},\boldsymbol{X}_{1},N_{1}\right);v\right)$
with the first argument fixed at $\left(\boldsymbol{b}.,\boldsymbol{z},m\right)$.
The expression for $\mathcal{M}_{1}^{n}$ is: 
\begin{eqnarray*}
 &  & \mathcal{M}_{1}^{n}\left(\boldsymbol{b}.,\boldsymbol{z},m;v\right)\\
 & = & -\mathbbm{1}\left(m=n\right)\frac{1}{m}\sum_{i=1}^{m}\frac{1}{h^{2+d}}K_{f}'\left(\frac{\xi\left(b_{i},\boldsymbol{z},m\right)-v}{h},\frac{\boldsymbol{z}-\boldsymbol{x}}{h}\right)\frac{G\left(b_{i},\boldsymbol{z},m\right)}{g\left(b_{i},\boldsymbol{z},m\right)^{2}}\\
 &  & \times\int\sum_{m'\in\mathcal{N}}\int\cdots\int\left(\mathbbm{1}\left(m'=n\right)\frac{1}{m'}\sum_{j=1}^{m'}\frac{1}{h^{1+d}}K_{g}\left(\frac{b_{j}'-b_{i}}{h}\right)K_{\boldsymbol{X}}\left(\frac{\boldsymbol{z}'-\boldsymbol{z}}{h}\right)-g\left(b_{i},\boldsymbol{z},m\right)\right)\\
 &  & \times\left(\prod_{j=1}^{m'}g\left(b_{j}'|\boldsymbol{z}',m'\right)\right)\pi\left(m'|\boldsymbol{z}'\right)\varphi\left(\boldsymbol{z}'\right)\mathrm{d}b_{1}'\cdots\mathrm{d}b_{m'}'\mathrm{d}\boldsymbol{z}'.
\end{eqnarray*}
As in the case of the simplified model, the contribution of $\mathcal{M}_{1}^{n}$
is asymptotically negligible. This happens for the same reason as
in Section \ref{sec:Asymptotic Normality}: $\mathcal{M}_{1}^{n}$
depends on the difference between the expectation of the kernel function
and the true density. Hence, the asymptotic distribution of the GPV
estimator is driven solely by the $\mathcal{M}_{2}^{n}$ term, which
is the expectation of the kernel $\mathcal{M}^{n}\left(\left(\boldsymbol{B}_{\cdot1},\boldsymbol{X}_{1},N_{1}\right),\left(\boldsymbol{b}.,\boldsymbol{z},m\right);v\right)$
with the second argument fixed at $\left(\boldsymbol{b}.,\boldsymbol{z},m\right)$. 

We show in the supplement that a generalized version of (\ref{eq:E_m2^2 expansion})
holds:
\begin{align}
 & \mathrm{E}\left[Lh^{3+d}\left\{ \frac{1}{L}\sum_{l=1}^{L}\left(\mathcal{M}_{2}^{n}\left(\boldsymbol{B}_{\cdot l},\boldsymbol{X}_{l},N_{l};v\right)-\mu_{\mathcal{M}^{n}}\left(v\right)\right)\right\} ^{2}\right]\nonumber \\
= & \frac{1}{n\left(n-1\right)^{2}}\frac{1}{h^{3\left(1+d\right)}}\int\int\left(\int_{\mathcal{X}}\int_{\underline{b}\left(\boldsymbol{z}'\right)}^{\overline{b}\left(\boldsymbol{z}',n\right)}K_{f}'\left(\frac{\xi\left(b',\boldsymbol{z}',n\right)-v}{h},\frac{\boldsymbol{z}'-\boldsymbol{x}}{h}\right)\frac{G\left(b',\boldsymbol{z}',n\right)}{g\left(b',\boldsymbol{z}',n\right)}\right.\nonumber \\
 & \left.\times K_{g}\left(\frac{b-b'}{h}\right)K_{\boldsymbol{X}}\left(\frac{\boldsymbol{z}-\boldsymbol{z}'}{h}\right)\mathrm{d}b'\mathrm{d}\boldsymbol{z}'\right)^{2}g\left(b,\boldsymbol{z},n\right)\mathrm{d}b\mathrm{d}\boldsymbol{z}+O\left(h^{3}\right)\nonumber \\
\eqqcolon & \mathrm{V}_{\mathcal{M}}\left(v|\boldsymbol{x},n\right)+O\left(h^{3}\right),\label{eq:EM_2  heterogeneity}
\end{align}
where the remainder term is uniform in $v\in I\left(\boldsymbol{x}\right)$.
Moreover, similarly to (\ref{eq:variance limit simple}),
\begin{align}
 & \underset{h\downarrow0}{\mathrm{lim}}\frac{1}{h^{3\left(1+d\right)}}\int\int\left(\int_{\mathcal{X}}\int_{\underline{b}\left(\boldsymbol{z}'\right)}^{\overline{b}\left(\boldsymbol{z}',n\right)}K_{f}'\left(\frac{\xi\left(b',\boldsymbol{z}',n\right)-v}{h},\frac{\boldsymbol{z}'-\boldsymbol{x}}{h}\right)\frac{G\left(b',\boldsymbol{z}',n\right)}{g\left(b',\boldsymbol{z}',n\right)}\right.\nonumber \\
 & \left.\times K_{g}\left(\frac{b-b'}{h}\right)K_{\boldsymbol{X}}\left(\frac{\boldsymbol{z}-\boldsymbol{z}'}{h}\right)\mathrm{d}b'\mathrm{d}\boldsymbol{z}'\right)^{2}g\left(b,\boldsymbol{z},n\right)\mathrm{d}b\mathrm{d}\boldsymbol{z}\nonumber \\
= & \frac{G(s(v,\boldsymbol{x},n),\boldsymbol{x},n)^{2}s_{v}(v,\boldsymbol{x},n)^{2}}{g(s(v,\boldsymbol{x},n),\boldsymbol{x},n)}\nonumber \\
 & \times\int\int\left\{ \int\int K_{f}'\left(w,\boldsymbol{y}\right)K_{\boldsymbol{X}}\left(\boldsymbol{y}-\boldsymbol{z}\right)K_{g}\left(u-s_{v}w-s_{\boldsymbol{x}}^{\mathrm{T}}\boldsymbol{y}\right)\mathrm{d}w\mathrm{d}\boldsymbol{y}\right\} ^{2}\mathrm{d}u\mathrm{d}\boldsymbol{z},\label{eq:variance limit general}
\end{align}
where $s_{v}$ and $s_{\boldsymbol{x}}$ denote the partial derivatives
of the bidding function:
\begin{equation}
s_{v}\coloneqq\left.\frac{\partial s\left(u,\boldsymbol{z},n\right)}{\partial u}\right|_{\left(u,\boldsymbol{z}\right)=\left(v,\boldsymbol{x}\right)}\textrm{ and }s_{\boldsymbol{x}}\coloneqq\left.\frac{\partial s\left(u,\boldsymbol{z},n\right)}{\partial\boldsymbol{z}}\right|_{\left(u,\boldsymbol{z}\right)=\left(v,\boldsymbol{x}\right)}.\label{eq:derivatives}
\end{equation}

The asymptotic variance of the GPV estimator is the limit of $\left(\pi\left(n|\boldsymbol{x}\right)\varphi\left(\boldsymbol{x}\right)\right)^{-2}\mathrm{V}_{\mathcal{M}}\left(v|\boldsymbol{x},n\right)$.
After using a change of variable argument and (\ref{eq:variance limit general}),
the variance is shown to be
\begin{multline}
\mathrm{V}_{GPV}\left(v|\boldsymbol{x},n\right)\coloneqq\frac{1}{n\left(n-1\right)^{2}}\frac{F\left(v|\boldsymbol{x}\right)^{2}f\left(v|\boldsymbol{x}\right)^{2}}{\pi\left(n|\boldsymbol{x}\right)\varphi\left(\boldsymbol{x}\right)g\left(s\left(v,\boldsymbol{x},n\right)|\boldsymbol{x},n\right)^{3}}\\
\times\int\int\left\{ \int\int K_{f}'\left(w,\boldsymbol{y}\right)K_{\boldsymbol{X}}\left(\boldsymbol{y}-\boldsymbol{z}\right)K_{g}\left(u-s_{v}w-s_{\boldsymbol{x}}^{\mathrm{T}}\boldsymbol{y}\right)\mathrm{d}w\mathrm{d}\boldsymbol{y}\right\} ^{2}\mathrm{d}u\mathrm{d}\boldsymbol{z}.\label{eq:V_GPV heterogeneity definition}
\end{multline}

The following theorem is a generalization of Theorem \ref{thm:Asymptotic Normality GPV}.
The proof of the theorem as well as the proofs of all other results
provided in Section \ref{sec:Auction-Specific-Heterogeneity} are
in the Supplement.
\begin{thm}
\label{thm:heterogeneity} Suppose Assumptions \ref{assu:DGP} - \ref{assu: rate of bandwidth}
hold. Then, for any interior point $\left(v,\boldsymbol{x}\right)\in\mathcal{S}_{V,\boldsymbol{X}},$
\[
\left(Lh^{3+d}\right)^{\nicefrac{1}{2}}\left(\widehat{f}_{GPV}\left(v|\boldsymbol{x},n\right)-f\left(v|\boldsymbol{x}\right)\right)\rightarrow_{d}\mathrm{N}\left(0,\mathrm{V}_{GPV}\left(v|\boldsymbol{x},n\right)\right).
\]
Moreover, $\left\{ \widehat{f}_{GPV}\left(v|\boldsymbol{x},n\right):n\in\mathcal{N}\right\} $
are asymptotically independent.
\end{thm}
\begin{rembold}As in the simplified model, the asymptotic variance
of the GPV estimator depends on a convoluted integral transformation
involving the kernel function, its derivative and the derivatives
of the bidding strategy. Auction-specific heterogeneity complicates
the expression in two ways. Firstly, the dimension of the integral
depends on the number of auction characteristics (covariates), and
analytical calculation of the integral becomes cumbersome when there
are many covariates. Secondly, the expression now contains the derivatives
of the bidding strategy with respect to the covariates. The reason
for that is apparent from the expression for the second moment of
$\mathcal{M}_{2}^{n}\left(\boldsymbol{B}_{\cdot1},\boldsymbol{X}_{1},N_{1};v\right)$
in (\ref{eq:EM_2  heterogeneity}) as it involves averaging of the
inverse bidding function $\xi(u,\boldsymbol{z}',n)$ over the auction
characteristics $\boldsymbol{z}'$ in a shrinking neighborhood of
$\boldsymbol{x}$.

While the GPV estimator and the quantile-based estimator have the
same rate of convergence (see \citealp[Theorem 2]{Marmer_Shneyerov_Quantile_Auctions}
for the rate of convergence and the expression of the asymptotic variance
$\mathrm{V}_{QB}\left(v|\boldsymbol{x},n\right)$ of the quantile-based
estimator in the general model), one can show that the GPV estimator
has a smaller asymptotic variance than that of the quantile-based
estimator, as long as the two estimators use the same second-order
kernel function. It can be shown that the ratio $\nicefrac{\mathrm{V}_{GPV}\left(v|\boldsymbol{x},n\right)}{\mathrm{V}_{QB}\left(v|\boldsymbol{x},n\right)}\leq1$
by applying Jensen's inequality and standard techniques for multi-dimensional
integration. A detailed proof can be found in the Supplement.\end{rembold}

\begin{rembold} [Optimal Weights]Since the distribution of valuations
does not depend on the number of bidders, i.e., $f\left(v|\boldsymbol{x},n\right)=f\left(v|\boldsymbol{x}\right)$
for all $\left(v,\boldsymbol{x},n\right)\in\mathcal{S}_{V,\boldsymbol{X}}\times\mathcal{N}$,
one can average the estimators $\left\{ \widehat{f}_{GPV}\left(v|\boldsymbol{x},n\right):n\in\mathcal{N}\right\} $
to obtain a weighted estimator of the density $f\left(v|\boldsymbol{x}\right)$:
\[
\widehat{f}_{GPV}^{w}\left(v|\boldsymbol{x}\right)\coloneqq\sum_{n\in\mathcal{N}}\widehat{w}\left(n,v,\boldsymbol{x}\right)\widehat{f}_{GPV}\left(v|\boldsymbol{x},n\right),
\]
where the weights $\widehat{w}\left(n,v,\boldsymbol{x}\right)$, $n\in\mathcal{N}$
should satisfy $\widehat{w}\left(n,v,\boldsymbol{x}\right)\rightarrow_{p}w\left(n,v,\boldsymbol{x}\right)$,
$n\in\mathcal{N}$ and $\sum_{n\in\mathcal{N}}w\left(n,v,\boldsymbol{x}\right)=1$.
As in \citet{Marmer_Shneyerov_Quantile_Auctions}, the optimal weights
that minimize the asymptotic variance of of the resulting weighted
estimator are inversely related to $\mathrm{V}_{GPV}\left(v|\boldsymbol{x},n\right)$
and given by
\[
w^{\mathrm{opt}}\left(n,v,\boldsymbol{x}\right)\coloneqq\frac{\frac{1}{\mathrm{V}_{GPV}\left(v|\boldsymbol{x},n\right)}}{\sum_{n\in\mathcal{N}}\frac{1}{\mathrm{V}_{GPV}\left(v|\boldsymbol{x},n\right)}}.
\]
These weights can be consistently estimated by the plug-in principle
using an estimator of $\mathrm{V}_{GPV}\left(v|\boldsymbol{x},n\right)$.
The original GPV estimator uses the weights $\widehat{w}\left(v,\boldsymbol{x},n\right)=\widehat{\pi}(n|\boldsymbol{x})$,
$n\in\mathcal{N}$. See (\ref{eq:f_hat_GPV definition}) and (\ref{eq:f_hat_GPV heterogeneity}).
Note, however, that such weights would be sub-optimal from the point
of view of minimizing the asymptotic variance. We provide the asymptotic
normality of $\widehat{f}_{GPV}\left(v|\boldsymbol{x}\right)$ in
the corollary below.\end{rembold}
\begin{cor}
\label{cor:asymptotic normality heterogeneity}Suppose Assumptions
\ref{assu:DGP} - \ref{assu: rate of bandwidth} hold. Then, for any
interior point $\left(v,\boldsymbol{x}\right)\in\mathcal{S}_{V,\boldsymbol{X}}$,
\[
\left(Lh^{3+d}\right)^{\nicefrac{1}{2}}\left(\widehat{f}_{GPV}\left(v|\boldsymbol{x}\right)-f\left(v|\boldsymbol{x}\right)\right)\rightarrow_{d}\mathrm{N}\left(0,\mathrm{V}_{GPV}\left(v|\boldsymbol{x}\right)\right),
\]
where $\mathrm{V}_{GPV}\left(v|\boldsymbol{x}\right)\coloneqq\sum_{n\in\mathcal{N}}\pi\left(n|\boldsymbol{x}\right)^{2}\mathrm{V}_{GPV}\left(v|\boldsymbol{x},n\right)$.
\end{cor}
For practical purposes, it is important to have a consistent estimator
of the asymptotic variance $\mathrm{V}_{GPV}\left(v|\boldsymbol{x}\right)$
that avoids estimation of the bidding strategy $s\left(\cdot,\cdot,n\right)$
and its derivatives. It is also highly desirable to avoid analytical
or numerical evaluation of a multidimensional integral in the definition
of the asymptotic variance. Following the same approach we used in
the case of the simplified model (see (\ref{eq:definition estimator of asymptotic variance})),
we rely on the sample analogue of (\ref{eq:EM_2  heterogeneity}):
\begin{multline}
\widehat{\mathrm{V}}_{GPV}\left(v|\boldsymbol{x},n\right)\coloneqq\frac{1}{n(n-1)^{2}}\frac{1}{\widehat{\pi}\left(n|\boldsymbol{x}\right)^{2}\widehat{\varphi}\left(\boldsymbol{x}\right)^{2}h^{3\left(1+d\right)}}\frac{1}{L\left(L-1\right)\left(L-2\right)}\\
\times\sum_{l=1}^{L}\sum_{k\neq l}\sum_{k'\neq k,k'\neq l}\mathbbm{1}\left(N_{l}=n,N_{k}=n,N_{k'}=n\right)\frac{1}{N_{l}}\sum_{i=1}^{N_{l}}\eta_{il,k}(v,\boldsymbol{x})\eta_{il,k'}(v,\boldsymbol{x}),\label{eq:V_GPV estimator heterogeneity}
\end{multline}
where 
\begin{multline*}
\eta_{il,k}(v,\boldsymbol{x})\coloneqq\frac{1}{N_{k}}\\
\times\sum_{j=1}^{N_{k}}\mathbb{T}_{jk}K_{f}'\left(\frac{\widehat{V}_{jk}-v}{h},\frac{\boldsymbol{X}_{k}-\boldsymbol{x}}{h}\right)\frac{\widehat{G}\left(B_{jk},\boldsymbol{X}_{k},N_{k}\right)}{\widehat{g}\left(B_{jk},\boldsymbol{X}_{k},N_{k}\right)^{2}}K_{g}\left(\frac{B_{il}-B_{jk}}{h}\right)K_{\boldsymbol{X}}\left(\frac{\boldsymbol{X}_{l}-\boldsymbol{X}_{k}}{h}\right).
\end{multline*}

The following result is a generalization of Theorem \ref{thm:variance estimator}.
\begin{thm}
\label{thm:variance estimator convergence rate heterogeneity}Suppose
Assumptions \ref{assu:DGP} - \ref{assu: rate of bandwidth} hold.
Then, for any interior point $\boldsymbol{x}$, 
\[
\underset{v\in I\left(\boldsymbol{x}\right)}{\mathrm{sup}}\left|\widehat{\mathrm{V}}_{GPV}\left(v|\boldsymbol{x},n\right)-\left(\pi\left(n|\boldsymbol{x}\right)\varphi\left(\boldsymbol{x}\right)\right)^{-2}\mathrm{V}_{\mathcal{M}}\left(v|\boldsymbol{x},n\right)\right|=O_{p}\left(\left(\frac{\mathrm{log}\left(L\right)}{Lh^{3+d}}\right)^{\nicefrac{1}{2}}+h^{R}\right).
\]
\end{thm}
\begin{rembold} An estimator for the asymptotic variance $\mathrm{V}_{GPV}(v|\boldsymbol{x})$
of the estimator $\widehat{f}_{GPV}\left(v|\boldsymbol{x}\right)$
in Corollary \ref{cor:asymptotic normality heterogeneity} can be
constructed using the plug-in approach:
\[
\mathrm{\widehat{V}}_{GPV}\left(v|\boldsymbol{x}\right)\coloneqq\sum_{n\in\mathcal{N}}\widehat{\pi}\left(n|\boldsymbol{x}\right)^{2}\widehat{\mathrm{V}}_{GPV}\left(v|\boldsymbol{x},n\right).
\]
Its rate of convergence is the same as in Theorem \ref{thm:variance estimator convergence rate heterogeneity}.
\end{rembold} 

\subsection{Bootstrap-based Inference and Uniform Confidence Bands}

To generate bootstrap samples, we apply the same resampling procedure
as that proposed in \citet[Section 4]{Marmer_Shneyerov_Quantile_Auctions}.
First, we randomly draw $L$ observations from $\left\{ \left(\boldsymbol{X}_{l},N_{l}\right):l=1,...,L\right\} $
(i.e., the auction-specific characteristics) with replacement. Next,
we randomly draw bids with replacement from the bids corresponding
to each selected auction. Given $\left(\boldsymbol{X}_{l}^{*},N_{l}^{*}\right)=\left(\boldsymbol{X}_{l'},N_{l'}\right)$
in the first step, in the second step $\{B_{il}^{*}:i=1,...,N_{l}^{*}\}$
is generated as an empirical bootstrap sample drawn from $\{B_{il'}:i=1,...,N_{l'}\}$.
Let $\widehat{\xi}^{*}\left(\cdot,\cdot,\cdot\right)$ and $\widehat{\varphi}^{*}\left(\cdot\right)$
be the bootstrap analogues of $\widehat{\xi}\left(\cdot,\cdot,\cdot\right)$
and $\widehat{\varphi}\left(\cdot\right)$ respectively. Let $\widehat{f}_{GPV}^{*}\left(v|\boldsymbol{x}\right)$
denote the bootstrap version of the GPV estimator: 
\[
\widehat{f}_{GPV}^{*}\left(v|\boldsymbol{x}\right)\coloneqq\frac{1}{\widehat{\varphi}^{*}\left(\boldsymbol{x}\right)L}\sum_{l=1}^{L}\frac{1}{N_{l}^{*}}\sum_{i=1}^{N_{l}^{*}}\mathbb{T}_{il}^{*}\frac{1}{h^{1+d}}K_{f}\left(\frac{\widehat{V}_{il}^{*}-v}{h},\frac{\boldsymbol{X}_{l}^{*}-\boldsymbol{x}}{h}\right),
\]
where $\widehat{V}_{il}^{*}\coloneqq\widehat{\xi}^{*}\left(B_{il}^{*},\boldsymbol{X}_{l}^{*},N_{l}^{*}\right)$
and the bootstrap version of the trimming factor is given by
\[
\mathbb{T}_{il}^{*}\coloneqq\mathbbm{1}\left(\mathbb{H}\left(\left(B_{il}^{*},\boldsymbol{X}_{l}^{*}\right),2h\right)\subseteq\mathcal{\widehat{S}}_{B,\boldsymbol{X}}^{N_{l}}\right).
\]

Consider the scaled deviation of the GPV estimator from the true PDF,
and its bootstrap analogue: 
\begin{gather*}
S\left(v|\boldsymbol{x}\right)\coloneqq\left(Lh^{3+d}\right)^{\nicefrac{1}{2}}\left(\widehat{f}_{GPV}\left(v|\boldsymbol{x}\right)-f\left(v|\boldsymbol{x}\right)\right)\textrm{ and}\\
S^{*}\left(v|\boldsymbol{x}\right)\coloneqq\left(Lh^{3+d}\right)^{\nicefrac{1}{2}}\left(\widehat{f}_{GPV}^{*}\left(v|\boldsymbol{x}\right)-\widehat{f}_{GPV}\left(v|\boldsymbol{x}\right)\right).
\end{gather*}
The following result, which is a generalization of Theorem \ref{thm: bootstrap consistency},
establishes the validity of the percentile bootstrap for $f\left(v|\boldsymbol{x}\right)$.
\begin{thm}
Suppose Assumptions \ref{assu:DGP} - \ref{assu: rate of bandwidth}
hold. Then, for any interior point $\left(v,\boldsymbol{x}\right)\in\mathcal{S}_{V,\boldsymbol{X}}$,
\[
\underset{z\in\mathbb{R}}{\mathrm{sup}}\left|\mathrm{P}^{*}\left[S^{*}\left(v|\boldsymbol{x}\right)\leq z\right]-\mathrm{P}\left[S\left(v|\boldsymbol{x}\right)\leq z\right]\right|\rightarrow_{p}0,\,\textrm{as \ensuremath{L\uparrow\infty}}.
\]
\end{thm}
We now turn to construction of uniform confidence bands for $\left\{ f\left(v|\boldsymbol{x}\right):v\in I\left(\boldsymbol{x}\right)\right\} $
given a fixed interior point $\boldsymbol{x}$. Consider the following
processes: 
\[
Z\left(v|\boldsymbol{x}\right)\coloneqq\frac{\widehat{f}_{GPV}\left(v|\boldsymbol{x}\right)-f\left(v|\boldsymbol{x}\right)}{\left(Lh^{3+d}\right)^{-\nicefrac{1}{2}}\widehat{\mathrm{V}}_{GPV}\left(v|\boldsymbol{x}\right)^{\nicefrac{1}{2}}}\textrm{ and }Z^{*}\left(v|\boldsymbol{x}\right)\coloneqq\frac{\widehat{f}_{GPV}^{*}\left(v|\boldsymbol{x}\right)-\widehat{f}_{GPV}\left(v|\boldsymbol{x}\right)}{\left(Lh^{3+d}\right)^{-\nicefrac{1}{2}}\widehat{\mathrm{V}}_{GPV}\left(v|\boldsymbol{x}\right)^{\nicefrac{1}{2}}},\textrm{ \ensuremath{v\in I\left(\boldsymbol{x}\right)}}.
\]
Similarly to the simplified model, the distribution of $\left\Vert Z\left(\cdot|\boldsymbol{x}\right)\right\Vert _{I\left(\boldsymbol{x}\right)}$
can be approximated by the conditional distribution of $\left\Vert Z^{*}\left(\cdot|\boldsymbol{x}\right)\right\Vert _{I\left(\boldsymbol{x}\right)}$.
Let $\zeta_{L,\alpha}^{*}$ be the $\left(1-\alpha\right)$-quantile
of the conditional distribution of $\left\Vert Z^{*}\left(\cdot|\boldsymbol{x}\right)\right\Vert _{I\left(\boldsymbol{x}\right)}$
given the original sample. Consider the following confidence band:
for $\ensuremath{v\in I\left(\boldsymbol{x}\right)}$,
\[
CB^{*}\left(v|\boldsymbol{x}\right)\coloneqq\left[\widehat{f}_{GPV}\left(v|\boldsymbol{x}\right)-\zeta_{L,\alpha}^{*}\sqrt{\frac{\widehat{\mathrm{V}}_{GPV}\left(v|\boldsymbol{x}\right)}{Lh^{3+d}}},\,\widehat{f}_{GPV}\left(v|\boldsymbol{x}\right)+\zeta_{L,\alpha}^{*}\sqrt{\frac{\widehat{\mathrm{V}}_{GPV}\left(v|\boldsymbol{x}\right)}{Lh^{3+d}}}\right].
\]
The following result, which is a generalization of Corollary \ref{cor:consistency of IGA bootstrap},
establishes the validity of $CB^{*}\left(\cdot|\boldsymbol{x}\right)$. 
\begin{thm}
Suppose Assumptions \ref{assu:DGP} - \ref{assu: rate of bandwidth}
hold. Then,
\[
\mathrm{P}\left[f\left(v|\boldsymbol{x}\right)\in CB^{*}\left(v|\boldsymbol{x}\right),\,\textrm{for all \ensuremath{v\in I\left(\boldsymbol{x}\right)}}\right]\rightarrow1-\alpha,\,\textrm{as \ensuremath{L\uparrow\infty}}.
\]
\end{thm}

\section{Binding Reserve Price\label{sec:Binding-Reserve-Price}}

Section 4 of GPV shows how to modify their identification and estimation
strategy when there is a binding reserve price. Here, we discuss how
our approach can be applied in that case.

When there is a binding reserve price, it is assumed that only bidders
with valuations exceeding the reserve price submit bids. Thus, one
has to distinguish between the numbers of potential and actual (active)
bidders. Let $N$ denote the number of potential bidders, which is
assumed to be known to players. The bidding strategy depends on the
number of potential bidders instead of the number of active bidders.
As discussed in GPV, $N$ can be estimated by taking the maximum of
the observed numbers of actual bidders across the auctions: $\widehat{N}\coloneqq\mathrm{max}\left\{ N_{l}:l=1,\ldots,L\right\} ,$
where $N_{l}$ is the number of actual bidders in auction $l$. 

GPV assume that the reserve price in auction $l$, denoted $P_{0l}$,
is some unknown deterministic function of the auction characteristics
$\boldsymbol{X}_{l}$: $P_{0l}=p_{0}\left(\boldsymbol{X}_{l}\right)$.
The probability of drawing a valuation below the reserve price is
given by $\Phi\left(\boldsymbol{X}_{l}\right)\coloneqq F\left(P_{0l}|\boldsymbol{X}_{l}\right)$.
The conditional CDF and PDF of the distribution of valuations given
participation (submitting a bid) are
\begin{equation}
F^{\star}\left(v|\boldsymbol{x}\right)\coloneqq\frac{F\left(v|\boldsymbol{x}\right)-\Phi\left(\boldsymbol{x}\right)}{1-\Phi\left(\boldsymbol{x}\right)}\text{ and }f^{\star}(v|\boldsymbol{x})\coloneqq\frac{f\left(v|\boldsymbol{x}\right)}{1-\Phi\left(\boldsymbol{x}\right)}\label{eq:binding CDF PDF}
\end{equation}
respectively. The third displayed equation on page 550 of GPV shows
that $\Phi(\cdot)$ can be estimated using a nonparametric regression
of the number of actual bidders:
\[
\widehat{\Phi}\left(\boldsymbol{x}\right)\coloneqq1-\frac{1}{\widehat{N}\widehat{\varphi}\left(\boldsymbol{x}\right)L}\sum_{l=1}^{L}\frac{1}{h^{d}}N_{l}K_{\boldsymbol{X}}\left(\frac{\boldsymbol{X}_{l}-\boldsymbol{x}}{h}\right).
\]

GPV point out that the density of bids is unbounded at the reserve
price $p_{0}\left(\boldsymbol{x}\right)$ and, in its neighborhood,
behaves as $\nicefrac{1}{\sqrt{b-p_{0}(\boldsymbol{x})}}$ . To avoid
technical problems due to the unbounded density, they propose to transform
the bids as
\[
B_{\dagger il}=\left(B_{il}-P_{0l}\right)^{\nicefrac{1}{2}}.
\]
The support of $\left(B_{\dagger11},\boldsymbol{X}_{1}\right)$ is
given by $\mathcal{S}_{B_{\dagger},\boldsymbol{X}}\coloneqq\left\{ \left(b,\boldsymbol{x}\right):\boldsymbol{x}\in\mathcal{X},\,b\in\left[0,\overline{b}_{\dagger}\left(\boldsymbol{x}\right)\right]\right\} $,
where $\overline{b}_{\dagger}\left(\boldsymbol{z}\right)\coloneqq\left(\overline{b}_{\dagger}\left(\boldsymbol{z}\right)-p_{0}\left(\boldsymbol{z}\right)\right)^{\nicefrac{1}{2}}$.
The support can be estimated by $\mathcal{\widehat{S}}_{B_{\dagger},\boldsymbol{X}}\coloneqq\left\{ \left(b,\boldsymbol{x}\right):\boldsymbol{z}\in\mathcal{X},\,b\in\left[0,\widehat{\overline{b}}_{\dagger}\left(\boldsymbol{x}\right)\right]\right\} $,
where $\widehat{\overline{b}}_{\dagger}\left(\boldsymbol{x}\right)\coloneqq\mathrm{\mathrm{max}}\left\{ B_{\dagger pl}:p=1,...,N_{l},\,\boldsymbol{X}_{l}\in\Pi_{h_{\partial}}\left(\boldsymbol{x}\right),\,l=1,...,L\right\} $,
see page 550 in GPV.

Let $G_{\dagger}\left(\cdot|\cdot\right)$ and $g_{\dagger}\left(\cdot|\cdot\right)$
denote respectively the conditional CDF and PDF of the transformed
bids $B_{\dagger11}$ given $\boldsymbol{X}_{1}$. Let $G_{\dagger}\left(b_{\dagger},\boldsymbol{z}\right)\coloneqq G_{\dagger}\left(b_{\dagger}|\boldsymbol{z}\right)\varphi\left(\boldsymbol{z}\right)$
and $g_{\dagger}\left(b_{\dagger},\boldsymbol{z}\right)\coloneqq g_{\dagger}\left(b_{\dagger}|\boldsymbol{z}\right)\varphi\left(\boldsymbol{z}\right)$.
GPV show that $g_{\dagger}\left(\cdot,\cdot\right)$ is bounded on
its support, and that latent valuations can be recovered using
\[
V_{il}=\xi_{\dagger}\left(B_{\dagger il},\boldsymbol{X}_{l}\right)\coloneqq P_{0l}+B_{\dagger il}^{2}+\frac{2B_{\dagger il}}{N-1}\frac{\left(1-\Phi\left(\boldsymbol{X}_{l}\right)\right)G_{\dagger}\left(B_{\dagger il},\boldsymbol{X}_{l}\right)+\Phi\left(\boldsymbol{X}_{l}\right)\varphi\left(\boldsymbol{X}_{l}\right)}{\left(1-\Phi\left(\boldsymbol{X}_{l}\right)\right)g_{\dagger}\left(B_{\dagger il},\boldsymbol{X}_{l}\right)}.
\]

In the modified GPV procedure, one first estimates $\xi_{\dagger}\left(\cdot,\cdot\right)$
by replacing $N$, $\Phi\left(\cdot\right)$, $G_{\dagger}\left(\cdot,\cdot\right)$,
$g_{\dagger}\left(\cdot,\cdot\right)$ and $\varphi\left(\cdot\right)$
with their estimators. $G_{\dagger}\left(\cdot,\cdot\right)$ and
$g_{\dagger}\left(\cdot,\cdot\right)$ can be estimated using the
transformed bids $B_{\dagger il}$: 
\begin{gather*}
\widehat{G}_{\dagger}(b_{\dagger},\boldsymbol{x})\coloneqq\frac{1}{L}\sum_{l=1}^{L}\frac{1}{N_{l}}\sum_{i=1}^{N_{l}}\mathbbm{1}\left(B_{\dagger il}\leq b_{\dagger}\right)\frac{1}{h^{d}}K_{\boldsymbol{X}}\left(\frac{\boldsymbol{X}_{l}-\boldsymbol{x}}{h}\right),\\
\widehat{g}_{\dagger}(b_{\dagger},\boldsymbol{x})\coloneqq\frac{1}{L}\sum_{l=1}^{L}\frac{1}{N_{l}}\sum_{i=1}^{N_{l}}\frac{1}{h^{1+d}}K_{g}\left(\frac{B_{\dagger il}-b_{\dagger}}{h}\right)K_{\boldsymbol{X}}\left(\frac{\boldsymbol{X}_{l}-\boldsymbol{x}}{h}\right).
\end{gather*}
In the second step of the modified GPV procedure, one uses the pseudo
valuations 
\[
\left\{ \widehat{V}_{il}\coloneqq\widehat{\xi}_{\dagger}\left(B_{\dagger il},\boldsymbol{X}_{l}\right):i=1,\dots,N_{l},l=1,\ldots,L\right\} ,
\]
where $\widehat{\xi}_{\dagger}\left(\cdot,\cdot\right)$ is the estimated
version of $\xi_{\dagger}(\cdot,\cdot)$, in place of latent valuations
to construct a kernel density estimator of $f^{\star}\left(v|\boldsymbol{x}\right)$:
\[
\widehat{f}_{GPV}^{\star}(v|\boldsymbol{x})\coloneqq\frac{1}{\widehat{\varphi}(\boldsymbol{x})L}\sum_{l=1}^{L}\frac{1}{N_{l}}\sum_{i=1}^{N_{l}}\mathbb{T}_{il}\frac{1}{h^{1+d}}K_{f}\left(\frac{\widehat{V}_{il}-v}{h},\frac{\boldsymbol{X}_{l}-\boldsymbol{x}}{h}\right),
\]
where the trimming factors $\mathbb{T}_{il}$ can be defined analogously
to the case with no binding reserve price: $\mathbb{T}_{il}\coloneqq\mathbbm{1}\left(\mathbb{H}\left(\left(B_{\dagger il},\boldsymbol{X}_{l}\right),2h\right)\subseteq\mathcal{\widehat{S}}_{B_{\dagger},\boldsymbol{X}}\right).$

Our approach can be used to obtain the asymptotic distribution of
the modified GPV estimator as follows. In view of the definitions
of $\xi_{\dagger}\left(\cdot.\cdot\right)$ and its estimator, the
$\widehat{V}_{il}-V_{il}$ term in the analogue of (\ref{eq:the first stochastic approximation result})
can be expanded as
\[
\widehat{V}_{il}-V_{il}=\frac{2B_{\dagger il}}{N-1}\frac{\left(1-\Phi\left(\boldsymbol{X}_{l}\right)\right)G_{\dagger}\left(B_{\dagger il},\boldsymbol{X}_{l}\right)+\Phi\left(\boldsymbol{X}_{l}\right)\varphi\left(\boldsymbol{X}_{l}\right)}{\left(1-\Phi\left(\boldsymbol{X}_{l}\right)\right)g_{\dagger}\left(B_{\dagger il},\boldsymbol{X}_{l}\right)^{2}}\left(\widehat{g}_{\dagger}\left(B_{\dagger il},\boldsymbol{X}_{l}\right)-g_{\dagger}\left(B_{\dagger il},\boldsymbol{X}_{l}\right)\right)+s.o.,
\]
where ``$s.o.$'' stands for smaller order terms. Hence, the GPV
estimator of $f^{\star}\left(v|\boldsymbol{x}\right)$ still has a
representation of the same form as in (\ref{eq:auction heterogeneity V statistic approximation}):
\begin{multline*}
\widehat{f}_{GPV}^{\star}\left(v|\boldsymbol{x}\right)-f^{\star}\left(v|\boldsymbol{x}\right)\\
=\frac{1}{\widehat{\varphi}\left(\boldsymbol{x}\right)L^{2}}\sum_{l=1}^{L}\sum_{k=1}^{L}\mathcal{M}_{\dagger}\left(\left(\boldsymbol{B}_{\dagger\cdot l},\boldsymbol{X}_{l},N_{l}\right),\left(\boldsymbol{B}_{\dagger\cdot k},\boldsymbol{X}_{k},N_{k}\right);v\right)+o_{p}\left(\left(Lh^{3+d}\right)^{-\nicefrac{1}{2}}\right),
\end{multline*}
where $\boldsymbol{B}_{\dagger\cdot l}\coloneqq\left(B_{\dagger1l},...,B_{\dagger N_{l}l}\right)$,
and 
\begin{eqnarray*}
 &  & \mathcal{M}_{\dagger}\left(\left(\boldsymbol{b}_{\cdot},\boldsymbol{z},m\right),\left(\boldsymbol{b}_{\cdot}',\boldsymbol{z}',m'\right);v\right)\\
 & \coloneqq & -\frac{1}{m}\sum_{i=1}^{m}\frac{1}{h^{2+d}}K_{f}'\left(\frac{\xi_{\dagger}\left(b_{i},\boldsymbol{z}\right)-v}{h},\frac{\boldsymbol{z}-\boldsymbol{x}}{h}\right)\frac{2b_{i}\left(\left(1-\Phi\left(\boldsymbol{z}\right)\right)G_{\dagger}\left(b_{i},\boldsymbol{z}\right)+\Phi\left(\boldsymbol{z}\right)\varphi\left(\boldsymbol{z}\right)\right)}{\left(N-1\right)\left(1-\Phi\left(\boldsymbol{z}\right)\right)g_{\dagger}\left(b_{i},\boldsymbol{z}\right)^{2}}\\
 &  & \times\left(\frac{1}{m'}\sum_{j=1}^{m'}\frac{1}{h^{1+d}}K_{g}\left(\frac{b_{j}'-b_{i}}{h}\right)K_{\boldsymbol{X}}\left(\frac{\boldsymbol{z}'-\boldsymbol{z}}{h}\right)-g_{\dagger}\left(b_{i},\boldsymbol{z}\right)\right).
\end{eqnarray*}

Similarly to the case with no reserve price, one can apply the Hoeffding
decomposition with only $\mathcal{M}_{\dagger2}\left(\boldsymbol{b}.,\boldsymbol{z},m;v\right)\coloneqq\mathrm{E}\left[\mathcal{M}_{\dagger}\left(\left(\boldsymbol{B}_{\dagger\cdot1},\boldsymbol{X}_{1},N_{1}\right),\left(\boldsymbol{b}.,\boldsymbol{z},m\right);v\right)\right]$
contributing to the asymptotic variance:
\begin{eqnarray*}
 &  & \mathcal{M}_{\dagger2}\left(\boldsymbol{b}.,\boldsymbol{z},m;v\right)\\
 & = & -\frac{1}{h^{2+d}}\int\int K_{f}'\left(\frac{\xi_{\dagger}\left(b',\boldsymbol{z}'\right)-v}{h},\frac{\boldsymbol{z}'-\boldsymbol{x}}{h}\right)\frac{2b'\left((1-\Phi\left(\boldsymbol{z}'\right))G_{\dagger}\left(b',\boldsymbol{z}'\right)+\Phi\left(\boldsymbol{z}'\right)\varphi\left(\boldsymbol{z}'\right)\right)}{\left(N-1\right)\left(1-\Phi\left(\boldsymbol{z}'\right)\right)g_{\dagger}\left(b',\boldsymbol{z}'\right)^{2}}\\
 &  & \times\left(\frac{1}{m}\sum_{j=1}^{m}\frac{1}{h^{1+d}}K_{g}\left(\frac{b_{j}-b'}{h}\right)K_{\boldsymbol{X}}\left(\frac{\boldsymbol{z}-\boldsymbol{z}'}{h}\right)-g_{\dagger}\left(b',\boldsymbol{z}'\right)\right)g_{\dagger}\left(b',\boldsymbol{z}'\right)\mathrm{d}b'\mathrm{d}\boldsymbol{z}'.
\end{eqnarray*}
 Similarly to (\ref{eq:EM_2  heterogeneity}), the asymptotic variance
of the GPV estimator is now given by the limit of 
\begin{align}
 & \frac{\overline{\pi}\left(\boldsymbol{x}\right)}{\varphi\left(\boldsymbol{x}\right)^{2}\left(N-1\right)^{2}h^{3(1+d)}}\int\int\left\{ \int_{\mathcal{X}}\int_{0}^{\overline{b}_{\dagger}\left(\boldsymbol{z}'\right)}K_{f}'\left(\frac{\xi_{\dagger}\left(b',\boldsymbol{z}'\right)-v}{h},\frac{\boldsymbol{z}'-\boldsymbol{x}}{h}\right)\right.\nonumber \\
 & \times\frac{2b'\left((1-\Phi\left(\boldsymbol{z}'\right))G_{\dagger}\left(b',\boldsymbol{z}'\right)+\Phi\left(\boldsymbol{z}'\right)\varphi\left(\boldsymbol{z}'\right)\right)}{1-\Phi\left(\boldsymbol{z}'\right)g_{\dagger}\left(b',\boldsymbol{z}'\right)}\nonumber \\
 & \left.\times K_{g}\left(\frac{b-b'}{h}\right)K_{\boldsymbol{X}}\left(\frac{\boldsymbol{z}-\boldsymbol{z}'}{h}\right)\mathrm{d}b'\mathrm{d}\boldsymbol{z}'\right\} ^{2}g_{\dagger}\left(b,\boldsymbol{z}\right)\mathrm{d}b\mathrm{d}\boldsymbol{z}\label{eq:Em_2^2 reserve price}
\end{align}
as $h\downarrow0$, where $\overline{\pi}\left(\boldsymbol{x}\right)\coloneqq\mathrm{E}\left[N_{1}^{-1}\mid\boldsymbol{X}_{1}=\boldsymbol{x}\right]$.
Note that conditionally on $\boldsymbol{X}_{1}$, the number of active
bidders $N_{1}$ has a binomial distribution with parameters $N$
and $1-\Phi\left(\boldsymbol{X}_{1}\right)$. Lastly, similarly to
Theorem \ref{thm:heterogeneity}, the expression in (\ref{eq:variance limit general}),
and for any interior point $v\in\left(p_{0}\left(\boldsymbol{x}\right),\bar{v}\left(\boldsymbol{x}\right)\right)$,
the GPV estimator of $f^{\star}\left(v\mid\boldsymbol{x}\right)$
is asymptotically normal with the asymptotic variance given by 
\begin{eqnarray*}
 &  & \mathrm{V}_{\dagger GPV}\left(v,\boldsymbol{x}\right)\\
 & \coloneqq & \frac{\overline{\pi}\left(\boldsymbol{x}\right)}{\varphi\left(\boldsymbol{x}\right)^{2}\left(N-1\right)^{2}}\frac{\left(2s_{\dagger}\left(v,\boldsymbol{x}\right)s_{\dagger v}\left(v,\boldsymbol{x}\right)\left(\left(1-\Phi\left(\boldsymbol{x}\right)\right)G_{\dagger}\left(s_{\dagger}\left(v,\boldsymbol{x}\right),\boldsymbol{x}\right)+\Phi\left(\boldsymbol{x}\right)\varphi\left(\boldsymbol{x}\right)\right)\right)^{2}}{\left(1-\Phi\left(\boldsymbol{x}\right)\right)^{2}g_{\dagger}\left(s_{\dagger}\left(v,\boldsymbol{x}\right),\boldsymbol{x}\right)}\\
 &  & \times\int\int\left\{ \int\int K_{f}'\left(w,\boldsymbol{y}\right)K_{\boldsymbol{X}}\left(\boldsymbol{y}-\boldsymbol{z}\right)K_{g}\left(u-s_{\dagger v}w-s_{\dagger\boldsymbol{x}}^{\mathrm{T}}\boldsymbol{y}\right)\mathrm{d}w\mathrm{d}\boldsymbol{y}\right\} ^{2}\mathrm{d}u\mathrm{d}\boldsymbol{z},
\end{eqnarray*}
where $s_{\dagger}(\cdot,\boldsymbol{x})\coloneqq\xi_{\dagger}^{-1}(\cdot,\boldsymbol{x})$,
and the partial derivatives $s_{\dagger v}$ and $s_{\dagger\boldsymbol{x}}$
are defined similarly to $s_{v}$ and $s_{\boldsymbol{x}}$ in (\ref{eq:derivatives}).
Similarly to Corollary \ref{cor:asymptotic normality heterogeneity},
one can show:
\[
\left(Lh^{3+d}\right)^{\nicefrac{1}{2}}\left(\widehat{f}^{\star}\left(v\mid\boldsymbol{x}\right)-f^{\star}\left(v\mid\boldsymbol{x}\right)\right)\rightarrow_{d}\mathrm{N}\left(0,\mathrm{V}_{\dagger GPV}\left(v,\boldsymbol{x}\right)\right).
\]

To estimate the asymptotic variance $\mathrm{V}_{\dagger GPV}(v,\boldsymbol{x})$,
one can use the sample analogue of (\ref{eq:Em_2^2 reserve price})
in the same way as that used to construct the estimator of $\mathrm{V}_{GPV}(v,\boldsymbol{x})$
defined by (\ref{eq:V_GPV estimator heterogeneity}) from (\ref{eq:EM_2  heterogeneity}).
As before, the approach does not require estimation of the bidding
strategy $s_{\dagger}(v,\boldsymbol{x})$ or its derivatives. The
analogue estimator is given by
\begin{multline*}
\mathrm{\widehat{V}}_{\dagger GPV}\left(v,\boldsymbol{x}\right)\coloneqq\frac{\widehat{\overline{\pi}}\left(\boldsymbol{x}\right)}{\widehat{\varphi}\left(\boldsymbol{x}\right)^{2}\left(\widehat{N}-1\right)^{2}h^{3(1+d)}}\frac{1}{L\left(L-1\right)\left(L-2\right)}\\
\times\sum_{l=1}^{L}\sum_{k\neq l}\sum_{k'\neq k,k'\neq l}\frac{1}{N_{l}}\sum_{i=1}^{N_{l}}\eta_{\dagger il,k}(v,\boldsymbol{x})\eta_{\dagger il,k'}(v,\boldsymbol{x}),
\end{multline*}
where $\widehat{\overline{\pi}}\left(\boldsymbol{x}\right)$ is the
Nadaraya-Watson estimator of $\overline{\pi}\left(\boldsymbol{x}\right)$,
and
\begin{multline*}
\eta_{\dagger il,k}\left(v,\boldsymbol{x}\right)\coloneqq\frac{1}{N_{k}}\sum_{j=1}^{N_{k}}\mathbb{T}_{jk}K_{f}'\left(\frac{\widehat{V}_{jk}-v}{h},\frac{\boldsymbol{X}_{k}-\boldsymbol{x}}{h}\right)K_{g}\left(\frac{B_{\dagger il}-B_{\dagger jk}}{h}\right)K_{\boldsymbol{X}}\left(\frac{\boldsymbol{X}_{l}-\boldsymbol{X}_{k}}{h}\right)\\
\times\frac{2B_{\dagger jk}\left((1-\widehat{\Phi}\left(\boldsymbol{X}_{k}\right))\widehat{G}_{\dagger}\left(B_{\dagger jk},\boldsymbol{X}_{k}\right)+\widehat{\Phi}\left(\boldsymbol{X}_{k}\right)\widehat{\varphi}\left(\boldsymbol{X}_{k}\right)\right)}{1-\widehat{\Phi}\left(\boldsymbol{X}_{k}\right)\widehat{g}_{\dagger}\left(B_{\dagger jk},\boldsymbol{X}_{k}\right)^{2}}.
\end{multline*}

Suppose $I\left(\boldsymbol{x}\right)$ is an inner closed sub-interval
of $\left[p_{0}\left(\boldsymbol{x}\right),\bar{v}\left(\boldsymbol{x}\right)\right]$.
The uniform convergence rate of $\mathrm{\widehat{V}}_{\dagger GPV}\left(v,\boldsymbol{x}\right)$
to (\ref{eq:Em_2^2 reserve price}) can be shown to be the same as
that in the statement of Theorem \ref{thm:variance estimator convergence rate heterogeneity}.
In view of the definitions in (\ref{eq:binding CDF PDF}), the nonparametric
estimator for $f\left(v|\boldsymbol{x}\right)$ is $\left(1-\widehat{\Phi}\left(\boldsymbol{x}\right)\right)\widehat{f}^{\star}\left(v|\boldsymbol{x}\right)$.
Since $\widehat{\Phi}\left(\boldsymbol{x}\right)$ converges at a
faster rate than the PDF estimator $\widehat{f}^{\star}\left(v\mid\boldsymbol{x}\right)$,
one can see that 
\[
\left(Lh^{3+d}\right)^{\nicefrac{1}{2}}\left(\left(1-\widehat{\Phi}\left(\boldsymbol{x}\right)\right)\widehat{f}^{\star}\left(v\mid\boldsymbol{x}\right)-f\left(v\mid\boldsymbol{x}\right)\right)\rightarrow_{d}\mathrm{N}\left(0,\left(1-\Phi\left(\boldsymbol{x}\right)\right)^{2}\mathrm{V}_{\dagger GPV}\left(v,\boldsymbol{x}\right)\right).
\]

A valid uniform confidence band of $\left\{ f\left(v|\boldsymbol{x}\right):v\in I\left(\boldsymbol{x}\right)\right\} $
can be constructed by adapting the methods described in Section \ref{sec:Auction-Specific-Heterogeneity}.

\section{Monte Carlo Simulations\label{sec:Monte-Carlo-Simulations}}

In this section, we assess the finite-sample coverage accuracy of
the uniform confidence bands. Our simulation design follows \citet*{Marmer_Shneyerov_Quantile_Auctions},
and the DGP is described in Remark \ref{rem:comparison QB}. We consider
$\theta\in\left\{ 1,2\right\} $ and draw valuations from $f_{\theta}$.
We choose the triweight kernel when implementing the two-step estimator.
We used the second-order triweight kernel in the second step and used
the fourth-order triweight kernel in the first step. 

We need to choose the bandwidths in the first step when we construct
the pseudo valuations and the second step when we implement kernel
density estimation using the pseudo valuations. We follow GPV (see
Section 2.4) and use $h_{g}=3.72\cdot\widehat{\sigma}_{b}\cdot\left(N\cdot L\right)^{-\nicefrac{1}{5}}$as
the first-step bandwidth, where $\widehat{\sigma}_{b}$ is the estimated
standard deviation of the observed bids. We use $h_{f}=3.15\cdot\widehat{\sigma}_{v}\cdot\left(\left(N\cdot L\right)_{\mathbb{T}}\right)^{-\nicefrac{1}{5}}$as
the second-step bandwidth, where $\widehat{\sigma}_{v}$ is the estimated
standard deviation of the trimmed pseudo valuations and $\left(N\cdot L\right)_{\mathbb{T}}$
is the number of bids remaining after the trimming. The constants
$3.72$ and $3.15$ are Silverman's rule-of-thumb constants corresponding
to fourth-order and second-order triweight kernels.\footnote{See \citet{li2007net} for a description of the Silverman approach;
see also \citet[Section 3.2]{li2003semiparametric}.} We consider different numbers of bidders $N\in\left\{ 3,5,7\right\} $,
and also the density function over different ranges: $v\in\left[0.2,0.8\right]$
and $v\in\left[0.3,0.7\right]$.\footnote{We use grid maximization, where the grid is chosen as $[v_{l}:0.001:v_{u}]$.
We have also tried a finer grid $[v_{l}:0.0001:v_{u}],$ which produced
similar results.} The number of auctions $L$ is chosen so the total number of observations
is fixed as $N\cdot L=2100$. 

\begin{table}[t]
\caption{Coverage probabilities of the uniform confidence band $CB^{*}$ for
the number of bidders $N=3,5,7$, the distribution parameter $\theta=1,2$,
different ranges of valuations $v$, and the nominal coverage probability
$=0.90,0.95,0.99$. The number of auctions $L$ is determined by $N\cdot L=2100$\label{tab:Coverage-probabilities}\medskip{}
}

$\,$\thispagestyle{empty}
\centering{}%
\begin{tabular}{c>{\centering}p{2cm}>{\centering}p{2cm}>{\centering}p{2cm}>{\centering}p{2cm}>{\centering}p{2cm}>{\centering}p{2cm}}
\toprule 
$N$ & 0.90 & 0.95 & 0.99 & 0.90 & 0.95 & 0.99\tabularnewline
\midrule
 & \multicolumn{3}{c}{\uline{\mbox{$\theta=1,\quad v\in[0.3,0.7]$}}} & \multicolumn{3}{c}{\uline{\mbox{$\theta=1,\quad v\in[0.2,0.8]$}}}\tabularnewline
 &  &  &  &  &  & \tabularnewline
$3$ & 0.880 & 0.924 & 0.988 & 0.880 & 0.930 & 0.982\tabularnewline
$5$ & 0.922 & 0.962 & 0.998 & 0.918 & 0.962 & 0.994\tabularnewline
$7$ & 0.888 & 0.946 & 0.990 & 0.898 & 0.948 & 0.984\tabularnewline
 &  &  &  &  &  & \tabularnewline
 & \multicolumn{3}{c}{\uline{\mbox{$\theta=2,\quad v\in[0.3,0.7]$}}} & \multicolumn{3}{c}{\uline{\mbox{$\theta=2,\quad v\in[0.2,0.8]$}}}\tabularnewline
 &  &  &  &  &  & \tabularnewline
$3$ & 0.912 & 0.954 & 0.994 & 0.900 & 0.944 & 0.988\tabularnewline
$5$ & 0.882 & 0.938 & 0.990 & 0.872 & 0.932 & 0.990\tabularnewline
$7$ & 0.916 & 0.950 & 0.986 & 0.914 & 0.964 & 0.994\tabularnewline
 &  &  &  &  &  & \tabularnewline
\bottomrule
\end{tabular}
\end{table}

In Table \ref{tab:Coverage-probabilities}, we report our simulation
results for the bootstrap-based IGA uniform confidence band $CB^{*}$.
We find that the IGA bootstrap approach provides accurate coverage
probabilities. Additional simulation results are reported in the Supplement.

\section{Conclusion\label{sec:Conclusions}}

The GPV estimator has proven to be the essential input in virtually
all nonparametric structural auction models. By proving the asymptotic
normality and the first-order validity of the bootstrap uniform confidence
bands, this paper completes the econometric theory of the GPV estimator
and opens way to new applications.

Our pointwise asymptotic normality results can be used for inference
on an important policy variable: the optimal reserve price. As discussed,
e.g., in \citet{haile2003iim}, the optimal reserve price $r(\boldsymbol{x})$
in auctions with $\boldsymbol{X}_{l}=\boldsymbol{x}$ satisfies the
following equation:
\[
r(\boldsymbol{x})-\frac{1-F(r(\boldsymbol{x})|\boldsymbol{x})}{f(r(\boldsymbol{x})|\boldsymbol{x})}=c(\boldsymbol{x}),
\]
where $c(\boldsymbol{x})$ is the seller's own valuation. Suppose
that the estimator $\widehat{r}(\boldsymbol{x})$ is constructed by
solving an estimated version of the above equation with $f(\cdot|\boldsymbol{x})$
replaced by its GPV estimator $\widehat{f}_{GPV}\left(\cdot|\boldsymbol{x}\right)$.
In that case, our pointwise normality results imply the asymptotic
normality of the estimated optimal reserve price:
\begin{multline*}
\left(Lh^{3+d}\right)^{\nicefrac{1}{2}}\left(\widehat{r}\left(\boldsymbol{x}\right)-r(\boldsymbol{x})\right)\\
\rightarrow_{d}\mathrm{N}\left(0,\left(\frac{1-F(r(\boldsymbol{x})|\boldsymbol{x})}{2f(r(\boldsymbol{x})|\boldsymbol{x})^{2}+f'(r(\boldsymbol{x})|\boldsymbol{x})\left(1-F(r(\boldsymbol{x})|\boldsymbol{x})\right)}\right)^{2}\mathrm{V}_{GPV}\left(r(\boldsymbol{x})|\boldsymbol{x}\right)\right).
\end{multline*}
Our results for the validity of the percentile bootstrap of the GPV
estimator naturally carry over to the above estimator of the optimal
reserve price. Thus in practice, one can use the percentile bootstrap
for inference on the optimal reserve price. 

Our uniform confidence bands can be used for specification of the
density of valuations.

In future research, our results could be extended in several directions.
Below, we briefly describe some of potentially interesting extensions.
These extensions would address the limitations of the independent
private values model that underlies the GPV estimator.

First, in GPV the bidders are treated symmetrically. Empirically this
is not always the case. See, e.g., \citet{flambard2006asymmetry}
for an application to snow removal contracts. Second, we abstract
from the empirically relevant issue of unobserved heterogeneity, as
in \citet{krasnokutskaya2011identification}, \citet{hu2013identification}
and \citet{roberts2013unobserved}. Third, correlated values may also
be important empirically. \citet{li2002structural} and \citet{hubbard2012semiparametric}
extend the GPV estimator to the affiliated value environment. Fourth,
\citet{guerre2009nonparametric} and \citet{zincenko2018nonparametric}
provide extensions to an environment with risk-averse bidders. Fifth,
the literature on endogenous entry in auctions has developed rapidly.
See, e.g., \citet{li2009entry}, \citet{krasnokutskaya2011bid}, \citet{marmer2013model},
\citet{roberts2013should} and \citet{gentry2014identification}. 

In the above models, the basic GPV estimator is often adapted to suit
the needs of a particular application. But most of these estimators
share the underlying two-step structure of the GPV estimator. We conjecture
that our main results will also prove useful for establishing the
asymptotic normality and validity of certain uniform confidence bands
for these GPV-like estimators. 

A remaining unresolved important practical issue is bandwidth selection
for the GPV estimator. It is possible that the recent advances in
that area (e.g., \citealp{calonico2014robust} and \citealp{armstrong2016simple})
can be adapted to the framework of GPV.

%

\bibliographystyle{elsarticle-harv}
\bibliography{vgpv_art,Jun_Ma_auctions}

\newpage
\begin{appendices}
\part*{Appendix}

Let $\apprle$ denote an inequality up to a universal constant that
does not depend on the sample size $L$. Denote $\left\Vert f\right\Vert _{Q,2}\coloneqq\left(\int\left|f\right|^{2}\mathrm{d}Q\right)^{\nicefrac{1}{2}}$.
For a sequence of classes of functions $\mathscr{F}_{L}$ (that may
depend on the sample size) defined on $\left[\underline{b},\overline{b}\right]^{d}$,
let $N\left(\epsilon,\mathscr{F}_{L},\left\Vert \cdot\right\Vert _{Q,2}\right)$
denote the $\epsilon-$covering number, i.e., the smallest integer
$m$ such that there are $m$ balls of radius $\epsilon$ (with respect
to the metric induced by the norm $\left\Vert \cdot\right\Vert _{Q,2}$)
centered at points in $\mathscr{F}_{L}$ whose union covers $\mathscr{F}_{L}$.
A function $F_{L}:\left[\underline{b},\overline{b}\right]^{d}\rightarrow\mathbb{R}_{+}$
is an envelope of $\mathscr{F}_{L}$ if $F_{L}\geq\underset{f\in\mathscr{F}_{L}}{\mathrm{sup}}\left|f\right|$.
We say that $\mathscr{F}_{L}$ is a (uniform) Vapnik-Chervonenkis-type
(VC-type) class with respect to the envelope $F_{L}$ (see, e.g.,
\citealp[Definition 2.1]{chernozhukov2014gaussian} and \citealp[Definition 3.6.10]{gine2015mathematical})
if there exist some positive constants $A$ and $V$ that are independent
of $L$ such that
\[
N\left(\epsilon\left\Vert F_{L}\right\Vert _{Q,2},\mathscr{F}_{L},\left\Vert \cdot\right\Vert _{Q,2}\right)\leq\left(\frac{A}{\epsilon}\right)^{V},\,\textrm{for all \ensuremath{\epsilon\in\left(0,1\right]},}
\]
for all finitely discrete probability measure $Q$ on $\left[\underline{b},\overline{b}\right]^{d}$.
Note that the all function classes that appear later are dependent
on $L$. We suppress the dependence for notational simplicity.

Denote $\overline{C}_{K_{f}}\coloneqq\underset{u\in\mathbb{R}}{\mathrm{sup}}\left|K_{f}\left(u\right)\right|$
and $\overline{C}_{K_{g}}\coloneqq\underset{u\in\mathbb{R}}{\mathrm{sup}}\left|K_{g}\left(u\right)\right|$.
Since $\underset{u\in\mathbb{R}}{\mathrm{sup}}\left|K_{f}^{''}\left(u\right)\right|<\infty$,
the restriction of $K_{f}^{'}$ on $\left[-1,1\right]$ is of bounded
variation. We can decompose $K_{f}'=D_{1}-D_{2}$, where $D_{1}$
and $D_{2}$ are non-decreasing and bounded. Similarly, denote $\overline{C}_{D_{s}}\coloneqq\underset{u\in\mathbb{R}}{\mathrm{sup}}\left|D_{s}\left(u\right)\right|$,
for $s\in\left\{ 1,2\right\} $. Let $\overline{C}_{s'}\coloneqq\underset{u\in\left[\underline{v},\overline{v}\right]}{\mathrm{sup}}s'\left(u\right)$.
Let $C_{k}$, $k=1,2,...$ denote positive universal constants that
are independent of the sample size and whose values may change in
different places. $\delta_{k,L}$, $k=1,2,3,...$ denote positive
null sequences of real numbers (i.e., $\delta_{k,L}\downarrow0$,
as $L\uparrow\infty$). 

Let $\mathrm{E}^{*}\left[\cdot\right]$ denote the expectation under
$\mathrm{P}^{*}$, the conditional probability distribution given
the original sample. We use the following notation for bootstrap asymptotics.
We say that $\xi_{L}=o_{p}^{*}\left(\lambda_{L}\right)$ if for all
$\epsilon>0$, $\mathrm{P}^{*}\left[\left|\nicefrac{\xi_{L}}{\lambda_{L}}\right|>\epsilon\right]\rightarrow_{p}0$
as $L\uparrow\infty$. We say that $\xi_{L}=O_{p}^{*}\left(\lambda_{L}\right)$
if for all $\epsilon>0$, there is $M_{\epsilon}>0$ and some $L_{\epsilon}\in\mathbb{N}$
such that $\mathrm{P}\left[\mathrm{P}^{*}\left[\left|\nicefrac{\xi_{L}}{\lambda_{L}}\right|\geq M_{\epsilon}\right]>\epsilon\right]<\epsilon$
for all $L\geq L_{\epsilon}$. Properties of $O_{p}$ and $o_{p}$
notations carry over to $O_{p}^{*}$ and $o_{p}^{*}$ naturally. 

``With probability approaching 1'' is abbreviated as ``w.p.a.1''.
For notational simplicity, in the proofs, $\underset{i,l}{\mathrm{max}}$
is understood as $\underset{\left(i,l\right)\in\left\{ 1,...,N\right\} \times\left\{ 1,...,L\right\} }{\mathrm{max}}$.
$\underset{\left(2\right)}{\sum}$ is understood as $\sum_{\left(j,k\right)\neq\left(i,l\right)}$
and $\underset{\left(3\right)}{\sum}$ is understood as 
\[
\sum_{i,l}\sum_{\left(j,k\right)\neq\left(i,l\right)}\sum_{\left(j',k'\right)\neq\left(i,l\right),\,\left(j',k'\right)\neq\left(j,k\right)},
\]
i.e., summing over all distinct indices. We also adopt the following
notation: $\left(N\cdot L\right)_{2}$ is understood as $\left(N\cdot L\right)\left(N\cdot L-1\right)$
and $\left(N\cdot L\right)_{3}$ is understood as $\left(N\cdot L\right)\left(N\cdot L-1\right)\left(N\cdot L-2\right)$. 

In the proofs, we will often invoke maximal inequalities for empirical
processes. A sharper one is \citet[Corollary 5.1]{chernozhukov2014gaussian}.
It is more convenient to apply a modified version of this inequality,
i.e., \citet[Corollary 5.5]{Chen_Kato_U_Process} (with $r=k=1$).
To avoid ambiguity, we will still refer to this inequality as the
``CCK inequality'' in the proofs. In some applications, a less sharp
but simpler maximal inequality for empirical processes suffices. See,
e.g., \citet[Theorem 2.14.1]{van1996weak}. We will refer to this
inequality as the ``VW inequality''. We will also often invoke a
(less sharp) maximal inequality for U-processes, i.e., \citet[Corollary 5.6]{Chen_Kato_U_Process}.
We will refer to this inequality as the ``CK inequality''. The following
lemmas provide asymptotic expansions that are crucial for proving
the distributional results provided in Theorems \ref{thm:Asymptotic Normality GPV}
and \ref{thm:Gaussian coupling} and also the bootstrap consistency
results. Their proofs are relegated to the Supplement.
\section{Proofs for Sections \ref{sec:Asymptotic Normality}--\ref{sec:Uniform-Confidence-Bands} }\label{sec:Appendix_main}
\begin{lem}
\label{lem:lemma 2}Suppose Assumptions \ref{assu:DGP} - \ref{assu: rate of bandwidth}
hold. Then,
\begin{multline*}
\widehat{f}_{GPV}\left(v\right)-f\left(v\right)=\frac{1}{\left(N-1\right)}\frac{1}{\left(N\cdot L\right)^{2}}\sum_{i,l}\sum_{j,k}\mathcal{M}\left(B_{il},B_{jk};v\right)\\
+\frac{1}{R!}f^{\left(R\right)}\left(v\right)\left(\int K_{f}\left(u\right)u^{R}\mathrm{d}u\right)h^{R}+O_{p}\left(\left(\frac{\mathrm{log}\left(L\right)}{Lh}\right)^{\nicefrac{1}{2}}+\frac{\mathrm{log}\left(L\right)}{Lh^{3}}\right)+o\left(h^{R}\right),
\end{multline*}
where the remainder term is uniform in $v\in I$ and $\mathcal{M}$
is defined by (\ref{eq:V_stat kernel definition}).
\end{lem}
\begin{lem}
\label{lem:lemma 3}Suppose Assumptions \ref{assu:DGP} - \ref{assu: rate of bandwidth}
hold. Then,
\begin{multline*}
\widehat{f}_{GPV}\left(v\right)-f\left(v\right)=\frac{1}{N-1}\frac{1}{N\cdot L}\sum_{i,l}\left(\mathcal{M}_{2}\left(B_{il};v\right)-\mu_{\mathcal{M}}\left(v\right)\right)\\
+\frac{1}{R!}f^{\left(R\right)}\left(v\right)\left(\int K_{f}\left(u\right)u^{R}\mathrm{d}u\right)h^{R}+O_{p}\left(\left(\frac{\mathrm{log}\left(L\right)}{Lh}\right)^{\nicefrac{1}{2}}+\frac{\mathrm{log}\left(L\right)}{Lh^{3}}\right)+o\left(h^{R}\right),
\end{multline*}
where the remainder term is uniform in $v\in I$.
\end{lem}
\begin{proof}[Proof of Theorem \ref{thm:Asymptotic Normality GPV}]
It follows from Lemma \ref{lem:lemma 3} that
\[
\left(Lh^{3}\right)^{\nicefrac{1}{2}}\left(\widehat{f}_{GPV}\left(v\right)-f\left(v\right)\right)=\frac{1}{N^{\nicefrac{1}{2}}\left(N-1\right)}\frac{1}{\left(N\cdot L\right)^{\nicefrac{1}{2}}}\sum_{i,l}h^{\nicefrac{3}{2}}\left(\mathcal{M}_{2}\left(B_{il};v\right)-\mu_{\mathcal{M}}\left(v\right)\right)+o_{p}\left(1\right).
\]
Next, we show that a central limit theorem for triangular arrays can
be applied to the leading term on the right-hand side. 

Let 
\begin{gather*}
\mathcal{M}^{\ddagger}\left(b,b';v\right)\coloneqq-\frac{1}{h^{3}}K_{f}'\left(\frac{\xi\left(b\right)-v}{h}\right)\frac{G\left(b\right)}{g\left(b\right)^{2}}K_{g}\left(\frac{b'-b}{h}\right),\\
\mathcal{M}_{2}^{\ddagger}\left(b;v\right)\coloneqq-\int_{\underline{b}}^{\overline{b}}\frac{1}{h^{3}}K_{f}'\left(\frac{\xi\left(b'\right)-v}{h}\right)\frac{G\left(b'\right)}{g\left(b'\right)}K_{g}\left(\frac{b-b'}{h}\right)\mathrm{d}b'\textrm{ and}\\
\mu_{\mathcal{M}^{\ddagger}}\left(v\right)\coloneqq\int\int\mathcal{M}^{\ddagger}\left(b,b';v\right)\mathrm{d}G\left(b'\right)\mathrm{d}G\left(b\right)=\int\mathcal{M}_{2}^{\ddagger}\left(b;v\right)\mathrm{d}G\left(b\right).
\end{gather*}
It is easy to see that 
\begin{equation}
\mathcal{M}_{2}\left(B_{il};v\right)-\mu_{\mathcal{M}}\left(v\right)=\mathcal{M}_{2}^{\ddagger}\left(B_{il};v\right)-\mu_{\mathcal{M}^{\ddagger}}\left(v\right),\textrm{ for all \ensuremath{i=1,...,N} and \ensuremath{l=1,...,L}}.\label{eq:M_2 - miu_M =00003D M_2_ddagger - miu_M_ddagger}
\end{equation}

Define
\[
U_{il}\left(v\right)\coloneqq\frac{1}{N^{\nicefrac{1}{2}}\left(N-1\right)}\frac{1}{\left(N\cdot L\right)^{\nicefrac{1}{2}}}h^{\nicefrac{3}{2}}\left(\mathcal{M}_{2}^{\ddagger}\left(B_{il};v\right)-\mu_{\mathcal{M}^{\ddagger}}\left(v\right)\right)
\]
and
\begin{equation}
\sigma\left(v\right)\coloneqq\left(\sum_{i,l}\mathrm{E}\left[U_{il}\left(v\right)^{2}\right]\right)^{\nicefrac{1}{2}}=\left(\frac{1}{N\left(N-1\right)^{2}}h^{3}\mathrm{E}\left[\left(\mathcal{M}_{2}^{\ddagger}\left(B_{11};v\right)-\mu_{\mathcal{M}^{\ddagger}}\left(v\right)\right)^{2}\right]\right)^{\nicefrac{1}{2}}.\label{eq:sigma_L definition}
\end{equation}
By the definition of $U_{il}\left(v\right)$, 
\begin{equation}
\left(Lh^{3}\right)^{\nicefrac{1}{2}}\left(\widehat{f}_{GPV}\left(v\right)-f\left(v\right)\right)=\sum_{i,l}U_{il}\left(v\right)+o_{p}\left(1\right).\label{eq:f_hat - f U's}
\end{equation}

Next, we show that
\begin{equation}
\frac{1}{N\left(N-1\right)^{2}}h^{3}\mathrm{E}\left[\left(\mathcal{M}_{2}^{\ddagger}\left(B_{11};v\right)-\mu_{\mathcal{M}^{\ddagger}}\left(v\right)\right)^{2}\right]-\mathrm{V}_{\mathcal{M}}\left(v\right)=O\left(h^{3}\right)\label{eq:M_2^ddagger variance - V_M}
\end{equation}
uniformly in $v\in I$. It is easy to check that 
\[
\mu_{\mathcal{M}^{\ddagger}}\left(v\right)=\mu_{\mathcal{M}}\left(v\right)+\left(\int_{\underline{b}}^{\overline{b}}-\frac{1}{h^{2}}K_{f}'\left(\frac{\xi\left(b\right)-v}{h}\right)G\left(b\right)\mathrm{d}b\right).
\]
By change of variable and a mean value expansion, 
\begin{align}
 & \int_{\underline{b}}^{\overline{b}}K_{f}'\left(\frac{\xi\left(b\right)-v}{h}\right)G\left(b\right)\mathrm{d}b\nonumber \\
= & h\int_{\frac{\underline{v}-v}{h}}^{\frac{\overline{v}-v}{h}}K'_{f}\left(u\right)G\left(s\left(hu+v\right)\right)s'\left(hu+v\right)\mathrm{d}u\nonumber \\
= & h\int_{\frac{\underline{v}-v}{h}}^{\frac{\overline{v}-v}{h}}K_{f}'\left(u\right)\left\{ G\left(s\left(v\right)\right)s'\left(v\right)+\left(g\left(s\left(\dot{v}\right)\right)s'\left(\dot{v}\right)^{2}+G\left(s\left(\dot{v}\right)\right)s''\left(\dot{v}\right)\right)hu\right\} \mathrm{d}u\label{eq:E_m_2^2 expansion second term}
\end{align}
for some mean value $\dot{v}$ (depending on $u$) such that $\left|\dot{v}-v\right|\leq h\left|u\right|$.
Since $\int K_{f}'\left(u\right)\mathrm{d}u=0$ and $K_{f}'$ is supported
on $\left[-1,1\right]$, we have
\begin{align*}
 & \underset{v\in I}{\mathrm{sup}}\left|\int_{\frac{\underline{v}-v}{h}}^{\frac{\overline{v}-v}{h}}K_{f}'\left(u\right)G\left(s\left(hu+v\right)\right)s'\left(hu+v\right)\mathrm{d}u\right|\\
\leq & h\left(\int\left|K_{f}'\left(u\right)u\right|\mathrm{d}u\right)\left(\underset{u\in\left[\underline{v},\overline{v}\right]}{\mathrm{sup}}\left|g\left(s\left(u\right)\right)s'\left(u\right)^{2}+G\left(s\left(u\right)\right)s''\left(u\right)\right|\right).
\end{align*}
By the above result, (\ref{eq:E_m_2^2 expansion second term}), the
continuity of $s'$ and $s''$ and the continuity of $g$ and $G$,
we have
\begin{equation}
\underset{v\in I}{\mathrm{sup}}\left|\int_{\underline{b}}^{\overline{b}}K_{f}'\left(\frac{\xi\left(b\right)-v}{h}\right)G\left(b\right)\mathrm{d}b\right|=O\left(h^{2}\right).\footnotemark\label{eq:K_prime G s_prime integral bound}
\end{equation}
\footnotetext{See Proposition 1 and Lemma A1 of GPV.}It is shown
in the proof of Lemma \ref{lem:lemma 3} that $\underset{v\in I}{\mathrm{sup}}\left|\mu_{\mathcal{M}}\left(v\right)\right|=o\left(h^{R}\right)$.
Therefore we have 
\begin{equation}
\underset{v\in I}{\mathrm{sup}}\left|\mu_{\mathcal{M}^{\ddagger}}\left(v\right)\right|=O\left(1\right).\label{eq:miu_ddagger sup rate}
\end{equation}
It is clear that
\begin{eqnarray*}
\mathrm{E}\left[h^{3}\mathcal{M}_{2}^{\ddagger}\left(B_{11};v\right)^{2}\right] & = & h^{-3}\int\left\{ \int_{\underline{b}}^{\overline{b}}K_{f}'\left(\frac{\xi\left(b\right)-v}{h}\right)\frac{G\left(b\right)}{g\left(b\right)}K_{g}\left(\frac{b'-b}{h}\right)\mathrm{d}b\right\} ^{2}\mathrm{d}G\left(b'\right)\\
 & = & N\left(N-1\right)^{2}\mathrm{V}_{\mathcal{M}}\left(v\right).
\end{eqnarray*}
Now (\ref{eq:M_2^ddagger variance - V_M}) follows from the above
result and (\ref{eq:miu_ddagger sup rate}).

By change of variables $u=\nicefrac{\left(\xi\left(b\right)-v\right)}{h}$
and $w=\nicefrac{\left(b-s\left(v\right)\right)}{h}$, 
\begin{equation}
\mathrm{E}\left[h^{3}\mathcal{M}_{2}^{\ddagger}\left(B_{11};v\right)^{2}\right]=\int_{\frac{\underline{b}-s\left(v\right)}{h}}^{\frac{\overline{b}-s\left(v\right)}{h}}\left\{ \int_{\frac{\underline{v}-v}{h}}^{\frac{\overline{v}-v}{h}}\rho\left(u,w;v\right)\mathrm{d}u\right\} ^{2}g\left(hw+s\left(v\right)\right)\mathrm{d}w,\label{eq:E=00005BM_2(B)^2=00005D approximation 1}
\end{equation}
where $\psi\left(z\right)\coloneqq\nicefrac{G\left(s\left(z\right)\right)s'\left(z\right)}{g\left(s\left(z\right)\right)}$
and 
\[
\rho\left(u,w;v\right)\coloneqq K_{f}'\left(u\right)\psi\left(hu+v\right)K_{g}\left(w-\frac{s\left(hu+v\right)-s\left(v\right)}{h}\right).
\]
Denote 
\[
\overline{\rho}\left(w;v\right)\coloneqq\psi\left(v\right)\left\{ \int K_{f}'\left(u\right)K_{g}\left(w-s'\left(v\right)u\right)\mathrm{d}u\right\} .
\]

Next, we show that 
\begin{align}
 & \mathrm{E}\left[h^{3}\mathcal{M}_{2}^{\ddagger}\left(B_{11};v\right)^{2}\right]\nonumber \\
= & g\left(s\left(v\right)\right)\int\overline{\rho}\left(w;v\right)^{2}\mathrm{d}w+o\left(1\right)\nonumber \\
= & \frac{F\left(v\right)^{2}f\left(v\right)^{2}}{g\left(s\left(v\right)\right)^{3}}\int\left\{ \int K_{f}'\left(u\right)K_{g}\left(w-s'\left(v\right)u\right)\mathrm{d}u\right\} ^{2}\mathrm{d}w+o\left(1\right),\label{eq:Em_2^2 as integral}
\end{align}
where the remainder term is uniform in $v\in I$ and we applied the
equality $f\left(v\right)=g\left(s\left(v\right)\right)s'\left(v\right)$
to obtain the second equality. Since $s'$ is continuous and $K_{g}$
and $K_{f}'$ are supported on $\left[-1,1\right]$ and bounded, it
follows from (\ref{eq:density of bids bounded  from 0}) and the reverse
triangle inequality that
\begin{eqnarray}
\underset{v\in I}{\mathrm{sup}}\left\{ \mathbbm{1}\left(u\in\left[\frac{\underline{v}-v}{h},\frac{\overline{v}-v}{h}\right]\right)\left|\rho\left(u,w;v\right)\right|\right\}  & \lesssim & \mathbbm{1}\left(\left|u\right|\leq1\right)\mathbbm{1}\left(\left|w-\frac{s\left(hu+v\right)-s\left(v\right)}{h}\right|\leq1\right)\nonumber \\
 & \lesssim & \mathbbm{1}\left(\left|u\right|\leq1\right)\mathbbm{1}\left(\left|w\right|\leq1+\overline{C}_{s'}\right),\label{eq:theorem 1 domination condition}
\end{eqnarray}
for all $\left(u,w\right)\in\mathbb{R}^{^{2}}$. Similarly, 
\begin{equation}
\underset{v\in I}{\mathrm{sup}}\left|\overline{\rho}\left(w;v\right)\right|\apprle\mathbbm{1}\left(\left|w\right|\leq1+\overline{C}_{s'}\right).\label{eq:rho_bar sup bound}
\end{equation}

Next, by the triangle inequality, 
\begin{align}
 & \left|\int_{\frac{\underline{b}-s\left(v\right)}{h}}^{\frac{\overline{b}-s\left(v\right)}{h}}\left\{ \int_{\frac{\underline{v}-v}{h}}^{\frac{\overline{v}-v}{h}}\rho\left(u,w;v\right)\mathrm{d}u\right\} ^{2}g\left(hw+s\left(v\right)\right)\mathrm{d}w-g\left(s\left(v\right)\right)\int\overline{\rho}\left(w;v\right)^{2}\mathrm{d}w\right|\nonumber \\
\leq & \left|\int_{\frac{\underline{b}-s\left(v\right)}{h}}^{\frac{\overline{b}-s\left(v\right)}{h}}\left\{ \int_{\frac{\underline{v}-v}{h}}^{\frac{\overline{v}-v}{h}}\rho\left(u,w;v\right)\mathrm{d}u\right\} ^{2}\left\{ g\left(hw+s\left(v\right)\right)-g\left(s\left(v\right)\right)\right\} \mathrm{d}w\right|\nonumber \\
 & +\left|\int_{\frac{\underline{b}-s\left(v\right)}{h}}^{\frac{\overline{b}-s\left(v\right)}{h}}\left(\left\{ \int_{\frac{\underline{v}-v}{h}}^{\frac{\overline{v}-v}{h}}\rho\left(u,w;v\right)\mathrm{d}u\right\} ^{2}-\overline{\rho}\left(w;v\right)^{2}\right)g\left(s\left(v\right)\right)\mathrm{d}w\right|\nonumber \\
 & +g\left(s\left(v\right)\right)\left|\int_{\frac{\underline{b}-s\left(v\right)}{h}}^{\frac{\overline{b}-s\left(v\right)}{h}}\overline{\rho}\left(w;v\right)^{2}\mathrm{d}w-\int\overline{\rho}\left(w;v\right)^{2}\mathrm{d}w\right|.\label{eq:proof of theorem 1, triangle bound}
\end{align}
Equation (\ref{eq:rho_bar sup bound}) implies that the last term
of the right-hand side of (\ref{eq:proof of theorem 1, triangle bound})
is zero for all $v\in I$, when $h$ is sufficiently small. Now (\ref{eq:theorem 1 domination condition})
implies
\begin{align}
 & \underset{v\in I}{\mathrm{sup}}\left|\int_{\frac{\underline{b}-s\left(v\right)}{h}}^{\frac{\overline{b}-s\left(v\right)}{h}}\left\{ \int_{\frac{\underline{v}-v}{h}}^{\frac{\overline{v}-v}{h}}\rho\left(u,w;v\right)\mathrm{d}u\right\} ^{2}\left\{ g\left(hw+s\left(v\right)\right)-g\left(s\left(v\right)\right)\right\} \mathrm{d}w\right|\nonumber \\
\apprle & \underset{v\in I}{\mathrm{sup}}\int\mathbbm{1}\left(\left|w\right|\leq1+\overline{C}_{s'}\right)\left|g\left(hw+s\left(v\right)\right)-g\left(s\left(v\right)\right)\right|\mathrm{d}w\nonumber \\
\apprle & \mathrm{sup}\left\{ \left|g\left(b'\right)-g\left(b\right)\right|:b\in\left[s\left(v_{l}\right),s\left(v_{u}\right)\right],\,\left|b'-b\right|\leq\left(1+\overline{C}_{s'}\right)h\right\} \nonumber \\
= & o\left(1\right),\label{eq:proof of theorem 1, sup bound 1}
\end{align}
where the inequalities hold when $h$ is sufficiently small and the
equality follows from the fact that $g$ is uniformly continuous on
any inner closed subinterval of $\left[\underline{b},\overline{b}\right]$. 

By (\ref{eq:theorem 1 domination condition}) and (\ref{eq:rho_bar sup bound}),
\begin{align}
 & \underset{v\in I}{\mathrm{sup}}\left|\int_{\frac{\underline{b}-s\left(v\right)}{h}}^{\frac{\overline{b}-s\left(v\right)}{h}}\left(\left\{ \int_{\frac{\underline{v}-v}{h}}^{\frac{\overline{v}-v}{h}}\rho\left(u,w;v\right)\mathrm{d}u\right\} ^{2}-\overline{\rho}\left(w;v\right)^{2}\right)\mathrm{d}w\right|\nonumber \\
\apprle & \underset{v\in I}{\mathrm{sup}}\int_{\frac{\underline{b}-s\left(v\right)}{h}}^{\frac{\overline{b}-s\left(v\right)}{h}}\mathbbm{1}\left(\left|w\right|\leq1+\overline{C}_{s'}\right)\left|\int_{\frac{\underline{v}-v}{h}}^{\frac{\overline{v}-v}{h}}\rho\left(u,w;v\right)\mathrm{d}u-\overline{\rho}\left(w;v\right)\right|\mathrm{d}w.\label{eq:proof of theorem 1, sup bound 2}
\end{align}
It follows from the uniform continuity of $\psi$, which is implied
by GPV Proposition 1 and Lemma A1, that 
\[
\underset{w\in\mathbb{R}}{\mathrm{sup}}\,\underset{v\in I}{\mathrm{sup}}\left|\int_{\frac{\underline{v}-v}{h}}^{\frac{\overline{v}-v}{h}}\rho\left(u,w;v\right)\mathrm{d}u-\psi\left(v\right)\int_{\frac{\underline{v}-v}{h}}^{\frac{\overline{v}-v}{h}}K_{f}'\left(u\right)K_{g}\left(w-\frac{s\left(hu+v\right)-s\left(v\right)}{h}\right)\mathrm{d}u\right|=o\left(1\right).
\]
By the mean value theorem, since $K_{f}'$ and $K_{g}$ are both supported
on $\left[-1,1\right]$ and bounded, 
\begin{align}
 & \underset{w\in\mathbb{R}}{\mathrm{sup}}\,\underset{v\in I}{\mathrm{sup}}\left|\psi\left(v\right)\int_{\frac{\underline{v}-v}{h}}^{\frac{\overline{v}-v}{h}}K_{f}'\left(u\right)K_{g}\left(w-\frac{s\left(hu+v\right)-s\left(v\right)}{h}\right)\mathrm{d}u-\overline{\rho}\left(w;v\right)\right|\nonumber \\
\apprle & \underset{v\in I}{\mathrm{sup}}\int\left|K_{f}'\left(u\right)u\right|\left|\frac{s\left(hu+v\right)-s\left(v\right)}{hu}-s'\left(v\right)\right|\mathrm{d}u\nonumber \\
\apprle & \mathrm{sup}\left\{ \left|s'\left(v'\right)-s'\left(v\right)\right|:v\in I,\left|v-v'\right|\leq h\right\} \nonumber \\
= & o\left(1\right),\label{eq:proof of theorem 1, sup bound 3}
\end{align}
where the inequalities hold when $h$ is sufficiently small and the
equality holds since $s'$ is uniformly continuous. Now it follows
that 
\[
\underset{w\in\mathbb{R}}{\mathrm{sup}}\,\underset{v\in I}{\mathrm{sup}}\left|\int_{\frac{\underline{v}-v}{h}}^{\frac{\overline{v}-v}{h}}\rho\left(u,w;v\right)\mathrm{d}u-\overline{\rho}\left(w;v\right)\right|=o\left(1\right).
\]
It follows from the above result and (\ref{eq:proof of theorem 1, sup bound 2})
that
\[
\underset{v\in I}{\mathrm{sup}}\left|\int_{\frac{\underline{b}-s\left(v\right)}{h}}^{\frac{\overline{b}-s\left(v\right)}{h}}\left(\left\{ \int_{\frac{\underline{v}-v}{h}}^{\frac{\overline{v}-v}{h}}\rho\left(u,w;v\right)\mathrm{d}u\right\} ^{2}-\overline{\rho}\left(w;v\right)^{2}\right)\mathrm{d}w\right|=o\left(1\right).
\]
Then (\ref{eq:Em_2^2 as integral}) follows from the above result,
(\ref{eq:proof of theorem 1, triangle bound}) and (\ref{eq:proof of theorem 1, sup bound 1}).

We have
\begin{align}
 & \mathrm{E}\left[h^{3}\left(\mathcal{M}_{2}^{\ddagger}\left(B_{11};v\right)-\mu_{\mathcal{M}^{\ddagger}}\left(v\right)\right)^{2}\right]\nonumber \\
= & \mathrm{E}\left[h^{3}\mathcal{M}_{2}^{\ddagger}\left(B_{11};v\right)^{2}\right]-h^{3}\mu_{\mathcal{M}_{2}^{\ddagger}}\left(v\right)^{2}\nonumber \\
= & \frac{F\left(v\right)^{2}f\left(v\right)^{2}}{g\left(s\left(v\right)\right)^{3}}\int\left\{ \int K_{f}'\left(u\right)K_{g}\left(w-s'\left(v\right)u\right)\mathrm{d}u\right\} ^{2}\mathrm{d}w+o\left(1\right),\label{eq:Em_2^2}
\end{align}
where the remainder term is uniform in $v\in I$.

Lastly, we verify Lyapunov's condition for each fixed $v\in I$. It
is clear from the definition of $\sigma\left(v\right)$ (see (\ref{eq:sigma_L definition}))
and (\ref{eq:Em_2^2}) that
\begin{equation}
\sigma\left(v\right)=\frac{1}{N^{\nicefrac{1}{2}}\left(N-1\right)}\left\{ \frac{F\left(v\right)^{2}f\left(v\right)^{2}}{g\left(s\left(v\right)\right)^{3}}\int\left\{ \int K_{f}'\left(u\right)K_{g}\left(w-s'\left(v\right)u\right)\mathrm{d}u\right\} ^{2}\mathrm{d}w\right\} ^{\nicefrac{1}{2}}+o\left(1\right).\label{eq:s_L limit}
\end{equation}
By Loève's $c_{r}$ inequality, 
\begin{eqnarray}
\sum_{i,l}\mathrm{E}\left[\left|\frac{U_{il}\left(v\right)}{\sigma\left(v\right)}\right|^{3}\right] & = & \sigma\left(v\right)^{-3}\left(N-1\right)^{-3}\left(N\cdot L\right)^{-\nicefrac{1}{2}}\mathrm{E}\left[h^{\nicefrac{9}{2}}\left|\left(\mathcal{M}_{2}^{\ddagger}\left(B_{11};v\right)-\mu_{\mathcal{M}^{\ddagger}}\left(v\right)\right)\right|^{3}\right]\nonumber \\
 & \apprle & \sigma\left(v\right)^{-3}\left(N\cdot L\right)^{-\nicefrac{1}{2}}\left(h^{\nicefrac{9}{2}}\mathrm{E}\left[\left|\mathcal{M}_{2}^{\ddagger}\left(B_{11};v\right)\right|^{3}\right]+h^{\nicefrac{9}{2}}\left|\mu_{\mathcal{M}^{\ddagger}}\left(v\right)\right|^{3}\right).\nonumber \\
\label{eq:Liapunov's condition sum}
\end{eqnarray}
It follows from the $c_{r}$ inequality and change of variables that
\[
\mathrm{E}\left[h^{\nicefrac{9}{2}}\left|\mathcal{M}_{2}^{\ddagger}\left(B_{11};v\right)\right|^{3}\right]\apprle h^{-\nicefrac{1}{2}}\int_{\frac{\underline{b}-s\left(v\right)}{h}}^{\frac{\overline{b}-s\left(v\right)}{h}}\left|\int_{\frac{\underline{v}-v}{h}}^{\frac{\overline{v}-v}{h}}\rho\left(u,w;v\right)\mathrm{d}u\right|^{3}g\left(hw+s\left(v\right)\right)\mathrm{d}w.
\]
Now it is clear from (\ref{eq:theorem 1 domination condition}) that
\begin{equation}
\underset{v\in I}{\mathrm{sup}}\,\mathrm{E}\left[h^{\nicefrac{9}{2}}\left|\mathcal{M}_{2}^{\ddagger}\left(B_{11};v\right)\right|^{3}\right]=O\left(h^{-\nicefrac{1}{2}}\right).\label{eq:E=00005B|M_2_ddagger|^3=00005D rate}
\end{equation}
It now follows from the above result, (\ref{eq:miu_ddagger sup rate}),
(\ref{eq:s_L limit}) and (\ref{eq:Liapunov's condition sum}) that
\begin{equation}
\sum_{i,l}\mathrm{E}\left[\left|\frac{U_{il}\left(v\right)}{\sigma\left(v\right)}\right|^{3}\right]\downarrow0,\textrm{ as \ensuremath{L\uparrow\infty}}.\label{eq:Liapunov's condition}
\end{equation}

Hence, by Lyapunov's central limit theorem, 
\[
\sum_{i,l}\frac{U_{il}\left(v\right)}{\sigma\left(v\right)}\rightarrow_{d}\mathrm{N}\left(0,1\right),\textrm{ as \ensuremath{L\uparrow\infty.}}
\]
The conclusion follows from the above result, (\ref{eq:f_hat - f U's}),
(\ref{eq:s_L limit}) and Slutsky's lemma. \end{proof}

\begin{proof}[Proof of Theorem \ref{thm:variance estimator}]Write
\[
N\left(N-1\right)^{2}\widehat{\mathrm{V}}_{GPV}\left(v\right)=\mathit{\Delta}_{1}^{\dagger}\left(v\right)+\mathit{\Delta}_{2}^{\dagger}\left(v\right)+\mathit{\Delta}_{3}^{\dagger}\left(v\right),
\]
where 
\begin{eqnarray*}
\mathit{\Delta}_{1}^{\dagger}\left(v\right) & \coloneqq & \frac{1}{\left(N\cdot L\right)_{3}}\underset{\left(3\right)}{\sum}\frac{1}{h^{3}}\mathbb{T}_{jk}K_{f}'\left(\frac{\widehat{V}_{jk}-v}{h}\right)\frac{G\left(B_{jk}\right)}{g\left(B_{jk}\right)^{2}}K_{g}\left(\frac{B_{il}-B_{jk}}{h}\right)\\
 &  & \times\mathbb{T}_{j'k'}K_{f}'\left(\frac{\widehat{V}_{j'k'}-v}{h}\right)\frac{G\left(B_{j'k'}\right)}{g\left(B_{j'k'}\right)^{2}}K_{g}\left(\frac{B_{il}-B_{j'k'}}{h}\right),
\end{eqnarray*}
\begin{eqnarray*}
\mathit{\Delta}_{2}^{\dagger}\left(v\right) & \coloneqq & \frac{2}{\left(N\cdot L\right)_{3}}\underset{\left(3\right)}{\sum}\frac{1}{h^{3}}\mathbb{T}_{jk}K_{f}'\left(\frac{\widehat{V}_{jk}-v}{h}\right)\frac{G\left(B_{jk}\right)}{g\left(B_{jk}\right)^{2}}K_{g}\left(\frac{B_{il}-B_{jk}}{h}\right)\\
 &  & \times\mathbb{T}_{j'k'}K_{f}'\left(\frac{\widehat{V}_{j'k'}-v}{h}\right)\left(\frac{\widehat{G}\left(B_{j'k'}\right)}{\widehat{g}\left(B_{j'k'}\right)^{2}}-\frac{G\left(B_{j'k'}\right)}{g\left(B_{j'k'}\right)^{2}}\right)K_{g}\left(\frac{B_{il}-B_{j'k'}}{h}\right)
\end{eqnarray*}
and
\begin{eqnarray*}
\mathit{\Delta}_{3}^{\dagger}\left(v\right) & \coloneqq & \frac{1}{\left(N\cdot L\right)_{3}}\underset{\left(3\right)}{\sum}\frac{1}{h^{3}}\mathbb{T}_{jk}K_{f}'\left(\frac{\widehat{V}_{jk}-v}{h}\right)\left(\frac{\widehat{G}\left(B_{jk}\right)}{\widehat{g}\left(B_{jk}\right)^{2}}-\frac{G\left(B_{jk}\right)}{g\left(B_{jk}\right)^{2}}\right)K_{g}\left(\frac{B_{il}-B_{jk}}{h}\right)\\
 &  & \times\mathbb{T}_{j'k'}K_{f}'\left(\frac{\widehat{V}_{j'k'}-v}{h}\right)\left(\frac{\widehat{G}\left(B_{j'k'}\right)}{\widehat{g}\left(B_{j'k'}\right)^{2}}-\frac{G\left(B_{j'k'}\right)}{g\left(B_{j'k'}\right)^{2}}\right)K_{g}\left(\frac{B_{il}-B_{j'k'}}{h}\right).
\end{eqnarray*}

By standard arguments (see, e.g., \citealp*[Lemma 1]{Marmer_Shneyerov_Quantile_Auctions}),
\begin{gather}
\underset{b\in\left[\underline{b},\overline{b}\right]}{\mathrm{sup}}\left|\widehat{G}\left(b\right)-G\left(b\right)\right|=O_{p}\left(\left(\frac{\mathrm{log}\left(L\right)}{L}\right)^{\nicefrac{1}{2}}\right)\nonumber \\
\underset{b\in\left[\underline{b}+h,\overline{b}-h\right]}{\mathrm{sup}}\left|\widehat{g}\left(b\right)-g\left(b\right)\right|=O_{p}\left(\left(\frac{\mathrm{log}\left(L\right)}{Lh}\right)^{\nicefrac{1}{2}}+h^{1+R}\right).\label{eq:uniform rates}
\end{gather}
Then by the triangle inequality, (\ref{eq:density of bids bounded  from 0}),
(\ref{eq:uniform rates}) and the identity 
\[
\frac{a}{b}=\frac{a}{c}-\frac{a\left(b-c\right)}{c^{2}}+\frac{a\left(b-c\right)^{2}}{bc^{2}},
\]
\begin{eqnarray}
\left|\mathit{\Delta}_{2}^{\dagger}\left(v\right)\right| & \apprle & \left\{ \underset{j',k'}{\mathrm{max}}\,\mathbb{T}_{j'k'}\left|\frac{\widehat{G}\left(B_{j'k'}\right)}{\widehat{g}\left(B_{j'k'}\right)^{2}}-\frac{G\left(B_{j'k'}\right)}{g\left(B_{j'k'}\right)^{2}}\right|\right\} \left\{ \frac{1}{\left(N\cdot L\right)_{3}}\underset{\left(3\right)}{\sum}\frac{1}{h^{3}}\mathbb{T}_{jk}\left|K_{f}'\left(\frac{\widehat{V}_{jk}-v}{h}\right)\right|\right.\nonumber \\
 &  & \left.\times\left|K_{g}\left(\frac{B_{il}-B_{jk}}{h}\right)\right|\mathbb{T}_{j'k'}\left|K_{f}'\left(\frac{\widehat{V}_{j'k'}-v}{h}\right)\right|\left|K_{g}\left(\frac{B_{il}-B_{j'k'}}{h}\right)\right|\right\} \nonumber \\
 & \apprle & O_{p}\left(\left(\frac{\mathrm{log}\left(L\right)}{Lh}\right)^{\nicefrac{1}{2}}+h^{1+R}\right)\left\{ \frac{1}{\left(N\cdot L\right)_{3}}\underset{\left(3\right)}{\sum}h^{-3}\mathbb{T}_{jk}\mathbbm{1}\left(\left|\widehat{V}_{jk}-v\right|\leq h\right)\right.\nonumber \\
 &  & \left.\times\left|K_{g}\left(\frac{B_{il}-B_{jk}}{h}\right)\right|\mathbb{T}_{j'k'}\mathbbm{1}\left(\left|\widehat{V}_{j'k'}-v\right|\leq h\right)\left|K_{g}\left(\frac{B_{il}-B_{j'k'}}{h}\right)\right|\right\} .\label{eq:I_2 decomposition}
\end{eqnarray}
Since it is shown in the proof of Lemma \ref{lem:lemma 2} that
\begin{equation}
\underset{i,l}{\mathrm{max}}\,\mathbb{T}_{il}\left|\widehat{V}_{il}-V_{il}\right|=O_{p}\left(\left(\frac{\mathrm{log}\left(L\right)}{Lh}\right)^{\nicefrac{1}{2}}+h^{1+R}\right)=o_{p}\left(h\right),\label{eq:max difference v_hat v feasible}
\end{equation}
by the triangle inequality, 
\begin{align}
 & \frac{1}{\left(N\cdot L\right)_{3}}\underset{\left(3\right)}{\sum}h^{-3}\mathbb{T}_{jk}\mathbbm{1}\left(\left|\widehat{V}_{jk}-v\right|\leq h\right)\left|K_{g}\left(\frac{B_{il}-B_{jk}}{h}\right)\right|\mathbb{T}_{j'k'}\mathbbm{1}\left(\left|\widehat{V}_{j'k'}-v\right|\leq h\right)\nonumber \\
 & \hspace{6em}\times\left|K_{g}\left(\frac{B_{il}-B_{j'k'}}{h}\right)\right|\nonumber \\
\leq & \frac{1}{\left(N\cdot L\right)_{3}}\underset{\left(3\right)}{\sum}h^{-3}\mathbbm{1}\left(\left|\xi\left(B_{jk}\right)-v\right|\leq2h\right)\left|K_{g}\left(\frac{B_{il}-B_{jk}}{h}\right)\right|\mathbbm{1}\left(\left|\xi\left(B_{j'k'}\right)-v\right|\leq2h\right)\nonumber \\
 & \hspace{6em}\times\left|K_{g}\left(\frac{B_{il}-B_{j'k'}}{h}\right)\right|\nonumber \\
\eqqcolon & \frac{1}{\left(N\cdot L\right)_{3}}\underset{\left(3\right)}{\sum}\mathcal{K}\left(B_{il},B_{jk},B_{j'k'};v\right),\label{eq:K U-stat upper bound}
\end{align}
where the inequality holds w.p.a.1. 

Define 
\begin{gather}
\mathcal{K}_{1}^{\left(1\right)}\left(b;v\right)\coloneqq\int\int\mathcal{K}\left(b,b',b'';v\right)\mathrm{d}G\left(b'\right)\mathrm{d}G\left(b''\right),\nonumber \\
\mathcal{K}_{2}^{\left(1\right)}\left(b;v\right)\coloneqq\int\int\mathcal{K}\left(b',b,b'';v\right)\mathrm{d}G\left(b'\right)\mathrm{d}G\left(b''\right),\nonumber \\
\mathcal{K}_{3}^{\left(1\right)}\left(b;v\right)\coloneqq\int\int\mathcal{K}\left(b',b'',b;v\right)\mathrm{d}G\left(b'\right)\mathrm{d}G\left(b''\right),\label{eq:K Hoeffding decomposition def 1}
\end{gather}
\begin{gather}
\mathcal{K}_{1}^{\left(2\right)}\left(b,b';v\right)\coloneqq\int\mathcal{K}\left(b,b',b'';v\right)\mathrm{d}G\left(b''\right),\nonumber \\
\mathcal{K}_{2}^{\left(2\right)}\left(b,b';v\right)\coloneqq\int\mathcal{K}\left(b,b'',b';v\right)\mathrm{d}G\left(b''\right),\nonumber \\
\mathcal{K}_{3}^{\left(2\right)}\left(b,b';v\right)\coloneqq\int\mathcal{K}\left(b'',b,b';v\right)\mathrm{d}G\left(b''\right),\label{eq:K Hoeffding decomposition def 2}
\end{gather}
and 
\begin{equation}
\mu_{\mathcal{K}}\left(v\right)\coloneqq\int\int\int\mathcal{K}\left(b,b',b'';v\right)\mathrm{d}G\left(b\right)\mathrm{d}G\left(b'\right)\mathrm{d}G\left(b''\right).\label{eq:K Hoeffding decomposition def 3}
\end{equation}

The Hoeffding decomposition yields 
\begin{align}
 & \frac{1}{\left(N\cdot L\right)_{3}}\underset{\left(3\right)}{\sum}\mathcal{K}\left(B_{il},B_{jk},B_{j'k'};v\right)\nonumber \\
= & \mu_{\mathcal{K}}\left(v\right)+\frac{1}{N\cdot L}\sum_{i,l}\left(\mathcal{K}_{1}^{\left(1\right)}\left(B_{il};v\right)-\mu_{\mathcal{K}}\left(v\right)\right)+\frac{1}{N\cdot L}\sum_{i,l}\left(\mathcal{K}_{2}^{\left(1\right)}\left(B_{il};v\right)-\mu_{\mathcal{K}}\left(v\right)\right)\nonumber \\
 & +\frac{1}{N\cdot L}\sum_{i,l}\left(\mathcal{K}_{3}^{\left(1\right)}\left(B_{il};v\right)-\mu_{\mathcal{K}}\left(v\right)\right)+\Upsilon_{\mathcal{K}}^{1}\left(v\right)+\Upsilon_{\mathcal{K}}^{2}\left(v\right)+\Upsilon_{\mathcal{K}}^{3}\left(v\right)+\Psi_{\mathcal{K}}\left(v\right),\label{eq:K Hoeffding decomposition}
\end{align}
where $\Upsilon_{\mathcal{K}}^{1}\left(v\right)$, $\Upsilon_{\mathcal{K}}^{2}\left(v\right)$
and $\Upsilon_{\mathcal{K}}^{3}\left(v\right)$ are degenerate U-statistics
of order two and $\Psi_{\mathcal{K}}\left(v\right)$ is a degenerate
U-statistic of order three:
\begin{gather}
\Upsilon_{\mathcal{K}}^{1}\left(v\right)\coloneqq\frac{1}{\left(N\cdot L\right)_{2}}\sum_{\left(2\right)}\left\{ \mathcal{K}_{1}^{\left(2\right)}\left(B_{il},B_{jk};v\right)-\mathcal{K}_{1}^{\left(1\right)}\left(B_{il};v\right)-\mathcal{K}_{2}^{\left(1\right)}\left(B_{jk};v\right)+\mu_{\mathcal{K}}\left(v\right)\right\} ,\nonumber \\
\Upsilon_{\mathcal{K}}^{2}\left(v\right)\coloneqq\frac{1}{\left(N\cdot L\right)_{2}}\sum_{\left(2\right)}\left\{ \mathcal{K}_{2}^{\left(2\right)}\left(B_{il},B_{jk};v\right)-\mathcal{K}_{1}^{\left(1\right)}\left(B_{il};v\right)-\mathcal{K}_{3}^{\left(1\right)}\left(B_{jk};v\right)+\mu_{\mathcal{K}}\left(v\right)\right\} ,\nonumber \\
\Upsilon_{\mathcal{K}}^{3}\left(v\right)\coloneqq\frac{1}{\left(N\cdot L\right)_{2}}\sum_{\left(2\right)}\left\{ \mathcal{K}_{3}^{\left(2\right)}\left(B_{il},B_{jk};v\right)-\mathcal{K}_{2}^{\left(1\right)}\left(B_{il};v\right)-\mathcal{K}_{3}^{\left(1\right)}\left(B_{jk};v\right)+\mu_{\mathcal{K}}\left(v\right)\right\} ,\label{eq:K Hoeffding decomposition def 4}
\end{gather}
and
\begin{eqnarray}
\Psi_{\mathcal{K}}\left(v\right) & \coloneqq & \frac{1}{\left(N\cdot L\right)_{3}}\sum_{\left(3\right)}\left\{ \mathcal{K}\left(B_{il},B_{jk},B_{j'k'};v\right)-\mathcal{K}_{1}^{\left(2\right)}\left(B_{il},B_{jk};v\right)\right.\nonumber \\
 &  & -\mathcal{K}_{2}^{\left(2\right)}\left(B_{il},B_{j'k'};v\right)-\mathcal{K}_{3}^{\left(2\right)}\left(B_{jk},B_{j'k'};v\right)\nonumber \\
 &  & \left.+\mathcal{K}_{1}^{\left(1\right)}\left(B_{il};v\right)+\mathcal{K}_{2}^{\left(1\right)}\left(B_{jk};v\right)+\mathcal{K}_{3}^{\left(1\right)}\left(B_{j'k'};v\right)-\mu_{\mathcal{K}}\left(v\right)\right\} .\label{eq:K Hoeffding decomposition def 5}
\end{eqnarray}

By change of variables, the expression for $\mu_{\mathcal{K}}\left(v\right)$
is given by
\begin{align*}
 & \int_{\frac{\underline{b}-s\left(v\right)}{h}}^{\frac{\overline{b}-s\left(v\right)}{h}}\left\{ \int_{\frac{\underline{v}-v}{h}}^{\frac{\overline{v}-v}{h}}\mathbbm{1}\left(\left|w\right|\leq2\right)\left|K_{g}\left(u-\frac{s\left(hw+v\right)-s\left(v\right)}{h}\right)\right|s'\left(hw+v\right)g\left(s\left(hw+v\right)\right)\mathrm{d}w\right\} ^{2}\\
 & \hspace{6em}\times g\left(hu+s\left(v\right)\right)\mathrm{d}u\\
 & \leq\overline{C}_{K_{g}}^{2}\int_{\frac{\underline{b}-s\left(v\right)}{h}}^{\frac{\overline{b}-s\left(v\right)}{h}}\left\{ \int_{\frac{\underline{v}-v}{h}}^{\frac{\overline{v}-v}{h}}\mathbbm{1}\left(\left|w\right|\leq2\right)\mathbbm{1}\left(\left|u\right|\leq1+2\overline{C}_{s'}\right)s'\left(hw+v\right)g\left(s\left(hw+v\right)\right)\mathrm{d}w\right\} ^{2}\\
 & \times g\left(hu+s\left(v\right)\right)\mathrm{d}u,
\end{align*}
and therefore 
\begin{equation}
\underset{v\in I}{\mathrm{sup}}\left|\mu_{\mathcal{K}}\left(v\right)\right|=O\left(1\right).\label{eq:miu_K sup bound}
\end{equation}

Now we derive a bound for the order of 
\[
\underset{v\in I}{\mathrm{sup}}\left|\frac{1}{\left(N\cdot L\right)_{3}}\underset{\left(3\right)}{\sum}\mathcal{K}\left(B_{il},B_{jk},B_{j'k'};v\right)-\mu_{\mathcal{K}}\left(v\right)\right|
\]
by deriving a bound for the order of each of the terms in the Hoeffding
decomposition (\ref{eq:K Hoeffding decomposition}).

Since $\mathbbm{1}\left(\left|\cdot\right|\leq2\right)$ is the difference
of two non-decreasing functions, following standard arguments (see
the proof of Lemma \ref{lem:lemma 2} in the Supplement), we
can easily check that the function class 
\[
\left\{ \left(b,b'\right)\mapsto h^{-\nicefrac{3}{2}}\mathbbm{1}\left(\left|\xi\left(b\right)-v\right|\leq2h\right)\left|K_{g}\left(\frac{b-b'}{h}\right)\right|:v\in I\right\} 
\]
is (uniformly) VC-type with respect to the constant envelope $h^{-\nicefrac{3}{2}}\overline{C}_{K_{g}}$
by using \citet[Lemma 2.6.16]{van1996weak}, \citet[Lemma 9.9 (vi, vii)]{kosorok2007introduction},
\citet[Theorem 3.6.9]{gine2015mathematical} and \citet[Lemma 16]{nolan1987u}.
Now $\mathscr{K}\coloneqq\left\{ \mathcal{K}\left(\cdot,\cdot,\cdot;v\right):v\in I\right\} $
can be viewed as a subset of a pointwise product of two VC-type classes.
\citet[Corollary A.1]{chernozhukov2014gaussian} gives that $\mathscr{K}$
is also (uniformly) VC-type with respect to the constant envelope
$h^{-3}\overline{C}_{K_{g}}^{2}$. The CK inequality yields 
\begin{equation}
\mathrm{E}\left[\underset{v\in I}{\mathrm{sup}}\left|\Upsilon_{\mathcal{K}}^{k}\left(v\right)\right|\right]\apprle\left(Lh^{3}\right)^{-1},\textrm{ for }k=1,2,3\textrm{ and }\mathrm{E}\left[\underset{v\in I}{\mathrm{sup}}\left|\Psi_{\mathcal{K}}\left(v\right)\right|\right]\apprle L^{-\nicefrac{3}{2}}h^{-3}.\label{eq:K U-processes expectation bounds}
\end{equation}

Since $\mathscr{K}$ is VC-type, it follows from \citet[Lemma 5.4]{Chen_Kato_U_Process}
that the function classes $\left\{ \mathcal{K}_{k}^{\left(1\right)}\left(\cdot;v\right):v\in I\right\} $,
for $k=1,2,3$ are all VC-type with respect to the constant envelope
$h^{-3}\overline{C}_{K_{g}}^{2}$. By Jensen's inequality, 
\[
\underset{v\in I}{\mathrm{sup}}\,\mathrm{E}\left[\mathcal{K}_{k}^{\left(1\right)}\left(B_{11};v\right)^{2}\right]\leq\underset{v\in I}{\mathrm{sup}}\int\int\int\mathcal{K}\left(b,b',b'';v\right)^{2}\mathrm{d}G\left(b\right)\mathrm{d}G\left(b'\right)\mathrm{d}G\left(b''\right),\textrm{ for }k=1,2,3.
\]
By change of variables, 
\begin{align*}
 & \underset{v\in I}{\mathrm{sup}}\int\int\int\mathcal{K}\left(b,b',b'';v\right)^{2}\mathrm{d}G\left(b\right)\mathrm{d}G\left(b'\right)\mathrm{d}G\left(b''\right)\\
\leq & h^{-3}\overline{C}_{K_{g}}^{2}\left\{ \underset{v\in I}{\mathrm{sup}}\int_{\frac{\underline{b}-s\left(v\right)}{h}}^{\frac{\overline{b}-s\left(v\right)}{h}}\left\{ \int_{\frac{\underline{v}-v}{h}}^{\frac{\overline{v}-v}{h}}\mathbbm{1}\left(\left|w\right|\leq2\right)\mathbbm{1}\left(\left|u\right|\leq1+2\overline{C}_{s'}\right)s'\left(hw+v\right)g\left(s\left(hw+v\right)\right)\mathrm{d}w\right\} ^{2}\right.\\
 & \left.\times g\left(hu+s\left(v\right)\right)\mathrm{d}u\right\} 
\end{align*}
and hence,
\begin{equation}
\underset{v\in I}{\mathrm{sup}}\,\mathrm{E}\left[\mathcal{K}_{k}^{\left(1\right)}\left(B_{11};v\right)^{2}\right]\apprle h^{-3},\textrm{ for }k=1,2,3.\label{eq:sup E=00005BK^(1)=00005D bound}
\end{equation}
Now we apply the CCK inequality with $\sigma^{2}$ being the left-hand
side of (\ref{eq:sup E=00005BK^(1)=00005D bound}) and $F$ being
$h^{-3}\overline{C}_{K_{g}}^{2}$. It follows that
\begin{eqnarray}
 &  & \mathrm{E}\left[\underset{v\in I}{\mathrm{sup}}\left|\frac{1}{N\cdot L}\sum_{i,l}\left(\mathcal{K}_{k}^{\left(1\right)}\left(B_{il};v\right)-\mu_{\mathcal{K}}\left(v\right)\right)\right|\right]\nonumber \\
 & \leq & C_{1}\left\{ \left(Lh^{3}\right)^{-\nicefrac{1}{2}}\mathrm{log}\left(C_{2}L\right)^{\nicefrac{1}{2}}+\left(Lh^{3}\right)^{-1}\mathrm{log}\left(C_{2}L\right)\right\} \nonumber \\
 & = & O\left(\left(\frac{\mathrm{log}\left(L\right)}{Lh^{3}}\right)^{\nicefrac{1}{2}}\right),\textrm{ for \ensuremath{k=1,2,3.}}\label{eq:K empirical process expectation bound}
\end{eqnarray}

Now, 
\begin{equation}
\underset{v\in I}{\mathrm{sup}}\,\frac{1}{\left(N\cdot L\right)_{3}}\underset{\left(3\right)}{\sum}\mathcal{K}\left(B_{il},B_{jk},B_{j'k'};v\right)=O_{p}\left(1\right)\label{eq:K U-process supremum rate}
\end{equation}
follows from (\ref{eq:K Hoeffding decomposition}), (\ref{eq:miu_K sup bound}),
(\ref{eq:K empirical process expectation bound}) and (\ref{eq:K U-processes expectation bounds})
and 
\[
\underset{v\in I}{\mathrm{sup}}\left|\mathit{\Delta}_{2}^{\dagger}\left(v\right)\right|=O_{p}\left(\left(\frac{\mathrm{log}\left(L\right)}{Lh}\right)^{\nicefrac{1}{2}}+h^{1+R}\right)
\]
follows from (\ref{eq:K U-process supremum rate}) and (\ref{eq:I_2 decomposition}).
Similarly, we can show 
\[
\underset{v\in I}{\mathrm{sup}}\left|\mathit{\Delta}_{3}^{\dagger}\left(v\right)\right|=O_{p}\left(\frac{\mathrm{log}\left(L\right)}{Lh}+h^{2+2R}\right).
\]

Using mean-value expansion arguments, 
\begin{eqnarray*}
\mathit{\Delta}_{1}^{\dagger}\left(v\right) & = & \frac{1}{\left(N\cdot L\right)_{3}}\underset{\left(3\right)}{\sum}h^{-3}\mathbb{T}_{jk}K_{f}'\left(\frac{V_{jk}-v}{h}\right)\frac{G\left(B_{jk}\right)}{g\left(B_{jk}\right)^{2}}K_{g}\left(\frac{B_{il}-B_{jk}}{h}\right)\\
 &  & \times\mathbb{T}_{j'k'}K_{f}'\left(\frac{V_{j'k'}-v}{h}\right)\frac{G\left(B_{j'k'}\right)}{g\left(B_{j'k'}\right)^{2}}K_{g}\left(\frac{B_{il}-B_{j'k'}}{h}\right)\\
 &  & +\frac{2}{\left(N\cdot L\right)_{3}}\underset{\left(3\right)}{\sum}h^{-4}\mathbb{T}_{jk}K_{f}''\left(\frac{\dot{V}_{jk}-v}{h}\right)\left(\widehat{V}_{jk}-V_{jk}\right)\\
 &  & \times\frac{G\left(B_{jk}\right)}{g\left(B_{jk}\right)^{2}}K_{g}\left(\frac{B_{il}-B_{jk}}{h}\right)\mathbb{T}_{j'k'}K_{f}'\left(\frac{V_{j'k'}-v}{h}\right)\frac{G\left(B_{j'k'}\right)}{g\left(B_{j'k'}\right)^{2}}K_{g}\left(\frac{B_{il}-B_{j'k'}}{h}\right)\\
 &  & +\frac{1}{\left(N\cdot L\right)_{3}}\underset{\left(3\right)}{\sum}h^{-5}\mathbb{T}_{jk}K_{f}''\left(\frac{\dot{V}_{jk}-v}{h}\right)\left(\widehat{V}_{jk}-V_{jk}\right)\frac{G\left(B_{jk}\right)}{g\left(B_{jk}\right)^{2}}K_{g}\left(\frac{B_{il}-B_{jk}}{h}\right)\mathbb{T}_{j'k'}\\
 &  & \times K_{f}''\left(\frac{\dot{V}_{j'k'}-v}{h}\right)\left(\widehat{V}_{j'k'}-V_{j'k'}\right)\frac{G\left(B_{j'k'}\right)}{g\left(B_{j'k'}\right)^{2}}K_{g}\left(\frac{B_{il}-B_{j'k'}}{h}\right)\\
 & \eqqcolon & \mathit{\Delta}_{4}^{\dagger}\left(v\right)+2\cdot\mathit{\Delta}_{5}^{\dagger}\left(v\right)+\mathit{\Delta}_{6}^{\dagger}\left(v\right),
\end{eqnarray*}
where $\dot{V}_{jk}$ ($\dot{V}_{j'k'}$) is the mean value that lies
between $V_{jk}$ ($V_{j'k'}$) and $\widehat{V}_{jk}$ ($\widehat{V}_{j'k'}$).
Now by (\ref{eq:density of bids bounded  from 0}), the triangle inequality,
the fact $\underset{i,l}{\mathrm{max}}\,\mathbb{T}_{il}\left|\widehat{V}_{il}-V_{il}\right|=o_{p}\left(h\right)$
and the fact that $K_{f}'$ and $K_{f}''$ are both compactly supported
on $\left[-1,1\right]$, 
\begin{eqnarray*}
\left|\mathit{\Delta}_{5}^{\dagger}\left(v\right)\right| & \lesssim & h^{-1}\left\{ \underset{j,k}{\mathrm{max}}\,\mathbb{T}_{jk}\left|\widehat{V}_{jk}-V_{jk}\right|\right\} \left\{ \frac{1}{\left(N\cdot L\right)_{3}}\underset{\left(3\right)}{\sum}\mathcal{K}\left(B_{il},B_{jk},B_{j'k'};v\right)\right\} \\
 & = & O_{p}\left(\left(\frac{\mathrm{log}\left(L\right)}{Lh^{3}}\right)^{\nicefrac{1}{2}}+h^{R}\right),
\end{eqnarray*}
where the inequality holds w.p.a.1 and the equality is uniform in
$v\in I$, and also 
\begin{eqnarray*}
\left|\mathit{\Delta}_{6}^{\dagger}\left(v\right)\right| & \lesssim & h^{-2}\left\{ \underset{j,k}{\mathrm{max}}\,\mathbb{T}_{jk}\left|\widehat{V}_{jk}-V_{jk}\right|\right\} ^{2}\left\{ \frac{1}{\left(N\cdot L\right)_{3}}\underset{\left(3\right)}{\sum}\mathcal{K}\left(B_{il},B_{jk},B_{j'k'};v\right)\right\} \\
 & = & O_{p}\left(\frac{\mathrm{log}\left(L\right)}{Lh^{3}}+h^{2R}\right),
\end{eqnarray*}
where the inequality holds w.p.a.1 and the equality is uniform in
$v\in I$. 

Denote 
\[
\mathcal{H}\left(b,b',b'';v\right)\coloneqq\frac{1}{h^{3}}K_{f}'\left(\frac{\xi\left(b'\right)-v}{h}\right)K_{g}\left(\frac{b-b'}{h}\right)\frac{G\left(b'\right)}{g\left(b'\right)^{2}}K_{f}'\left(\frac{\xi\left(b''\right)-v}{h}\right)K_{g}\left(\frac{b-b''}{h}\right)\frac{G\left(b''\right)}{g\left(b''\right)^{2}}.
\]
Since $K_{f}'$ is compactly supported on $\left[-1,1\right]$, the
trimming is asymptotically negligible:
\[
\mathit{\Delta}_{4}^{\dagger}\left(v\right)=\frac{1}{\left(N\cdot L\right)_{3}}\underset{\left(3\right)}{\sum}\mathcal{H}\left(B_{il},B_{jk},B_{j'k'};v\right),\,\textrm{for all \ensuremath{v\in I,}}
\]
w.p.a.1.\footnote{Formally, we can show that the supremum (over $v\in I$) of the difference
between $\mathit{\Delta}_{4}^{\dagger}\left(v\right)$ and the same
term with $\mathbb{T}_{jk}$ ($\mathbb{T}_{j'k'}$) replaced by $\widetilde{\mathbb{T}}_{jk}$
($\widetilde{\mathbb{T}}_{j'k'}$) is equal to zero w.p.a.1. Then
since $K_{f}'$ is compactly supported on $\left[-1,1\right]$, it
is straightforward to see that the contribution of the trimming factors
$\widetilde{\mathbb{T}}_{jk}$ and $\widetilde{\mathbb{T}}_{j'k'}$
is negligible. } The Hoeffding decomposition yields
\begin{align}
 & \frac{1}{\left(N\cdot L\right)_{3}}\underset{\left(3\right)}{\sum}\mathcal{H}\left(B_{il},B_{jk},B_{j'k'};v\right)\nonumber \\
= & \mu_{\mathcal{\mathcal{H}}}\left(v\right)+\frac{1}{N\cdot L}\sum_{i,l}\left(\mathcal{\mathcal{H}}_{1}^{\left(1\right)}\left(B_{il};v\right)-\mu_{\mathcal{\mathcal{H}}}\left(v\right)\right)+\frac{1}{N\cdot L}\sum_{i,l}\left(\mathcal{\mathcal{H}}_{2}^{\left(1\right)}\left(B_{il};v\right)-\mu_{\mathcal{\mathcal{H}}}\left(v\right)\right)\nonumber \\
 & +\frac{1}{N\cdot L}\sum_{i,l}\left(\mathcal{\mathcal{H}}_{3}^{\left(1\right)}\left(B_{il};v\right)-\mu_{\mathcal{\mathcal{H}}}\left(v\right)\right)+\Upsilon_{\mathcal{\mathcal{H}}}^{1}\left(v\right)+\Upsilon_{\mathcal{\mathcal{H}}}^{2}\left(v\right)+\Upsilon_{\mathcal{\mathcal{H}}}^{3}\left(v\right)+\Psi_{\mathcal{\mathcal{H}}}\left(v\right),\label{eq:H Hoeffding decomposition}
\end{align}
where the terms in the decomposition are defined by (\ref{eq:K Hoeffding decomposition def 1})
to (\ref{eq:K Hoeffding decomposition def 5}) with $\mathcal{K}$
replaced by $\mathcal{H}$. Note that we have $\mu_{\mathcal{\mathcal{H}}}\left(v\right)=N\left(N-1\right)^{2}\mathrm{V}_{\mathcal{M}}\left(v\right)$.
Also define $\mathscr{H}\coloneqq\left\{ \mathcal{H}\left(\cdot,\cdot,\cdot;v\right):v\in I\right\} $.
By the arguments used to show that $\mathscr{K}$ is (uniformly) VC-type,
we can show that $\mathscr{H}$ is (uniformly) VC-type with respect
to the constant envelope 
\begin{equation}
h^{-3}\left(\overline{C}_{D_{1}}+\overline{C}_{D_{2}}\right)^{2}\overline{C}_{K_{g}}^{2}\underline{C}_{g}^{-4}.\label{eq:H class envelope}
\end{equation}
Since $\mathscr{H}$ is VC-type, it again follows from \citet[Lemma 5.4]{Chen_Kato_U_Process}
that the function classes $\left\{ \mathcal{H}_{k}^{\left(1\right)}\left(\cdot;v\right):v\in I\right\} $,
for $k=1,2,3$ are all VC-type with respect to the constant envelope
(\ref{eq:H class envelope}).

Similarly, by Jensen's inequality and change of variables, for $k=1,2,3$,
\begin{eqnarray*}
\mathrm{E}\left[\mathcal{\mathcal{H}}_{k}^{\left(1\right)}\left(B_{11};v\right)^{2}\right] & \leq & \int\int\int\mathcal{H}\left(b,b',b'';v\right)^{2}\mathrm{d}G\left(b\right)\mathrm{d}G\left(b'\right)\mathrm{d}G\left(b''\right)\\
 & = & h^{-6}\int\left\{ \int_{\underline{b}}^{\overline{b}}K_{f}'\left(\frac{\xi\left(b'\right)-v}{h}\right)^{2}K_{g}\left(\frac{b-b'}{h}\right)^{2}\frac{G\left(b'\right)^{2}}{g\left(b'\right)^{3}}\mathrm{d}b'\right\} ^{2}g\left(b\right)\mathrm{d}b\\
 & \apprle & h^{-3}\int\left\{ \int\mathbbm{1}\left(\left|u\right|\leq1\right)\mathbbm{1}\left(\left|w\right|\leq1+\overline{C}_{s'}\right)\mathrm{d}u\right\} ^{2}\mathrm{d}w,
\end{eqnarray*}
for all $v\in I$. Therefore,
\begin{equation}
\underset{v\in I}{\mathrm{sup}}\,\mathrm{E}\left[\mathcal{\mathcal{H}}_{k}^{\left(1\right)}\left(B_{11};v\right)^{2}\right]\apprle h^{-3},\,k=1,2,3.\label{eq:sup E=00005BH^(1)=00005D bound}
\end{equation}
We apply the CCK inequality again with $\sigma^{2}$ being the left-hand
side of (\ref{eq:sup E=00005BH^(1)=00005D bound}) and $F$ being
(\ref{eq:H class envelope}) and also the CK inequality. It now follows
that the bounds (\ref{eq:K U-processes expectation bounds}) and (\ref{eq:K empirical process expectation bound})
also hold as we replace $\mathcal{K}$ by $\mathcal{H}$.\end{proof}

Let $\widetilde{f}$ denote the infeasible estimator that uses the
unobserved true valuations:
\[
\widetilde{f}\left(v\right)=\frac{1}{N\cdot L}\sum_{i,l}\frac{1}{h}K_{f}\left(\frac{V_{il}-v}{h}\right).
\]
Let 
\[
\widetilde{f}^{*}\left(v\right)\coloneqq\frac{1}{N\cdot L}\sum_{i,l}\frac{1}{h}K_{f}\left(\frac{V_{il}^{*}-v}{h}\right)
\]
with $V_{il}^{*}\coloneqq\xi\left(B_{il}^{*}\right)$ be the empirical
bootstrap analogue of $\widetilde{f}\left(v\right)$. The following
lemma is used in the the proofs of bootstrap consistency results (Theorems
\ref{thm: bootstrap consistency} and \ref{thm:Gaussian coupling for empirical bootstrap process}).
Its proof can be found in the Supplement.
\begin{lem}
\label{lem:bootstrap lemma 0}Suppose Assumptions \ref{assu:DGP}
- \ref{assu: rate of bandwidth} hold. Then,
\[
\underset{v\in I}{\mathrm{sup}}\left|\widetilde{f}^{*}\left(v\right)-\widetilde{f}\left(v\right)\right|=O_{p}^{*}\left(\left(\frac{\mathrm{log}\left(L\right)}{Lh}\right)^{\nicefrac{1}{2}}\right).
\]
\end{lem}
The proofs of the bootstrap consistency results also hinge on an asymptotic
expansion for $\widehat{f}_{GPV}^{*}\left(v\right)-\widehat{f}_{GPV}\left(v\right)$,
which is the empirical bootstrap analogue of $\widehat{f}_{GPV}\left(v\right)-f\left(v\right)$.
The following lemma provides a crucial asymptotic expansion result
that is invoked in the proof of bootstrap consistency. Its proof is
relegated to the Supplement.
\begin{lem}
\label{lem:bootstrap Lemma 3}Suppose that Assumptions \ref{assu:DGP}
- \ref{assu: rate of bandwidth} hold. Then 
\begin{eqnarray*}
\widehat{f}_{GPV}^{*}\left(v\right)-\widetilde{f}^{*}\left(v\right) & = & \frac{1}{\left(N-1\right)}\frac{1}{\left(N\cdot L\right)^{2}}\sum_{i,l}\sum_{j,k}\mathcal{M}\left(B_{il},B_{jk};v\right)\\
 &  & +\frac{1}{N-1}\left\{ \frac{1}{N\cdot L}\sum_{i,l}\mathcal{M}_{2}\left(B_{il}^{*};v\right)-\frac{1}{N\cdot L}\sum_{i,l}\mathcal{M}_{2}\left(B_{il};v\right)\right\} \\
 &  & +O_{p}^{*}\left(\left(\frac{\mathrm{log}\left(L\right)}{Lh}\right)^{\nicefrac{1}{2}}+\frac{\mathrm{log}\left(L\right)}{Lh^{3}}+h^{R}\right),
\end{eqnarray*}
where the remainder term is uniform in $v\in I$.
\end{lem}
\begin{proof}[Proof of Theorem \ref{thm: bootstrap consistency}]Write
\begin{equation}
\widehat{f}_{GPV}^{*}\left(v\right)-\widehat{f}_{GPV}\left(v\right)=\left(\widehat{f}_{GPV}^{*}\left(v\right)-\widetilde{f}^{*}\left(v\right)\right)+\left(\widetilde{f}^{*}\left(v\right)-\widetilde{f}\left(v\right)\right)-\left(\widehat{f}_{GPV}\left(v\right)-\widetilde{f}\left(v\right)\right).\label{eq:bootstrap decomposition}
\end{equation}
It follows from Lemma \ref{lem:lemma 2} and the fact $\widetilde{f}\left(v\right)-f\left(v\right)=O_{p}\left(\left(\frac{\mathrm{log}\left(L\right)}{Lh}\right)^{\nicefrac{1}{2}}+h^{R}\right)$
that 
\begin{multline*}
\widehat{f}_{GPV}\left(v\right)-\widetilde{f}\left(v\right)=\frac{1}{\left(N-1\right)}\frac{1}{\left(N\cdot L\right)^{2}}\sum_{i,l}\sum_{j,k}\mathcal{M}\left(B_{il},B_{jk};v\right)\\
+O_{p}\left(\left(\frac{\mathrm{log}\left(L\right)}{Lh}\right)^{\nicefrac{1}{2}}+\frac{\mathrm{log}\left(L\right)}{Lh^{3}}+h^{R}\right),
\end{multline*}
where the remainder term is uniform in $v\in I$. It follows from
the above result, Lemmas \ref{lem:bootstrap lemma 0}, \ref{lem:bootstrap Lemma 3}
and \citet[online supplement, Lemma S.1]{Marmer_Shneyerov_Quantile_Auctions}
that for any fixed $v\in\left(\underline{v},\overline{v}\right)$,
\begin{align*}
 & \left(Lh^{3}\right)^{\nicefrac{1}{2}}\left(\widehat{f}_{GPV}^{*}\left(v\right)-\widehat{f}_{GPV}\left(v\right)\right)\\
= & \frac{1}{N^{\nicefrac{1}{2}}\left(N\cdot L\right)^{\nicefrac{1}{2}}}\sum_{i,l}h^{\nicefrac{3}{2}}\left(\mathcal{M}_{2}\left(B_{il}^{*};v\right)-\frac{1}{NL}\sum_{i,l}\mathcal{M}_{2}\left(B_{il};v\right)\right)+o_{p}^{*}\left(1\right),
\end{align*}
where the leading term of the right-hand side is an empirical bootstrap
analogue of
\[
\frac{1}{N^{\nicefrac{1}{2}}\left(N\cdot L\right)^{\nicefrac{1}{2}}}\sum_{i,l}h^{\nicefrac{3}{2}}\left(\mathcal{M}_{2}\left(B_{il};v\right)-\mu_{\mathcal{M}}\left(v\right)\right),
\]
which was shown to converge in distribution to a normal random variable
in the proof of Theorem \ref{thm:Asymptotic Normality GPV}. The conclusion
now follows from Theorem 1 of \citet*{Mammen_Bootstrap_and_Normality_1992},
P\'{o}lya's theorem and the conditional Slutsky's lemma (see, e.g., \citealp[Lemma 4.1]{Lahiri_Dependent_Bootstrap}).
\end{proof}

The following lemma is invoked in the proof of Theorem \ref{thm:Gaussian coupling}.
It essentially follows from Lemma \ref{lem:lemma 3} and Theorem \ref{thm:variance estimator}.
Its proof is relegated to the Supplement. It is easy to check
that 
\[
\mathit{\Gamma}\left(v\right)=\frac{1}{\left(N\cdot L\right)^{\nicefrac{1}{2}}}\sum_{i,l}\frac{\mathcal{M}_{2}^{\ddagger}\left(B_{il};v\right)-\mu_{\mathcal{M}^{\ddagger}}\left(v\right)}{\mathrm{Var}\left[\mathcal{M}_{2}^{\ddagger}\left(B_{11};v\right)\right]^{\nicefrac{1}{2}}},\,\textrm{ for all \ensuremath{v\in I}}.
\]

\begin{lem}
\label{lem:lemma 4 (approximation error rate of the linearization)}Suppose
Assumptions \ref{assu:DGP} - \ref{assu: rate of bandwidth} hold.
Then,
\[
\underset{v\in I}{\mathrm{sup}}\left|Z\left(v\right)-\mathit{\Gamma}\left(v\right)\right|=O_{p}\left(\mathrm{log}\left(L\right)^{\nicefrac{1}{2}}h+\frac{\mathrm{log}\left(L\right)}{\left(Lh^{3}\right)^{\nicefrac{1}{2}}}+L^{\nicefrac{1}{2}}h^{\nicefrac{3}{2}+R}\right),
\]
where $\left\{ Z\left(v\right):v\in I\right\} $ was defined by (\ref{eq:Z definition})
and $\left\{ \varGamma\left(v\right):v\in I\right\} $ was defined
by (\ref{eq:GAMMA defnition}).
\end{lem}
The following result is useful \citep[see ][Lemma 2.1]{chernozhukov2016empirical}. 
\begin{lem}
\label{lem:auxiliary 1}Let $V$ and $W$ be random variables such
that $\mathrm{P}\left[\left|V-W\right|>r_{1}\right]\leq r_{2}$, for
some positive constants $r_{1}$ and $r_{2}$. Then,
\[
\left|\mathrm{P}\left[V\leq t\right]-\mathrm{P}\left[W\leq t\right]\right|\leq\mathrm{P}\left[\left|W-t\right|\leq r_{1}\right]+r_{2},\textrm{ for all \ensuremath{t\in\mathbb{R}}}.
\]
\end{lem}
\begin{proof}[Proof of Theorem \ref{thm:Gaussian coupling}]Recall
that $\mathrm{Var}\left[h^{\nicefrac{3}{2}}\mathcal{M}_{2}^{\ddagger}\left(B_{11};v\right)\right]$
converges to $N\left(N-1\right)^{2}\mathrm{V}_{GPV}\left(v\right)$
uniformly in $v\in I$ as $h\downarrow0$, see the proof of Theorem
\ref{thm:Asymptotic Normality GPV}. Therefore, when $h$ is sufficiently
small,
\[
\underset{v\in I}{\mathrm{inf}}\,\mathrm{Var}\left[h^{\nicefrac{3}{2}}\mathcal{M}_{2}^{\ddagger}\left(B_{11};v\right)\right]>C_{1}>0.
\]
By standard arguments (see the proof of Lemma A.1), we can easily
verify that $\mathscr{M}^{\ddagger}\coloneqq\left\{ \mathcal{M}^{\ddagger}\left(\cdot,\cdot;v\right):v\in I\right\} $
is (uniformly) VC-type with respect to the constant envelope 
\begin{equation}
h^{-3}\left(\overline{C}_{D_{1}}+\overline{C}_{D_{1}}\right)\underline{C}_{g}^{-2}\overline{C}_{K_{g}}.\label{eq:M_ddagger envelope}
\end{equation}
It follows from the fact that $\mathscr{M}^{\ddagger}$ is a VC-type
class with respect to the constant envelope (\ref{eq:M_ddagger envelope})
and \citet[Lemma 5.4]{Chen_Kato_U_Process} that $\left\{ \mathcal{M}_{2}^{\ddagger}\left(\cdot;v\right):v\in I\right\} $
is also VC-type with respect to the constant envelope (\ref{eq:M_ddagger envelope}).
It follows from this result and \citet[Corollary A.1]{chernozhukov2014gaussian}
that the function class
\[
\mathscr{S}\coloneqq\left\{ \frac{h^{\nicefrac{3}{2}}\left(\mathcal{M}_{2}\left(\cdot;v\right)-\mu_{\mathcal{M}}\left(v\right)\right)}{\mathrm{Var}\left[h^{\nicefrac{3}{2}}\mathcal{M}_{2}\left(B_{11};v\right)\right]^{\nicefrac{1}{2}}}:v\in I\right\} =\left\{ \frac{h^{\nicefrac{3}{2}}\left(\mathcal{M}_{2}^{\ddagger}\left(\cdot;v\right)-\mu_{\mathcal{M}^{\ddagger}}\left(v\right)\right)}{\mathrm{Var}\left[h^{\nicefrac{3}{2}}\mathcal{M}_{2}^{\ddagger}\left(B_{11};v\right)\right]^{\nicefrac{1}{2}}}:v\in I\right\} 
\]
is (uniformly) VC-type with respect to a constant envelope that is
a multiple of $h^{-\nicefrac{3}{2}}$ when $h$ is sufficiently small. 

Note that $\left\{ \mathit{\Gamma}\left(v\right):v\in I\right\} $
is an empirical process indexed by $\mathscr{S}$. \citet[Lemma 2.1]{chernozhukov2014gaussian}
implies the existence of a tight Gaussian element in $\ell^{\infty}\left(\mathscr{S}\right)$,
denoted by $S_{G}$ with mean zero and covariance function 
\[
\mathrm{E}\left[S_{G}\left(f\right)S_{G}\left(g\right)\right]=\mathrm{E}\left[f\left(B_{11}\right)g\left(B_{11}\right)\right],\,\textrm{for all \ensuremath{\left(f,g\right)\in\mathscr{S}^{2}}.}
\]

Define another mean-zero Gaussian process 
\[
\mathit{\Gamma}_{G}\left(v\right)\coloneqq S_{G}\left(\frac{h^{\nicefrac{3}{2}}\left(\mathcal{M}_{2}^{\ddagger}\left(\cdot;v\right)-\mu_{\mathcal{M}^{\ddagger}}\left(v\right)\right)}{\mathrm{Var}\left[h^{\nicefrac{3}{2}}\mathcal{M}_{2}^{\ddagger}\left(B_{11};v\right)\right]^{\nicefrac{1}{2}}}\right),\,v\in I.
\]
It is easy to check that the process $\left\{ \mathit{\Gamma}_{G}\left(v\right):v\in I\right\} $
is a tight Borel measurable random element in $\ell^{\infty}\left(I\right)$
(for any fixed $L$) by referring to the definitions. See, e.g., \citet[Page 105]{kosorok2007introduction}.
Note that $\left\{ \mathit{\Gamma}_{G}\left(v\right):v\in I\right\} $
has the same covariance structure as $\left\{ \mathit{\Gamma}\left(v\right):v\in I\right\} $.
\citet[Lemma 7.2 and Lemma 7.4]{kosorok2007introduction} yield that
$I$ is totally bounded if endowed with the intrinsic pseudo metric
$\left(v,v'\right)\mapsto\mathrm{E}\left[\left(\mathit{\Gamma}_{G}\left(v\right)-\mathit{\Gamma}_{G}\left(v'\right)\right)^{2}\right]^{\nicefrac{1}{2}}$
and $\left\{ \mathit{\Gamma}_{G}\left(v\right):v\in I\right\} $ is
separable as a stochastic process. Application of \citet[Corollary 2.2]{chernozhukov2014gaussian}
with $q=\infty$, $b\apprle h^{-\nicefrac{3}{2}}$, $\gamma=\mathrm{log}\left(L\right)^{-1}$
and $\sigma=1$ yields that there exists a sequence of random variables
$W_{L}$ with $W_{L}\overset{d}{=}\left\Vert S_{G}\right\Vert _{\mathscr{S}}=\left\Vert \mathit{\Gamma}_{G}\right\Vert _{I}$
satisfying
\begin{equation}
\left|\left\Vert \mathit{\Gamma}\right\Vert _{I}-W_{L}\right|=O_{p}\left(\frac{\mathrm{log}\left(L\right)}{\left(Lh^{3}\right)^{\nicefrac{1}{6}}}\right).\label{eq:sup GAMMA - W_L order bound}
\end{equation}

Note that the distribution of $W_{L}$ (also that of $\left\Vert \mathit{\Gamma}_{G}\right\Vert _{I}$)
changes with $L$. Since $\mathrm{E}\left[\mathit{\Gamma}_{G}\left(v\right)^{2}\right]=1$
for all $v\in I$, the diameter of $I$ under the intrinsic metric
is less than or equal to 2. By the calculations used in the proof
of \citet[Corollary 5.1]{chernozhukov2014gaussian} and approximation
based on the strong law of large numbers (see \citealp[Problem 2.5.1]{van1996weak}),
we have 
\[
\int_{0}^{1}\sqrt{\mathrm{log}\left(2N\left(\epsilon,\mathscr{S},\left\Vert \cdot\right\Vert _{G,2}\right)\right)}\mathrm{d}\epsilon\apprle\int_{0}^{1}\sqrt{1+\mathrm{log}\left(\frac{h^{-\nicefrac{3}{2}}}{\epsilon}\right)}\mathrm{d}\epsilon\apprle\mathrm{log}\left(h^{-1}\right)^{\nicefrac{1}{2}}.
\]
Then in view of the fact that $\mathrm{E}\left[\mathit{\Gamma}_{G}\left(v\right)^{2}\right]=1$
for all $v\in I$, Dudley's bound (see, e.g., \citealp[Theorem 2.3.7]{gine2015mathematical})
yields 
\begin{equation}
\mathrm{E}\left[\left\Vert \mathit{\Gamma}_{G}\right\Vert _{I}\right]=\mathrm{E}\left[\left\Vert S_{G}\right\Vert _{\mathscr{S}}\right]=O\left(\mathrm{log}\left(h^{-1}\right)^{\nicefrac{1}{2}}\right).\label{eq:Dudley's bound}
\end{equation}

Since $\left\{ \mathit{\Gamma}_{G}\left(v\right):v\in I\right\} $
is a centered separable Gaussian process with $\mathrm{E}\left[\mathit{\Gamma}_{G}\left(v\right)^{2}\right]=1$
for all $v\in I$, the Gaussian anti-concentration inequality \citet[Corollary 2.1]{chernozhukov2014anti}
yields 
\begin{equation}
\underset{z\in\mathbb{R}}{\mathrm{sup}}\,\mathrm{P}\left[\left|\left\Vert \mathit{\Gamma}_{G}\right\Vert _{I}-z\right|\leq\epsilon\right]\leq4\epsilon\left(\mathrm{E}\left[\left\Vert \mathit{\Gamma}_{G}\right\Vert _{I}\right]+1\right),\,\textrm{for all \ensuremath{\epsilon\geq0}}.\label{eq:anti-concentration Gaussian process}
\end{equation}

Note that Lemma \ref{lem:lemma 4 (approximation error rate of the linearization)}
and (\ref{eq:sup GAMMA - W_L order bound}) yield $\left|\left\Vert Z\right\Vert _{I}-W_{L}\right|=o_{p}\left(\mathrm{log}\left(h^{-1}\right)^{-\nicefrac{1}{2}}\right).$
This result and \citet[Theorem 9.2.2]{dudley2002real} imply that
there exists some null sequence $\delta_{1,L}\downarrow0$, 
\[
\mathrm{P}\left[\left|\left\Vert Z\right\Vert _{I}-W_{L}\right|>\mathrm{log}\left(h^{-1}\right)^{-\nicefrac{1}{2}}\delta_{1,L}\right]<\delta_{1,L}.
\]
By the above result, Lemma \ref{lem:auxiliary 1}, the fact $W_{L}\overset{d}{=}\left\Vert \mathit{\Gamma}_{G}\right\Vert _{I}$,
(\ref{eq:Dudley's bound}) and (\ref{eq:anti-concentration Gaussian process}),
\[
\underset{z\in\mathbb{R}}{\mathrm{sup}}\left|\mathrm{P}\left[\left\Vert Z\right\Vert _{I}\leq z\right]-\mathrm{P}\left[\left\Vert \mathit{\Gamma}_{G}\right\Vert _{I}\leq z\right]\right|\leq\underset{z\in\mathbb{R}}{\mathrm{sup}}\,\mathrm{P}\left[\left|\left\Vert \mathit{\Gamma}_{G}\right\Vert _{I}-z\right|\leq\mathrm{log}\left(h^{-1}\right)^{-\nicefrac{1}{2}}\delta_{1,L}\right]+\delta_{1,L}=O\left(\delta_{1,L}\right).
\]

\noindent \end{proof}

Denote $\widehat{\mu}_{\mathcal{M}_{2}^{\ddagger}}\left(v\right)\coloneqq\left(N\cdot L\right)^{-1}\sum_{i,l}\mathcal{M}_{2}^{\ddagger}\left(B_{il};v\right)$
and consider the following bootstrap analogue of $\mathit{\Gamma}$:
\begin{equation}
\mathit{\Gamma}^{*}\left(v\right)\coloneqq\frac{1}{\left(N\cdot L\right)^{\nicefrac{1}{2}}}\sum_{i,l}\frac{\mathcal{M}_{2}^{\ddagger}\left(B_{il}^{*};v\right)-\widehat{\mu}_{\mathcal{M}_{2}^{\ddagger}}\left(v\right)}{\mathrm{Var}\left[\mathcal{M}_{2}^{\ddagger}\left(B_{11};v\right)\right]^{\nicefrac{1}{2}}},\,\textrm{for }v\in I.\label{eq:GAMMA_star definition}
\end{equation}
The following lemma is invoked in the proof of Theorem \ref{thm:Gaussian coupling for empirical bootstrap process}.
Its proof is similar to that of Lemma \ref{lem:lemma 4 (approximation error rate of the linearization)}.
It essentially follows from Lemmas \ref{lem:lemma 2}, \ref{lem:bootstrap lemma 0},
\ref{lem:bootstrap Lemma 3} and Theorem \ref{thm:variance estimator}.
We relegate its proof to the Supplement.
\begin{lem}
\label{lem:bootstrap lemma 4 (approximation error rate of the linearization)}Suppose
that Assumptions \ref{assu:DGP} - \ref{assu: rate of bandwidth}
hold. Then
\[
\underset{v\in I}{\mathrm{sup}}\left|Z^{*}\left(v\right)-\mathit{\Gamma}^{*}\left(v\right)\right|=O_{p}^{*}\left(\mathrm{log}\left(L\right)^{\nicefrac{1}{2}}h+\frac{\mathrm{log}\left(L\right)}{\left(Lh^{3}\right)^{\nicefrac{1}{2}}}+L^{\nicefrac{1}{2}}h^{\nicefrac{3}{2}+R}\right),
\]
where $\left\{ Z^{*}\left(v\right):v\in I\right\} $ is defined by
(\ref{eq:Z_star process}) and $\left\{ \mathit{\Gamma}^{*}\left(v\right):v\in I\right\} $
is defined by (\ref{eq:GAMMA_star definition}).
\end{lem}
We also need the following technical lemma in the proof of Theorem
\ref{thm:Gaussian coupling for empirical bootstrap process}. Its
proof can be found in the Supplement.
\begin{lem}
\label{lem:auxiliary 4}Let $V_{L}^{*}$ and $W_{L}^{*}$ be statistics
computed using the bootstrap sample with $\left|V_{L}^{*}-W_{L}^{*}\right|=O_{p}^{*}\left(\lambda_{L}\right)$
for some $\lambda_{L}\downarrow0$. Suppose for any fixed $C_{1}>0$,
\[
\underset{z\in\mathbb{R}}{\mathrm{sup}}\,\mathrm{P}^{*}\left[\left|W_{L}^{*}-z\right|\leq C_{1}\lambda_{L}\right]\rightarrow_{p}0,\,\textrm{as \ensuremath{L\uparrow\infty}}.
\]
Then, 
\[
\underset{z\in\mathbb{R}}{\mathrm{sup}}\left|\mathrm{P}^{*}\left[V_{L}^{*}\leq z\right]-\mathrm{P}^{*}\left[W_{L}^{*}\leq z\right]\right|\rightarrow_{p}0,\,\textrm{as \ensuremath{L\uparrow\infty}}.
\]
\end{lem}
\begin{proof}[Proof of Theorem \ref{thm:Gaussian coupling for empirical bootstrap process}]Application
of \citet[Theorem 2.3]{chernozhukov2016empirical} with $B\left(f\right)=0$,
$q=\infty$, $b\apprle h^{-\nicefrac{3}{2}}$, $\gamma=\mathrm{log}\left(L\right)^{-1}$
and $\sigma=1$ yields that there exists a sequence of random variables,
$W_{L}^{*}$, with the property that the conditional distribution
of $W_{L}^{*}$ given the original sample is the same as the (marginal)
distribution of $\left\Vert S_{G}\right\Vert _{\mathscr{S}}=\left\Vert \mathit{\Gamma}_{G}\right\Vert _{I}$
almost surely, and 
\[
\left|\left\Vert \mathit{\Gamma}^{*}\right\Vert _{I}-W_{L}^{*}\right|=O_{p}\left(\frac{\mathrm{log}\left(L\right)}{\left(Lh^{3}\right)^{\nicefrac{1}{6}}}\right).
\]
This result, Lemma \ref{lem:bootstrap lemma 4 (approximation error rate of the linearization)}
and Markov's inequality yield 
\[
\left|\left\Vert Z^{*}\right\Vert _{I}-W_{L}^{*}\right|=O_{p}^{*}\left(\lambda_{L}^{*}\right),
\]
where
\[
\lambda_{L}^{*}\coloneqq\mathrm{log}\left(L\right)^{\nicefrac{1}{2}}h+\frac{\mathrm{log}\left(L\right)}{\left(Lh^{3}\right)^{\nicefrac{1}{6}}}+L^{\nicefrac{1}{2}}h^{\nicefrac{3}{2}+R}.
\]

Since the conditional distribution of $W_{L}^{*}$ under $\mathrm{P}^{*}$
is the same as the distribution of $\left\Vert S_{G}\right\Vert _{\mathscr{S}}=\left\Vert \mathit{\Gamma}_{G}\right\Vert _{I}$,
(\ref{eq:Dudley's bound}) and the anti-concentration bound (\ref{eq:anti-concentration Gaussian process})
now yield 
\[
\underset{z\in\mathbb{R}}{\mathrm{sup}}\,\mathrm{P}^{*}\left[\left|W_{L}^{*}-z\right|\leq C_{1}\lambda_{L}^{*}\right]\leq4C_{1}\lambda_{L}^{*}\left(\mathrm{E}\left[\left\Vert \mathit{\Gamma}_{G}\right\Vert _{I}\right]+1\right)=o\left(1\right),
\]
for any $C_{1}>0$. The conclusion now follows from the fact that
$\mathrm{P}^{*}\left[W_{L}^{*}\leq z\right]=\mathrm{P}\left[\left\Vert \mathit{\Gamma}_{G}\right\Vert _{I}\leq z\right]$
for all $z\in\mathbb{R}$ almost surely, and Lemma \ref{lem:auxiliary 4}
(with $V_{L}^{*}=\left\Vert Z^{*}\right\Vert _{I}$).\end{proof}

\begin{proof}[Proof of Corollary \ref{cor:consistency of IGA bootstrap}]Note
that (\ref{eq:anti-concentration Gaussian process}) implies that
the CDF of $\left\Vert \mathit{\Gamma}_{G}\right\Vert _{I}$ is Lipschitz
continuous. Let $\zeta\left(1-\tau\right)$ denote the $\left(1-\tau\right)$-th
quantile of $\left\Vert \mathit{\Gamma}_{G}\right\Vert _{I}$. Now
Theorem \ref{thm:Gaussian coupling for empirical bootstrap process}
and \citet[Theorem 9.2.2]{dudley2002real} imply that there exists
some null sequence $\delta_{2,L}\downarrow0$ such that 
\[
\mathrm{P}\left[\left|\mathrm{P}^{*}\left[\left\Vert Z^{*}\right\Vert _{I}\leq\zeta\left(1-\alpha+\delta_{2,L}\right)\right]-\mathrm{P}\left[\left\Vert \mathit{\Gamma}_{G}\right\Vert _{I}\leq\zeta\left(1-\alpha+\delta_{2,L}\right)\right]\right|>\delta_{2,L}\right]<\delta_{2,L},
\]
which clearly implies $\mathrm{P}\left[\mathrm{P}^{*}\left[\left\Vert Z^{*}\right\Vert _{I}\leq\zeta\left(1-\alpha+\delta_{2,L}\right)\right]<1-\alpha\right]<\delta_{2,L}.$
By this result and the definition of $\zeta_{L,\alpha}^{*}$ (see
(\ref{eq:zeta_L^* definition})), we have 
\begin{equation}
\mathrm{P}\left[\zeta_{L,\alpha}^{*}>\zeta\left(1-\alpha+\delta_{2,L}\right)\right]<\delta_{2,L}.\label{eq:zeta_L^*}
\end{equation}
This further implies 
\begin{eqnarray*}
\mathrm{P}\left[\left\Vert Z\right\Vert _{I}\leq\zeta_{L,\alpha}^{*}\right] & \leq & \mathrm{P}\left[\left\Vert Z\right\Vert _{I}\leq\zeta\left(1-\alpha+\delta_{2,L}\right)\right]+\delta_{2,L}\\
 & \leq & \mathrm{P}\left[\left\Vert \mathit{\Gamma}_{G}\right\Vert _{I}\leq\zeta\left(1-\alpha+\delta_{2,L}\right)\right]+\delta_{2,L}+\delta_{3,L}\\
 & = & \left(1-\alpha\right)+2\delta_{2,L}+\delta_{3,L},
\end{eqnarray*}
for some null sequence $\delta_{3,L}\downarrow0$, where the second
inequality follows from Theorem \ref{thm:Gaussian coupling} and the
equality follows from \citet[Lemma 21.1(ii)]{VanDerVaart_Asymptotic_Statistics_Book}
and the continuity of the CDF of $\left\Vert \mathit{\Gamma}_{G}\right\Vert _{I}$.
Similarly, by using 
\[
\mathrm{P}\left[\left|\mathrm{P}^{*}\left[\left\Vert Z^{*}\right\Vert _{I}\leq\zeta\left(1-\alpha-\delta_{2,L}\right)\right]-\mathrm{P}\left[\left\Vert \mathit{\Gamma}_{G}\right\Vert _{I}\leq\zeta\left(1-\alpha-\delta_{2,L}\right)\right]\right|>\delta_{2,L}\right]<\delta_{2,L}
\]
which implies $\mathrm{P}\left[\zeta_{L,\alpha}^{*}\leq\zeta\left(1-\alpha-\delta_{2,L}\right)\right]<\delta_{2,L}$,
we can show that a lower bound holds: 
\[
\mathrm{P}\left[\left\Vert Z\right\Vert _{I}\leq\zeta_{L,\alpha}^{*}\right]\geq\left(1-\alpha\right)-\left(2\delta_{2,L}+\delta_{3,L}\right).
\]
Now we have 
\[
\mathrm{P}\left[\left\Vert Z\right\Vert _{I}\leq\zeta_{L,\alpha}^{*}\right]\rightarrow1-\alpha,\,\textrm{as \ensuremath{L\uparrow\infty}},
\]
which is exactly the first conclusion.

By the definition of $\zeta\left(\cdot\right)$ and the Borell-Sudakov-Tsirelson
inequality (see \citealp[Theorems 2.2.7 and 2.5.8]{gine2015mathematical}),
we have 
\[
\zeta\left(1-\alpha+\delta_{2,L}\right)\leq\mathrm{E}\left[\left\Vert \mathit{\Gamma}_{G}\right\Vert _{I}\right]+\left(2\mathrm{log}\left(\frac{1}{\alpha-\delta_{2,L}}\right)\right)^{\nicefrac{1}{2}}=O\left(\mathrm{log}\left(h^{-1}\right)^{\nicefrac{1}{2}}\right).
\]
By (\ref{eq:zeta_L^*}), $\zeta_{L,\alpha}^{*}\leq\zeta\left(1-\alpha+\delta_{2,L}\right)$
w.p.a.1, and the second conclusion follows.\end{proof}

\end{appendices}

\input{2nd_Revision_Supplement_V5}

\end{document}

%% file: 2nd_Revision_Supplement_V5.tex


\allowdisplaybreaks
\setcounter{equation}{0}
\setcounter{table}{0}

\renewcommand{\thesection}{S.\arabic{section}}  
\renewcommand*{\theHsection}{S.\the\value{section}}

\renewcommand{\thethm}{\thesection.\arabic{thm}}
\renewcommand{\thetable}{S.\arabic{table}}   
\renewcommand{\thefigure}{S.\arabic{figure}}
\renewcommand{\theequation}{\thesection.\arabic{equation}}
\renewcommand{\thelem}{\thesection.\arabic{lem}}
\renewcommand{\thetable}{S.\arabic{table}}
\renewcommand{\theassumption}{\thesection.\arabic{assumption}}
\renewcommand{\thecor}{\thesection.\arabic{cor}}

%

\newgeometry{verbose,tmargin=2cm,bmargin=2cm,lmargin=1.6cm,rmargin=1.6cm}

\fontsize{8}{10}\selectfont
\setcounter{page}{1}
\setcounter{section}{0}
\part*{Supplement}
\addcontentsline{toc}{part}{Supplement}

\section{Proofs of the Lemmas in Appendix \ref{sec:Appendix_main}}
\renewcommand{\thepage}{S.\arabic{page} }

To prove Lemmas A.1 and A.2, we derive the following intermediate
asymptotic expansion first.

\begin{lem}
\label{lem:lemma 1 Stochastic Expansion} Suppose Assumptions 1 -
3 hold. Let $\widetilde{\mathbb{T}}_{il}\coloneqq\mathbbm{1}\left(\left|V_{il}-v\right|\leq\overline{\delta}\right)$
be an infeasible trimming factor. Then,
\begin{eqnarray*}
\widehat{f}_{GPV}\left(v\right)-f\left(v\right) & = & \frac{1}{N\cdot L}\sum_{i,l}\widetilde{\mathbb{T}}_{il}\frac{1}{h^{2}}K_{f}'\left(\frac{V_{il}-v}{h}\right)\left(\widehat{V}_{il}-V_{il}\right)+\frac{1}{R!}f^{\left(R\right)}\left(v\right)\left(\int K_{f}\left(u\right)u^{R}\mathrm{d}u\right)h^{R}\\
 &  & +O_{p}\left(\left(\frac{\mathrm{log}\left(L\right)}{Lh}\right)^{\nicefrac{1}{2}}+\frac{\mathrm{log}\left(L\right)}{Lh^{3}}\right)+o\left(h^{R}\right),
\end{eqnarray*}
where the remainder term is uniform in $v\in I$.
\end{lem}
\begin{proof}[Proof of Lemma \ref{lem:lemma 1 Stochastic Expansion}]
By standard arguments (see, e.g., \citealp*[Lemma 1]{Marmer_Shneyerov_Quantile_Auctions}),
\begin{equation}
\underset{b\in\left[\underline{b},\overline{b}\right]}{\mathrm{sup}}\left|\widehat{G}\left(b\right)-G\left(b\right)\right|=O_{p}\left(\left(\frac{\mathrm{log}\left(L\right)}{L}\right)^{\nicefrac{1}{2}}\right)\textrm{ and }\underset{b\in\left[\underline{b}+h,\overline{b}-h\right]}{\mathrm{sup}}\left|\widehat{g}\left(b\right)-g\left(b\right)\right|=O_{p}\left(\left(\frac{\mathrm{log}\left(L\right)}{Lh}\right)^{\nicefrac{1}{2}}+h^{1+R}\right).\label{eq:uniform rates}
\end{equation}

It follows from the definitions of $\widehat{V}_{il}$ and $V_{il}$
and the identity 
\begin{equation}
\frac{a}{b}=\frac{a}{c}-\frac{a\left(b-c\right)}{c^{2}}+\frac{a\left(b-c\right)^{2}}{bc^{2}}\label{eq:abc_identity}
\end{equation}
 that
\begin{equation}
\widehat{V}_{il}-V_{il}=\frac{1}{N-1}\left\{ \frac{\widehat{G}\left(B_{il}\right)-G\left(B_{il}\right)}{g\left(B_{il}\right)}-\frac{\widehat{G}\left(B_{il}\right)\left(\widehat{g}\left(B_{il}\right)-g\left(B_{il}\right)\right)}{g\left(B_{il}\right)^{2}}+\frac{\widehat{G}\left(B_{il}\right)}{\widehat{g}\left(B_{il}\right)}\frac{\left(\widehat{g}\left(B_{il}\right)-g\left(B_{il}\right)\right)^{2}}{g\left(B_{il}\right)^{2}}\right\} .\label{eq:v_hat v difference algebra}
\end{equation}
Then by the triangle inequality, 
\begin{equation}
\underset{i,l}{\mathrm{max}}\,\mathbb{T}_{il}\left|\widehat{V}_{il}-V_{il}\right|\leq\underset{i,l}{\mathrm{max}}\,\mathbb{T}_{il}\left|\frac{G\left(B_{il}\right)-\widehat{G}\left(B_{il}\right)}{g\left(B_{il}\right)}\right|+\underset{i,l}{\mathrm{max}}\,\mathbb{T}_{il}\left|\frac{\widehat{g}\left(B_{il}\right)-g\left(B_{il}\right)}{g\left(B_{il}\right)^{2}}\right|+\underset{i,l}{\mathrm{max}}\,\mathbb{T}_{il}\left|\frac{\left(\widehat{g}\left(B_{il}\right)-g\left(B_{il}\right)\right)^{2}}{\widehat{g}\left(B_{il}\right)g\left(B_{il}\right)^{2}}\right|.\label{eq:v_hat v difference}
\end{equation}
The order of magnitude of the first and second terms of the right
hand side of (\ref{eq:v_hat v difference}) is easily obtained by
using 
\begin{equation}
\underline{C}_{g}\coloneqq\underset{b\in\left[\underline{b},\overline{b}\right]}{\mathrm{inf}}g\left(b\right)>0,\label{eq:density of bids bounded  from 0}
\end{equation}
(\ref{eq:uniform rates}) and the fact that $\widehat{\underline{b}}\geq\underline{b}$
and $\widehat{\overline{b}}\leq\overline{b}$. 

For the third term, since $\underset{i,l}{\mathrm{max}}\,\mathbb{T}_{il}\left|\widehat{g}\left(B_{il}\right)-g\left(B_{il}\right)\right|=o_{p}\left(1\right)$,
we have $\underset{i,l}{\mathrm{max}}\,\mathbb{T}_{il}\left|\widehat{g}\left(B_{il}\right)^{-1}\right|\leq\left(\nicefrac{\underline{C}_{g}}{2}\right)^{-1}$
w.p.a.1 and consequently,
\begin{equation}
\underset{i,l}{\mathrm{max}}\,\mathbb{T}_{il}\left|\frac{\left(\widehat{g}\left(B_{il}\right)-g\left(B_{il}\right)\right)^{2}}{\widehat{g}\left(B_{il}\right)g\left(B_{il}\right)^{2}}\right|\leq2\underline{C}_{g}^{-3}\left\{ \underset{i,l}{\mathrm{max}}\,\mathbb{T}_{il}\left(\widehat{g}\left(B_{il}\right)-g\left(B_{il}\right)\right)^{2}\right\} =O_{p}\left(\frac{\mathrm{log}\left(L\right)}{Lh}+h^{2+2R}\right),\label{eq:difference bw g_hat g bound}
\end{equation}
where the inequality holds w.p.a.1. 

Now it follows that
\begin{equation}
\underset{i,l}{\mathrm{max}}\,\mathbb{T}_{il}\left|\widehat{V}_{il}-V_{il}\right|=O_{p}\left(\left(\frac{\mathrm{log}\left(L\right)}{Lh}\right)^{\nicefrac{1}{2}}+h^{1+R}\right).\label{eq:max difference v_hat v feasible}
\end{equation}
Similarly, we can also obtain
\begin{equation}
\underset{v\in I}{\mathrm{\mathrm{sup}}}\,\underset{i,l}{\mathrm{max}}\,\mathbb{\widetilde{T}}_{il}\left|\widehat{V}_{il}-V_{il}\right|=O_{p}\left(\left(\frac{\mathrm{log}\left(L\right)}{Lh}\right)^{\nicefrac{1}{2}}+h^{1+R}\right)\label{eq:max difference v_hat v infeasible-1}
\end{equation}
by observing that $\mathbb{\widetilde{T}}_{il}=\mathbbm{1}\left(B_{il}\in\left[s\left(v-\overline{\delta}\right),s\left(v+\overline{\delta}\right)\right]\right)$
and using (\ref{eq:uniform rates}) and (\ref{eq:v_hat v difference})
(with $\mathbb{T}_{il}$ replaced by $\mathbb{\widetilde{T}}_{il}$). 

Write
\begin{eqnarray*}
\widehat{f}_{GPV}\left(v\right) & = & \frac{1}{N\cdot L}\sum_{i,l}\left\{ \mathbb{\widetilde{T}}_{il}\frac{1}{h}K_{f}\left(\frac{\widehat{V}_{il}-v}{h}\right)+\mathbb{T}_{il}\left(1-\mathbb{\widetilde{T}}_{il}\right)\frac{1}{h}K_{f}\left(\frac{\widehat{V}_{il}-v}{h}\right)+\left(\mathbb{T}_{il}-1\right)\mathbb{\widetilde{T}}_{il}\frac{1}{h}K_{f}\left(\frac{\widehat{V}_{il}-v}{h}\right)\right\} \\
 & \eqqcolon & \frac{1}{N\cdot L}\sum_{i,l}\mathbb{\widetilde{T}}_{il}\frac{1}{h}K_{f}\left(\frac{\widehat{V}_{il}-v}{h}\right)+\kappa_{1}^{\dagger}\left(v\right)+\kappa_{2}^{\dagger}\left(v\right).
\end{eqnarray*}
Since $K_{f}$ is compactly supported on $\left[-1,1\right]$, $K_{f}\left(\nicefrac{\left(\widehat{V}_{il}-v\right)}{h}\right)$
is zero if $\widehat{V}_{il}$ is outside of an $h-$neighborhood
of $v$. By the triangle inequality, we have 
\begin{eqnarray*}
\left|\kappa_{1}^{\dagger}\left(v\right)\right| & \apprle & \frac{1}{N\cdot L}\sum_{i,l}h^{-1}\mathbb{T}_{il}\left(1-\mathbb{\widetilde{T}}_{il}\right)\mathbbm{1}\left(\left|\frac{\widehat{V}_{il}-v}{h}\right|\leq1\right)\\
 & \apprle & \frac{1}{N\cdot L}\sum_{i,l}h^{-1}\mathbb{T}_{il}\left(1-\mathbb{\widetilde{T}}_{il}\right)\mathbbm{1}\left(\left|V_{il}-v\right|\leq h+\underset{j,k}{\mathrm{max}}\,\mathbb{T}_{jk}\left|\widehat{V}_{jk}-V_{jk}\right|\right).
\end{eqnarray*}
Therefore,
\[
\mathrm{P}\left[\underset{v\in I}{\mathrm{sup}}\left|\kappa_{1}^{\dagger}\left(v\right)\right|=0\right]\geq\mathrm{P}\left[\underset{i,l}{\mathrm{max}}\,\mathbb{T}_{il}\left|\widehat{V}_{il}-V_{il}\right|<\frac{\overline{\delta}}{2}\right],
\]
when $h$ is sufficiently small. The right hand side of the above
inequality tends to one as $L\uparrow\infty$ (see (\ref{eq:max difference v_hat v feasible})).
Therefore, we have $\underset{v\in I}{\mathrm{sup}}\left|\kappa_{1}^{\dagger}\left(v\right)\right|=0$
w.p.a.1. 

Similarly, we have
\begin{eqnarray*}
\left|\kappa_{2}^{\dagger}\left(v\right)\right| & \apprle & \frac{1}{N\cdot L}\sum_{i,l}h^{-1}\left(1-\mathbb{T}_{il}\right)\mathbb{\widetilde{T}}_{il}\\
 & \apprle & \frac{1}{N\cdot L}\sum_{i,l}h^{-1}\mathbbm{1}\left(\left|\widehat{\underline{b}}-\underline{b}\right|+\underline{b}+h>B_{il}\right)\mathbb{\widetilde{T}}_{il}+\frac{1}{N\cdot L}\sum_{i,l}h^{-1}\mathbbm{1}\left(B_{il}>\overline{b}-\left|\widehat{\overline{b}}-\overline{b}\right|-h\right)\mathbb{\widetilde{T}}_{il}\\
 & \leq & h^{-1}\mathbbm{1}\left(\left|\widehat{\underline{b}}-\underline{b}\right|+\underline{b}+h>s\left(v_{l}-\overline{\delta}\right)\right)+h^{-1}\mathbbm{1}\left(\overline{b}-\left|\widehat{\overline{b}}-\overline{b}\right|-h<s\left(v_{u}+\overline{\delta}\right)\right)
\end{eqnarray*}
and hence
\[
\mathrm{P}\left[\underset{v\in I}{\mathrm{sup}}\left|\kappa_{2}^{\dagger}\left(v\right)\right|>0\right]\leq\mathrm{P}\left[\left|\widehat{\underline{b}}-\underline{b}\right|+\underline{b}+h>s\left(v_{l}-\overline{\delta}\right)\right]+\mathrm{P}\left[\overline{b}-\left|\widehat{\overline{b}}-\overline{b}\right|-h<s\left(v_{u}+\overline{\delta}\right)\right].
\]
The right-hand side of the above inequality tends to zero as $L\uparrow\infty$
since by the Borel-Cantelli lemma (see also GPV Proposition 2) and
(\ref{eq:density of bids bounded  from 0}), we have
\[
\left|\widehat{\underline{b}}-\underline{b}\right|=O_{p}\left(\frac{\mathrm{\mathrm{log}}\left(L\right)}{L}\right),\,\left|\widehat{\overline{b}}-\overline{b}\right|=O_{p}\left(\frac{\mathrm{log}\left(L\right)}{L}\right).
\]
Therefore now we have
\[
\widehat{f}_{GPV}\left(v\right)=\frac{1}{N\cdot L}\sum_{i,l}\mathbb{\widetilde{T}}_{il}\frac{1}{h}K_{f}\left(\frac{\widehat{V}_{il}-v}{h}\right),\textrm{ for all \ensuremath{v\in I}, w.p.a.1.}
\]

Since $K_{f}$ is compactly supported on $\left[-1,1\right]$, 
\begin{equation}
\underset{v\in I}{\mathrm{sup}}\,\underset{i,l}{\mathrm{max}}\left|\left(\widetilde{\mathbb{T}}_{il}-1\right)\frac{1}{h}K_{f}\left(\frac{V_{il}-v}{h}\right)\right|=0,\,\textrm{ when \ensuremath{h<\overline{\delta}}}.\label{eq:negligible trimming}
\end{equation}
It now follows that
\[
\widehat{f}_{GPV}\left(v\right)-\widetilde{f}\left(v\right)=\frac{1}{N\cdot L}\sum_{i,l}\mathbb{\widetilde{T}}_{il}\frac{1}{h}\left(K_{f}\left(\frac{\widehat{V}_{il}-v}{h}\right)-K_{f}\left(\frac{V_{il}-v}{h}\right)\right),\textrm{ for all \ensuremath{v\in I}, w.p.a.1.}
\]
By a second-order Taylor expansion of the right-hand side of the above
equality, 
\begin{equation}
\widehat{f}_{GPV}\left(v\right)-\widetilde{f}\left(v\right)=\frac{1}{N\cdot L}\sum_{i,l}\mathbb{\widetilde{T}}_{il}\frac{1}{h^{2}}K_{f}'\left(\frac{V_{il}-v}{h}\right)\left(\widehat{V}_{il}-V_{il}\right)+\frac{1}{2}\cdot\frac{1}{N\cdot L}\sum_{i,l}\mathbb{\widetilde{T}}_{il}\frac{1}{h^{3}}K_{f}''\left(\frac{\dot{V}_{il}-v}{h}\right)\left(\widehat{V}_{il}-V_{il}\right)^{2},\label{eq:difference f_hat f_til 1}
\end{equation}
for some mean value $\dot{V}_{il}$ that lies on the line joining
$\widehat{V}_{il}$ and $V_{il}$. 

Since $K_{f}''$ is compactly supported on $\left[-1,1\right]$ and
bounded, by the triangle inequality, we have
\begin{align}
 & \underset{v\in I}{\mathrm{sup}}\left|\frac{1}{N\cdot L}\sum_{i,l}\mathbb{\widetilde{T}}_{il}\frac{1}{h^{3}}K_{f}''\left(\frac{\dot{V}_{il}-v}{h}\right)\left(\widehat{V}_{il}-V_{il}\right)^{2}\right|\nonumber \\
\apprle & \left\{ \underset{v\in I}{\mathrm{sup}}\,\frac{1}{N\cdot L}\sum_{i,l}\mathbb{\widetilde{T}}_{il}h^{-3}\mathbbm{1}\left(\left|\dot{V}_{il}-v\right|\leq h\right)\right\} \left\{ \underset{v\in I}{\mathrm{sup}}\,\underset{i,l}{\mathrm{max}}\,\mathbb{\widetilde{T}}_{il}\left(\widehat{V}_{il}-V_{il}\right)^{2}\right\} \nonumber \\
\leq & \left\{ \underset{v\in I}{\mathrm{sup}}\,\frac{1}{N\cdot L}\sum_{i,l}\mathbb{\widetilde{T}}_{il}h^{-3}\mathbbm{1}\left(\left|V_{il}-v\right|\leq2h\right)\right\} \left\{ \underset{v\in I}{\mathrm{sup}}\,\underset{i,l}{\mathrm{max}}\,\mathbb{\widetilde{T}}_{il}\left(\widehat{V}_{il}-V_{il}\right)^{2}\right\} ,\label{eq:K_double_prime * (V_hat - V)^2 bound}
\end{align}
where the last inequality holds w.p.a.1, since $\underset{v\in I}{\mathrm{sup}}\,\underset{i,l}{\mathrm{max}}\,\mathbb{\widetilde{T}}_{il}\left|\dot{V}_{il}-V_{il}\right|=o_{p}\left(h\right)$
(see (\ref{eq:max difference v_hat v infeasible-1})).

Denote $\reallywidecheck{\mathbb{T}}_{il}\coloneqq\mathbbm{1}\left(\left|V_{il}-v\right|\leq2h\right)$.
The CCK inequality and Markov's inequality yield 
\begin{equation}
\underset{v\in I}{\mathrm{sup}}\left|\frac{1}{N\cdot L}\sum_{i,l}h^{-1}\reallywidecheck{\mathbb{T}}_{il}-\mathrm{E}\left[h^{-1}\reallywidecheck{\mathbb{T}}_{il}\right]\right|=O_{p}\left(\left(\frac{\mathrm{log}\left(L\right)}{Lh}\right)^{\nicefrac{1}{2}}\right).\label{eq:T_check sup bound 1}
\end{equation}
Since $\underset{v\in I}{\mathrm{sup}}\,\mathrm{E}\left[h^{-1}\reallywidecheck{\mathbb{T}}_{il}\right]\leq4\underset{u\in\left[\underline{v},\overline{v}\right]}{\mathrm{sup}}f\left(u\right)$,
it follows that
\begin{equation}
\underset{v\in I}{\mathrm{sup}}\,\frac{1}{N\cdot L}\sum_{i,l}h^{-1}\reallywidecheck{\mathbb{T}}_{il}=O_{p}\left(1\right).\label{eq:T_check sup bound 2}
\end{equation}
It follows from the above result, (\ref{eq:max difference v_hat v infeasible-1}),
(\ref{eq:K_double_prime * (V_hat - V)^2 bound}) and (\ref{eq:T_check sup bound 1})
that
\begin{equation}
\underset{v\in I}{\mathrm{sup}}\left|\frac{1}{N\cdot L}\sum_{i,l}\mathbb{\widetilde{T}}_{il}\frac{1}{h^{3}}K_{f}''\left(\frac{\dot{V}_{il}-v}{h}\right)\left(\widehat{V}_{il}-V_{il}\right)^{2}\right|=O_{p}\left(\frac{\mathrm{log}\left(L\right)}{Lh^{3}}+h^{2R}\right).\label{eq:difference f_hat f_til 2}
\end{equation}

By standard arguments for kernel density estimation (see, e.g., \citealp{Newey_Kernel_ET_1994}),
\begin{equation}
\widetilde{f}\left(v\right)-\mathrm{E}\left[\widetilde{f}\left(v\right)\right]=O_{p}\left(\left(\frac{\mathrm{log}\left(L\right)}{Lh}\right)^{\nicefrac{1}{2}}\right)\textrm{ and }\mathrm{E}\left[\widetilde{f}\left(v\right)\right]-f\left(v\right)=\frac{1}{R!}f^{\left(R\right)}\left(v\right)\left(\int K_{f}\left(u\right)u^{R}\mathrm{d}u\right)h^{R}+o\left(h^{R}\right),\label{eq:infeasible PDF estimate error rate}
\end{equation}
where the remainder terms are uniform in $v\in I$. The conclusion
now follows from (\ref{eq:difference f_hat f_til 1}), (\ref{eq:difference f_hat f_til 2})
and (\ref{eq:infeasible PDF estimate error rate}).\end{proof} 

\begin{proof}[Proof of Lemma A.1] By using (\ref{eq:v_hat v difference algebra}),
\begin{align}
 & \frac{1}{N\cdot L}\sum_{i,l}\mathbb{\widetilde{T}}_{il}\frac{1}{h^{2}}K_{f}'\left(\frac{V_{il}-v}{h}\right)\left(\widehat{V}_{il}-V_{il}\right)\nonumber \\
= & -\frac{1}{\left(N-1\right)}\frac{1}{N\cdot L}\sum_{i,l}\mathbb{\widetilde{T}}_{il}\frac{1}{h^{2}}K_{f}'\left(\frac{V_{il}-v}{h}\right)\frac{G\left(B_{il}\right)}{g\left(B_{il}\right)^{2}}\left(\widehat{g}\left(B_{il}\right)-g\left(B_{il}\right)\right)+\mathit{\Delta}_{1}^{\ddagger}\left(v\right)+\mathit{\Delta}_{2}^{\ddagger}\left(v\right)+\mathit{\Delta}_{3}^{\ddagger}\left(v\right),\label{eq:lemma 1 leading term decomposition}
\end{align}
where
\begin{gather*}
\mathit{\Delta}_{1}^{\ddagger}\left(v\right)\coloneqq\frac{1}{\left(N-1\right)}\frac{1}{N\cdot L}\sum_{i,l}\mathbb{\widetilde{T}}_{il}\frac{1}{h^{2}}K_{f}'\left(\frac{V_{il}-v}{h}\right)\frac{\left(\widehat{G}\left(B_{il}\right)-G\left(B_{il}\right)\right)}{g\left(B_{il}\right)},\\
\mathit{\Delta}_{2}^{\ddagger}\left(v\right)\coloneqq-\frac{1}{\left(N-1\right)}\frac{1}{N\cdot L}\sum_{i,l}\mathbb{\widetilde{T}}_{il}\frac{1}{h^{2}}K_{f}'\left(\frac{V_{il}-v}{h}\right)\frac{\left(\widehat{G}\left(B_{il}\right)-G\left(B_{il}\right)\right)\left(\widehat{g}\left(B_{il}\right)-g\left(B_{il}\right)\right)}{g\left(B_{il}\right)^{2}}\textrm{ and}\\
\mathit{\Delta}_{3}^{\ddagger}\left(v\right)\coloneqq-\frac{1}{\left(N-1\right)}\frac{1}{N\cdot L}\sum_{i,l}\mathbb{\widetilde{T}}_{il}\frac{1}{h^{2}}K_{f}'\left(\frac{V_{il}-v}{h}\right)\frac{\widehat{G}\left(B_{il}\right)}{\widehat{g}\left(B_{il}\right)}\frac{\left(\widehat{g}\left(B_{il}\right)-g\left(B_{il}\right)\right)^{2}}{g\left(B_{il}\right)^{2}}.
\end{gather*}

We have
\begin{equation}
\underset{v\in I}{\mathrm{sup}}\left|\mathit{\Delta}_{2}^{\ddagger}\left(v\right)\right|=O_{p}\left(\frac{\mathrm{log}\left(L\right)}{Lh^{\nicefrac{3}{2}}}+h^{R}\left(\frac{\mathrm{log}\left(L\right)}{L}\right)^{\nicefrac{1}{2}}\right)\label{eq:lemma 2 I_2 order bound}
\end{equation}
by using the triangle inequality, (\ref{eq:density of bids bounded  from 0}),
(\ref{eq:uniform rates}) and the fact that
\[
\underset{v\in I}{\mathrm{sup}}\,\frac{1}{N\cdot L}\sum_{i,l}\left|\frac{1}{h}K_{f}'\left(\frac{V_{il}-v}{h}\right)\right|\apprle\underset{v\in I}{\mathrm{sup}}\,\frac{1}{N\cdot L}\sum_{i,l}h^{-1}\reallywidecheck{\mathbb{T}}_{il}=O_{p}\left(1\right).
\]
It also follows from the above result, the triangle inequality and
(\ref{eq:difference bw g_hat g bound}) that
\begin{equation}
\underset{v\in I}{\mathrm{sup}}\left|\mathit{\Delta}_{3}^{\ddagger}\left(v\right)\right|=O_{p}\left(\frac{\mathrm{log}\left(L\right)}{Lh^{2}}+h^{2R+1}\right).\label{eq:lemma 2 I_3 order bound}
\end{equation}

Next, we apply the maximal inequalities for empirical processes and
degenerate U-processes to obtain the order bound for $\underset{v\in I}{\mathrm{sup}}\left|\mathit{\Delta}_{1}^{\ddagger}\left(v\right)\right|$.
Since $K_{f}'$ is compactly supported on $\left[-1,1\right]$, the
contribution of the trimmed observations is asymptotically negligible:
\begin{equation}
\underset{v\in I}{\mathrm{sup}}\,\underset{i,l}{\mathrm{max}}\left|\left(\mathbb{\widetilde{T}}_{il}-1\right)K_{f}'\left(\frac{V_{il}-v}{h}\right)\frac{\left(\widehat{G}\left(B_{il}\right)-G\left(B_{il}\right)\right)}{g\left(B_{il}\right)}\right|=0,\textrm{ when \ensuremath{h\leq\overline{\delta}}}.\label{eq:trimming negligible G}
\end{equation}

Define
\[
\mathcal{G}\left(b,b';v\right)\coloneqq\frac{1}{h^{2}}K_{f}'\left(\frac{\xi\left(b\right)-v}{h}\right)\frac{\left(\mathbbm{1}\left(b'\leq b\right)-G\left(b\right)\right)}{g\left(b\right)}.
\]
By the definition of $\widehat{G}$ and (\ref{eq:trimming negligible G}),
\begin{equation}
\mathit{\Delta}_{1}^{\ddagger}\left(v\right)=\frac{1}{\left(N-1\right)}\frac{1}{\left(N\cdot L\right)^{2}}\sum_{\left(2\right)}\mathcal{G}\left(B_{il},B_{jk};v\right)+\frac{1}{\left(N-1\right)}\frac{1}{\left(N\cdot L\right)^{2}}\sum_{i,l}\mathcal{G}\left(B_{il},B_{il};v\right),\,\textrm{for all \ensuremath{v\in I}},\label{eq:I_1 decomposition}
\end{equation}
when $h$ is sufficiently small. The kernel $\mathcal{G}$ satisfies
\[
\mathcal{G}_{1}\left(b;v\right)\coloneqq\int\mathcal{G}\left(b,b';v\right)\mathrm{d}G\left(b'\right)=0\textrm{ and }\mu_{\mathcal{G}}\left(v\right)\coloneqq\int\int\mathcal{G}\left(b,b';v\right)\mathrm{d}G\left(b'\right)\mathrm{d}G\left(b\right)=0.
\]
Also define 
\[
\mathcal{G}_{2}\left(b;v\right)\coloneqq\int\mathcal{G}\left(b',b;v\right)\mathrm{d}G\left(b'\right).
\]

The Hoeffding decomposition yields
\begin{equation}
\frac{1}{\left(N\cdot L\right)_{2}}\sum_{\left(2\right)}\mathcal{G}\left(B_{il},B_{jk};v\right)=\frac{1}{N\cdot L}\sum_{i,l}\mathcal{G}_{2}\left(B_{il};v\right)+\frac{1}{\left(N\cdot L\right)_{2}}\sum_{\left(2\right)}\left\{ \mathcal{G}\left(B_{il},B_{jk};v\right)-\mathcal{G}_{2}\left(B_{il};v\right)\right\} .\label{eq:G Hoeffding decomposition}
\end{equation}

By \citet[Lemma 2.6.16]{van1996weak}, for any positive $h$, the
class $\left\{ K_{f}'\left(\frac{\cdot-v}{h}\right):v\in I\right\} $
is the (pointwise) difference of two Vapnik-Chervonenkis (VC) subgraph
classes of functions on $\left[\underline{v},\overline{v}\right]$:
$\left\{ D_{s}\left(\frac{\cdot-v}{h}\right):v\in I\right\} $, $s\in\left\{ 1,2\right\} $.
Each of these classes has VC index less than or equal to two. Let
\[
\mathscr{D}_{s}\coloneqq\left\{ D_{s}\left(\frac{\xi\left(\cdot\right)-v}{h}\right):v\in I\right\} ,\,s\in\left\{ 1,2\right\} .
\]
By \citet[Lemma 9.9(vii)]{kosorok2007introduction}, each of $\mathscr{D}_{1}$
and $\mathscr{D}_{2}$ has VC index less than or equal to two. Let
$\reallywidecheck{g}\left(b,b'\right)\coloneqq h^{-2}g\left(b\right)^{-1}\left(\mathbbm{1}\left(b'\leq b\right)-G\left(b\right)\right)$.
It then follows from \citet[Lemma 9.9 (vi)]{kosorok2007introduction}
that for any positive $h$, each of the classes $\mathscr{D}_{s}\cdot\reallywidecheck{g}$
($s\in\left\{ 1,2\right\} $) is VC-subgraph with VC index less than
or equal to three. They have constant envelopes $h^{-2}2\overline{C}_{D_{1}}\underline{C}_{g}^{-1}$
and $h^{-2}2\overline{C}_{D_{2}}\underline{C}_{g}^{-1}$ respectively.
\citet[Theorem 3.6.9]{gine2015mathematical} (see also \citealp[Theorem 9.3]{kosorok2007introduction})
yields the following non-asymptotic bound:
\begin{equation}
N\left(\epsilon\left(h^{-2}2\overline{C}_{D_{1}}\underline{C}_{g}^{-1}\right),\mathscr{D}_{s}\cdot\reallywidecheck{g},\left\Vert \cdot\right\Vert _{Q,2}\right)\leq\left(\frac{A}{\epsilon}\right)^{V},\,\textrm{for \ensuremath{\epsilon\in\left(0,1\right]}},\label{eq:D_s*g covering number}
\end{equation}
for any (not necessarily discrete) probability measure $Q$, where
$A>1$ and $V>1$ are universal constants that are independent of
$L$. Now $\mathscr{G}\coloneqq\left\{ \mathcal{G}\left(\cdot,\cdot;v\right):v\in I\right\} $
is a subset of the pointwise difference of $\mathscr{D}_{1}\cdot\reallywidecheck{g}$
and $\mathscr{D}_{2}\cdot\reallywidecheck{g}$. It follows from (\ref{eq:D_s*g covering number})
and \citet[Lemma 16]{nolan1987u} that $\mathscr{G}$ is a (uniform)
VC-type class with respect to the constant envelope 
\begin{equation}
h^{-2}2\left(\overline{C}_{D_{1}}+\overline{C}_{D_{2}}\right)\underline{C}_{g}^{-1}.\label{eq:G constant envelope}
\end{equation}

For the higher-order term in the Hoeffding decomposition (\ref{eq:G Hoeffding decomposition}),
the CK inequality suffices to yield 
\begin{equation}
\mathrm{E}\left[\underset{v\in I}{\mathrm{sup}}\left|\frac{1}{\left(N\cdot L\right)_{2}}\sum_{\left(2\right)}\left\{ \mathcal{G}\left(B_{il},B_{jk};v\right)-\mathcal{G}_{2}\left(B_{il};v\right)\right\} \right|\right]\apprle\left(Lh^{2}\right)^{-1}.\label{eq:G U process expectation bound}
\end{equation}
To obtain a bound for the order of the supremum of the first term
in the Hoeffding decomposition (\ref{eq:G Hoeffding decomposition}),
a sharper maximal inequality for empirical processes is needed. 

Compute
\begin{eqnarray}
\mathrm{E}\left[\mathcal{G}_{2}\left(B_{11};v\right)^{2}\right] & = & \int\left(\int_{\underline{b}}^{\overline{b}}\frac{1}{h^{2}}K_{f}'\left(\frac{\xi\left(b\right)-v}{h}\right)\left(\mathbbm{1}\left(b'\leq b\right)-G\left(b\right)\right)\mathrm{d}b\right)^{2}g\left(b'\right)\mathrm{d}b'\nonumber \\
 & = & \int\left(\int_{\underline{b}}^{\overline{b}}\frac{1}{h^{2}}K_{f}'\left(\frac{\xi\left(b\right)-v}{h}\right)\mathbbm{1}\left(b'\leq b\right)\mathrm{d}b\right)^{2}g\left(b'\right)\mathrm{d}b'-\left(\int_{\underline{b}}^{\overline{b}}\frac{1}{h^{2}}K_{f}'\left(\frac{\xi\left(b\right)-v}{h}\right)G\left(b\right)\mathrm{d}b\right)^{2},\nonumber \\
\label{eq:E=00005BG_2(B)^2=00005D decomposition}
\end{eqnarray}
where we applied the Fubini-Tonelli theorem to obtain the second equality.
Applying the Fubini-Tonelli theorem again, we have
\begin{equation}
\int\left(\int_{\underline{b}}^{\overline{b}}K_{f}'\left(\frac{\xi\left(b\right)-v}{h}\right)\mathbbm{1}\left(b'\leq b\right)\mathrm{d}b\right)^{2}g\left(b'\right)\mathrm{d}b'=\int_{\underline{b}}^{\overline{b}}\int_{\underline{b}}^{\overline{b}}K_{f}'\left(\frac{\xi\left(b'\right)-v}{h}\right)K_{f}'\left(\frac{\xi\left(b\right)-v}{h}\right)G\left(\mathrm{min}\left\{ b',b\right\} \right)\mathrm{d}b'\mathrm{d}b.\label{eq:G(min(b',b)) double integral equality}
\end{equation}
By change of variables $u=\nicefrac{\left(\xi\left(b'\right)-v\right)}{h}$
and $w=\nicefrac{\left(\xi\left(b\right)-v\right)}{h}$, and since
$K_{f}'$ is supported on $\left[-1,1\right]$, we have
\begin{align*}
 & \int_{\underline{b}}^{\overline{b}}\int_{\underline{b}}^{\overline{b}}K_{f}'\left(\frac{\xi\left(b'\right)-v}{h}\right)K_{f}'\left(\frac{\xi\left(b\right)-v}{h}\right)G\left(\mathrm{min}\left\{ b',b\right\} \right)\mathrm{d}b'\mathrm{d}b\\
= & h^{2}\int_{\frac{\underline{v}-v}{h}}^{\frac{\overline{v}-v}{h}}\int_{\frac{\underline{v}-v}{h}}^{\frac{\overline{v}-v}{h}}K_{f}'\left(u\right)K_{f}'\left(w\right)G\left(\mathrm{min}\left\{ s\left(hu+v\right),s\left(hw+v\right)\right\} \right)s'\left(hu+v\right)s'\left(hw+v\right)\mathrm{d}u\mathrm{d}w\\
= & 2h^{2}\int K_{f}'\left(w\right)s'\left(hw+v\right)\left(\int_{-\infty}^{w}K_{f}'\left(u\right)G\left(s\left(hu+v\right)\right)s'\left(hu+v\right)\mathrm{d}u\right)\mathrm{d}w,
\end{align*}
when $h$ is sufficiently small, where the last equality follows from
symmetry. By a mean value expansion, 
\[
\int_{-\infty}^{w}K_{f}'\left(u\right)G\left(s\left(hu+v\right)\right)s'\left(hu+v\right)\mathrm{d}u=\int_{-\infty}^{w}K_{f}'\left(u\right)\left\{ G\left(s\left(v\right)\right)s'\left(v\right)+\left(g\left(s\left(\dot{v}\right)\right)s'\left(\dot{v}\right)^{2}+G\left(s\left(\dot{v}\right)\right)s''\left(\dot{v}\right)\right)hu\right\} \mathrm{d}u
\]
for some mean value $\dot{v}$ (depending on $u$) such that $\left|\dot{v}-v\right|\leq h\left|u\right|$.
Now it follows that
\begin{align}
 & \int\left(\int_{\underline{b}}^{\overline{b}}K_{f}'\left(\frac{\xi\left(b\right)-v}{h}\right)\mathbbm{1}\left(b'\leq b\right)\mathrm{d}b\right)^{2}g\left(b'\right)\mathrm{d}b'\nonumber \\
= & 2G\left(s\left(v\right)\right)s'\left(v\right)h^{2}\left(\int K_{f}'\left(w\right)s'\left(hw+v\right)\int_{-\infty}^{w}K_{f}'\left(u\right)\mathrm{d}u\mathrm{d}w\right)\nonumber \\
 & +2h^{3}\left(\int K_{f}'\left(w\right)s'\left(hw+v\right)\int_{-\infty}^{w}K_{f}'\left(u\right)\left(g\left(s\left(\dot{v}\right)\right)s'\left(\dot{v}\right)^{2}+G\left(s\left(\dot{v}\right)\right)s''\left(\dot{v}\right)\right)u\mathrm{d}u\mathrm{d}w\right).\label{eq:V stat G integral bound 1}
\end{align}
Another mean value expansion with some mean value $\ddot{v}$ (depending
on $w$) such that $\left|\ddot{v}-v\right|\leq h\left|w\right|$
yields
\begin{eqnarray}
\underset{v\in I}{\mathrm{sup}}\left|\int K_{f}'\left(w\right)s'\left(hw+v\right)\int_{-\infty}^{w}K_{f}'\left(u\right)\mathrm{d}u\mathrm{d}w\right| & = & \underset{v\in I}{\mathrm{sup}}\left|\int K_{f}'\left(w\right)\left(s'\left(v\right)+s''\left(\ddot{v}\right)hw\right)\int_{-\infty}^{w}K_{f}'\left(u\right)\mathrm{d}u\mathrm{d}w\right|\nonumber \\
 & \leq & h\left(\underset{u\in\left[\underline{v},\overline{v}\right]}{\mathrm{sup}}\left|s''\left(u\right)\right|\right)\left(\int\left|K_{f}'\left(w\right)K_{f}\left(w\right)w\right|\mathrm{d}w\right),\label{eq:V stat G integral bound 2}
\end{eqnarray}
where the inequality holds when $h$ is sufficiently small and we
used the fact
\[
\int K_{f}'\left(w\right)\int_{-\infty}^{w}K_{f}'\left(u\right)\mathrm{d}u\mathrm{d}w=\int K_{f}'\left(w\right)K_{f}\left(w\right)\mathrm{d}w=0
\]
which holds under our assumption imposed on the kernel functions.
We also have
\begin{align*}
 & \underset{v\in I}{\mathrm{sup}}\left|\int K_{f}'\left(w\right)s'\left(hw+v\right)\int_{-\infty}^{w}K_{f}'\left(u\right)\left(g\left(s\left(\dot{v}\right)\right)s'\left(\dot{v}\right)^{2}+G\left(s\left(\dot{v}\right)\right)s''\left(\dot{v}\right)\right)u\mathrm{d}u\mathrm{d}w\right|\\
\leq & \left(\int\left|K_{f}'\left(w\right)\right|\mathrm{d}w\right)\left(\int\left|K_{f}'\left(u\right)u\right|\mathrm{d}u\right)\overline{C}_{s'}\left(\underset{u\in\left[\underline{v},\overline{v}\right]}{\mathrm{sup}}\left|g\left(s\left(u\right)\right)s'\left(u\right)^{2}+G\left(s\left(u\right)\right)s''\left(u\right)\right|\right),
\end{align*}
when $h$ is sufficiently small, by the definition of $\dot{v}$ and
the fact that $K_{f}'$ is supported on $\left[-1,1\right]$. It follows
from the above result, (\ref{eq:V stat G integral bound 1}), (\ref{eq:V stat G integral bound 2}),
our assumptions imposed on the kernel functions, the continuity of
$s$, $s'$ and $s''$ and the continuity of $g$ and $G$ that
\begin{equation}
\underset{v\in I}{\mathrm{sup}}\int\left(\int_{\underline{b}}^{\overline{b}}K_{f}'\left(\frac{\xi\left(b\right)-v}{h}\right)\mathbbm{1}\left(b'\leq b\right)\mathrm{d}b\right)^{2}g\left(b'\right)\mathrm{d}b'\apprle h^{3},\label{eq:G(min(b',b)) double integral sup bound}
\end{equation}
when $h$ is sufficiently small. The above result and (\ref{eq:E=00005BG_2(B)^2=00005D decomposition})
imply 
\begin{equation}
\underset{v\in I}{\mathrm{sup}}\,\mathrm{E}\left[\mathcal{G}_{2}\left(B_{11};v\right)^{2}\right]\apprle h^{-1},\label{eq:uniform bound E=00005BG_2^2=00005D}
\end{equation}
when $h$ is sufficiently small.

It follows from the fact that $\mathscr{G}$ is a VC-type class with
respect to the constant envelope (\ref{eq:G constant envelope}) and
\citet[Lemma 5.4]{Chen_Kato_U_Process} that $\left\{ \mathcal{G}_{2}\left(\cdot;v\right):v\in I\right\} $
is also VC-type with respect to the constant envelope (\ref{eq:G constant envelope}).
Now an application of the CCK inequality with $\sigma^{2}$ being
the left-hand side of (\ref{eq:uniform bound E=00005BG_2^2=00005D})
and $F$ being (\ref{eq:G constant envelope}) yields
\[
\mathrm{E}\left[\underset{v\in I}{\mathrm{sup}}\left|\frac{1}{N\cdot L}\sum_{i,l}\mathcal{G}_{2}\left(B_{il};v\right)\right|\right]\leq C_{1}\left\{ \left(Lh\right)^{-\nicefrac{1}{2}}\mathrm{log}\left(C_{2}L\right)^{\nicefrac{1}{2}}+\left(Lh^{2}\right)^{-1}\mathrm{log}\left(C_{2}L\right)\right\} =O\left(\left(\frac{\mathrm{log}\left(L\right)}{Lh}\right)^{\nicefrac{1}{2}}\right),
\]
where the inequality is non-asymptotic and holds when $h$ is sufficiently
small. Now the above result, (\ref{eq:G Hoeffding decomposition}),
(\ref{eq:G U process expectation bound}) and Markov's inequality
yield
\[
\underset{v\in I}{\mathrm{sup}}\left|\frac{1}{\left(N\cdot L\right)^{2}}\sum_{\left(2\right)}\mathcal{G}\left(B_{il},B_{jk};v\right)\right|=O_{p}\left(\frac{\mathrm{log}\left(L\right)}{Lh^{2}}+\left(\frac{\mathrm{log}\left(L\right)}{Lh}\right)^{\nicefrac{1}{2}}\right).
\]
Since $\mathscr{G}$ is uniformly bounded by (\ref{eq:G constant envelope}),
\[
\underset{v\in I}{\mathrm{sup}}\left|\frac{1}{\left(N\cdot L\right)^{2}}\sum_{i,l}\mathcal{G}\left(B_{il},B_{il};v\right)\right|\apprle\left(Lh^{2}\right)^{-1}.
\]
Now (\ref{eq:I_1 decomposition}) and these results yield 
\begin{equation}
\underset{v\in I}{\mathrm{sup}}\left|\mathit{\Delta}_{1}^{\ddagger}\left(v\right)\right|=O_{p}\left(\frac{\mathrm{log}\left(L\right)}{Lh^{2}}+\left(\frac{\mathrm{log}\left(L\right)}{Lh}\right)^{\nicefrac{1}{2}}\right).\label{eq:I_1 uniform bound}
\end{equation}

Since $K_{f}'$ is compactly supported on $\left[-1,1\right]$,
\[
\underset{v\in I}{\mathrm{sup}}\,\underset{i,l}{\mathrm{max}}\left|\left(\widetilde{\mathbb{T}}_{il}-1\right)K_{f}'\left(\frac{\xi\left(B_{il}\right)-v}{h}\right)\frac{G\left(B_{il}\right)}{g\left(B_{il}\right)^{2}}\left(\widehat{g}\left(B_{il}\right)-g\left(B_{il}\right)\right)\right|=0,\textrm{ when \ensuremath{h<\overline{\delta}}.}
\]
By the above result, Lemma \ref{lem:lemma 1 Stochastic Expansion},
(\ref{eq:lemma 1 leading term decomposition}), (\ref{eq:lemma 2 I_2 order bound}),
(\ref{eq:lemma 2 I_3 order bound}) and (\ref{eq:I_1 uniform bound}),
we have
\begin{eqnarray*}
\widehat{f}_{GPV}\left(v\right)-f\left(v\right) & = & -\frac{1}{\left(N-1\right)}\frac{1}{N\cdot L}\sum_{i,l}\frac{1}{h^{2}}K_{f}'\left(\frac{\xi\left(B_{il}\right)-v}{h}\right)\frac{G\left(B_{il}\right)}{g\left(B_{il}\right)^{2}}\left(\widehat{g}\left(B_{il}\right)-g\left(B_{il}\right)\right)\\
 &  & +O_{p}\left(\left(\frac{\mathrm{log}\left(L\right)}{Lh}\right)^{\nicefrac{1}{2}}+h^{R}+\frac{\mathrm{log}\left(L\right)}{Lh^{3}}\right),
\end{eqnarray*}
where the remainder term is uniform in $v\in I$. The conclusion now
follows from the definition of $\widehat{g}$. \end{proof} 

\begin{proof}[Proof of Lemma A.2] Again we observe that the maximal
inequalities provided in \citet{chernozhukov2014gaussian}, \citet[Section 5]{Chen_Kato_U_Process}
and \citet{van1996weak} yield non-asymptotic bounds for the suprema
of the absolute values of the terms in the Hoeffding decomposition
\begin{align*}
 & \frac{1}{\left(N\cdot L\right)^{2}}\sum_{i,l}\sum_{j,k}\mathcal{M}\left(B_{il},B_{jk};v\right)\\
= & \mu_{\mathcal{M}}\left(v\right)+\left\{ \frac{1}{N\cdot L}\sum_{i,l}\mathcal{M}_{1}\left(B_{il};v\right)-\mu_{\mathcal{M}}\left(v\right)\right\} +\left\{ \frac{1}{N\cdot L}\sum_{i,l}\mathcal{M}_{2}\left(B_{il};v\right)-\mu_{\mathcal{M}}\left(v\right)\right\} \\
 & +\frac{1}{\left(N\cdot L\right)\left(N\cdot L-1\right)}\sum_{\left(i,l\right)\neq\left(j,k\right)}\left\{ \mathcal{M}\left(B_{il},B_{jk};v\right)-\mathcal{M}_{1}\left(B_{il};v\right)-\mathcal{M}_{2}\left(B_{jk};v\right)+\mu_{\mathcal{M}}\left(v\right)\right\} \\
 & +\frac{1}{\left(N\cdot L\right)^{2}}\sum_{i,l}\mathcal{M}\left(B_{il},B_{il};v\right)-\frac{1}{\left(N\cdot L\right)^{2}\left(N\cdot L-1\right)}\sum_{\left(i,l\right)\neq\left(j,k\right)}\mathcal{M}\left(B_{il},B_{jk};v\right).
\end{align*}

We use the same arguments as those showing that $\mathscr{G}$ is
(uniformly) VC-type to show that the class $\mathscr{M}\coloneqq\left\{ \mathcal{M}\left(\cdot,\cdot;v\right):v\in I\right\} $
is (uniformly) VC-type with respect to the constant envelope
\begin{equation}
h^{-3}\left(\overline{C}_{D_{1}}+\overline{C}_{D_{2}}\right)\underline{C}_{g}^{-2}\overline{C}_{K_{g}}+h^{-2}\left(\overline{C}_{D_{1}}+\overline{C}_{D_{2}}\right)\underline{C}_{g}^{-1}.\label{eq:M constant envelope}
\end{equation}
Then the CK inequality yields
\begin{equation}
\mathrm{E}\left[\underset{v\in I}{\mathrm{sup}}\left|\frac{1}{\left(N\cdot L\right)_{2}}\sum_{\left(2\right)}\left\{ \mathcal{M}\left(B_{il},B_{jk};v\right)-\mathcal{M}_{1}\left(B_{il};v\right)-\mathcal{M}_{2}\left(B_{jk};v\right)+\mu_{\mathcal{M}}\left(v\right)\right\} \right|\right]\apprle\left(Lh^{3}\right)^{-1}.\label{eq:Lemma 3 bound 1}
\end{equation}

Since $\mathscr{M}$ is uniformly bounded by (\ref{eq:M constant envelope}),
we have
\begin{equation}
\underset{v\in I}{\mathrm{sup}}\left|\frac{1}{\left(N\cdot L\right)^{2}}\sum_{i,l}\mathcal{M}\left(B_{il},B_{il};v\right)\right|=O\left(\left(Lh^{3}\right)^{-1}\right)\label{eq:Lemma 3 bound 2}
\end{equation}
and
\begin{equation}
\underset{v\in I}{\mathrm{sup}}\left|\frac{1}{\left(N\cdot L\right)^{2}\left(N\cdot L-1\right)}\sum_{\left(2\right)}\mathcal{M}\left(B_{il},B_{jk};v\right)\right|=O\left(\left(Lh^{3}\right)^{-1}\right).\label{eq:Lemma 3 bound 3}
\end{equation}

By the definition, $\mu_{\mathcal{M}}\left(v\right)$ is given by
\[
\mu_{\mathcal{M}}\left(v\right)\coloneqq\int\int\mathcal{M}\left(b,b';v\right)\mathrm{d}G\left(b\right)\mathrm{d}G\left(b'\right)=-\int_{\underline{b}}^{\overline{b}}\frac{1}{h^{2}}K_{f}'\left(\frac{\xi\left(b\right)-v}{h}\right)\frac{G\left(b\right)\beta\left(b\right)}{g\left(b\right)}\mathrm{d}b
\]
where
\begin{equation}
\beta\left(b\right)\coloneqq\int\left(\frac{1}{h}K_{g}\left(\frac{b'-b}{h}\right)-g\left(b\right)\right)g\left(b'\right)\mathrm{d}b'\label{eq:definition beta}
\end{equation}
denotes the bias of the kernel density estimator for $g\left(b\right)$.
Since we assume that $K_{g}$ is supported on $\left[-1,1\right]$
and differentiable everywhere on $\mathbb{R}$, it is straightforward
to verify that $\beta$ is differentiable on $\left[s\left(v_{l}-\overline{\delta}\right),s\left(v_{u}+\overline{\delta}\right)\right]$
and
\[
\beta'\left(b\right)=\int\left(-\frac{1}{h^{2}}K_{g}'\left(\frac{b'-b}{h}\right)-g'\left(b\right)\right)g\left(b'\right)\mathrm{d}b'
\]
which is the bias of the kernel estimator for the density derivative
$g'\left(b\right)$. By Proposition 1(iv) of GPV, $g$ has $1+R$
continuous derivatives instead of $R$. By a standard argument for
the bias of kernel estimators for the density (see, e.g., \citealp{Newey_Kernel_ET_1994}),
since $K_{g}$ is supported on $\left[-1,1\right]$, for each $b\in\left[s\left(v_{l}-\overline{\delta}\right),s\left(v_{u}+\overline{\delta}\right)\right]$,
\begin{equation}
\left|\beta\left(b\right)\right|\leq\frac{h^{1+R}}{\left(1+R\right)!}\underset{b'\in\left[b-h,b+h\right]}{\mathrm{sup}}\left\vert g^{(1+R)}\left(b'\right)\right\vert \int\left\vert u^{1+R}K_{g}\left(u\right)\right\vert \mathrm{d}u,\label{eq:beta bound-1}
\end{equation}
when $h$ is sufficiently small (so that $\left[b-h,b+h\right]$ is
an inner closed subset of $\left[\underline{b},\overline{b}\right]$).
By change of variable and Taylor expansion, 
\begin{eqnarray}
\underset{b\in\left[s\left(v_{l}-\overline{\delta}\right),s\left(v_{u}+\overline{\delta}\right)\right]}{\mathrm{sup}}\left|\beta'\left(b\right)\right| & = & \underset{b\in\left[s\left(v_{l}-\overline{\delta}\right),s\left(v_{u}+\overline{\delta}\right)\right]}{\mathrm{sup}}\left|\int\frac{1}{h}K_{g}\left(\frac{b'-b}{h}\right)g'\left(b'\right)\mathrm{d}b'-g'\left(b\right)\right|\nonumber \\
 & \leq & \underset{b\in\left[s\left(v_{l}-\overline{\delta}\right),s\left(v_{u}+\overline{\delta}\right)\right]}{\mathrm{sup}}\frac{h^{R}}{R!}\left|\int K_{g}\left(u\right)u^{R}\left(g^{\left(1+R\right)}\left(\dot{b}\right)-g^{\left(1+R\right)}\left(b\right)\right)\mathrm{d}u\right|,\label{eq:sup beta_prime bound}
\end{eqnarray}
when $h$ is sufficiently small, where $\dot{b}$ is the mean value
depending on $u$ with $\left|\dot{b}-b\right|\leq h\left|u\right|$.
Since $g^{\left(1+R\right)}$ is uniformly continuous on any inner
closed subset of $\left[\underline{b},\overline{b}\right]$, the assumption
that $K_{g}$ is supported on $\left[-1,1\right]$ and (\ref{eq:sup beta_prime bound})
imply that $\beta'\left(b\right)=o\left(h^{R}\right)$ uniformly in
$b\in\left[s\left(v_{l}-\overline{\delta}\right),s\left(v_{u}+\overline{\delta}\right)\right]$. 

By change of variable,
\[
\int_{\underline{b}}^{\overline{b}}K_{f}'\left(\frac{\xi\left(b\right)-v}{h}\right)\frac{G\left(b\right)\beta\left(b\right)}{g\left(b\right)}\mathrm{d}b=h\int_{\frac{\underline{v}-v}{h}}^{\frac{\overline{v}-v}{h}}K_{f}'\left(u\right)\frac{G\left(s\left(hu+v\right)\right)\beta\left(s\left(hu+v\right)\right)s'\left(hu+v\right)}{g\left(s\left(hu+v\right)\right)}\mathrm{d}u.
\]
Let $\psi\left(z\right)\coloneqq\nicefrac{G\left(s\left(z\right)\right)s'\left(z\right)}{g\left(s\left(z\right)\right)}$.
By Lemma A1 and Proposition 1 of GPV, both $\psi$ and $\psi'$ are
uniformly continuous on $\left[\underline{v},\overline{v}\right]$.
By a mean value expansion,
\begin{eqnarray}
\int_{\frac{\underline{v}-v}{h}}^{\frac{\overline{v}-v}{h}}K_{f}'\left(u\right)\psi\left(hu+v\right)\beta\left(s\left(hu+v\right)\right)\mathrm{d}u & = & \int_{\frac{\underline{v}-v}{h}}^{\frac{\overline{v}-v}{h}}K_{f}'\left(u\right)\psi\left(v\right)\beta\left(s\left(v\right)\right)\mathrm{d}u\nonumber \\
 &  & +\int_{\frac{\underline{v}-v}{h}}^{\frac{\overline{v}-v}{h}}K_{f}'\left(u\right)\left\{ \psi'\left(\dot{v}\right)\beta\left(s\left(\dot{v}\right)\right)+\psi\left(\dot{v}\right)\beta'\left(s\left(\dot{v}\right)\right)s'\left(\dot{v}\right)\right\} hu\mathrm{d}u,\label{eq:mu mean value expansion}
\end{eqnarray}
where $\dot{v}$ is the mean value depending on $u$ with $\left|\dot{v}-v\right|\leq h\left|u\right|$.
Since $\int K_{f}'\left(u\right)\mathrm{d}u=0$ by symmetry of the
kernel and $K_{f}'$ is compactly supported on $\left[-1,1\right]$,
the first term of the right-hand side of (\ref{eq:mu mean value expansion})
vanishes when $h$ is sufficiently small and for the second term,
we have
\begin{align}
 & \left|\int_{\frac{\underline{v}-v}{h}}^{\frac{\overline{v}-v}{h}}K_{f}'\left(u\right)\left\{ \psi'\left(\dot{v}\right)\beta\left(s\left(\dot{v}\right)\right)+\psi\left(\dot{v}\right)\beta'\left(s\left(\dot{v}\right)\right)s'\left(\dot{v}\right)\right\} u\mathrm{d}u\right|\nonumber \\
\leq & \left(\int\left|K_{f}'\left(u\right)u\right|\mathrm{d}u\right)\left(\underset{u\in\left[v_{l}-\overline{\delta},v_{u}+\overline{\delta}\right]}{\mathrm{sup}}\left|\psi'\left(u\right)\beta\left(s\left(u\right)\right)+\psi\left(u\right)\beta'\left(s\left(u\right)\right)s'\left(u\right)\right|\right),\label{eq:order of mean of u stat}
\end{align}
when $h$ is sufficiently small. 

Since $\beta\left(b\right)$ and $\beta'\left(b\right)$ are $o\left(h^{R}\right)$
uniformly in $b\in\left[s\left(v_{l}-\overline{\delta}\right),s\left(v_{u}+\overline{\delta}\right)\right]$.
It follows from (\ref{eq:mu mean value expansion}) and (\ref{eq:order of mean of u stat})
that
\begin{equation}
\underset{v\in I}{\mathrm{sup}}\left|\mu_{\mathcal{M}}\left(v\right)\right|=\underset{v\in I}{\mathrm{sup}}\left|\int_{\underline{b}}^{\overline{b}}\frac{1}{h^{2}}K_{f}'\left(\frac{\xi\left(b\right)-v}{h}\right)\frac{G\left(b\right)\beta\left(b\right)}{g\left(b\right)}\mathrm{d}b\right|=o\left(h^{R}\right).\label{eq:miu_M uniform rate}
\end{equation}

It is clear from the definition of $\mathcal{M}_{1}$ and the arguments
used in the proof of Lemma A.1 that for any positive $h$ the class
$\left\{ \mathcal{M}_{1}\left(\cdot;v\right):v\in I\right\} $ is
a subset of the difference of two VC-subgraph classes, each of which
has VC index less than or equal to three. \citet[Theorem 3.6.9]{gine2015mathematical}
and \citet[Lemma 16]{nolan1987u} imply that $\left\{ \mathcal{M}_{1}\left(\cdot;v\right):v\in I\right\} $
is (uniformly) VC-type with respect to the constant envelope: 
\begin{equation}
h^{-2}\left(\overline{C}_{D_{1}}+\overline{C}_{D_{2}}\right)\underline{C}_{g}^{-2}\left\{ \underset{b\in\left[s\left(v_{l}-\overline{\delta}\right),s\left(v_{u}+\overline{\delta}\right)\right]}{\mathrm{sup}}\left|\beta\left(b\right)\right|\right\} =O\left(h^{R-1}\right),\label{eq:M_1 constant envelope}
\end{equation}
when $h$ is sufficiently small. The VW inequality suffices to yield
\[
\mathrm{E}\left[\underset{v\in I}{\mathrm{sup}}\left|\frac{1}{N\cdot L}\sum_{i,l}\mathcal{M}_{1}\left(B_{il};v\right)-\mu_{\mathcal{M}}\left(v\right)\right|\right]\apprle\left(Lh^{2}\right)^{-\nicefrac{1}{2}}h^{R},
\]
when $h$ is sufficiently small. The conclusion follows from Lemma
A.1, (\ref{eq:Lemma 3 bound 1}), (\ref{eq:Lemma 3 bound 2}), (\ref{eq:Lemma 3 bound 3}),
(\ref{eq:miu_M uniform rate}), the above inequality and Markov's
inequality. \end{proof}

The following lemma is essentially the same as \citet[online supplement, Lemma S.2]{Marmer_Shneyerov_Quantile_Auctions}.
It is clear from its proof that it suffices to bound the first (absolute)
moment. The proof of the following lemma is almost the same as that
of \citet[Lemma S.2]{Marmer_Shneyerov_Quantile_Auctions} and hence
omitted. 
\begin{lem}
\label{lem:auxiliary 3}Let $\widehat{\vartheta}^{*}$ be a statistic
computed using the bootstrap sample satisfying $\mathrm{E}^{*}\left[\left|\widehat{\vartheta}^{*}\right|\right]=O_{p}\left(\epsilon_{L}\right)$
for some null sequence $\epsilon_{L}\downarrow0$. Then $\widehat{\vartheta}^{*}=O_{p}^{*}\left(\epsilon_{L}\right)$. 
\end{lem}
\begin{proof}[Proof of Lemma A.3]Note that we have $\widetilde{f}\left(v\right)=\mathrm{E}^{*}\left[\widetilde{f}^{*}\left(v\right)\right]$
and also the function class $\left\{ h^{-1}K_{f}\left(\frac{\cdot-v}{h}\right):v\in I\right\} $
is (uniformly) VC-type with respect to the constant envelope $h^{-1}\overline{C}_{K_{f}}$.
Note that
\[
\widehat{\sigma}_{V}^{2}\coloneqq\underset{v\in I}{\mathrm{sup}}\,\mathrm{E}^{*}\left[\frac{1}{h^{2}}K_{f}\left(\frac{V_{11}^{*}-v}{h}\right)^{2}\right]=\underset{v\in I}{\mathrm{sup}}\,\frac{1}{N\cdot L}\sum_{i,l}\frac{1}{h^{2}}K_{f}\left(\frac{V_{il}-v}{h}\right)^{2}\apprle\underset{v\in I}{\mathrm{sup}}\,\frac{1}{N\cdot L}\sum_{i,l}h^{-2}\reallywidecheck{\mathbb{T}}_{il}.
\]
Now it is clear that $\widehat{\sigma}_{V}=O_{p}\left(h^{-\nicefrac{1}{2}}\right)$
follows from (\ref{eq:T_check sup bound 2}). 

Next we apply the CCK inequality with $\sigma=\widehat{\sigma}_{V}$
and $F$ being the constant envelope $h^{-1}\overline{C}_{K_{f}}$.
Observing the non-asymptotic nature of the CCK inequality and applying
it, we have
\[
\mathrm{E}^{*}\left[\underset{v\in I}{\mathrm{sup}}\left|\widetilde{f}^{*}\left(v\right)-\widetilde{f}\left(v\right)\right|\right]\leq C_{1}\left\{ L^{-\nicefrac{1}{2}}\widehat{\sigma}_{V}\mathrm{log}\left(C_{2}L\right)^{\nicefrac{1}{2}}+\left(Lh\right)^{-1}\mathrm{log}\left(L\right)\right\} =O_{p}\left(\left(\frac{\mathrm{log}\left(L\right)}{Lh}\right)^{\nicefrac{1}{2}}\right),
\]
where the inequality is non-asymptotic. The conclusion follows from
the above result and Lemma \ref{lem:auxiliary 3}.\end{proof}

To prove Lemma A.4, we derive intermediate asymptotic expansions that
are empirical bootstrap analogues of those provided in Lemmas \ref{lem:lemma 1 Stochastic Expansion}
and A.1 first. These expansions are given in the following two lemmas.
\begin{lem}
\label{lem:bootstrap lemma 1}Suppose that Assumptions 1 - 3 hold.
Let $\textrm{ }\mathbb{\widetilde{T}}_{il}^{*}\coloneqq\mathbbm{1}\left(\left|V_{il}^{*}-v\right|\leq\overline{\delta}\right)$.
Then
\[
\widehat{f}_{GPV}^{*}\left(v\right)-\widetilde{f}^{*}\left(v\right)=\frac{1}{N\cdot L}\sum_{i,l}\mathbb{\widetilde{T}}_{il}^{*}\frac{1}{h^{2}}K_{f}'\left(\frac{V_{il}^{*}-v}{h}\right)\left(\widehat{V}_{il}^{*}-V_{il}^{*}\right)+O_{p}^{*}\left(\frac{\mathrm{log}\left(L\right)}{Lh^{3}}+h^{R}\right),
\]
where the remainder term is uniform in $v\in I$.
\end{lem}
\begin{proof}[Proof of Lemma \ref{lem:bootstrap lemma 1}]By \citet*[Lemmas 1, S.1 and S.4]{Marmer_Shneyerov_Quantile_Auctions},
we have
\begin{equation}
\underset{b\in\left[\underline{b},\overline{b}\right]}{\mathrm{sup}}\left|\widehat{G}^{*}\left(b\right)-G\left(b\right)\right|=O_{p}^{*}\left(\left(\frac{\mathrm{log}\left(L\right)}{L}\right)^{\nicefrac{1}{2}}\right)\textrm{ and }\underset{b\in\left[\underline{b}+h,\overline{b}-h\right]}{\mathrm{sup}}\left|\widehat{g}^{*}\left(b\right)-g\left(b\right)\right|=O_{p}^{*}\left(\left(\frac{\mathrm{log}\left(L\right)}{Lh}\right)^{\nicefrac{1}{2}}+h^{1+R}\right).\label{eq:bootstrap world uniform convergence}
\end{equation}
Note that it is straightforward to verify that (\ref{eq:density of bids bounded  from 0})
and (\ref{eq:bootstrap world uniform convergence}) imply 
\[
\mathrm{P}^{*}\left[\underset{i,l}{\mathrm{max}}\,\mathbb{T}_{il}^{*}\left|\widehat{g}^{*}\left(B_{il}^{*}\right)^{-1}\right|\leq\left(\frac{\underline{C}_{g}}{2}\right)^{-1}\right]\rightarrow_{p}1,\,\textrm{as \ensuremath{L\uparrow\infty}},
\]
which further implies $\underset{i,l}{\mathrm{max}}\,\mathbb{T}_{il}^{*}\left|\widehat{g}^{*}\left(B_{il}^{*}\right)^{-1}\right|=O_{p}^{*}\left(1\right)$.
Now bootstrap analogues of (\ref{eq:max difference v_hat v feasible})
and (\ref{eq:max difference v_hat v infeasible-1}) can be easily
obtained by using (\ref{eq:bootstrap world uniform convergence})
and this result. We have
\begin{equation}
\underset{i,l}{\mathrm{max}}\,\mathbb{T}_{il}^{*}\left|\widehat{V}_{il}^{*}-V_{il}^{*}\right|=O_{p}^{*}\left(\left(\frac{\mathrm{log}\left(L\right)}{Lh}\right)^{\nicefrac{1}{2}}+h^{1+R}\right)\label{eq:v_star uniform 1 infeasible}
\end{equation}
and
\begin{equation}
\underset{v\in I}{\mathrm{sup}}\,\underset{i,l}{\mathrm{max}}\,\mathbb{\widetilde{T}}_{il}^{*}\left|\widehat{V}_{il}^{*}-V_{il}^{*}\right|=O_{p}^{*}\left(\left(\frac{\mathrm{log}\left(L\right)}{Lh}\right)^{\nicefrac{1}{2}}+h^{1+R}\right).\label{eq:v_star uniform 1 feasible}
\end{equation}

Write
\begin{eqnarray*}
\widehat{f}_{GPV}^{*}\left(v\right) & = & \frac{1}{N\cdot L}\sum_{i,l}\left\{ \mathbb{\widetilde{T}}_{il}^{*}\frac{1}{h}K_{f}\left(\frac{\widehat{V}_{il}^{*}-v}{h}\right)+\mathbb{T}_{il}^{*}\left(1-\mathbb{\widetilde{T}}_{il}^{*}\right)\frac{1}{h}K_{f}\left(\frac{\widehat{V}_{il}^{*}-v}{h}\right)+\left(\mathbb{T}_{il}^{*}-1\right)\mathbb{\widetilde{T}}_{il}^{*}\frac{1}{h}K_{f}\left(\frac{\widehat{V}_{il}^{*}-v}{h}\right)\right\} \\
 & \eqqcolon & \frac{1}{N\cdot L}\sum_{i,l}\mathbb{\widetilde{T}}_{il}^{*}\frac{1}{h}K_{f}\left(\frac{\widehat{V}_{il}^{*}-v}{h}\right)+\kappa_{1}^{*}\left(v\right)+\kappa_{2}^{*}\left(v\right).
\end{eqnarray*}
By the arguments used in the proof of Lemma \ref{lem:lemma 1 Stochastic Expansion},
\[
\mathrm{P}^{*}\left[\underset{v\in I}{\mathrm{sup}}\left|\kappa_{1}^{*}\left(v\right)\right|>0\right]\leq\mathrm{P}^{*}\left[\underset{i,l}{\mathrm{max}}\,\mathbb{T}_{il}^{*}\left|\widehat{V}_{il}^{*}-V_{il}^{*}\right|\geq\frac{\overline{\delta}}{2}\right]=o_{p}\left(1\right),
\]
where the inequality holds when $h$ is sufficiently small. 

Since for all $v\in I$,
\begin{eqnarray*}
\left|\kappa_{2}^{*}\left(v\right)\right| & \apprle & \underset{v\in I}{\mathrm{sup}}\frac{1}{N\cdot L}\sum_{i,l}h^{-1}\left(1-\mathbb{T}_{il}^{*}\right)\mathbb{\widetilde{T}}_{il}^{*}\\
 & \leq & h^{-1}\mathbbm{1}\left(\left|\widehat{\underline{b}}-\underline{b}\right|+\underline{b}+h>s\left(v_{l}-\overline{\delta}\right)\right)+h^{-1}\mathbbm{1}\left(\overline{b}-\left|\widehat{\overline{b}}-\overline{b}\right|-h<s\left(v_{u}+\overline{\delta}\right)\right),
\end{eqnarray*}
by Markov's inequality, 
\begin{eqnarray*}
\mathrm{P}^{*}\left[\underset{v\in I}{\mathrm{sup}}\left|\kappa_{2}^{*}\left(v\right)\right|>0\right] & \leq & \mathbbm{1}\left(\left|\widehat{\underline{b}}-\underline{b}\right|+\underline{b}+h>s\left(v_{l}-\overline{\delta}\right)\right)+\mathbbm{1}\left(\overline{b}-\left|\widehat{\overline{b}}-\overline{b}\right|-h<s\left(v_{u}+\overline{\delta}\right)\right)\\
 & = & o_{p}\left(1\right).
\end{eqnarray*}

Thus we have 
\[
\widehat{f}_{GPV}^{*}\left(v\right)=\frac{1}{N\cdot L}\sum_{i,l}\mathbb{\widetilde{T}}_{il}^{*}\frac{1}{h}K_{f}\left(\frac{\widehat{V}_{il}^{*}-v}{h}\right)+o_{p}^{*}\left(\epsilon_{L}\right),
\]
where the remainder term is uniform in $v\in I$, for any null sequence
$\epsilon_{L}\downarrow0$, and hence is negligible. 

Then by a Taylor expansion and (\ref{eq:negligible trimming}) with
$\widetilde{\mathbb{T}}_{il}$ ($V_{il}$) replaced by their bootstrap
counterparts $\widetilde{\mathbb{T}}_{il}^{*}$ ($V_{il}^{*}$), 
\begin{equation}
\widehat{f}_{GPV}^{*}\left(v\right)-\widetilde{f}^{*}\left(v\right)=\frac{1}{N\cdot L}\sum_{i,l}\mathbb{\widetilde{T}}_{il}^{*}\frac{1}{h^{2}}K_{f}'\left(\frac{V_{il}^{*}-v}{h}\right)\left(\widehat{V}_{il}^{*}-V_{il}^{*}\right)+\frac{1}{2}\cdot\frac{1}{N\cdot L}\sum_{i,l}\mathbb{\widetilde{T}}_{il}^{*}\frac{1}{h^{3}}K_{f}''\left(\frac{\dot{V}_{il}^{*}-v}{h}\right)\left(\widehat{V}_{il}^{*}-V_{il}^{*}\right)^{2}\label{eq:difference f_hat_star f_til_star  bootstrap}
\end{equation}
for some mean value $\dot{V}_{il}^{*}$ that lies on the line joining
$\widehat{V}_{il}^{*}$ and $V_{il}^{*}$, with some remainder error
term that is $o_{p}^{*}\left(\epsilon_{L}\right)$ for any null sequence
$\epsilon_{L}\downarrow0$. 

Since $K_{f}''$ is supported on $\left[-1,1\right]$, by the triangle
inequality, 
\begin{align}
 & \underset{v\in I}{\mathrm{sup}}\left|\frac{1}{N\cdot L}\sum_{i,l}\mathbb{\widetilde{T}}_{il}^{*}\frac{1}{h^{3}}K_{f}''\left(\frac{\dot{V}_{il}^{*}-v}{h}\right)\left(\widehat{V}_{il}^{*}-V_{il}^{*}\right)^{2}\right|\nonumber \\
\apprle & \left\{ \underset{v\in I}{\mathrm{sup}}\,\frac{1}{N\cdot L}\sum_{i,l}h^{-3}\reallywidecheck{\mathbb{T}}_{il}^{*}+\underset{v\in I}{\mathrm{sup}}\left|\kappa_{3}^{*}\left(v\right)\right|\right\} \left\{ \underset{v\in I}{\mathrm{sup}}\,\underset{i,l}{\mathrm{max}}\,\mathbb{\widetilde{T}}_{il}^{*}\left(\widehat{V}_{il}^{*}-V_{il}^{*}\right)^{2}\right\} ,\label{eq:K_double_prime * (V_hat - V)^2 bound bootstrap}
\end{align}
where $\reallywidecheck{\mathbb{T}}_{il}^{*}\coloneqq\mathbbm{1}\left(\left|V_{il}^{*}-v\right|\leq2h\right),$
and
\[
\kappa_{3}^{*}\left(v\right)\coloneqq\frac{1}{N\cdot L}\sum_{i,l}\mathbb{\widetilde{T}}_{il}^{*}h^{-3}\mathbbm{1}\left(\left|V_{il}^{*}-v\right|>2h\right)\mathbbm{1}\left(\left|V_{il}^{*}-v\right|\leq h+\underset{i,l}{\mathrm{max}}\,\mathbb{\widetilde{T}}_{il}^{*}\left|\dot{V}_{il}^{*}-V_{il}^{*}\right|\right).
\]
Then it is clear that
\[
\mathrm{P}^{*}\left[\underset{v\in I}{\mathrm{sup}}\left|\kappa_{3}^{*}\left(v\right)\right|>0\right]\leq\mathrm{P}^{*}\left[\underset{v\in I}{\mathrm{sup}}\,\underset{i,l}{\mathrm{max}}\,\mathbb{\widetilde{T}}_{il}^{*}\left|\dot{V}_{il}^{*}-V_{il}^{*}\right|>h\right]=o_{p}\left(1\right),
\]
where the equality follows from (\ref{eq:v_star uniform 1 feasible}).
Therefore, $\underset{v\in I}{\mathrm{sup}}\left|\kappa_{3}^{*}\left(v\right)\right|=o_{p}^{*}\left(\epsilon_{L}\right)$,
for any null sequence $\epsilon_{L}\downarrow0$. A application of
the CCK inequality and Lemma \ref{lem:auxiliary 3} yields 
\begin{equation}
\underset{v\in I}{\mathrm{sup}}\left|\frac{1}{N\cdot L}\sum_{i,l}h^{-1}\reallywidecheck{\mathbb{T}}_{il}^{*}-\mathrm{E}^{*}\left[h^{-1}\reallywidecheck{\mathbb{T}}_{il}^{*}\right]\right|=O_{p}^{*}\left(\left(\frac{\mathrm{log}\left(L\right)}{Lh}\right)^{\nicefrac{1}{2}}\right).\label{eq:T_check sup bound bootstrap}
\end{equation}
It is argued in the proof of Lemma \ref{lem:lemma 1 Stochastic Expansion}
that $\underset{v\in I}{\mathrm{sup}}\,\mathrm{E}^{*}\left[h^{-1}\reallywidecheck{\mathbb{T}}_{il}^{*}\right]=O_{p}\left(1\right)$.
It then follows from this result, (\ref{eq:v_star uniform 1 feasible}),
(\ref{eq:K_double_prime * (V_hat - V)^2 bound bootstrap}), (\ref{eq:T_check sup bound bootstrap})
and \citet[Lemma S.1]{Marmer_Shneyerov_Quantile_Auctions} that
\[
\underset{v\in I}{\mathrm{sup}}\left|\frac{1}{N\cdot L}\sum_{i,l}\mathbb{\widetilde{T}}_{il}^{*}\frac{1}{h^{3}}K_{f}''\left(\frac{\dot{V}_{il}^{*}-v}{h}\right)\left(\widehat{V}_{il}^{*}-V_{il}^{*}\right)^{2}\right|=O_{p}^{*}\left(\frac{\mathrm{log}\left(L\right)}{Lh^{3}}+h^{2R}\right).
\]
The conclusion follows from the above result and (\ref{eq:difference f_hat_star f_til_star  bootstrap}).
\end{proof}

\begin{lem}
\label{lem:bootstrap lemma 2}Suppose that Assumptions 1 - 3 hold.
Then 
\[
\widehat{f}_{GPV}^{*}\left(v\right)-\widetilde{f}^{*}\left(v\right)=\frac{1}{\left(N-1\right)}\frac{1}{\left(N\cdot L\right)^{2}}\sum_{i,l}\sum_{j,k}\mathcal{M}\left(B_{il}^{*},B_{jk}^{*};v\right)+O_{p}^{*}\left(\left(\frac{\mathrm{log}\left(L\right)}{Lh}\right)^{\nicefrac{1}{2}}+\frac{\mathrm{log}\left(L\right)}{Lh^{3}}+h^{R}\right),
\]
where the remainder term is uniform in $v\in I$.
\end{lem}
\begin{proof}[Proof of Lemma \ref{lem:bootstrap lemma 2}] By using
(\ref{eq:v_hat v difference algebra}) with all objects replaced by
their bootstrap counterparts, we have
\begin{align}
 & \frac{1}{N\cdot L}\sum_{i,l}\mathbb{\widetilde{T}}_{il}^{*}\frac{1}{h^{2}}K_{f}'\left(\frac{V_{il}^{*}-v}{h}\right)\left(\widehat{V}_{il}^{*}-V_{il}^{*}\right)\nonumber \\
= & -\frac{1}{\left(N-1\right)}\frac{1}{N\cdot L}\sum_{i,l}\mathbb{\widetilde{T}}_{il}^{*}\frac{1}{h^{2}}K_{f}'\left(\frac{V_{il}^{*}-v}{h}\right)\frac{G\left(B_{il}^{*}\right)}{g\left(B_{il}^{*}\right)^{2}}\left(\widehat{g}^{*}\left(B_{il}^{*}\right)-g\left(B_{il}^{*}\right)\right)+\mathit{\Delta}_{1}^{*}\left(v\right)+\mathit{\Delta}_{2}^{*}\left(v\right)+\mathit{\Delta}_{3}^{*}\left(v\right),\label{eq:lemma 1 leading term decomposition bootstrap}
\end{align}
where
\[
\mathit{\Delta}_{1}^{*}\left(v\right)\coloneqq\frac{1}{\left(N-1\right)}\frac{1}{N\cdot L}\sum_{i,l}\mathbb{\widetilde{T}}_{il}^{*}\frac{1}{h^{2}}K_{f}'\left(\frac{V_{il}^{*}-v}{h}\right)\frac{\left(\widehat{G}^{*}\left(B_{il}^{*}\right)-G\left(B_{il}^{*}\right)\right)}{g\left(B_{il}^{*}\right)},
\]
and the order bounds
\[
\underset{v\in I}{\mathrm{sup}}\left|\mathit{\Delta}_{2}^{*}\left(v\right)\right|=O_{p}^{*}\left(\frac{\mathrm{log}\left(L\right)}{Lh^{\nicefrac{3}{2}}}+h^{R}\left(\frac{\mathrm{log}\left(L\right)}{L}\right)^{\nicefrac{1}{2}}\right)\textrm{ and }\underset{v\in I}{\mathrm{sup}}\left|\mathit{\Delta}_{3}^{*}\left(v\right)\right|=O_{p}^{*}\left(\frac{\mathrm{log}\left(L\right)}{Lh^{2}}+h^{2R+1}\right)
\]
can be easily obtained by using (\ref{eq:bootstrap world uniform convergence})
and the fact 
\[
\underset{v\in I}{\mathrm{sup}}\frac{1}{N\cdot L}\sum_{i,l}\left|\frac{1}{h}K_{f}'\left(\frac{V_{il}^{*}-v}{h}\right)\right|\apprle\underset{v\in I}{\mathrm{sup}}\frac{1}{N\cdot L}\sum_{i,l}h^{-1}\reallywidecheck{\mathbb{T}}_{il}^{*}=O_{p}^{*}\left(1\right).
\]

Since $K_{f}'$ is supported on $\left[-1,1\right]$, by (\ref{eq:trimming negligible G})
with all objects replaced by their bootstrap counterparts, we have
\[
\mathit{\Delta}_{1}^{*}\left(v\right)=\frac{1}{\left(N-1\right)}\frac{1}{\left(N\cdot L\right)^{2}}\sum_{i,l}\sum_{j,k}\mathcal{G}\left(B_{il}^{*},B_{jk}^{*};v\right),\,\textrm{for all \ensuremath{v\in I}},
\]
when $h$ is sufficiently small. Note that the conditional distribution
of each of $\left\{ B_{il}^{*}:i=1,...,N,\,l=1,...,L\right\} $ is
$\widehat{G}$. Let 
\begin{equation}
\widehat{\mu}_{\mathcal{G}}\left(v\right)\coloneqq\int\int\mathcal{G}\left(b,b';v\right)\mathrm{d}\widehat{G}\left(b\right)\mathrm{d}\widehat{G}\left(b'\right)=\mathit{\Delta}_{1}^{\ddagger}\left(v\right)=O_{p}\left(\frac{\mathrm{log}\left(L\right)}{Lh^{2}}+\left(\frac{\mathrm{log}\left(L\right)}{Lh}\right)^{\nicefrac{1}{2}}\right),\label{eq:miu_hat_G stochastic bound}
\end{equation}
where the last equality is uniform in $v\in I$ and shown in the proof
of Lemma A.1. Let 
\[
\mathcal{\widehat{G}}_{1}\left(b;v\right)\coloneqq\int\mathcal{G}\left(b,b';v\right)\mathrm{d}\widehat{G}\left(b'\right)=\frac{1}{h^{2}}K_{f}'\left(\frac{\xi\left(b\right)-v}{h}\right)\frac{\widehat{G}\left(b\right)-G\left(b\right)}{g\left(b\right)}
\]
and 
\[
\mathcal{\widehat{G}}_{2}\left(b;v\right)=\frac{1}{N\cdot L}\sum_{j,k}\frac{1}{h^{2}}K_{f}'\left(\frac{\xi\left(B_{jk}\right)-v}{h}\right)\frac{\mathbbm{1}\left(b\leq B_{jk}\right)-G\left(B_{jk}\right)}{g\left(B_{jk}\right)}.
\]

The Hoeffding decomposition gives 
\begin{eqnarray}
\frac{1}{\left(N\cdot L\right)^{2}}\sum_{i,l}\sum_{j,k}\mathcal{G}\left(B_{il}^{*},B_{jk}^{*};v\right) & = & \widehat{\mu}_{\mathcal{G}}\left(v\right)+\left\{ \frac{1}{N\cdot L}\sum_{i,l}\mathcal{\widehat{G}}_{1}\left(B_{il}^{*};v\right)-\widehat{\mu}_{\mathcal{G}}\left(v\right)\right\} +\left\{ \frac{1}{N\cdot L}\sum_{i,l}\mathcal{\widehat{G}}_{2}\left(B_{il}^{*};v\right)-\widehat{\mu}_{\mathcal{G}}\left(v\right)\right\} \nonumber \\
 &  & +\frac{1}{\left(N\cdot L\right)_{2}}\sum_{\left(2\right)}\left\{ \mathcal{G}\left(B_{il}^{*},B_{jk}^{*};v\right)-\mathcal{\widehat{G}}_{1}\left(B_{il}^{*};v\right)-\mathcal{\widehat{G}}_{2}\left(B_{jk}^{*};v\right)+\widehat{\mu}_{\mathcal{G}}\left(v\right)\right\} \nonumber \\
 &  & +\frac{1}{\left(N\cdot L\right)^{2}}\sum_{i,l}\mathcal{G}\left(B_{il}^{*},B_{il}^{*};v\right)-\frac{1}{\left(N\cdot L\right)^{2}\left(N\cdot L-1\right)}\sum_{\left(2\right)}\mathcal{G}\left(B_{il}^{*},B_{jk}^{*};v\right).\label{eq:G Hoeffding decomposition bootstrap world}
\end{eqnarray}

We argued that $\mathscr{G}$ is a (uniform) VC-type class with respect
to the constant envelope (\ref{eq:G constant envelope}). Due to the
non-asymptotic nature of the maximal inequalities we invoked in the
proofs, we can apply the same inequalities in the ``bootstrap world''
combined with Lemma \ref{lem:auxiliary 3} to obtain the desired result.
The CK inequality yields the following non-asymptotic bound:
\begin{equation}
\mathrm{E}^{*}\left[\underset{v\in I}{\mathrm{sup}}\left|\frac{1}{\left(N\cdot L\right)_{2}}\sum_{\left(2\right)}\left\{ \mathcal{G}\left(B_{il}^{*},B_{jk}^{*};v\right)-\mathcal{\widehat{G}}_{1}\left(B_{il}^{*};v\right)-\mathcal{\widehat{G}}_{2}\left(B_{jk}^{*};v\right)+\widehat{\mu}_{\mathcal{G}}\left(v\right)\right\} \right|\right]\apprle\left(Lh^{2}\right)^{-1}.\label{eq:G U-process supremum bound bootstrap world}
\end{equation}

It is clear from the definition of $\mathcal{\widehat{G}}_{1}$ and
the standard arguments that (conditionally on the original sample,
in the bootstrap world) for any positive $h$ the (non-random) class
$\left\{ \mathcal{\widehat{G}}_{1}\left(\cdot;v\right):v\in I\right\} $
is a subset of the difference of two VC-subgraph classes, each of
which has VC index less than or equal to three. \citet[Theorem 3.6.9]{gine2015mathematical}
and \citet[Lemma 16]{nolan1987u} imply that $\left\{ \mathcal{\widehat{G}}_{1}\left(\cdot;v\right):v\in I\right\} $
is (uniformly) VC-type with respect to the (conditionally) constant
envelope: 
\[
h^{-2}\left(\overline{C}_{D_{1}}+\overline{C}_{D_{2}}\right)\underline{C}_{g}^{-1}\left\{ \underset{b\in\mathbb{R}}{\mathrm{sup}}\left|\widehat{G}\left(b\right)-G\left(b\right)\right|\right\} .
\]
Now the VW inequality yields:
\begin{equation}
\mathrm{E}^{*}\left[\underset{v\in I}{\mathrm{sup}}\left|\frac{1}{N\cdot L}\sum_{i,l}\mathcal{\widehat{G}}_{1}\left(B_{il}^{*};v\right)-\widehat{\mu}_{\mathcal{G}}\left(v\right)\right|\right]\apprle\frac{1}{L^{\nicefrac{1}{2}}h^{2}}\left\{ \underset{b\in\mathbb{R}}{\mathrm{sup}}\left|\widehat{G}\left(b\right)-G\left(b\right)\right|\right\} =O_{p}\left(\frac{\mathrm{log}\left(L\right)^{\nicefrac{1}{2}}}{Lh^{2}}\right),\label{eq:G_hat_1 empirical process supremum expectation bound}
\end{equation}
where the inequality is non-asymptotic.

Note that 
\[
\frac{1}{N\cdot L}\sum_{i,l}\mathcal{\widehat{G}}_{2}\left(B_{il}^{*};v\right)-\widehat{\mu}_{\mathcal{G}}\left(v\right)=\frac{1}{N\cdot L}\sum_{i,l}\mathcal{\widehat{G}}_{2}^{\ddagger}\left(B_{il}^{*};v\right)-\mathrm{E}^{*}\left[\mathcal{\widehat{G}}_{2}^{\ddagger}\left(B_{11}^{*};v\right)\right],
\]
where 
\[
\mathcal{\widehat{G}}_{2}^{\ddagger}\left(b;v\right)\coloneqq\frac{1}{N\cdot L}\sum_{j,k}\frac{1}{h^{2}}K_{f}'\left(\frac{\xi\left(B_{jk}\right)-v}{h}\right)\frac{\mathbbm{1}\left(b\leq B_{jk}\right)}{g\left(B_{jk}\right)},
\]
since 
\[
\mathcal{\widehat{G}}_{2}^{\ddagger}\left(B_{il}^{*};v\right)-\widehat{\mu}_{\mathcal{G}}\left(v\right)=\mathcal{\widehat{G}}_{2}^{\ddagger}\left(B_{il}^{*};v\right)-\mathrm{E}^{*}\left[\mathcal{\widehat{G}}_{2}^{\ddagger}\left(B_{11}^{*};v\right)\right],\textrm{ for all \ensuremath{i=1,...,N} and \ensuremath{l=1,...,L}}.
\]
It is easy to check that the class $\left\{ \mathcal{\widehat{G}}_{2}^{\ddagger}\left(\cdot;v\right):v\in I\right\} $
is non-random conditionally on the original sample and VC-type with
respect to the constant envelope (\ref{eq:G constant envelope}) by
\citet[Lemma 5.4]{Chen_Kato_U_Process}. Compute
\begin{eqnarray*}
\mathrm{E}^{*}\left[\mathcal{\widehat{G}}_{2}^{\ddagger}\left(B_{11}^{*};v\right)^{2}\right] & = & \frac{1}{\left(N\cdot L\right)^{3}}\sum_{i,l}\sum_{j,k}\sum_{j',k'}\frac{1}{h^{4}}K_{f}'\left(\frac{\xi\left(B_{jk}\right)-v}{h}\right)\frac{\mathbbm{1}\left(B_{il}\leq B_{jk}\right)}{g\left(B_{jk}\right)}K_{f}'\left(\frac{\xi\left(B_{j'k'}\right)-v}{h}\right)\frac{\mathbbm{1}\left(B_{il}\leq B_{j'k'}\right)}{g\left(B_{j'k'}\right)}\\
 & \eqqcolon & \frac{1}{\left(N\cdot L\right)^{3}}\sum_{i,l}\sum_{j,k}\sum_{j',k'}\mathcal{J}\left(B_{il},B_{jk},B_{j'k'};v\right).
\end{eqnarray*}

Now by observing that $\mathcal{J}$ is symmetric with respect to
the second the the third arguments and the V-statistic decomposition
argument of \citet[5.7.3]{Serfling_Approximation_Theorems}, 
\begin{align}
 & \frac{1}{\left(N\cdot L\right)^{3}}\sum_{i,l}\sum_{j,k}\sum_{j',k'}\mathcal{J}\left(B_{il},B_{jk},B_{j'k'};v\right)\nonumber \\
= & \frac{1}{\left(N\cdot L\right)_{3}}\sum_{\left(3\right)}\mathcal{J}\left(B_{il},B_{jk},B_{j'k'};v\right)\nonumber \\
 & +\frac{O\left(L^{-1}\right)}{3\left(N\cdot L\right)^{2}-2\left(N\cdot L\right)}\left\{ \sum_{\left(2\right)}\left(2\mathcal{J}\left(B_{il},B_{il},B_{jk};v\right)+\mathcal{J}\left(B_{jk},B_{il},B_{il};v\right)\right)+\sum_{i,l}\mathcal{J}\left(B_{il},B_{il},B_{il};v\right)\right\} .\label{eq:E=00005BG_hat_2(B_11^*)=00005D V-stat decomposition}
\end{align}

Define 
\begin{gather}
\mathcal{J}_{1}^{\left(1\right)}\left(b;v\right)\coloneqq\int\int\mathcal{J}\left(b,b',b'';v\right)\mathrm{d}G\left(b'\right)\mathrm{d}G\left(b''\right),\nonumber \\
\mathcal{J}_{2}^{\left(1\right)}\left(b;v\right)\coloneqq\int\int\mathcal{J}\left(b',b,b'';v\right)\mathrm{d}G\left(b'\right)\mathrm{d}G\left(b''\right),\nonumber \\
\mathcal{J}_{3}^{\left(1\right)}\left(b;v\right)\coloneqq\int\int\mathcal{J}\left(b',b'',b;v\right)\mathrm{d}G\left(b'\right)\mathrm{d}G\left(b''\right),\label{eq:K Hoeffding decomposition def 1}
\end{gather}
\begin{gather}
\mathcal{J}_{1}^{\left(2\right)}\left(b,b';v\right)\coloneqq\int\mathcal{J}\left(b,b',b'';v\right)\mathrm{d}G\left(b''\right),\nonumber \\
\mathcal{J}_{2}^{\left(2\right)}\left(b,b';v\right)\coloneqq\int\mathcal{J}\left(b,b'',b';v\right)\mathrm{d}G\left(b''\right),\nonumber \\
\mathcal{J}_{3}^{\left(2\right)}\left(b,b';v\right)\coloneqq\int\mathcal{J}\left(b'',b,b';v\right)\mathrm{d}G\left(b''\right)\label{eq:K Hoeffding decomposition def 2}
\end{gather}
and 
\begin{equation}
\mu_{\mathcal{J}}\left(v\right)\coloneqq\int\int\int\mathcal{J}\left(b,b',b'';v\right)\mathrm{d}G\left(b\right)\mathrm{d}G\left(b'\right)\mathrm{d}G\left(b''\right).\label{eq:K Hoeffding decomposition def 3}
\end{equation}

The Hoeffding decomposition yields 
\begin{eqnarray}
\frac{1}{\left(N\cdot L\right)_{3}}\underset{\left(3\right)}{\sum}\mathcal{J}\left(B_{il},B_{jk},B_{j'k'};v\right) & = & \mu_{\mathcal{J}}\left(v\right)+\frac{1}{N\cdot L}\sum_{i,l}\left(\mathcal{\mathcal{J}}_{1}^{\left(1\right)}\left(B_{il};v\right)-\mu_{\mathcal{\mathcal{J}}}\left(v\right)\right)+\frac{1}{N\cdot L}\sum_{i,l}\left(\mathcal{\mathcal{J}}_{2}^{\left(1\right)}\left(B_{il};v\right)-\mu_{\mathcal{\mathcal{J}}}\left(v\right)\right)\nonumber \\
 &  & +\frac{1}{N\cdot L}\sum_{i,l}\left(\mathcal{\mathcal{J}}_{3}^{\left(1\right)}\left(B_{il};v\right)-\mu_{\mathcal{\mathcal{J}}}\left(v\right)\right)+\Upsilon_{\mathcal{\mathcal{J}}}^{1}\left(v\right)+\Upsilon_{\mathcal{\mathcal{J}}}^{2}\left(v\right)+\Upsilon_{\mathcal{\mathcal{J}}}^{3}\left(v\right)+\Psi_{\mathcal{\mathcal{J}}}\left(v\right),\label{eq:J Hoeffding decomposition-1}
\end{eqnarray}
where $\Upsilon_{\mathcal{\mathcal{J}}}^{1}\left(v\right)$, $\Upsilon_{\mathcal{\mathcal{J}}}^{2}\left(v\right)$
and $\Upsilon_{\mathcal{\mathcal{J}}}^{3}\left(v\right)$ are degenerate
U-statistics of order two and $\Psi_{\mathcal{\mathcal{J}}}\left(v\right)$
is a degenerate U-statistic of order three:
\begin{gather}
\Upsilon_{\mathcal{\mathcal{J}}}^{1}\left(v\right)\coloneqq\frac{1}{\left(N\cdot L\right)_{2}}\sum_{\left(2\right)}\left\{ \mathcal{\mathcal{J}}_{1}^{\left(2\right)}\left(B_{il},B_{jk};v\right)-\mathcal{\mathcal{J}}_{1}^{\left(1\right)}\left(B_{il};v\right)-\mathcal{\mathcal{J}}_{2}^{\left(1\right)}\left(B_{jk};v\right)+\mu_{\mathcal{\mathcal{J}}}\left(v\right)\right\} ,\nonumber \\
\Upsilon_{\mathcal{\mathcal{J}}}^{2}\left(v\right)\coloneqq\frac{1}{\left(N\cdot L\right)_{2}}\sum_{\left(2\right)}\left\{ \mathcal{\mathcal{J}}_{2}^{\left(2\right)}\left(B_{il},B_{jk};v\right)-\mathcal{\mathcal{J}}_{1}^{\left(1\right)}\left(B_{il};v\right)-\mathcal{\mathcal{J}}_{3}^{\left(1\right)}\left(B_{jk};v\right)+\mu_{\mathcal{\mathcal{J}}}\left(v\right)\right\} ,\nonumber \\
\Upsilon_{\mathcal{\mathcal{J}}}^{3}\left(v\right)\coloneqq\frac{1}{\left(N\cdot L\right)_{2}}\sum_{\left(2\right)}\left\{ \mathcal{\mathcal{J}}_{3}^{\left(2\right)}\left(B_{il},B_{jk};v\right)-\mathcal{\mathcal{J}}_{2}^{\left(1\right)}\left(B_{il};v\right)-\mathcal{\mathcal{J}}_{3}^{\left(1\right)}\left(B_{jk};v\right)+\mu_{\mathcal{\mathcal{J}}}\left(v\right)\right\} \label{eq:K Hoeffding decomposition def 4}
\end{gather}
and
\begin{eqnarray}
\Psi_{\mathcal{\mathcal{J}}}\left(v\right) & \coloneqq & \frac{1}{\left(N\cdot L\right)_{3}}\sum_{\left(3\right)}\left\{ \mathcal{\mathcal{J}}\left(B_{il},B_{jk},B_{j'k'};v\right)-\mathcal{\mathcal{J}}_{1}^{\left(2\right)}\left(B_{il},B_{jk};v\right)-\mathcal{\mathcal{J}}_{2}^{\left(2\right)}\left(B_{il},B_{j'k'};v\right)-\mathcal{\mathcal{J}}_{3}^{\left(2\right)}\left(B_{jk},B_{j'k'};v\right)\right.\nonumber \\
 &  & \left.+\mathcal{\mathcal{J}}_{1}^{\left(1\right)}\left(B_{il};v\right)+\mathcal{\mathcal{J}}_{2}^{\left(1\right)}\left(B_{jk};v\right)+\mathcal{\mathcal{J}}_{3}^{\left(1\right)}\left(B_{j'k'};v\right)-\mu_{\mathcal{\mathcal{J}}}\left(v\right)\right\} .\label{eq:K Hoeffding decomposition def 5}
\end{eqnarray}

It is easy to check that 
\[
\mu_{\mathcal{J}}\left(v\right)=\int_{\underline{b}}^{\overline{b}}\int_{\underline{b}}^{\overline{b}}\frac{1}{h^{4}}K_{f}'\left(\frac{\xi\left(b\right)-v}{h}\right)K_{f}'\left(\frac{\xi\left(b'\right)-v}{h}\right)G\left(\mathrm{min}\left\{ b,b'\right\} \right)\mathrm{d}b\mathrm{d}b'.
\]
It is shown in the proof of Lemma A.1 that
\begin{equation}
\underset{v\in I}{\mathrm{sup}}\left|\mu_{\mathcal{J}}\left(v\right)\right|=O\left(h^{-1}\right),\label{eq:miu_J sup bound}
\end{equation}
see (\ref{eq:G(min(b',b)) double integral equality}) and (\ref{eq:G(min(b',b)) double integral sup bound}).

Next, we obtain the order bounds for the suprema of the absolute values
of the remainder terms in the Hoeffding decomposition (\ref{eq:J Hoeffding decomposition-1}).
Firstly, we observe that $\mathscr{J}\coloneqq\left\{ \mathcal{J}\left(\cdot,\cdot,\cdot;v\right):v\in I\right\} $
is (uniformly) VC-type with respect to the constant envelope $h^{-4}\underline{C}_{g}^{-2}\left(\overline{C}_{D_{1}}+\overline{C}_{D_{2}}\right)^{2}$
by using the same arguments applied to show that $\mathscr{K}$ is
VC-type in the proof of Theorem 4.1. The CK inequality yields 
\begin{equation}
\mathrm{E}\left[\underset{v\in I}{\mathrm{sup}}\left|\Upsilon_{\mathcal{J}}^{k}\left(v\right)\right|\right]\apprle\left(Lh^{4}\right)^{-1},\textrm{ for }k=1,2,3\textrm{ and }\mathrm{E}\left[\underset{v\in I}{\mathrm{sup}}\left|\Psi_{\mathcal{J}}\left(v\right)\right|\right]\apprle L^{-\nicefrac{3}{2}}h^{-4}.\label{eq:UPSILON PSI J rate}
\end{equation}

Since $\mathscr{J}$ is VC-type with respect to the constant envelope
$h^{-4}\underline{C}_{g}^{-2}\left(\overline{C}_{D_{1}}+\overline{C}_{D_{2}}\right)^{2}$,
the fact that $\left\{ \mathcal{\mathcal{J}}_{k}^{\left(1\right)}\left(\cdot;v\right):v\in I\right\} $,
$k=1,2,3$ are all VC-type with respect to the same constant envelope
follows from \citet[Lemma 5.4]{Chen_Kato_U_Process}. Now by change
of variable and using the fact that $K_{f}'$ is compactly supported
on $\left[-1,1\right]$, 
\begin{eqnarray*}
\underset{v\in I}{\mathrm{sup}}\,\mathrm{E}\left[\mathcal{\mathcal{J}}_{1}^{\left(1\right)}\left(B_{11};v\right)^{2}\right] & = & \underset{v\in I}{\mathrm{sup}}\int\left\{ \int_{\underline{b}}^{\overline{b}}\frac{1}{h^{2}}K_{f}'\left(\frac{\xi\left(b'\right)-v}{h}\right)\mathbbm{1}\left(b\leq b'\right)\mathrm{d}b'\right\} ^{4}\mathrm{d}G\left(b\right)\\
 & \leq & \underset{v\in I}{\mathrm{sup}}\left\{ \int_{\underline{b}}^{\overline{b}}\frac{1}{h^{2}}\left|K_{f}'\left(\frac{\xi\left(b'\right)-v}{h}\right)\right|\mathrm{d}b'\right\} ^{4}\\
 & \apprle & h^{-4}
\end{eqnarray*}
and 
\begin{eqnarray*}
\underset{v\in I}{\mathrm{sup}}\,\mathrm{E}\left[\mathcal{\mathcal{J}}_{2}^{\left(1\right)}\left(B_{11};v\right)^{2}\right] & = & \underset{v\in I}{\mathrm{sup}}\int_{\underline{b}}^{\overline{b}}K_{f}'\left(\frac{\xi\left(b\right)-v}{h}\right)^{2}\frac{1}{g\left(b\right)}\left\{ \int_{\underline{b}}^{\overline{b}}\frac{1}{h^{4}}K_{f}'\left(\frac{\xi\left(b'\right)-v}{h}\right)G\left(\mathrm{min}\left\{ b,b'\right\} \right)\mathrm{d}b'\right\} ^{2}\mathrm{d}b\\
 & \apprle & h^{-5},
\end{eqnarray*}
where the last inequality holds when $h$ is sufficiently small. Now
the CCK inequality with $\sigma^{2}$ being 
\[
\underset{v\in I}{\mathrm{sup}}\,\mathrm{E}\left[\mathcal{\mathcal{J}}_{1}^{\left(1\right)}\left(B_{11};v\right)^{2}\right]\textrm{ or }\underset{v\in I}{\mathrm{sup}}\,\mathrm{E}\left[\mathcal{\mathcal{J}}_{2}^{\left(1\right)}\left(B_{11};v\right)^{2}\right]
\]
and $F$ being $h^{-4}\underline{C}_{g}^{-2}\left(\overline{C}_{D_{1}}+\overline{C}_{D_{2}}\right)^{2}$
yield 
\begin{gather*}
\mathrm{E}\left[\underset{v\in I}{\mathrm{sup}}\left|\frac{1}{N\cdot L}\sum_{i,l}\left(\mathcal{J}_{1}^{\left(1\right)}\left(B_{il};v\right)-\mu_{\mathcal{J}}\left(v\right)\right)\right|\right]\leq C_{1}\left\{ \left(Lh^{4}\right)^{-\nicefrac{1}{2}}\mathrm{log}\left(C_{2}L\right)^{\nicefrac{1}{2}}+\left(Lh^{4}\right)^{-1}\mathrm{log}\left(C_{2}L\right)\right\} \textrm{ and}\\
\mathrm{E}\left[\underset{v\in I}{\mathrm{sup}}\left|\frac{1}{N\cdot L}\sum_{i,l}\left(\mathcal{J}_{2}^{\left(1\right)}\left(B_{il};v\right)-\mu_{\mathcal{J}}\left(v\right)\right)\right|\right]\leq C_{1}\left\{ \left(Lh^{5}\right)^{-\nicefrac{1}{2}}\mathrm{log}\left(C_{2}L\right)^{\nicefrac{1}{2}}+\left(Lh^{4}\right)^{-1}\mathrm{log}\left(C_{2}L\right)\right\} ,
\end{gather*}
when $h$ is sufficiently small. Note that the second inequality holds
if $\mathcal{J}_{2}^{\left(1\right)}$ is replaced by $\mathcal{J}_{3}^{\left(1\right)}$
since $\mathcal{J}\left(\cdot,\cdot,\cdot;v\right)$ is symmetric
with respect to the second and the third arguments. Now 
\begin{equation}
\underset{v\in I}{\mathrm{sup}}\left|\frac{1}{\left(N\cdot L\right)_{3}}\underset{\left(3\right)}{\sum}\mathcal{J}\left(B_{il},B_{jk},B_{j'k'};v\right)\right|=O_{p}\left(h^{-1}+\left(\frac{\mathrm{log}\left(L\right)}{Lh^{5}}\right)^{\nicefrac{1}{2}}+\frac{\mathrm{log}\left(L\right)}{Lh^{4}}\right)\label{eq:J U-stat stochastic order}
\end{equation}
follows from these inequalities, Markov's inequality, (\ref{eq:J Hoeffding decomposition-1}),
(\ref{eq:miu_J sup bound}) and (\ref{eq:UPSILON PSI J rate}). 

\[
\underset{v\in I}{\mathrm{sup}}\left|\frac{1}{3\left(N\cdot L\right)^{2}-2\left(N\cdot L\right)}\sum_{\left(2\right)}\left(2\mathcal{J}\left(B_{il},B_{il},B_{jk};v\right)+\mathcal{J}\left(B_{jk},B_{il},B_{il};v\right)\right)\right|\apprle h^{-4}
\]
and 
\[
\underset{v\in I}{\mathrm{sup}}\left|\frac{1}{N\cdot L}\sum_{i,l}\mathcal{J}\left(B_{il},B_{il},B_{il};v\right)\right|\apprle h^{-4}
\]
follow from the fact that $\mathscr{J}$ is uniformly bounded by $h^{-4}\underline{C}_{g}^{-2}\left(\overline{C}_{D_{1}}+\overline{C}_{D_{2}}\right)^{2}$.
Now it is clear from these inequalities, (\ref{eq:E=00005BG_hat_2(B_11^*)=00005D V-stat decomposition})
and (\ref{eq:J U-stat stochastic order}) that 
\[
\widehat{\sigma}_{\mathcal{\widehat{G}}_{2}^{\ddagger}}^{2}\coloneqq\underset{v\in I}{\mathrm{sup}}\,\mathrm{E}^{*}\left[\mathcal{\widehat{G}}_{2}^{\ddagger}\left(B_{11}^{*};v\right)^{2}\right]=O_{p}\left(h^{-1}+\left(\frac{\mathrm{log}\left(L\right)}{Lh^{5}}\right)^{\nicefrac{1}{2}}+\frac{\mathrm{log}\left(L\right)}{Lh^{4}}\right).
\]
The CCK inequality with $\sigma=\widehat{\sigma}_{\mathcal{\widehat{G}}_{2}^{\ddagger}}$
and $F$ being the constant envelope (\ref{eq:G constant envelope})
yields
\begin{eqnarray*}
\mathrm{E}^{*}\left[\underset{v\in I}{\mathrm{sup}}\left|\frac{1}{N\cdot L}\sum_{i,l}\mathcal{\widehat{G}}_{2}^{\ddagger}\left(B_{il}^{*};v\right)-\mathrm{E}^{*}\left[\mathcal{\widehat{G}}_{2}^{\ddagger}\left(B_{11}^{*};v\right)\right]\right|\right] & \leq & C_{1}\left\{ L^{-\nicefrac{1}{2}}\widehat{\sigma}_{\mathcal{\widehat{G}}_{2}^{\ddagger}}\mathrm{log}\left(C_{2}L\right)^{\nicefrac{1}{2}}+\left(Lh^{2}\right)^{-1}\mathrm{log}\left(L\right)\right\} \\
 & = & O_{p}\left(\left(\frac{\mathrm{log}\left(L\right)}{Lh}\right)^{\nicefrac{1}{2}}+\frac{\mathrm{log}\left(L\right)}{Lh^{2}}\right),
\end{eqnarray*}
where the inequality is non-asymptotic. Now 
\[
\underset{v\in I}{\mathrm{sup}}\left|\mathit{\Delta}_{1}^{*}\left(v\right)\right|=O_{p}^{*}\left(\left(\frac{\mathrm{log}\left(L\right)}{Lh}\right)^{\nicefrac{1}{2}}+\frac{\mathrm{log}\left(L\right)}{Lh^{2}}\right)
\]
follows from the above result, Lemma \ref{lem:auxiliary 3}, (\ref{eq:miu_hat_G stochastic bound}),
(\ref{eq:G Hoeffding decomposition bootstrap world}), (\ref{eq:G U-process supremum bound bootstrap world}),
(\ref{eq:G_hat_1 empirical process supremum expectation bound}),
\citet[Lemma S.1]{Marmer_Shneyerov_Quantile_Auctions} and uniform
boundedness of $\mathscr{G}$ which implies that the last two terms
in the decomposition (\ref{eq:G Hoeffding decomposition bootstrap world})
are both $O\left(\left(Lh^{2}\right)^{-1}\right)$ uniformly in $v\in I$.
The conclusion follows from Lemma \ref{lem:bootstrap lemma 1}, the
order bounds for the suprema of $\mathit{\Delta}_{k}^{*}$, $k=1,2,3$
and also the definition of $\widehat{g}^{*}$. \end{proof}

\begin{proof}[Proof of Lemma A.4] Define 
\[
\widehat{\mathcal{M}}_{1}\left(b;v\right)\coloneqq\int\mathcal{M}\left(b,b';v\right)\mathrm{d}\widehat{G}\left(b'\right)\textrm{, }\widehat{\mathcal{M}}_{2}\left(b;v\right)\coloneqq\int\mathcal{M}\left(b',b;v\right)\mathrm{d}\widehat{G}\left(b'\right)\textrm{ and }\widehat{\mu}_{\mathcal{M}}\left(v\right)\coloneqq\int\int\mathcal{M}\left(b,b';v\right)\mathrm{d}\widehat{G}\left(b\right)\mathrm{d}\widehat{G}\left(b'\right).
\]
Note that we have 
\[
\widehat{\mu}_{\mathcal{M}}\left(v\right)=\frac{1}{\left(N\cdot L\right)^{2}}\sum_{i,l}\sum_{j,k}\mathcal{M}\left(B_{il},B_{jk};v\right)
\]
by definition. The Hoeffding decomposition yields 
\begin{eqnarray}
\frac{1}{\left(N\cdot L\right)^{2}}\sum_{i,l}\sum_{j,k}\mathcal{M}\left(B_{il}^{*},B_{jk}^{*};v\right) & = & \widehat{\mu}_{\mathcal{M}}\left(v\right)+\left\{ \frac{1}{N\cdot L}\sum_{i,l}\widehat{\mathcal{M}}_{1}\left(B_{il}^{*};v\right)-\widehat{\mu}_{\mathcal{M}}\left(v\right)\right\} +\left\{ \frac{1}{N\cdot L}\sum_{i,l}\mathcal{\widehat{M}}_{2}\left(B_{il}^{*};v\right)-\widehat{\mu}_{\mathcal{M}}\left(v\right)\right\} \nonumber \\
 &  & +\frac{1}{\left(N\cdot L\right)_{2}}\sum_{\left(2\right)}\left\{ \mathcal{M}\left(B_{il}^{*},B_{jk}^{*};v\right)-\mathcal{\widehat{M}}_{1}\left(B_{il}^{*};v\right)-\mathcal{\widehat{M}}_{2}\left(B_{jk}^{*};v\right)+\widehat{\mu}_{\mathcal{M}}\left(v\right)\right\} \nonumber \\
 &  & +\frac{1}{\left(N\cdot L\right)^{2}}\sum_{i,l}\mathcal{M}\left(B_{il}^{*},B_{il}^{*};v\right)-\frac{1}{\left(N\cdot L\right)^{2}\left(N\cdot L-1\right)}\sum_{\left(2\right)}\mathcal{M}\left(B_{il}^{*},B_{jk}^{*};v\right).\label{eq:M Hoeffding decomposition bootstrap}
\end{eqnarray}

It is argued in the proof of Lemma A.2 that $\mathscr{M}$ is VC-type
with respect to the constant envelope (\ref{eq:M constant envelope}).
Now the CK inequality yields 
\begin{equation}
\mathrm{E}^{*}\left[\underset{v\in I}{\mathrm{sup}}\left|\frac{1}{\left(N\cdot L\right)_{2}}\sum_{\left(2\right)}\left\{ \mathcal{M}\left(B_{il}^{*},B_{jk}^{*};v\right)-\mathcal{\widehat{M}}_{1}\left(B_{il}^{*};v\right)-\mathcal{\widehat{M}}_{2}\left(B_{jk}^{*};v\right)+\widehat{\mu}_{\mathcal{M}}\left(v\right)\right\} \right|\right]\apprle\left(Lh^{3}\right)^{-1}+\left(Lh^{2}\right)^{-1}.\label{eq:M U-process supremum bound bootstrap world}
\end{equation}

Note that by definition, 
\[
\widehat{\mathcal{M}}_{1}\left(b;v\right)=-\frac{1}{h^{2}}K_{f}'\left(\frac{\xi\left(b\right)-v}{h}\right)\frac{G\left(b\right)}{g\left(b\right)^{2}}\left\{ \widehat{g}\left(b\right)-g\left(b\right)\right\} .
\]
By the arguments used to show that $\left\{ \widehat{\mathcal{G}}_{1}\left(\cdot;v\right):v\in I\right\} $
is VC-type, we can show that $\left\{ \widehat{\mathcal{M}}_{1}\left(\cdot;v\right):v\in I\right\} $
is (non-random) VC-type (conditionally on the original sample) with
respect to the (conditionally) constant envelope: 
\[
h^{-2}\left(\overline{C}_{D_{1}}+\overline{C}_{D_{2}}\right)\underline{C}_{g}^{-2}\underset{b\in\left[s\left(v_{l}-h\right),s\left(v_{u}+h\right)\right]}{\mathrm{sup}}\left|\widehat{g}\left(b\right)-g\left(b\right)\right|,
\]
when $h$ is sufficiently small. The VW inequality yields:
\begin{equation}
\mathrm{E}^{*}\left[\underset{v\in I}{\mathrm{sup}}\left|\frac{1}{N\cdot L}\sum_{i,l}\widehat{\mathcal{M}}_{1}\left(B_{il}^{*};v\right)-\widehat{\mu}_{\mathcal{M}}\left(v\right)\right|\right]\apprle L^{-\nicefrac{1}{2}}h^{-2}\underset{b\in\left[s\left(v_{l}-h\right),s\left(v_{u}+h\right)\right]}{\mathrm{sup}}\left|\widehat{g}\left(b\right)-g\left(b\right)\right|=O_{p}\left(\frac{\mathrm{log}\left(L\right)^{\nicefrac{1}{2}}}{Lh^{\nicefrac{5}{2}}}+\frac{h^{R-1}}{L^{\nicefrac{1}{2}}}\right),\label{eq:M_hat_1 sup bound bootstrap}
\end{equation}
where the inequality is non-asymptotic.

Consider the following process: 
\[
\varDelta^{**}\left(v\right)\coloneqq\frac{1}{N\cdot L}\sum_{i,l}\left\{ \left(\widehat{\mathcal{M}}_{2}\left(B_{il}^{*};v\right)-\widehat{\mu}_{\mathcal{M}}\left(v\right)\right)-\left(\mathcal{M}_{2}\left(B_{il}^{*};v\right)-\frac{1}{N\cdot L}\sum_{j,k}\mathcal{M}_{2}\left(B_{jk};v\right)\right)\right\} ,\,v\in I.
\]
Let 
\[
\widehat{\mathcal{M}}_{2}^{\ddagger}\left(b;v\right)\coloneqq\int\mathcal{M}^{\ddagger}\left(b',b;v\right)\mathrm{d}\widehat{G}\left(b'\right)\textrm{ and }\widehat{\mu}_{\mathcal{M}^{\ddagger}}\left(v\right)\coloneqq\int\int\mathcal{M}^{\ddagger}\left(b',b;v\right)\mathrm{d}\widehat{G}\left(b'\right)\mathrm{d}\widehat{G}\left(b\right).
\]
It follows from 
\[
\mathcal{M}_{2}\left(B_{il}^{*};v\right)-\frac{1}{N\cdot L}\sum_{j,k}\mathcal{M}_{2}\left(B_{jk};v\right)=\mathcal{M}_{2}^{\ddagger}\left(B_{il}^{*};v\right)-\frac{1}{N\cdot L}\sum_{j,k}\mathcal{M}_{2}^{\ddagger}\left(B_{jk};v\right),\textrm{ for all \ensuremath{i=1,...,N} and \ensuremath{l=1,...,L}}
\]
and
\[
\widehat{\mathcal{M}}_{2}\left(B_{il}^{*};v\right)-\widehat{\mu}_{\mathcal{M}}\left(v\right)=\widehat{\mathcal{M}}_{2}^{\ddagger}\left(B_{il}^{*};v\right)-\widehat{\mu}_{\mathcal{M}^{\ddagger}}\left(v\right),\textrm{ for all \ensuremath{i=1,...,N} and \ensuremath{l=1,...,L}}
\]
that
\[
\varDelta^{**}\left(v\right)=\frac{1}{N\cdot L}\sum_{i,l}\left\{ \left(\widehat{\mathcal{M}}_{2}^{\ddagger}\left(B_{il}^{*};v\right)-\widehat{\mu}_{\mathcal{M}^{\ddagger}}\left(v\right)\right)-\left(\mathcal{M}_{2}^{\ddagger}\left(B_{il}^{*};v\right)-\frac{1}{N\cdot L}\sum_{j,k}\mathcal{M}_{2}^{\ddagger}\left(B_{jk};v\right)\right)\right\} ,\,\textrm{for all }v\in I.
\]

\noindent Simple algebra yields
\begin{equation}
\mathrm{E}^{*}\left[\left(\widehat{\mathcal{M}}_{2}^{\ddagger}\left(B_{11}^{*};v\right)-\mathcal{M}_{2}^{\ddagger}\left(B_{11}^{*};v\right)\right)^{2}\right]=\frac{1}{\left(N\cdot L\right)^{3}}\sum_{i,l}\sum_{j,k}\sum_{j',k'}\mathcal{L}\left(B_{il},B_{jk},B_{j'k'};v\right),\label{eq:E*=00005BDELTA**^2=00005D}
\end{equation}
where
\[
\mathcal{L}\left(b,b',b'';v\right)\coloneqq\left(\mathcal{M}^{\ddagger}\left(b',b;v\right)-\mathcal{M}_{2}^{\ddagger}\left(b;v\right)\right)\left(\mathcal{M}^{\ddagger}\left(b'',b;v\right)-\mathcal{M}_{2}^{\ddagger}\left(b;v\right)\right).
\]

By the V-statistic decomposition argument of \citet[5.7.3]{Serfling_Approximation_Theorems},
since the kernel $\mathcal{L}$ is symmetric with respect to the second
and the third arguments, we have 
\begin{align}
 & \frac{1}{\left(N\cdot L\right)^{3}}\sum_{i,l}\sum_{j,k}\sum_{j',k'}\mathcal{L}\left(B_{il},B_{jk},B_{j'k'};v\right)\nonumber \\
= & \frac{1}{\left(N\cdot L\right)_{3}}\sum_{\left(3\right)}\mathcal{L}\left(B_{il},B_{jk},B_{j'k'};v\right)\nonumber \\
 & +\frac{O\left(L^{-1}\right)}{3\left(N\cdot L\right)^{2}-2\left(N\cdot L\right)}\left\{ \sum_{\left(2\right)}\left(2\mathcal{L}\left(B_{il},B_{il},B_{jk};v\right)+\mathcal{L}\left(B_{jk},B_{il},B_{il};v\right)\right)+\sum_{i,l}\mathcal{L}\left(B_{il},B_{il},B_{il};v\right)\right\} .\label{eq:L V-stat decomposition}
\end{align}

Since we argued that $\mathscr{M}^{\ddagger}$ and $\left\{ \mathcal{M}_{2}^{\ddagger}\left(\cdot;v\right):v\in I\right\} $
are both VC-type with respect to the constant envelope 
\begin{equation}
h^{-3}\left(\overline{C}_{D_{1}}+\overline{C}_{D_{1}}\right)\underline{C}_{g}^{-2}\overline{C}_{K_{g}},\label{eq:M_ddagger envelope}
\end{equation}
it follows from \citet[Lemma 16]{nolan1987u} and \citet[Lemma A.1]{chernozhukov2014gaussian}
that $\mathscr{L}\coloneqq\left\{ \mathcal{L}\left(\cdot,\cdot,\cdot;v\right):v\in I\right\} $
is VC-type with respect to the constant envelope
\begin{equation}
4\left\{ h^{-3}\left(\overline{C}_{D_{1}}+\overline{C}_{D_{2}}\right)\underline{C}_{g}^{-2}\overline{C}_{K_{g}}\right\} ^{2}.\label{eq:L constant envelope}
\end{equation}

Note that the leading U-process is first-order degenerate:
\[
\mu_{\mathcal{L}}\left(v\right)=\mathcal{L}_{1}^{\left(1\right)}\left(b;v\right)=\mathcal{L}_{2}^{\left(1\right)}\left(b;v\right)=\mathcal{L}_{3}^{\left(1\right)}\left(b;v\right)=0,\textrm{ for all \ensuremath{\left(b,v\right)\in\left[\underline{b},\overline{b}\right]\times I}}
\]
and it is also clear that
\[
\mathcal{L}_{1}^{\left(2\right)}\left(b,b';v\right)=\mathcal{L}_{2}^{\left(2\right)}\left(b,b';v\right)=0,\textrm{ for all \ensuremath{\left(b,b',v\right)\in\left[\underline{b},\overline{b}\right]^{2}\times I}},
\]
where these terms in the Hoeffding decomposition are defined by (\ref{eq:K Hoeffding decomposition def 1})
to (\ref{eq:K Hoeffding decomposition def 3}) with $\mathcal{K}$
replaced by $\mathcal{L}$. Now the Hoeffding decomposition is simply
\begin{align*}
 & \frac{1}{\left(N\cdot L\right)_{3}}\sum_{\left(3\right)}\mathcal{L}\left(B_{il},B_{jk},B_{j'k'};v\right)\\
= & \frac{1}{\left(N\cdot L\right)_{2}}\sum_{\left(2\right)}\mathcal{L}_{3}^{\left(2\right)}\left(B_{il},B_{jk};v\right)+\frac{1}{\left(N\cdot L\right)_{3}}\sum_{\left(3\right)}\left\{ \mathcal{L}\left(B_{il},B_{jk},B_{j'k'};v\right)-\mathcal{L}_{3}^{\left(2\right)}\left(B_{jk},B_{j'k'};v\right)\right\} .
\end{align*}

The CK inequality yields 
\[
\mathrm{E}\left[\underset{v\in I}{\mathrm{sup}}\left|\frac{1}{\left(N\cdot L\right)_{2}}\sum_{\left(2\right)}\mathcal{L}_{3}^{\left(2\right)}\left(B_{il},B_{jk};v\right)\right|\right]=O\left(\left(Lh^{6}\right)^{-1}\right)
\]
and 
\[
\mathrm{E}\left[\underset{v\in I}{\mathrm{sup}}\left|\frac{1}{\left(N\cdot L\right)_{3}}\sum_{\left(3\right)}\left\{ \mathcal{L}\left(B_{il},B_{jk},B_{j'k'};v\right)-\mathcal{L}_{3}^{\left(2\right)}\left(B_{jk},B_{j'k'};v\right)\right\} \right|\right]=O\left(L^{-\nicefrac{3}{2}}h^{-6}\right).
\]
It follows from these inequalities, (\ref{eq:L V-stat decomposition}),
the fact that $\mathscr{L}$ is uniformly bounded by (\ref{eq:L constant envelope})
and Markov's inequality that 
\[
\underset{v\in I}{\mathrm{sup}}\left|\frac{1}{\left(N\cdot L\right)^{3}}\sum_{i,l}\sum_{j,k}\sum_{j',k'}\mathcal{L}\left(B_{il},B_{jk},B_{j'k'};v\right)\right|=O_{p}\left(\left(Lh^{6}\right)^{-1}\right).
\]
It then follows from the above result and (\ref{eq:E*=00005BDELTA**^2=00005D})
that
\[
\widehat{\sigma}_{\varDelta}^{2}\coloneqq\underset{v\in I}{\mathrm{sup}}\,\mathrm{E}^{*}\left[\left(\widehat{\mathcal{M}}_{2}^{\ddagger}\left(B_{11}^{*};v\right)-\mathcal{M}_{2}^{\ddagger}\left(B_{11}^{*};v\right)\right)^{2}\right]=O_{p}\left(\left(Lh^{6}\right)^{-1}\right).
\]

Since $\mathscr{M}^{\ddagger}$ is VC-type with respect to the constant
envelope (\ref{eq:M_ddagger envelope}), the fact that both $\left\{ \mathcal{M}_{2}^{\ddagger}\left(\cdot;v\right):v\in I\right\} $
and $\left\{ \widehat{\mathcal{M}}_{2}^{\ddagger}\left(\cdot;v\right):v\in I\right\} $
are VC-type classes (conditionally on the original sample) with respect
to the same constant envelope follows from \citet[Lemma 5.4]{Chen_Kato_U_Process}.
Now \citet[Lemma 16]{nolan1987u} implies that the class $\left\{ \widehat{\mathcal{M}}_{2}^{\ddagger}\left(\cdot;v\right)-\mathcal{M}_{2}^{\ddagger}\left(\cdot;v\right):v\in I\right\} $
is also VC-type (conditionally on the original sample) with respect
to a constant envelope that is twice of (\ref{eq:M_ddagger envelope}).
The CCK inequality with $\sigma=\widehat{\sigma}_{\varDelta}$ and
$F$ being this constant envelope yields: 
\[
\mathrm{E}^{*}\left[\underset{v\in I}{\mathrm{sup}}\left|\varDelta^{**}\left(v\right)\right|\right]\leq C_{1}\left\{ L^{-\nicefrac{1}{2}}\widehat{\sigma}_{\varDelta}\mathrm{log}\left(C_{2}L\right)^{\nicefrac{1}{2}}+L^{-1}h^{-3}\mathrm{log}\left(C_{2}L\right)\right\} =O_{p}\left(\frac{\mathrm{log}\left(L\right)}{Lh^{3}}\right),
\]
where the inequality is non-asymptotic. The conclusion follows from
the above result, Lemma \ref{lem:bootstrap lemma 2}, (\ref{eq:M Hoeffding decomposition bootstrap}),
(\ref{eq:M U-process supremum bound bootstrap world}), (\ref{eq:M_hat_1 sup bound bootstrap}),
Lemma \ref{lem:auxiliary 3} and the fact that $\mathscr{M}$ is uniformly
bounded by (\ref{eq:M constant envelope}). \end{proof}

\begin{proof}[Proof of Lemma A.5]We first show
\begin{equation}
\underset{v\in I}{\mathrm{sup}}\left|\frac{\widehat{f}_{GPV}\left(v\right)-f\left(v\right)}{\left(Lh^{3}\right)^{-\nicefrac{1}{2}}\mathrm{V}_{\mathcal{M}}\left(v\right)^{\nicefrac{1}{2}}}-\mathit{\Gamma}\left(v\right)\right|=O_{p}\left(\mathrm{log}\left(L\right)^{\nicefrac{1}{2}}h+\frac{\mathrm{log}\left(L\right)}{\left(Lh^{3}\right)^{\nicefrac{1}{2}}}+L^{\nicefrac{1}{2}}h^{\nicefrac{3}{2}+R}\right).\label{eq:studentized_by_V_M - GAMMA uniform rate}
\end{equation}

Lemma A.2 showed that 
\[
\vartheta_{1}\coloneqq\underset{v\in I}{\mathrm{sup}}\left|\widehat{f}_{GPV}\left(v\right)-f\left(v\right)-\frac{1}{\left(N-1\right)}\frac{1}{N\cdot L}\sum_{i,l}\left(\mathcal{M}_{2}\left(B_{il};v\right)-\mu_{\mathcal{M}}\left(v\right)\right)\right|=O_{p}\left(\left(\frac{\mathrm{log}\left(L\right)}{Lh}\right)^{\nicefrac{1}{2}}+\frac{\mathrm{log}\left(L\right)}{Lh^{3}}+h^{R}\right),
\]
where the remainder is uniform in $v\in I$. 

Since $\mathrm{V}_{\mathcal{M}}\left(v\right)$ uniformly converges
to $\mathrm{V}_{GPV}\left(v\right)$ as $h\downarrow0$, we have 
\begin{equation}
\underline{\mathrm{V}}_{\mathcal{M}}\coloneqq\underset{v\in I}{\mathrm{inf}}\,\mathrm{V}_{\mathcal{M}}\left(v\right)>C_{1}>0,\label{eq:V_M bounded away from zero}
\end{equation}
when $h$ is sufficiently small. It is shown in the proof of Theorem
3.1 (see (A.8) and (A.10)) that 
\begin{equation}
\underset{v\in I}{\mathrm{sup}}\,\mathrm{E}\left[\mathcal{M}_{2}^{\ddagger}\left(B_{11};v\right)^{2}\right]\apprle h^{-3}.\label{eq:uniform bound E=00005BM_2^2=00005D}
\end{equation}
An application of the CCK inequality with $\sigma^{2}$ being the
left-hand side of (\ref{eq:uniform bound E=00005BM_2^2=00005D}) and
$F$ being (A.38) gives 
\begin{equation}
\mathrm{E}\left[\underset{v\in I}{\mathrm{sup}}\left|\frac{1}{N\cdot L}\sum_{i,l}\left(\mathcal{M}_{2}^{\ddagger}\left(B_{il};v\right)-\mu_{\mathcal{M}^{\ddagger}}\left(v\right)\right)\right|\right]\leq C_{1}\left\{ \left(Lh^{3}\right)^{-\nicefrac{1}{2}}\mathrm{log}\left(C_{2}L\right)^{\nicefrac{1}{2}}+\left(Lh^{3}\right)^{-1}\mathrm{log}\left(C_{2}L\right)\right\} ,\label{eq:M_2 - miu supremum expectation bound}
\end{equation}
where the inequality is non-asymptotic. Then Markov's inequality yields
\[
\vartheta_{2}\coloneqq\underset{v\in I}{\mathrm{sup}}\left|\frac{1}{N-1}\frac{1}{N\cdot L}\sum_{i,l}\left(\mathcal{M}_{2}\left(B_{il};v\right)-\mu_{\mathcal{M}}\left(v\right)\right)\right|=O_{p}\left(\left(\frac{\mathrm{log}\left(L\right)}{Lh^{3}}\right)^{\nicefrac{1}{2}}\right).
\]

In the proof of Theorem 3.1, we showed 
\begin{equation}
\vartheta_{3}\coloneqq\underset{v\in I}{\mathrm{sup}}\left|\mathrm{V}_{\mathcal{M}}\left(v\right)^{\nicefrac{1}{2}}-\mathrm{Var}\left[N^{-\nicefrac{1}{2}}\left(N-1\right)^{-1}h^{\nicefrac{3}{2}}\mathcal{M}_{2}\left(B_{11};v\right)\right]^{\nicefrac{1}{2}}\right|=O\left(h^{3}\right).\label{eq:theta_3 definition}
\end{equation}
Then by the triangle inequality and 
\begin{equation}
\frac{a}{b}=\frac{a}{c}-\frac{a\left(b-c\right)}{c^{2}}+\frac{a\left(b-c\right)^{2}}{bc^{2}},\label{eq:abc identity}
\end{equation}
we have
\begin{align*}
 & \underset{v\in I}{\mathrm{sup}}\left|\frac{\widehat{f}_{GPV}\left(v\right)-f\left(v\right)}{\left(Lh^{3}\right)^{-\nicefrac{1}{2}}\mathrm{V}_{\mathcal{M}}\left(v\right)^{\nicefrac{1}{2}}}-\mathit{\Gamma}\left(v\right)\right|\\
\leq & \left(Lh^{3}\right)^{\nicefrac{1}{2}}\left\{ \underline{\mathrm{V}}_{\mathcal{M}}^{-\nicefrac{1}{2}}\vartheta_{1}+\underline{\mathrm{V}}_{\mathcal{M}}^{-1}\vartheta_{2}\vartheta_{3}+\mathrm{Var}\left[N^{-\nicefrac{1}{2}}\left(N-1\right)^{-1}h^{\nicefrac{3}{2}}\mathcal{M}_{2}\left(B_{11};v\right)\right]^{-\nicefrac{1}{2}}\underline{\mathrm{V}}_{\mathcal{M}}^{-1}\vartheta_{2}\vartheta_{3}^{2}\right\} .
\end{align*}
Now (\ref{eq:studentized_by_V_M - GAMMA uniform rate}) follows from
the above result and the the rates of convergence of $\vartheta_{1}$,
$\vartheta_{2}$ and $\vartheta_{3}$.

By Theorem 4.1 and (\ref{eq:V_M bounded away from zero}), we have
\begin{eqnarray}
\underset{v\in I}{\mathrm{sup}}\left|\widehat{\mathrm{V}}_{GPV}\left(v\right)^{\nicefrac{1}{2}}-\mathrm{V}_{\mathcal{M}}\left(v\right)^{\nicefrac{1}{2}}\right| & \leq & \frac{1}{2}\underset{v\in I}{\mathrm{sup}}\left\{ \left(\mathrm{max}\left\{ \widehat{\mathrm{V}}_{GPV}\left(v\right)^{-1},\mathrm{V}_{\mathcal{M}}\left(v\right)^{-1}\right\} \right)^{\nicefrac{1}{2}}\left|\widehat{\mathrm{V}}_{GPV}\left(v\right)-\mathrm{V}_{\mathcal{M}}\left(v\right)\right|\right\} \nonumber \\
 & = & O_{p}\left(\left(\frac{\mathrm{log}\left(L\right)}{Lh^{3}}\right)^{\nicefrac{1}{2}}+h^{R}\right).\label{eq:sqrt(V_GPV) - sqrt(V_M) sup rate}
\end{eqnarray}
It then follows from the above result, (\ref{eq:abc identity}), (\ref{eq:V_M bounded away from zero})
and the rates of convergence of $\vartheta_{1}$ and $\vartheta_{2}$
that
\[
\underset{v\in I}{\mathrm{sup}}\left|Z\left(v\right)-\frac{\widehat{f}_{GPV}\left(v\right)-f\left(v\right)}{\left(Lh^{3}\right)^{-\nicefrac{1}{2}}\mathrm{V}_{\mathcal{M}}\left(v\right)^{\nicefrac{1}{2}}}\right|=O_{p}\left(\frac{\mathrm{log}\left(L\right)}{\left(Lh^{3}\right)^{\nicefrac{1}{2}}}+\mathrm{log}\left(L\right)^{\nicefrac{1}{2}}h^{R}\right).
\]
The conclusion follows from the above result and (\ref{eq:studentized_by_V_M - GAMMA uniform rate}).\end{proof}

\begin{proof}[Proof of Lemma A.7]By decomposition (see (A.37)), Lemmas
A.1, A.3, A.4 and \citet[online supplement, Lemma S.1]{Marmer_Shneyerov_Quantile_Auctions},
we have 
\begin{eqnarray*}
\vartheta_{1}^{*} & \coloneqq & \underset{v\in I}{\mathrm{sup}}\left|\widehat{f}_{GPV}^{*}\left(v\right)-\widehat{f}_{GPV}\left(v\right)-\left\{ \frac{1}{N-1}\frac{1}{N\cdot L}\sum_{i,l}\left(\mathcal{M}_{2}\left(B_{il}^{*};v\right)-\frac{1}{N\cdot L}\sum_{j,k}\mathcal{M}_{2}\left(B_{jk};v\right)\right)\right\} \right|\\
 & = & O_{p}^{*}\left(\left(\frac{\mathrm{log}\left(L\right)}{Lh}\right)^{\nicefrac{1}{2}}+\frac{\mathrm{log}\left(L\right)}{Lh^{3}}+h^{R}\right).
\end{eqnarray*}

Let 
\begin{eqnarray*}
\vartheta_{2}^{*} & \coloneqq & \underset{v\in I}{\mathrm{sup}}\left|\frac{1}{N-1}\frac{1}{N\cdot L}\sum_{i,l}\left(\mathcal{M}_{2}\left(B_{il}^{*};v\right)-\frac{1}{N\cdot L}\sum_{j,k}\mathcal{M}_{2}\left(B_{jk};v\right)\right)\right|\\
 & = & \underset{v\in I}{\mathrm{sup}}\left|\frac{1}{N-1}\frac{1}{N\cdot L}\sum_{i,l}\left(\mathcal{M}_{2}^{\ddagger}\left(B_{il}^{*};v\right)-\frac{1}{N\cdot L}\sum_{j,k}\mathcal{M}_{2}^{\ddagger}\left(B_{jk};v\right)\right)\right|,
\end{eqnarray*}
where the second equality follows since it is easy to check that 
\begin{equation}
\mathcal{M}_{2}\left(B_{il}^{*};v\right)-\frac{1}{N\cdot L}\sum_{j,k}\mathcal{M}_{2}\left(B_{jk};v\right)=\mathcal{M}_{2}^{\ddagger}\left(B_{il}^{*};v\right)-\frac{1}{N\cdot L}\sum_{j,k}\mathcal{M}_{2}^{\ddagger}\left(B_{jk};v\right),\textrm{ for all \ensuremath{i=1,...,N} and \ensuremath{l=1,...,L}}.\label{eq:M_2 - miu_hat_M_2 - M_2_ddagger - miu_hat_M_2_ddagger}
\end{equation}
Now let
\begin{eqnarray}
\widehat{\sigma}_{\mathcal{M}_{2}^{\ddagger}}^{2} & \coloneqq & \underset{v\in I}{\mathrm{sup}}\,\mathrm{E}^{*}\left[\mathcal{M}_{2}^{\ddagger}\left(B_{11}^{*};v\right)^{2}\right]\nonumber \\
 & \leq & h^{-3}\left\{ \underset{v\in I}{\mathrm{sup}}\,\mathrm{E}\left[h^{3}\mathcal{M}_{2}^{\ddagger}\left(B_{11};v\right)^{2}\right]+\underset{v\in I}{\mathrm{sup}}\left|\frac{1}{N\cdot L}\sum_{i,l}h^{3}\mathcal{M}_{2}^{\ddagger}\left(B_{il};v\right)^{2}-\mathrm{E}\left[h^{3}\mathcal{M}_{2}^{\ddagger}\left(B_{11};v\right)^{2}\right]\right|\right\} .\label{eq:sigma_M_2_til^2 bound}
\end{eqnarray}
Since it was shown that $\mathrm{E}\left[h^{3}\mathcal{M}_{2}^{\ddagger}\left(B_{11};v\right)^{2}\right]$
converges to $N\left(N-1\right)^{2}\mathrm{V}_{GPV}\left(v\right)$
uniformly in $v\in I$, we have 
\begin{equation}
\underset{v\in I}{\mathrm{sup}}\,\mathrm{E}\left[h^{3}\mathcal{M}_{2}^{\ddagger}\left(B_{11};v\right)^{2}\right]=O\left(1\right).\label{eq:sigma_M_2_til^2 expectation bound}
\end{equation}

It follows from \citet[Corollary A.1]{chernozhukov2014gaussian} that
$\left\{ h^{3}\mathcal{M}_{2}^{\ddagger}\left(\cdot;v\right)^{2}:v\in I\right\} $
is VC-type with respect to the constant envelope $h^{-3}\left(\overline{C}_{D_{1}}+\overline{C}_{D_{1}}\right)^{2}\underline{C}_{g}^{-4}\overline{C}_{K_{g}}^{2}$.
Now by change of variables we have 
\begin{equation}
\underset{v\in I}{\mathrm{sup}}\mathrm{E}\left[h^{6}\mathcal{M}_{2}^{\ddagger}\left(B_{11};v\right)^{4}\right]=h^{-1}\underset{v\in I}{\mathrm{sup}}\int_{\frac{\underline{b}-s\left(v\right)}{h}}^{\frac{\overline{b}-s\left(v\right)}{h}}\left(\int_{\frac{\underline{v}-v}{h}}^{\frac{\overline{v}-v}{h}}\rho\left(u,w;v\right)\mathrm{d}u\right)^{4}g\left(hw+s\left(v\right)\right)\mathrm{d}w\apprle h^{-1}.\label{eq:E=00005BM_2_ddagger^4=00005D sup rate}
\end{equation}
Applying the CCK inequality with $\sigma^{2}$ being the left-hand
side of the inequality of (\ref{eq:E=00005BM_2_ddagger^4=00005D sup rate})
and $F$ being this constant envelope yields 
\begin{eqnarray*}
\mathrm{E}\left[\underset{v\in I}{\mathrm{sup}}\left|\frac{1}{N\cdot L}\sum_{i,l}h^{3}\mathcal{M}_{2}^{\ddagger}\left(B_{il};v\right)^{2}-\mathrm{E}\left[h^{3}\mathcal{M}_{2}^{\ddagger}\left(B_{11};v\right)^{2}\right]\right|\right] & \leq & C_{1}\left\{ \left(Lh\right)^{-\nicefrac{1}{2}}\mathrm{log}\left(C_{2}L\right)^{\nicefrac{1}{2}}+\left(Lh^{3}\right)^{-1}\mathrm{log}\left(C_{2}L\right)\right\} \\
 & = & O\left(\left(\frac{\mathrm{log}\left(L\right)}{Lh}\right)^{\nicefrac{1}{2}}+\frac{\mathrm{log}\left(L\right)}{Lh^{3}}\right),
\end{eqnarray*}
where the inequality is non-asymptotic. It follows from the above
result, Markov's inequality, (\ref{eq:sigma_M_2_til^2 bound}) and
(\ref{eq:sigma_M_2_til^2 expectation bound}) that $\widehat{\sigma}_{\mathcal{M}_{2}^{\ddagger}}^{2}=O_{p}\left(h^{-3}\right)$.
It now follows from the CCK inequality that 
\[
\mathrm{E}^{*}\left[\vartheta_{2}^{*}\right]\leq C_{1}\left\{ L^{-\nicefrac{1}{2}}\widehat{\sigma}_{\mathcal{M}_{2}^{\ddagger}}\mathrm{log}\left(C_{2}L\right)^{\nicefrac{1}{2}}+\left(Lh^{3}\right)^{-1}\mathrm{log}\left(C_{2}L\right)\right\} =O_{p}\left(\left(\frac{\mathrm{log}\left(L\right)}{Lh^{3}}\right)^{\nicefrac{1}{2}}\right).
\]
Now by the above result, Lemma \ref{lem:auxiliary 3}, the triangle
inequality, (\ref{eq:abc identity}) and the rates of convergence
of $\vartheta_{1}^{*}$ and $\vartheta_{3}$, we have
\begin{align*}
 & \underset{v\in I}{\mathrm{sup}}\left|\frac{\widehat{f}_{GPV}^{*}\left(v\right)-\widehat{f}_{GPV}\left(v\right)}{\left(Lh^{3}\right)^{-\nicefrac{1}{2}}\mathrm{V}_{\mathcal{M}}\left(v\right)^{\nicefrac{1}{2}}}-\mathit{\Gamma}^{*}\left(v\right)\right|\\
\leq & \left(Lh^{3}\right)^{\nicefrac{1}{2}}\left\{ \underline{\mathrm{V}}_{\mathcal{M}}^{-\nicefrac{1}{2}}\vartheta_{1}^{*}+\underline{\mathrm{V}}_{\mathcal{M}}^{-1}\vartheta_{2}^{*}\vartheta_{3}+\mathrm{Var}\left[N^{-\nicefrac{1}{2}}\left(N-1\right)^{-1}h^{\nicefrac{3}{2}}\mathcal{M}_{2}\left(B_{11};v\right)\right]^{-\nicefrac{1}{2}}\underline{\mathrm{V}}_{\mathcal{M}}^{-1}\vartheta_{2}^{*}\vartheta_{3}^{2}\right\} \\
= & O_{p}^{*}\left(\mathrm{log}\left(L\right)^{\nicefrac{1}{2}}h+\frac{\mathrm{log}\left(L\right)}{\left(Lh^{3}\right)^{\nicefrac{1}{2}}}+L^{\nicefrac{1}{2}}h^{\nicefrac{3}{2}+R}\right).
\end{align*}
The conclusion follows from the above result, (\ref{eq:sqrt(V_GPV) - sqrt(V_M) sup rate}),
the rates of convergence of $\vartheta_{1}^{*}$ and $\vartheta_{2}^{*}$.\end{proof}

\begin{proof}[Proof of Lemma A.8]By definition, fix any $\epsilon>0$,
there exists some $M_{\epsilon}>0$ and also some $L_{\epsilon}\in\mathbb{N}$,
such that 
\[
\mathrm{P}\left[\mathrm{P}^{*}\left[\left|V_{L}^{*}-W_{L}^{*}\right|>\lambda_{L}M_{\epsilon}\right]>\frac{\epsilon}{2}\right]<\frac{\epsilon}{2},
\]
when $L\geq L_{\epsilon}$. By Lemma A.6, 
\[
\underset{z\in\mathbb{R}}{\mathrm{sup}}\left|\mathrm{P}^{*}\left[V_{L}^{*}\leq z\right]-\mathrm{P}^{*}\left[W_{L}^{*}\leq z\right]\right|\leq\underset{z\in\mathbb{R}}{\mathrm{sup}}\,\mathrm{P}^{*}\left[\left|W_{L}^{*}-z\right|\leq\lambda_{L}M_{\epsilon}\right]+\frac{\epsilon}{2},
\]
when $\mathrm{P}^{*}\left[\left|V_{L}^{*}-W_{L}^{*}\right|>\lambda_{L}M_{\epsilon}\right]\leq\nicefrac{\epsilon}{2}$. 

Since 
\[
\underset{z\in\mathbb{R}}{\mathrm{sup}}\,\mathrm{P}^{*}\left[\left|W_{L}^{*}-z\right|\leq C_{1}\lambda_{L}\right]\rightarrow_{p}0,\,\textrm{as \ensuremath{L\uparrow\infty}},
\]
there exists some $L_{\epsilon}'\in\mathbb{N}$ such that
\[
\mathrm{P}\left[\underset{z\in\mathbb{R}}{\mathrm{sup}}\,\mathrm{P}^{*}\left[\left|W_{L}^{*}-z\right|\leq\lambda_{L}M_{\epsilon}\right]\geq\frac{\epsilon}{2}\right]<\frac{\epsilon}{2},
\]
when $L\geq L_{\epsilon}'$. Since the three events satisfy
\[
\left\{ \mathrm{P}^{*}\left[\left|V_{L}^{*}-W_{L}^{*}\right|>\lambda_{L}M_{\epsilon}\right]\leq\frac{\epsilon}{2}\right\} \cap\left\{ \underset{z\in\mathbb{R}}{\mathrm{sup}}\,\mathrm{P}^{*}\left[\left|W_{L}^{*}-z\right|\leq\lambda_{L}M_{\epsilon}\right]<\frac{\epsilon}{2}\right\} \subseteq\left\{ \underset{z\in\mathbb{R}}{\mathrm{sup}}\left|\mathrm{P}^{*}\left[V_{L}^{*}\leq z\right]-\mathrm{P}^{*}\left[W_{L}^{*}\leq z\right]\right|<\epsilon\right\} ,
\]
it is clear that 
\begin{eqnarray*}
\mathrm{P}\left[\underset{z\in\mathbb{R}}{\mathrm{sup}}\left|\mathrm{P}^{*}\left[V_{L}^{*}\leq z\right]-\mathrm{P}^{*}\left[W_{L}^{*}\leq z\right]\right|\geq\epsilon\right] & \leq & \mathrm{P}\left[\mathrm{P}^{*}\left[\left|V_{L}^{*}-W_{L}^{*}\right|>\lambda_{L}M_{\epsilon}\right]>\frac{\epsilon}{2}\right]+\mathrm{P}\left[\underset{z\in\mathbb{R}}{\mathrm{sup}}\,\mathrm{P}^{*}\left[\left|W_{L}^{*}-z\right|\leq\lambda_{L}M_{\epsilon}\right]\geq\frac{\epsilon}{2}\right]\\
 & < & \epsilon,
\end{eqnarray*}
when $L\geq\mathrm{max}\left\{ L_{\epsilon},L_{\epsilon}'\right\} $.
The conclusion now follows.\end{proof}

\section{Limiting Distribution of the Uniform Error}

Denote 
\[
\widetilde{\mathcal{M}}_{2}^{\ddagger}\left(b;v\right)\coloneqq\frac{-\frac{G\left(s\left(v\right)\right)s'\left(v\right)}{g\left(s\left(v\right)\right)}\int\frac{1}{h^{2}}K_{f}'\left(u\right)K_{g}\left(\frac{b-s\left(v\right)}{h}-s'\left(v\right)u\right)\mathrm{d}u}{\sqrt{\frac{g\left(b\right)}{h^{3}}\int\left\{ \int K_{f}'\left(u\right)K_{g}\left(w-s'\left(v\right)u\right)\mathrm{d}u\right\} ^{2}\mathrm{d}w}}
\]
and 
\[
\mu_{\widetilde{\mathcal{M}}_{2}^{\ddagger}}\left(v\right)\coloneqq\int\widetilde{\mathcal{M}}_{2}^{\ddagger}\left(b;v\right)\mathrm{d}G\left(b\right).
\]
Since $K_{g}$ is of bounded variation, it follows from \citet[Lemma 22(ii)]{nolan1987u}
that the function class 
\[
\left\{ \left(b,u\right)\mapsto K_{g}\left(\frac{b-s\left(v\right)}{h}-s'\left(v\right)u\right):v\in I\right\} 
\]
is uniformly VC-type with respect to some constant envelope. Since
$K_{f}'$ is supported on $\left[-1,1\right]$, it follows from \citet[Lemma A.2]{ghosal2000testing}
the function class 
\[
\left\{ b\mapsto\int K_{f}'\left(u\right)K_{g}\left(\frac{b-s\left(v\right)}{h}-s'\left(v\right)u\right)\mathrm{d}u:v\in I\right\} 
\]
is uniformly VC-type with respect to some constant envelope. Then
by standard arguments, we can verify that the function class $\left\{ \widetilde{\mathcal{M}}_{2}^{\ddagger}\left(\cdot;v\right):v\in I\right\} $
is uniformly VC-type with respect to some constant envelope that is
a multiple of $h^{-\nicefrac{1}{2}}$.

Consider the following empirical processes:
\[
\widetilde{\varGamma}\left(v\right)\coloneqq\frac{1}{\left(N\cdot L\right)^{\nicefrac{1}{2}}}\sum_{i,l}\left(\widetilde{\mathcal{M}}_{2}^{\ddagger}\left(B_{il};v\right)-\mu_{\widetilde{\mathcal{M}}_{2}^{\ddagger}}\left(v\right)\right)
\]
and 
\begin{equation}
\widetilde{\varDelta}\left(v\right)\coloneqq\frac{1}{\left(N\cdot L\right)^{\nicefrac{1}{2}}}\sum_{i,l}\left(\widetilde{\mathcal{M}}_{2}^{\ddagger}\left(B_{il};v\right)-\frac{\mathcal{M}_{2}^{\ddagger}\left(B_{il};v\right)}{\sqrt{\mathrm{Var}\left[\mathcal{M}_{2}^{\ddagger}\left(B_{11};v\right)\right]}}-\mu_{\widetilde{\mathcal{M}}_{2}^{\ddagger}}\left(v\right)+\frac{\mu_{\mathcal{M}^{\ddagger}}\left(v\right)}{\sqrt{\mathrm{Var}\left[\mathcal{M}_{2}^{\ddagger}\left(B_{11};v\right)\right]}}\right),\label{eq:delta_tilta definition}
\end{equation}
for $v\in I$. It follows from \citet[Lemma 16]{nolan1987u} that
\[
\left\{ \widetilde{\mathcal{M}}_{2}^{\ddagger}\left(\cdot;v\right)-\frac{\mathcal{M}_{2}^{\ddagger}\left(\cdot;v\right)}{\sqrt{\mathrm{Var}\left[\mathcal{M}_{2}^{\ddagger}\left(B_{11};v\right)\right]}}:v\in I\right\} 
\]
is uniformly VC-type with respect to some constant envelope that is
a multiple of $h^{-\nicefrac{3}{2}}+h^{-\nicefrac{1}{2}}$, when $h$
is sufficiently small. 

By tedious and lengthy calculations, we have
\begin{align*}
 & \mathrm{E}\left[\left(\widetilde{\mathcal{M}}_{2}^{\ddagger}\left(B_{11};v\right)-\frac{\mathcal{M}_{2}^{\ddagger}\left(B_{il};v\right)}{\sqrt{\mathrm{Var}\left[\mathcal{M}_{2}^{\ddagger}\left(B_{11};v\right)\right]}}\right)^{2}\right]\\
= & \int_{\underline{b}}^{\overline{b}}\left\{ \frac{\frac{G\left(s\left(v\right)\right)s'\left(v\right)}{g\left(s\left(v\right)\right)}\int\frac{1}{h^{2}}K_{f}'\left(u\right)K_{g}\left(\frac{b-s\left(v\right)}{h}-s'\left(v\right)u\right)\mathrm{d}u}{\sqrt{\frac{g\left(b\right)}{h^{3}}\int\left\{ \int K_{f}'\left(u\right)K_{g}\left(w-s'\left(v\right)u\right)\mathrm{d}u\right\} ^{2}\mathrm{d}w}}-\frac{\int_{\underline{b}}^{\overline{b}}\frac{1}{h^{3}}K_{f}'\left(\frac{\xi\left(b'\right)-v}{h}\right)\frac{G\left(b'\right)}{g\left(b'\right)}K_{g}\left(\frac{b-b'}{h}\right)\mathrm{d}b'\sqrt{g\left(b\right)}}{\sqrt{\int\left\{ \int_{\underline{b}}^{\overline{b}}\frac{1}{h^{3}}K_{f}'\left(\frac{\xi\left(b'\right)-v}{h}\right)\frac{G\left(b'\right)}{g\left(b'\right)}K_{g}\left(\frac{b-b'}{h}\right)\mathrm{d}b'\right\} ^{2}g\left(b\right)\mathrm{d}b-\mu_{\mathcal{M}^{\ddagger}}\left(v\right)^{2}}}\right\} \mathrm{d}b\\
= & O\left(h^{2}\right),
\end{align*}
uniformly in $v\in I$. Then by the CCK inequality and Markov's inequality,
\begin{equation}
\underset{v\in I}{\mathrm{sup}}\left|\widetilde{\varDelta}\left(v\right)\right|=O_{p}\left(\mathrm{log}\left(L\right)^{\nicefrac{1}{2}}h+\mathrm{log}\left(L\right)\left(Lh^{3}\right)^{-\nicefrac{1}{2}}\right).\label{eq:delta EP rate}
\end{equation}

Again, by tedious and lengthy calculations, it can be shown that 
\begin{equation}
\underset{v\in I}{\mathrm{sup}}\left|\mu_{\widetilde{\mathcal{M}}_{2}^{\ddagger}}\left(v\right)\right|=O\left(h^{\nicefrac{3}{2}}\right).\label{eq:sup miu_M_til_2 rate}
\end{equation}
By the arguments used in the proof of Theorem 5.1, we can easily verify
that there exists a centered Gaussian process $\left\{ \widetilde{\varGamma}_{G}\left(v\right):v\in I\right\} $
which is a tight random element in $\ell^{\infty}\left(I\right)$
and has the following covariance function:
\[
\mathrm{E}\left[\widetilde{\varGamma}_{G}\left(v\right)\widetilde{\varGamma}_{G}\left(v'\right)\right]=\mathrm{E}\left[\left(\widetilde{\mathcal{M}}_{2}^{\ddagger}\left(B_{11};v\right)-\mu_{\widetilde{\mathcal{M}}_{2}^{\ddagger}}\left(v\right)\right)\left(\widetilde{\mathcal{M}}_{2}^{\ddagger}\left(B_{11};v'\right)-\mu_{\widetilde{\mathcal{M}}_{2}^{\ddagger}}\left(v'\right)\right)\right],\textrm{ for all \ensuremath{\left(v,v'\right)\in I^{2}}}.
\]
Application of the coupling theorem \citet[Corollary 2.2]{chernozhukov2014gaussian}
with $q=\infty$, $b\apprle h^{-\nicefrac{1}{2}}$, $\gamma=\mathrm{log}\left(L\right)^{-1}$
and $\sigma=1$ yields that there exists a sequence of random variables
$\widetilde{W}_{L}$ with $\widetilde{W}_{L}\overset{d}{=}\left\Vert \widetilde{\varGamma}_{G}\right\Vert _{I}$
satisfying 
\begin{equation}
\left|\left\Vert \widetilde{\varGamma}\right\Vert _{I}-\widetilde{W}_{L}\right|=O_{p}\left(\frac{\mathrm{log}\left(L\right)}{\left(Lh\right)^{\nicefrac{1}{6}}}\right).\label{eq:coupling}
\end{equation}
It follows from Lemma A.5, (\ref{eq:delta_tilta definition}), (\ref{eq:delta EP rate})
and (\ref{eq:coupling}) that 
\begin{equation}
\left\Vert Z\right\Vert _{I}-\widetilde{W}_{L}=o_{p}\left(\text{log}\left(L\right)^{-1}\right).\label{eq:Z - W}
\end{equation}

Let $Z$ be a standard normal random variable that is independent
of $\left\{ \widetilde{\varGamma}_{G}\left(v\right):v\in I\right\} $.
Define 
\[
\widetilde{\varGamma}_{G}^{BR}\left(v\right)\coloneqq\widetilde{\varGamma}_{G}\left(v\right)+Z\cdot\mu_{\widetilde{\mathcal{M}}_{2}^{\ddagger}}\left(v\right),\,v\in I.
\]
Note that $\left\{ \widetilde{\varGamma}_{G}^{BR}\left(v\right):v\in I\right\} $
is a centered Gaussian process with covariance function 
\begin{eqnarray*}
\mathrm{E}\left[\widetilde{\varGamma}_{G}^{BR}\left(v\right)\widetilde{\varGamma}_{G}^{BR}\left(v'\right)\right] & = & \mathrm{E}\left[\widetilde{\mathcal{M}}_{2}^{\ddagger}\left(B_{11};v\right)\widetilde{\mathcal{M}}_{2}^{\ddagger}\left(B_{11};v'\right)\right]\\
 & = & \frac{\int_{\underline{b}}^{\overline{b}}\frac{1}{h}\int K_{f}'\left(u\right)K_{g}\left(\frac{w-s\left(v\right)}{h}-s'\left(v\right)u\right)\mathrm{d}u\int K_{f}'\left(u'\right)K_{g}\left(\frac{w-s\left(v'\right)}{h}-s'\left(v'\right)u'\right)\mathrm{d}u'\mathrm{d}w}{\sqrt{\int\left\{ \int K_{f}'\left(u\right)K_{g}\left(w-s'\left(v\right)u\right)\mathrm{d}u\right\} ^{2}\mathrm{d}w}\sqrt{\int\left\{ \int K_{f}'\left(u\right)K_{g}\left(w-s'\left(v'\right)u\right)\mathrm{d}u\right\} ^{2}\mathrm{d}w}}\\
 & = & \frac{\int\int K_{f}'\left(u\right)K_{g}\left(y-s'\left(v\right)u\right)\mathrm{d}u\int K_{f}'\left(u'\right)K_{g}\left(y-s'\left(v'\right)u'+\frac{s\left(v\right)-s\left(v'\right)}{h}\right)\mathrm{d}u'\mathrm{d}y}{\sqrt{\int\left\{ \int K_{f}'\left(u\right)K_{g}\left(w-s'\left(v\right)u\right)\mathrm{d}u\right\} ^{2}\mathrm{d}w}\sqrt{\int\left\{ \int K_{f}'\left(u\right)K_{g}\left(w-s'\left(v'\right)u\right)\mathrm{d}u\right\} ^{2}\mathrm{d}w}},
\end{eqnarray*}
where the third equality holds when $h$ is sufficiently small. Moreover,
it follows from (\ref{eq:sup miu_M_til_2 rate}) that
\begin{equation}
\left\Vert \widetilde{\varGamma}_{G}\right\Vert _{I}=\left\Vert \widetilde{\varGamma}_{G}^{BR}\right\Vert _{I}+O_{p}\left(h^{\nicefrac{3}{2}}\right).\label{eq:Gamma_G Gamma_G_BR}
\end{equation}
$\left\{ \widetilde{\varGamma}_{G}^{BR}\left(v\right):v\in I\right\} $
is a Bickel-Rosenblatt-type Gaussian approximation. 

Suppose that $s$ is a linear function so that $s'\left(v\right)=\overline{\gamma}$
for all $v\in I$, for some positive constant $\overline{\gamma}$.
It is easy to see that in this special case, the covariance function
is 
\[
\mathrm{E}\left[\widetilde{\varGamma}_{G}^{BR}\left(v\right)\widetilde{\varGamma}_{G}^{BR}\left(v'\right)\right]=\frac{\int\int K_{f}'\left(u\right)K_{g}\left(y-\overline{\gamma}u\right)\mathrm{d}u\int K_{f}'\left(u'\right)K_{g}\left(y-\overline{\gamma}u'+\frac{\overline{\gamma}}{h}\left(v-v'\right)\right)\mathrm{d}u'\mathrm{d}y}{\int\left\{ \int K_{f}'\left(u\right)K_{g}\left(w-\overline{\gamma}u\right)\mathrm{d}u\right\} ^{2}\mathrm{d}w},
\]
which is a function of $\left(v-v'\right)$. So the process $\left\{ \widetilde{\varGamma}_{G}^{BR}\left(v\right):v\in I\right\} $
is stationary. 

Denote 
\[
\rho\left(t\right)=\frac{\int\int K_{f}'\left(u\right)K_{g}\left(y-\overline{\gamma}u\right)\mathrm{d}u\int K_{f}'\left(u'\right)K_{g}\left(y-\overline{\gamma}u'+\overline{\gamma}t\right)\mathrm{d}u'\mathrm{d}y}{\int\left\{ \int K_{f}'\left(u\right)K_{g}\left(w-\overline{\gamma}u\right)\mathrm{d}u\right\} ^{2}\mathrm{d}w}.
\]
It is easy to check that 
\begin{gather*}
\rho\left(0\right)=1\\
\rho'\left(0\right)=\frac{\overline{\gamma}\int\int K_{f}'\left(u\right)K_{g}\left(y-\overline{\gamma}u\right)\mathrm{d}u\int K_{f}'\left(u'\right)K_{g}'\left(y-\overline{\gamma}u'\right)\mathrm{d}u'\mathrm{d}y}{\int\left\{ \int K_{f}'\left(u\right)K_{g}\left(w-\overline{\gamma}u\right)\mathrm{d}u\right\} ^{2}\mathrm{d}w}=0
\end{gather*}
and 
\begin{eqnarray*}
\rho''\left(0\right) & = & \frac{\overline{\gamma}^{2}\int\int K_{f}'\left(u\right)K_{g}\left(y-\overline{\gamma}u\right)\mathrm{d}u\int K_{f}'\left(u'\right)K_{g}''\left(y-\overline{\gamma}u'\right)\mathrm{d}u'\mathrm{d}y}{\int\left\{ \int K_{f}'\left(u\right)K_{g}\left(w-\overline{\gamma}u\right)\mathrm{d}u\right\} ^{2}\mathrm{d}w}\\
 & = & -\frac{\overline{\gamma}^{2}\int\left\{ \int K_{f}'\left(u\right)K_{g}'\left(y-\overline{\gamma}u\right)\mathrm{d}u\right\} ^{2}\mathrm{d}y}{\int\left\{ \int K_{f}'\left(u\right)K_{g}\left(w-\overline{\gamma}u\right)\mathrm{d}u\right\} ^{2}\mathrm{d}w},
\end{eqnarray*}
where the second equality follows from integration by parts. 

Define $\ddot{\varGamma}_{G}^{BR}\left(y\right)\coloneqq\widetilde{\varGamma}_{G}^{BR}\left(v_{l}+h\cdot y\right)$,
for $y\in\left[0,\frac{v_{u}-v_{l}}{h}\right]$. It is easy to see
that $\mathrm{E}\left[\ddot{\varGamma}_{G}^{BR}\left(y\right)\ddot{\varGamma}_{G}^{BR}\left(y'\right)\right]=\rho\left(y-y'\right)$
and
\[
\left\Vert \widetilde{\varGamma}_{G}^{BR}\right\Vert _{I}=\underset{y\in\left[0,\frac{v_{u}-v_{l}}{h}\right]}{\mathrm{sup}}\left|\ddot{\varGamma}_{G}^{BR}\left(y\right)\right|.
\]
Note that $\mathrm{E}\left[\ddot{\varGamma}_{G}^{BR}\left(y\right)\right]=0$
and $\mathrm{E}\left[\ddot{\varGamma}_{G}^{BR}\left(y\right)^{2}\right]=1$.
$\left\{ \ddot{\varGamma}_{G}^{BR}\left(y\right):y\in\mathbb{R}_{+}\right\} $
is a centered, normalized and stationary Gaussian process. It is easy
to check that since $K_{f}'$ and $K_{g}$ are both supported on $\left[-1,1\right]$,
$\rho\left(t\right)=0$ when $\left|t\right|>2$. Since $\rho\left(0\right)=1$
and $\rho'\left(0\right)=0$, we also have $\rho\left(t\right)=1-\lambda t^{2}+o\left(t^{2}\right)$,
as $t\downarrow0$, where $\lambda=-\rho''\left(0\right)$. Therefore
the conditions in the statement of \citet[Theorem 2.7.9]{gine2015mathematical}
are all satisfied. Let 
\begin{gather*}
a_{L}\coloneqq\left(2\cdot\mathrm{log}\left(\frac{v_{u}-v_{l}}{h}\right)\right)^{\nicefrac{1}{2}}\\
b_{L}\coloneqq\left(2\cdot\mathrm{log}\left(\frac{v_{u}-v_{l}}{h}\right)\right)^{\nicefrac{1}{2}}+\frac{\mathrm{log}\left(\frac{\lambda^{\nicefrac{1}{2}}}{2\pi}\right)}{\left(2\cdot\mathrm{log}\left(\frac{v_{u}-v_{l}}{h}\right)\right)^{\nicefrac{1}{2}}}.
\end{gather*}
By \citet[Theorem 2.7.9]{gine2015mathematical}, 
\[
\underset{L\uparrow\infty}{\mathrm{lim}}\mathrm{P}\left[a_{L}\left(\left\Vert \widetilde{\varGamma}_{G}^{BR}\right\Vert _{I}-b_{L}\right)\leq x\right]=\mathrm{exp}\left(-\mathrm{exp}\left(-x\right)\right),\textrm{ for all \ensuremath{x\in\mathbb{R}}},
\]
where $x\mapsto\mathrm{exp}\left(-\mathrm{exp}\left(-x\right)\right)$
is the standard Gumbel CDF. We note that $a_{L}=O\left(\mathrm{log}\left(L\right)^{\nicefrac{1}{2}}\right)$.
It then follows from (\ref{eq:Gamma_G Gamma_G_BR}) and Slutsky's
lemma that
\[
\underset{L\uparrow\infty}{\mathrm{lim}}\mathrm{P}\left[a_{L}\left(\left\Vert \widetilde{\varGamma}_{G}\right\Vert _{I}-b_{L}\right)\leq x\right]=\mathrm{exp}\left(-\mathrm{exp}\left(-x\right)\right),\textrm{ for all \ensuremath{x\in\mathbb{R}}}.
\]
Since $\widetilde{W}_{L}\overset{d}{=}\left\Vert \widetilde{\varGamma}_{G}\right\Vert _{I}$,
we also have 
\[
\underset{L\uparrow\infty}{\mathrm{lim}}\mathrm{P}\left[a_{L}\left(\widetilde{W}_{L}-b_{L}\right)\leq x\right]=\mathrm{exp}\left(-\mathrm{exp}\left(-x\right)\right),\textrm{ for all \ensuremath{x\in\mathbb{R}}},
\]
and therefore by (\ref{eq:Z - W}) and Slutsky's lemma, 
\[
\underset{L\uparrow\infty}{\mathrm{lim}}\mathrm{P}\left[a_{L}\left(\left\Vert Z\right\Vert _{I}-b_{L}\right)\leq x\right]=\mathrm{exp}\left(-\mathrm{exp}\left(-x\right)\right),\textrm{ for all \ensuremath{x\in\mathbb{R}}}.
\]
Note that unlike many other asymptotic results for limiting distributions
of uniform errors in the literature (see, e.g., \citealp{bickel1973some}
and \citealp{ghosal2000testing}), the normalizing constants for $\left\Vert Z\right\Vert _{I}$
depend on the unknown slope $\overline{\gamma}$.

For the general case, the difficulty is that the approximating Gaussian
process $\left\{ \widetilde{\varGamma}_{G}^{BR}\left(v\right):v\in I\right\} $
is non-stationary. Deriving the limiting distribution of $\left\Vert \widetilde{\varGamma}_{G}^{BR}\right\Vert _{I}$
(and $\left\Vert Z\right\Vert _{I}$) requires non-standard techniques
and is beyond the scope of this paper.

\section{Proofs of the Results in Section 6}

\subsection{Preliminaries and Notation}

For fixed $\boldsymbol{x}\in\mathrm{int}\left(\mathcal{X}\right)$,
let $I\left(\boldsymbol{x}\right)\coloneqq\left[v_{l}\left(\boldsymbol{x}\right),v_{u}\left(\boldsymbol{x}\right)\right]$
be an inner closed subset of $\left[\underline{v}\left(\boldsymbol{x}\right),\overline{v}\left(\boldsymbol{x}\right)\right]$.
Fix 
\[
\delta_{0}\coloneqq\mathrm{min}\left\{ \nicefrac{\left(\overline{v}\left(\boldsymbol{x}\right)-v_{u}\left(\boldsymbol{x}\right)\right)}{2},\nicefrac{\left(v_{l}\left(\boldsymbol{x}\right)-\underline{v}\left(\boldsymbol{x}\right)\right)}{2}\right\} .
\]
Then for any $n'\in\mathcal{N}$, by the strict monotonicity of $s\left(\cdot,\boldsymbol{x},n'\right)$
we have 
\[
s\left(v_{l}\left(\boldsymbol{x}\right)-\delta_{0},\boldsymbol{x},n'\right)>s\left(\underline{v}\left(\boldsymbol{x}\right),\boldsymbol{x},n'\right)=\underline{b}\left(\boldsymbol{x}\right).
\]

Since $\boldsymbol{x}$ is an interior point, by the continuity of
$s\left(\cdot,\cdot,n'\right)$ and $\underline{b}\left(\cdot\right)$,
there exists a neighborhood $\mathbb{H}\left(\boldsymbol{x},\underline{\delta}_{n'}\right)$
for some $\underline{\delta}_{n'}>0$ and some $\underline{\delta}_{n'}^{\dagger}>0$
such that $\mathbb{H}\left(\boldsymbol{x},\underline{\delta}_{n'}\right)\subseteq\mathrm{int}\left(\mathcal{X}\right)$
and
\[
v_{l}\left(\boldsymbol{x}\right)-\delta_{0}>\underline{v}\left(\boldsymbol{x}'\right),\,\textrm{for all \ensuremath{\boldsymbol{x}'\in\mathbb{H}\left(\boldsymbol{x},\underline{\delta}_{n'}\right)}}
\]
and 
\[
\underset{\boldsymbol{x}'\in\mathbb{H}\left(\boldsymbol{x},\underline{\delta}_{n'}\right)}{\mathrm{inf}}s\left(v_{l}\left(\boldsymbol{x}\right)-\delta_{0},\boldsymbol{x}',n'\right)>\underset{\boldsymbol{x}''\in\mathbb{H}\left(\boldsymbol{x},\underline{\delta}_{n'}\right)}{\mathrm{sup}}\underline{b}\left(\boldsymbol{x}''\right)+\underline{\delta}_{n'}^{\dagger}.
\]
Similarly, we can find some $\overline{\delta}_{n'}>0$ and some $\overline{\delta}_{n'}^{\dagger}>0$
such that $\mathbb{H}\left(\boldsymbol{x},\overline{\delta}_{n'}\right)\subseteq\mathrm{int}\left(\mathcal{X}\right)$
and
\[
v_{u}\left(\boldsymbol{x}\right)+\delta_{0}<\overline{v}\left(\boldsymbol{x}'\right),\,\textrm{for all \ensuremath{\boldsymbol{x}'\in\mathbb{H}\left(\boldsymbol{x},\overline{\delta}_{n'}\right)}}
\]
and
\[
\underset{\boldsymbol{x}'\in\mathbb{H}\left(\boldsymbol{x},\overline{\delta}_{n'}\right)}{\mathrm{sup}}s\left(v_{u}\left(\boldsymbol{x}\right)+\delta_{0},\boldsymbol{x}',n'\right)+\overline{\delta}_{n'}^{\dagger}<\underset{\boldsymbol{x}''\in\mathbb{H}\left(\boldsymbol{x},\overline{\delta}_{n'}\right)}{\mathrm{inf}}\overline{b}\left(\boldsymbol{x}'',n'\right).
\]
Let 

\[
\overline{\delta}\coloneqq\mathrm{min}\left\{ \delta_{0},\nicefrac{\underline{\delta}_{\underline{n}}}{2},...,\nicefrac{\underline{\delta}_{\overline{n}}}{2},\nicefrac{\overline{\delta}_{\underline{n}}}{2},...,\nicefrac{\overline{\delta}_{\overline{n}}}{2}\right\} .
\]
Note that now 
\[
\mathcal{C}_{V,\boldsymbol{X}}\coloneqq\left[v_{l}\left(\boldsymbol{x}\right)-\overline{\delta},v_{u}\left(\boldsymbol{x}\right)+\overline{\delta}\right]\times\mathbb{H}\left(\boldsymbol{x},\overline{\delta}\right)
\]
is a inner closed subset of $\mathcal{S}_{V,\boldsymbol{X}}$, when
$\boldsymbol{x}$ is an interior point of $\mathcal{X}$. Denote
\begin{equation}
\mathcal{C}_{B,\boldsymbol{X}}^{n'}\coloneqq\left\{ \left(s\left(v',\boldsymbol{x}',n'\right),\boldsymbol{x}''\right):\left(v',\boldsymbol{x}'\right)\in\mathcal{C}_{V,\boldsymbol{X}},\,\boldsymbol{x}''\in\mathbb{H}\left(\boldsymbol{x},\overline{\delta}\right)\right\} \label{eq:C_n definition}
\end{equation}
for each $n'\in\mathcal{N}$. By the continuity of $s\left(\cdot,\cdot,n'\right)$
(see Lemma A2 of GPV), $\mathcal{C}_{B,\boldsymbol{X}}^{n'}$ is a
compact inner subset of $\mathcal{S}_{B,\boldsymbol{X}}^{n'}$. 

Proposition 1(ii) of GPV gives that 
\begin{equation}
\underset{n'\in\mathcal{N}}{\mathrm{min}}\underset{\left(b,\boldsymbol{x}'\right)\in\mathcal{S}_{B,\boldsymbol{X}}^{n'}}{\mathrm{inf}}g\left(b,\boldsymbol{x}',n'\right)>\underline{C}_{g}>0\label{eq:bids density bounded away from zero}
\end{equation}
for some constant $\underline{C}_{g}$. Also denote 
\[
\overline{\varphi}\coloneqq\underset{\boldsymbol{z}\in\mathcal{X}}{\mathrm{sup}}\varphi\left(\boldsymbol{z}\right).
\]

Denote 
\[
G\left(b,\boldsymbol{x}',n'\right)\coloneqq G\left(b|\boldsymbol{x}',n'\right)\pi\left(n'|\boldsymbol{x}'\right)\varphi\left(\boldsymbol{x}'\right),\,g\left(b,\boldsymbol{x}',n'\right)\coloneqq g\left(b|\boldsymbol{x}',n'\right)\pi\left(n'|\boldsymbol{x}'\right)\varphi\left(\boldsymbol{x}'\right)
\]
and 
\[
\widehat{G}\left(b,\boldsymbol{x}',n'\right)\coloneqq\widehat{G}\left(b|\boldsymbol{x}',n'\right)\widehat{\pi}\left(n'|\boldsymbol{x}'\right)\widehat{\varphi}\left(\boldsymbol{x}'\right),\,\widehat{g}\left(b,\boldsymbol{x}',n'\right)\coloneqq\widehat{g}\left(b|\boldsymbol{x}',n'\right)\widehat{\pi}\left(n'|\boldsymbol{x}'\right)\widehat{\varphi}\left(\boldsymbol{x}'\right).
\]

Denote
\[
\varGamma\left(v|\boldsymbol{x}\right)\coloneqq\frac{1}{L^{\nicefrac{1}{2}}}\sum_{l=1}^{L}\frac{\sum_{n\in\mathcal{N}}\left\{ \mathcal{M}_{2}^{n}\left(\boldsymbol{B}_{\cdot l},\boldsymbol{X}_{l},N_{l};v\right)-\mu_{\mathcal{M}^{n}}\left(v\right)\right\} }{\mathrm{Var}\left[\sum_{n\in\mathcal{N}}\mathcal{M}_{2}^{n}\left(\boldsymbol{B}_{\cdot1},\boldsymbol{X}_{1},N_{1};v\right)\right]^{\nicefrac{1}{2}}},\textrm{ }v\in I\left(\boldsymbol{x}\right)
\]
and its bootstrap analogue 
\[
\varGamma^{*}\left(v|\boldsymbol{x}\right)\coloneqq\frac{1}{L^{\nicefrac{1}{2}}}\sum_{l=1}^{L}\frac{\sum_{n\in\mathcal{N}}\left\{ \mathcal{M}_{2}^{n}\left(\boldsymbol{B}_{\cdot l}^{*},\boldsymbol{X}_{l}^{*},N_{l}^{*};v\right)-\widehat{\mu}_{\mathcal{M}_{2}^{n}}\left(v\right)\right\} }{\mathrm{Var}\left[\sum_{n\in\mathcal{N}}\mathcal{M}_{2}^{n}\left(\boldsymbol{B}_{\cdot1},\boldsymbol{X}_{1},N_{1};v\right)\right]^{\nicefrac{1}{2}}},\textrm{ }v\in I\left(\boldsymbol{x}\right),
\]
where 
\begin{eqnarray*}
\widehat{\mu}_{\mathcal{M}_{2}^{n}}\left(v\right) & \coloneqq & \mathrm{E}^{*}\left[\mathcal{M}_{2}^{n}\left(\boldsymbol{B}_{\cdot1}^{*},\boldsymbol{X}_{1}^{*},N_{1}^{*};v\right)\right]\\
 & = & \mathrm{E}^{*}\left[\mathrm{E}^{*}\left[\mathcal{M}_{2}^{n}\left(\boldsymbol{B}_{\cdot1}^{*},\boldsymbol{X}_{1}^{*},N_{1}^{*};v\right)|\boldsymbol{X}_{1}^{*},N_{1}^{*}\right]\right]\\
 & = & \frac{1}{L}\sum_{l=1}^{L}\mathcal{M}_{2}^{n}\left(\boldsymbol{B}_{\cdot l},\boldsymbol{X}_{l},N_{l};v\right),
\end{eqnarray*}
where the second equality follows from LIE and the third equality
can be verified using the fact that the bids in the bootstrap sample
are conditionally i.i.d..

Denote

\[
\widetilde{f}\left(v,\boldsymbol{x},n\right)\coloneqq\frac{1}{L}\sum_{l=1}^{L}\mathbbm{1}\left(N_{l}=n\right)\frac{1}{N_{l}}\sum_{i=1}^{N_{l}}\frac{1}{h^{1+d}}K_{f}\left(\frac{V_{il}-v}{h},\frac{\boldsymbol{X}_{l}-\boldsymbol{x}}{h}\right),
\]
\[
\widetilde{f}\left(v|\boldsymbol{x}\right)\coloneqq\frac{1}{\widehat{\varphi}\left(\boldsymbol{x}\right)}\sum_{n\in\mathcal{N}}\widetilde{f}\left(v,\boldsymbol{x},n\right)
\]
and
\begin{eqnarray*}
\widehat{f}_{GPV}\left(v,\boldsymbol{x},n\right) & \coloneqq & \widehat{f}_{GPV}\left(v|\boldsymbol{x},n\right)\widehat{\pi}\left(n|\boldsymbol{x}\right)\widehat{\varphi}\left(\boldsymbol{x}\right)\\
 & = & \frac{1}{L}\sum_{l=1}^{L}\mathbbm{1}\left(N_{l}=n\right)\frac{1}{N_{l}}\sum_{i=1}^{N_{l}}\mathbb{T}_{il}\frac{1}{h^{1+d}}K_{f}\left(\frac{\widehat{V}_{il}-v}{h},\frac{\boldsymbol{X}_{l}-\boldsymbol{x}}{h}\right).
\end{eqnarray*}
Note that now we have 
\[
\widehat{f}_{GPV}\left(v|\boldsymbol{x}\right)=\frac{1}{\widehat{\varphi}\left(\boldsymbol{x}\right)}\sum_{n\in\mathcal{N}}\widehat{f}_{GPV}\left(v,\boldsymbol{x},n\right).
\]

Let 
\[
\widehat{G}^{*}\left(b',\boldsymbol{x}',n'\right)\coloneqq\frac{1}{L}\sum_{l=1}^{L}\mathbbm{1}\left(N_{l}^{*}=n'\right)\frac{1}{N_{l}^{*}}\sum_{i=1}^{N_{l}^{*}}\mathbbm{1}\left(B_{il}^{*}\leq b'\right)\frac{1}{h^{d}}K_{\boldsymbol{X}}\left(\frac{\boldsymbol{X}_{l}^{*}-\boldsymbol{x}'}{h}\right)
\]
and 
\[
\widehat{g}^{*}\left(b',\boldsymbol{x}',n'\right)\coloneqq\frac{1}{L}\sum_{l=1}^{L}\mathbbm{1}\left(N_{l}^{*}=n'\right)\frac{1}{N_{l}^{*}}\sum_{i=1}^{N_{l}^{*}}\mathbbm{1}\left(B_{il}^{*}\leq b'\right)\frac{1}{h^{1+d}}K_{g}\left(\frac{B_{il}^{*}-b'}{h}\right)K_{\boldsymbol{X}}\left(\frac{\boldsymbol{X}_{l}^{*}-\boldsymbol{x}'}{h}\right).
\]
Note that now we have 
\[
\widehat{V}_{il}^{*}=B_{il}^{*}+\frac{1}{N_{l}^{*}-1}\frac{\widehat{G}^{*}\left(B_{il}^{*},\boldsymbol{X}_{l}^{*},N_{l}^{*}\right)}{\widehat{g}^{*}\left(B_{il}^{*},\boldsymbol{X}_{l}^{*},N_{l}^{*}\right)}.
\]
Let 
\[
\widehat{f}_{GPV}^{*}\left(v|\boldsymbol{x},n\right)\coloneqq\frac{1}{\widehat{\pi}^{*}\left(n|\boldsymbol{x}\right)\widehat{\varphi}^{*}\left(\boldsymbol{x}\right)L}\sum_{l=1}^{L}\mathbbm{1}\left(N_{l}^{*}=n\right)\frac{1}{N_{l}^{*}}\sum_{i=1}^{N_{l}^{*}}\mathbb{T}_{il}^{*}\frac{1}{h^{1+d}}K_{f}\left(\frac{\widehat{V}_{il}^{*}-v}{h},\frac{\boldsymbol{X}_{l}^{*}-\boldsymbol{x}}{h}\right)
\]
and 
\[
\widehat{f}_{GPV}^{*}\left(v,\boldsymbol{x},n\right)\coloneqq\widehat{f}_{GPV}^{*}\left(v|\boldsymbol{x},n\right)\widehat{\pi}^{*}\left(n|\boldsymbol{x}\right)\widehat{\varphi}^{*}\left(\boldsymbol{x}\right).
\]
Note that we have 
\[
\widehat{f}_{GPV}^{*}\left(v|\boldsymbol{x}\right)=\frac{1}{\widehat{\varphi}^{*}\left(\boldsymbol{x}\right)}\sum_{n\in\mathcal{N}}\widehat{f}_{GPV}^{*}\left(v,\boldsymbol{x},n\right).
\]
Denote
\[
\widetilde{f}^{*}\left(v,\boldsymbol{x},n\right)\coloneqq\frac{1}{L}\sum_{l=1}^{L}\mathbbm{1}\left(N_{l}^{*}=n\right)\frac{1}{N_{l}^{*}}\sum_{i=1}^{N_{l}^{*}}\frac{1}{h^{1+d}}K_{f}\left(\frac{V_{il}^{*}-v}{h},\frac{\boldsymbol{X}_{l}^{*}-\boldsymbol{x}}{h}\right),
\]
where $V_{il}^{*}\coloneqq\xi\left(B_{il}^{*},\boldsymbol{X}_{l}^{*},N_{l}^{*}\right)$. 

For a function $\phi:\mathbb{R}^{1+d}\rightarrow\mathbb{R}$, denote
\[
D_{j}^{\alpha}\phi\left(x_{1},...,x_{1+d}\right)\coloneqq\left.\frac{\partial^{\alpha}\phi\left(z_{1},...,z_{1+d}\right)}{\partial z_{j}^{\alpha}}\right|_{\left(z_{1},...,z_{1+d}\right)=\left(x_{1},...,x_{1+d}\right)},
\]
for $\alpha\in\mathbb{Z}_{+}$ and $j=1,...,1+d$. It is also convenient
to denote
\[
\phi'\left(x_{1},...,x_{1+d}\right)\coloneqq\left.\frac{\partial\phi\left(z_{1},...,z_{1+d}\right)}{\partial z_{1}}\right|_{\left(z_{1},...,z_{1+d}\right)=\left(x_{1},...,x_{1+d}\right)}\textrm{ and }\phi''\left(x_{1},...,x_{1+d}\right)\coloneqq\left.\frac{\partial^{2}\phi\left(z_{1},...,z_{1+d}\right)}{\partial z_{1}^{2}}\right|_{\left(z_{1},...,z_{1+d}\right)=\left(x_{1},...,x_{1+d}\right)}
\]
to be the partial derivatives with respect to the first argument. 

Let $\left\Vert \cdot\right\Vert _{1}$ denote the $L_{1}$ norm on
$\mathbb{R}^{k}$: $\left\Vert \left(z_{1},...,z_{k}\right)\right\Vert _{1}=\left|z_{1}\right|+\cdots+\left|z_{k}\right|$.
Also denote 
\[
K_{\boldsymbol{X}}^{0}\left(\boldsymbol{x}\right)\coloneqq\prod_{k=1}^{d}K_{0}\left(x_{k}\right),\textrm{ for \ensuremath{\boldsymbol{x}=\left(x_{1},...,x_{d}\right)\in\mathbb{R}^{d}}}
\]
and
\[
\mathbb{S}_{R}\coloneqq\left\{ \left(\alpha_{1},...,\alpha_{1+d}\right)\in\mathbb{Z}_{+}^{1+d}:\sum_{j=1}^{1+d}\alpha_{j}=R\right\} .
\]

Law of iterated expectations is abbreviated as ``LIE''. $\underset{i,l}{\mathrm{max}}$
is understood as $\underset{l\in\left\{ 1,...,L\right\} }{\mathrm{max}}\underset{i\in\left\{ 1,...,N_{l}\right\} }{\mathrm{max}}$.
$\underset{\left(2\right)}{\sum}$ is understood as $\sum_{k\neq l}$
and $\underset{\left(3\right)}{\sum}$ is understood as $\sum_{l}\sum_{k\neq l}\sum_{k'\neq l,\,k'\neq k},$
i.e., summing over all distinct indices. $\left(L\right)_{2}$ is
understood as $\left(L\right)\left(L-1\right)$ and $\left(L\right)_{3}$
is understood as $L\left(L-1\right)\left(L-2\right)$. 

\subsection{Proofs of the Results in Section 6}

\begin{proof}[Proof of Theorem 6.1]Now it follows from Lemma \ref{Lemma 3}
that
\begin{equation}
\left(Lh^{3+d}\right)^{\nicefrac{1}{2}}\left(\widehat{f}_{GPV}\left(v|\boldsymbol{x},n\right)-f\left(v|\boldsymbol{x}\right)\right)=\frac{1}{\widehat{\pi}\left(n|\boldsymbol{x}\right)\widehat{\varphi}\left(\boldsymbol{x}\right)}\frac{1}{L^{\nicefrac{1}{2}}}\sum_{l=1}^{L}h^{\nicefrac{\left(3+d\right)}{2}}\left(\mathcal{M}_{2}^{n}\left(\boldsymbol{B}_{\cdot l},\boldsymbol{X}_{l},N_{l};v\right)-\mu_{\mathcal{M}^{n}}\left(v\right)\right)+o_{p}\left(1\right).\label{eq:f_hat - f linearization 2}
\end{equation}
We now show that a central limit theorem for triangular arrays can
be applied to the leading term of the right hand side of (\ref{eq:f_hat - f linearization 2}). 

Let 
\begin{align*}
 & \mathcal{M}^{\ddagger,n}\left(\left(\boldsymbol{b}.,\boldsymbol{z},m\right),\left(\boldsymbol{b}.',\boldsymbol{z}',m'\right);v\right)\\
\coloneqq & -\mathbbm{1}\left(m=n\right)\frac{1}{m}\sum_{i=1}^{m}\frac{1}{h^{3+2d}}K_{f}'\left(\frac{\xi\left(b_{i},\boldsymbol{z},m\right)-v}{h},\frac{\boldsymbol{z}-\boldsymbol{x}}{h}\right)\frac{G\left(b_{i},\boldsymbol{z},m\right)}{\left(m-1\right)g\left(b_{i},\boldsymbol{z},m\right)^{2}}\mathbbm{1}\left(m'=m\right)\frac{1}{m'}\sum_{j=1}^{m'}K_{g}\left(\frac{b_{j}'-b_{i}}{h}\right)K_{\boldsymbol{X}}\left(\frac{\boldsymbol{z}'-\boldsymbol{z}}{h}\right),
\end{align*}
\[
\mathcal{M}_{2}^{\ddagger,n}\left(\boldsymbol{b}.,\boldsymbol{z},m;v\right)\coloneqq\mathrm{E}\left[\mathcal{M}^{\ddagger,n}\left(\left(\boldsymbol{B}_{\cdot1},\boldsymbol{X}_{1},N_{1}\right),\left(\boldsymbol{b}.,\boldsymbol{z},m\right);v\right)\right]
\]
and 
\[
\mu_{\mathcal{M}^{\ddagger,n}}\left(v\right)\coloneqq\mathrm{E}\left[\mathcal{M}^{\ddagger,n}\left(\left(\boldsymbol{B}_{\cdot1},\boldsymbol{X}_{1},N_{1}\right),\left(\boldsymbol{B}_{\cdot2},\boldsymbol{X}_{2},N_{2}\right);v\right)\right].
\]
By the LIE, we have 
\begin{align*}
 & \mathcal{M}_{2}^{\ddagger,n}\left(\boldsymbol{b}.,\boldsymbol{z},m;v\right)\\
= & \mathrm{E}\left[\mathrm{E}\left[\mathcal{M}^{\ddagger,n}\left(\left(\boldsymbol{B}_{\cdot1},\boldsymbol{X}_{1},N_{1}\right),\left(\boldsymbol{b}.,\boldsymbol{z},m\right);v\right)|\boldsymbol{X}_{1},N_{1}\right]\right]\\
= & -\mathbbm{1}\left(m=n\right)\frac{1}{m}\sum_{i=1}^{m}\int_{\mathcal{X}}\int_{\underline{b}\left(\boldsymbol{z}'\right)}^{\overline{b}\left(\boldsymbol{z}',n\right)}\frac{1}{h^{3+2d}}K_{f}'\left(\frac{\xi\left(b',\boldsymbol{z}',n\right)-v}{h},\frac{\boldsymbol{z}'-\boldsymbol{x}}{h}\right)\frac{G\left(b',\boldsymbol{z}',n\right)}{\left(n-1\right)g\left(b',\boldsymbol{z}',n\right)}K_{g}\left(\frac{b'-b_{i}}{h}\right)K_{\boldsymbol{X}}\left(\frac{\boldsymbol{z}'-\boldsymbol{z}}{h}\right)\mathrm{d}b'\mathrm{d}\boldsymbol{z}'
\end{align*}
and
\begin{align*}
 & \mathcal{M}_{2}^{n}\left(\boldsymbol{b}.,\boldsymbol{z},m;v\right)\\
= & \mathrm{E}\left[-\mathbbm{1}\left(N_{1}=n\right)\frac{1}{N_{1}}\sum_{i=1}^{N_{1}}\frac{1}{h^{2+d}}K_{f}'\left(\frac{\xi\left(B_{i1},\boldsymbol{X}_{1},N_{1}\right)-v}{h},\frac{\boldsymbol{X}_{1}-\boldsymbol{x}}{h}\right)\frac{G\left(B_{i1},\boldsymbol{X}_{1},N_{1}\right)}{\left(N_{1}-1\right)g\left(B_{i1},\boldsymbol{X}_{1},N_{1}\right)^{2}}\right.\\
 & \left.\times\mathbbm{1}\left(m=N_{1}\right)\frac{1}{m}\sum_{j=1}^{m}\frac{1}{h^{1+d}}K_{g}\left(\frac{b_{j}-B_{i1}}{h}\right)K_{\boldsymbol{X}}\left(\frac{\boldsymbol{X}_{1}-\boldsymbol{z}}{h}\right)\right]\\
 & +\mathrm{E}\left[\mathbbm{1}\left(N_{1}=n\right)\frac{1}{N_{1}}\sum_{i=1}^{N_{1}}\frac{1}{h^{2+d}}K_{f}'\left(\frac{\xi\left(B_{i1},\boldsymbol{X}_{1},N_{1}\right)-v}{h},\frac{\boldsymbol{X}_{1}-\boldsymbol{x}}{h}\right)\frac{G\left(B_{i1},\boldsymbol{X}_{1},N_{1}\right)}{\left(N_{1}-1\right)g\left(B_{i1},\boldsymbol{X}_{1},N_{1}\right)}\right]\\
= & -\mathbbm{1}\left(m=n\right)\frac{1}{m}\sum_{i=1}^{m}\int_{\mathcal{X}}\int_{\underline{b}\left(\boldsymbol{z}'\right)}^{\overline{b}\left(\boldsymbol{z}',n\right)}\frac{1}{h^{3+2d}}K_{f}'\left(\frac{\xi\left(b',\boldsymbol{z}',n\right)-v}{h},\frac{\boldsymbol{z}'-\boldsymbol{x}}{h}\right)\frac{G\left(b',\boldsymbol{z}',n\right)}{\left(n-1\right)g\left(b',\boldsymbol{z}',n\right)}K_{g}\left(\frac{b'-b_{i}}{h}\right)K_{\boldsymbol{X}}\left(\frac{\boldsymbol{z}'-\boldsymbol{z}}{h}\right)\mathrm{d}b'\mathrm{d}\boldsymbol{z}'\\
 & +\frac{1}{n-1}\int_{\mathcal{X}}\int_{\underline{b}\left(\boldsymbol{z}'\right)}^{\overline{b}\left(\boldsymbol{z}',n\right)}\frac{1}{h^{2+d}}K_{f}'\left(\frac{\xi\left(b',\boldsymbol{z}',n\right)-v}{h},\frac{\boldsymbol{z}'-\boldsymbol{x}}{h}\right)G\left(b',\boldsymbol{z}',n\right)\mathrm{d}b'\mathrm{d}\boldsymbol{z}'.
\end{align*}
It is straightforward to check 
\begin{equation}
\mathcal{M}_{2}^{n}\left(\boldsymbol{B}_{\cdot l},\boldsymbol{X}_{l},N_{l};v\right)-\mu_{\mathcal{M}^{n}}\left(v\right)=\mathcal{M}_{2}^{\ddagger,n}\left(\boldsymbol{B}_{\cdot l},\boldsymbol{X}_{l},N_{l};v\right)-\mu_{\mathcal{M}^{\ddagger,n}}\left(v\right),\textrm{ for all \ensuremath{l=1,...,L}}.\label{eq:M_2_n - miu_M_n =00003D M_2_ddagger - miu_ddagger}
\end{equation}

Denote
\[
U_{l}^{n}\left(v\right)\coloneqq L^{-\nicefrac{1}{2}}h^{\nicefrac{\left(3+d\right)}{2}}\left(\mathcal{M}_{2}^{\ddagger,n}\left(\boldsymbol{B}_{\cdot l},\boldsymbol{X}_{l},N_{l};v\right)-\mu_{\mathcal{M}^{\ddagger,n}}\left(v\right)\right)
\]
and 
\begin{equation}
\sigma^{n}\left(v\right)\coloneqq\left(\sum_{l=1}^{L}\mathrm{E}\left[U_{l}^{n}\left(v\right)^{2}\right]\right)^{\nicefrac{1}{2}}=\left(\mathrm{E}\left[h^{3+d}\left(\mathcal{M}_{2}^{\ddagger,n}\left(\boldsymbol{B}_{\cdot1},\boldsymbol{X}_{1},N_{1};v\right)-\mu_{\mathcal{M}^{\ddagger,n}}\left(v\right)\right)^{2}\right]\right)^{\nicefrac{1}{2}}.\label{eq:sigma_L definition}
\end{equation}
By the definition of $U_{l}^{n}\left(v\right)$'s and (\ref{eq:f_hat - f linearization 2}),
\begin{equation}
\left(Lh^{3+d}\right)^{\nicefrac{1}{2}}\left(\widehat{f}_{GPV}\left(v|\boldsymbol{x},n\right)-f\left(v|\boldsymbol{x}\right)\right)=\frac{1}{\widehat{\pi}\left(n|\boldsymbol{x}\right)\widehat{\varphi}\left(\boldsymbol{x}\right)}\sum_{l=1}^{L}U_{l}^{n}\left(v\right)+o_{p}\left(1\right).\label{eq:f_hat - f linearization 3}
\end{equation}
It is easy to check that 
\[
\mu_{\mathcal{M}^{\ddagger,n}}\left(v\right)=\mu_{\mathcal{M}}\left(v\right)+\left(-\frac{1}{n-1}\int_{\mathcal{X}}\int_{\underline{b}\left(\boldsymbol{z}'\right)}^{\overline{b}\left(\boldsymbol{z}',n\right)}\frac{1}{h^{2+d}}K_{f}'\left(\frac{\xi\left(b',\boldsymbol{z}',n\right)-v}{h},\frac{\boldsymbol{z}'-\boldsymbol{x}}{h}\right)G\left(b',\boldsymbol{z}',n\right)\mathrm{d}b'\mathrm{d}\boldsymbol{z}'\right).
\]

By change of variables and a mean value expansion, we have
\begin{align}
 & \int_{\mathcal{X}}\int_{\underline{b}\left(\boldsymbol{z}'\right)}^{\overline{b}\left(\boldsymbol{z}',n\right)}\frac{1}{h^{2+d}}K_{f}'\left(\frac{\xi\left(b',\boldsymbol{z}',n\right)-v}{h},\frac{\boldsymbol{z}'-\boldsymbol{x}}{h}\right)G\left(b',\boldsymbol{z}',n\right)\mathrm{d}b'\mathrm{d}\boldsymbol{z}'\nonumber \\
= & \int_{\mathcal{X}}\int_{\frac{\underline{v}\left(\boldsymbol{z}'\right)-v}{h}}^{\frac{\overline{v}\left(\boldsymbol{z}'\right)-v}{h}}\frac{1}{h^{1+d}}K_{f}'\left(u,\frac{\boldsymbol{z}'-\boldsymbol{x}}{h}\right)G\left(s\left(hu+v,\boldsymbol{z}',n\right),\boldsymbol{z}',n\right)s'\left(hu+v,\boldsymbol{z}',n\right)\mathrm{d}u\mathrm{d}\boldsymbol{z}'\nonumber \\
= & \int_{\mathcal{X}}\frac{1}{h^{1+d}}K_{\boldsymbol{X}}^{0}\left(\frac{\boldsymbol{z}'-\boldsymbol{x}}{h}\right)\left(G\left(s\left(v,\boldsymbol{z}',n\right),\boldsymbol{z}',n\right)\right)s'\left(v,\boldsymbol{z}',n\right)\left\{ \int_{\frac{\underline{v}\left(\boldsymbol{z}'\right)-v}{h}}^{\frac{\overline{v}\left(\boldsymbol{z}'\right)-v}{h}}K_{0}'\left(u\right)\mathrm{d}u\right\} \mathrm{d}\boldsymbol{z}'\nonumber \\
 & +\int_{\mathcal{X}}\frac{1}{h^{d}}K_{\boldsymbol{X}}^{0}\left(\frac{\boldsymbol{z}'-\boldsymbol{x}}{h}\right)\int_{\frac{\underline{v}\left(\boldsymbol{z}'\right)-v}{h}}^{\frac{\overline{v}\left(\boldsymbol{z}'\right)-v}{h}}K_{0}'\left(u\right)u\left\{ g\left(s\left(\dot{v},\boldsymbol{z}',n\right),\boldsymbol{z}',n\right)s'\left(\dot{v},\boldsymbol{z}',n\right)^{2}+G\left(s\left(\dot{v},\boldsymbol{z}',n\right),\boldsymbol{z}',n\right)s''\left(\dot{v},\boldsymbol{z}',n\right)\right\} \mathrm{d}u\mathrm{d}\boldsymbol{z}',\label{eq:K_f'*G integral expansion}
\end{align}
where $\dot{v}$ is the mean value (depending on $u$ and $\boldsymbol{z}'$)
with $\left|\dot{v}-v\right|\leq h\left|u\right|$. When $h$ is sufficiently
small, the term on the third line of (\ref{eq:K_f'*G integral expansion})
vanishes since $K_{0}'$ is odd. And also we have
\begin{align*}
 & \underset{v\in I\left(\boldsymbol{x}\right)}{\mathrm{sup}}\left|\int_{\mathcal{X}}\frac{1}{h^{d}}K_{\boldsymbol{X}}^{0}\left(\frac{\boldsymbol{z}'-\boldsymbol{x}}{h}\right)\int_{\frac{\underline{v}\left(\boldsymbol{z}'\right)-v}{h}}^{\frac{\overline{v}\left(\boldsymbol{z}'\right)-v}{h}}K_{0}'\left(u\right)u\left\{ g\left(s\left(\dot{v},\boldsymbol{z}',n\right),\boldsymbol{z}',n\right)s'\left(\dot{v},\boldsymbol{z}',n\right)^{2}+G\left(s\left(\dot{v},\boldsymbol{z}',n\right),\boldsymbol{z}',n\right)s''\left(\dot{v},\boldsymbol{z}',n\right)\right\} \mathrm{d}u\mathrm{d}\boldsymbol{z}'\right|\\
\apprle & \left(\int_{\mathcal{X}}\frac{1}{h^{d}}\left|K_{\boldsymbol{X}}^{0}\left(\frac{\boldsymbol{z}'-\boldsymbol{x}}{h}\right)\right|\mathrm{d}\boldsymbol{z}'\right)\left\{ \underset{\left(u,\boldsymbol{z}'\right)\in\mathcal{C}_{V,\boldsymbol{X}}}{\mathrm{sup}}\left|g\left(s\left(u,\boldsymbol{z}',n\right),\boldsymbol{z}',n\right)s'\left(u,\boldsymbol{z}',n\right)^{2}+G\left(s\left(u,\boldsymbol{z}',n\right),\boldsymbol{z}',n\right)s''\left(u,\boldsymbol{z}',n\right)\right|\right\} \\
= & O\left(1\right),
\end{align*}
where the inequality holds when $h$ is sufficiently small and the
equality follows from the fact that $\mathcal{C}_{V,\boldsymbol{X}}$
is an inner closed subset of $\mathcal{S}_{V,\boldsymbol{X}}$, the
continuity of $g\left(\cdot,\cdot,n\right)$, $G\left(\cdot,\cdot,n\right)$,
$s'\left(\cdot,\cdot,n\right)$ and $s''\left(\cdot,\cdot,n\right)$
and the fact 
\begin{equation}
\int_{\mathcal{X}}\frac{1}{h^{d}}\left|K_{\boldsymbol{X}}^{0}\left(\frac{\boldsymbol{z}'-\boldsymbol{x}}{h}\right)\right|\mathrm{d}\boldsymbol{z}'=O\left(1\right),\label{eq:|K_X| integral order}
\end{equation}
which follows from change of variables. Now it follows that 
\begin{equation}
\underset{v\in I\left(\boldsymbol{x}\right)}{\mathrm{sup}}\left|\int_{\mathcal{X}}\int_{\underline{b}\left(\boldsymbol{z}'\right)}^{\overline{b}\left(\boldsymbol{z}',n\right)}\frac{1}{h^{2+d}}K_{f}'\left(\frac{\xi\left(b',\boldsymbol{z}',n\right)-v}{h},\frac{\boldsymbol{z}'-\boldsymbol{x}}{h}\right)G\left(b',\boldsymbol{z}',n\right)\mathrm{d}b'\mathrm{d}\boldsymbol{z}'\right|=O\left(1\right).\label{eq:sup integral K_f'*G bound}
\end{equation}
Therefore it follows from the fact $\underset{v\in I\left(\boldsymbol{x}\right)}{\mathrm{sup}}\left|\mu_{\mathcal{M}^{n}}\left(v\right)\right|=O\left(h^{R}\right)$
which is shown in the proof of Lemma \ref{Lemma 3} and (\ref{eq:sup integral K_f'*G bound})
that
\begin{equation}
\underset{v\in I\left(\boldsymbol{x}\right)}{\mathrm{sup}}\left|\mu_{\mathcal{M}^{\ddagger,n}}\left(v\right)\right|=O\left(1\right),\textrm{ for all \ensuremath{n\in\mathcal{N}}}.\label{eq:sup miu_M_ddagger rate}
\end{equation}

By the LIE, we have
\begin{align}
 & \mathrm{E}\left[h^{3+d}\mathcal{M}_{2}^{\ddagger,n}\left(\boldsymbol{B}_{\cdot1},\boldsymbol{X}_{1},N_{1};v\right)^{2}\right]\nonumber \\
= & \mathrm{E}\left[\frac{1}{h^{1+d}}\left\{ \mathbbm{1}\left(N_{1}=n\right)\frac{1}{N_{1}}\sum_{i=1}^{N_{1}}\int_{\mathcal{X}}\int_{\underline{b}\left(\boldsymbol{z}'\right)}^{\overline{b}\left(\boldsymbol{z}',n\right)}\frac{1}{h^{1+d}}K_{f}'\left(\frac{\xi\left(b',\boldsymbol{z}',n\right)-v}{h},\frac{\boldsymbol{z}'-\boldsymbol{x}}{h}\right)\right.\right.\nonumber \\
 & \left.\left.\times\frac{G\left(b',\boldsymbol{z}',n\right)}{\left(n-1\right)g\left(b',\boldsymbol{z}',n\right)}K_{g}\left(\frac{B_{i1}-b'}{h}\right)K_{\boldsymbol{X}}\left(\frac{\boldsymbol{X}_{1}-\boldsymbol{z}'}{h}\right)\mathrm{d}b'\mathrm{d}\boldsymbol{z}'\right\} ^{2}\right]\nonumber \\
= & \frac{1}{n\left(n-1\right)^{2}}\frac{1}{h^{1+d}}\int_{\mathcal{X}}\int_{\underline{b}\left(\boldsymbol{z}\right)}^{\overline{b}\left(\boldsymbol{z},n\right)}\left\{ \int_{\mathcal{X}}\int_{\underline{b}\left(\boldsymbol{z}'\right)}^{\overline{b}\left(\boldsymbol{z}',n\right)}\frac{1}{h^{1+d}}K_{f}'\left(\frac{\xi\left(b',\boldsymbol{z}',n\right)-v}{h},\frac{\boldsymbol{z}'-\boldsymbol{x}}{h}\right)\frac{G\left(b',\boldsymbol{z}',n\right)}{g\left(b',\boldsymbol{z}',n\right)}K_{g}\left(\frac{b-b'}{h}\right)K_{\boldsymbol{X}}\left(\frac{\boldsymbol{z}-\boldsymbol{z}'}{h}\right)\mathrm{d}b'\mathrm{d}\boldsymbol{z}'\right\} ^{2}\nonumber \\
 & \times g\left(b,\boldsymbol{z},n\right)\mathrm{d}b\mathrm{d}\boldsymbol{z}\nonumber \\
 & +\frac{1}{n\left(n-1\right)}\frac{1}{h^{1+d}}\int_{\mathcal{X}}\left\{ \int_{\mathcal{X}}\int_{\underline{b}\left(\boldsymbol{z}'\right)}^{\overline{b}\left(\boldsymbol{z}',n\right)}\int_{\underline{b}\left(\boldsymbol{z}\right)}^{\overline{b}\left(\boldsymbol{z},n\right)}\frac{1}{h^{1+d}}K_{f}'\left(\frac{\xi\left(b',\boldsymbol{z}',n\right)-v}{h},\frac{\boldsymbol{z}'-\boldsymbol{x}}{h}\right)\frac{G\left(b',\boldsymbol{z}',n\right)}{g\left(b',\boldsymbol{z}',n\right)}K_{g}\left(\frac{b-b'}{h}\right)K_{\boldsymbol{X}}\left(\frac{\boldsymbol{z}-\boldsymbol{z}'}{h}\right)\right.\nonumber \\
 & \left.\times g\left(b|\boldsymbol{z},n\right)\mathrm{d}b\mathrm{d}b'\mathrm{d}\boldsymbol{z}'\right\} ^{2}\pi\left(n|\boldsymbol{z}\right)\varphi\left(\boldsymbol{z}\right)\mathrm{d}\boldsymbol{z}.\label{eq:E=00005BM_2^2=00005D expansion 2}
\end{align}

By change of variables, we have 
\begin{align*}
 & \frac{1}{n\left(n-1\right)}\frac{1}{h^{1+d}}\int_{\mathcal{X}}\left\{ \int_{\mathcal{X}}\int_{\underline{b}\left(\boldsymbol{z}'\right)}^{\overline{b}\left(\boldsymbol{z}',n\right)}\int_{\underline{b}\left(\boldsymbol{z}\right)}^{\overline{b}\left(\boldsymbol{z},n\right)}\frac{1}{h^{1+d}}K_{f}'\left(\frac{\xi\left(b',\boldsymbol{z}',n\right)-v}{h},\frac{\boldsymbol{z}'-\boldsymbol{x}}{h}\right)\frac{G\left(b',\boldsymbol{z}',n\right)}{g\left(b',\boldsymbol{z}',n\right)}K_{g}\left(\frac{b-b'}{h}\right)K_{\boldsymbol{X}}\left(\frac{\boldsymbol{z}-\boldsymbol{z}'}{h}\right)\right.\\
 & \left.\times g\left(b|\boldsymbol{z},n\right)\mathrm{d}b\mathrm{d}b'\mathrm{d}\boldsymbol{z}'\right\} ^{2}\pi\left(n|\boldsymbol{z}\right)\varphi\left(\boldsymbol{z}\right)\mathrm{d}\boldsymbol{z}\\
= & h\frac{1}{n\left(n-1\right)}\int_{\mathcal{Y}}\left\{ \int_{\mathcal{Y}}\int_{\frac{\underline{v}\left(h\boldsymbol{y}'+\boldsymbol{x}\right)-v}{h}}^{\frac{\overline{v}\left(h\boldsymbol{y}'+\boldsymbol{x}\right)-v}{h}}\int_{\frac{\underline{b}\left(h\boldsymbol{y}+\boldsymbol{x}\right)-s\left(v,\boldsymbol{x},n\right)}{h}}^{\frac{\overline{b}\left(h\boldsymbol{y}+\boldsymbol{x},n\right)-s\left(v,\boldsymbol{x},n\right)}{h}}K_{f}'\left(u,\boldsymbol{y}'\right)\frac{G\left(s\left(hu+v,h\boldsymbol{y}'+\boldsymbol{x},n\right),h\boldsymbol{y}'+\boldsymbol{x},n\right)s'\left(hu+v,h\boldsymbol{y}'+\boldsymbol{x}\right)}{g\left(s\left(hu+v,h\boldsymbol{y}'+\boldsymbol{x},n\right),h\boldsymbol{y}'+\boldsymbol{x},n\right)}\right.\\
 & \left.\times K_{g}\left(w+\frac{s\left(v,\boldsymbol{x},n\right)-s\left(hu+v,h\boldsymbol{y}'+\boldsymbol{x},n\right)}{h}\right)K_{\boldsymbol{X}}\left(\boldsymbol{y}-\boldsymbol{y}'\right)g\left(hw+s\left(v,\boldsymbol{x},n\right)|h\boldsymbol{y}+\boldsymbol{x},n\right)\mathrm{d}w\mathrm{d}u\mathrm{d}\boldsymbol{y}'\right\} ^{2}\pi\left(n|h\boldsymbol{y}+\boldsymbol{x}\right)\varphi\left(h\boldsymbol{y}+\boldsymbol{x}\right)\mathrm{d}\boldsymbol{y}\\
= & h\frac{1}{n\left(n-1\right)}\int_{\mathcal{Y}}\left\{ \int_{\mathcal{Y}}\int_{\frac{\underline{v}\left(h\boldsymbol{y}'+\boldsymbol{x}\right)-v}{h}}^{\frac{\overline{v}\left(h\boldsymbol{y}'+\boldsymbol{x}\right)-v}{h}}K_{f}'\left(u,\boldsymbol{y}'\right)\frac{G\left(s\left(hu+v,h\boldsymbol{y}'+\boldsymbol{x},n\right),h\boldsymbol{y}'+\boldsymbol{x},n\right)s'\left(hu+v,h\boldsymbol{y}'+\boldsymbol{x}\right)}{g\left(s\left(hu+v,h\boldsymbol{y}'+\boldsymbol{x},n\right),h\boldsymbol{y}'+\boldsymbol{x},n\right)}K_{\boldsymbol{X}}\left(\boldsymbol{y}-\boldsymbol{y}'\right)\right.\\
 & \left.\times\left(\int_{\frac{\underline{b}\left(h\boldsymbol{y}+\boldsymbol{x}\right)-s\left(hu+v,h\boldsymbol{y}'+\boldsymbol{x},n\right)}{h}}^{\frac{\overline{b}\left(h\boldsymbol{y}+\boldsymbol{x},n\right)-s\left(hu+v,h\boldsymbol{y}'+\boldsymbol{x},n\right)}{h}}K_{g}\left(w\right)g\left(hw+s\left(hu+v,h\boldsymbol{y}'+\boldsymbol{x},n\right)|h\boldsymbol{y}+\boldsymbol{x},n\right)\mathrm{d}w\right)\mathrm{d}u\mathrm{d}\boldsymbol{y}'\right\} ^{2}\pi\left(n|h\boldsymbol{y}+\boldsymbol{x}\right)\varphi\left(h\boldsymbol{y}+\boldsymbol{x}\right)\mathrm{d}\boldsymbol{y},
\end{align*}
where
\[
\mathcal{Y}\coloneqq\left[\frac{\underline{x}-x_{1}}{h},\frac{\overline{x}-x_{1}}{h}\right]\times\cdots\times\left[\frac{\underline{x}-x_{d}}{h},\frac{\overline{x}-x_{d}}{h}\right].
\]

Denote 
\[
\psi\left(u,\boldsymbol{z},n\right)\coloneqq\frac{G\left(s\left(u,\boldsymbol{z},n\right),\boldsymbol{z},n\right)s'\left(u,\boldsymbol{z},n\right)}{g\left(s\left(u,\boldsymbol{z},n\right),\boldsymbol{z},n\right)}.
\]
By the $c_{r}$ inequality, we have 
\begin{align*}
 & \frac{1}{n\left(n-1\right)}\frac{1}{h^{1+d}}\int_{\mathcal{X}}\left\{ \int_{\mathcal{X}}\int_{\underline{b}\left(\boldsymbol{z}'\right)}^{\overline{b}\left(\boldsymbol{z}',n\right)}\int_{\underline{b}\left(\boldsymbol{z}\right)}^{\overline{b}\left(\boldsymbol{z},n\right)}\frac{1}{h^{1+d}}K_{f}'\left(\frac{\xi\left(b',\boldsymbol{z}',n\right)-v}{h},\frac{\boldsymbol{z}'-\boldsymbol{x}}{h}\right)\frac{G\left(b',\boldsymbol{z}',n\right)}{g\left(b',\boldsymbol{z}',n\right)}K_{g}\left(\frac{b-b'}{h}\right)K_{\boldsymbol{X}}\left(\frac{\boldsymbol{z}-\boldsymbol{z}'}{h}\right)\right.\\
 & \left.\times g\left(b|\boldsymbol{z},n\right)\mathrm{d}b\mathrm{d}b'\mathrm{d}\boldsymbol{z}'\right\} ^{2}\pi\left(n|\boldsymbol{z}\right)\varphi\left(\boldsymbol{z}\right)\mathrm{d}\boldsymbol{z}\\
\apprle & h\int_{\mathcal{Y}}\left\{ \int_{\mathcal{Y}}\int_{\frac{\underline{v}\left(h\boldsymbol{y}'+\boldsymbol{x}\right)-v}{h}}^{\frac{\overline{v}\left(h\boldsymbol{y}'+\boldsymbol{x}\right)-v}{h}}K_{f}'\left(u,\boldsymbol{y}'\right)\psi\left(hu+v,h\boldsymbol{y}'+\boldsymbol{x},n\right)K_{\boldsymbol{X}}\left(\boldsymbol{y}-\boldsymbol{y}'\right)g\left(s\left(hu+v,h\boldsymbol{y}'+\boldsymbol{x},n\right)|h\boldsymbol{y}+\boldsymbol{x},n\right)\right.\\
 & \left.\times\left(\int_{\frac{\underline{b}\left(h\boldsymbol{y}+\boldsymbol{x}\right)-s\left(hu+v,h\boldsymbol{y}'+\boldsymbol{x},n\right)}{h}}^{\frac{\overline{b}\left(h\boldsymbol{y}+\boldsymbol{x},n\right)-s\left(hu+v,h\boldsymbol{y}'+\boldsymbol{x},n\right)}{h}}K_{g}\left(w\right)\mathrm{d}w\right)\mathrm{d}u\mathrm{d}\boldsymbol{y}'\right\} ^{2}\pi\left(n|h\boldsymbol{y}+\boldsymbol{x}\right)\varphi\left(h\boldsymbol{y}+\boldsymbol{x}\right)\mathrm{d}\boldsymbol{y}\\
 & +h\int_{\mathcal{Y}}\left\{ \int_{\mathcal{Y}}\int_{\frac{\underline{v}\left(h\boldsymbol{y}'+\boldsymbol{x}\right)-v}{h}}^{\frac{\overline{v}\left(h\boldsymbol{y}'+\boldsymbol{x}\right)-v}{h}}K_{f}'\left(u,\boldsymbol{y}'\right)\psi\left(hu+v,h\boldsymbol{y}'+\boldsymbol{x},n\right)K_{\boldsymbol{X}}\left(\boldsymbol{y}-\boldsymbol{y}'\right)\left(\int_{\frac{\underline{b}\left(h\boldsymbol{y}+\boldsymbol{x}\right)-s\left(hu+v,h\boldsymbol{y}'+\boldsymbol{x},n\right)}{h}}^{\frac{\overline{b}\left(h\boldsymbol{y}+\boldsymbol{x},n\right)-s\left(hu+v,h\boldsymbol{y}'+\boldsymbol{x},n\right)}{h}}K_{g}\left(w\right)\right.\right.\\
 & \left.\left.\times\left(g\left(hw+s\left(hu+v,h\boldsymbol{y}'+\boldsymbol{x},n\right)|h\boldsymbol{y}+\boldsymbol{x},n\right)-g\left(s\left(hu+v,h\boldsymbol{y}'+\boldsymbol{x},n\right)|h\boldsymbol{y}+\boldsymbol{x},n\right)\right)\right)\mathrm{d}w\mathrm{d}u\mathrm{d}\boldsymbol{y}'\right\} ^{2}\pi\left(n|h\boldsymbol{y}+\boldsymbol{x}\right)\varphi\left(h\boldsymbol{y}+\boldsymbol{x}\right)\mathrm{d}\boldsymbol{y}.
\end{align*}
When $h$ is sufficiently small ($h\leq\mathrm{min}\left\{ \overline{\delta},\underline{\delta}_{\underline{n}}^{\dagger},...,\underline{\delta}_{\overline{n}}^{\dagger},\overline{\delta}_{\underline{n}}^{\dagger},...,\overline{\delta}_{\overline{n}}^{\dagger}\right\} $),
\begin{equation}
\int_{\frac{\underline{b}\left(h\boldsymbol{y}+\boldsymbol{x}\right)-s\left(hu+v,h\boldsymbol{y}'+\boldsymbol{x},n\right)}{h}}^{\frac{\overline{b}\left(h\boldsymbol{y}+\boldsymbol{x},n\right)-s\left(hu+v,h\boldsymbol{y}'+\boldsymbol{x},n\right)}{h}}K_{g}\left(w\right)\mathrm{d}w=1,\textrm{ for all \ensuremath{\boldsymbol{y}'\in\mathbb{H}\left(\boldsymbol{0},1\right),} \ensuremath{\left|u\right|\leq1}, \ensuremath{\boldsymbol{y}\in\mathbb{H}\left(\boldsymbol{0},2\right)} and \ensuremath{v\in I\left(\boldsymbol{x}\right)}}.\label{eq:K_g integral =00003D1}
\end{equation}

By a mean value expansion, for any $\boldsymbol{y}\in\mathcal{Y}$,
\begin{align}
 & \int_{\mathcal{Y}}\int_{\frac{\underline{v}\left(h\boldsymbol{y}'+\boldsymbol{x}\right)-v}{h}}^{\frac{\overline{v}\left(h\boldsymbol{y}'+\boldsymbol{x}\right)-v}{h}}K_{f}'\left(u,\boldsymbol{y}'\right)\psi\left(hu+v,h\boldsymbol{y}'+\boldsymbol{x},n\right)K_{\boldsymbol{X}}\left(\boldsymbol{y}-\boldsymbol{y}'\right)g\left(s\left(hu+v,h\boldsymbol{y}'+\boldsymbol{x},n\right)|h\boldsymbol{y}+\boldsymbol{x},n\right)\mathrm{d}u\mathrm{d}\boldsymbol{y}'\nonumber \\
= & \int_{\mathcal{Y}}K_{\boldsymbol{X}}\left(\boldsymbol{y}-\boldsymbol{y}'\right)\int_{\frac{\underline{v}\left(h\boldsymbol{y}'+\boldsymbol{x}\right)-v}{h}}^{\frac{\overline{v}\left(h\boldsymbol{y}'+\boldsymbol{x}\right)-v}{h}}K_{f}'\left(u,\boldsymbol{y}'\right)\left\{ \psi\left(v,h\boldsymbol{y}'+\boldsymbol{x},n\right)g\left(s\left(v,h\boldsymbol{y}'+\boldsymbol{x},n\right)|h\boldsymbol{y}+\boldsymbol{x},n\right)\right.\nonumber \\
 & \left.+hu\left(\psi'\left(\dot{v},h\boldsymbol{y}'+\boldsymbol{x},n\right)g\left(s\left(\dot{v},h\boldsymbol{y}'+\boldsymbol{x},n\right)|h\boldsymbol{y}+\boldsymbol{x},n\right)+\psi\left(\dot{v},h\boldsymbol{y}'+\boldsymbol{x},n\right)g'\left(s\left(\dot{v},h\boldsymbol{y}'+\boldsymbol{x},n\right)|h\boldsymbol{y}+\boldsymbol{x},n\right)s'\left(\dot{v},h\boldsymbol{y}'+\boldsymbol{x},n\right)\right)\right\} \mathrm{d}u\mathrm{d}\boldsymbol{y}'\nonumber \\
= & \int_{\mathcal{Y}}K_{\boldsymbol{X}}\left(\boldsymbol{y}-\boldsymbol{y}'\right)\psi\left(v,h\boldsymbol{y}'+\boldsymbol{x},n\right)g\left(s\left(v,h\boldsymbol{y}'+\boldsymbol{x},n\right)|h\boldsymbol{y}+\boldsymbol{x},n\right)K_{\boldsymbol{X}}^{0}\left(\boldsymbol{y}'\right)\left(\int_{\frac{\underline{v}\left(h\boldsymbol{y}'+\boldsymbol{x}\right)-v}{h}}^{\frac{\overline{v}\left(h\boldsymbol{y}'+\boldsymbol{x}\right)-v}{h}}K_{0}'\left(u\right)\mathrm{d}u\right)\mathrm{d}\boldsymbol{y}'\nonumber \\
 & +h\int_{\mathcal{Y}}K_{\boldsymbol{X}}\left(\boldsymbol{y}-\boldsymbol{y}'\right)\int_{\frac{\underline{v}\left(h\boldsymbol{y}'+\boldsymbol{x}\right)-v}{h}}^{\frac{\overline{v}\left(h\boldsymbol{y}'+\boldsymbol{x}\right)-v}{h}}K_{f}'\left(u,\boldsymbol{y}'\right)u\left\{ \psi'\left(\dot{v},h\boldsymbol{y}'+\boldsymbol{x},n\right)g\left(s\left(\dot{v},h\boldsymbol{y}'+\boldsymbol{x},n\right)|h\boldsymbol{y}+\boldsymbol{x},n\right)\right.\nonumber \\
 & \left.+\psi\left(\dot{v},h\boldsymbol{y}'+\boldsymbol{x},n\right)g'\left(s\left(\dot{v},h\boldsymbol{y}'+\boldsymbol{x},n\right)|h\boldsymbol{y}+\boldsymbol{x},n\right)s'\left(\dot{v},h\boldsymbol{y}'+\boldsymbol{x},n\right)\right\} \mathrm{d}u\mathrm{d}\boldsymbol{y}',\label{eq:theorem 1 expansion 1}
\end{align}
where $\dot{v}$ is the mean value with $\left|\dot{v}-v\right|\leq h\left|u\right|$.
Since $K_{0}'$ is odd, 
\[
\int_{\frac{\underline{v}\left(h\boldsymbol{y}'+\boldsymbol{x}\right)-v}{h}}^{\frac{\overline{v}\left(h\boldsymbol{y}'+\boldsymbol{x}\right)-v}{h}}K_{0}'\left(u\right)\mathrm{d}u=0\textrm{, for all \ensuremath{\ensuremath{\boldsymbol{y}'\in\mathbb{H}\left(\boldsymbol{0},1\right)}} and \ensuremath{v\in I\left(\boldsymbol{x}\right)},}
\]
when $h$ is sufficiently small. It follows from this fact, (\ref{eq:K_g integral =00003D1}),
(\ref{eq:theorem 1 expansion 1}), continuity of $\psi\left(\cdot,\cdot,n\right)$,
$\psi'\left(\cdot,\cdot,n\right)$, $s\left(\cdot,\cdot,n\right)$,
$s'\left(\cdot,\cdot,n\right)$, $g\left(\cdot|\cdot,n\right)$ and
$g'\left(\cdot|\cdot,n\right)$ that 
\begin{align*}
 & \underset{v\in I\left(\boldsymbol{x}\right)}{\mathrm{sup}}h\int_{\mathcal{Y}}\left\{ \int_{\mathcal{Y}}\int_{\frac{\underline{v}\left(h\boldsymbol{y}'+\boldsymbol{x}\right)-v}{h}}^{\frac{\overline{v}\left(h\boldsymbol{y}'+\boldsymbol{x}\right)-v}{h}}K_{f}'\left(u,\boldsymbol{y}'\right)\psi\left(hu+v,h\boldsymbol{y}'+\boldsymbol{x},n\right)K_{\boldsymbol{X}}\left(\boldsymbol{y}-\boldsymbol{y}'\right)g\left(s\left(hu+v,h\boldsymbol{y}'+\boldsymbol{x},n\right)|h\boldsymbol{y}+\boldsymbol{x},n\right)\right.\\
 & \left.\times\left(\int_{\frac{\underline{b}\left(h\boldsymbol{y}+\boldsymbol{x}\right)-s\left(hu+v,h\boldsymbol{y}'+\boldsymbol{x},n\right)}{h}}^{\frac{\overline{b}\left(h\boldsymbol{y}+\boldsymbol{x},n\right)-s\left(hu+v,h\boldsymbol{y}'+\boldsymbol{x},n\right)}{h}}K_{g}\left(w\right)\mathrm{d}w\right)\mathrm{d}u\mathrm{d}\boldsymbol{y}'\right\} ^{2}\pi\left(n|h\boldsymbol{y}+\boldsymbol{x}\right)\varphi\left(h\boldsymbol{y}+\boldsymbol{x}\right)\mathrm{d}\boldsymbol{y}\\
= & O\left(h^{3}\right).
\end{align*}

By a second-order Taylor expansion and the fact 
\[
\int_{\frac{\underline{b}\left(h\boldsymbol{y}+\boldsymbol{x}\right)-s\left(hu+v,h\boldsymbol{y}'+\boldsymbol{x},n\right)}{h}}^{\frac{\overline{b}\left(h\boldsymbol{y}+\boldsymbol{x},n\right)-s\left(hu+v,h\boldsymbol{y}'+\boldsymbol{x},n\right)}{h}}K_{g}\left(w\right)w\mathrm{d}w=0,\textrm{ for all \ensuremath{\boldsymbol{y}'\in\mathbb{H}\left(\boldsymbol{0},1\right),} \ensuremath{\boldsymbol{y}\in\mathbb{H}\left(\boldsymbol{0},2\right)}, \ensuremath{\left|u\right|\leq1} and \ensuremath{v\in I\left(\boldsymbol{x}\right)}},
\]
when $h$ is sufficiently small, we have 
\begin{align*}
 & \underset{v\in I\left(\boldsymbol{x}\right)}{\mathrm{sup}}h\int_{\mathcal{Y}}\left\{ \int_{\mathcal{Y}}\int_{\frac{\underline{v}\left(h\boldsymbol{y}'+\boldsymbol{x}\right)-v}{h}}^{\frac{\overline{v}\left(h\boldsymbol{y}'+\boldsymbol{x}\right)-v}{h}}K_{f}'\left(u,\boldsymbol{y}'\right)\psi\left(hu+v,h\boldsymbol{y}'+\boldsymbol{x},n\right)K_{\boldsymbol{X}}\left(\boldsymbol{y}-\boldsymbol{y}'\right)\left(\int_{\frac{\underline{b}\left(h\boldsymbol{y}+\boldsymbol{x}\right)-s\left(hu+v,h\boldsymbol{y}'+\boldsymbol{x},n\right)}{h}}^{\frac{\overline{b}\left(h\boldsymbol{y}+\boldsymbol{x},n\right)-s\left(hu+v,h\boldsymbol{y}'+\boldsymbol{x},n\right)}{h}}K_{g}\left(w\right)\right.\right.\\
 & \left.\left.\times\left(g\left(hw+s\left(hu+v,h\boldsymbol{y}'+\boldsymbol{x},n\right)|h\boldsymbol{y}+\boldsymbol{x},n\right)-g\left(s\left(hu+v,h\boldsymbol{y}'+\boldsymbol{x},n\right)|h\boldsymbol{y}+\boldsymbol{x},n\right)\right)\right)\mathrm{d}w\mathrm{d}u\mathrm{d}\boldsymbol{y}'\right\} ^{2}\pi\left(n|h\boldsymbol{y}+\boldsymbol{x}\right)\varphi\left(h\boldsymbol{y}+\boldsymbol{x}\right)\mathrm{d}\boldsymbol{y}\\
= & O\left(h^{5}\right).
\end{align*}
Now it follows that 
\begin{align*}
 & \underset{v\in I\left(\boldsymbol{x}\right)}{\mathrm{sup}}\frac{1}{n\left(n-1\right)}\frac{1}{h^{1+d}}\int_{\mathcal{X}}\left\{ \int_{\mathcal{X}}\int_{\underline{b}\left(\boldsymbol{z}'\right)}^{\overline{b}\left(\boldsymbol{z}',n\right)}\int_{\underline{b}\left(\boldsymbol{z}\right)}^{\overline{b}\left(\boldsymbol{z},n\right)}\frac{1}{h^{1+d}}K_{f}'\left(\frac{\xi\left(b',\boldsymbol{z}',n\right)-v}{h},\frac{\boldsymbol{z}'-\boldsymbol{x}}{h}\right)\frac{G\left(b',\boldsymbol{z}',n\right)}{g\left(b',\boldsymbol{z}',n\right)}K_{g}\left(\frac{b-b'}{h}\right)K_{\boldsymbol{X}}\left(\frac{\boldsymbol{z}-\boldsymbol{z}'}{h}\right)\right.\\
 & \left.\times g\left(b|\boldsymbol{z},n\right)\mathrm{d}b\mathrm{d}b'\mathrm{d}\boldsymbol{z}'\right\} ^{2}\pi\left(n|\boldsymbol{z}\right)\varphi\left(\boldsymbol{z}\right)\mathrm{d}\boldsymbol{z}\\
= & O\left(h^{3}\right).
\end{align*}
Now by this result, (\ref{eq:E=00005BM_2^2=00005D expansion 2}) and
change of variables, we have 
\begin{align*}
 & \mathrm{E}\left[h^{3+d}\mathcal{M}_{2}^{\ddagger,n}\left(\boldsymbol{B}_{\cdot1},\boldsymbol{X}_{1},N_{1};v\right)^{2}\right]\\
= & \frac{1}{n\left(n-1\right)^{2}}\int_{\mathcal{Y}}\int_{\frac{\underline{b}\left(h\boldsymbol{y}+\boldsymbol{x}\right)-s\left(v,\boldsymbol{x},n\right)}{h}}^{\frac{\overline{b}\left(h\boldsymbol{y}+\boldsymbol{x},n\right)-s\left(v,\boldsymbol{x},n\right)}{h}}\left\{ \int_{\mathcal{Y}}\int_{\frac{\underline{v}\left(h\boldsymbol{y}'+\boldsymbol{x}\right)-v}{h}}^{\frac{\overline{v}\left(h\boldsymbol{y}'+\boldsymbol{x}\right)-v}{h}}K_{f}'\left(u,\boldsymbol{y}'\right)\psi\left(hu+v,h\boldsymbol{y}'+\boldsymbol{x},n\right)\right.\\
 & \left.\times K_{g}\left(w-\frac{s\left(hu+v,h\boldsymbol{y}'+\boldsymbol{x},n\right)-s\left(v,\boldsymbol{x},n\right)}{h}\right)K_{\boldsymbol{X}}\left(\boldsymbol{y}-\boldsymbol{y}'\right)\mathrm{d}u\mathrm{d}\boldsymbol{y}'\right\} ^{2}g\left(hw+s\left(v,\boldsymbol{x},n\right),h\boldsymbol{y}+\boldsymbol{x},n\right)\mathrm{d}w\mathrm{d}\boldsymbol{y}+O\left(h^{3}\right),
\end{align*}
where the remainder term is uniform in $v\in I\left(\boldsymbol{x}\right)$.

Denote 
\[
\rho^{n}\left(u,\boldsymbol{y}',w,\boldsymbol{y};v\right)\coloneqq K_{f}'\left(u,\boldsymbol{y}'\right)\psi\left(hu+v,h\boldsymbol{y}'+\boldsymbol{x},n\right)K_{g}\left(w-\frac{s\left(hu+v,h\boldsymbol{y}'+\boldsymbol{x},n\right)-s\left(v,\boldsymbol{x},n\right)}{h}\right)K_{\boldsymbol{X}}\left(\boldsymbol{y}-\boldsymbol{y}'\right)
\]
and 
\[
\overline{\rho}^{n}\left(w,\boldsymbol{y};v\right)\coloneqq\psi\left(v,\boldsymbol{x},n\right)\int\int K_{f}'\left(u,\boldsymbol{y}'\right)K_{g}\left(w-s_{v}u-s_{\boldsymbol{x}}^{\mathrm{T}}\boldsymbol{y}'\right)K_{\boldsymbol{X}}\left(\boldsymbol{y}-\boldsymbol{y}'\right)\mathrm{d}u\mathrm{d}\boldsymbol{y}'.
\]
Now we have 
\begin{align}
 & \mathrm{E}\left[h^{3+d}\mathcal{M}_{2}^{\ddagger,n}\left(\boldsymbol{B}_{\cdot1},\boldsymbol{X}_{1},N_{1};v\right)^{2}\right]\nonumber \\
= & \frac{1}{n\left(n-1\right)^{2}}\int_{\mathcal{Y}}\int_{\frac{\underline{b}\left(h\boldsymbol{y}+\boldsymbol{x}\right)-s\left(v,\boldsymbol{x},n\right)}{h}}^{\frac{\overline{b}\left(h\boldsymbol{y}+\boldsymbol{x},n\right)-s\left(v,\boldsymbol{x},n\right)}{h}}\left\{ \int_{\mathcal{Y}}\int_{\frac{\underline{v}\left(h\boldsymbol{y}'+\boldsymbol{x}\right)-v}{h}}^{\frac{\overline{v}\left(h\boldsymbol{y}'+\boldsymbol{x}\right)-v}{h}}\rho^{n}\left(u,\boldsymbol{y}',w,\boldsymbol{y};v\right)\mathrm{d}u\mathrm{d}\boldsymbol{y}'\right\} ^{2}g\left(hw+s\left(v,\boldsymbol{x},n\right),h\boldsymbol{y}+\boldsymbol{x},n\right)\mathrm{d}w\mathrm{d}\boldsymbol{y}+O\left(h^{3}\right).\label{eq:Thm 5.1 (1)}
\end{align}

Let 
\[
\overline{\tau}\coloneqq\underset{\left(u,\boldsymbol{z}\right)\in\mathcal{C}_{V,\boldsymbol{X}}}{\mathrm{sup}}\sum_{j=1}^{1+d}\left|D_{j}s\left(u,\boldsymbol{z},n\right)\right|.
\]
When $h$ is sufficiently small, 
\begin{equation}
\underset{v\in I\left(\boldsymbol{x}\right)}{\mathrm{sup}}\left|\rho^{n}\left(u,\boldsymbol{y}',w,\boldsymbol{y};v\right)\right|\apprle\mathbbm{1}\left(\boldsymbol{y}'\in\mathbb{H}\left(\boldsymbol{0},1\right)\right)\mathbbm{1}\left(\boldsymbol{y}\in\mathbb{H}\left(\boldsymbol{0},2\right)\right)\mathbbm{1}\left(\left|w\right|\leq1+\overline{\tau}\right)\mathbbm{1}\left(\left|u\right|\leq1\right)\label{eq:rho_n bound}
\end{equation}
and

\begin{equation}
\underset{v\in I\left(\boldsymbol{x}\right)}{\mathrm{sup}}\left|\overline{\rho}^{n}\left(w,\boldsymbol{y};v\right)\right|\apprle\mathbbm{1}\left(\boldsymbol{y}\in\mathbb{H}\left(\boldsymbol{0},2\right)\right)\mathbbm{1}\left(\left|w\right|\leq1+\overline{\tau}\right).\label{eq:rho bound}
\end{equation}

By the triangle inequality, we have
\begin{align}
 & \left|\int_{\mathcal{Y}}\int_{\frac{\underline{b}\left(h\boldsymbol{y}+\boldsymbol{x}\right)-s\left(v,\boldsymbol{x},n\right)}{h}}^{\frac{\overline{b}\left(h\boldsymbol{y}+\boldsymbol{x},n\right)-s\left(v,\boldsymbol{x},n\right)}{h}}\left\{ \int_{\mathcal{Y}}\int_{\frac{\underline{v}\left(h\boldsymbol{y}'+\boldsymbol{x}\right)-v}{h}}^{\frac{\overline{v}\left(h\boldsymbol{y}'+\boldsymbol{x}\right)-v}{h}}\rho^{n}\left(u,\boldsymbol{y}',w,\boldsymbol{y};v\right)\mathrm{d}u\mathrm{d}\boldsymbol{y}'\right\} ^{2}g\left(hw+s\left(v,\boldsymbol{x},n\right),h\boldsymbol{y}+\boldsymbol{x},n\right)\mathrm{d}w\mathrm{d}\boldsymbol{y}\right.\nonumber \\
 & \left.-g\left(s\left(v,\boldsymbol{x},n\right),\boldsymbol{x},n\right)\int\int\overline{\rho}^{n}\left(w,\boldsymbol{y};v\right)^{2}\mathrm{d}w\mathrm{d}\boldsymbol{y}\right|\nonumber \\
\leq & \left|\int_{\mathcal{Y}}\int_{\frac{\underline{b}\left(h\boldsymbol{y}+\boldsymbol{x}\right)-s\left(v,\boldsymbol{x},n\right)}{h}}^{\frac{\overline{b}\left(h\boldsymbol{y}+\boldsymbol{x},n\right)-s\left(v,\boldsymbol{x},n\right)}{h}}\left\{ \int_{\mathcal{Y}}\int_{\frac{\underline{v}\left(h\boldsymbol{y}'+\boldsymbol{x}\right)-v}{h}}^{\frac{\overline{v}\left(h\boldsymbol{y}'+\boldsymbol{x}\right)-v}{h}}\rho^{n}\left(u,\boldsymbol{y}',w,\boldsymbol{y};v\right)\mathrm{d}u\mathrm{d}\boldsymbol{y}'\right\} ^{2}\left\{ g\left(hw+s\left(v,\boldsymbol{x},n\right),h\boldsymbol{y}+\boldsymbol{x},n\right)\right.\right.\nonumber \\
 & \left.\left.-g\left(s\left(v,\boldsymbol{x},n\right),\boldsymbol{x},n\right)\right\} \mathrm{d}w\mathrm{d}\boldsymbol{y}\right|\nonumber \\
 & +g\left(s\left(v,\boldsymbol{x},n\right),\boldsymbol{x},n\right)\left|\int_{\mathcal{Y}}\int_{\frac{\underline{b}\left(h\boldsymbol{y}+\boldsymbol{x}\right)-s\left(v,\boldsymbol{x},n\right)}{h}}^{\frac{\overline{b}\left(h\boldsymbol{y}+\boldsymbol{x},n\right)-s\left(v,\boldsymbol{x},n\right)}{h}}\left(\left\{ \int_{\mathcal{Y}}\int_{\frac{\underline{v}\left(h\boldsymbol{y}'+\boldsymbol{x}\right)-v}{h}}^{\frac{\overline{v}\left(h\boldsymbol{y}'+\boldsymbol{x}\right)-v}{h}}\rho^{n}\left(u,\boldsymbol{y}',w,\boldsymbol{y};v\right)\mathrm{d}u\mathrm{d}\boldsymbol{y}'\right\} ^{2}-\overline{\rho}^{n}\left(w,\boldsymbol{y};v\right)^{2}\right)\mathrm{d}w\mathrm{d}\boldsymbol{y}\right|\nonumber \\
 & +g\left(s\left(v,\boldsymbol{x},n\right),\boldsymbol{x},n\right)\left|\int_{\mathcal{Y}}\int_{\frac{\underline{b}\left(h\boldsymbol{y}+\boldsymbol{x}\right)-s\left(v,\boldsymbol{x},n\right)}{h}}^{\frac{\overline{b}\left(h\boldsymbol{y}+\boldsymbol{x},n\right)-s\left(v,\boldsymbol{x},n\right)}{h}}\overline{\rho}^{n}\left(w,\boldsymbol{y};v\right)^{2}\mathrm{d}w\mathrm{d}\boldsymbol{y}-\int\int\overline{\rho}^{n}\left(w,\boldsymbol{y};v\right)^{2}\mathrm{d}w\mathrm{d}\boldsymbol{y}\right|.\label{eq:Thm 5.1 (2)}
\end{align}

It is clear that when $h$ is sufficiently small,

\begin{equation}
g\left(s\left(v,\boldsymbol{x},n\right),\boldsymbol{x},n\right)\left|\int_{\mathcal{Y}}\int_{\frac{\underline{b}\left(h\boldsymbol{y}+\boldsymbol{x}\right)-s\left(v,\boldsymbol{x},n\right)}{h}}^{\frac{\overline{b}\left(h\boldsymbol{y}+\boldsymbol{x},n\right)-s\left(v,\boldsymbol{x},n\right)}{h}}\overline{\rho}^{n}\left(w,\boldsymbol{y};v\right)^{2}\mathrm{d}w\mathrm{d}\boldsymbol{y}-\int\int\overline{\rho}^{n}\left(w,\boldsymbol{y};v\right)^{2}\mathrm{d}w\mathrm{d}\boldsymbol{y}\right|=0,\textrm{ for all \ensuremath{v\in I\left(\boldsymbol{x}\right)}}.\label{eq:Thm 5.1 (3)}
\end{equation}
And
\begin{align}
 & \underset{v\in I\left(\boldsymbol{x}\right)}{\mathrm{sup}}\left|\int_{\mathcal{Y}}\int_{\frac{\underline{b}\left(h\boldsymbol{y}+\boldsymbol{x}\right)-s\left(v,\boldsymbol{x},n\right)}{h}}^{\frac{\overline{b}\left(h\boldsymbol{y}+\boldsymbol{x},n\right)-s\left(v,\boldsymbol{x},n\right)}{h}}\left\{ \int_{\mathcal{Y}}\int_{\frac{\underline{v}\left(h\boldsymbol{y}'+\boldsymbol{x}\right)-v}{h}}^{\frac{\overline{v}\left(h\boldsymbol{y}'+\boldsymbol{x}\right)-v}{h}}\rho^{n}\left(u,\boldsymbol{y}',w,\boldsymbol{y};v\right)\mathrm{d}u\mathrm{d}\boldsymbol{y}'\right\} ^{2}\left\{ g\left(hw+s\left(v,\boldsymbol{x},n\right),h\boldsymbol{y}+\boldsymbol{x},n\right)-g\left(s\left(v,\boldsymbol{x},n\right),\boldsymbol{x},n\right)\right\} \right|\mathrm{d}w\mathrm{d}\boldsymbol{y}\nonumber \\
= & o\left(1\right)\label{eq:Thm 5.1 (4)}
\end{align}
follows from uniform continuity of $g\left(\cdot,\cdot,n\right)$
and $s\left(\cdot,\cdot,n\right)$. It is also clear that
\begin{align*}
 & \underset{v\in I\left(\boldsymbol{x}\right)}{\mathrm{sup}}\underset{\left(w,\boldsymbol{y}\right)\in\mathbb{R}^{1+d}}{\mathrm{sup}}\left|\psi\left(v,\boldsymbol{x},n\right)\int_{\mathcal{Y}}\int_{\frac{\underline{v}\left(h\boldsymbol{y}'+\boldsymbol{x}\right)-v}{h}}^{\frac{\overline{v}\left(h\boldsymbol{y}'+\boldsymbol{x}\right)-v}{h}}K_{f}'\left(u,\boldsymbol{y}'\right)K_{g}\left(w-\frac{s\left(hu+v,h\boldsymbol{y}'+\boldsymbol{x},n\right)-s\left(v,\boldsymbol{x},n\right)}{h}\right)K_{\boldsymbol{X}}\left(\boldsymbol{y}-\boldsymbol{y}'\right)\mathrm{d}u\mathrm{d}\boldsymbol{y}'\right.\\
 & \left.-\psi\left(v,\boldsymbol{x},n\right)\int\int K_{f}'\left(u,\boldsymbol{y}'\right)K_{g}\left(w-s_{v}u-s_{\boldsymbol{x}}^{\mathrm{T}}\boldsymbol{y}'\right)K_{\boldsymbol{X}}\left(\boldsymbol{y}-\boldsymbol{y}'\right)\mathrm{d}u\mathrm{d}\boldsymbol{y}'\right|\\
\apprle & \underset{v\in I\left(\boldsymbol{x}\right)}{\mathrm{sup}}\underset{\left(u,\boldsymbol{z}\right)\in\mathbb{H}\left(\left(v,\boldsymbol{x}\right),h\right)}{\mathrm{sup}}\sum_{j=1}^{1+d}\left|D_{j}s\left(u,\boldsymbol{z},n\right)-D_{j}s\left(v,\boldsymbol{x},n\right)\right|\\
= & o\left(1\right),
\end{align*}
where the equality follows from uniform continuity of the partial
derivatives of $s\left(\cdot,\cdot,n\right)$ and now 
\begin{equation}
\underset{v\in I\left(\boldsymbol{x}\right)}{\mathrm{sup}}\left|\int_{\mathcal{Y}}\int_{\frac{\underline{b}\left(h\boldsymbol{y}+\boldsymbol{x}\right)-s\left(v,\boldsymbol{x},n\right)}{h}}^{\frac{\overline{b}\left(h\boldsymbol{y}+\boldsymbol{x},n\right)-s\left(v,\boldsymbol{x},n\right)}{h}}\left(\left\{ \int_{\mathcal{Y}}\int_{\frac{\underline{v}\left(h\boldsymbol{y}'+\boldsymbol{x}\right)-v}{h}}^{\frac{\overline{v}\left(h\boldsymbol{y}'+\boldsymbol{x}\right)-v}{h}}\rho^{n}\left(u,\boldsymbol{y}',w,\boldsymbol{y};v\right)\mathrm{d}u\mathrm{d}\boldsymbol{y}'\right\} ^{2}-\overline{\rho}^{n}\left(w,\boldsymbol{y};v\right)^{2}\right)\mathrm{d}w\mathrm{d}\boldsymbol{y}\right|=o\left(1\right)\label{eq:Thm 5.1 (5)}
\end{equation}
follows from this result, continuity of $\psi\left(\cdot,\cdot,n\right)$,
(\ref{eq:rho_n bound}) and (\ref{eq:rho bound}).

Now by (\ref{eq:sup miu_M_ddagger rate}), (\ref{eq:Thm 5.1 (1)}),
(\ref{eq:Thm 5.1 (2)}), (\ref{eq:Thm 5.1 (3)}), (\ref{eq:Thm 5.1 (4)})
and (\ref{eq:Thm 5.1 (5)}), we have
\begin{align*}
 & \mathrm{E}\left[h^{3+d}\left(\mathcal{M}_{2}^{\ddagger,n}\left(\boldsymbol{B}_{\cdot1},\boldsymbol{X}_{1},N_{1};v\right)-\mu_{\mathcal{M}^{\ddagger,n}}\left(v\right)\right)^{2}\right]\\
= & \mathrm{E}\left[h^{3+d}\mathcal{M}_{2}^{\ddagger,n}\left(\boldsymbol{B}_{\cdot1},\boldsymbol{X}_{1},N_{1};v\right)^{2}\right]-h^{3+d}\mu_{\mathcal{M}^{\ddagger,n}}\left(v\right)^{2}\\
= & \frac{1}{n\left(n-1\right)^{2}}\frac{F\left(v|\boldsymbol{x}\right)^{2}f\left(v|\boldsymbol{x}\right)^{2}}{\pi\left(n|\boldsymbol{x}\right)\varphi\left(\boldsymbol{x}\right)g\left(s\left(v,\boldsymbol{x},n\right)|\boldsymbol{x},n\right)^{3}}\int\int\left\{ \int\int K_{f}'\left(u,\boldsymbol{y}\right)K_{g}\left(w-s_{v}u-s_{\boldsymbol{x}}^{\mathrm{T}}\boldsymbol{y}\right)K_{\boldsymbol{X}}\left(\boldsymbol{z}-\boldsymbol{y}\right)\mathrm{d}u\mathrm{d}\boldsymbol{y}\right\} ^{2}\mathrm{d}w\mathrm{d}\boldsymbol{z}+o\left(1\right)
\end{align*}
and also 
\begin{equation}
\sigma^{n}\left(v\right)=\frac{1}{n^{\nicefrac{1}{2}}\left(n-1\right)}\left(\frac{F\left(v|\boldsymbol{x}\right)^{2}f\left(v|\boldsymbol{x}\right)^{2}}{\pi\left(n|\boldsymbol{x}\right)\varphi\left(\boldsymbol{x}\right)g\left(s\left(v,\boldsymbol{x},n\right)|\boldsymbol{x},n\right)^{3}}\int\int\left\{ \int\int K_{f}'\left(u,\boldsymbol{y}\right)K_{g}\left(w-s_{v}u-s_{\boldsymbol{x}}^{\mathrm{T}}\boldsymbol{y}\right)K_{\boldsymbol{X}}\left(\boldsymbol{z}-\boldsymbol{y}\right)\mathrm{d}u\mathrm{d}\boldsymbol{y}\right\} ^{2}\mathrm{d}w\mathrm{d}\boldsymbol{z}\right)^{\nicefrac{1}{2}}+o\left(1\right),\label{eq:sigma limit}
\end{equation}
where the remainder terms are uniform in $v\in I\left(\boldsymbol{x}\right)$.

By the $c_{r}$ inequality, we have
\begin{eqnarray}
\sum_{l=1}^{L}\mathrm{E}\left[\left|\frac{U_{l}^{n}\left(v\right)}{\sigma^{n}\left(v\right)}\right|^{3}\right] & = & \sigma^{n}\left(v\right)^{-3}L^{-\nicefrac{1}{2}}\mathrm{E}\left[h^{\nicefrac{3\left(3+d\right)}{2}}\left|\mathcal{M}_{2}^{\ddagger,n}\left(\boldsymbol{B}_{\cdot1},\boldsymbol{X}_{1},N_{1};v\right)-\mu_{\mathcal{M}^{\ddagger,n}}\left(v\right)\right|^{3}\right]\nonumber \\
 & \apprle & \sigma^{n}\left(v\right)^{-3}L^{-\nicefrac{1}{2}}\left(\mathrm{E}\left[h^{\nicefrac{3\left(3+d\right)}{2}}\left|\mathcal{M}_{2}^{\ddagger,n}\left(\boldsymbol{B}_{\cdot1},\boldsymbol{X}_{1},N_{1};v\right)\right|^{3}\right]+h^{\nicefrac{3\left(3+d\right)}{2}}\left|\mu_{\mathcal{M}^{\ddagger,n}}\left(v\right)\right|^{3}\right).\label{eq:lyapounov condition inequality}
\end{eqnarray}
By the $c_{r}$ inequality, Jensen's inequality and change of variables,
we have
\begin{align*}
 & \mathrm{E}\left[h^{\nicefrac{3\left(3+d\right)}{2}}\left|\mathcal{M}_{2}^{\ddagger,n}\left(\boldsymbol{B}_{\cdot1},\boldsymbol{X}_{1},N_{1};v\right)\right|^{3}\right]\\
\apprle & \mathrm{E}\left[h^{\nicefrac{3\left(3+d\right)}{2}}\left|\mathbbm{1}\left(N_{1}=n\right)\frac{1}{N_{1}}\sum_{j=1}^{N_{1}}\int_{\mathcal{X}}\int_{\underline{b}\left(\boldsymbol{z}'\right)}^{\overline{b}\left(\boldsymbol{z}',n\right)}\frac{1}{h^{3+2d}}K_{f}'\left(\frac{\xi\left(b',\boldsymbol{z}',n\right)-v}{h},\frac{\boldsymbol{z}'-\boldsymbol{x}}{h}\right)\frac{G\left(b',\boldsymbol{z}',n\right)}{\left(n-1\right)g\left(b',\boldsymbol{z}',n\right)}\right.\right.\\
 & \left.\left.\times K_{g}\left(\frac{b'-B_{j1}}{h}\right)K_{\boldsymbol{X}}\left(\frac{\boldsymbol{z}'-\boldsymbol{X}_{1}}{h}\right)\mathrm{d}b'\mathrm{d}\boldsymbol{z}'\right|^{3}\right]\\
\apprle & h^{\nicefrac{3\left(3+d\right)}{2}}\int\int\left|\int_{\mathcal{X}}\int_{\underline{b}\left(\boldsymbol{z}'\right)}^{\overline{b}\left(\boldsymbol{z}',n\right)}\frac{1}{h^{3+2d}}K_{f}'\left(\frac{\xi\left(b',\boldsymbol{z}',n\right)-v}{h},\frac{\boldsymbol{z}'-\boldsymbol{x}}{h}\right)\frac{G\left(b',\boldsymbol{z}',n\right)}{g\left(b',\boldsymbol{z}',n\right)}K_{g}\left(\frac{b'-b}{h}\right)K_{\boldsymbol{X}}\left(\frac{\boldsymbol{z}'-\boldsymbol{z}}{h}\right)\mathrm{d}b'\mathrm{d}\boldsymbol{z}'\right|^{3}g\left(b,\boldsymbol{z},n\right)\mathrm{d}b\mathrm{d}\boldsymbol{z}\\
= & h^{-\nicefrac{\left(1+d\right)}{2}}\int_{\mathcal{Y}}\int_{\frac{\underline{b}\left(h\boldsymbol{y}+\boldsymbol{x}\right)-s\left(v,\boldsymbol{x},n\right)}{h}}^{\frac{\overline{b}\left(h\boldsymbol{y}+\boldsymbol{x},n\right)-s\left(v,\boldsymbol{x},n\right)}{h}}\left|\int_{\mathcal{Y}}\int_{\frac{\underline{v}\left(h\boldsymbol{y}'+\boldsymbol{x}\right)-v}{h}}^{\frac{\overline{v}\left(h\boldsymbol{y}'+\boldsymbol{x}\right)-v}{h}}\rho^{n}\left(u,\boldsymbol{y}',w,\boldsymbol{y};v\right)\mathrm{d}u\mathrm{d}\boldsymbol{y}'\right|^{3}g\left(hw+s\left(v,\boldsymbol{x},n\right),h\boldsymbol{y}+\boldsymbol{x},n\right)\mathrm{d}w\mathrm{d}\boldsymbol{y}.
\end{align*}
Now it follows from this result and (\ref{eq:rho_n bound}) that
\begin{equation}
\underset{v\in I\left(\boldsymbol{x}\right)}{\mathrm{sup}}\mathrm{E}\left[h^{\nicefrac{3\left(3+d\right)}{2}}\left|\mathcal{M}_{2}^{\ddagger,n}\left(\boldsymbol{B}_{\cdot1},\boldsymbol{X}_{1},N_{1};v\right)\right|^{3}\right]=O\left(h^{-\nicefrac{\left(1+d\right)}{2}}\right).\label{eq:E_M_2^3 order}
\end{equation}
It now follows from this result, (\ref{eq:sup miu_M_ddagger rate}),
(\ref{eq:sigma limit}) and (\ref{eq:lyapounov condition inequality})
that
\begin{equation}
\sum_{l=1}^{L}\mathrm{E}\left[\left|\frac{U_{l}^{n}\left(v\right)}{\sigma^{n}\left(v\right)}\right|^{3}\right]\downarrow0,\textrm{ as \ensuremath{L\uparrow\infty}}.\label{eq:Lyapunov condition}
\end{equation}

By the Lyapunov's central limit theorem, we have 
\begin{equation}
\sum_{l=1}^{L}\frac{U_{l}^{n}\left(v\right)}{\sigma^{n}\left(v\right)}\rightarrow_{d}\mathrm{N}\left(0,1\right),\textrm{ as \ensuremath{L\uparrow\infty}}.\label{eq:U's standard normal limit}
\end{equation}
Standard arguments for kernel-smoothing based nonparametric estimation
shows that $\widehat{\pi}\left(n|\boldsymbol{x}\right)\widehat{\varphi}\left(\boldsymbol{x}\right)$
is consistent for $\pi\left(n|\boldsymbol{x}\right)\varphi\left(\boldsymbol{x}\right)$.
Now the asymptotic normality follows from this result, (\ref{eq:f_hat - f linearization 3}),
(\ref{eq:sigma limit}), (\ref{eq:U's standard normal limit}), the
above displayed result and Slutsky's lemma.

Fixing any $n_{1},n_{2}\in\mathcal{N}$ with $n_{1}\neq n_{2}$, we
have
\begin{gather}
\left(Lh^{3+d}\right)^{\nicefrac{1}{2}}\left(\widehat{f}_{GPV}\left(v|\boldsymbol{x},n_{1}\right)-f\left(v|\boldsymbol{x}\right)\right)=\frac{1}{\widehat{\pi}\left(n_{1}|\boldsymbol{x}\right)\widehat{\varphi}\left(\boldsymbol{x}\right)L^{\nicefrac{1}{2}}}\sum_{l=1}^{L}h^{\nicefrac{\left(3+d\right)}{2}}\left(\mathcal{M}_{2}^{\ddagger,n_{1}}\left(\boldsymbol{B}_{\cdot l},\boldsymbol{X}_{l},N_{l};v\right)-\mu_{\mathcal{M}^{\ddagger,n_{1}}}\left(v\right)\right)+o_{p}\left(1\right)\nonumber \\
\left(Lh^{3+d}\right)^{\nicefrac{1}{2}}\left(\widehat{f}_{GPV}\left(v|\boldsymbol{x},n_{2}\right)-f\left(v|\boldsymbol{x}\right)\right)=\frac{1}{\widehat{\pi}\left(n_{2}|\boldsymbol{x}\right)\widehat{\varphi}\left(\boldsymbol{x}\right)L^{\nicefrac{1}{2}}}\sum_{l=1}^{L}h^{\nicefrac{\left(3+d\right)}{2}}\left(\mathcal{M}_{2}^{\ddagger,n_{2}}\left(\boldsymbol{B}_{\cdot l},\boldsymbol{X}_{l},N_{l};v\right)-\mu_{\mathcal{M}^{\ddagger,n_{2}}}\left(v\right)\right)+o_{p}\left(1\right)\label{eq:multi dimensional  asymptotic equivalence}
\end{gather}
Denote
\[
\boldsymbol{Y}_{l}\left(v\right)\coloneqq\left(L^{-\nicefrac{1}{2}}h^{\nicefrac{\left(3+d\right)}{2}}\left(\mathcal{M}_{2}^{\ddagger,n_{1}}\left(\boldsymbol{B}_{\cdot l},\boldsymbol{X}_{l},N_{l};v\right)-\mu_{\mathcal{M}^{\ddagger,n_{1}}}\left(v\right)\right),L^{-\nicefrac{1}{2}}h^{\nicefrac{\left(3+d\right)}{2}}\left(\mathcal{M}_{2}^{\ddagger,n_{2}}\left(\boldsymbol{B}_{\cdot l},\boldsymbol{X}_{l},N_{l};v\right)-\mu_{\mathcal{M}^{\ddagger,n_{2}}}\left(v\right)\right)\right)^{\mathrm{T}}.
\]
It follows from (\ref{eq:sup miu_M_ddagger rate}) that
\[
\mathrm{E}\left[h^{3+d}\left(\mathcal{M}_{2}^{\ddagger,n_{1}}\left(\boldsymbol{B}_{\cdot1},\boldsymbol{X}_{1},N_{1};v\right)-\mu_{\mathcal{M}^{\ddagger,n_{1}}}\left(v\right)\right)\left(\mathcal{M}_{2}^{\ddagger,n_{2}}\left(\boldsymbol{B}_{\cdot1},\boldsymbol{X}_{1},N_{1};v\right)-\mu_{\mathcal{M}^{\ddagger,n_{2}}}\left(v\right)\right)\right]=O\left(h^{3+d}\right),
\]
uniformly in $v\in I\left(\boldsymbol{x}\right)$. It follows from
the above result and (\ref{eq:sigma limit}) that 
\[
\sum_{l=1}^{L}\mathrm{Var}\left[\boldsymbol{Y}_{l}\left(v\right)\right]\rightarrow\left[\begin{array}{cc}
\mathrm{V}_{GPV}\left(v,\boldsymbol{x},n_{1}\right) & 0\\
0 & \mathrm{V}_{GPV}\left(v,\boldsymbol{x},n_{2}\right)
\end{array}\right],\textrm{ as }L\uparrow\infty.
\]

It is straightforward to verify that given the above result, 
\begin{equation}
\sum_{l=1}^{L}\mathrm{E}\left[\left\Vert \boldsymbol{Y}_{l}\left(v\right)\right\Vert ^{3}\right]\downarrow0,\textrm{ as }L\uparrow\infty\label{eq:multi dimensional Lyapounov condition}
\end{equation}
is sufficient for the Lyapunov condition for the the multi-dimensional
Lyapunov central limit theorem which can be established as a consequence
of the Cramer-Wold device and the one-dimensional central limit theorem
for triangular arrays. (\ref{eq:multi dimensional Lyapounov condition})
can be established by using the $c_{r}$ inequality and (\ref{eq:E_M_2^3 order}).
The joint asymptotic normality of $\sum_{l=1}^{L}\boldsymbol{Y}_{l}\left(v\right)$
and asymptotic independence of 
\[
\left(\left(Lh^{3+d}\right)^{\nicefrac{1}{2}}\left(\widehat{f}_{GPV}\left(v|\boldsymbol{x},n_{1}\right)-f\left(v|\boldsymbol{x}\right)\right),\,\left(Lh^{3+d}\right)^{\nicefrac{1}{2}}\left(\widehat{f}_{GPV}\left(v|\boldsymbol{x},n_{2}\right)-f\left(v|\boldsymbol{x}\right)\right)\right)^{\mathrm{T}}
\]
follows from (\ref{eq:multi dimensional  asymptotic equivalence}),
(\ref{eq:multi dimensional Lyapounov condition}), the consistency
of $\widehat{\pi}\left(n_{1}|\boldsymbol{x}\right)\widehat{\varphi}\left(\boldsymbol{x}\right)$
and $\widehat{\pi}\left(n_{2}|\boldsymbol{x}\right)\widehat{\varphi}\left(\boldsymbol{x}\right)$
and Slutsky's lemma.\end{proof}

\begin{proof}[Proof of Corollary 6.1]The conclusion of this corollary
follows from Theorem 6.1, consistency of $\widehat{\pi}\left(n|\boldsymbol{x}\right)$,
$n\in\mathcal{N}$ and the continuous mapping theorem.\end{proof}

Note that the estimator for the asymptotic variance is numerically
equivalent to 
\begin{eqnarray*}
\widehat{\mathrm{V}}_{GPV}\left(v|\boldsymbol{x},n\right) & = & \frac{1}{\widehat{\pi}\left(n|\boldsymbol{x}\right)^{2}\widehat{\varphi}\left(\boldsymbol{x}\right)^{2}h^{3\left(1+d\right)}\left(L\right)_{3}}\sum_{\left(3\right)}\mathbbm{1}\left(N_{l}=n\right)\frac{1}{N_{l}^{2}}\sum_{i=1}^{N_{l}}\frac{1}{N_{k}}\sum_{j=1}^{N_{k}}\frac{1}{N_{k}-1}\mathbb{T}_{jk}K_{f}'\left(\frac{\widehat{V}_{jk}-v}{h},\frac{\boldsymbol{X}_{k}-\boldsymbol{x}}{h}\right)\frac{\widehat{G}\left(B_{jk},\boldsymbol{X}_{k},N_{k}\right)}{\widehat{g}\left(B_{jk},\boldsymbol{X}_{k},N_{k}\right)^{2}}\\
 &  & \times\mathbbm{1}\left(N_{l}=N_{k}\right)K_{g}\left(\frac{B_{il}-B_{jk}}{h}\right)K_{\boldsymbol{X}}\left(\frac{\boldsymbol{X}_{l}-\boldsymbol{X}_{k}}{h}\right)\frac{1}{N_{k'}}\sum_{j'=1}^{N_{k'}}\frac{1}{N_{k'}-1}\mathbb{T}_{j'k'}K_{f}'\left(\frac{\widehat{V}_{j'k'}-v}{h},\frac{\boldsymbol{X}_{k'}-\boldsymbol{x}}{h}\right)\\
 &  & \times\frac{\widehat{G}\left(B_{j'k'},\boldsymbol{X}_{k'},N_{k'}\right)}{\widehat{g}\left(B_{j'k'},\boldsymbol{X}_{k'},N_{k'}\right)^{2}}\mathbbm{1}\left(N_{l}=N_{k'}\right)K_{g}\left(\frac{B_{il}-B_{j'k'}}{h}\right)K_{\boldsymbol{X}}\left(\frac{\boldsymbol{X}_{l}-\boldsymbol{X}_{k'}}{h}\right).
\end{eqnarray*}
And it is clear from this expression that 
\begin{eqnarray*}
\widehat{\mathrm{V}}_{GPV}\left(v|\boldsymbol{x}\right) & = & \sum_{n\in\mathcal{N}}\widehat{\pi}\left(n|\boldsymbol{x}\right)^{2}\widehat{\mathrm{V}}_{GPV}\left(v|\boldsymbol{x},n\right)\\
 & = & \frac{1}{\widehat{\varphi}\left(\boldsymbol{x}\right)^{2}h^{3\left(1+d\right)}\left(L\right)_{3}}\sum_{\left(3\right)}\frac{1}{N_{l}^{2}}\sum_{i=1}^{N_{l}}\frac{1}{N_{k}}\sum_{j=1}^{N_{k}}\frac{1}{N_{k}-1}\mathbb{T}_{jk}K_{f}'\left(\frac{\widehat{V}_{jk}-v}{h},\frac{\boldsymbol{X}_{k}-\boldsymbol{x}}{h}\right)\frac{\widehat{G}\left(B_{jk},\boldsymbol{X}_{k},N_{k}\right)}{\widehat{g}\left(B_{jk},\boldsymbol{X}_{k},N_{k}\right)^{2}}\\
 &  & \times\mathbbm{1}\left(N_{l}=N_{k}\right)K_{g}\left(\frac{B_{il}-B_{jk}}{h}\right)K_{\boldsymbol{X}}\left(\frac{\boldsymbol{X}_{l}-\boldsymbol{X}_{k}}{h}\right)\frac{1}{N_{k'}}\sum_{j'=1}^{N_{k'}}\frac{1}{N_{k'}-1}\mathbb{T}_{j'k'}K_{f}'\left(\frac{\widehat{V}_{j'k'}-v}{h},\frac{\boldsymbol{X}_{k'}-\boldsymbol{x}}{h}\right)\\
 &  & \times\frac{\widehat{G}\left(B_{j'k'},\boldsymbol{X}_{k'},N_{k'}\right)}{\widehat{g}\left(B_{j'k'},\boldsymbol{X}_{k'},N_{k'}\right)^{2}}\mathbbm{1}\left(N_{l}=N_{k'}\right)K_{g}\left(\frac{B_{il}-B_{j'k'}}{h}\right)K_{\boldsymbol{X}}\left(\frac{\boldsymbol{X}_{l}-\boldsymbol{X}_{k'}}{h}\right).
\end{eqnarray*}

\begin{proof}[Proof of Theorem 6.2]Write 
\[
\widehat{\pi}\left(n|\boldsymbol{x}\right)^{2}\widehat{\varphi}\left(\boldsymbol{x}\right)^{2}\widehat{\mathrm{V}}_{GPV}\left(v|\boldsymbol{x},n\right)=\varDelta_{1}^{\dagger}\left(v\right)+\varDelta_{2}^{\dagger}\left(v\right)+\varDelta_{3}^{\dagger}\left(v\right),
\]
where
\begin{eqnarray*}
\varDelta_{1}^{\dagger}\left(v\right) & \coloneqq & \frac{1}{h^{3\left(1+d\right)}\left(L\right)_{3}}\sum_{\left(3\right)}\mathbbm{1}\left(N_{l}=n\right)\frac{1}{N_{l}^{2}}\sum_{i=1}^{N_{l}}\frac{1}{N_{k}}\sum_{j=1}^{N_{k}}\frac{1}{N_{k}-1}\mathbb{T}_{jk}K_{f}'\left(\frac{\widehat{V}_{jk}-v}{h},\frac{\boldsymbol{X}_{k}-\boldsymbol{x}}{h}\right)\frac{G\left(B_{jk},\boldsymbol{X}_{k},N_{k}\right)}{g\left(B_{jk},\boldsymbol{X}_{k},N_{k}\right)^{2}}\\
 &  & \times\mathbbm{1}\left(N_{l}=N_{k}\right)K_{g}\left(\frac{B_{il}-B_{jk}}{h}\right)K_{\boldsymbol{X}}\left(\frac{\boldsymbol{X}_{l}-\boldsymbol{X}_{k}}{h}\right)\frac{1}{N_{k'}}\sum_{j'=1}^{N_{k'}}\frac{1}{N_{k'}-1}\mathbb{T}_{j'k'}K_{f}'\left(\frac{\widehat{V}_{j'k'}-v}{h},\frac{\boldsymbol{X}_{k'}-\boldsymbol{x}}{h}\right)\\
 &  & \times\frac{G\left(B_{j'k},\boldsymbol{X}_{k'},N_{k'}\right)}{g\left(B_{j'k},\boldsymbol{X}_{k'},N_{k'}\right)^{2}}\mathbbm{1}\left(N_{l}=N_{k'}\right)K_{g}\left(\frac{B_{il}-B_{j'k'}}{h}\right)K_{\boldsymbol{X}}\left(\frac{\boldsymbol{X}_{l}-\boldsymbol{X}_{k'}}{h}\right),
\end{eqnarray*}
\begin{eqnarray*}
\varDelta_{2}^{\dagger}\left(v\right) & \coloneqq & \frac{2}{h^{3\left(1+d\right)}\left(L\right)_{3}}\sum_{\left(3\right)}\mathbbm{1}\left(N_{l}=n\right)\frac{1}{N_{l}^{2}}\sum_{i=1}^{N_{l}}\frac{1}{N_{k}}\sum_{j=1}^{N_{k}}\frac{1}{N_{k}-1}\mathbb{T}_{jk}K_{f}'\left(\frac{\widehat{V}_{jk}-v}{h},\frac{\boldsymbol{X}_{k}-\boldsymbol{x}}{h}\right)\frac{G\left(B_{jk},\boldsymbol{X}_{k},N_{k}\right)}{g\left(B_{jk},\boldsymbol{X}_{k},N_{k}\right)^{2}}\\
 &  & \times\mathbbm{1}\left(N_{l}=N_{k}\right)K_{g}\left(\frac{B_{il}-B_{jk}}{h}\right)K_{\boldsymbol{X}}\left(\frac{\boldsymbol{X}_{l}-\boldsymbol{X}_{k}}{h}\right)\frac{1}{N_{k'}}\sum_{j'=1}^{N_{k'}}\frac{1}{N_{k'}-1}\mathbb{T}_{j'k'}K_{f}'\left(\frac{\widehat{V}_{j'k'}-v}{h},\frac{\boldsymbol{X}_{k'}-\boldsymbol{x}}{h}\right)\\
 &  & \times\left\{ \frac{\widehat{G}\left(B_{j'k},\boldsymbol{X}_{k'},N_{k'}\right)}{\widehat{g}\left(B_{j'k},\boldsymbol{X}_{k'},N_{k'}\right)^{2}}-\frac{G\left(B_{j'k},\boldsymbol{X}_{k'},N_{k'}\right)}{g\left(B_{j'k},\boldsymbol{X}_{k'},N_{k'}\right)^{2}}\right\} \mathbbm{1}\left(N_{l}=N_{k'}\right)K_{g}\left(\frac{B_{il}-B_{j'k'}}{h}\right)K_{\boldsymbol{X}}\left(\frac{\boldsymbol{X}_{l}-\boldsymbol{X}_{k'}}{h}\right)
\end{eqnarray*}
and
\begin{eqnarray*}
\varDelta_{3}^{\dagger}\left(v\right) & \coloneqq & \frac{1}{h^{3\left(1+d\right)}\left(L\right)_{3}}\sum_{\left(3\right)}\mathbbm{1}\left(N_{l}=n\right)\frac{1}{N_{l}^{2}}\sum_{i=1}^{N_{l}}\frac{1}{N_{k}}\sum_{j=1}^{N_{k}}\frac{1}{N_{k}-1}\mathbb{T}_{jk}K_{f}'\left(\frac{\widehat{V}_{jk}-v}{h},\frac{\boldsymbol{X}_{k}-\boldsymbol{x}}{h}\right)\\
 &  & \times\left\{ \frac{\widehat{G}\left(B_{jk},\boldsymbol{X}_{k},N_{k}\right)}{\widehat{g}\left(B_{jk},\boldsymbol{X}_{k},N_{k}\right)^{2}}-\frac{G\left(B_{jk},\boldsymbol{X}_{k},N_{k}\right)}{g\left(B_{jk},\boldsymbol{X}_{k},N_{k}\right)^{2}}\right\} \mathbbm{1}\left(N_{l}=N_{k}\right)K_{g}\left(\frac{B_{il}-B_{jk}}{h}\right)K_{\boldsymbol{X}}\left(\frac{\boldsymbol{X}_{l}-\boldsymbol{X}_{k}}{h}\right)\\
 &  & \times\frac{1}{N_{k'}}\sum_{j'=1}^{N_{k'}}\frac{1}{N_{k'}-1}\mathbb{T}_{j'k'}K_{f}'\left(\frac{\widehat{V}_{j'k'}-v}{h},\frac{\boldsymbol{X}_{k'}-\boldsymbol{x}}{h}\right)\left\{ \frac{\widehat{G}\left(B_{j'k},\boldsymbol{X}_{k'},N_{k'}\right)}{\widehat{g}\left(B_{j'k},\boldsymbol{X}_{k'},N_{k'}\right)^{2}}-\frac{G\left(B_{j'k},\boldsymbol{X}_{k'},N_{k'}\right)}{g\left(B_{j'k},\boldsymbol{X}_{k'},N_{k'}\right)^{2}}\right\} \\
 &  & \times\mathbbm{1}\left(N_{l}=N_{k'}\right)K_{g}\left(\frac{B_{il}-B_{j'k'}}{h}\right)K_{\boldsymbol{X}}\left(\frac{\boldsymbol{X}_{l}-\boldsymbol{X}_{k'}}{h}\right).
\end{eqnarray*}

Then we have 
\begin{eqnarray}
\left|\varDelta_{2}^{\dagger}\left(v\right)\right| & \apprle & \left\{ \underset{j',k'}{\mathrm{max}}\,\mathbb{T}_{j'k'}\left|\frac{\widehat{G}\left(B_{j'k},\boldsymbol{X}_{k'},N_{k'}\right)}{\widehat{g}\left(B_{j'k},\boldsymbol{X}_{k'},N_{k'}\right)^{2}}-\frac{G\left(B_{j'k},\boldsymbol{X}_{k'},N_{k'}\right)}{g\left(B_{j'k},\boldsymbol{X}_{k'},N_{k'}\right)^{2}}\right|\right\} \nonumber \\
 &  & \times\frac{1}{h^{3\left(1+d\right)}\left(L\right)_{3}}\sum_{\left(3\right)}\mathbbm{1}\left(N_{l}=n\right)\frac{1}{N_{l}^{2}}\sum_{i=1}^{N_{l}}\frac{1}{N_{k}}\sum_{j=1}^{N_{k}}\frac{1}{N_{k}-1}\mathbb{T}_{jk}\left|K_{f}'\left(\frac{\widehat{V}_{jk}-v}{h},\frac{\boldsymbol{X}_{k}-\boldsymbol{x}}{h}\right)\right|\nonumber \\
 &  & \times\mathbbm{1}\left(N_{l}=N_{k}\right)\left|K_{g}\left(\frac{B_{il}-B_{jk}}{h}\right)\right|\left|K_{\boldsymbol{X}}\left(\frac{\boldsymbol{X}_{l}-\boldsymbol{X}_{k}}{h}\right)\right|\frac{1}{N_{k'}}\sum_{j'=1}^{N_{k'}}\frac{1}{N_{k'}-1}\mathbb{T}_{j'k'}\left|K_{f}'\left(\frac{\widehat{V}_{j'k'}-v}{h},\frac{\boldsymbol{X}_{k'}-\boldsymbol{x}}{h}\right)\right|\nonumber \\
 &  & \times\mathbbm{1}\left(N_{l}=N_{k'}\right)\left|K_{g}\left(\frac{B_{il}-B_{j'k'}}{h}\right)\right|\left|K_{\boldsymbol{X}}\left(\frac{\boldsymbol{X}_{l}-\boldsymbol{X}_{k'}}{h}\right)\right|\nonumber \\
 & \apprle & \left\{ \underset{j',k'}{\mathrm{max}}\,\mathbb{T}_{j'k'}\left|\frac{\widehat{G}\left(B_{j'k},\boldsymbol{X}_{k'},N_{k'}\right)}{\widehat{g}\left(B_{j'k},\boldsymbol{X}_{k'},N_{k'}\right)^{2}}-\frac{G\left(B_{j'k},\boldsymbol{X}_{k'},N_{k'}\right)}{g\left(B_{j'k},\boldsymbol{X}_{k'},N_{k'}\right)^{2}}\right|\right\} \nonumber \\
 &  & \times\frac{1}{h^{3\left(1+d\right)}\left(L\right)_{3}}\sum_{\left(3\right)}\mathbbm{1}\left(N_{l}=n\right)\frac{1}{N_{l}^{2}}\sum_{i=1}^{N_{l}}\frac{1}{N_{k}}\sum_{j=1}^{N_{k}}\frac{1}{N_{k}-1}\mathbbm{1}\left(\left|V_{jk}-v\right|\leq2h\right)\left|K_{\boldsymbol{X}}^{0}\left(\frac{\boldsymbol{X}_{k}-\boldsymbol{x}}{h}\right)\right|\nonumber \\
 &  & \times\mathbbm{1}\left(N_{l}=N_{k}\right)\left|K_{g}\left(\frac{B_{il}-B_{jk}}{h}\right)\right|\left|K_{\boldsymbol{X}}\left(\frac{\boldsymbol{X}_{l}-\boldsymbol{X}_{k}}{h}\right)\right|\frac{1}{N_{k'}}\sum_{j'=1}^{N_{k'}}\frac{1}{N_{k'}-1}\mathbbm{1}\left(\left|V_{j'k'}-v\right|\leq2h\right)\left|K_{\boldsymbol{X}}^{0}\left(\frac{\boldsymbol{X}_{k'}-\boldsymbol{x}}{h}\right)\right|\nonumber \\
 &  & \times\mathbbm{1}\left(N_{l}=N_{k'}\right)\left|K_{g}\left(\frac{B_{il}-B_{j'k'}}{h}\right)\right|\left|K_{\boldsymbol{X}}\left(\frac{\boldsymbol{X}_{l}-\boldsymbol{X}_{k'}}{h}\right)\right|,\label{eq:DELTA_2 bound}
\end{eqnarray}
where the second inequality holds w.p.a.1. Let 
\begin{align*}
 & \mathcal{K}\left(\left(\boldsymbol{b}.,\boldsymbol{z},m\right),\left(\boldsymbol{b}.',\boldsymbol{z}',m'\right),\left(\boldsymbol{b}.'',\boldsymbol{z}'',m''\right);v\right)\\
\coloneqq & h^{-3\left(1+d\right)}\mathbbm{1}\left(m=n\right)\frac{1}{m^{2}}\sum_{i=1}^{m}\frac{1}{m'}\sum_{j=1}^{m'}\frac{1}{m'-1}\mathbbm{1}\left(\left|\xi\left(\boldsymbol{b}.',\boldsymbol{z}',m'\right)-v\right|\leq2h\right)\left|K_{\boldsymbol{X}}^{0}\left(\frac{\boldsymbol{z}'-\boldsymbol{x}}{h}\right)\right|\\
 & \times\mathbbm{1}\left(m=m'\right)\left|K_{g}\left(\frac{b_{i}-b_{j}'}{h}\right)\right|\left|K_{\boldsymbol{X}}\left(\frac{\boldsymbol{z}-\boldsymbol{z}'}{h}\right)\right|\frac{1}{m''}\sum_{j'=1}^{m''}\frac{1}{m''-1}\mathbbm{1}\left(\left|\xi\left(\boldsymbol{b}.'',\boldsymbol{z}'',m''\right)-v\right|\leq2h\right)\\
 & \times\left|K_{\boldsymbol{X}}^{0}\left(\frac{\boldsymbol{z}''-\boldsymbol{x}}{h}\right)\right|\mathbbm{1}\left(m=m''\right)\left|K_{g}\left(\frac{b_{i}-b_{j'}''}{h}\right)\right|\left|K_{\boldsymbol{X}}\left(\frac{\boldsymbol{z}-\boldsymbol{z}''}{h}\right)\right|.
\end{align*}
(\ref{eq:DELTA_2 bound}) can be written as
\begin{eqnarray*}
\left|\varDelta_{2}^{\dagger}\left(v\right)\right| & \apprle & \left\{ \underset{j',k'}{\mathrm{max}}\,\mathbb{T}_{j'k'}\left|\frac{\widehat{G}\left(B_{j'k},\boldsymbol{X}_{k'},N_{k'}\right)}{\widehat{g}\left(B_{j'k},\boldsymbol{X}_{k'},N_{k'}\right)^{2}}-\frac{G\left(B_{j'k},\boldsymbol{X}_{k'},N_{k'}\right)}{g\left(B_{j'k},\boldsymbol{X}_{k'},N_{k'}\right)^{2}}\right|\right\} \\
 &  & \times\left\{ \frac{1}{\left(L\right)_{3}}\sum_{\left(3\right)}\mathcal{K}\left(\left(\boldsymbol{B}_{\cdot l},\boldsymbol{X}_{l},N_{l}\right),\left(\boldsymbol{B}_{\cdot k},\boldsymbol{X}_{k},N_{k}\right),\left(\boldsymbol{B}_{\cdot k'},\boldsymbol{X}_{k'},N_{k'}\right);v\right)\right\} .
\end{eqnarray*}
It can be verified that $\left\{ \mathcal{K}\left(\cdot,\cdot,\cdot;v\right):v\in I\left(\boldsymbol{x}\right)\right\} $
is uniformly VC-type with respect to the envelope 
\begin{equation}
F_{\mathcal{K}}\left(\boldsymbol{z},\boldsymbol{z}',\boldsymbol{z}''\right)\coloneqq\frac{\overline{C}_{K_{g}}^{2}}{n\left(n-1\right)^{2}h^{3\left(1+d\right)}}\left|K_{\boldsymbol{X}}^{0}\left(\frac{\boldsymbol{z}'-\boldsymbol{x}}{h}\right)\right|\left|K_{\boldsymbol{X}}\left(\frac{\boldsymbol{z}-\boldsymbol{z}'}{h}\right)\right|\left|K_{\boldsymbol{X}}^{0}\left(\frac{\boldsymbol{z}''-\boldsymbol{x}}{h}\right)\right|\left|K_{\boldsymbol{X}}\left(\frac{\boldsymbol{z}-\boldsymbol{z}''}{h}\right)\right|.\label{eq:K envelope}
\end{equation}

Define 
\begin{equation}
\mu_{\mathcal{K}}\left(v\right)\coloneqq\mathrm{E}\left[\mathcal{K}\left(\left(\boldsymbol{B}_{\cdot1},\boldsymbol{X}_{1},N_{1}\right),\left(\boldsymbol{B}_{\cdot2},\boldsymbol{X}_{2},N_{2}\right),\left(\boldsymbol{B}_{\cdot3},\boldsymbol{X}_{3},N_{3}\right);v\right)\right],\label{eq:K Hoeffding term 1}
\end{equation}
\begin{gather}
\mathcal{K}_{1}^{\left(1\right)}\left(\boldsymbol{b}.,\boldsymbol{z},m;v\right)\coloneqq\mathrm{E}\left[\mathcal{K}\left(\left(\boldsymbol{b}.,\boldsymbol{z},m\right),\left(\boldsymbol{B}_{\cdot1},\boldsymbol{X}_{1},N_{1}\right),\left(\boldsymbol{B}_{\cdot2},\boldsymbol{X}_{2},N_{2}\right);v\right)\right]\nonumber \\
\mathcal{K}_{2}^{\left(1\right)}\left(\boldsymbol{b}.,\boldsymbol{z},m;v\right)\coloneqq\mathrm{E}\left[\mathcal{K}\left(\left(\boldsymbol{B}_{\cdot1},\boldsymbol{X}_{1},N_{1}\right),\left(\boldsymbol{b}.,\boldsymbol{z},m\right),\left(\boldsymbol{B}_{\cdot2},\boldsymbol{X}_{2},N_{2}\right);v\right)\right]\nonumber \\
\mathcal{K}_{3}^{\left(1\right)}\left(\boldsymbol{b}.,\boldsymbol{z},m;v\right)\coloneqq\mathrm{E}\left[\mathcal{K}\left(\left(\boldsymbol{B}_{\cdot1},\boldsymbol{X}_{1},N_{1}\right),\left(\boldsymbol{B}_{\cdot2},\boldsymbol{X}_{2},N_{2}\right),\left(\boldsymbol{b}.,\boldsymbol{z},m\right);v\right)\right]\label{eq:K Hoeffding term 2}
\end{gather}
and 
\begin{gather}
\mathcal{K}_{1}^{\left(2\right)}\left(\left(\boldsymbol{b}.,\boldsymbol{z},m\right),\left(\boldsymbol{b}.',\boldsymbol{z}',m'\right);v\right)\coloneqq\mathrm{E}\left[\mathcal{K}\left(\left(\boldsymbol{b}.,\boldsymbol{z},m\right),\left(\boldsymbol{b}.',\boldsymbol{z}',m'\right),\left(\boldsymbol{B}_{\cdot1},\boldsymbol{X}_{1},N_{1}\right);v\right)\right]\nonumber \\
\mathcal{K}_{2}^{\left(2\right)}\left(\left(\boldsymbol{b}.,\boldsymbol{z},m\right),\left(\boldsymbol{b}.',\boldsymbol{z}',m'\right);v\right)\coloneqq\mathrm{E}\left[\mathcal{K}\left(\left(\boldsymbol{b}.,\boldsymbol{z},m\right),\left(\boldsymbol{B}_{\cdot1},\boldsymbol{X}_{1},N_{1}\right),\left(\boldsymbol{b}.',\boldsymbol{z}',m'\right);v\right)\right]\nonumber \\
\mathcal{K}_{3}^{\left(2\right)}\left(\left(\boldsymbol{b}.,\boldsymbol{z},m\right),\left(\boldsymbol{b}.',\boldsymbol{z}',m'\right);v\right)\coloneqq\mathrm{E}\left[\mathcal{K}\left(\left(\boldsymbol{B}_{\cdot1},\boldsymbol{X}_{1},N_{1}\right),\left(\boldsymbol{b}.,\boldsymbol{z},m\right),\left(\boldsymbol{b}.',\boldsymbol{z}',m'\right);v\right)\right].\label{eq:K Hoeffding term 3}
\end{gather}

The Hoeffding decomposition yields 
\begin{align*}
 & \frac{1}{\left(L\right)_{3}}\sum_{\left(3\right)}\mathcal{K}\left(\left(\boldsymbol{B}_{\cdot l},\boldsymbol{X}_{l},N_{l}\right),\left(\boldsymbol{B}_{\cdot k},\boldsymbol{X}_{k},N_{k}\right),\left(\boldsymbol{B}_{\cdot k'},\boldsymbol{X}_{k'},N_{k'}\right);v\right)\\
= & \mu_{\mathcal{K}}\left(v\right)+\frac{1}{L}\sum_{l=1}^{L}\left(\mathcal{K}_{1}^{\left(1\right)}\left(\boldsymbol{B}_{\cdot l},\boldsymbol{X}_{l},N_{l};v\right)-\mu_{\mathcal{K}}\left(v\right)\right)+\frac{1}{L}\sum_{l=1}^{L}\left(\mathcal{K}_{2}^{\left(1\right)}\left(\boldsymbol{B}_{\cdot l},\boldsymbol{X}_{l},N_{l};v\right)-\mu_{\mathcal{K}}\left(v\right)\right)+\frac{1}{L}\sum_{l=1}^{L}\left(\mathcal{K}_{3}^{\left(1\right)}\left(\boldsymbol{B}_{\cdot l},\boldsymbol{X}_{l},N_{l};v\right)-\mu_{\mathcal{K}}\left(v\right)\right)\\
 & +\Upsilon_{\mathcal{K}}^{1}\left(v\right)+\Upsilon_{\mathcal{K}}^{2}\left(v\right)+\Upsilon_{\mathcal{K}}^{3}\left(v\right)+\Psi_{\mathcal{K}}\left(v\right),
\end{align*}
where $\Upsilon_{\mathcal{K}}^{1}\left(v\right)$, $\Upsilon_{\mathcal{K}}^{2}\left(v\right)$
and $\Upsilon_{\mathcal{K}}^{3}\left(v\right)$ are degenerate U-statistics
of order two and $\Psi_{\mathcal{K}}\left(v\right)$ is a degenerate
U-statistic of order three:
\begin{gather}
\Upsilon_{\mathcal{K}}^{1}\left(v\right)\coloneqq\frac{1}{\left(L\right)_{2}}\sum_{\left(2\right)}\left\{ \mathcal{K}_{1}^{\left(2\right)}\left(\left(\boldsymbol{B}_{\cdot l},\boldsymbol{X}_{l},N_{l}\right),\left(\boldsymbol{B}_{\cdot k},\boldsymbol{X}_{k},N_{k}\right);v\right)-\mathcal{K}_{1}^{\left(1\right)}\left(\boldsymbol{B}_{\cdot l},\boldsymbol{X}_{l},N_{l};v\right)-\mathcal{K}_{2}^{\left(1\right)}\left(\boldsymbol{B}_{\cdot k},\boldsymbol{X}_{k},N_{k};v\right)+\mu_{\mathcal{K}}\left(v\right)\right\} ,\nonumber \\
\Upsilon_{\mathcal{K}}^{2}\left(v\right)\coloneqq\frac{1}{\left(L\right)_{2}}\sum_{\left(2\right)}\left\{ \mathcal{K}_{2}^{\left(2\right)}\left(\left(\boldsymbol{B}_{\cdot l},\boldsymbol{X}_{l},N_{l}\right),\left(\boldsymbol{B}_{\cdot k},\boldsymbol{X}_{k},N_{k}\right);v\right)-\mathcal{K}_{1}^{\left(1\right)}\left(\boldsymbol{B}_{\cdot l},\boldsymbol{X}_{l},N_{l};v\right)-\mathcal{K}_{3}^{\left(1\right)}\left(\boldsymbol{B}_{\cdot k},\boldsymbol{X}_{k},N_{k};v\right)+\mu_{\mathcal{K}}\left(v\right)\right\} ,\nonumber \\
\Upsilon_{\mathcal{K}}^{3}\left(v\right)\coloneqq\frac{1}{\left(L\right)_{2}}\sum_{\left(2\right)}\left\{ \mathcal{K}_{3}^{\left(2\right)}\left(\left(\boldsymbol{B}_{\cdot l},\boldsymbol{X}_{l},N_{l}\right),\left(\boldsymbol{B}_{\cdot k},\boldsymbol{X}_{k},N_{k}\right);v\right)-\mathcal{K}_{2}^{\left(1\right)}\left(\boldsymbol{B}_{\cdot l},\boldsymbol{X}_{l},N_{l};v\right)-\mathcal{K}_{3}^{\left(1\right)}\left(\boldsymbol{B}_{\cdot k},\boldsymbol{X}_{k},N_{k};v\right)+\mu_{\mathcal{K}}\left(v\right)\right\} \label{eq:K Hoeffding term 4}
\end{gather}
and
\begin{eqnarray}
\Psi_{\mathcal{K}}\left(v\right) & \coloneqq & \frac{1}{\left(L\right)_{3}}\sum_{\left(3\right)}\left\{ \mathcal{K}\left(\left(\boldsymbol{B}_{\cdot l},\boldsymbol{X}_{l},N_{l}\right),\left(\boldsymbol{B}_{\cdot k},\boldsymbol{X}_{k},N_{k}\right),\left(\boldsymbol{B}_{\cdot k'},\boldsymbol{X}_{k'},N_{k'}\right);v\right)-\mathcal{K}_{1}^{\left(2\right)}\left(\left(\boldsymbol{B}_{\cdot l},\boldsymbol{X}_{l},N_{l}\right),\left(\boldsymbol{B}_{\cdot k},\boldsymbol{X}_{k},N_{k}\right);v\right)\right.\nonumber \\
 &  & -\mathcal{K}_{2}^{\left(2\right)}\left(\left(\boldsymbol{B}_{\cdot l},\boldsymbol{X}_{l},N_{l}\right),\left(\boldsymbol{B}_{\cdot k'},\boldsymbol{X}_{k'},N_{k'}\right);v\right)-\mathcal{K}_{3}^{\left(2\right)}\left(\left(\boldsymbol{B}_{\cdot k},\boldsymbol{X}_{k},N_{k}\right),\left(\boldsymbol{B}_{\cdot k'},\boldsymbol{X}_{k'},N_{k'}\right);v\right)\nonumber \\
 &  & \left.+\mathcal{K}_{1}^{\left(1\right)}\left(\boldsymbol{B}_{\cdot l},\boldsymbol{X}_{l},N_{l};v\right)+\mathcal{K}_{2}^{\left(1\right)}\left(\boldsymbol{B}_{\cdot k},\boldsymbol{X}_{k},N_{k};v\right)+\mathcal{K}_{3}^{\left(1\right)}\left(\boldsymbol{B}_{\cdot k'},\boldsymbol{X}_{k'},N_{k'};v\right)-\mu_{\mathcal{K}}\left(v\right)\right\} .\label{eq:K Hoeffding term 5}
\end{eqnarray}

By the LIE, we have 
\begin{eqnarray*}
\mu_{\mathcal{K}}\left(v\right) & = & \int_{\mathcal{X}}\sum_{m\in\mathcal{N}}\int_{\mathcal{X}}\sum_{m'\in\mathcal{N}}\int_{\mathcal{X}}\sum_{m''\in\mathcal{N}}\int_{\underline{b}\left(\boldsymbol{z}''\right)}^{\overline{b}\left(\boldsymbol{z}'',m''\right)}\int_{\underline{b}\left(\boldsymbol{z}'\right)}^{\overline{b}\left(\boldsymbol{z}',m'\right)}\int_{\underline{b}\left(\boldsymbol{z}\right)}^{\overline{b}\left(\boldsymbol{z},m\right)}\frac{1}{h^{3\left(1+d\right)}}\mathbbm{1}\left(m=n\right)\frac{1}{m}\frac{1}{m'-1}\\
 &  & \times\mathbbm{1}\left(\left|\xi\left(b',\boldsymbol{z}',m'\right)-v\right|\leq2h\right)\left|K_{\boldsymbol{X}}^{0}\left(\frac{\boldsymbol{z}'-\boldsymbol{x}}{h}\right)\right|\mathbbm{1}\left(m=m'\right)\left|K_{g}\left(\frac{b-b'}{h}\right)\right|\left|K_{\boldsymbol{X}}\left(\frac{\boldsymbol{z}'-\boldsymbol{z}}{h}\right)\right|\\
 &  & \times\frac{1}{m''-1}\mathbbm{1}\left(\left|\xi\left(b'',\boldsymbol{z}'',m''\right)-v\right|\leq2h\right)\left|K_{\boldsymbol{X}}^{0}\left(\frac{\boldsymbol{z}''-\boldsymbol{x}}{h}\right)\right|\mathbbm{1}\left(m=m''\right)\left|K_{g}\left(\frac{b-b''}{h}\right)\right|\\
 &  & \times\left|K_{\boldsymbol{X}}\left(\frac{\boldsymbol{z}''-\boldsymbol{z}}{h}\right)\right|g\left(b,\boldsymbol{z},m\right)g\left(b',\boldsymbol{z}',m'\right)g\left(b'',\boldsymbol{z}'',m''\right)\mathrm{d}b\mathrm{d}b'\mathrm{d}b''\mathrm{d}\boldsymbol{z}''\mathrm{d}\boldsymbol{z}'\mathrm{d}\boldsymbol{z}\\
 & = & \frac{1}{h^{3\left(1+d\right)}}\int_{\mathcal{X}}\int_{\underline{b}\left(\boldsymbol{z}\right)}^{\overline{b}\left(\boldsymbol{z},n\right)}\left\{ \int_{\mathcal{X}}\int_{\underline{b}\left(\boldsymbol{z}'\right)}^{\overline{b}\left(\boldsymbol{z}',n\right)}\frac{1}{n\left(n-1\right)^{2}}\mathbbm{1}\left(\left|\xi\left(b',\boldsymbol{z}',m'\right)-v\right|\leq2h\right)\left|K_{\boldsymbol{X}}^{0}\left(\frac{\boldsymbol{z}'-\boldsymbol{x}}{h}\right)\right|\left|K_{g}\left(\frac{b-b'}{h}\right)\right|\right.\\
 &  & \left.\times\left|K_{\boldsymbol{X}}\left(\frac{\boldsymbol{z}-\boldsymbol{z}'}{h}\right)\right|g\left(b',\boldsymbol{z}',n\right)\mathrm{d}b'\mathrm{d}\boldsymbol{z}'\right\} ^{2}g\left(b,\boldsymbol{z},n\right)\mathrm{d}b\mathrm{d}\boldsymbol{z}.
\end{eqnarray*}
Then, by change of variables, we have
\begin{eqnarray*}
\mu_{\mathcal{K}}\left(v\right) & = & \int_{\mathcal{Y}}\int_{\frac{\underline{b}\left(h\boldsymbol{y}+\boldsymbol{x}\right)-s\left(v,\boldsymbol{x},n\right)}{h}}^{\frac{\overline{b}\left(h\boldsymbol{y}+\boldsymbol{x},n\right)-s\left(v,\boldsymbol{x},n\right)}{h}}\left\{ \int_{\mathcal{Y}}\int_{\frac{\underline{v}\left(h\boldsymbol{y}'+\boldsymbol{x}\right)-v}{h}}^{\frac{\overline{v}\left(h\boldsymbol{y}'+\boldsymbol{x}\right)-v}{h}}\frac{1}{n\left(n-1\right)^{2}}\mathbbm{1}\left(\left|u\right|\leq2\right)\left|K_{\boldsymbol{X}}^{0}\left(\boldsymbol{y}'\right)\right|\right.\\
 &  & \times\left|K_{g}\left(w-\frac{s\left(hu+v,h\boldsymbol{y}'+\boldsymbol{x},n\right)-s\left(v,\boldsymbol{x},n\right)}{h}\right)\right|\left|K_{\boldsymbol{X}}\left(\boldsymbol{y}-\boldsymbol{y}'\right)\right|\\
 &  & \left.\times g\left(s\left(hu+v,h\boldsymbol{y}'+\boldsymbol{x},n\right),h\boldsymbol{y}'+\boldsymbol{x},n\right)s'\left(hu+v,h\boldsymbol{y}'+\boldsymbol{x},n\right)\mathrm{d}u\mathrm{d}\boldsymbol{y}'\right\} ^{2}g\left(hw+s\left(v,\boldsymbol{x},n\right),h\boldsymbol{y}+\boldsymbol{x},n\right)\mathrm{d}w\mathrm{d}\boldsymbol{y}.
\end{eqnarray*}
It is now clear that 
\begin{eqnarray*}
\mu_{\mathcal{K}}\left(v\right) & \apprle & \int_{\mathcal{Y}}\int_{\frac{\underline{b}\left(h\boldsymbol{y}+\boldsymbol{x}\right)-s\left(v,\boldsymbol{x},n\right)}{h}}^{\frac{\overline{b}\left(h\boldsymbol{y}+\boldsymbol{x},n\right)-s\left(v,\boldsymbol{x},n\right)}{h}}\left\{ \int_{\mathcal{Y}}\int_{\frac{\underline{v}\left(h\boldsymbol{y}'+\boldsymbol{x}\right)-v}{h}}^{\frac{\overline{v}\left(h\boldsymbol{y}'+\boldsymbol{x}\right)-v}{h}}\mathbbm{1}\left(\boldsymbol{y}'\in\mathbb{H}\left(\boldsymbol{0},1\right)\right)\mathbbm{1}\left(\boldsymbol{y}\in\mathbb{H}\left(\boldsymbol{0},2\right)\right)\mathbbm{1}\left(\left|w\right|\leq1+\overline{\tau}\right)\mathbbm{1}\left(\left|u\right|\leq1\right)\right.\\
 &  & \left.\times g\left(s\left(hu+v,h\boldsymbol{y}'+\boldsymbol{x},n\right),h\boldsymbol{y}'+\boldsymbol{x},n\right)s'\left(hu+v,h\boldsymbol{y}'+\boldsymbol{x},n\right)\mathrm{d}u\mathrm{d}\boldsymbol{y}'\right\} ^{2}g\left(hw+s\left(v,\boldsymbol{x},n\right),h\boldsymbol{y}+\boldsymbol{x},n\right)\mathrm{d}w\mathrm{d}\boldsymbol{y},
\end{eqnarray*}
when $h$ is sufficiently small. Now it is clear that $\underset{v\in I\left(\boldsymbol{x}\right)}{\mathrm{sup}}\mu_{\mathcal{K}}\left(v\right)=O\left(1\right)$. 

By the LIE, we have
\begin{eqnarray*}
\mathcal{K}_{1}^{\left(1\right)}\left(\boldsymbol{b}.,\boldsymbol{z},m;v\right) & = & \frac{1}{\left(m-1\right)^{2}m^{2}}\sum_{i=1}^{m}\mathbbm{1}\left(m=n\right)\int_{\mathcal{X}}\int_{\underline{b}\left(\boldsymbol{z}''\right)}^{\overline{b}\left(\boldsymbol{z}'',m\right)}\int_{\mathcal{X}}\int_{\underline{b}\left(\boldsymbol{z}'\right)}^{\overline{b}\left(\boldsymbol{z}',m\right)}\frac{1}{h^{3\left(1+d\right)}}\mathbbm{1}\left(\left|\xi\left(b',\boldsymbol{z}',m\right)-v\right|\leq2h\right)\\
 &  & \times\left|K_{\boldsymbol{X}}^{0}\left(\frac{\boldsymbol{z}'-\boldsymbol{x}}{h}\right)\right|\left|K_{g}\left(\frac{b_{i}-b'}{h}\right)\right|\left|K_{\boldsymbol{X}}\left(\frac{\boldsymbol{z}-\boldsymbol{z}'}{h}\right)\right|\mathbbm{1}\left(\left|\xi\left(b'',\boldsymbol{z}'',m\right)-v\right|\leq2h\right)\left|K_{\boldsymbol{X}}^{0}\left(\frac{\boldsymbol{z}''-\boldsymbol{x}}{h}\right)\right|\\
 &  & \times\left|K_{g}\left(\frac{b_{i}-b''}{h}\right)\right|g\left(b',\boldsymbol{z}',m\right)g\left(b'',\boldsymbol{z}'',m\right)\mathrm{d}b'\mathrm{d}\boldsymbol{z}'\mathrm{d}b''\mathrm{d}\boldsymbol{z}''
\end{eqnarray*}
and 
\begin{eqnarray*}
\mathcal{K}_{2}^{\left(1\right)}\left(\boldsymbol{b}.,\boldsymbol{z},m;v\right) & = & \frac{1}{n\left(n-1\right)}\mathbbm{1}\left(m=n\right)\frac{1}{m}\sum_{i=1}^{m}\int_{\mathcal{X}}\int_{\mathcal{X}}\int_{\underline{b}\left(\boldsymbol{z}''\right)}^{\overline{b}\left(\boldsymbol{z}'',n\right)}\int_{\underline{b}\left(\boldsymbol{z}'\right)}^{\overline{b}\left(\boldsymbol{z}',n\right)}\mathbbm{1}\left(\left|\xi\left(b_{i},\boldsymbol{z},m\right)-v\right|\leq2h\right)\left|K_{\boldsymbol{X}}^{0}\left(\frac{\boldsymbol{z}-\boldsymbol{x}}{h}\right)\right|\\
 &  & \times\left|K_{g}\left(\frac{b'-b_{i}}{h}\right)\right|\left|K_{\boldsymbol{X}}\left(\frac{\boldsymbol{z}'-\boldsymbol{z}}{h}\right)\right|\mathbbm{1}\left(\left|\xi\left(b'',\boldsymbol{z}'',n\right)-v\right|\leq2h\right)\left|K_{\boldsymbol{X}}^{0}\left(\frac{\boldsymbol{z}''-\boldsymbol{x}}{h}\right)\right|\left|K_{g}\left(\frac{b'-b''}{h}\right)\right|\\
 &  & \times\left|K_{\boldsymbol{X}}^{0}\left(\frac{\boldsymbol{z}'-\boldsymbol{z}''}{h}\right)\right|g\left(b',\boldsymbol{z}',n\right)g\left(b'',\boldsymbol{z}'',n\right)\mathrm{d}b'\mathrm{d}b''\mathrm{d}\boldsymbol{z}'\mathrm{d}\boldsymbol{z}''.
\end{eqnarray*}

By Jensen's inequality, the LIE and change of variables, we have 
\begin{align*}
 & \mathrm{E}\left[\mathcal{K}_{1}^{\left(1\right)}\left(\boldsymbol{B}_{\cdot1},\boldsymbol{X}_{1},N_{1};v\right)^{2}\right]\\
\leq & \mathrm{E}\left[\frac{1}{N_{1}^{2}\left(N_{1}-1\right)^{4}}\mathbbm{1}\left(N_{1}=n\right)\frac{1}{N_{1}}\sum_{i=1}^{N_{1}}\left\{ \int_{\mathcal{X}}\int_{\underline{b}\left(\boldsymbol{z}''\right)}^{\overline{b}\left(\boldsymbol{z}'',n\right)}\int_{\mathcal{X}}\int_{\underline{b}\left(\boldsymbol{z}'\right)}^{\overline{b}\left(\boldsymbol{z}',n\right)}\frac{1}{h^{3\left(1+d\right)}}\mathbbm{1}\left(\left|\xi\left(b',\boldsymbol{z}',N_{1}\right)-v\right|\leq2h\right)\right.\right.\\
 & \times\left|K_{\boldsymbol{X}}^{0}\left(\frac{\boldsymbol{z}'-\boldsymbol{x}}{h}\right)\right|\left|K_{g}\left(\frac{B_{i1}-b'}{h}\right)\right|\left|K_{\boldsymbol{X}}\left(\frac{X_{1}-\boldsymbol{z}'}{h}\right)\right|\mathbbm{1}\left(\left|\xi\left(b'',\boldsymbol{z}'',N_{1}\right)-v\right|\leq2h\right)\left|K_{\boldsymbol{X}}^{0}\left(\frac{\boldsymbol{z}''-\boldsymbol{x}}{h}\right)\right|\\
 & \left.\left.\times g\left(b',\boldsymbol{z}',N_{1}\right)g\left(b'',\boldsymbol{z}'',N_{1}\right)\mathrm{d}b'\mathrm{d}\boldsymbol{z}'\mathrm{d}b''\mathrm{d}\boldsymbol{z}''\right\} ^{2}\right]\\
= & \frac{1}{n^{2}\left(n-1\right)^{4}}\frac{1}{h^{1+d}}\int_{\mathcal{Y}}\int_{\frac{\underline{b}\left(h\boldsymbol{y}+\boldsymbol{x}\right)-s\left(v,\boldsymbol{x},n\right)}{h}}^{\frac{\overline{b}\left(h\boldsymbol{y}+\boldsymbol{x},n\right)-s\left(v,\boldsymbol{x},n\right)}{h}}\left\{ \int_{\mathcal{Y}}\int_{\frac{\underline{v}\left(h\boldsymbol{y}''+\boldsymbol{x}\right)-v}{h}}^{\frac{\overline{v}\left(h\boldsymbol{y}''+\boldsymbol{x}\right)-v}{h}}\int_{\mathcal{Y}}\int_{\frac{\underline{v}\left(h\boldsymbol{y}'+\boldsymbol{x}\right)-v}{h}}^{\frac{\overline{v}\left(h\boldsymbol{y}'+\boldsymbol{x}\right)-v}{h}}\mathbbm{1}\left(\left|u\right|\leq2\right)\left|K_{\boldsymbol{X}}^{0}\left(\boldsymbol{y}'\right)\right|\right.\\
 & \times\left|K_{g}\left(w-\frac{s\left(hu+v,h\boldsymbol{y}'+\boldsymbol{x},n\right)-s\left(v,\boldsymbol{x},n\right)}{h}\right)\right|\left|K_{\boldsymbol{X}}\left(\boldsymbol{y}-\boldsymbol{y}'\right)\right|\mathbbm{1}\left(\left|u'\right|\leq2\right)\left|K_{\boldsymbol{X}}^{0}\left(\boldsymbol{y}''\right)\right|\\
 & \times\left|K_{g}\left(w-\frac{s\left(hu'+v,h\boldsymbol{y}''+\boldsymbol{x},n\right)-s\left(v,\boldsymbol{x},n\right)}{h}\right)\right|g\left(s\left(hu+v,h\boldsymbol{y}'+\boldsymbol{x},n\right),h\boldsymbol{y}'+\boldsymbol{x},n\right)g\left(s\left(hu'+v,h\boldsymbol{y}''+\boldsymbol{x},n\right),h\boldsymbol{y}''+\boldsymbol{x},n\right)\\
 & \left.\times s'\left(hu+v,h\boldsymbol{y}'+\boldsymbol{x},n\right)s'\left(hu'+v,h\boldsymbol{y}''+\boldsymbol{x},n\right)\mathrm{d}u\mathrm{d}\boldsymbol{y}'\mathrm{d}u'\mathrm{d}\boldsymbol{y}''\right\} ^{2}g\left(hw+s\left(v,\boldsymbol{x},n\right),h\boldsymbol{y}+\boldsymbol{x},n\right)\mathrm{d}w\mathrm{d}\boldsymbol{y}
\end{align*}
and 
\begin{align*}
 & \mathrm{E}\left[\mathcal{K}_{2}^{\left(1\right)}\left(\boldsymbol{B}_{\cdot1},\boldsymbol{X}_{1},N_{1};v\right)^{2}\right]\\
\leq & \mathrm{E}\left[\frac{1}{\left(N_{1}-1\right)^{2}}\frac{1}{N_{1}}\sum_{i=1}^{N_{1}}\left\{ \int_{\mathcal{X}}\int_{\mathcal{X}}\int_{\underline{b}\left(\boldsymbol{z}''\right)}^{\overline{b}\left(\boldsymbol{z}'',n\right)}\int_{\underline{b}\left(\boldsymbol{z}'\right)}^{\overline{b}\left(\boldsymbol{z}',n\right)}\frac{1}{n\left(n-1\right)}\mathbbm{1}\left(N_{1}=n\right)\frac{1}{h^{3\left(1+d\right)}}\mathbbm{1}\left(\left|\xi\left(B_{i1},\boldsymbol{X}_{1},N_{1}\right)-v\right|\leq2h\right)\right.\right.\\
 & \times\left|K_{\boldsymbol{X}}^{0}\left(\frac{\boldsymbol{X}_{1}-\boldsymbol{x}}{h}\right)\right|\left|K_{g}\left(\frac{b'-B_{i1}}{h}\right)\right|\left|K_{\boldsymbol{X}}\left(\frac{\boldsymbol{z}-\boldsymbol{X}_{1}}{h}\right)\right|\mathbbm{1}\left(\left|\xi\left(b'',\boldsymbol{z}'',n\right)-v\right|\leq2h\right)\left|K_{\boldsymbol{X}}^{0}\left(\frac{\boldsymbol{z}''-\boldsymbol{x}}{h}\right)\right|\left|K_{g}\left(\frac{b'-b''}{h}\right)\right|\\
 & \left.\left.\times\left|K_{\boldsymbol{X}}\left(\frac{\boldsymbol{z}'-\boldsymbol{z}''}{h}\right)\right|g\left(b',\boldsymbol{z}',n\right)g\left(b'',\boldsymbol{z}'',n\right)\mathrm{d}b'\mathrm{d}b''\mathrm{d}\boldsymbol{z}'\mathrm{d}\boldsymbol{z}''\right\} ^{2}\right]\\
= & \frac{1}{n^{2}\left(n-1\right)^{4}}\frac{1}{h^{1+d}}\int_{\mathcal{Y}}\int_{\frac{\underline{v}\left(h\boldsymbol{y}+\boldsymbol{x}\right)-v}{h}}^{\frac{\overline{v}\left(h\boldsymbol{y}+\boldsymbol{x}\right)-v}{h}}\left\{ \int_{\mathcal{Y}}\int_{\mathcal{Y}}\int_{\frac{\underline{v}\left(h\boldsymbol{y}''+\boldsymbol{x}\right)-v}{h}}^{\frac{\overline{v}\left(h\boldsymbol{y}''+\boldsymbol{x}\right)-v}{h}}\int_{\frac{\underline{b}\left(h\boldsymbol{y}'+\boldsymbol{x}\right)-s\left(v,\boldsymbol{x},n\right)}{h}}^{\frac{\overline{b}\left(h\boldsymbol{y}'+\boldsymbol{x},n\right)-s\left(v,\boldsymbol{x},n\right)}{h}}\mathbbm{1}\left(\left|u\right|\leq2\right)\left|K_{\boldsymbol{X}}^{0}\left(\boldsymbol{y}\right)\right|\right.\\
 & \times\left|K_{g}\left(w-\frac{s\left(hu+v,h\boldsymbol{y}+\boldsymbol{x},n\right)-s\left(v,\boldsymbol{x},n\right)}{h}\right)\right|\left|K_{\boldsymbol{X}}\left(\boldsymbol{y}'-\boldsymbol{y}\right)\right|\mathbbm{1}\left(\left|w'\right|\leq2\right)\left|K_{\boldsymbol{X}}^{0}\left(\boldsymbol{y}''\right)\right|\\
 & \times\left|K_{g}\left(w-\frac{s\left(hw'+v,h\boldsymbol{y}''+\boldsymbol{x},n\right)-s\left(v,\boldsymbol{x},n\right)}{h}\right)\right|g\left(hw+s\left(v,\boldsymbol{x},n\right),h\boldsymbol{y}'+\boldsymbol{x},n\right)g\left(s\left(hw'+v,h\boldsymbol{y}''+\boldsymbol{x},n\right),h\boldsymbol{y}''+\boldsymbol{x},n\right)\\
 & \left.\times s'\left(hw'+v,h\boldsymbol{y}''+\boldsymbol{x},n\right)\mathrm{d}w\mathrm{d}w'\mathrm{d}\boldsymbol{y}'\mathrm{d}\boldsymbol{y}''\right\} ^{2}g\left(s\left(hu+v,h\boldsymbol{y}+\boldsymbol{x},n\right),h\boldsymbol{y}+\boldsymbol{x},n\right)s'\left(hu+v,h\boldsymbol{y}+\boldsymbol{x},n\right)\mathrm{d}u\mathrm{d}\boldsymbol{y}.
\end{align*}
Now it is clear that 
\begin{equation}
\sigma_{\mathcal{K}_{1}^{\left(1\right)}}^{2}\coloneqq\underset{v\in I\left(\boldsymbol{x}\right)}{\mathrm{sup}}\mathrm{E}\left[\mathcal{K}_{1}^{\left(1\right)}\left(\boldsymbol{B}_{\cdot1},\boldsymbol{X}_{1},N_{1};v\right)^{2}\right]=O\left(h^{-\left(1+d\right)}\right).\label{eq:E=00005BK^(1)_1^2=00005D bound}
\end{equation}

Since $\left\{ \mathcal{K}\left(\cdot,\cdot,\cdot;v\right):v\in I\left(\boldsymbol{x}\right)\right\} $
is (uniformly) VC-type with respect to the envelope (\ref{eq:K envelope}),
it follows from \citet[Lemma 5.4]{Chen_Kato_U_Process} that the class
$\left\{ \mathcal{K}_{1}^{\left(1\right)}\left(\cdot;v\right):v\in I\left(\boldsymbol{x}\right)\right\} $
is VC-type with respect to the envelope 
\begin{eqnarray*}
F_{\mathcal{K}_{1}^{\left(1\right)}}\left(\boldsymbol{z}\right) & \coloneqq & \frac{\overline{C}_{K_{g}}^{2}}{n\left(n-1\right)^{2}h^{3\left(1+d\right)}}\left(\int_{\mathcal{X}}\left|K_{\boldsymbol{X}}^{0}\left(\frac{\boldsymbol{z}'-\boldsymbol{x}}{h}\right)\right|\left|K_{\boldsymbol{X}}\left(\frac{\boldsymbol{z}-\boldsymbol{z}'}{h}\right)\right|\varphi\left(\boldsymbol{z}'\right)\mathrm{d}\boldsymbol{z}'\right)^{2}\\
 & \apprle & h^{-\left(3+d\right)}\left(\int_{\mathcal{Y}}\left|K_{\boldsymbol{X}}^{0}\left(\boldsymbol{y}'\right)\right|\left|K_{\boldsymbol{X}}\left(\frac{\boldsymbol{z}-\boldsymbol{x}}{h}-\boldsymbol{y}'\right)\right|\varphi\left(h\boldsymbol{y}'+\boldsymbol{x}\right)\mathrm{d}\boldsymbol{y}'\right)^{2}
\end{eqnarray*}
and it is clear that $\left\Vert F_{\mathcal{K}_{1}^{\left(1\right)}}\right\Vert _{\mathcal{X}}=O\left(h^{-\left(3+d\right)}\right)$. 

Now applying the CCK inequality with $\sigma=\sigma_{\mathcal{K}_{1}^{\left(1\right)}}$
and $F=F_{\mathcal{K}_{1}^{\left(1\right)}}$, we have
\begin{eqnarray}
\mathrm{E}\left[\underset{v\in I\left(\boldsymbol{x}\right)}{\mathrm{sup}}\left|\frac{1}{L}\sum_{l=1}^{L}\mathcal{K}_{1}^{\left(1\right)}\left(\boldsymbol{B}_{\cdot l},\boldsymbol{X}_{l},N_{l};v\right)-\mu_{\mathcal{K}}\left(v\right)\right|\right] & \leq & C_{1}\left\{ L^{-\nicefrac{1}{2}}\sigma_{\mathcal{K}_{1}^{\left(1\right)}}\mathrm{log}\left(C_{2}L\right)^{\nicefrac{1}{2}}+L^{-1}\left\Vert F_{\mathcal{K}_{1}^{\left(1\right)}}\right\Vert _{\mathcal{X}}\mathrm{log}\left(C_{2}L\right)\right\} \nonumber \\
 & = & O\left(\left(\frac{\mathrm{log}\left(L\right)}{Lh^{1+d}}\right)^{\nicefrac{1}{2}}+\frac{\mathrm{log}\left(L\right)}{Lh^{3+d}}\right).\label{eq:K bound 1}
\end{eqnarray}
Similarly, 
\begin{equation}
\mathrm{E}\left[\underset{v\in I\left(\boldsymbol{x}\right)}{\mathrm{sup}}\left|\frac{1}{L}\sum_{l=1}^{L}\mathcal{K}_{2}^{\left(1\right)}\left(\boldsymbol{B}_{\cdot l},\boldsymbol{X}_{l},N_{l};v\right)-\mu_{\mathcal{K}}\left(v\right)\right|\right]=O\left(\left(\frac{\mathrm{log}\left(L\right)}{Lh^{1+d}}\right)^{\nicefrac{1}{2}}+\frac{\mathrm{log}\left(L\right)}{Lh^{3+d}}\right)\label{eq:K bound 2}
\end{equation}
follows from 
\[
\underset{v\in I\left(\boldsymbol{x}\right)}{\mathrm{sup}}\mathrm{E}\left[\mathcal{K}_{2}^{\left(1\right)}\left(\boldsymbol{B}_{\cdot1},\boldsymbol{X}_{1},N_{1};v\right)^{2}\right]=O\left(h^{-\left(1+d\right)}\right).
\]
We also have
\begin{equation}
\mathrm{E}\left[\underset{v\in I\left(\boldsymbol{x}\right)}{\mathrm{sup}}\left|\frac{1}{L}\sum_{l=1}^{L}\mathcal{K}_{3}^{\left(1\right)}\left(\boldsymbol{B}_{\cdot l},\boldsymbol{X}_{l},N_{l};v\right)-\mu_{\mathcal{K}}\left(v\right)\right|\right]=O\left(\left(\frac{\mathrm{log}\left(L\right)}{Lh^{1+d}}\right)^{\nicefrac{1}{2}}+\frac{\mathrm{log}\left(L\right)}{Lh^{3+d}}\right)\label{eq:K bound 3}
\end{equation}
since $\mathcal{K}$ is symmetric with respect to the second and the
third arguments.

The CK inequality yields 
\begin{eqnarray}
\mathrm{E}\left[\underset{v\in I\left(\boldsymbol{x}\right)}{\mathrm{sup}}\left|\Upsilon_{\mathcal{K}}^{1}\left(v\right)\right|\right] & \apprle & L^{-1}\left\{ \int_{\mathcal{X}}\int_{\mathcal{X}}\left(\int_{\mathcal{X}}\frac{1}{h^{3\left(1+d\right)}}\left|K_{\boldsymbol{X}}^{0}\left(\frac{\boldsymbol{z}'-\boldsymbol{x}}{h}\right)\right|\left|K_{\boldsymbol{X}}\left(\frac{\boldsymbol{z}-\boldsymbol{z}'}{h}\right)\right|\left|K_{\boldsymbol{X}}^{0}\left(\frac{\boldsymbol{z}''-\boldsymbol{x}}{h}\right)\right|\left|K_{\boldsymbol{X}}\left(\frac{\boldsymbol{z}-\boldsymbol{z}''}{h}\right)\right|\varphi\left(\boldsymbol{z}''\right)\mathrm{d}\boldsymbol{z}''\right)^{2}\right.\nonumber \\
 &  & \left.\times\varphi\left(\boldsymbol{z}\right)\varphi\left(\boldsymbol{z}'\right)\mathrm{d}\boldsymbol{z}\mathrm{d}\boldsymbol{z}'\right\} ^{\nicefrac{1}{2}}\nonumber \\
 & = & \left(Lh^{3+d}\right)^{-1}\left\{ \int_{\mathcal{Y}}\int_{\mathcal{Y}}\left(\int_{\mathcal{Y}}\left|K_{\boldsymbol{X}}^{0}\left(\boldsymbol{y}'\right)\right|\left|K_{\boldsymbol{X}}\left(\boldsymbol{y}-\boldsymbol{y}''\right)\right|\left|K_{\boldsymbol{X}}^{0}\left(\boldsymbol{y}''\right)\right|\left|K_{\boldsymbol{X}}\left(\boldsymbol{y}-\boldsymbol{y}''\right)\right|\varphi\left(h\boldsymbol{y}''+\boldsymbol{x}\right)\mathrm{d}\boldsymbol{y}''\right)^{2}\right.\nonumber \\
 &  & \left.\times\varphi\left(h\boldsymbol{y}+\boldsymbol{x}\right)\varphi\left(h\boldsymbol{y}'+\boldsymbol{x}\right)\mathrm{d}\boldsymbol{y}\mathrm{d}\boldsymbol{y}'\right\} ^{\nicefrac{1}{2}}\nonumber \\
 & = & O\left(\left(Lh^{3+d}\right)^{-1}\right),\label{eq:K bound 4}
\end{eqnarray}
\begin{eqnarray}
\mathrm{E}\left[\underset{v\in I\left(\boldsymbol{x}\right)}{\mathrm{sup}}\left|\Upsilon_{\mathcal{K}}^{3}\left(v\right)\right|\right] & \apprle & L^{-1}\left\{ \int_{\mathcal{X}}\int_{\mathcal{X}}\left(\int_{\mathcal{X}}\frac{1}{h^{3\left(1+d\right)}}\left|K_{\boldsymbol{X}}^{0}\left(\frac{\boldsymbol{z}'-\boldsymbol{x}}{h}\right)\right|\left|K_{\boldsymbol{X}}\left(\frac{\boldsymbol{z}-\boldsymbol{z}'}{h}\right)\right|\left|K_{\boldsymbol{X}}^{0}\left(\frac{\boldsymbol{z}''-\boldsymbol{x}}{h}\right)\right|\left|K_{\boldsymbol{X}}\left(\frac{\boldsymbol{z}-\boldsymbol{z}''}{h}\right)\right|\varphi\left(\boldsymbol{z}\right)\mathrm{d}\boldsymbol{z}\right)^{2}\right.\nonumber \\
 &  & \left.\times\varphi\left(\boldsymbol{z}'\right)\varphi\left(\boldsymbol{z}''\right)\mathrm{d}\boldsymbol{z}'\mathrm{d}\boldsymbol{z}''\right\} ^{\nicefrac{1}{2}}\nonumber \\
 & = & \left(Lh^{3+d}\right)^{-1}\left\{ \int_{\mathcal{Y}}\int_{\mathcal{Y}}\left(\int_{\mathcal{Y}}\left|K_{\boldsymbol{X}}^{0}\left(\boldsymbol{y}'\right)\right|\left|K_{\boldsymbol{X}}\left(\boldsymbol{y}-\boldsymbol{y}'\right)\right|\left|K_{\boldsymbol{X}}^{0}\left(\boldsymbol{y}''\right)\right|\left|K_{\boldsymbol{X}}\left(\boldsymbol{y}-\boldsymbol{y}''\right)\right|\varphi\left(h\boldsymbol{y}+\boldsymbol{x}\right)\mathrm{d}\boldsymbol{y}\right)^{2}\right.\nonumber \\
 &  & \left.\times\varphi\left(h\boldsymbol{y}'+\boldsymbol{x}\right)\varphi\left(h\boldsymbol{y}''+\boldsymbol{x}\right)\mathrm{d}\boldsymbol{y}'\mathrm{d}\boldsymbol{y}''\right\} ^{\nicefrac{1}{2}}\nonumber \\
 & = & O\left(\left(Lh^{3+d}\right)^{-1}\right)\label{eq:K bound 5}
\end{eqnarray}
and 
\begin{eqnarray}
\mathrm{E}\left[\underset{v\in I\left(\boldsymbol{x}\right)}{\mathrm{sup}}\left|\Psi_{\mathcal{K}}\left(v\right)\right|\right] & \apprle & L^{-\nicefrac{3}{2}}\left\{ \int_{\mathcal{X}}\int_{\mathcal{X}}\int_{\mathcal{X}}\frac{1}{h^{6\left(1+d\right)}}K_{\boldsymbol{X}}^{0}\left(\frac{\boldsymbol{z}'-\boldsymbol{x}}{h}\right)^{2}K_{\boldsymbol{X}}\left(\frac{\boldsymbol{z}-\boldsymbol{z}'}{h}\right)^{2}K_{\boldsymbol{X}}^{0}\left(\frac{\boldsymbol{z}''-\boldsymbol{x}}{h}\right)^{2}K_{\boldsymbol{X}}\left(\frac{\boldsymbol{z}-\boldsymbol{z}''}{h}\right)^{2}\varphi\left(\boldsymbol{z}\right)\right.\nonumber \\
 &  & \left.\times\varphi\left(\boldsymbol{z}'\right)\varphi\left(\boldsymbol{z}''\right)\mathrm{d}\boldsymbol{z}\mathrm{d}\boldsymbol{z}'\mathrm{d}\boldsymbol{z}''\right\} ^{\nicefrac{1}{2}}\nonumber \\
 & = & L^{-\nicefrac{3}{2}}h^{-\left(3+\nicefrac{3}{2}d\right)}\left\{ \int_{\mathcal{Y}}\int_{\mathcal{Y}}\int_{\mathcal{Y}}K_{\boldsymbol{X}}^{0}\left(\boldsymbol{y}'\right)^{2}K_{\boldsymbol{X}}\left(\boldsymbol{y}-\boldsymbol{y}'\right)^{2}K_{\boldsymbol{X}}^{0}\left(\boldsymbol{y}''\right)^{2}K_{\boldsymbol{X}}\left(\boldsymbol{y}-\boldsymbol{y}''\right)^{2}\varphi\left(h\boldsymbol{y}+\boldsymbol{x}\right)\right.\nonumber \\
 &  & \left.\times\varphi\left(h\boldsymbol{y}'+\boldsymbol{x}\right)\varphi\left(h\boldsymbol{y}''+\boldsymbol{x}\right)\mathrm{d}\boldsymbol{y}\mathrm{d}\boldsymbol{y}'\mathrm{d}\boldsymbol{y}''\right\} ^{\nicefrac{1}{2}}\nonumber \\
 & = & o\left(\left(Lh^{3+d}\right)^{-1}\right).\label{eq:K bound 6}
\end{eqnarray}
Since $\mathcal{K}$ is symmetric with respect to the second and the
third arguments, we have
\begin{equation}
\mathrm{E}\left[\underset{v\in I\left(\boldsymbol{x}\right)}{\mathrm{sup}}\left|\Upsilon_{\mathcal{K}}^{2}\left(v\right)\right|\right]=O\left(\left(Lh^{3+d}\right)^{-1}\right).\label{eq:K bound 7}
\end{equation}

Write 
\[
\varDelta_{1}^{\dagger}\left(v\right)=\varDelta_{4}^{\dagger}\left(v\right)+2\varDelta_{5}^{\dagger}\left(v\right)+\varDelta_{6}^{\dagger}\left(v\right),
\]
where
\begin{eqnarray*}
\varDelta_{4}^{\dagger}\left(v\right) & \coloneqq & \frac{1}{h^{3\left(1+d\right)}\left(L\right)_{3}}\sum_{\left(3\right)}\mathbbm{1}\left(N_{l}=n\right)\frac{1}{N_{l}^{2}}\sum_{i=1}^{N_{l}}\frac{1}{N_{k}}\sum_{j=1}^{N_{k}}\frac{1}{N_{k}-1}\mathbb{T}_{jk}K_{f}'\left(\frac{V_{jk}-v}{h},\frac{\boldsymbol{X}_{k}-\boldsymbol{x}}{h}\right)\frac{G\left(B_{jk},\boldsymbol{X}_{k},N_{k}\right)}{g\left(B_{jk},\boldsymbol{X}_{k},N_{k}\right)^{2}}\\
 &  & \times\mathbbm{1}\left(N_{l}=N_{k}\right)K_{g}\left(\frac{B_{il}-B_{jk}}{h}\right)K_{\boldsymbol{X}}\left(\frac{\boldsymbol{X}_{l}-\boldsymbol{X}_{k}}{h}\right)\frac{1}{N_{k'}}\sum_{j'=1}^{N_{k'}}\frac{1}{N_{k'}-1}\mathbb{T}_{j'k'}K_{f}'\left(\frac{V_{j'k'}-v}{h},\frac{\boldsymbol{X}_{k'}-\boldsymbol{x}}{h}\right)\\
 &  & \times\frac{G\left(B_{j'k},\boldsymbol{X}_{k'},N_{k'}\right)}{g\left(B_{j'k},\boldsymbol{X}_{k'},N_{k'}\right)^{2}}\mathbbm{1}\left(N_{l}=N_{k'}\right)K_{g}\left(\frac{B_{il}-B_{j'k'}}{h}\right)K_{\boldsymbol{X}}\left(\frac{\boldsymbol{X}_{l}-\boldsymbol{X}_{k'}}{h}\right),
\end{eqnarray*}
\begin{eqnarray*}
\varDelta_{5}^{\dagger}\left(v\right) & \coloneqq & \frac{1}{h^{4+3d}\left(L\right)_{3}}\sum_{\left(3\right)}\mathbbm{1}\left(N_{l}=n\right)\frac{1}{N_{l}^{2}}\sum_{i=1}^{N_{l}}\frac{1}{N_{k}}\sum_{j=1}^{N_{k}}\frac{1}{N_{k}-1}\mathbb{T}_{jk}K_{f}''\left(\frac{\dot{V}_{jk}-v}{h},\frac{\boldsymbol{X}_{k}-\boldsymbol{x}}{h}\right)\left(\widehat{V}_{jk}-V_{jk}\right)\frac{G\left(B_{jk},\boldsymbol{X}_{k},N_{k}\right)}{g\left(B_{jk},\boldsymbol{X}_{k},N_{k}\right)^{2}}\\
 &  & \times\mathbbm{1}\left(N_{l}=N_{k}\right)K_{g}\left(\frac{B_{il}-B_{jk}}{h}\right)K_{\boldsymbol{X}}\left(\frac{\boldsymbol{X}_{l}-\boldsymbol{X}_{k}}{h}\right)\frac{1}{N_{k'}}\sum_{j'=1}^{N_{k'}}\frac{1}{N_{k'}-1}\mathbb{T}_{j'k'}K_{f}'\left(\frac{V_{j'k'}-v}{h},\frac{\boldsymbol{X}_{k'}-\boldsymbol{x}}{h}\right)\\
 &  & \times\frac{G\left(B_{j'k},\boldsymbol{X}_{k'},N_{k'}\right)}{g\left(B_{j'k},\boldsymbol{X}_{k'},N_{k'}\right)^{2}}\mathbbm{1}\left(N_{l}=N_{k'}\right)K_{g}\left(\frac{B_{il}-B_{j'k'}}{h}\right)K_{\boldsymbol{X}}\left(\frac{\boldsymbol{X}_{l}-\boldsymbol{X}_{k'}}{h}\right)
\end{eqnarray*}
and 
\begin{eqnarray*}
\varDelta_{6}^{\dagger}\left(v\right) & \coloneqq & \frac{1}{h^{5+3d}\left(L\right)_{3}}\sum_{\left(3\right)}\mathbbm{1}\left(N_{l}=n\right)\frac{1}{N_{l}^{2}}\sum_{i=1}^{N_{l}}\frac{1}{N_{k}}\sum_{j=1}^{N_{k}}\frac{1}{N_{k}-1}\mathbb{T}_{jk}\left(\frac{\dot{V}_{jk}-v}{h},\frac{\boldsymbol{X}_{k}-\boldsymbol{x}}{h}\right)\left(\widehat{V}_{jk}-V_{jk}\right)\frac{G\left(B_{jk},\boldsymbol{X}_{k},N_{k}\right)}{g\left(B_{jk},\boldsymbol{X}_{k},N_{k}\right)^{2}}\\
 &  & \times\mathbbm{1}\left(N_{l}=N_{k}\right)K_{g}\left(\frac{B_{il}-B_{jk}}{h}\right)K_{\boldsymbol{X}}\left(\frac{\boldsymbol{X}_{l}-\boldsymbol{X}_{k}}{h}\right)\frac{1}{N_{k'}}\sum_{j'=1}^{N_{k'}}\frac{1}{N_{k'}-1}\mathbb{T}_{j'k'}K_{f}'\left(\frac{\dot{V}_{j'k'}-v}{h},\frac{\boldsymbol{X}_{k'}-\boldsymbol{x}}{h}\right)\left(\widehat{V}_{j'k'}-V_{j'k'}\right)\\
 &  & \times\frac{G\left(B_{j'k},\boldsymbol{X}_{k'},N_{k'}\right)}{g\left(B_{j'k},\boldsymbol{X}_{k'},N_{k'}\right)^{2}}\mathbbm{1}\left(N_{l}=N_{k'}\right)K_{g}\left(\frac{B_{il}-B_{j'k'}}{h}\right)K_{\boldsymbol{X}}\left(\frac{\boldsymbol{X}_{l}-\boldsymbol{X}_{k'}}{h}\right),
\end{eqnarray*}
where $\dot{V}_{jk}$ ($\dot{V}_{j'k'}$) is the mean value that lies
between $V_{jk}$ ($V_{j'k'}$) and $\widehat{V}_{jk}$ ($\widehat{V}_{j'k'}$). 

Now by the fact $\underset{i,l}{\mathrm{max}}\,\mathbb{T}_{il}\left|\widehat{V}_{il}-V_{il}\right|=o_{p}\left(h\right)$
and the fact that $K_{0}'$ and $K_{0}''$ are both compactly supported
on $\left[-1,1\right]$, we have 
\[
\left|\mathit{\Delta}_{5}^{\dagger}\left(v\right)\right|\lesssim h^{-1}\left\{ \underset{j,k}{\mathrm{max}}\,\mathbb{T}_{jk}\left|\widehat{V}_{jk}-V_{jk}\right|\right\} \left\{ \frac{1}{\left(L\right)_{3}}\sum_{\left(3\right)}\mathcal{K}\left(\left(\boldsymbol{B}_{\cdot l},\boldsymbol{X}_{l},N_{l}\right),\left(\boldsymbol{B}_{\cdot k},\boldsymbol{X}_{k},N_{k}\right),\left(\boldsymbol{B}_{\cdot k'},\boldsymbol{X}_{k'},N_{k'}\right);v\right)\right\} =O_{p}\left(\left(\frac{\mathrm{log}\left(L\right)}{Lh^{3+d}}\right)^{\nicefrac{1}{2}}+h^{R}\right),
\]
where equality holds w.p.a.1. and the equality is uniform in $v\in I\left(\boldsymbol{x}\right)$
and also 
\[
\left|\mathit{\Delta}_{6}^{\dagger}\left(v\right)\right|\lesssim h^{-2}\left\{ \underset{j,k}{\mathrm{max}}\,\mathbb{T}_{jk}\left|\widehat{V}_{jk}-V_{jk}\right|\right\} ^{2}\left\{ \frac{1}{\left(L\right)_{3}}\sum_{\left(3\right)}\mathcal{K}\left(\left(\boldsymbol{B}_{\cdot l},\boldsymbol{X}_{l},N_{l}\right),\left(\boldsymbol{B}_{\cdot k},\boldsymbol{X}_{k},N_{k}\right),\left(\boldsymbol{B}_{\cdot k'},\boldsymbol{X}_{k'},N_{k'}\right);v\right)\right\} =O_{p}\left(\frac{\mathrm{log}\left(L\right)}{Lh^{3+d}}+h^{2R}\right)
\]
where the equality holds w.p.a.1 and the equality is uniform in $v\in I\left(\boldsymbol{x}\right)$. 

Since $K_{0}'$ is compactly supported on $\left[-1,1\right]$, the
trimming is asymptotically negligible:

\[
\varDelta_{4}^{\dagger}\left(v\right)=\frac{1}{\left(L\right)_{3}}\sum_{\left(3\right)}\mathcal{H}\left(\left(\boldsymbol{B}_{\cdot l},\boldsymbol{X}_{l},N_{l}\right),\left(\boldsymbol{B}_{\cdot k},\boldsymbol{X}_{k},N_{k}\right),\left(\boldsymbol{B}_{\cdot k'},\boldsymbol{X}_{k'},N_{k'}\right);v\right),\textrm{ for all \ensuremath{v\in I\left(\boldsymbol{x}\right)}, w.p.a.1,}
\]
where

\begin{align*}
 & \mathcal{H}\left(\left(\boldsymbol{b}.,\boldsymbol{z},m\right),\left(\boldsymbol{b}'.,\boldsymbol{z}',m'\right),\left(\boldsymbol{b}.'',\boldsymbol{z}'',m''\right);v\right)\\
\coloneqq & h^{-3\left(1+d\right)}\mathbbm{1}\left(m=n\right)\frac{1}{m^{2}}\sum_{i=1}^{m}\frac{1}{m'}\sum_{j=1}^{m'}\frac{1}{m'-1}K_{f}'\left(\frac{\xi\left(b_{j}',\boldsymbol{z}',m'\right)-v}{h},\frac{\boldsymbol{z}'-\boldsymbol{x}}{h}\right)\\
 & \times\frac{G\left(b_{j}',\boldsymbol{z}',m'\right)}{g\left(b_{j}',\boldsymbol{z}',m'\right)^{2}}\mathbbm{1}\left(m=m'\right)K_{g}\left(\frac{b_{i}-b_{j}'}{h}\right)K_{\boldsymbol{X}}\left(\frac{\boldsymbol{z}-\boldsymbol{z}'}{h}\right)\frac{1}{m''}\sum_{j'=1}^{m''}\frac{1}{m''-1}\\
 & \times K_{f}'\left(\frac{\xi\left(b_{j'}'',\boldsymbol{z}'',m''\right)-v}{h},\frac{\boldsymbol{z}''-\boldsymbol{x}}{h}\right)\frac{G\left(b_{j'}'',\boldsymbol{z}'',m''\right)}{g\left(b_{j'}'',\boldsymbol{z}'',m''\right)^{2}}\mathbbm{1}\left(m=m''\right)K_{g}\left(\frac{b_{i}-b_{j'}''}{h}\right)K_{\boldsymbol{X}}\left(\frac{\boldsymbol{z}-\boldsymbol{z}''}{h}\right).
\end{align*}

The Hoeffding decomposition yields 
\begin{align*}
 & \frac{1}{\left(L\right)_{3}}\sum_{\left(3\right)}\mathcal{H}\left(\left(\boldsymbol{B}_{\cdot l},\boldsymbol{X}_{l},N_{l}\right),\left(\boldsymbol{B}_{\cdot k},\boldsymbol{X}_{k},N_{k}\right),\left(\boldsymbol{B}_{\cdot k'},\boldsymbol{X}_{k'},N_{k'}\right);v\right)\\
= & \mu_{\mathcal{H}}\left(v\right)+\frac{1}{L}\sum_{l=1}^{L}\left(\mathcal{H}_{1}^{\left(1\right)}\left(\boldsymbol{B}_{\cdot l},\boldsymbol{X}_{l},N_{l};v\right)-\mu_{\mathcal{K}}\left(v\right)\right)+\frac{1}{L}\sum_{l=1}^{L}\left(\mathcal{H}_{2}^{\left(1\right)}\left(\boldsymbol{B}_{\cdot l},\boldsymbol{X}_{l},N_{l};v\right)-\mu_{\mathcal{K}}\left(v\right)\right)\\
 & +\frac{1}{L}\sum_{l=1}^{L}\left(\mathcal{H}_{3}^{\left(1\right)}\left(\boldsymbol{B}_{\cdot l},\boldsymbol{X}_{l},N_{l};v\right)-\mu_{\mathcal{K}}\left(v\right)\right)+\Upsilon_{\mathcal{H}}^{1}\left(v\right)+\Upsilon_{\mathcal{H}}^{2}\left(v\right)+\Upsilon_{\mathcal{H}}^{3}\left(v\right)+\Psi_{\mathcal{H}}\left(v\right),
\end{align*}
where the terms in the decomposition are defined by (\ref{eq:K Hoeffding term 1})
to (\ref{eq:K Hoeffding term 5}) with $\mathcal{K}$ replaced by
$\mathcal{H}$. By the LIE, we can easily check that 
\[
\mu_{\mathcal{H}}\left(v\right)=\mathrm{V}_{\mathcal{M}}\left(v|\boldsymbol{x},n\right),
\]
\begin{eqnarray*}
\mathcal{H}_{1}^{\left(1\right)}\left(\boldsymbol{b}.,\boldsymbol{z},m;v\right) & \coloneqq & \frac{1}{\left(n-1\right)^{2}}\frac{\text{1}}{h^{3\left(1+d\right)}}\mathbbm{1}\left(m=n\right)\frac{1}{m^{2}}\sum_{i=1}^{m}\left\{ \int_{\mathcal{X}}\int_{\underline{b}\left(\boldsymbol{z}'\right)}^{\overline{b}\left(\boldsymbol{z}',n\right)}K_{f}'\left(\frac{\xi\left(b',\boldsymbol{z}',n\right)-v}{h},\frac{\boldsymbol{z}'-\boldsymbol{x}}{h}\right)\frac{G\left(b',\boldsymbol{z}',n\right)}{g\left(b',\boldsymbol{z}',n\right)}\right.\\
 &  & \left.\times K_{g}\left(\frac{b_{i}-b'}{h}\right)K_{\boldsymbol{X}}\left(\frac{\boldsymbol{z}-\boldsymbol{z}'}{h}\right)\mathrm{d}b'\mathrm{d}\boldsymbol{z}'\right\} ^{2}
\end{eqnarray*}
and
\begin{eqnarray*}
\mathcal{H}_{2}^{\left(1\right)}\left(\boldsymbol{b}.,\boldsymbol{z},m;v\right) & \coloneqq & \frac{1}{m}\sum_{i=1}^{m}\frac{1}{h^{3\left(1+d\right)}}\frac{1}{m-1}K_{f}'\left(\frac{\xi\left(b_{i},\boldsymbol{z},m\right)-v}{h},\frac{\boldsymbol{z}-\boldsymbol{x}}{h}\right)\frac{G\left(b_{i},\boldsymbol{z},m\right)}{g\left(b_{i},\boldsymbol{z},m\right)^{2}}\mathbbm{1}\left(m=n\right)\int_{\mathcal{X}}\int_{\mathcal{X}}\int_{\underline{b}\left(\boldsymbol{z}''\right)}^{\overline{b}\left(\boldsymbol{z}'',n\right)}\int_{\underline{b}\left(\boldsymbol{z}'\right)}^{\overline{b}\left(\boldsymbol{z}',n\right)}K_{g}\left(\frac{b'-b_{i}}{h}\right)\\
 &  & \times K_{\boldsymbol{X}}\left(\frac{\boldsymbol{z}'-\boldsymbol{z}}{h}\right)\frac{1}{n-1}K_{f}'\left(\frac{\xi\left(b'',\boldsymbol{z}'',n\right)-v}{h},\frac{\boldsymbol{z}''-\boldsymbol{x}}{h}\right)\frac{G\left(b'',\boldsymbol{z}'',n\right)}{g\left(b'',\boldsymbol{z}'',n\right)}K_{g}\left(\frac{b'-b''}{h}\right)K_{\boldsymbol{X}}\left(\frac{\boldsymbol{z}'-\boldsymbol{z}''}{h}\right)\\
 &  & \times g\left(b',\boldsymbol{z}',n\right)\mathrm{d}b'\mathrm{d}b''\mathrm{d}\boldsymbol{z}'\mathrm{d}\boldsymbol{z}''.
\end{eqnarray*}
We have $\mathcal{H}_{3}^{\left(1\right)}=\mathcal{H}_{2}^{\left(1\right)}$
since $\mathcal{H}$ is symmetric with respect to the second and the
third arguments.

Then it can be easily verified that 
\[
\underset{v\in I\left(\boldsymbol{x}\right)}{\mathrm{sup}}\mathrm{E}\left[\mathcal{H}_{1}^{\left(1\right)}\left(\boldsymbol{B}_{\cdot1},\boldsymbol{X}_{1},N_{1};v\right)^{2}\right]=O\left(h^{-\left(1+d\right)}\right)
\]
and 
\[
\underset{v\in I\left(\boldsymbol{x}\right)}{\mathrm{sup}}\mathrm{E}\left[\mathcal{H}_{2}^{\left(1\right)}\left(\boldsymbol{B}_{\cdot1},\boldsymbol{X}_{1},N_{1};v\right)^{2}\right]=O\left(h^{-\left(1+d\right)}\right).
\]
The bounds in (\ref{eq:K bound 1}) to (\ref{eq:K bound 7}) also
hold if $\mathcal{K}$ is replaced by $\mathcal{H}$.

Now we have shown that
\[
\underset{v\in I\left(\boldsymbol{x}\right)}{\mathrm{sup}}\left|\widehat{\pi}\left(n|\boldsymbol{x}\right)^{2}\widehat{\varphi}\left(\boldsymbol{x}\right)^{2}\widehat{\mathrm{V}}_{GPV}\left(v|\boldsymbol{x},n\right)-\mathrm{V}_{\mathcal{M}}\left(v|\boldsymbol{x},n\right)\right|=O_{p}\left(\left(\frac{\mathrm{log}\left(L\right)}{Lh^{3+d}}\right)^{\nicefrac{1}{2}}+h^{R}\right).
\]
The first conclusion of the theorem follows from the above fact, 
\begin{eqnarray*}
\left|\widehat{\mathrm{V}}_{GPV}\left(v|\boldsymbol{x},n\right)-\left(\pi\left(n|\boldsymbol{x}\right)\varphi\left(\boldsymbol{x}\right)\right)^{-2}\mathrm{V}_{\mathcal{M}}\left(v|\boldsymbol{x},n\right)\right| & \leq & \left|\frac{1}{\widehat{\pi}\left(n|\boldsymbol{x}\right)^{2}\widehat{\varphi}\left(\boldsymbol{x}\right)^{2}}-\frac{1}{\pi\left(n|\boldsymbol{x}\right)^{2}\varphi\left(\boldsymbol{x}\right)^{2}}\right|\left|\widehat{\pi}\left(n|\boldsymbol{x}\right)^{2}\widehat{\varphi}\left(\boldsymbol{x}\right)^{2}\widehat{\mathrm{V}}_{GPV}\left(v|\boldsymbol{x},n\right)\right|\\
 &  & +\left|\frac{\widehat{\pi}\left(n|\boldsymbol{x}\right)^{2}\widehat{\varphi}\left(\boldsymbol{x}\right)^{2}\widehat{\mathrm{V}}_{GPV}\left(v|\boldsymbol{x},n\right)-\mathrm{V}_{\mathcal{M}}\left(v|\boldsymbol{x},n\right)}{\pi\left(n|\boldsymbol{x}\right)^{2}\varphi\left(\boldsymbol{x}\right)^{2}}\right|
\end{eqnarray*}
and the fact
\begin{equation}
\widehat{\pi}\left(n|\boldsymbol{x}\right)-\pi\left(n|\boldsymbol{x}\right)=O_{p}\left(\left(\frac{\mathrm{log}\left(L\right)}{Lh^{d}}\right)^{\nicefrac{1}{2}}+h^{1+R}\right)\textrm{ and }\widehat{\varphi}\left(\boldsymbol{x}\right)-\varphi\left(\boldsymbol{x}\right)=O_{p}\left(\left(\frac{\mathrm{log}\left(L\right)}{Lh^{d}}\right)^{\nicefrac{1}{2}}+h^{1+R}\right).\label{eq:phi_hat - phi rate}
\end{equation}

The second conclusion of the theorem follows similarly.\end{proof}

\begin{proof}[Proof of Theorem 6.3]Note
\begin{align}
 & \sum_{n\in\mathcal{N}}\widehat{f}_{GPV}^{*}\left(v,\boldsymbol{x},n\right)-\sum_{n\in\mathcal{N}}\widehat{f}_{GPV}\left(v,\boldsymbol{x},n\right)\nonumber \\
= & \left(\sum_{n\in\mathcal{N}}\widehat{f}_{GPV}^{*}\left(v,\boldsymbol{x},n\right)-\sum_{n\in\mathcal{N}}\widetilde{f}^{*}\left(v,\boldsymbol{x},n\right)\right)+\left(\sum_{n\in\mathcal{N}}\widetilde{f}^{*}\left(v,\boldsymbol{x},n\right)-\sum_{n\in\mathcal{N}}\widetilde{f}\left(v,\boldsymbol{x},n\right)\right)-\left(\sum_{n\in\mathcal{N}}\widehat{f}_{GPV}\left(v,\boldsymbol{x},n\right)-\sum_{n\in\mathcal{N}}\widetilde{f}\left(v,\boldsymbol{x},n\right)\right).\label{eq:f_GPV_hat_star - f_GPV_hat}
\end{align}
It is shown in the proofs of Lemmas \ref{lem:Lemma 1 Stochastic Expansion}
and \ref{Lemma 2} that
\[
\widehat{f}_{GPV}\left(v,\boldsymbol{x},n\right)-\widetilde{f}\left(v,\boldsymbol{x},n\right)=\frac{1}{L^{2}}\sum_{l=1}^{L}\sum_{k=1}^{L}\mathcal{M}^{n}\left(\left(\boldsymbol{B}_{\cdot l},\boldsymbol{X}_{l},N_{l}\right),\left(\boldsymbol{B}_{\cdot k},\boldsymbol{X}_{k},N_{k}\right);v\right)+O_{p}\left(\left(\frac{\mathrm{log}\left(L\right)}{Lh^{1+d}}\right)^{\nicefrac{1}{2}}+\frac{\mathrm{log}\left(L\right)}{Lh^{3+d}}+h^{R}\right).
\]
It follows from this result, (\ref{eq:f_GPV_hat_star - f_GPV_hat}),
Lemmas \ref{Lemma 5} and \ref{Lemma 9} and \citet[Lemma S.1]{Marmer_Shneyerov_Quantile_Auctions}
that 
\begin{eqnarray*}
\sum_{n\in\mathcal{N}}\widehat{f}_{GPV}^{*}\left(v,\boldsymbol{x},n\right)-\sum_{n\in\mathcal{N}}\widehat{f}_{GPV}\left(v,\boldsymbol{x},n\right) & = & \frac{1}{L}\sum_{l=1}^{L}\sum_{n\in\mathcal{N}}\left\{ \mathcal{M}_{2}^{n}\left(\boldsymbol{B}_{\cdot l}^{*},\boldsymbol{X}_{l}^{*},N_{l}^{*};v\right)-\frac{1}{L}\sum_{k=1}^{L}\mathcal{M}_{2}^{n}\left(\boldsymbol{B}_{\cdot k},\boldsymbol{X}_{k},N_{k};v\right)\right\} \\
 &  & +O_{p}^{*}\left(\left(\frac{\mathrm{log}\left(L\right)}{Lh^{1+d}}\right)^{\nicefrac{1}{2}}+\frac{\mathrm{log}\left(L\right)}{Lh^{3+d}}+h^{R}\right).
\end{eqnarray*}
It is also shown in the proof of Lemma \ref{Lemma 10} that 
\[
\underset{v\in I\left(\boldsymbol{x}\right)}{\mathrm{sup}}\left|\frac{1}{L}\sum_{l=1}^{L}\sum_{n\in\mathcal{N}}\left\{ \mathcal{M}_{2}^{n}\left(\boldsymbol{B}_{\cdot l}^{*},\boldsymbol{X}_{l}^{*},N_{l}^{*};v\right)-\frac{1}{L}\sum_{k=1}^{L}\mathcal{M}_{2}^{n}\left(\boldsymbol{B}_{\cdot k},\boldsymbol{X}_{k},N_{k};v\right)\right\} \right|=O_{p}^{*}\left(\left(\frac{\mathrm{log}\left(L\right)}{Lh^{3+d}}\right)^{\nicefrac{1}{2}}\right).
\]
Note that by \citet[Lemmas S.1 and S.4]{Marmer_Shneyerov_Quantile_Auctions},
we have
\[
\widehat{\varphi}^{*}\left(\boldsymbol{x}\right)-\varphi\left(\boldsymbol{x}\right)=O_{p}^{*}\left(\left(\frac{\mathrm{log}\left(L\right)}{Lh^{d}}\right)^{\nicefrac{1}{2}}+h^{1+R}\right).
\]

Now using these results, (\ref{eq:abc identity}), (\ref{eq:phi_hat - phi rate})
and the fact
\[
\sum_{n\in\mathcal{N}}\widehat{f}_{GPV}\left(v,\boldsymbol{x},n\right)\rightarrow_{p}f\left(v|\boldsymbol{x}\right)\varphi\left(\boldsymbol{x}\right),\textrm{ uniformly in \ensuremath{v\in I\left(\boldsymbol{x}\right)}},
\]
we can obtain 
\begin{eqnarray*}
\widehat{f}_{GPV}^{*}\left(v|\boldsymbol{x}\right)-\widehat{f}_{GPV}\left(v|\boldsymbol{x}\right) & = & \frac{1}{\widehat{\varphi}^{*}\left(\boldsymbol{x}\right)}\sum_{n\in\mathcal{N}}\widehat{f}_{GPV}^{*}\left(v,\boldsymbol{x},n\right)-\frac{1}{\widehat{\varphi}\left(\boldsymbol{x}\right)}\sum_{n\in\mathcal{N}}\widehat{f}_{GPV}\left(v,\boldsymbol{x},n\right)\\
 & = & \frac{1}{L}\sum_{l=1}^{L}\frac{1}{\varphi\left(\boldsymbol{x}\right)}\sum_{n\in\mathcal{N}}\left\{ \mathcal{M}_{2}^{n}\left(\boldsymbol{B}_{\cdot l}^{*},\boldsymbol{X}_{l}^{*},N_{l}^{*};v\right)-\frac{1}{L}\sum_{k=1}^{L}\mathcal{M}_{2}^{n}\left(\boldsymbol{B}_{\cdot k},\boldsymbol{X}_{k},N_{k};v\right)\right\} \\
 &  & +O_{p}^{*}\left(\left(\frac{\mathrm{log}\left(L\right)}{Lh^{1+d}}\right)^{\nicefrac{1}{2}}+\frac{\mathrm{log}\left(L\right)}{Lh^{3+d}}+h^{R}\right).
\end{eqnarray*}
Similarly, we have 
\[
\widehat{f}_{GPV}\left(v|\boldsymbol{x}\right)-f\left(v|\boldsymbol{x}\right)=\frac{1}{L}\sum_{l=1}^{L}\frac{1}{\varphi\left(\boldsymbol{x}\right)}\sum_{n\in\mathcal{N}}\left\{ \mathcal{M}_{2}^{n}\left(\boldsymbol{B}_{\cdot l},\boldsymbol{X}_{l},N_{l};v\right)-\mu_{\mathcal{M}^{n}}\left(v\right)\right\} +O_{p}\left(\left(\frac{\mathrm{log}\left(L\right)}{Lh^{1+d}}\right)^{\nicefrac{1}{2}}+\frac{\mathrm{log}\left(L\right)}{Lh^{3+d}}+h^{R}\right).
\]

Now it follows that
\begin{equation}
\left(Lh^{3+d}\right)^{\nicefrac{1}{2}}\left(\widehat{f}_{GPV}\left(v|\boldsymbol{x}\right)-f\left(v|\boldsymbol{x}\right)\right)=\frac{1}{L^{\nicefrac{1}{2}}}\sum_{l=1}^{L}\frac{1}{\varphi\left(\boldsymbol{x}\right)}\sum_{n\in\mathcal{N}}h^{\nicefrac{\left(3+d\right)}{2}}\left\{ \mathcal{M}_{2}^{n}\left(\boldsymbol{B}_{\cdot l},\boldsymbol{X}_{l},N_{l};v\right)-\mu_{\mathcal{M}^{n}}\left(v\right)\right\} +o_{p}\left(1\right),\label{eq:(Lh^3+d)(f_hat - f) leading term}
\end{equation}
where the leading term converges in distribution to $N\left(0,\mathrm{V}_{GPV}\left(v|\boldsymbol{x}\right)\right)$
and 
\[
\left(Lh^{3+d}\right)^{\nicefrac{1}{2}}\left(\widehat{f}_{GPV}^{*}\left(v|\boldsymbol{x}\right)-\widehat{f}_{GPV}\left(v|\boldsymbol{x}\right)\right)=\frac{1}{L^{\nicefrac{1}{2}}}\sum_{l=1}^{L}\frac{1}{\varphi\left(\boldsymbol{x}\right)}\sum_{n\in\mathcal{N}}h^{\nicefrac{\left(3+d\right)}{2}}\left\{ \mathcal{M}_{2}^{n}\left(\boldsymbol{B}_{\cdot l}^{*},\boldsymbol{X}_{l}^{*},N_{l}^{*};v\right)-\frac{1}{L}\sum_{k=1}^{L}\mathcal{M}_{2}^{n}\left(\boldsymbol{B}_{\cdot k},\boldsymbol{X}_{k},N_{k};v\right)\right\} +o_{p}^{*}\left(1\right),
\]
where the leading term is the bootstrap analogue of the leading term
on the right hand side of (\ref{eq:(Lh^3+d)(f_hat - f) leading term}).
The desired result follows from this observation and the arguments
used in the proof of Theorem 4.2. \end{proof}

\begin{proof}[Proof of Theorem 6.4]It is shown in the proof of Theorem
6.1 that when $h$ is sufficiently small,
\begin{equation}
\underset{v\in I\left(\boldsymbol{x}\right)}{\mathrm{inf}}\mathrm{Var}\left[h^{\nicefrac{\left(3+d\right)}{2}}\sum_{n\in\mathcal{N}}\mathcal{M}_{2}^{\ddagger,n}\left(\boldsymbol{B}_{\cdot1},\boldsymbol{X}_{1},N_{1};v\right)\right]>C_{1}>0.\label{eq:Variance M_ddagger bounded awary from zero}
\end{equation}
Since it is argued in the proof of Lemma \ref{Lemma 4} that $\left\{ \mathcal{M}_{2}^{\ddagger,n}\left(\cdot;v\right):v\in I\left(\boldsymbol{x}\right)\right\} $
is uniformly VC-type with respect to the envelope (\ref{eq:M_ddagger_2 envelope})
which satisfies $\left\Vert F_{\mathcal{M}_{2}^{\ddagger,n}}\right\Vert _{\mathcal{X}}=O\left(h^{-\left(3+d\right)}\right)$,
it follows from \citet[Lemma 16]{nolan1987u} and (\ref{eq:Variance M_ddagger bounded awary from zero})
that the class 
\[
\left\{ \frac{h^{\nicefrac{\left(3+d\right)}{2}}\sum_{n\in\mathcal{N}}\left(\mathcal{M}_{2}^{\ddagger,n}\left(\cdot;v\right)-\mu_{\mathcal{M}^{\ddagger,n}}\left(v\right)\right)}{\mathrm{Var}\left[h^{\nicefrac{\left(3+d\right)}{2}}\sum_{n\in\mathcal{N}}\mathcal{M}_{2}^{\ddagger,n}\left(\boldsymbol{B}_{\cdot1},\boldsymbol{X}_{1},N_{1};v\right)\right]^{\nicefrac{1}{2}}}\right\} ,\,v\in I\left(\boldsymbol{x}\right),
\]
is uniformly VC-type with respect to an envelope that is a multiple
of $h^{\nicefrac{\left(3+d\right)}{2}}\sum_{n\in\mathcal{N}}F_{\mathcal{M}_{2}^{\ddagger,n}}$,
which satisfies 
\[
\left\Vert h^{\nicefrac{\left(3+d\right)}{2}}\sum_{n\in\mathcal{N}}F_{\mathcal{M}_{2}^{\ddagger,n}}\right\Vert _{\infty}=O\left(h^{-\nicefrac{\left(3+d\right)}{2}}\right),
\]
when $h$ is sufficiently small.

By the arguments used in the proof of Theorem 5.1, we can show the
existence of a tight Gaussian random element in $\ell^{\infty}\left(I\left(\boldsymbol{x}\right)\right)$,
denoted by $\left\{ \varGamma_{G}\left(v|\boldsymbol{x}\right):v\in I\left(\boldsymbol{x}\right)\right\} $
that has the same covariance structure as the empirical process $\left\{ \varGamma\left(v|\boldsymbol{x}\right):v\in I\left(\boldsymbol{x}\right)\right\} $.
Application of \citet[Corollary 2.2]{chernozhukov2014gaussian} with
$q=\infty$, $b\apprle h^{-\nicefrac{\left(3+d\right)}{2}}$, $\gamma=\mathrm{log}\left(L\right)^{-1}$
and $\sigma=1$ yields the existence of a coupling $W_{L}$ with $W_{L}\overset{d}{=}\left\Vert \varGamma_{G}\left(\cdot|\boldsymbol{x}\right)\right\Vert _{I\left(\boldsymbol{x}\right)}$
satisfying 
\[
\left|\left\Vert \varGamma\left(\cdot|\boldsymbol{x}\right)\right\Vert _{I\left(\boldsymbol{x}\right)}-W_{L}\right|=O_{p}\left(\frac{\mathrm{log}\left(L\right)}{\left(Lh^{3+d}\right)^{\nicefrac{1}{6}}}\right).
\]
This result and Lemma \ref{Lemma 4} yield 
\begin{equation}
\left|\left\Vert Z\left(\cdot|\boldsymbol{x}\right)\right\Vert _{I\left(\boldsymbol{x}\right)}-W_{L}\right|=O_{p}\left(\lambda_{L}^{*}\right),\label{eq:sup Z - W bound}
\end{equation}
where 
\[
\lambda_{L}^{*}\coloneqq\mathrm{log}\left(L\right)^{\nicefrac{1}{2}}h+\frac{\mathrm{log}\left(L\right)}{\left(Lh^{3+d}\right)^{\nicefrac{1}{6}}}+L^{\nicefrac{1}{2}}h^{\nicefrac{\left(3+d\right)}{2}+R}.
\]
Then by applying Lemma A.6, (\ref{eq:sup Z - W bound}) and the Gaussian
anti-concentration inequality of \citet{chernozhukov2014anti}, we
have
\begin{equation}
\underset{z\in\mathbb{R}}{\mathrm{sup}}\left|\mathrm{P}\left[\left\Vert Z\left(\cdot|\boldsymbol{x}\right)\right\Vert _{I\left(\boldsymbol{x}\right)}\leq z\right]-\mathrm{P}\left[\left\Vert \varGamma_{G}\left(\cdot|\boldsymbol{x}\right)\right\Vert _{I\left(\boldsymbol{x}\right)}\leq z\right]\right|=o\left(1\right).\label{eq:Theorem 5.5 Part 1}
\end{equation}

Application of \citet[Theorem 2.3]{chernozhukov2016empirical} with
$B\left(f\right)=0$, $q=\infty$, $b\apprle h^{-\nicefrac{\left(3+d\right)}{2}}$,
$\gamma=\mathrm{log}\left(L\right)^{-1}$ and $\sigma=1$ yields that
there exists a coupling $W_{L}^{*}$, with the property that the conditional
distribution of $W_{L}^{*}$ given the original sample is the same
as the marginal distribution of $\left\Vert \varGamma_{G}\left(\cdot|\boldsymbol{x}\right)\right\Vert _{I\left(\boldsymbol{x}\right)}$
almost surely, and 
\[
\left|\left\Vert \varGamma^{*}\left(\cdot|\boldsymbol{x}\right)\right\Vert _{I\left(\boldsymbol{x}\right)}-W_{L}^{*}\right|=O_{p}\left(\frac{\mathrm{log}\left(L\right)}{\left(Lh^{3+d}\right)^{\nicefrac{1}{6}}}\right).
\]
This result, Lemma \ref{Lemma 10} and Markov's inequality yield 
\[
\left|\left\Vert Z^{*}\left(\cdot|\boldsymbol{x}\right)\right\Vert _{I\left(\boldsymbol{x}\right)}-W_{L}^{*}\right|=O_{p}^{*}\left(\lambda_{L}^{*}\right).
\]

Next we apply the arguments used in the proof of Theorem 5.2. It follows
from the above displayed result, the Gaussian anti-concentration inequality
of \citet{chernozhukov2014anti} and Lemma A.8 that 
\begin{equation}
\underset{z\in\mathbb{R}}{\mathrm{sup}}\left|\mathrm{P}^{*}\left[\left\Vert Z^{*}\left(\cdot|\boldsymbol{x}\right)\right\Vert _{I\left(\boldsymbol{x}\right)}\leq z\right]-\mathrm{P}\left[\left\Vert \varGamma_{G}\left(\cdot|\boldsymbol{x}\right)\right\Vert _{I\left(\boldsymbol{x}\right)}\leq z\right]\right|=o_{p}\left(1\right).\label{eq:Theorem 5.5 Part 2}
\end{equation}

The conclusion follows from (\ref{eq:Theorem 5.5 Part 1}), (\ref{eq:Theorem 5.5 Part 2})
and the arguments used in the proof of Corollary 5.1.\end{proof}

\subsection{Lemmas}

\begin{lem}
\label{lem:Lemma 1 Stochastic Expansion}Suppose that Assumptions
1 - 3 hold. Let $\boldsymbol{x}$ be an interior point of $\mathcal{X}$
and $n\in\mathcal{N}$ be fixed. Let
\[
\widetilde{\mathbb{T}}_{il}\coloneqq\mathbbm{1}\left(\left(V_{il},\boldsymbol{X}_{l}\right)\in\mathbb{H}\left(\left(v,\boldsymbol{x}\right),\overline{\delta}\right)\right).
\]
Then
\begin{eqnarray*}
\widehat{f}_{GPV}\left(v,\boldsymbol{x},n\right)-f\left(v|\boldsymbol{x}\right)\varphi\left(\boldsymbol{x}\right)\pi\left(n|\boldsymbol{x}\right) & = & \frac{1}{L}\sum_{l=1}^{L}\mathbbm{1}\left(N_{l}=n\right)\frac{1}{N_{l}}\sum_{i=1}^{N_{l}}\widetilde{\mathbb{T}}_{il}\frac{1}{h^{2+d}}K_{f}^{\prime}\left(\frac{V_{il}-v}{h},\frac{\boldsymbol{X}_{l}-\boldsymbol{x}}{h}\right)\left(\widehat{V}_{il}-V_{il}\right)\\
 &  & +O_{p}\left(\left(\frac{\mathrm{log}\left(L\right)}{Lh^{1+d}}\right)^{\nicefrac{1}{2}}+\frac{\mathrm{log}\left(L\right)}{Lh^{3+d}}+h^{R}\right),
\end{eqnarray*}
where the remainder term is uniform in $v\in I\left(\boldsymbol{x}\right)$.
\end{lem}
\begin{proof}[Proof of Lemma \ref{lem:Lemma 1 Stochastic Expansion}]
It is clear from the proof of Lemma B2 of GPV that
\begin{gather}
\underset{n'\in\mathcal{N}}{\mathrm{max}}\underset{\mathbb{H}\left(\left(b',\boldsymbol{x}'\right),h\right)\subseteq\mathcal{S}_{B,\boldsymbol{X}}^{n'}}{\mathrm{sup}}\left|\widehat{G}\left(b',\boldsymbol{x}',n'\right)-G\left(b',\boldsymbol{x}',n'\right)\right|=O_{p}\left(\left(\frac{\mathrm{log}\left(L\right)}{Lh^{d}}\right)^{\nicefrac{1}{2}}+h^{1+R}\right)\nonumber \\
\underset{n'\in\mathcal{N}}{\mathrm{max}}\underset{\mathbb{H}\left(\left(b',\boldsymbol{x}'\right),h\right)\subseteq\mathcal{S}_{B,\boldsymbol{X}}^{n'}}{\mathrm{sup}}\left|\widehat{g}\left(b',\boldsymbol{x}',n'\right)-g\left(b',\boldsymbol{x}',n'\right)\right|=O_{p}\left(\left(\frac{\mathrm{log}\left(L\right)}{Lh^{1+d}}\right)^{\nicefrac{1}{2}}+h^{1+R}\right).\label{eq:uniform convergence rate}
\end{gather}

It follows from 
\begin{gather*}
V_{il}=\xi\left(B_{il},\boldsymbol{X}_{l},N_{l}\right)\coloneqq B_{il}+\frac{1}{N_{l}-1}\frac{G\left(B_{il},\boldsymbol{X}_{l},N_{l}\right)}{g\left(B_{il},\boldsymbol{X}_{l},N_{l}\right)}\\
\widehat{V}_{il}\coloneqq\widehat{\xi}\left(B_{il},\boldsymbol{X}_{l},N_{l}\right)\coloneqq B_{il}+\frac{1}{N_{l}-1}\frac{\widehat{G}\left(B_{il},\boldsymbol{X}_{l},N_{l}\right)}{\widehat{g}\left(B_{il},\boldsymbol{X}_{l},N_{l}\right)}
\end{gather*}
and (\ref{eq:abc identity}) that
\begin{eqnarray}
\widehat{V}_{il}-V_{il} & = & \frac{1}{N_{l}-1}\left(\frac{\widehat{G}\left(B_{il},\boldsymbol{X}_{l},N_{l}\right)-G\left(B_{il},\boldsymbol{X}_{l},N_{l}\right)}{g\left(B_{il},\boldsymbol{X}_{l},N_{l}\right)}-\frac{\widehat{G}\left(B_{il},\boldsymbol{X}_{l},N_{l}\right)\left(\widehat{g}\left(B_{il},\boldsymbol{X}_{l},N_{l}\right)-g\left(B_{il},\boldsymbol{X}_{l},N_{l}\right)\right)}{g\left(B_{il},\boldsymbol{X}_{l},N_{l}\right)^{2}}\right.\nonumber \\
 &  & \left.+\frac{\widehat{G}\left(B_{il},\boldsymbol{X}_{l},N_{l}\right)}{\widehat{g}\left(B_{il},\boldsymbol{X}_{l},N_{l}\right)}\frac{\left(\widehat{g}\left(B_{il},\boldsymbol{X}_{l},N_{l}\right)-g\left(B_{il},\boldsymbol{X}_{l},N_{l}\right)\right)^{2}}{g\left(B_{il},\boldsymbol{X}_{l},N_{l}\right)^{2}}\right).\label{eq:V_hat - V expansion}
\end{eqnarray}
It then follows from the triangle inequality that
\begin{eqnarray}
\underset{i,l}{\mathrm{max}}\,\mathbb{T}_{il}\left|\widehat{V}_{il}-V_{il}\right| & \leq & \underset{i,l}{\mathrm{max}}\,\mathbb{T}_{il}\left|\frac{\widehat{G}\left(B_{il},\boldsymbol{X}_{l},N_{l}\right)-G\left(B_{il},\boldsymbol{X}_{l},N_{l}\right)}{g\left(B_{il},\boldsymbol{X}_{l},N_{l}\right)}\right|+\underset{i,l}{\mathrm{max}}\,\mathbb{T}_{il}\left|\frac{\widehat{G}\left(B_{il},\boldsymbol{X}_{l},N_{l}\right)\left(\widehat{g}\left(B_{il},\boldsymbol{X}_{l},N_{l}\right)-g\left(B_{il},\boldsymbol{X}_{l},N_{l}\right)\right)}{g\left(B_{il},\boldsymbol{X}_{l},N_{l}\right)^{2}}\right|\nonumber \\
 &  & +\underset{i,l}{\mathrm{max}}\,\mathbb{T}_{il}\left|\frac{\widehat{G}\left(B_{il},\boldsymbol{X}_{l},N_{l}\right)}{\widehat{g}\left(B_{il},\boldsymbol{X}_{l},N_{l}\right)}\frac{\left(\widehat{g}\left(B_{il},\boldsymbol{X}_{l},N_{l}\right)-g\left(B_{il},\boldsymbol{X}_{l},N_{l}\right)\right)^{2}}{g\left(B_{il},\boldsymbol{X}_{l},N_{l}\right)^{2}}\right|.\label{eq:sup V_hat - V decomposition}
\end{eqnarray}

Denote
\[
\overline{\mathbb{T}}_{il}\coloneqq\mathbbm{1}\left(\mathbb{H}\left(\left(B_{il},\boldsymbol{X}_{l}\right),h\right)\subseteq\mathcal{S}_{B,\boldsymbol{X}}^{N_{l}}\right).
\]
For the first term of the right hand side of (\ref{eq:sup V_hat - V decomposition}),
we have
\begin{align}
 & \underset{i,l}{\mathrm{max}}\,\mathbb{T}_{il}\left|\frac{\widehat{G}\left(B_{il},\boldsymbol{X}_{l},N_{l}\right)-G\left(B_{il},\boldsymbol{X}_{l},N_{l}\right)}{g\left(B_{il},\boldsymbol{X}_{l},N_{l}\right)}\right|\nonumber \\
= & \underset{i,l}{\mathrm{max}}\,\mathbb{T}_{il}\overline{\mathbb{T}}_{il}\left|\frac{\widehat{G}\left(B_{il},\boldsymbol{X}_{l},N_{l}\right)-G\left(B_{il},\boldsymbol{X}_{l},N_{l}\right)}{g\left(B_{il},\boldsymbol{X}_{l},N_{l}\right)}\right|+\underset{i,l}{\mathrm{max}}\,\mathbb{T}_{il}\left(1-\overline{\mathbb{T}}_{il}\right)\left|\frac{\widehat{G}\left(B_{il},\boldsymbol{X}_{l},N_{l}\right)-G\left(B_{il},\boldsymbol{X}_{l},N_{l}\right)}{g\left(B_{il},\boldsymbol{X}_{l},N_{l}\right)}\right|\label{eq:Lemma 1 order bound 1}
\end{align}
and by (\ref{eq:bids density bounded away from zero}) and (\ref{eq:uniform convergence rate}),
we have
\begin{eqnarray*}
\underset{i,l}{\mathrm{max}}\,\mathbb{T}_{il}\overline{\mathbb{T}}_{il}\left|\frac{\widehat{G}\left(B_{il},\boldsymbol{X}_{l},N_{l}\right)-G\left(B_{il},\boldsymbol{X}_{l},N_{l}\right)}{g\left(B_{il},\boldsymbol{X}_{l},N_{l}\right)}\right| & \apprle & \underset{i,l}{\mathrm{max}}\,\overline{\mathbb{T}}_{il}\left|\widehat{G}\left(B_{il},\boldsymbol{X}_{l},N_{l}\right)-G\left(B_{il},\boldsymbol{X}_{l},N_{l}\right)\right|\\
 & = & O_{p}\left(\left(\frac{\mathrm{log}\left(L\right)}{Lh^{d}}\right)^{\nicefrac{1}{2}}+h^{1+R}\right).
\end{eqnarray*}

For the second term of the right hand side of (\ref{eq:Lemma 1 order bound 1}),
we first note that $\mathbb{H}\left(\left(B_{il},\boldsymbol{X}_{l}\right),2h\right)\subseteq\mathcal{\widehat{S}}_{B,\boldsymbol{X}}^{N_{l}}$
if and only if $\mathbb{H}\left(\boldsymbol{X}_{l},2h\right)\subseteq\mathcal{X}$
and for all $\boldsymbol{x}'\in\mathbb{H}\left(\boldsymbol{X}_{l},2h\right)$,
$B_{il}+2h\leq\widehat{\overline{b}}\left(\boldsymbol{x}',N_{l}\right)$
and $B_{il}-2h\geq\widehat{\underline{b}}\left(\boldsymbol{x}'\right)$.
Proposition 2 of GPV gives that
\begin{equation}
\underset{\left(\boldsymbol{x}',n'\right)\in\mathcal{X}\times\mathcal{N}}{\mathrm{sup}}\left|\widehat{\overline{b}}\left(\boldsymbol{x}',n'\right)-\overline{b}\left(\boldsymbol{x}',n'\right)\right|=o_{p}\left(h\right)\textrm{ and }\underset{\boldsymbol{x}'\in\mathcal{X}}{\mathrm{sup}}\left|\widehat{\underline{b}}\left(\boldsymbol{x}'\right)-\underline{b}\left(\boldsymbol{x}'\right)\right|=o_{p}\left(h\right).\label{eq:boundary estimator rate}
\end{equation}
For any $\left(i,l\right)$, if $\mathbb{H}\left(\left(B_{il},\boldsymbol{X}_{l}\right),2h\right)\subseteq\mathcal{\widehat{S}}_{B,\boldsymbol{X}}^{N_{l}}$,
we have for all $\boldsymbol{x}'\in\mathbb{H}\left(\boldsymbol{X}_{l},h\right)$,
\[
B_{il}+h\leq\overline{b}\left(\boldsymbol{x}',N_{l}\right)+\left(\underset{\left(\boldsymbol{z},n'\right)\in\mathcal{X}\times\mathcal{N}}{\mathrm{sup}}\left|\widehat{\overline{b}}\left(\boldsymbol{z},n'\right)-\overline{b}\left(\boldsymbol{z},n'\right)\right|-h\right)\textrm{ and }B_{il}-h\geq\underline{b}\left(\boldsymbol{x}'\right)-\left(\underset{\boldsymbol{z}\in\mathcal{X}}{\mathrm{sup}}\left|\widehat{\underline{b}}\left(\boldsymbol{z}\right)-\underline{b}\left(\boldsymbol{z}\right)\right|-h\right).
\]
Therefore, if $\underset{\left(\boldsymbol{z},n'\right)\in\mathcal{X}\times\mathcal{N}}{\mathrm{sup}}\left|\widehat{\overline{b}}\left(\boldsymbol{z},n'\right)-\overline{b}\left(\boldsymbol{z},n'\right)\right|\leq h$,
$\underset{\boldsymbol{z}\in\mathcal{X}}{\mathrm{sup}}\left|\widehat{\underline{b}}\left(\boldsymbol{z}\right)-\underline{b}\left(\boldsymbol{z}\right)\right|\leq h$
and $\mathbb{H}\left(\left(B_{il},\boldsymbol{X}_{l}\right),2h\right)\subseteq\mathcal{\widehat{S}}_{B,\boldsymbol{X}}^{N_{l}}$,
we must have $\mathbb{H}\left(\left(B_{il},\boldsymbol{X}_{l}\right),h\right)\subseteq\mathcal{S}_{B,X}^{N_{l}}$.
Now it is clear that for sufficiently small $h$,
\[
\mathrm{P}\left[\underset{i,l}{\mathrm{max}}\,\mathbb{T}_{il}\left(1-\overline{\mathbb{T}}_{il}\right)=0\right]\geq\mathrm{P}\left[\underset{\left(\boldsymbol{z},n'\right)\in\mathcal{X}\times\mathcal{N}}{\mathrm{sup}}\left|\widehat{\overline{b}}\left(\boldsymbol{z},n'\right)-\overline{b}\left(\boldsymbol{z},n'\right)\right|\leq h\textrm{ and }\underset{\boldsymbol{z}\in\mathcal{X}}{\mathrm{sup}}\left|\widehat{\underline{b}}\left(\boldsymbol{z}\right)-\underline{b}\left(\boldsymbol{z}\right)\right|\leq h\right].
\]
By using (\ref{eq:boundary estimator rate}), we have $\underset{i,l}{\mathrm{max}}\,\mathbb{T}_{il}\left(1-\overline{\mathbb{T}}_{il}\right)=0$,
w.p.a.1. Therefore we have
\begin{equation}
\underset{i,l}{\mathrm{max}}\,\mathbb{T}_{il}\left|\frac{\widehat{G}\left(B_{il},\boldsymbol{X}_{l},N_{l}\right)-G\left(B_{il},\boldsymbol{X}_{l},N_{l}\right)}{g\left(B_{il},\boldsymbol{X}_{l},N_{l}\right)}\right|=O_{p}\left(\left(\frac{\mathrm{log}\left(L\right)}{Lh^{d}}\right)^{\nicefrac{1}{2}}+h^{1+R}\right).\label{eq:sup V_hat - V first term order bound}
\end{equation}

Similarly, by using (\ref{eq:uniform convergence rate}) and the triangle
inequality, we have
\begin{align}
 & \underset{i,l}{\mathrm{max}}\,\mathbb{T}_{il}\left|\frac{\widehat{G}\left(B_{il},\boldsymbol{X}_{l},N_{l}\right)\left(\widehat{g}\left(B_{il},\boldsymbol{X}_{l},N_{l}\right)-g\left(B_{il},\boldsymbol{X}_{l},N_{l}\right)\right)}{g\left(B_{il},\boldsymbol{X}_{l},N_{l}\right)^{2}}\right|\nonumber \\
\apprle & \underset{i,l}{\mathrm{max}}\,\overline{\mathbb{T}}_{il}\left|\left(\widehat{G}\left(B_{il},\boldsymbol{X}_{l},N_{l}\right)-G\left(B_{il},\boldsymbol{X}_{l},N_{l}\right)\right)\left(\widehat{g}\left(B_{il},\boldsymbol{X}_{l},N_{l}\right)-g\left(B_{il},\boldsymbol{X}_{l},N_{l}\right)\right)\right|+\underset{i,l}{\mathrm{max}}\,\overline{\mathbb{T}}_{il}\left|\widehat{g}\left(B_{il},\boldsymbol{X}_{l},N_{l}\right)-g\left(B_{il},\boldsymbol{X}_{l},N_{l}\right)\right|\nonumber \\
= & O_{p}\left(\left(\frac{\mathrm{log}\left(L\right)}{Lh^{1+d}}\right)^{\nicefrac{1}{2}}+h^{1+R}\right),\label{eq:sup V_hat - V second term order bound}
\end{align}
where the inequality holds w.p.a.1.

For the third term of the right hand side of (\ref{eq:sup V_hat - V decomposition}),
it follows from
\[
\underset{i,l}{\mathrm{max}}\,\overline{\mathbb{T}}_{il}\left|\widehat{g}\left(B_{il},\boldsymbol{X}_{l},N_{l}\right)-g\left(B_{il},\boldsymbol{X}_{l},N_{l}\right)\right|=o_{p}\left(1\right)
\]
and (\ref{eq:bids density bounded away from zero}) that 
\[
\underset{i,l}{\mathrm{max}}\,\overline{\mathbb{T}}_{il}\widehat{g}\left(B_{il},\boldsymbol{X}_{l},N_{l}\right)^{-1}\leq\left(\frac{\underline{C}_{g}}{2}\right)^{-1}\textrm{ w.p.a.1}
\]
and therefore we have
\begin{align*}
 & \underset{i,l}{\mathrm{max}}\,\mathbb{T}_{il}\left|\frac{\widehat{G}\left(B_{il},\boldsymbol{X}_{l},N_{l}\right)}{\widehat{g}\left(B_{il},\boldsymbol{X}_{l},N_{l}\right)}\frac{\left(\widehat{g}\left(B_{il},\boldsymbol{X}_{l},N_{l}\right)-g\left(B_{il},\boldsymbol{X}_{l},N_{l}\right)\right)^{2}}{g\left(B_{il},\boldsymbol{X}_{l},N_{l}\right)^{2}}\right|\\
\apprle & \underset{i,l}{\mathrm{max}}\,\overline{\mathbb{T}}_{il}\left|\left(\widehat{G}\left(B_{il},\boldsymbol{X}_{l},N_{l}\right)-G\left(B_{il},\boldsymbol{X}_{l},N_{l}\right)\right)\left(\widehat{g}\left(B_{il},\boldsymbol{X}_{l},N_{l}\right)-g\left(B_{il},\boldsymbol{X}_{l},N_{l}\right)\right)^{2}\right|+\underset{i,l}{\mathrm{max}}\,\overline{\mathbb{T}}_{il}\left(\widehat{g}\left(B_{il},\boldsymbol{X}_{l},N_{l}\right)-g\left(B_{il},\boldsymbol{X}_{l},N_{l}\right)\right)^{2}\\
= & O_{p}\left(\frac{\mathrm{log}\left(L\right)}{Lh^{1+d}}+h^{2+2R}\right),
\end{align*}
where the inequality holds w.p.a.1. Therefore it follows from (\ref{eq:sup V_hat - V decomposition}),
(\ref{eq:sup V_hat - V first term order bound}), (\ref{eq:sup V_hat - V second term order bound})
and the above result that
\begin{equation}
\underset{i,l}{\mathrm{max}}\,\mathbb{T}_{il}\left|\widehat{V}_{il}-V_{il}\right|=O_{p}\left(\left(\frac{\mathrm{log}\left(L\right)}{Lh^{1+d}}\right)^{\nicefrac{1}{2}}+h^{1+R}\right).\label{eq:V_hat - V Feasible trimming}
\end{equation}

Note that $\mathbb{H}\left(\left(B_{il},\boldsymbol{X}_{l}\right),h\right)\subseteq\mathcal{S}_{B,\boldsymbol{X}}^{N_{l}}$
if and only if $\mathbb{H}\left(\boldsymbol{X}_{l},h\right)\subseteq\mathcal{X}$
and for all $\boldsymbol{x}'\in\mathbb{H}\left(\boldsymbol{X}_{l},h\right)$,
$B_{il}+h\leq\overline{b}\left(\boldsymbol{x}',N_{l}\right)$ and
$B_{il}-h\geq\underline{b}\left(\boldsymbol{x}'\right)$. Now we can
show that when $h$ is sufficiently small, for every $v\in I\left(\boldsymbol{x}\right)$,
$\left(V_{il},\boldsymbol{X}_{l}\right)\in\mathbb{H}\left(\left(v,\boldsymbol{x}\right),\overline{\delta}\right)$
implies $\mathbb{H}\left(\left(B_{il},\boldsymbol{X}_{l}\right),h\right)\subseteq\mathcal{S}_{B,\boldsymbol{X}}^{N_{l}}$.
When $h<\overline{\delta}$ and $\left(V_{il},\boldsymbol{X}_{l}\right)\in\mathbb{H}\left(\left(v,\boldsymbol{x}\right),\overline{\delta}\right)$,
clearly, for any $n'\in\mathcal{N}$ we have $\mathbb{H}\left(\boldsymbol{X}_{l},h\right)\subseteq\mathbb{H}\left(\boldsymbol{x},\underline{\delta}_{n'}\right)\subseteq\mathcal{X}$
since for any $\boldsymbol{x}'\in\mathbb{H}\left(\boldsymbol{X}_{l},h\right)$
we have $X_{l}\in\mathbb{H}\left(\boldsymbol{x},\nicefrac{\underline{\delta}_{n'}}{2}\right)$
by assumption and thus $\boldsymbol{x}'\in\mathbb{H}\left(\boldsymbol{x},\underline{\delta}_{n'}\right)$
by the triangle inequality. By the definition of $\underline{\delta}_{n'}$,
since $\left|V_{il}-v\right|\leq\delta_{0}$ and $v\in I\left(\boldsymbol{x}\right)$,
we have 
\[
\underline{b}\left(\boldsymbol{x}'\right)+\underline{\delta}_{n'}^{\dagger}<s\left(V_{il},\boldsymbol{X}_{l},n'\right),
\]
for all $\boldsymbol{x}'\in\mathbb{H}\left(\boldsymbol{X}_{l},h\right)$.
Since this holds for all $n'\in\mathcal{N}$, we have $B_{il}>\underline{b}\left(\boldsymbol{x}'\right)+h$,
for all $\boldsymbol{x}'\in\mathbb{H}\left(\boldsymbol{X}_{l},h\right)$,
when $h<\mathrm{min}\left\{ \underline{\delta}_{\underline{n}}^{\dagger},...,\overline{\delta}_{\overline{n}}^{\dagger}\right\} $.
Similarly, we have $B_{il}+h<\overline{b}\left(\boldsymbol{x}',N_{l}\right)$
when $h$ is sufficiently small. Therefore we have
\begin{equation}
\underset{v\in I\left(\boldsymbol{x}\right)}{\mathrm{sup}}\underset{i,l}{\mathrm{max}}\,\widetilde{\mathbb{T}}_{il}\left|\widehat{V}_{il}-V_{il}\right|=O_{p}\left(\left(\frac{\mathrm{log}\left(L\right)}{Lh^{1+d}}\right)^{\nicefrac{1}{2}}+h^{1+R}\right).\label{eq:V_hat - V Infeasible Trimming}
\end{equation}

Write
\begin{align*}
 & \frac{1}{L}\sum_{l=1}^{L}\mathbbm{1}\left(N_{l}=n\right)\frac{1}{N_{l}}\sum_{i=1}^{N_{l}}\mathbb{T}_{il}\frac{1}{h^{1+d}}K_{f}\left(\frac{\widehat{V}_{il}-v}{h},\frac{\boldsymbol{X}_{l}-\boldsymbol{x}}{h}\right)\\
= & \frac{1}{L}\sum_{l=1}^{L}\mathbbm{1}\left(N_{l}=n\right)\frac{1}{N_{l}}\sum_{i=1}^{N_{l}}\widetilde{\mathbb{T}}_{il}\frac{1}{h^{1+d}}K_{f}\left(\frac{\widehat{V}_{il}-v}{h},\frac{\boldsymbol{X}_{l}-\boldsymbol{x}}{h}\right)+\kappa_{1}^{\dagger}\left(v\right)+\kappa_{2}^{\dagger}\left(v\right)
\end{align*}
where
\begin{gather*}
\kappa_{1}^{\dagger}\left(v\right)\coloneqq\frac{1}{L}\sum_{l=1}^{L}\mathbbm{1}\left(N_{l}=n\right)\frac{1}{N_{l}}\sum_{i=1}^{N_{l}}\mathbb{T}_{il}\left(1-\widetilde{\mathbb{T}}_{il}\right)\frac{1}{h^{1+d}}K_{f}\left(\frac{\widehat{V}_{il}-v}{h},\frac{\boldsymbol{X}_{l}-\boldsymbol{x}}{h}\right)\\
\kappa_{2}^{\dagger}\left(v\right)\coloneqq\frac{1}{L}\sum_{l=1}^{L}\mathbbm{1}\left(N_{l}=n\right)\frac{1}{N_{l}}\sum_{i=1}^{N_{l}}\widetilde{\mathbb{T}}_{il}\left(\mathbb{T}_{il}-1\right)\frac{1}{h^{1+d}}K_{f}\left(\frac{\widehat{V}_{il}-v}{h},\frac{\boldsymbol{X}_{l}-\boldsymbol{x}}{h}\right).
\end{gather*}

Since $K_{0}$ is supported on $\left[-1,1\right]$, $K_{0}\left(\nicefrac{\left(\widehat{V}_{il}-v\right)}{h}\right)$
is zero if $\widehat{V}_{il}$ is outside of a $h-$neighborhood of
$v$. By the triangle inequality, we have
\begin{eqnarray*}
\left|\kappa_{1}^{\dagger}\left(v\right)\right| & \apprle & \frac{1}{L}\sum_{l=1}^{L}\mathbbm{1}\left(N_{l}=n\right)\frac{1}{N_{l}}\sum_{i=1}^{N_{l}}\mathbb{T}_{il}\left(1-\widetilde{\mathbb{T}}_{il}\right)h^{-\left(1+d\right)}\mathbbm{1}\left(\left|\widehat{V}_{il}-v\right|\leq h\right)\mathbbm{1}\left(\boldsymbol{X}_{l}\in\mathbb{H}\left(\boldsymbol{x},h\right)\right)\\
 & \apprle & \frac{1}{L}\sum_{l=1}^{L}\mathbbm{1}\left(N_{l}=n\right)\frac{1}{N_{l}}\sum_{i=1}^{N_{l}}\mathbb{T}_{il}\left(1-\widetilde{\mathbb{T}}_{il}\right)h^{-\left(1+d\right)}\mathbbm{1}\left(\left|V_{il}-v\right|\leq h+\underset{j,k}{\mathrm{max}}\,\mathbb{T}_{jk}\left|\widehat{V}_{jk}-V_{jk}\right|\right)\mathbbm{1}\left(\boldsymbol{X}_{l}\in\mathbb{H}\left(\boldsymbol{x},h\right)\right).
\end{eqnarray*}
Therefore it is clear that
\[
\mathrm{P}\left[\underset{v\in I\left(\boldsymbol{x}\right)}{\mathrm{sup}}\left|\kappa_{1}^{\dagger}\left(v\right)\right|=0\right]\geq\mathrm{P}\left[\underset{j,k}{\mathrm{max}}\,\mathbb{T}_{jk}\left|\widehat{V}_{jk}-V_{jk}\right|<\frac{\overline{\delta}}{2}\right],
\]
when $h$ is sufficiently small. Therefore, we have $\underset{v\in I\left(\boldsymbol{x}\right)}{\mathrm{sup}}\left|\kappa_{1}^{\dagger}\left(v\right)\right|=0$,
w.p.a.1. 

It is clear that
\begin{eqnarray*}
\underset{v\in I\left(\boldsymbol{x}\right)}{\mathrm{sup}}\left|\kappa_{2}^{\dagger}\left(v\right)\right| & \apprle & \underset{v\in I\left(\boldsymbol{x}\right)}{\mathrm{sup}}\frac{1}{L}\sum_{l=1}^{L}\mathbbm{1}\left(N_{l}=n\right)\frac{1}{N_{l}}\sum_{i=1}^{N_{l}}h^{-\left(1+d\right)}\widetilde{\mathbb{T}}_{il}\mathbbm{1}\left(B_{il}+2h>\underset{\boldsymbol{x}'\in\mathbb{H}\left(\boldsymbol{X}_{l},2h\right)\cap\mathcal{X}}{\mathrm{inf}}\widehat{\overline{b}}\left(\boldsymbol{x}',N_{l}\right)\right)\mathbbm{1}\left(\boldsymbol{X}_{l}\in\mathbb{H}\left(\boldsymbol{x},h\right)\right)\\
 &  & +\underset{v\in I\left(\boldsymbol{x}\right)}{\mathrm{sup}}\frac{1}{L}\sum_{l=1}^{L}\mathbbm{1}\left(N_{l}=n\right)\frac{1}{N_{l}}\sum_{i=1}^{N_{l}}h^{-\left(1+d\right)}\widetilde{\mathbb{T}}_{il}\mathbbm{1}\left(B_{il}-2h<\underset{\boldsymbol{x}'\in\mathbb{H}\left(\boldsymbol{X}_{l},2h\right)\cap\mathcal{X}}{\mathrm{sup}}\widehat{\underline{b}}\left(\boldsymbol{x}'\right)\right)\mathbbm{1}\left(\boldsymbol{X}_{l}\in\mathbb{H}\left(\boldsymbol{x},h\right)\right)\\
 & \leq & h^{-\left(1+d\right)}\mathbbm{1}\left(\underset{\boldsymbol{x}'\in\mathbb{H}\left(\boldsymbol{x},3h\right)}{\mathrm{sup}}s\left(v_{u}\left(\boldsymbol{x}\right)+\overline{\delta},\boldsymbol{x}',n\right)+2h>\underset{\boldsymbol{x}'\in\mathbb{H}\left(\boldsymbol{x},3h\right)}{\mathrm{inf}}\widehat{\overline{b}}\left(\boldsymbol{x}',n\right)\right)\\
 &  & +h^{-\left(1+d\right)}\mathbbm{1}\left(\underset{\boldsymbol{x}'\in\mathbb{H}\left(\boldsymbol{x},3h\right)}{\mathrm{inf}}s\left(v_{l}\left(\boldsymbol{x}\right)-\overline{\delta},\boldsymbol{x}',n\right)-2h<\underset{\boldsymbol{x}'\in\mathbb{H}\left(\boldsymbol{x},3h\right)}{\mathrm{sup}}\widehat{\underline{b}}\left(\boldsymbol{x}'\right)\right)\\
 & \leq & h^{-\left(1+d\right)}\mathbbm{1}\left(\underset{\boldsymbol{x}'\in\mathbb{H}\left(\boldsymbol{x},3h\right)}{\mathrm{sup}}s\left(v_{u}\left(\boldsymbol{x}\right)+\overline{\delta},\boldsymbol{x}',n\right)+2h>\underset{\boldsymbol{x}'\in\mathbb{H}\left(\boldsymbol{x},3h\right)}{\mathrm{inf}}\overline{b}\left(\boldsymbol{x}',n\right)-\underset{\boldsymbol{x}'\in\mathcal{X}}{\mathrm{sup}}\left|\widehat{\overline{b}}\left(\boldsymbol{x}',n\right)-\overline{b}\left(\boldsymbol{x}',n\right)\right|\right)\\
 &  & +h^{-\left(1+d\right)}\mathbbm{1}\left(\underset{\boldsymbol{x}'\in\mathbb{H}\left(\boldsymbol{x},3h\right)}{\mathrm{inf}}s\left(v_{l}\left(\boldsymbol{x}\right)-\overline{\delta},\boldsymbol{x}',n\right)-2h<\underset{\boldsymbol{x}'\in\mathbb{H}\left(\boldsymbol{x},3h\right)}{\mathrm{sup}}\underline{b}\left(\boldsymbol{x}'\right)+\underset{\boldsymbol{x}'\in\mathcal{X}}{\mathrm{sup}}\left|\widehat{\underline{b}}\left(\boldsymbol{x}'\right)-\underline{b}\left(\boldsymbol{x}'\right)\right|\right),
\end{eqnarray*}
where the second inequality holds when $h$ is sufficiently small
and follows from the definition of $\widetilde{\mathbb{T}}_{il}$.
Now it follows that 
\begin{eqnarray*}
\mathrm{P}\left[\underset{v\in I\left(\boldsymbol{x}\right)}{\mathrm{sup}}\left|\kappa_{2}^{\dagger}\left(v\right)\right|>0\right] & \leq & \mathrm{P}\left[\underset{\boldsymbol{x}'\in\mathbb{H}\left(\boldsymbol{x},3h\right)}{\mathrm{sup}}s\left(v_{u}\left(\boldsymbol{x}\right)+\overline{\delta},\boldsymbol{x}',n\right)+2h>\underset{\boldsymbol{x}'\in\mathbb{H}\left(\boldsymbol{x},3h\right)}{\mathrm{inf}}\overline{b}\left(\boldsymbol{x}',n\right)-\underset{\boldsymbol{x}'\in\mathcal{X}}{\mathrm{sup}}\left|\widehat{\overline{b}}\left(\boldsymbol{x}',n\right)-\overline{b}\left(\boldsymbol{x}',n\right)\right|\right]\\
 &  & +\mathrm{P}\left[\underset{\boldsymbol{x}'\in\mathbb{H}\left(\boldsymbol{x},3h\right)}{\mathrm{inf}}s\left(v_{l}\left(\boldsymbol{x}\right)-\overline{\delta},\boldsymbol{x}',n\right)-2h<\underset{\boldsymbol{x}'\in\mathbb{H}\left(\boldsymbol{x},3h\right)}{\mathrm{sup}}\underline{b}\left(\boldsymbol{x}'\right)+\underset{\boldsymbol{x}'\in\mathcal{X}}{\mathrm{sup}}\left|\widehat{\underline{b}}\left(\boldsymbol{x}'\right)-\underline{b}\left(\boldsymbol{x}'\right)\right|\right],
\end{eqnarray*}
when $h$ is sufficiently small. It follows from (\ref{eq:boundary estimator rate})
that the right-hand side of the inequality tends to zero as $L\uparrow\infty$.

Now we have
\begin{align}
 & \widehat{f}_{GPV}\left(v,\boldsymbol{x},n\right)-\widetilde{f}\left(v,\boldsymbol{x},n\right)\nonumber \\
= & \frac{1}{L}\sum_{l=1}^{L}\mathbbm{1}\left(N_{l}=n\right)\frac{1}{N_{l}}\sum_{i=1}^{N_{l}}\widetilde{\mathbb{T}}_{il}\frac{1}{h^{1+d}}\left(K_{f}\left(\frac{\widehat{V}_{il}-v}{h},\frac{\boldsymbol{X}_{l}-\boldsymbol{x}}{h}\right)-K_{f}\left(\frac{V_{il}-v}{h},\frac{\boldsymbol{X}_{l}-\boldsymbol{x}}{h}\right)\right)\label{eq:f_hat - f_til}
\end{align}
for all $v\in I\left(\boldsymbol{x}\right)$, w.p.a.1. Then a second-order
Taylor expansion of the right-hand side of (\ref{eq:f_hat - f_til})
gives
\begin{eqnarray}
\widehat{f}_{GPV}\left(v,\boldsymbol{x},n\right)-\widetilde{f}\left(v,\boldsymbol{x},n\right) & = & \frac{1}{L}\sum_{l=1}^{L}\mathbbm{1}\left(N_{l}=n\right)\frac{1}{N_{l}}\sum_{i=1}^{N_{l}}\widetilde{\mathbb{T}}_{il}\frac{1}{h^{2+d}}K_{f}'\left(\frac{V_{il}-v}{h},\frac{\boldsymbol{X}_{l}-\boldsymbol{x}}{h}\right)\left(\widehat{V}_{il}-V_{il}\right)\nonumber \\
 &  & +\frac{1}{2}\frac{1}{L}\sum_{l=1}^{L}\mathbbm{1}\left(N_{l}=n\right)\frac{1}{N_{l}}\sum_{i=1}^{N_{l}}\widetilde{\mathbb{T}}_{il}\frac{1}{h^{3+d}}K_{f}''\left(\frac{\dot{V}_{il}-v}{h},\frac{\boldsymbol{X}_{l}-\boldsymbol{x}}{h}\right)\left(\widehat{V}_{il}-V_{il}\right)^{2},\label{eq:Lemma 1 approximation error 1}
\end{eqnarray}
for some mean value $\dot{V}_{il}$ that lies on the line joining
$\widehat{V}_{il}$ and $V_{il}$. It follows from triangle inequality
and the Lipschitz condition imposed on the kernel that
\begin{align}
 & \left|\frac{1}{L}\sum_{l=1}^{L}\mathbbm{1}\left(N_{l}=n\right)\frac{1}{N_{l}}\sum_{i=1}^{N_{l}}\widetilde{\mathbb{T}}_{il}\frac{1}{h^{3+d}}K_{f}''\left(\frac{\dot{V}_{il}-v}{h},\frac{\boldsymbol{X}_{l}-\boldsymbol{x}}{h}\right)\left(\hat{V}_{il}-V_{il}\right)^{2}\right|\nonumber \\
\apprle & \left(\frac{1}{L}\sum_{l=1}^{L}\mathbbm{1}\left(N_{l}=n\right)\frac{1}{N_{l}}\sum_{i=1}^{N_{l}}\widetilde{\mathbb{T}}_{il}\frac{1}{h^{3+d}}\left|K_{\boldsymbol{X}}^{0}\left(\frac{\boldsymbol{X}_{l}-\boldsymbol{x}}{h}\right)\right|\mathbbm{1}\left(\left|\dot{V}_{il}-v\right|\leq h\right)\right)\left(\underset{i,l}{\mathrm{max}}\,\widetilde{\mathbb{T}}_{il}\left(\widehat{V}_{il}-V_{il}\right)^{2}\right).\label{eq:Lemma 1 approximation error 2}
\end{align}
By the triangle inequality, we have
\begin{align}
 & \frac{1}{L}\sum_{l=1}^{L}\mathbbm{1}\left(N_{l}=n\right)\frac{1}{N_{l}}\sum_{i=1}^{N_{l}}\widetilde{\mathbb{T}}_{il}\frac{1}{h^{3+d}}\left|K_{\boldsymbol{X}}^{0}\left(\frac{\boldsymbol{X}_{l}-\boldsymbol{x}}{h}\right)\right|\mathbbm{1}\left(\left|\dot{V}_{il}-v\right|\leq h\right)\nonumber \\
\leq & \frac{1}{L}\sum_{l=1}^{L}\mathbbm{1}\left(N_{l}=n\right)\frac{1}{N_{l}}\sum_{i=1}^{N_{l}}\widetilde{\mathbb{T}}_{il}\left(1-\reallywidecheck{\mathbb{T}}_{il}+\reallywidecheck{\mathbb{T}}_{il}\right)\frac{1}{h^{3+d}}\left|K_{\boldsymbol{X}}^{0}\left(\frac{\boldsymbol{X}_{l}-\boldsymbol{x}}{h}\right)\right|\mathbbm{1}\left(\left|V_{il}-v\right|\leq h+\underset{j,k}{\mathrm{max}}\,\widetilde{\mathbb{T}}_{jk}\left|\dot{V}_{jk}-V_{jk}\right|\right)\nonumber \\
\leq & \frac{1}{L}\sum_{l=1}^{L}\mathbbm{1}\left(N_{l}=n\right)\frac{1}{N_{l}}\sum_{i=1}^{N_{l}}\reallywidecheck{\mathbb{T}}_{il}\frac{1}{h^{3+d}}\left|K_{\boldsymbol{X}}\left(\frac{\boldsymbol{X}_{l}-\boldsymbol{x}}{h}\right)\right|+\kappa_{3}^{\dagger}\left(v\right),\label{eq:Lemma 1 approximation error indicator}
\end{align}
where
\[
\reallywidecheck{\mathbb{T}}_{il}\coloneqq\mathbbm{1}\left(\left|V_{il}-v\right|\leq2h\right)
\]
and
\[
\kappa_{3}^{\dagger}\left(v\right)\coloneqq\frac{1}{L}\sum_{l=1}^{L}\mathbbm{1}\left(N_{l}=n\right)\frac{1}{N_{l}}\sum_{i=1}^{N_{l}}\widetilde{\mathbb{T}}_{il}\mathbbm{1}\left(\left|V_{il}-v\right|>2h\right)\frac{1}{h^{3+d}}\left|K_{\boldsymbol{X}}^{0}\left(\frac{\boldsymbol{X}_{l}-\boldsymbol{x}}{h}\right)\right|\mathbbm{1}\left(\left|V_{il}-v\right|\leq h+\underset{j,k}{\mathrm{max}}\,\widetilde{\mathbb{T}}_{jk}\left|\dot{V}_{jk}-V_{jk}\right|\right).
\]
Clearly we have 
\[
\mathrm{P}\left[\underset{v\in I\left(\boldsymbol{x}\right)}{\mathrm{sup}}\left|\kappa_{3}^{\dagger}\left(v\right)\right|=0\right]\geq\mathrm{P}\left[\underset{v\in I\left(\boldsymbol{x}\right)}{\mathrm{sup}}\underset{j,k}{\mathrm{max}}\,\widetilde{\mathbb{T}}_{jk}\left|\dot{V}_{jk}-V_{jk}\right|\leq h\right]
\]
where the right hand side tends to 1 as $L\uparrow\infty$ since $\underset{v\in I\left(\boldsymbol{x}\right)}{\mathrm{sup}}\underset{j,k}{\mathrm{max}}\,\widetilde{\mathbb{T}}_{jk}\left|\dot{V}_{jk}-V_{jk}\right|=o_{p}\left(h\right)$
(see (\ref{eq:V_hat - V Infeasible Trimming})). Therefore, we have
$\underset{v\in I\left(\boldsymbol{x}\right)}{\mathrm{sup}}\left|\kappa_{3}^{\dagger}\left(v\right)\right|=0$
w.p.a.1. 

By the LIE and change of variables, we have
\begin{align}
 & \underset{v\in I\left(\boldsymbol{x}\right)}{\mathrm{sup}}\mathrm{E}\left[\mathbbm{1}\left(N_{1}=n\right)\frac{1}{N_{1}}\sum_{i=1}^{N_{1}}\reallywidecheck{\mathbb{T}}_{i1}\frac{1}{h^{1+d}}\left|K_{\boldsymbol{X}}^{0}\left(\frac{\boldsymbol{X}_{1}-\boldsymbol{x}}{h}\right)\right|\right]\nonumber \\
= & \underset{v\in I\left(\boldsymbol{x}\right)}{\mathrm{sup}}\mathrm{E}\left[\mathbbm{1}\left(N_{1}=n\right)\frac{1}{h^{1+d}}\left|K_{\boldsymbol{X}}^{0}\left(\frac{\boldsymbol{X}_{1}-\boldsymbol{x}}{h}\right)\right|\mathrm{E}\left[\reallywidecheck{\mathbb{T}}_{11}|\boldsymbol{X}_{1},N_{1}\right]\right]\nonumber \\
= & \underset{v\in I\left(\boldsymbol{x}\right)}{\mathrm{sup}}\int_{\mathcal{Y}}\int_{\underline{v}\left(h\boldsymbol{z}+\boldsymbol{x}\right)}^{\overline{v}\left(h\boldsymbol{z}+\boldsymbol{x}\right)}h^{-1}\mathbbm{1}\left(\left|w-v\right|\leq2h\right)\left|K_{\boldsymbol{X}}^{0}\left(\boldsymbol{z}\right)\right|f\left(w|h\boldsymbol{z}+\boldsymbol{x}\right)\pi\left(n|h\boldsymbol{z}+\boldsymbol{x}\right)\varphi\left(h\boldsymbol{z}+\boldsymbol{x}\right)\mathrm{d}w\mathrm{d}\boldsymbol{z}\nonumber \\
\apprle & \left(\underset{\left(w,\boldsymbol{z}\right)\in\mathcal{C}_{V,\boldsymbol{X}}}{\mathrm{sup}}f\left(w|\boldsymbol{z},n\right)\right)\left(\underset{\boldsymbol{z}\in\mathbb{H}\left(\boldsymbol{x},\overline{\delta}\right)}{\mathrm{sup}}\pi\left(n|\boldsymbol{z}\right)\varphi\left(\boldsymbol{z}\right)\right),\label{eq:Lemma 1 error change of variable}
\end{align}
where the inequality holds when $h$ is sufficiently small, since
$K_{0}$ is supported on $\left[-1,1\right]$.

By Jensen's inequality and LIE, we have 
\begin{eqnarray}
\sigma_{\mathcal{F}^{n}}^{2} & \coloneqq & \underset{v\in I\left(\boldsymbol{x}\right)}{\mathrm{sup}}\mathrm{E}\left[\mathbbm{1}\left(N_{1}=n\right)\left\{ \frac{1}{N_{1}}\sum_{i=1}^{N_{1}}\reallywidecheck{\mathbb{T}}_{i1}\frac{1}{h^{1+d}}K_{\boldsymbol{X}}\left(\frac{\boldsymbol{X}_{1}-\boldsymbol{x}}{h}\right)\right\} ^{2}\right]\nonumber \\
 & \leq & \underset{v\in I\left(\boldsymbol{x}\right)}{\mathrm{sup}}\mathrm{E}\left[\mathbbm{1}\left(N_{1}=n\right)\reallywidecheck{\mathbb{T}}_{11}\frac{1}{h^{2+2d}}K_{\boldsymbol{X}}\left(\frac{\boldsymbol{X}_{1}-\boldsymbol{x}}{h}\right)^{2}\right]\nonumber \\
 & \apprle & h^{-\left(1+d\right)},\label{eq:sigma sup bound 1}
\end{eqnarray}
where the second inequality holds when $h$ is sufficiently small.
Let $\boldsymbol{u}.\coloneqq\left(u_{1},...,u_{m}\right)$ and 
\[
\mathcal{F}^{n}\left(\boldsymbol{u},\boldsymbol{z},m,;v\right)\coloneqq\mathbbm{1}\left(m=n\right)\frac{1}{m}\sum_{i=1}^{m}\mathbbm{1}\left(\left|u_{i}-v\right|\leq2h\right)\frac{1}{h^{1+d}}\left|K_{\boldsymbol{X}}\left(\frac{\boldsymbol{z}-\boldsymbol{x}}{h}\right)\right|.
\]
By standard arguments, we can verify that the class $\left\{ \mathcal{F}^{n}\left(\cdot;v\right):v\in I\left(\boldsymbol{x}\right)\right\} $,
which implicitly depends on $L$, is uniformly VC-type with respect
to the envelope 
\[
F_{\mathcal{F}^{n}}\left(\boldsymbol{z}\right)\coloneqq\frac{1}{h^{1+d}}\left|K_{\boldsymbol{X}}\left(\frac{\boldsymbol{z}-\boldsymbol{x}}{h}\right)\right|.
\]
The CCK inequality yields 
\begin{eqnarray*}
\mathrm{E}\left[\underset{v\in I\left(\boldsymbol{x}\right)}{\mathrm{sup}}\left|\frac{1}{L}\sum_{l=1}^{L}\mathcal{F}^{n}\left(\boldsymbol{V}_{\cdot l},\boldsymbol{X}_{l},N_{l};v\right)-\mathrm{E}\left[\mathcal{F}^{n}\left(\boldsymbol{V}_{\cdot1},\boldsymbol{X}_{1},N_{1};v\right)\right]\right|\right] & \leq & C_{1}\left\{ L^{-\nicefrac{1}{2}}\sigma_{\mathcal{F}^{n}}\mathrm{log}\left(C_{2}L\right)^{\nicefrac{1}{2}}+L^{-1}\left\Vert F_{\mathcal{F}^{n}}\right\Vert _{\mathcal{X}}\mathrm{log}\left(C_{2}L\right)\right\} \\
 & = & O\left(\left(\frac{\mathrm{log}\left(L\right)}{Lh^{1+d}}\right)^{\nicefrac{1}{2}}\right)
\end{eqnarray*}
and
\begin{equation}
\underset{v\in I\left(\boldsymbol{x}\right)}{\mathrm{sup}}\left|\frac{1}{L}\sum_{l=1}^{L}\mathcal{F}^{n}\left(\boldsymbol{V}_{\cdot l},\boldsymbol{X}_{l},N_{l};v\right)-\mathrm{E}\left[\mathcal{F}^{n}\left(\boldsymbol{V}_{\cdot1},\boldsymbol{X}_{1},N_{1};v\right)\right]\right|=O_{p}\left(\left(\frac{\mathrm{log}\left(L\right)}{Lh^{1+d}}\right)^{\nicefrac{1}{2}}\right)\label{eq:EP sup bound 1}
\end{equation}
follows from Markov's inequality. It now follows that 
\begin{equation}
\underset{v\in I\left(\boldsymbol{x}\right)}{\mathrm{sup}}\frac{1}{L}\sum_{l=1}^{L}\mathcal{F}^{n}\left(\boldsymbol{V}_{\cdot l},\boldsymbol{X}_{l},N_{l},;v\right)=O_{p}\left(1\right).\label{eq:sup F average O_p(1)}
\end{equation}

It follows from the above result, (\ref{eq:V_hat - V Infeasible Trimming}),
(\ref{eq:Lemma 1 approximation error 2}), (\ref{eq:Lemma 1 approximation error indicator})
and (\ref{eq:Lemma 1 error change of variable}) that
\[
\underset{v\in I\left(\boldsymbol{x}\right)}{\mathrm{sup}}\left|\frac{1}{L}\sum_{l=1}^{L}\mathbbm{1}\left(N_{l}=n\right)\frac{1}{N_{l}}\sum_{i=1}^{N_{l}}\widetilde{\mathbb{T}}_{il}\frac{1}{h^{3+d}}K_{\boldsymbol{X}}\left(\frac{\boldsymbol{X}_{l}-\boldsymbol{x}}{h}\right)K_{f}''\left(\frac{\dot{V}_{il}-v}{h}\right)\left(\widehat{V}_{il}-V_{il}\right)^{2}\right|=O_{p}\left(\frac{\mathrm{log}\left(L\right)}{Lh^{3+d}}+h^{2R}\right).
\]
It follows from the above result, (\ref{eq:Lemma 1 approximation error 1})
and (\ref{eq:Lemma 1 approximation error 2}) that
\begin{eqnarray}
\widehat{f}_{GPV}\left(v,\boldsymbol{x},n\right)-\widetilde{f}\left(v,\boldsymbol{x},n\right) & = & \frac{1}{L}\sum_{l=1}^{L}\mathbbm{1}\left(N_{l}=n\right)\frac{1}{N_{l}}\sum_{i=1}^{N_{l}}\widetilde{\mathbb{T}}_{il}\frac{1}{h^{2+d}}K_{f}'\left(\frac{V_{il}-v}{h},\frac{\boldsymbol{X}_{l}-\boldsymbol{x}}{h}\right)\left(\widehat{V}_{il}-V_{il}\right)\nonumber \\
 &  & +O_{p}\left(\left(\frac{\mathrm{log}\left(L\right)}{Lh^{1+d}}\right)^{\nicefrac{1}{2}}+\frac{\mathrm{log}\left(L\right)}{Lh^{3+d}}+h^{2R}\right).\label{eq:f_hat - f_til linearization}
\end{eqnarray}

Standard arguments for the kernel density estimation for conditional
densities gives
\[
\widetilde{f}\left(v,\boldsymbol{x},n\right)=f\left(v|\boldsymbol{x}\right)\varphi\left(\boldsymbol{x}\right)\pi\left(n|\boldsymbol{x}\right)+O_{p}\left(\left(\frac{\mathrm{log}\left(L\right)}{Lh^{1+d}}\right)^{\nicefrac{1}{2}}+h^{R}\right),
\]
where the remainder term is uniform in $I\left(\boldsymbol{x}\right)$.
See, e.g., \citet[Lemma 1(f)]{Marmer_Shneyerov_Quantile_Auctions}.
Then the conclusion follows from (\ref{eq:f_hat - f_til linearization})
and the above result.\end{proof}
\begin{lem}
\label{Lemma 2}Suppose that Assumptions 1 - 3 hold. Let $\boldsymbol{x}$
be an interior point of $\mathcal{X}$ and $n\in\mathcal{N}$ be fixed.
Then
\[
\widehat{f}_{GPV}\left(v,\boldsymbol{x},n\right)-f\left(v|\boldsymbol{x}\right)\varphi\left(\boldsymbol{x}\right)\pi\left(n|\boldsymbol{x}\right)=\frac{1}{L^{2}}\sum_{l=1}^{L}\sum_{k=1}^{L}\mathcal{M}^{n}\left(\left(\boldsymbol{B}_{\cdot l},\boldsymbol{X}_{l},N_{l}\right),\left(\boldsymbol{B}_{\cdot k},\boldsymbol{X}_{k},N_{k}\right);v\right)+O_{p}\left(\left(\frac{\mathrm{log}\left(L\right)}{Lh^{1+d}}\right)^{\nicefrac{1}{2}}+\frac{\mathrm{log}\left(L\right)}{Lh^{3+d}}+h^{R}\right),
\]
where the remainder term is uniform in $v\in I\left(\boldsymbol{x}\right)$.
\end{lem}
\begin{proof}[Proof of Lemma \ref{Lemma 2}] Using (\ref{eq:V_hat - V expansion}),
we have
\begin{align}
 & \frac{1}{L}\sum_{l=1}^{L}\mathbbm{1}\left(N_{l}=n\right)\frac{1}{N_{l}}\sum_{i=1}^{N_{l}}\widetilde{\mathbb{T}}_{il}\frac{1}{h^{2+d}}K_{f}'\left(\frac{V_{il}-v}{h},\frac{\boldsymbol{X}_{l}-\boldsymbol{x}}{h}\right)\left(\widehat{V}_{il}-V_{il}\right)\nonumber \\
= & -\frac{1}{L}\sum_{l=1}^{L}\mathbbm{1}\left(N_{l}=n\right)\frac{1}{N_{l}}\sum_{i=1}^{N_{l}}\widetilde{\mathbb{T}}_{il}\frac{1}{h^{2+d}}K_{f}'\left(\frac{V_{il}-v}{h},\frac{\boldsymbol{X}_{l}-\boldsymbol{x}}{h}\right)\frac{1}{N_{l}-1}\frac{G\left(B_{il},\boldsymbol{X}_{l},N_{l}\right)}{g\left(B_{il},\boldsymbol{X}_{l},N_{l}\right)^{2}}\left(\widehat{g}\left(B_{il},\boldsymbol{X}_{l},N_{l}\right)-g\left(B_{il},\boldsymbol{X}_{l},N_{l}\right)\right)\nonumber \\
 & +\varDelta_{1}^{\ddagger}\left(v\right)+\varDelta_{2}^{\ddagger}\left(v\right)+\varDelta_{3}^{\ddagger}\left(v\right),\label{eq:lemma 2 decomposition}
\end{align}
where
\begin{eqnarray*}
\varDelta_{1}^{\ddagger}\left(v\right) & \coloneqq & \frac{1}{L}\sum_{l=1}^{L}\mathbbm{1}\left(N_{l}=n\right)\frac{1}{N_{l}}\sum_{i=1}^{N_{l}}\widetilde{\mathbb{T}}_{il}\frac{1}{h^{2+d}}K_{f}'\left(\frac{V_{il}-v}{h},\frac{\boldsymbol{X}_{l}-\boldsymbol{x}}{h}\right)\frac{1}{N_{l}-1}\frac{G\left(B_{il},\boldsymbol{X}_{l},N_{l}\right)-\widehat{G}\left(B_{il},\boldsymbol{X}_{l},N_{l}\right)}{g\left(B_{il},\boldsymbol{X}_{l},N_{l}\right)},\\
\varDelta_{2}^{\ddagger}\left(v\right) & \coloneqq & -\frac{1}{L}\sum_{l=1}^{L}\mathbbm{1}\left(N_{l}=n\right)\frac{1}{N_{l}}\sum_{i=1}^{N_{l}}\widetilde{\mathbb{T}}_{il}\frac{1}{h^{2+d}}K_{f}'\left(\frac{V_{il}-v}{h},\frac{\boldsymbol{X}_{l}-\boldsymbol{x}}{h}\right)\frac{1}{N_{l}-1}\frac{\widehat{G}\left(B_{il},\boldsymbol{X}_{l},N_{l}\right)-G\left(B_{il},\boldsymbol{X}_{l},N_{l}\right)}{g\left(B_{il},\boldsymbol{X}_{l},N_{l}\right)^{2}},\\
 &  & \times\left(\widehat{g}\left(B_{il},\boldsymbol{X}_{l},N_{l}\right)-g\left(B_{il},\boldsymbol{X}_{l},N_{l}\right)\right)\\
\varDelta_{3}^{\ddagger}\left(v\right) & \coloneqq & -\frac{1}{L}\sum_{l=1}^{L}\mathbbm{1}\left(N_{l}=n\right)\frac{1}{N_{l}}\sum_{i=1}^{N_{l}}\widetilde{\mathbb{T}}_{il}\frac{1}{h^{2+d}}K_{f}'\left(\frac{V_{il}-v}{h},\frac{\boldsymbol{X}_{l}-\boldsymbol{x}}{h}\right)\frac{1}{N_{l}-1}\frac{\widehat{G}\left(B_{il},\boldsymbol{X}_{l},N_{l}\right)}{\widehat{g}\left(B_{il},\boldsymbol{X}_{l},N_{l}\right)}\frac{\left(\widehat{g}\left(B_{il},\boldsymbol{X}_{l},N_{l}\right)-g\left(B_{il},\boldsymbol{X}_{l},N_{l}\right)\right)^{2}}{g\left(B_{il},\boldsymbol{X}_{l},N_{l}\right)^{2}}.
\end{eqnarray*}

By using the triangle inequality, (\ref{eq:bids density bounded away from zero}),
(\ref{eq:uniform convergence rate}) and
\begin{equation}
\underset{v\in I\left(\boldsymbol{x}\right)}{\mathrm{sup}}\frac{1}{L}\sum_{l=1}^{L}\mathbbm{1}\left(N_{l}=n\right)\frac{1}{N_{l}}\sum_{i=1}^{N_{l}}\frac{1}{N_{l}-1}\frac{1}{h^{1+d}}\left|K_{f}'\left(\frac{V_{il}-v}{h},\frac{\boldsymbol{X}_{l}-\boldsymbol{x}}{h}\right)\right|=O_{p}\left(1\right)\label{eq:K_d K_prime bound}
\end{equation}
(which follows from arguments used to show (\ref{eq:sup F average O_p(1)})),
we have 
\[
\underset{v\in I\left(\boldsymbol{x}\right)}{\mathrm{sup}}\left|\varDelta_{2}^{\ddagger}\left(v\right)\right|=O_{p}\left(\frac{\mathrm{log}\left(L\right)}{Lh^{\nicefrac{3}{2}+d}}+h^{2R+1}\right).
\]
It follows from the triangle inequality, (\ref{eq:bids density bounded away from zero})
and the inequality $G\left(b,\boldsymbol{x}',n'\right)\leq\pi\left(n'|\boldsymbol{x}'\right)\varphi\left(\boldsymbol{x}'\right)$
that
\begin{eqnarray*}
\left|\varDelta_{3}^{\ddagger}\left(v\right)\right| & \apprle & \left(\underset{i,l}{\mathrm{min}}\,\widetilde{\mathbb{T}}_{il}\widehat{g}\left(B_{il},\boldsymbol{X}_{l},N_{l}\right)\right)^{-1}\left\{ \frac{1}{L}\sum_{l=1}^{L}\mathbbm{1}\left(N_{l}=n\right)\pi\left(N_{l}|\boldsymbol{X}_{l}\right)\varphi\left(\boldsymbol{X}_{l}\right)\frac{1}{N_{l}\left(N_{l}-1\right)}\sum_{i=1}^{n}\widetilde{\mathbb{T}}_{il}\frac{1}{h^{2+d}}\left|K_{f}'\left(\frac{V_{il}-v}{h},\frac{\boldsymbol{X}_{l}-\boldsymbol{x}}{h}\right)\right|\right\} \\
 &  & \times\left\{ \underset{i,l}{\mathrm{max}}\,\widetilde{\mathbb{T}}_{il}\left(\widehat{g}\left(B_{il},\boldsymbol{X}_{l},N_{l}\right)-g\left(B_{il},\boldsymbol{X}_{l},N_{l}\right)\right)^{2}\right\} .
\end{eqnarray*}
It follows from (\ref{eq:uniform convergence rate}), (\ref{eq:K_d K_prime bound})
and the fact 
\[
\left(\underset{i,l}{\mathrm{min}}\,\widetilde{\mathbb{T}}_{il}\widehat{g}\left(B_{il},X_{l},N_{l}\right)\right)^{-1}<\left(\frac{\underline{C}_{g}}{2}\right)^{-1},\textrm{ w.p.a.1}
\]
that 
\[
\underset{v\in I\left(\boldsymbol{x}\right)}{\mathrm{sup}}\left|\varDelta_{3}^{\ddagger}\left(v\right)\right|=O_{p}\left(\frac{\mathrm{log}\left(L\right)}{Lh^{2+d}}+h^{2R-1}\right).
\]

For $\varDelta_{1}^{\ddagger}\left(v\right)$, firstly, the contribution
of the trimmed observations is asymptotically negligible, since $K_{0}'$
has a bounded support. When $h$ is sufficiently small,
\begin{equation}
\sum_{l=1}^{L}\mathbbm{1}\left(N_{l}=n\right)\frac{1}{N_{l}}\sum_{i=1}^{N_{l}}\left(\widetilde{\mathbb{T}}_{il}-1\right)K_{f}'\left(\frac{V_{il}-v}{h},\frac{\boldsymbol{X}_{l}-\boldsymbol{x}}{h}\right)\frac{G\left(B_{il},\boldsymbol{X}_{l},N_{l}\right)-\widehat{G}\left(B_{il},\boldsymbol{X}_{l},N_{l}\right)}{\left(N_{l}-1\right)g\left(B_{il},\boldsymbol{X}_{l},N_{l}\right)}=0.\label{eq:lemma 2 infeasible trimming negligible}
\end{equation}
Since
\begin{eqnarray*}
\varDelta_{1}^{\ddagger}\left(v\right) & = & -\frac{1}{L}\sum_{l=1}^{L}\mathbbm{1}\left(N_{l}=n\right)\frac{1}{N_{l}}\sum_{i=1}^{N_{l}}\frac{1}{h^{2+d}}K_{f}'\left(\frac{V_{il}-v}{h},\frac{\boldsymbol{X}_{l}-\boldsymbol{x}}{h}\right)\frac{1}{\left(N_{l}-1\right)g\left(B_{il},\boldsymbol{X}_{l},N_{l}\right)}\\
 &  & \times\left\{ \frac{1}{L}\sum_{k=1}^{L}\mathbbm{1}\left(N_{k}=N_{l}\right)\frac{1}{N_{k}}\sum_{j=1}^{N_{k}}\mathbbm{1}\left(B_{jk}\leq B_{il}\right)\frac{1}{h^{d}}K_{\boldsymbol{X}}\left(\frac{\boldsymbol{X}_{k}-\boldsymbol{X}_{l}}{h}\right)-G\left(B_{il},\boldsymbol{X}_{l},N_{l}\right)\right\} ,
\end{eqnarray*}
when $h$ is sufficiently small, it is clear that we can write
\begin{equation}
\varDelta_{1}^{\ddagger}\left(v\right)=\frac{1}{L^{2}}\sum_{l=1}^{L}\sum_{k=1}^{L}\mathcal{G}^{n}\left(\left(\boldsymbol{B}_{\cdot l},\boldsymbol{X}_{l},N_{l}\right),\left(\boldsymbol{B}_{\cdot k},\boldsymbol{X}_{k},N_{k}\right);v\right),\label{eq:I_1 V_statistic representation}
\end{equation}
where
\begin{eqnarray}
\mathcal{G}^{n}\left(\left(\boldsymbol{b}_{\cdot},\boldsymbol{z},m\right),\left(\boldsymbol{b}_{\cdot}',\boldsymbol{z}',m'\right);v\right) & \coloneqq & -\mathbbm{1}\left(m=n\right)\frac{1}{m}\sum_{i=1}^{m}\frac{1}{h^{2+d}}K_{f}'\left(\frac{\xi\left(b_{i},\boldsymbol{z},m\right)-v}{h},\frac{\boldsymbol{z}-\boldsymbol{x}}{h}\right)\frac{1}{\left(m-1\right)g\left(b_{i},\boldsymbol{z},m\right)}\nonumber \\
 &  & \times\left\{ \mathbbm{1}\left(m'=m\right)\frac{1}{m'}\sum_{j=1}^{m'}\mathbbm{1}\left(b_{j}'\leq b_{i}\right)\frac{1}{h^{d}}K_{\boldsymbol{X}}\left(\frac{\boldsymbol{z}'-\boldsymbol{z}}{h}\right)-G\left(b_{i},\boldsymbol{z},m\right)\right\} .\label{eq:G_n kernel definition}
\end{eqnarray}

Let
\begin{eqnarray*}
\mu_{\mathcal{G}^{n}}\left(v\right) & \coloneqq & \int\sum_{m\in\mathcal{N}}\int\cdots\int\mathcal{G}{}_{1}^{n}\left(\boldsymbol{b}_{\cdot},\boldsymbol{z},m;v\right)\prod_{j=1}^{m}g\left(b_{j}|\boldsymbol{z},m\right)\pi\left(m|\boldsymbol{z}\right)\varphi\left(\boldsymbol{z}\right)\mathrm{d}b_{1}\cdots\mathrm{d}b_{m}\mathrm{d}\boldsymbol{z}\\
 & = & \int\sum_{m\in\mathcal{N}}\int\cdots\int\mathcal{G}_{2}^{n}\left(\boldsymbol{b}_{\cdot},\boldsymbol{z},m;v\right)\prod_{j=1}^{m}g\left(b_{j}|\boldsymbol{z},m\right)\pi\left(m|\boldsymbol{z}\right)\varphi\left(\boldsymbol{z}\right)\mathrm{d}b_{1}\cdots\mathrm{d}b_{m}\mathrm{d}\boldsymbol{z}
\end{eqnarray*}
and
\[
\mathcal{G}_{1}^{n}\left(\boldsymbol{b}_{\cdot},\boldsymbol{z},m;v\right)\coloneqq\mathrm{E}\left[\mathcal{G}^{n}\left(\left(\boldsymbol{b}_{\cdot},\boldsymbol{z},m\right),\left(\boldsymbol{B}_{\cdot1},\boldsymbol{X}_{1},N_{1}\right);v\right)\right]\textrm{ and }\mathcal{G}_{2}^{n}\left(\boldsymbol{b}_{\cdot},\boldsymbol{z},m;v\right)\coloneqq\mathrm{E}\left[\mathcal{G}^{n}\left(\left(\boldsymbol{B}_{\cdot1},\boldsymbol{X}_{1},N_{1}\right),\left(\boldsymbol{b}_{\cdot},\boldsymbol{z},m\right);v\right)\right].
\]
The Hoeffding decomposition yields 
\begin{eqnarray*}
\varDelta_{1}^{\ddagger}\left(v\right) & = & \mu_{\mathcal{G}^{n}}\left(v\right)+\left\{ \frac{1}{L}\sum_{l=1}^{L}\mathcal{G}_{1}^{n}\left(\boldsymbol{B}_{\cdot l},\boldsymbol{X}_{l},N_{l};v\right)-\mu_{\mathcal{G}^{n}}\left(v\right)\right\} +\left\{ \frac{1}{L}\sum_{l=1}^{L}\mathcal{G}_{2}^{n}\left(\boldsymbol{B}_{\cdot l},\boldsymbol{X}_{l},N_{l};v\right)-\mu_{\mathcal{G}^{n}}\left(v\right)\right\} \\
 &  & +\frac{1}{L\left(L-1\right)}\sum_{l\neq k}\left\{ \mathcal{G}^{n}\left(\left(\boldsymbol{B}_{\cdot l},\boldsymbol{X}_{l},N_{l}\right),\left(\boldsymbol{B}_{\cdot k},\boldsymbol{X}_{k},N_{k}\right);v\right)-\mathcal{G}_{1}^{n}\left(\boldsymbol{B}_{\cdot l},\boldsymbol{X}_{l},N_{l};v\right)-\mathcal{G}_{2}^{n}\left(\boldsymbol{B}_{\cdot k},\boldsymbol{X}_{k},N_{k};v\right)+\mu_{\mathcal{G}^{n}}\left(v\right)\right\} \\
 &  & +\frac{1}{L^{2}}\sum_{l=1}^{L}\mathcal{G}^{n}\left(\left(\boldsymbol{B}_{\cdot l},\boldsymbol{X}_{l},N_{l}\right),\left(\boldsymbol{B}_{\cdot l},\boldsymbol{X}_{l},N_{l}\right);v\right)-\frac{1}{L^{2}\left(L-1\right)}\sum_{l\neq k}\mathcal{G}^{n}\left(\left(\boldsymbol{B}_{\cdot l},\boldsymbol{X}_{l},N_{l}\right),\left(\boldsymbol{B}_{\cdot k},\boldsymbol{X}_{k},N_{k}\right);v\right).
\end{eqnarray*}

By the LIE, we have
\begin{eqnarray*}
\mathcal{G}_{1}^{n}\left(\boldsymbol{b}_{\cdot},\boldsymbol{z},m;v\right) & = & -\mathbbm{1}\left(m=n\right)\frac{1}{m}\sum_{i=1}^{m}\frac{1}{h^{2+d}}K_{f}'\left(\frac{\xi\left(b_{i},\boldsymbol{z},m\right)-v}{h},\frac{\boldsymbol{z}-\boldsymbol{x}}{h}\right)\frac{1}{\left(m-1\right)g\left(b_{i},\boldsymbol{z},m\right)}\\
 &  & \times\left\{ \mathrm{E}\left[\mathbbm{1}\left(N_{1}=m\right)G\left(b_{i}|\boldsymbol{X}_{1},N_{1}\right)\frac{1}{h^{d}}K_{\boldsymbol{X}}\left(\frac{\boldsymbol{X}_{1}-\boldsymbol{z}}{h}\right)\right]-G\left(b_{i},\boldsymbol{z},m\right)\right\} 
\end{eqnarray*}
and
\begin{eqnarray*}
\mathcal{G}_{2}^{n}\left(\boldsymbol{b}_{\cdot},\boldsymbol{z},m;v\right) & = & \mathrm{E}\left[\mathrm{E}\left[-\mathbbm{1}\left(N_{1}=n\right)\frac{1}{h^{2+2d}}K_{f}'\left(\frac{\xi\left(B_{11},\boldsymbol{X}_{1},N_{1}\right)-v}{h},\frac{\boldsymbol{X}_{1}-\boldsymbol{x}}{h}\right)\frac{1}{\left(N_{1}-1\right)g\left(B_{11},\boldsymbol{X}_{1},N_{1}\right)}\right.\right.\\
 &  & \left.\left.\times\left\{ \mathbbm{1}\left(m=N_{1}\right)\frac{1}{m}\sum_{j=1}^{m}\mathbbm{1}\left(b_{j}\leq B_{11}\right)\frac{1}{h^{d}}K_{\boldsymbol{X}}\left(\frac{\boldsymbol{z}-\boldsymbol{X}_{1}}{h}\right)-G\left(B_{11},\boldsymbol{X}_{1},N_{1}\right)\right\} |\boldsymbol{X}_{1},N_{1}\right]\right]\\
 & = & -\int_{\mathcal{X}}\int_{\underline{b}\left(\boldsymbol{z}'\right)}^{\overline{b}\left(\boldsymbol{z}',n\right)}\frac{1}{h^{2+d}}K_{f}'\left(\frac{\xi\left(b',\boldsymbol{z}',n\right)-v}{h},\frac{\boldsymbol{z}'-\boldsymbol{x}}{h}\right)\frac{1}{n-1}\mathbbm{1}\left(m=n\right)\frac{1}{m}\sum_{j=1}^{m}\mathbbm{1}\left(b_{j}\leq b'\right)\frac{1}{h^{d}}K_{\boldsymbol{X}}\left(\frac{\boldsymbol{z}-\boldsymbol{z}'}{h}\right)\mathrm{d}b'\mathrm{d}\boldsymbol{z}'\\
 &  & +\int_{\mathcal{X}}\int_{\underline{b}\left(\boldsymbol{z}'\right)}^{\overline{b}\left(\boldsymbol{z}',n\right)}\frac{1}{h^{2+d}}K_{f}'\left(\frac{\xi\left(b',\boldsymbol{z}',n\right)-v}{h},\frac{\boldsymbol{z}'-\boldsymbol{x}}{h}\right)\frac{1}{n-1}G\left(b',\boldsymbol{z}',n\right)\mathrm{d}b'\mathrm{d}\boldsymbol{z}'.
\end{eqnarray*}

Let
\begin{eqnarray*}
\beta_{\mathcal{G}^{n}}\left(b,\boldsymbol{z}\right) & \coloneqq & \mathrm{E}\left[\mathbbm{1}\left(N_{1}=n\right)G\left(b|\boldsymbol{X}_{1},N_{1}\right)\frac{1}{h^{d}}K_{\boldsymbol{X}}\left(\frac{\boldsymbol{X}_{1}-\boldsymbol{z}}{h}\right)\right]-G\left(b,\boldsymbol{z},n\right)\\
 & = & \int\sum_{m'\in\mathcal{N}}\int\mathbbm{1}\left(m'=n\right)\mathbbm{1}\left(b'\leq b\right)\frac{1}{h^{d}}K_{\boldsymbol{X}}\left(\frac{\boldsymbol{z}'-\boldsymbol{z}}{h}\right)g\left(b'|\boldsymbol{z}',m'\right)\pi\left(m'|\boldsymbol{z}'\right)\varphi\left(\boldsymbol{z}'\right)\mathrm{d}b'\mathrm{d}\boldsymbol{z}'-G\left(b,\boldsymbol{z},n\right)\\
 & = & \int_{\mathcal{X}}G\left(b,\boldsymbol{z}',n\right)\frac{1}{h^{d}}K_{\boldsymbol{X}}\left(\frac{\boldsymbol{z}'-\boldsymbol{z}}{h}\right)\mathrm{d}\boldsymbol{z}'-G\left(b,\boldsymbol{z},n\right)
\end{eqnarray*}
be the bias of the kernel estimator of $G\left(b,\boldsymbol{z},n\right)$.
It is clear that 
\begin{eqnarray*}
\mu_{\mathcal{G}^{n}}\left(v\right) & = & -\frac{1}{n-1}\int_{\mathcal{X}}\int_{\underline{b}\left(\boldsymbol{z}\right)}^{\overline{b}\left(\boldsymbol{z},n\right)}\frac{1}{h^{2+d}}K_{f}'\left(\frac{\xi\left(b,\boldsymbol{z},n\right)-v}{h},\frac{\boldsymbol{z}-x}{h}\right)\beta_{\mathcal{G}^{n}}\left(b,\boldsymbol{z}\right)\mathrm{d}b\mathrm{d}\boldsymbol{z}.\\
 & = & -\frac{1}{n-1}\int_{\mathcal{Y}}\int_{\frac{\underline{v}\left(h\boldsymbol{y}+\boldsymbol{x}\right)-v}{h}}^{\frac{\overline{v}\left(h\boldsymbol{y}+\boldsymbol{x}\right)-v}{h}}\frac{1}{h}K_{f}'\left(u,\boldsymbol{y}\right)\beta_{\mathcal{G}^{n}}\left(s\left(hu+v,h\boldsymbol{y}+\boldsymbol{x},n\right),h\boldsymbol{y}+\boldsymbol{x}\right)s'\left(hu+v,h\boldsymbol{y}+\boldsymbol{x},n\right)\mathrm{d}u\mathrm{d}\boldsymbol{y},
\end{eqnarray*}
where the second equality follows from change of variables.

Since we assume that $K_{1}$ is supported on $\left[-1,1\right]$
and differentiable on $\mathbb{R}$, it is straightforward to verify
that for all $\boldsymbol{z}\in\mathbb{H}\left(\boldsymbol{x},h\right)$,
$\beta_{\mathcal{G}}\left(\cdot,\boldsymbol{z}\right)$ satisfies
the assumptions of Theorem 2.27 of \citet*{Folland_text_book} and
therefore $\beta_{\mathcal{G}^{n}}\left(\cdot,\boldsymbol{z}\right)$
is differentiable on $\left[s\left(v_{l}\left(\boldsymbol{x}\right)-h,\boldsymbol{z},n\right),s\left(v_{u}\left(\boldsymbol{x}\right)+h,\boldsymbol{z},n\right)\right]$,
which is well-defined when $h$ is sufficiently small, and 
\[
\beta_{\mathcal{G}^{n}}'\left(b,\boldsymbol{z}\right)=\int_{\mathcal{X}}\frac{1}{h^{d}}K_{\boldsymbol{X}}\left(\frac{\boldsymbol{z}'-\boldsymbol{z}}{h}\right)g\left(b,\boldsymbol{z}',n\right)\mathrm{d}\boldsymbol{z}'-g\left(b,\boldsymbol{z},n\right)
\]
for all $b\in\left[s\left(v_{l}\left(\boldsymbol{x}\right)-h,\boldsymbol{z},n\right),s\left(v_{u}\left(\boldsymbol{x}\right)+h,\boldsymbol{z},n\right)\right]$. 

By the usual argument for the bias of kernel estimators (see, e.g.,
\citet*{Newey_Kernel_ET_1994}) and the assumption that $K_{0}$ is
supported on $\left[-1,1\right]$, for each $\left(b,\boldsymbol{z}\right)\in\mathcal{C}_{B,\boldsymbol{X}}^{n}$,
\begin{equation}
\left|\beta_{\mathcal{G}^{n}}\left(b,\boldsymbol{z}\right)\right|\leq\frac{h^{R+1}}{\left(R+1\right)!}\left(\underset{\left(\alpha_{1},...,\alpha_{d}\right)\in\mathbb{S}_{R+1}}{\mathrm{sup}}\underset{\boldsymbol{z}'\in\mathbb{H}\left(\boldsymbol{z},h\right)}{\mathrm{sup}}\left|D_{2}^{\alpha_{1}}\cdots D_{1+d}^{\alpha_{d}}\,G\left(b,\boldsymbol{z}',n\right)\right|\right)\left(\int\left\Vert \boldsymbol{z}'\right\Vert _{1}^{R+1}\left|K_{\boldsymbol{X}}\left(\boldsymbol{z}'\right)\right|\mathrm{d}\boldsymbol{z}'\right)\label{eq:beta_G bound}
\end{equation}
and
\begin{equation}
\left|\beta_{\mathcal{G}^{n}}'\left(b,\boldsymbol{z}\right)\right|\leq\frac{h^{R+1}}{\left(R+1\right)!}\left(\underset{\left(\alpha_{1},...,\alpha_{d}\right)\in\mathbb{S}_{R+1}}{\mathrm{sup}}\underset{\boldsymbol{z}'\in\mathbb{H}\left(\boldsymbol{z},h\right)}{\mathrm{sup}}\left|D_{2}^{\alpha_{1}}\cdots D_{1+d}^{\alpha_{d}}\,g\left(b,\boldsymbol{z}',n\right)\right|\right)\left(\int\left\Vert \boldsymbol{z}'\right\Vert _{1}^{R+1}\left|K_{\boldsymbol{X}}\left(\boldsymbol{z}'\right)\right|\mathrm{d}\boldsymbol{z}'\right),\label{eq:beta_G_prime bound}
\end{equation}
when $h$ is sufficiently small. Since $\mathcal{C}_{B,\boldsymbol{X}}^{n}$
is a compact inner closed subset of $\mathcal{S}_{B,\boldsymbol{X}}^{n}$,
it is clear that $\beta_{\mathcal{G}^{n}}\left(b,\boldsymbol{z}\right)$
and $\beta_{\mathcal{G}^{n}}'\left(b,\boldsymbol{z}\right)$ are $O\left(h^{R+1}\right)$
uniformly in $\left(b,\boldsymbol{z}\right)\in\mathcal{C}_{B,\boldsymbol{X}}^{n}$. 

By change of variable, the fact that $s$ is twice continuously differentiable
(see Lemma A1 of GPV) and a mean value expansion, we have
\begin{eqnarray*}
\mu_{\mathcal{G}^{n}}\left(v\right) & = & -\frac{1}{n-1}\int_{\mathcal{X}}\frac{1}{h^{1+d}}K_{\boldsymbol{X}}^{0}\left(\frac{\boldsymbol{z}-\boldsymbol{x}}{h}\right)\left\{ \int_{\frac{\underline{v}\left(\boldsymbol{z}\right)-v}{h}}^{\frac{\overline{v}\left(\boldsymbol{z}\right)-v}{h}}K_{0}'\left(u\right)\beta_{\mathcal{G}^{n}}\left(s\left(hu+v,\boldsymbol{z},n\right),\boldsymbol{z}\right)s'\left(hu+v,\boldsymbol{z},n\right)\mathrm{d}u\right\} \mathrm{d}\boldsymbol{z}\\
 & = & -\frac{1}{n-1}\int_{\mathcal{X}}\frac{1}{h^{1+d}}K_{\boldsymbol{X}}^{0}\left(\frac{\boldsymbol{z}-\boldsymbol{x}}{h}\right)\left\{ \int_{\frac{\underline{v}\left(\boldsymbol{z}\right)-v}{h}}^{\frac{\overline{v}\left(\boldsymbol{z}\right)-v}{h}}K_{0}'\left(u\right)\left(\beta_{\mathcal{G}^{n}}\left(s\left(v,\boldsymbol{z},n\right),\boldsymbol{z}\right)s'\left(v,\boldsymbol{z},n\right)\right.\right.\\
 &  & \left.\left.+\left(\beta_{\mathcal{G}^{n}}'\left(s\left(\dot{v},\boldsymbol{z},n\right),\boldsymbol{z}\right)s'\left(\dot{v},\boldsymbol{z},n\right)^{2}+\beta_{\mathcal{G}^{n}}\left(s\left(\dot{v},\boldsymbol{z},n\right),\boldsymbol{z}\right)s''\left(\dot{v},\boldsymbol{z},n\right)\right)hu\right)\mathrm{d}u\right\} \mathrm{d}\boldsymbol{z},
\end{eqnarray*}
where $\dot{v}$ is the mean value (depending on $u$ and $\boldsymbol{z}$)
with $\left|\dot{v}-v\right|\leq h\left|u\right|$. By the triangle
inequality, we have 
\begin{eqnarray}
\left|\mu_{\mathcal{G}^{n}}\left(v\right)\right| & \apprle & \int_{\mathcal{X}}\frac{1}{h^{1+d}}\left|K_{\boldsymbol{X}}^{0}\left(\frac{\boldsymbol{z}-\boldsymbol{x}}{h}\right)\right|\left|\beta_{\mathcal{G}^{n}}\left(s\left(v,\boldsymbol{z},n\right),\boldsymbol{z}\right)s'\left(v,\boldsymbol{z},n\right)\int_{\frac{\underline{v}\left(\boldsymbol{z}\right)-v}{h}}^{\frac{\overline{v}\left(\boldsymbol{z}\right)-v}{h}}K_{0}'\left(u\right)\mathrm{d}u\right|\mathrm{d}\boldsymbol{z}\nonumber \\
 &  & +\int_{\mathcal{X}}\frac{1}{h^{1+d}}\left|K_{\boldsymbol{X}}^{0}\left(\frac{\boldsymbol{z}-\boldsymbol{x}}{h}\right)\right|\int_{\frac{\underline{v}\left(\boldsymbol{z}\right)-v}{h}}^{\frac{\overline{v}\left(\boldsymbol{z}\right)-v}{h}}\left|K_{0}'\left(u\right)u\right|\left|\beta_{\mathcal{G}^{n}}'\left(s\left(\dot{v},\boldsymbol{z},n\right),\boldsymbol{z}\right)s'\left(\dot{v},\boldsymbol{z},n\right)^{2}+\beta_{\mathcal{G}^{n}}\left(s\left(\dot{v},\boldsymbol{z},n\right),\boldsymbol{z}\right)s''\left(\dot{v},\boldsymbol{z},n\right)\right|\mathrm{d}u\mathrm{d}\boldsymbol{z}.\nonumber \\
\label{eq:mu_G order bound}
\end{eqnarray}
Since $K_{0}$ is assumed to be supported on $\left[-1,1\right]$,
$K_{\boldsymbol{X}}^{0}\left(\nicefrac{\left(\boldsymbol{z}-\boldsymbol{x}\right)}{h}\right)$
is zero for all $\boldsymbol{z}\notin\mathbb{H}\left(\boldsymbol{x},\overline{\delta}\right)$,
when $h$ is sufficiently small ($h<\overline{\delta}$). But if $h<\overline{\delta}$
and $\boldsymbol{z}\in\mathbb{H}\left(\boldsymbol{x},\overline{\delta}\right)$,
$\int_{\frac{\underline{v}\left(\boldsymbol{z}\right)-v}{h}}^{\frac{\overline{v}\left(\boldsymbol{z}\right)-v}{h}}K_{0}'\left(u\right)\mathrm{d}u=0$
for all $v\in I\left(\boldsymbol{x}\right)$ since $\mathcal{C}_{V,\boldsymbol{X}}$
is an inner closed subset of $\mathcal{S}_{V,\boldsymbol{X}}$ (i.e.,
for all $\boldsymbol{z}\in\mathbb{H}\left(\boldsymbol{x},\overline{\delta}\right)$,
$\left[v_{l}\left(\boldsymbol{x}\right)-\overline{\delta},v_{u}\left(\boldsymbol{x}\right)+\overline{\delta}\right]\subseteq\left(\underline{v}\left(\boldsymbol{z}\right),\overline{v}\left(\boldsymbol{z}\right)\right)$).
Therefore the first term on the right hand side of (\ref{eq:mu_G order bound})
vanishes when $h$ is sufficiently small. Therefore, 
\[
\underset{v\in I\left(\boldsymbol{x}\right)}{\mathrm{sup}}\left|\mu_{\mathcal{G}^{n}}\left(v\right)\right|\apprle\left(\int\frac{1}{h^{1+d}}\left|K_{\boldsymbol{X}}^{0}\left(\frac{\boldsymbol{z}-\boldsymbol{x}}{h}\right)\right|\mathrm{d}\boldsymbol{z}\right)\left\{ \underset{\left(u,\boldsymbol{z}\right)\in\mathcal{C}_{V,\boldsymbol{X}}}{\mathrm{sup}}\left|\beta_{\mathcal{G}^{n}}'\left(s\left(u,\boldsymbol{z},n\right),\boldsymbol{z}\right)s'\left(u,\boldsymbol{z},n\right)^{2}+\beta_{\mathcal{G}^{n}}\left(s\left(u,\boldsymbol{z},n\right),\boldsymbol{z}\right)s''\left(u,\boldsymbol{z},n\right)\right|\right\} .
\]
It follows from the fact that $\beta_{\mathcal{G}^{n}}\left(b,\boldsymbol{z}\right)$
and $\beta_{\mathcal{G}^{n}}'\left(b,\boldsymbol{z}\right)$ are $O\left(h^{R+1}\right)$
uniformly in $\left(b,\boldsymbol{z}\right)\in\mathcal{C}_{B,\boldsymbol{X}}^{n}$
and (\ref{eq:|K_X| integral order}) that 
\begin{equation}
\underset{v\in I\left(\boldsymbol{x}\right)}{\mathrm{sup}}\left|\mu_{\mathcal{G}}\left(v\right)\right|=O\left(h^{R}\right).\label{eq:sup miu_G rate}
\end{equation}

Standard arguments can be applied to verify that class $\left\{ \mathcal{G}^{n}\left(\cdot,\cdot,\cdot;v\right):v\in I\left(\boldsymbol{x}\right)\right\} $
is (uniformly) VC-type with respect to the envelope 
\begin{equation}
F_{\mathcal{G}^{n}}\left(\boldsymbol{z},\boldsymbol{z}'\right)\coloneqq\frac{1}{\left(n-1\right)\underline{C}_{g}h^{2+2d}}\left|K_{\boldsymbol{X}}^{0}\left(\frac{\boldsymbol{z}-\boldsymbol{x}}{h}\right)\right|\left|K_{\boldsymbol{X}}\left(\frac{\boldsymbol{z}'-\boldsymbol{z}}{h}\right)\right|+\frac{\underline{C}_{g}^{-1}\overline{\varphi}}{\left(n-1\right)h^{2+d}}\left|K_{\boldsymbol{X}}^{0}\left(\frac{\boldsymbol{z}-\boldsymbol{x}}{h}\right)\right|.\label{eq:F_G_n envelope}
\end{equation}
The CK inequality yields
\begin{align*}
 & \mathrm{E}\left[\underset{v\in I\left(\boldsymbol{x}\right)}{\mathrm{sup}}\left|\frac{1}{\left(L\right)_{2}}\sum_{\left(2\right)}\left\{ \mathcal{G}^{n}\left(\left(\boldsymbol{B}_{\cdot l},\boldsymbol{X}_{l},N_{l}\right),\left(\boldsymbol{B}_{\cdot k},\boldsymbol{X}_{k},N_{k}\right);v\right)-\mathcal{G}_{1}^{n}\left(\boldsymbol{B}_{\cdot l},\boldsymbol{X}_{l},N_{l};v\right)-\mathcal{G}_{2}^{n}\left(\boldsymbol{B}_{\cdot k},\boldsymbol{X}_{k},N_{k};v\right)+\mu_{\mathcal{G}^{n}}\left(v\right)\right\} \right|\right]\\
\apprle & L^{-1}\left(\mathrm{E}\left[F_{\mathcal{G}^{n}}\left(\boldsymbol{X}_{1},\boldsymbol{X}_{2}\right)^{2}\right]\right)^{\nicefrac{1}{2}}\\
= & O\left(\left(Lh^{2+d}\right)^{-1}\right).
\end{align*}

It is clear from the definition that when $h$ is sufficiently small,
the class $\left\{ \mathcal{G}_{1}^{n}\left(\cdot;v\right):v\in I\left(\boldsymbol{x}\right)\right\} $
is uniformly VC-type with respect to the envelope
\[
F_{\mathcal{G}_{1}^{n}}\left(\boldsymbol{z}\right)\coloneqq\frac{1}{h^{2+d}}\left|K_{\boldsymbol{X}}^{0}\left(\frac{\boldsymbol{z}-\boldsymbol{x}}{h}\right)\right|\left\{ \underset{\left(b',\boldsymbol{z}'\right)\in\mathcal{C}_{B,\boldsymbol{X}}^{n}}{\mathrm{sup}}\left|\beta_{\mathcal{G}^{n}}\left(b',\boldsymbol{z}'\right)\right|\right\} .
\]
The VW inequality yields 
\[
\mathrm{E}\left[\underset{v\in I\left(\boldsymbol{x}\right)}{\mathrm{sup}}\left|\frac{1}{L}\sum_{l=1}^{L}\mathcal{G}_{1}^{n}\left(\boldsymbol{B}_{\cdot l},\boldsymbol{X}_{l},N_{l};v\right)-\mu_{\mathcal{G}^{n}}\left(v\right)\right|\right]\leq L^{-\nicefrac{1}{2}}\mathrm{E}\left[F_{\mathcal{G}_{1}^{n}}\left(\boldsymbol{X}_{1}\right)^{2}\right]^{\nicefrac{1}{2}}=O\left(\frac{h^{R-1}}{\left(Lh^{1+d}\right)^{\nicefrac{1}{2}}}\right),
\]
where the inequality holds when $h$ is sufficiently small and the
equality follows from a standard change of variable argument and (\ref{eq:beta_G bound}).

Let 
\begin{eqnarray*}
\mathcal{G}^{\ddagger,n}\left(\left(\boldsymbol{b}.,\boldsymbol{z},m\right),\left(\boldsymbol{b}.',\boldsymbol{z}',m'\right);v\right) & \coloneqq & -\mathbbm{1}\left(m=n\right)\frac{1}{m}\sum_{i=1}^{m}\frac{1}{h^{2+2d}}K_{f}'\left(\frac{\xi\left(b_{i},\boldsymbol{z},m\right)-v}{h},\frac{\boldsymbol{z}-\boldsymbol{x}}{h}\right)\frac{1}{\left(m-1\right)g\left(b_{i},\boldsymbol{z},m\right)}\\
 &  & \times\mathbbm{1}\left(m'=m\right)\frac{1}{m'}\sum_{j=1}^{m'}\mathbbm{1}\left(b_{j}'\leq b_{i}\right)K_{\boldsymbol{X}}\left(\frac{\boldsymbol{z}'-\boldsymbol{z}}{h}\right),
\end{eqnarray*}
\begin{eqnarray*}
\mathcal{G}_{2}^{\ddagger,n}\left(\boldsymbol{b}.,\boldsymbol{z},m;v\right) & \coloneqq & \mathrm{E}\left[\mathcal{G}^{\ddagger,n}\left(\left(\boldsymbol{B}_{\cdot1},\boldsymbol{X}_{1},N_{1}\right),\left(\boldsymbol{b}.,\boldsymbol{z},m\right);v\right)\right]\\
 & = & -\mathbbm{1}\left(m=n\right)\frac{1}{m}\sum_{i=1}^{m}\int_{\mathcal{X}}\int_{\underline{b}\left(\boldsymbol{z}'\right)}^{\overline{b}\left(\boldsymbol{z}',n\right)}\frac{1}{h^{2+2d}}K_{f}'\left(\frac{\xi\left(b',\boldsymbol{z}',n\right)-v}{h},\frac{\boldsymbol{z}'-\boldsymbol{x}}{h}\right)\frac{1}{\left(n-1\right)}\mathbbm{1}\left(b_{i}\leq b'\right)K\left(\frac{\boldsymbol{z}-\boldsymbol{z}'}{h}\right)\mathrm{d}b'\mathrm{d}\boldsymbol{z}',
\end{eqnarray*}
where the second equality follows from LIE and 
\[
\mu_{\mathcal{G}^{\ddagger,n}}\left(v\right)\coloneqq\mathrm{E}\left[\mathcal{G}^{\ddagger,n}\left(\left(\boldsymbol{B}_{\cdot1},\boldsymbol{X}_{1},N_{1}\right),\left(\boldsymbol{B}_{\cdot2},\boldsymbol{X}_{2},N_{2}\right);v\right)\right].
\]
It is straightforward to check that 
\[
\mathcal{G}_{2}^{n}\left(\boldsymbol{B}_{\cdot l},\boldsymbol{X}_{l},N_{l};v\right)-\mu_{\mathcal{G}^{n}}\left(v\right)=\mathcal{G}_{2}^{\ddagger,n}\left(\boldsymbol{B}_{\cdot l},\boldsymbol{X}_{l},N_{l};v\right)-\mu_{\mathcal{G}^{\ddagger,n}}\left(v\right),\textrm{ for all \ensuremath{l=1,...,L}}
\]
and thus 
\[
\frac{1}{L}\sum_{l=1}^{L}\mathcal{G}_{2}^{n}\left(\boldsymbol{B}_{\cdot l},\boldsymbol{X}_{l},N_{l};v\right)-\mu_{\mathcal{G}^{n}}\left(v\right)=\frac{1}{L}\sum_{l=1}^{L}\mathcal{G}_{2}^{\ddagger,n}\left(\boldsymbol{B}_{\cdot l},\boldsymbol{X}_{l},N_{l};v\right)-\mu_{\mathcal{G}^{\ddagger,n}}\left(v\right).
\]
It is also easy to check that 
\[
\mu_{\mathcal{G}^{\ddagger,n}}\left(v\right)=\mu_{\mathcal{G}^{n}}\left(v\right)+\left(-\frac{1}{n-1}\int_{\mathcal{X}}\int_{\underline{b}\left(\boldsymbol{z}'\right)}^{\overline{b}\left(\boldsymbol{z}',n\right)}\frac{1}{h^{2+d}}K_{f}'\left(\frac{\xi\left(b',\boldsymbol{z}',n\right)-v}{h},\frac{\boldsymbol{z}'-\boldsymbol{x}}{h}\right)G\left(b',\boldsymbol{z}',n\right)\mathrm{d}b'\mathrm{d}\boldsymbol{z}'\right).
\]
And it follows from (\ref{eq:sup integral K_f'*G bound}) and (\ref{eq:sup miu_G rate})
that $\underset{v\in I\left(\boldsymbol{x}\right)}{\mathrm{sup}}\left|\mu_{\mathcal{G}^{\ddagger,n}}\left(v\right)\right|=O\left(1\right).$

By Jensen's inequality and LIE, we have
\begin{align*}
 & \mathrm{E}\left[\mathcal{G}_{2}^{\ddagger,n}\left(\boldsymbol{B}_{\cdot1},\boldsymbol{X}_{1},N_{1};v\right)^{2}\right]\\
\leq & \mathrm{E}\left[\mathbbm{1}\left(N_{1}=n\right)\frac{1}{N_{1}}\sum_{j=1}^{N_{1}}\left\{ \int_{\mathcal{X}}\int_{\underline{b}\left(\boldsymbol{z}'\right)}^{\overline{b}\left(\boldsymbol{z}',n\right)}\frac{1}{h^{2+d}}K_{f}'\left(\frac{\xi\left(b',\boldsymbol{z}',n\right)-v}{h},\frac{\boldsymbol{z}'-\boldsymbol{x}}{h}\right)\frac{1}{\left(n-1\right)}\mathbbm{1}\left(B_{j1}\leq b'\right)\frac{1}{h^{d}}K_{\boldsymbol{X}}\left(\frac{\boldsymbol{X}_{1}-\boldsymbol{z}'}{h}\right)\mathrm{d}b'\mathrm{d}\boldsymbol{z}'\right\} ^{2}\right]\\
= & \mathrm{E}\left[\frac{\mathbbm{1}\left(N_{1}=n\right)}{\left(n-1\right)^{2}}\int_{\mathcal{X}}\int_{\underline{b}\left(\boldsymbol{z}'\right)}^{\overline{b}\left(\boldsymbol{z}',n\right)}\int_{\mathcal{X}}\int_{\underline{b}\left(\boldsymbol{z}''\right)}^{\overline{b}\left(\boldsymbol{z}'',n\right)}\frac{1}{h^{4+4d}}K_{f}'\left(\frac{\xi\left(b',\boldsymbol{z}',n\right)-v}{h},\frac{\boldsymbol{z}'-\boldsymbol{x}}{h}\right)K_{\boldsymbol{X}}\left(\frac{\boldsymbol{X}_{1}-\boldsymbol{z}'}{h}\right)\right.\\
 & \left.\times K_{f}'\left(\frac{\xi\left(b'',\boldsymbol{z}'',n\right)-v}{h},\frac{\boldsymbol{z}''-\boldsymbol{x}}{h}\right)K_{\boldsymbol{X}}\left(\frac{\boldsymbol{X}_{1}-\boldsymbol{z}''}{h}\right)G\left(\mathrm{min}\left\{ b',b''\right\} |\boldsymbol{X}_{1},N_{1}\right)\mathrm{d}b''\mathrm{d}\boldsymbol{z}''\mathrm{d}b'\mathrm{d}\boldsymbol{z}'\right]\\
= & \int_{\mathcal{X}}\int_{\mathcal{X}}\int_{\underline{b}\left(\boldsymbol{z}'\right)}^{\overline{b}\left(\boldsymbol{z}',n\right)}\int_{\mathcal{X}}\int_{\underline{b}\left(\boldsymbol{z}''\right)}^{\overline{b}\left(\boldsymbol{z}'',n\right)}\frac{1}{\left(n-1\right)^{2}h^{4+4d}}K_{f}'\left(\frac{\xi\left(b',\boldsymbol{z}',n\right)-v}{h},\frac{\boldsymbol{z}'-\boldsymbol{x}}{h}\right)K_{\boldsymbol{X}}\left(\frac{\boldsymbol{z}-\boldsymbol{z}'}{h}\right)\\
 & \times K_{f}'\left(\frac{\xi\left(b'',\boldsymbol{z}'',n\right)-v}{h},\frac{\boldsymbol{z}''-\boldsymbol{x}}{h}\right)K_{\boldsymbol{X}}\left(\frac{\boldsymbol{z}-\boldsymbol{z}''}{h}\right)G\left(\mathrm{min}\left\{ b',b''\right\} ,\boldsymbol{z},n\right)\mathrm{d}b''\mathrm{d}\boldsymbol{z}''\mathrm{d}b'\mathrm{d}\boldsymbol{z}'\mathrm{d}\boldsymbol{z}.
\end{align*}
By the change of variables, we have
\begin{eqnarray*}
\mathrm{E}\left[\mathcal{G}_{2}^{\ddagger,n}\left(\boldsymbol{B}_{\cdot1},\boldsymbol{X}_{1},N_{1};v\right)^{2}\right] & = & \int_{\mathcal{X}}\int_{\mathcal{X}}\int_{\underline{b}\left(\boldsymbol{z}'\right)}^{\overline{b}\left(\boldsymbol{z}',n\right)}\int_{\mathcal{X}}\int_{\underline{b}\left(\boldsymbol{z}''\right)}^{\overline{b}\left(\boldsymbol{z}'',n\right)}\frac{1}{\left(n-1\right)^{2}h^{4+4d}}K_{f}'\left(\frac{\xi\left(b',\boldsymbol{z}',n\right)-v}{h},\frac{\boldsymbol{z}'-\boldsymbol{x}}{h}\right)K_{\boldsymbol{X}}\left(\frac{\boldsymbol{z}-\boldsymbol{z}'}{h}\right)\\
 &  & \times K_{f}'\left(\frac{\xi\left(b'',\boldsymbol{z}'',n\right)-v}{h},\frac{\boldsymbol{z}''-\boldsymbol{x}}{h}\right)K_{\boldsymbol{X}}\left(\frac{\boldsymbol{z}-\boldsymbol{z}''}{h}\right)G\left(\mathrm{min}\left\{ b',b''\right\} ,\boldsymbol{z},n\right)\mathrm{d}b''\mathrm{d}\boldsymbol{z}''\mathrm{d}b'\mathrm{d}\boldsymbol{z}'\mathrm{d}\boldsymbol{z}\\
 & = & \int_{\mathcal{Y}}\int_{\mathcal{Y}}\int_{\frac{\underline{v}\left(h\boldsymbol{y}''+\boldsymbol{x}\right)-v}{h}}^{\frac{\overline{v}\left(h\boldsymbol{y}''+\boldsymbol{x}\right)-v}{h}}\int_{\mathcal{Y}}\int_{\frac{\underline{v}\left(h\boldsymbol{y}'+\boldsymbol{x}\right)-v}{h}}^{\frac{\overline{v}\left(h\boldsymbol{y}'+\boldsymbol{x}\right)-v}{h}}\frac{1}{\left(n-1\right)^{2}h^{2+d}}K_{f}'\left(u,\boldsymbol{y}'\right)K_{\boldsymbol{X}}\left(\boldsymbol{y}'-\boldsymbol{y}\right)K_{f}'\left(w,\boldsymbol{y}''\right)K_{\boldsymbol{X}}\left(\boldsymbol{y}''-\boldsymbol{y}\right)\\
 &  & \times G\left(\mathrm{min}\left\{ s\left(hw+v,h\boldsymbol{y}''+\boldsymbol{x},n\right),s\left(hu+v,h\boldsymbol{y}'+\boldsymbol{x},n\right)\right\} ,h\boldsymbol{y}+\boldsymbol{x},n\right)\\
 &  & \left.\times s'\left(hu+v,h\boldsymbol{y}'+\boldsymbol{x},n\right)s'\left(hw+v,h\boldsymbol{y}''+\boldsymbol{x},n\right)\mathrm{d}u\mathrm{d}\boldsymbol{y}'\mathrm{d}w\mathrm{d}\boldsymbol{y}''\mathrm{d}\boldsymbol{y}\right\} .
\end{eqnarray*}
It follows from symmetry that
\begin{eqnarray*}
\mathrm{E}\left[\mathcal{G}_{2}^{\ddagger,n}\left(\boldsymbol{B}_{\cdot1},\boldsymbol{X}_{1},N_{1};v\right)^{2}\right] & = & 2\int_{\mathcal{Y}}\int_{\mathcal{Y}}\int_{\frac{\underline{v}\left(h\boldsymbol{y}'+\boldsymbol{x}\right)-v}{h}}^{\frac{\overline{v}\left(h\boldsymbol{y}'+\boldsymbol{x}\right)-v}{h}}\int_{\mathcal{Y}}\int_{\frac{\underline{v}\left(h\boldsymbol{y}''+\boldsymbol{x}\right)-v}{h}}^{\frac{\overline{v}\left(h\boldsymbol{y}''+\boldsymbol{x}\right)-v}{h}}\frac{1}{\left(n-1\right)^{2}h^{2+d}}K_{f}'\left(u,\boldsymbol{y}'\right)K_{\boldsymbol{X}}\left(\boldsymbol{y}'-\boldsymbol{y}\right)K_{f}'\left(w,\boldsymbol{y}''\right)K_{\boldsymbol{X}}\left(\boldsymbol{y}''-\boldsymbol{y}\right)\\
 &  & \times\mathbbm{1}\left(s\left(hu+v,h\boldsymbol{y}''+\boldsymbol{x},n\right)\leq s\left(hw+v,h\boldsymbol{y}'+\boldsymbol{x},n\right)\right)G\left(s\left(hu+v,h\boldsymbol{y}''+\boldsymbol{x},n\right),h\boldsymbol{y}+\boldsymbol{x},n\right)\\
 &  & \times s'\left(hu+v,h\boldsymbol{y}''+\boldsymbol{x},n\right)s'\left(hw+v,h\boldsymbol{y}'+\boldsymbol{x},n\right)\mathrm{d}u\mathrm{d}\boldsymbol{y}''\mathrm{d}w\mathrm{d}\boldsymbol{y}'\mathrm{d}\boldsymbol{y}.
\end{eqnarray*}
By a mean value expansion, we have 
\[
\mathrm{E}\left[\mathcal{G}_{2}^{\ddagger,n}\left(\boldsymbol{B}_{\cdot1},\boldsymbol{X}_{1},N_{1};v\right)^{2}\right]=\varrho_{1}\left(v\right)+\varrho_{2}\left(v\right)
\]
where
\begin{eqnarray*}
\varrho_{1}\left(v\right) & \coloneqq & 2\int_{\mathcal{Y}}\int_{\mathcal{Y}}\int_{\frac{\underline{v}\left(h\boldsymbol{y}''+\boldsymbol{x}\right)-v}{h}}^{\frac{\overline{v}\left(h\boldsymbol{y}''+\boldsymbol{x}\right)-v}{h}}\int_{\mathcal{Y}}\int_{\frac{\underline{v}\left(h\boldsymbol{y}'+\boldsymbol{x}\right)-v}{h}}^{\frac{\overline{v}\left(h\boldsymbol{y}'+\boldsymbol{x}\right)-v}{h}}\frac{1}{\left(n-1\right)^{2}h^{2+d}}K_{f}'\left(u,\boldsymbol{y}'\right)K_{\boldsymbol{X}}\left(\boldsymbol{y}'-\boldsymbol{y}\right)K_{f}'\left(w,\boldsymbol{y}''\right)K_{\boldsymbol{X}}\left(\boldsymbol{y}''-\boldsymbol{y}\right)\\
 &  & \times\mathbbm{1}\left(s\left(hu+v,h\boldsymbol{y}'+\boldsymbol{x},n\right)\leq s\left(hw+v,h\boldsymbol{y}''+\boldsymbol{x},n\right)\right)G\left(s\left(v,h\boldsymbol{y}''+\boldsymbol{x},n\right),h\boldsymbol{y}+\boldsymbol{x},n\right)\\
 &  & \left.\times s'\left(v,h\boldsymbol{y}''+\boldsymbol{x},n\right)s'\left(hw+v,h\boldsymbol{y}'+\boldsymbol{x},n\right)\mathrm{d}u\mathrm{d}\boldsymbol{y}'\mathrm{d}w\mathrm{d}\boldsymbol{y}''\mathrm{d}\boldsymbol{y}\right\} 
\end{eqnarray*}
and
\begin{eqnarray*}
\varrho_{2}\left(v\right) & \coloneqq & 2\int_{\mathcal{Y}}\int_{\mathcal{Y}}\int_{\frac{\underline{v}\left(h\boldsymbol{y}''+\boldsymbol{x}\right)-v}{h}}^{\frac{\overline{v}\left(h\boldsymbol{y}''+\boldsymbol{x}\right)-v}{h}}\int_{\mathcal{Y}}\int_{\frac{\underline{v}\left(h\boldsymbol{y}'+\boldsymbol{x}\right)-v}{h}}^{\frac{\overline{v}\left(h\boldsymbol{y}'+\boldsymbol{x}\right)-v}{h}}\frac{1}{\left(n-1\right)^{2}h^{1+d}}K_{f}'\left(u,\boldsymbol{y}'\right)K_{\boldsymbol{X}}\left(\boldsymbol{y}'-\boldsymbol{y}\right)K_{f}'\left(w,\boldsymbol{y}''\right)K_{\boldsymbol{X}}\left(\boldsymbol{y}''-\boldsymbol{y}\right)\\
 &  & \times\mathbbm{1}\left(s\left(hu+v,h\boldsymbol{y}'+\boldsymbol{x},n\right)\leq s\left(hw+v,h\boldsymbol{y}''+\boldsymbol{x},n\right)\right)u\left\{ g\left(s\left(\dot{v},h\boldsymbol{y}''+\boldsymbol{x},n\right),h\boldsymbol{y}+\boldsymbol{x},n\right)s\left(\dot{v},h\boldsymbol{y}''+\boldsymbol{x},n\right)^{2}\right.\\
 &  & \left.+G\left(s\left(\dot{v},h\boldsymbol{y}''+\boldsymbol{x},n\right),h\boldsymbol{y}+\boldsymbol{x},n\right)s''\left(\dot{v},h\boldsymbol{y}''+\boldsymbol{x},n\right)\right\} s'\left(hw+v,h\boldsymbol{y}'+\boldsymbol{x},n\right)\mathrm{d}u\mathrm{d}\boldsymbol{y}'\mathrm{d}w\mathrm{d}\boldsymbol{y}''\mathrm{d}\boldsymbol{y}
\end{eqnarray*}
for some mean value $\dot{v}$ with $\left|\dot{v}-v\right|\leq h\left|u\right|$.
It is clear that when $h$ is sufficiently small,
\begin{eqnarray}
\underset{v\in I\left(\boldsymbol{x}\right)}{\mathrm{sup}}\left|\varrho_{2}\left(v\right)\right| & \apprle & h^{-\left(1+d\right)}\left\{ \underset{v\in I\left(\boldsymbol{x}\right)}{\mathrm{sup}}\underset{\left(u,\boldsymbol{z}',w,\boldsymbol{z}''\right)\in\mathbb{H}\left(\left(v,\boldsymbol{x}\right),\overline{\delta}\right)^{2}}{\mathrm{sup}}s'\left(u,\boldsymbol{z}',n\right)g\left(s\left(w,\boldsymbol{z}'',n\right),\boldsymbol{z},n\right)s'\left(w,\boldsymbol{z}'',n\right)^{2}\right.\nonumber \\
 &  & \left.+\underset{v\in I\left(\boldsymbol{x}\right)}{\mathrm{sup}}\underset{\left(u,\boldsymbol{z}',w,\boldsymbol{z}''\right)\in\mathbb{H}\left(\left(v,\boldsymbol{x}\right),\overline{\delta}\right)^{2}}{\mathrm{sup}}s'\left(u,\boldsymbol{z}',n\right)G\left(s\left(w,\boldsymbol{z}'',n\right),\boldsymbol{z},n\right)s''\left(w,\boldsymbol{z}'',n\right)\right\} \label{eq:rho_2 order bound}
\end{eqnarray}
and therefore $\underset{v\in I\left(\boldsymbol{x}\right)}{\mathrm{sup}}\left|\varrho_{2}\left(v\right)\right|=O\left(h^{-\left(1+d\right)}\right)$. 

When $h$ is sufficiently small,
\begin{align*}
 & \int_{\frac{\underline{v}\left(h\boldsymbol{y}'+\boldsymbol{x}\right)-v}{h}}^{\frac{\overline{v}\left(h\boldsymbol{y}'+\boldsymbol{x}\right)-v}{h}}K_{0}'\left(u\right)\mathbbm{1}\left(s\left(hu+v,h\boldsymbol{y}'+\boldsymbol{x},n\right)\leq s\left(hw+v,h\boldsymbol{y}''+\boldsymbol{x},n\right)\right)\mathrm{d}u\\
= & K_{0}\left(\frac{\xi\left(s\left(hw+v,h\boldsymbol{y}''+\boldsymbol{x},n\right),h\boldsymbol{y}'+\boldsymbol{x},n\right)-v}{h}\right)
\end{align*}
for all $\boldsymbol{y}',\boldsymbol{y}''\in\mathbb{H}\left(\boldsymbol{0},1\right)$,
$\left|w\right|\leq1$ and $v\in I\left(\boldsymbol{x}\right)$ and
thus
\begin{eqnarray*}
\varrho_{1}\left(v\right) & = & 2\int_{\mathcal{Y}}\int_{\mathcal{Y}}\int_{\frac{\underline{v}\left(h\boldsymbol{y}''+\boldsymbol{x}\right)-v}{h}}^{\frac{\overline{v}\left(h\boldsymbol{y}''+\boldsymbol{x}\right)-v}{h}}\int_{\mathcal{Y}}\frac{1}{\left(n-1\right)^{2}h^{2+d}}K_{0}\left(\frac{\xi\left(s\left(hw+v,h\boldsymbol{y}''+\boldsymbol{x},n\right),h\boldsymbol{y}'+\boldsymbol{x},n\right)-v}{h}\right)K_{\boldsymbol{X}}^{0}\left(\boldsymbol{y}'\right)K_{\boldsymbol{X}}\left(\boldsymbol{y}'-\boldsymbol{y}\right)\\
 &  & \times K_{f}'\left(w,\boldsymbol{y}''\right)K_{\boldsymbol{X}}\left(\boldsymbol{y}''-\boldsymbol{y}\right)G\left(s\left(v,h\boldsymbol{y}''+\boldsymbol{x},n\right),h\boldsymbol{y}+\boldsymbol{x},n\right)s'\left(v,h\boldsymbol{y}''+\boldsymbol{x},n\right)s'\left(hw+v,h\boldsymbol{y}'+\boldsymbol{x},n\right)\mathrm{d}\boldsymbol{y}''\mathrm{d}w\mathrm{d}\boldsymbol{y}'\mathrm{d}\boldsymbol{y}
\end{eqnarray*}
for all $v\in I\left(\boldsymbol{x}\right)$. 

Let 
\[
\xi_{\boldsymbol{x}}\coloneqq\left.\frac{\partial\xi\left(u,\boldsymbol{z},n\right)}{\partial\boldsymbol{z}}\right|_{\left(u,\boldsymbol{z}\right)=\left(s\left(v,\boldsymbol{x},n\right),\boldsymbol{x}\right)}\textrm{ and }\tau_{1}\left(u,\boldsymbol{z}_{1},\boldsymbol{z}_{2}\right)\coloneqq\xi\left(s\left(u,\boldsymbol{z}_{1},n\right),\boldsymbol{z}_{2},n\right).
\]
For all $\boldsymbol{y}',\boldsymbol{y}''\in\mathbb{H}\left(\boldsymbol{0},1\right)$,
$\boldsymbol{y}\in\mathbb{H}\left(\boldsymbol{0},2\right)$ and $\left|w\right|\leq1$,
\begin{align*}
 & \underset{v\in I\left(\boldsymbol{x}\right)}{\mathrm{sup}}\left|K_{0}\left(\frac{\xi\left(s\left(hw+v,h\boldsymbol{y}''+\boldsymbol{x},n\right),h\boldsymbol{y}'+\boldsymbol{x},n\right)-v}{h}\right)-K_{0}\left(w+\xi'\left(s\left(v,\boldsymbol{x},n\right),\boldsymbol{x},n\right)s_{\boldsymbol{x}}^{\mathrm{T}}\boldsymbol{y}''+\xi_{\boldsymbol{x}}^{\mathrm{T}}\boldsymbol{y}'\right)\right|\\
\apprle & h\left(\underset{v\in I\left(\boldsymbol{x}\right)}{\mathrm{sup}}\underset{\left(u,\boldsymbol{z}_{1},\boldsymbol{z}_{2}\right)\in\mathbb{H}\left(\left(v,\boldsymbol{x},\boldsymbol{x}\right),\overline{\delta}\right)}{\mathrm{sup}}\sum_{\left(\alpha_{1},...,\alpha_{1+2d}\right)\in\mathbb{S}_{2}}\left|D_{1}^{\alpha_{1}}\cdots D_{1+2d}^{\alpha_{1+2d}}\tau_{1}\left(u,\boldsymbol{z}_{1},\boldsymbol{z}_{2}\right)\right|\right).
\end{align*}

Let
\[
\tau_{2}\left(u,\boldsymbol{z}_{1},\boldsymbol{z}_{2},\boldsymbol{z}_{3};v\right)\coloneqq s'\left(u,\boldsymbol{z}_{1},n\right)G\left(s\left(v,\boldsymbol{z}_{1},n\right),\boldsymbol{z}_{3},n\right)s'\left(v,\boldsymbol{z}_{2},n\right).
\]
For all $\boldsymbol{y}',\boldsymbol{y}''\in\mathbb{H}\left(\boldsymbol{0},1\right)$,
$\boldsymbol{y}\in\mathbb{H}\left(\boldsymbol{0},2\right)$ and $\left|w\right|\leq1$,
\begin{align*}
 & \underset{v\in I\left(\boldsymbol{x}\right)}{\mathrm{sup}}\left|s'\left(hw+v,h\boldsymbol{y}'+\boldsymbol{x},n\right)G\left(s\left(v,h\boldsymbol{y}'+\boldsymbol{x},n\right),h\boldsymbol{y}+\boldsymbol{x},n\right)s'\left(v,h\boldsymbol{y}''+\boldsymbol{x},n\right)-s'\left(v,\boldsymbol{x},n\right)^{2}G\left(s\left(v,\boldsymbol{x},n\right),\boldsymbol{x},n\right)\right|\\
\apprle & h\left(\underset{v\in I\left(\boldsymbol{x}\right)}{\mathrm{sup}}\underset{\left(u,\boldsymbol{z}_{1},\boldsymbol{z}_{2},\boldsymbol{z}_{3}\right)\in\mathbb{H}\left(\left(v,\boldsymbol{x},\boldsymbol{x}\right),\overline{\delta}\right)\times\mathbb{H}\left(\boldsymbol{x},2\overline{\delta}\right)}{\mathrm{sup}}\sum_{j=1}^{1+3d}\left|D_{j}\tau_{2}\left(u,\boldsymbol{z}_{1},\boldsymbol{z}_{2},\boldsymbol{z}_{3};v\right)\right|\right),
\end{align*}
when $h$ is sufficiently small. Therefore by the change of variable
argument, we have
\begin{eqnarray*}
\varrho_{1}\left(v\right) & = & \frac{2s'\left(v,\boldsymbol{x},n\right)^{2}G\left(s\left(v,\boldsymbol{x},n\right),\boldsymbol{x},n\right)}{h^{2+d}\left(n-1\right)^{2}}\int\int\int\int K_{\boldsymbol{X}}\left(\boldsymbol{y}'-\boldsymbol{y}\right)K_{f}'\left(w,\boldsymbol{y}''\right)K_{\boldsymbol{X}}\left(\boldsymbol{y}''-\boldsymbol{y}\right)\\
 &  & \times K_{f}\left(w+\xi'\left(s\left(v,\boldsymbol{x},n\right),\boldsymbol{x},n\right)s_{\boldsymbol{x}}^{\mathrm{T}}\boldsymbol{y}''+\xi_{\boldsymbol{x}}^{\mathrm{T}}\boldsymbol{y}'\right)\mathrm{d}w\mathrm{d}\boldsymbol{y}'\mathrm{d}\boldsymbol{y}''\mathrm{d}\boldsymbol{y}+O\left(h^{-\left(1+d\right)}\right),
\end{eqnarray*}
where the remainder term is uniform in $v\in I\left(\boldsymbol{x}\right)$,
when $h$ is sufficiently small. We note that the leading term on
the right hand side of the above displayed equation vanishes since
the integrand is an odd function. Therefore we have 
\begin{equation}
\sigma_{\mathcal{G}_{2}^{\ddagger,n}}^{2}\coloneqq\underset{v\in I\left(\boldsymbol{x}\right)}{\mathrm{sup}}\mathrm{E}\left[\mathcal{G}_{2}^{\ddagger,n}\left(\boldsymbol{B}_{\cdot1},\boldsymbol{X}_{1},N_{1};v\right)^{2}\right]\leq\underset{v\in I\left(\boldsymbol{x}\right)}{\mathrm{sup}}\left|\varrho_{1}\left(v\right)\right|+\underset{v\in I\left(\boldsymbol{x}\right)}{\mathrm{sup}}\left|\varrho_{2}\left(v\right)\right|=O\left(h^{-\left(1+d\right)}\right).\label{eq:sigma_G_ddagger_2 rate}
\end{equation}

Since standard arguments can be applied to verify that the class $\left\{ \mathcal{G}^{\ddagger,n}\left(\cdot,\cdot;v\right):v\in I\left(\boldsymbol{x}\right)\right\} $
is (uniformly) VC-type with respect to the envelope 
\begin{equation}
F_{\mathcal{G}^{\ddagger,n}}\left(\boldsymbol{z},\boldsymbol{z}'\right)\coloneqq\frac{1}{\left(n-1\right)\underline{C}_{g}h^{2+2d}}\left|K_{\boldsymbol{X}}^{0}\left(\frac{\boldsymbol{z}-\boldsymbol{x}}{h}\right)\right|\left|K_{\boldsymbol{X}}\left(\frac{\boldsymbol{z}'-\boldsymbol{z}}{h}\right)\right|,\label{eq:F_G_ddagger envelope}
\end{equation}
it follows from \citet[Lemma 5.4]{Chen_Kato_U_Process} that the class
$\left\{ \mathcal{G}_{2}^{\ddagger,n}\left(\cdot;v\right):v\in I\left(\boldsymbol{x}\right)\right\} $
is uniformly VC-type with respect to the envelope
\begin{eqnarray*}
F_{\mathcal{G}_{2}^{\ddagger,n}}\left(\boldsymbol{z}\right) & \coloneqq & \int F_{\mathcal{G}^{\ddagger,n}}\left(\boldsymbol{z}',\boldsymbol{z}\right)\varphi\left(\boldsymbol{z}'\right)\mathrm{d}\boldsymbol{z}'\\
 & = & \frac{1}{\left(n-1\right)\underline{C}_{g}}\int\frac{1}{h^{2+2d}}\left|K_{\boldsymbol{X}}^{0}\left(\frac{\boldsymbol{z}'-\boldsymbol{x}}{h}\right)\right|\left|K_{\boldsymbol{X}}\left(\frac{\boldsymbol{z}'-\boldsymbol{z}}{h}\right)\right|\varphi\left(\boldsymbol{z}'\right)\mathrm{d}\boldsymbol{z}'.
\end{eqnarray*}
The CCK inequality yields 
\begin{eqnarray*}
\mathrm{E}\left[\underset{v\in I\left(\boldsymbol{x}\right)}{\mathrm{sup}}\left|\frac{1}{L}\sum_{l=1}^{L}\mathcal{G}_{2}^{\ddagger,n}\left(\boldsymbol{B}_{\cdot l},\boldsymbol{X}_{l},N_{l};v\right)-\mu_{\mathcal{G}^{\ddagger,n}}\left(v\right)\right|\right] & \leq & C_{1}\left\{ L^{-\nicefrac{1}{2}}\sigma_{\mathcal{G}_{2}^{\ddagger,n}}\mathrm{log}\left(C_{2}L\right)^{\nicefrac{1}{2}}+L^{-1}\left\Vert F_{\mathcal{G}_{2}^{\ddagger,n}}\right\Vert _{\mathcal{X}}\mathrm{log}\left(C_{2}L\right)\right\} \\
 & = & O\left(\left(\frac{\mathrm{log}\left(L\right)}{Lh^{1+d}}\right)^{\nicefrac{1}{2}}+\frac{\mathrm{log}\left(L\right)}{Lh^{2+d}}\right),
\end{eqnarray*}
where the inequality is non-asymptotic and the equality follows from
(\ref{eq:sigma_G_ddagger_2 rate}) and $\left\Vert F_{\mathcal{G}_{2}^{n}}\right\Vert _{\mathcal{X}}=O\left(h^{-\left(2+d\right)}\right)$
(which follows from change of variables). 

It is easy to check 
\begin{align*}
 & \underset{v\in I\left(\boldsymbol{x}\right)}{\mathrm{sup}}\left|\frac{1}{L^{2}\left(L-1\right)}\sum_{l\neq k}\mathcal{G}^{n}\left(\left(\boldsymbol{B}_{\cdot l},\boldsymbol{X}_{l},N_{l}\right),\left(\boldsymbol{B}_{\cdot k},\boldsymbol{X}_{k},N_{k}\right);v\right)\right|\\
\apprle & \frac{1}{L^{2}\left(L-1\right)}\sum_{l\neq k}\left\{ \frac{1}{h^{2+2d}}\left|K_{\boldsymbol{X}}^{0}\left(\frac{\boldsymbol{X}_{l}-\boldsymbol{x}}{h}\right)\right|\left|K_{\boldsymbol{X}}\left(\frac{\boldsymbol{X}_{l}-\boldsymbol{X}_{k}}{h}\right)\right|+\frac{1}{h^{2+d}}\left|K_{\boldsymbol{X}}^{0}\left(\frac{\boldsymbol{X}_{l}-\boldsymbol{x}}{h}\right)\right|\right\} \\
= & O_{p}\left(\left(Lh^{2}\right)^{-1}\right),
\end{align*}
where the equality follows from change of variables and Markov's inequality
and 
\begin{eqnarray*}
\underset{v\in I\left(\boldsymbol{x}\right)}{\mathrm{sup}}\left|\frac{1}{L^{2}}\sum_{l=1}^{L}\mathcal{G}^{n}\left(\left(\boldsymbol{B}_{\cdot l},\boldsymbol{X}_{l},N_{l}\right),\left(\boldsymbol{B}_{\cdot l},\boldsymbol{X}_{l},N_{l}\right);v\right)\right| & \apprle & \frac{1}{L^{\text{2}}}\sum_{l=1}^{L}\left\{ \frac{1}{h^{2+2d}}\left|K_{\boldsymbol{X}}^{0}\left(\frac{\boldsymbol{X}_{l}-\boldsymbol{x}}{h}\right)\right|+\frac{1}{h^{2+d}}\left|K_{\boldsymbol{X}}^{0}\left(\frac{\boldsymbol{X}_{l}-\boldsymbol{x}}{h}\right)\right|\right\} \\
 & = & O_{p}\left(\left(Lh^{2+d}\right)^{-1}\right),
\end{eqnarray*}
where the equality follows from change of variables and Markov's inequality.
Now it follows that 
\[
\underset{v\in I\left(\boldsymbol{x}\right)}{\mathrm{sup}}\left|\varDelta_{1}^{\ddagger}\left(v\right)\right|=O_{p}\left(h^{R}+\left(\frac{\mathrm{log}\left(L\right)}{Lh^{1+d}}\right)^{\nicefrac{1}{2}}+\frac{\mathrm{log}\left(L\right)}{Lh^{2+d}}\right)
\]
and the conclusion follows.\end{proof}
\begin{lem}
\label{Lemma 3}Suppose that Assumptions 1 - 3 hold. Let $\boldsymbol{x}$
be an interior point of $\mathcal{X}$ and $n\in\mathcal{N}$ be fixed.
Then
\[
\widehat{f}_{GPV}\left(v,\boldsymbol{x},n\right)-f\left(v|\boldsymbol{x}\right)\varphi\left(\boldsymbol{x}\right)\pi\left(n|\boldsymbol{x}\right)=\frac{1}{L}\sum_{l=1}^{L}\left\{ \mathcal{M}_{2}^{n}\left(\boldsymbol{B}_{\cdot l},\boldsymbol{X}_{l},N_{l};v\right)-\mu_{\mathcal{M}^{n}}\left(v\right)\right\} +O_{p}\left(\left(\frac{\mathrm{log}\left(L\right)}{Lh^{1+d}}\right)^{\nicefrac{1}{2}}+\frac{\mathrm{log}\left(L\right)}{Lh^{3+d}}+h^{R}\right),
\]
where the remainder term is uniform in $v\in I\left(\boldsymbol{x}\right)$.
\end{lem}
\begin{proof}[Proof of Lemma \ref{Lemma 3}] The Hoeffding decomposition
yields
\begin{align*}
 & \frac{1}{L^{2}}\sum_{l=1}^{L}\sum_{k=1}^{L}\mathcal{M}^{n}\left(\left(\boldsymbol{B}_{\cdot l},\boldsymbol{X}_{l},N_{l}\right),\left(\boldsymbol{B}_{\cdot k},\boldsymbol{X}_{k},N_{k}\right);v\right)\\
= & \mu_{\mathcal{M}^{n}}\left(v\right)+\left\{ \frac{1}{L}\sum_{l=1}^{L}\mathcal{M}_{1}^{n}\left(\boldsymbol{B}_{\cdot l},\boldsymbol{X}_{l},N_{l};v\right)-\mu_{\mathcal{M}^{n}}\left(v\right)\right\} +\left\{ \frac{1}{L}\sum_{l=1}^{L}\mathcal{M}_{2}^{n}\left(\boldsymbol{B}_{\cdot l},\boldsymbol{X}_{l},N_{l};v\right)-\mu_{\mathcal{M}^{n}}\left(v\right)\right\} \\
 & +\frac{1}{L\left(L-1\right)}\sum_{l\neq k}\left\{ \mathcal{M}^{n}\left(\left(\boldsymbol{B}_{\cdot l},\boldsymbol{X}_{l},N_{l}\right),\left(\boldsymbol{B}_{\cdot k},\boldsymbol{X}_{k},N_{k}\right);v\right)-\mathcal{M}_{1}^{n}\left(\boldsymbol{B}_{\cdot l},\boldsymbol{X}_{l},N_{l};v\right)-\mathcal{M}_{2}^{n}\left(\boldsymbol{B}_{\cdot k},\boldsymbol{X}_{k},N_{k};v\right)+\mu_{\mathcal{M}^{n}}\left(v\right)\right\} \\
 & +\frac{1}{L^{2}}\sum_{l=1}^{L}\mathcal{M}^{n}\left(\left(\boldsymbol{B}_{\cdot l},\boldsymbol{X}_{l},N_{l}\right),\left(\boldsymbol{B}_{\cdot l},\boldsymbol{X}_{l},N_{l}\right);v\right)-\frac{1}{L^{2}\left(L-1\right)}\sum_{l\neq k}\mathcal{M}^{n}\left(\left(\boldsymbol{B}_{\cdot l},\boldsymbol{X}_{l},N_{l}\right),\left(\boldsymbol{B}_{\cdot k},\boldsymbol{X}_{k},N_{k}\right);v\right).
\end{align*}

Let 
\begin{eqnarray*}
\beta_{\mathcal{M}^{n}}\left(b,\boldsymbol{z}\right) & \coloneqq & \mathrm{E}\left[\mathbbm{1}\left(N_{1}=n\right)\frac{1}{N_{1}}\sum_{j=1}^{N_{1}}\frac{1}{h^{1+d}}K_{g}\left(\frac{B_{j1}-b}{h}\right)K_{\boldsymbol{X}}\left(\frac{\boldsymbol{X}_{1}-\boldsymbol{z}}{h}\right)\right]-g\left(b,\boldsymbol{z},n\right)\\
 & = & \int_{\mathcal{X}}\int_{\underline{b}\left(\boldsymbol{z}'\right)}^{\overline{b}\left(\boldsymbol{z}',n\right)}\frac{1}{h^{1+d}}K_{g}\left(\frac{b'-b}{h}\right)K_{\boldsymbol{X}}\left(\frac{\boldsymbol{z}'-\boldsymbol{z}}{h}\right)g\left(b',\boldsymbol{z}',n\right)\mathrm{d}b'\mathrm{d}\boldsymbol{z}'-g\left(b,\boldsymbol{z},n\right)
\end{eqnarray*}
where the second equality follows from LIE. 

By the definition of $\mathcal{M}^{n}\left(\cdot,\cdot\right)$, $\mu_{\mathcal{M}^{n}}\left(v\right)$
is given by
\begin{eqnarray*}
\mu_{\mathcal{M}^{n}}\left(v\right) & = & -\int_{\mathcal{X}}\sum_{m\in\mathcal{N}}\int\cdots\int\mathbbm{1}\left(m=n\right)\frac{1}{m}\sum_{i=1}^{m}\frac{1}{h^{2+d}}K_{f}'\left(\frac{\xi\left(b_{i},\boldsymbol{z},n\right)-v}{h},\frac{\boldsymbol{z}-\boldsymbol{x}}{h}\right)\frac{G\left(b_{i},\boldsymbol{z},n\right)\beta_{\mathcal{M}^{n}}\left(b_{i},\boldsymbol{z}\right)}{\left(m-1\right)g\left(b_{i},\boldsymbol{z},n\right)^{2}}\\
 &  & \times\left(\prod_{i=1}^{m}g\left(b_{i}|\boldsymbol{z},m\right)\right)\pi\left(m|\boldsymbol{z}\right)\varphi\left(\boldsymbol{z}\right)\mathrm{d}b_{1}\cdots\mathrm{d}b_{m}\mathrm{d}\boldsymbol{z}\\
 & = & -\frac{1}{n-1}\int_{\mathcal{X}}\int_{\underline{b}\left(\boldsymbol{z}\right)}^{\overline{b}\left(\boldsymbol{z},n\right)}\frac{1}{h^{2+d}}K_{f}'\left(\frac{\xi\left(b,\boldsymbol{z},n\right)-v}{h},\frac{\boldsymbol{z}-\boldsymbol{x}}{h}\right)\frac{G\left(b,\boldsymbol{z},n\right)\beta_{\mathcal{M}^{n}}\left(b,\boldsymbol{z}\right)}{g\left(b,\boldsymbol{z},n\right)}\mathrm{d}b\mathrm{d}\boldsymbol{z}.
\end{eqnarray*}
It is clear that $\beta_{\mathcal{M}^{n}}\left(b,\boldsymbol{z}\right)$
is the bias of the kernel estimator for $g\left(b,\boldsymbol{z},n\right)$.
Since we assume that $K_{0}$ is supported on $\left[-1,1\right]$
and differentiable everywhere on $\mathbb{R}$, it is straightforward
to verify that for $\beta_{\mathcal{M}^{n}}\left(b,\boldsymbol{z}\right)$,
the assumptions of Theorem 2.27 of \citet*{Folland_text_book} are
satisfied. Therefore $\beta_{\mathcal{M}^{n}}\left(\cdot,\boldsymbol{z}\right)$
is differentiable on $\left[s\left(v_{l}\left(\boldsymbol{x}\right)-h,\boldsymbol{z},n\right),s\left(v_{u}\left(\boldsymbol{x}\right)+h,\boldsymbol{z},n\right)\right]$
for all $\boldsymbol{z}\in\mathbb{H}\left(\boldsymbol{x},h\right)$
when $h$ is sufficiently small and
\[
\beta_{\mathcal{M}^{n}}'\left(b,\boldsymbol{z}\right)=\int_{\mathcal{X}}\int_{\underline{b}\left(\boldsymbol{z}'\right)}^{\overline{b}\left(\boldsymbol{z}',n\right)}-\frac{1}{h^{2+d}}K_{g}'\left(\frac{b'-b}{h}\right)K_{\boldsymbol{X}}\left(\frac{\boldsymbol{z}'-\boldsymbol{z}}{h}\right)g\left(b',\boldsymbol{z}',n\right)\mathrm{d}b'\mathrm{d}\boldsymbol{z}'-g'\left(b,\boldsymbol{z},n\right),
\]
which is the bias of the kernel estimator for the partial derivative
$g'\left(b,\boldsymbol{z},n\right)$. By the usual argument for the
bias of kernel estimators for the density (see, e.g., \citet{Newey_Kernel_ET_1994})
and the assumption that $K_{0}$ is supported on $\left[-1,1\right]$,
for each $\left(b,\boldsymbol{z}\right)\in\mathcal{C}_{B,\boldsymbol{X}}^{n}$,
\begin{eqnarray}
\left|\beta_{\mathcal{M}^{n}}\left(b,\boldsymbol{z}\right)\right| & \leq & \frac{h^{R+1}}{\left(R+1\right)!}\left\{ \underset{\left(\alpha_{1},...,\alpha_{1+d}\right)\in\mathbb{S}_{R+1}}{\mathrm{sup}}\underset{\left(b',\boldsymbol{z}'\right)\in\mathbb{H}\left(\left(b,\boldsymbol{z}\right),h\right)}{\mathrm{sup}}\left|D_{1}^{\alpha_{1}}\cdots D_{1+d}^{\alpha_{1+d}}g\left(b',\boldsymbol{z}',n\right)\right|\right\} \nonumber \\
 &  & \times\left\{ \int\int\left\Vert \left(b',\boldsymbol{z}'\right)\right\Vert _{1}^{R}\left|K_{g}\left(b'\right)K_{\boldsymbol{X}}\left(\boldsymbol{z}'\right)\right|\mathrm{d}b'\mathrm{d}\boldsymbol{z}'\right\} ,\label{eq:beta bound}
\end{eqnarray}
when $h$ is sufficiently small. It follows from Proposition 1(iv)
of GPV that $g\left(\cdot,\cdot,n\right)$ admits $R+1$ continuous
partial derivatives on the interior of $\mathcal{S}_{B,\boldsymbol{X}}^{n}$
for each $n\in\mathcal{N}$. By using the standard argument for the
bias of kernel estimators for the density derivatives (see, e.g.,
\citet{Newey_Kernel_ET_1994}), for each $\left(b,z\right)\in\mathcal{C}_{B,\boldsymbol{X}}^{n}$,
\begin{equation}
\left|\beta_{\mathcal{M}^{n}}'\left(b,\boldsymbol{z}\right)\right|\leq\frac{h^{R}}{R!}\left\{ \underset{\left(\alpha_{1},...,\alpha_{1+d}\right)\in\mathbb{S}_{R}}{\mathrm{sup}}\underset{\left(b',\boldsymbol{z}'\right)\in\mathbb{H}\left(\left(b,\boldsymbol{z}\right),h\right)}{\mathrm{sup}}\left|D_{1}^{1+\alpha_{1}}\cdots D_{1+d}^{\alpha_{1+d}}g'\left(b',\boldsymbol{z}',n\right)\right|\right\} \left\{ \int\int\left\Vert \left(b',\boldsymbol{z}'\right)\right\Vert _{1}^{R}\left|K_{g}\left(b'\right)K_{\boldsymbol{X}}\left(\boldsymbol{z}'\right)\right|\mathrm{d}b'\mathrm{d}\boldsymbol{z}'\right\} ,\label{eq:beta prime bound}
\end{equation}
when $h$ is sufficiently small. Since $\mathcal{C}_{B,\boldsymbol{X}}^{n}$
is an inner closed subset of $\mathcal{S}_{B,\boldsymbol{X}}^{n}$,
(\ref{eq:beta bound}) and (\ref{eq:beta prime bound}) imply that
$\beta_{\mathcal{M}^{n}}\left(b,\boldsymbol{z}\right)$ and $\beta_{\mathcal{M}^{n}}'\left(b,\boldsymbol{z}\right)$
are also $O\left(h^{R}\right)$ uniformly in $\left(b,\boldsymbol{z}\right)\in\mathcal{C}_{B,\boldsymbol{X}}^{n}$. 

By change of variables, we have
\[
\mu_{\mathcal{M}^{n}}\left(v\right)=-\frac{1}{n-1}\int_{\mathcal{X}}\int_{\frac{\underline{v}\left(\boldsymbol{z}\right)-v}{h}}^{\frac{\overline{v}\left(\boldsymbol{z}\right)-v}{h}}\frac{1}{h^{1+d}}K_{f}'\left(u,\frac{\boldsymbol{z}-\boldsymbol{x}}{h}\right)\frac{G\left(s\left(hu+v,\boldsymbol{z},n\right),\boldsymbol{z},n\right)s'\left(hu+v,\boldsymbol{z},n\right)}{g\left(s\left(hu+v,\boldsymbol{z},n\right),\boldsymbol{z},n\right)}\beta_{\mathcal{M}^{n}}\left(s\left(hu+v,\boldsymbol{z},n\right),\boldsymbol{z}\right)\mathrm{d}u\mathrm{d}\boldsymbol{z}.
\]
A mean value expansion gives
\begin{eqnarray*}
\mu_{\mathcal{M}^{n}}\left(v\right) & = & -\frac{1}{n-1}\int_{\mathcal{X}}\int_{\frac{\underline{v}\left(\boldsymbol{z}\right)-v}{h}}^{\frac{\overline{v}\left(\boldsymbol{z}\right)-v}{h}}\frac{1}{h^{1+d}}K_{f}'\left(u,\frac{\boldsymbol{z}-\boldsymbol{x}}{h}\right)\psi\left(hu+v,\boldsymbol{z},n\right)\beta_{\mathcal{M}^{n}}\left(s\left(hu+v,\boldsymbol{z},n\right),\boldsymbol{z}\right)\mathrm{d}u\mathrm{d}\boldsymbol{z}\\
 & = & -\frac{1}{n-1}\int_{\mathcal{X}}\int_{\frac{\underline{v}\left(\boldsymbol{z}\right)-v}{h}}^{\frac{\overline{v}\left(\boldsymbol{z}\right)-v}{h}}\frac{1}{h^{1+d}}K_{f}'\left(u,\frac{\boldsymbol{z}-\boldsymbol{x}}{h}\right)\left\{ \psi\left(v,\boldsymbol{z},n\right)\beta_{\mathcal{M}^{n}}\left(s\left(v,\boldsymbol{z},n\right),\boldsymbol{z}\right)\right.\\
 &  & \left.+\left(\psi'\left(\dot{v},\boldsymbol{z},n\right)\beta_{\mathcal{M}^{n}}\left(s\left(\dot{v},z,n\right),\boldsymbol{z}\right)+\psi\left(\dot{v},\boldsymbol{z},n\right)\beta_{\mathcal{M}^{n}}'\left(s\left(\dot{v},\boldsymbol{z},n\right),\boldsymbol{z}\right)s'\left(\dot{v},\boldsymbol{z},n\right)\right)hu\right\} \mathrm{d}u\mathrm{d}\boldsymbol{z},
\end{eqnarray*}
where $\dot{v}$ is the mean value (depending on $u$ and $\boldsymbol{z}$)
with $\left|\dot{v}-v\right|\leq h\left|u\right|$. By the triangle
inequality, we have
\begin{eqnarray}
\left|\mu_{\mathcal{M}^{n}}\left(v\right)\right| & \leq & \int_{\mathcal{X}}\frac{1}{h^{1+d}}\left|K_{\boldsymbol{X}}^{0}\left(\frac{\boldsymbol{z}-\boldsymbol{x}}{h}\right)\right|\left|\psi\left(v,\boldsymbol{z},n\right)\beta_{\mathcal{M}^{n}}\left(s\left(v,\boldsymbol{z},n\right),\boldsymbol{z}\right)\int_{\frac{\underline{v}\left(\boldsymbol{z}\right)-v}{h}}^{\frac{\overline{v}\left(\boldsymbol{z}\right)-v}{h}}K_{0}'\left(u\right)\mathrm{d}u\right|\mathrm{d}\boldsymbol{z}\nonumber \\
 &  & +\int_{\mathcal{X}}\frac{1}{h^{d}}\left|K_{\boldsymbol{X}}^{0}\left(\frac{\boldsymbol{z}-\boldsymbol{x}}{h}\right)\right|\int_{\frac{\underline{v}\left(\boldsymbol{z}\right)-v}{h}}^{\frac{\overline{v}\left(\boldsymbol{z}\right)-v}{h}}\left|K_{0}'\left(u\right)u\right|\nonumber \\
 &  & \times\left|\left(\psi'\left(\dot{v},\boldsymbol{z},n\right)\beta_{\mathcal{M}^{n}}\left(s\left(\dot{v},\boldsymbol{z},n\right),z\right)+\psi\left(\dot{v},\boldsymbol{z},n\right)\beta_{\mathcal{M}^{n}}'\left(s\left(\dot{v},\boldsymbol{z},n\right),\boldsymbol{z}\right)s'\left(\dot{v},\boldsymbol{z},n\right)\right)\right|\mathrm{d}u\mathrm{d}\boldsymbol{z}.\label{eq:mu bound 1}
\end{eqnarray}
By the argument used in the proof of Lemma \ref{Lemma 2}, the first
term on the right hand side of (\ref{eq:mu bound 1}) vanishes for
all sufficiently small $h$. Therefore now we have
\begin{eqnarray}
\underset{v\in I\left(\boldsymbol{x}\right)}{\mathrm{sup}}\left|\mu_{\mathcal{M}^{n}}\left(v\right)\right| & \leq & \left\{ \int\frac{1}{h^{d}}\left|K_{\boldsymbol{X}}^{0}\left(\frac{\boldsymbol{z}-\boldsymbol{x}}{h}\right)\right|\mathrm{d}\boldsymbol{z}\right\} \nonumber \\
 &  & \times\left\{ \underset{\left(u,\boldsymbol{z}\right)\in\mathcal{C}_{V,\boldsymbol{X}}}{\mathrm{sup}}\left|\psi'\left(u,\boldsymbol{z}\right)\beta_{\mathcal{M}^{n}}\left(s\left(u,\boldsymbol{z},n\right),\boldsymbol{z}\right)+\psi\left(u,\boldsymbol{z}\right)\beta_{\mathcal{M}^{n}}'\left(s\left(u,\boldsymbol{z},n\right),\boldsymbol{z}\right)s'\left(u,\boldsymbol{z},n\right)\right|\right\} ,\label{eq:mu bound 2}
\end{eqnarray}
when $h$ is sufficiently small. It follows from this result, the
fact that $\beta_{\mathcal{M}^{n}}\left(b,\boldsymbol{z}\right)$
and $\beta_{\mathcal{M}^{n}}'\left(b,\boldsymbol{z}\right)$ are both
$O\left(h^{R}\right)$ uniformly in $\left(b,\boldsymbol{z}\right)\in\mathcal{C}_{B,\boldsymbol{X}}^{n}$
and (\ref{eq:|K_X| integral order}) that $\underset{v\in I\left(\boldsymbol{x}\right)}{\mathrm{sup}}\left|\mu_{\mathcal{M}^{n}}\left(v\right)\right|=O\left(h^{R}\right)$. 

Standard arguments can be applied to verify that class $\left\{ \mathcal{M}^{n}\left(\cdot,\cdot;v\right):v\in I\left(\boldsymbol{x}\right)\right\} $
is (uniformly) VC-type with respect to the envelope 
\begin{equation}
F_{\mathcal{M}^{n}}\left(\boldsymbol{z},\boldsymbol{z}'\right)\coloneqq\frac{\overline{\varphi}}{\left(n-1\right)\underline{C}_{g}^{2}h^{3+2d}}\left|K_{\boldsymbol{X}}^{0}\left(\frac{\boldsymbol{z}-\boldsymbol{x}}{h}\right)\right|\left|K_{\boldsymbol{X}}\left(\frac{\boldsymbol{z}'-\boldsymbol{z}}{h}\right)\right|+\frac{\overline{\varphi}}{\left(n-1\right)\underline{C}_{g}h^{2+d}}\left|K_{\boldsymbol{X}}^{0}\left(\frac{\boldsymbol{z}-\boldsymbol{x}}{h}\right)\right|.\label{eq:F_M_n envelope}
\end{equation}
The CK inequality gives 
\begin{align*}
 & \mathrm{E}\left[\underset{v\in I\left(\boldsymbol{x}\right)}{\mathrm{sup}}\left|\frac{1}{\left(L\right)_{2}}\sum_{\left(2\right)}\left\{ \mathcal{M}^{n}\left(\left(\boldsymbol{B}_{\cdot l},\boldsymbol{X}_{l},N_{l}\right),\left(\boldsymbol{B}_{\cdot k},\boldsymbol{X}_{k},N_{k}\right);v\right)-\mathcal{M}_{1}^{n}\left(\boldsymbol{B}_{\cdot l},\boldsymbol{X}_{l},N_{l};v\right)-\mathcal{M}_{2}^{n}\left(\boldsymbol{B}_{\cdot k},\boldsymbol{X}_{k},N_{k};v\right)+\mu_{\mathcal{M}^{n}}\left(v\right)\right\} \right|\right]\\
\apprle & L^{-1}\left(\mathrm{E}\left[F_{\mathcal{M}^{n}}\left(\boldsymbol{X}_{1},\boldsymbol{X}_{2}\right)^{2}\right]\right)^{\nicefrac{1}{2}}\\
= & O\left(\left(Lh^{3+d}\right)^{-1}\right)
\end{align*}
where the equality follows from change of variables.

Next, we have

\begin{eqnarray*}
\mathcal{M}_{1}^{n}\left(\boldsymbol{b}_{\cdot},\boldsymbol{z},m;v\right) & = & \mathrm{E}\left[\mathcal{M}^{n}\left(\left(\boldsymbol{b}_{\cdot},\boldsymbol{z},m\right),\left(\boldsymbol{B}_{\cdot1},\boldsymbol{X}_{1},N_{1}\right);v\right)\right]\\
 & = & -\mathbbm{1}\left(m=n\right)\frac{1}{m}\sum_{i=1}^{m}\frac{1}{h^{2+d}}K_{f}'\left(\frac{\xi\left(b_{i},\boldsymbol{z},m\right)-v}{h},\frac{\boldsymbol{z}-\boldsymbol{x}}{h}\right)\frac{G\left(b_{i},\boldsymbol{z},m\right)}{\left(m-1\right)g\left(b_{i},\boldsymbol{z},m\right)^{2}}\\
 &  & \times\left(\mathrm{E}\left[\mathbbm{1}\left(N_{1}=m\right)\frac{1}{N_{1}}\sum_{j=1}^{N_{1}}\frac{1}{h^{1+d}}K_{g}\left(\frac{B_{j1}-b_{i}}{h}\right)K_{\boldsymbol{X}}\left(\frac{\boldsymbol{X}_{1}-\boldsymbol{z}}{h}\right)\right]-g\left(b_{i},\boldsymbol{z},m\right)\right)\\
 & = & -\mathbbm{1}\left(m=n\right)\frac{1}{m}\sum_{i=1}^{m}\frac{1}{h^{2+d}}K_{f}'\left(\frac{\xi\left(b_{i},\boldsymbol{z},m\right)-v}{h},\frac{\boldsymbol{z}-\boldsymbol{x}}{h}\right)\frac{G\left(b_{i},\boldsymbol{z},m\right)\beta_{\mathcal{M}^{n}}\left(b_{i},\boldsymbol{z}\right)}{\left(m-1\right)g\left(b_{i},\boldsymbol{z},m\right)^{2}}.
\end{eqnarray*}
It is clear from the definition that when $h$ is sufficiently small,
the class $\left\{ \mathcal{M}_{1}^{n}\left(\cdot;v\right):v\in I\left(\boldsymbol{x}\right)\right\} $
is uniformly VC-type with respect to the envelope
\[
F_{\mathcal{M}_{1}^{n}}\left(\boldsymbol{z}\right)\coloneqq\frac{\left(\overline{C}_{D_{1}}+\overline{C}_{D_{2}}\right)\overline{\varphi}}{\left(n-1\right)\underline{C}_{g}^{2}h^{2+d}}\left|K_{\boldsymbol{X}}^{0}\left(\frac{\boldsymbol{z}-\boldsymbol{x}}{h}\right)\right|\left\{ \underset{\left(b',\boldsymbol{z}'\right)\in\mathcal{C}_{B,\boldsymbol{X}}^{n}}{\mathrm{sup}}\left|\beta_{\mathcal{M}^{n}}\left(b',\boldsymbol{z}'\right)\right|\right\} .
\]
The VW inequality yields 
\[
\mathrm{E}\left[\underset{v\in I\left(\boldsymbol{x}\right)}{\mathrm{sup}}\left|\frac{1}{L}\sum_{l=1}^{L}\mathcal{M}_{1}^{n}\left(\boldsymbol{B}_{\cdot l},\boldsymbol{X}_{l},N_{l};v\right)-\mu_{\mathcal{M}^{n}}\left(v\right)\right|\right]\leq L^{-\nicefrac{1}{2}}\mathrm{E}\left[F_{\mathcal{M}_{1}^{n}}\left(\boldsymbol{X}_{1}\right)^{2}\right]^{\nicefrac{1}{2}}=O\left(\frac{h^{R-1}}{\left(Lh^{1+d}\right)^{\nicefrac{1}{2}}}\right),
\]
where the inequality holds when $h$ is sufficiently small and the
equality follows from a standard change of variable argument and (\ref{eq:beta bound}).

It follows from Markov's inequality and change of variables that
\begin{eqnarray*}
\underset{v\in I\left(\boldsymbol{x}\right)}{\mathrm{sup}}\left|\frac{1}{L^{2}}\sum_{l=1}^{L}\mathcal{M}^{n}\left(\left(\boldsymbol{B}_{\cdot l},\boldsymbol{X}_{l},N_{l}\right),\left(\boldsymbol{B}_{\cdot l},\boldsymbol{X}_{l},N_{l}\right);v\right)\right| & \apprle & \frac{1}{L^{2}}\sum_{l=1}^{L}\frac{1}{h^{3+2d}}\left|K_{\boldsymbol{X}}^{0}\left(\frac{\boldsymbol{X}_{l}-\boldsymbol{x}}{h}\right)\right|+\frac{1}{L^{2}}\sum_{l=1}^{L}\frac{1}{h^{2+d}}\left|K_{\boldsymbol{X}}^{0}\left(\frac{\boldsymbol{X}_{l}-\boldsymbol{x}}{h}\right)\right|\\
 & = & O_{p}\left(\left(Lh^{3+d}\right)^{-1}\right)
\end{eqnarray*}
and similarly, we have
\begin{align*}
 & \underset{v\in I\left(\boldsymbol{x}\right)}{\mathrm{sup}}\left|\frac{1}{L^{2}\left(L-1\right)}\sum_{l\neq k}\mathcal{M}^{n}\left(\left(\boldsymbol{B}_{\cdot l},\boldsymbol{X}_{l},N_{l}\right),\left(\boldsymbol{B}_{\cdot k},\boldsymbol{X}_{k},N_{k}\right);v\right)\right|\\
\apprle & \frac{1}{L^{2}\left(L-1\right)}\sum_{l\neq k}\frac{1}{h^{3+2d}}\left|K_{\boldsymbol{X}}^{0}\left(\frac{\boldsymbol{X}_{l}-\boldsymbol{x}}{h}\right)\right|\left|K_{\boldsymbol{X}}\left(\frac{\boldsymbol{X}_{k}-\boldsymbol{X}_{l}}{h}\right)\right|+\frac{1}{L^{2}}\sum_{l=1}^{L}\frac{1}{h^{2+d}}\left|K_{\boldsymbol{X}}^{0}\left(\frac{\boldsymbol{X}_{l}-\boldsymbol{x}}{h}\right)\right|\\
= & O_{p}\left(\left(Lh^{3}\right)^{-1}\right).
\end{align*}
The conclusion follows from these bounds. \end{proof}
\begin{lem}
\label{Lemma 4}Suppose Assumptions 1 - 3 hold. Then 
\[
\underset{v\in I\left(\boldsymbol{x}\right)}{\mathrm{sup}}\left|Z\left(v|\boldsymbol{x}\right)-\varGamma\left(v|\boldsymbol{x}\right)\right|=O_{p}\left(\mathrm{log}\left(L\right)^{\nicefrac{1}{2}}h+\frac{\mathrm{log}\left(L\right)}{\left(Lh^{3+d}\right)^{\nicefrac{1}{2}}}+L^{\nicefrac{1}{2}}h^{\nicefrac{\left(3+d\right)}{2}+R}\right).
\]
\end{lem}
\begin{proof}[Proof of Lemma \ref{Lemma 4}]It is straightforward
to check that 
\[
\varGamma\left(v|\boldsymbol{x}\right)=\frac{1}{L^{\nicefrac{1}{2}}}\sum_{l=1}^{L}\frac{\sum_{n\in\mathcal{N}}\left\{ \mathcal{M}_{2}^{\ddagger,n}\left(\boldsymbol{B}_{\cdot l},\boldsymbol{X}_{l},N_{l};v\right)-\mu_{\mathcal{M}^{\ddagger,n}}\left(v\right)\right\} }{\mathrm{Var}\left[\sum_{n\in\mathcal{N}}\mathcal{M}_{2}^{\ddagger,n}\left(\boldsymbol{B}_{\cdot1},\boldsymbol{X}_{1},N_{1}\right)\right]^{\nicefrac{1}{2}}}
\]
and 
\[
\frac{1}{L}\sum_{l=1}^{L}\mathcal{M}_{2}^{n}\left(\boldsymbol{B}_{\cdot l},\boldsymbol{X}_{l},N_{l};v\right)-\mu_{\mathcal{M}^{n}}\left(v\right)=\frac{1}{L}\sum_{l=1}^{L}\mathcal{M}_{2}^{\ddagger,n}\left(\boldsymbol{B}_{\cdot l},\boldsymbol{X}_{l},N_{l};v\right)-\mu_{\mathcal{M}^{\ddagger,n}}\left(v\right)
\]
for all $v\in I\left(\boldsymbol{x}\right)$, since 
\[
\mathcal{M}_{2}^{n}\left(\boldsymbol{B}_{\cdot l},\boldsymbol{X}_{l},N_{l}\right)-\mu_{\mathcal{M}^{n}}\left(v\right)=\mathcal{M}_{2}^{\ddagger,n}\left(\boldsymbol{B}_{\cdot l},\boldsymbol{X}_{l},N_{l}\right)-\mu_{\mathcal{M}^{\ddagger,n}}\left(v\right),\textrm{ for all \ensuremath{l=1,...,L}}.
\]

It follows from standard arguments that $\left\{ \mathcal{M}^{\ddagger,n}\left(\cdot,\cdot;v\right):v\in I\left(\boldsymbol{x}\right)\right\} $
is uniformly VC-type with respect to the envelope 
\begin{equation}
F_{\mathcal{M}^{\ddagger,n}}\left(\boldsymbol{z},\boldsymbol{z}'\right)\coloneqq\frac{\overline{\varphi}\left(\overline{C}_{D_{1}}+\overline{C}_{D_{2}}\right)\overline{C}_{K_{g}}}{\left(n-1\right)\underline{C}_{g}h^{3+2d}}\left|K_{\boldsymbol{X}}^{0}\left(\frac{\boldsymbol{z}-\boldsymbol{x}}{h}\right)\right|\left|K_{\boldsymbol{X}}\left(\frac{\boldsymbol{z}'-\boldsymbol{z}}{h}\right)\right|.\label{eq:F_M_ddagger}
\end{equation}
Then it follows from \citet[Lemma 5.4]{Chen_Kato_U_Process} that$\left\{ \mathcal{M}_{2}^{\ddagger,n}\left(\cdot;v\right):v\in I\left(\boldsymbol{x}\right)\right\} $
is uniformly VC-type with respect to the envelope 
\begin{equation}
F_{\mathcal{M}_{2}^{\ddagger,n}}\left(\boldsymbol{z}\right)\coloneqq\frac{\overline{\varphi}\left(\overline{C}_{D_{1}}+\overline{C}_{D_{2}}\right)\overline{C}_{K_{g}}}{\left(n-1\right)\underline{C}_{g}}\int\frac{1}{h^{3+2d}}\left|K_{\boldsymbol{X}}^{0}\left(\frac{\boldsymbol{z}'-\boldsymbol{x}}{h}\right)\right|\left|K_{\boldsymbol{X}}\left(\frac{\boldsymbol{z}'-\boldsymbol{z}}{h}\right)\right|\varphi\left(\boldsymbol{z}'\right)\mathrm{d}\boldsymbol{z}'.\label{eq:M_ddagger_2 envelope}
\end{equation}
Let
\begin{equation}
\sigma_{\mathcal{M}_{2}^{\ddagger,n}}^{2}\coloneqq\underset{v\in I\left(\boldsymbol{x}\right)}{\mathrm{sup}}\mathrm{E}\left[\mathcal{M}_{2}^{\ddagger,n}\left(\boldsymbol{B}_{\cdot1},\boldsymbol{X}_{1},N_{1};v\right)^{2}\right]=O\left(h^{-\left(3+d\right)}\right).\label{eq:sigma_M_ddagger_2 rate}
\end{equation}
The CCK inequality yields
\begin{eqnarray*}
\mathrm{E}\left[\underset{v\in I\left(\boldsymbol{x}\right)}{\mathrm{sup}}\left|\frac{1}{L}\sum_{l=1}^{L}\mathcal{M}_{2}^{\ddagger,n}\left(\boldsymbol{B}_{\cdot l},\boldsymbol{X}_{l},N_{l};v\right)-\mu_{\mathcal{M}^{\ddagger,n}}\left(v\right)\right|\right] & \leq & C_{1}\left\{ L^{-\nicefrac{1}{2}}\sigma_{\mathcal{M}_{2}^{\ddagger,n}}\mathrm{log}\left(C_{2}L\right)^{\nicefrac{1}{2}}+L^{-1}\left\Vert F_{\mathcal{M}_{2}^{\ddagger,n}}\right\Vert _{\mathcal{X}}\mathrm{log}\left(C_{2}L\right)\right\} .\\
 & = & O\left(\left(\frac{\mathrm{log}\left(L\right)}{Lh^{3+d}}\right)^{\nicefrac{1}{2}}\right),
\end{eqnarray*}
where the equality follows from (\ref{eq:sigma_M_ddagger_2 rate})
and $\left\Vert F_{\mathcal{M}_{2}^{\ddagger,n}}\right\Vert _{\mathcal{X}}=O\left(h^{-\left(3+d\right)}\right)$
(which follows from change of variables). By Markov's inequality,
we have 
\begin{equation}
\underset{v\in I\left(\boldsymbol{x}\right)}{\mathrm{sup}}\left|\frac{1}{L}\sum_{l=1}^{L}\mathcal{M}_{2}^{n}\left(\boldsymbol{B}_{\cdot l},\boldsymbol{X}_{l},N_{l};v\right)-\mu_{\mathcal{M}^{n}}\left(v\right)\right|=\underset{v\in I\left(\boldsymbol{x}\right)}{\mathrm{sup}}\left|\frac{1}{L}\sum_{l=1}^{L}\mathcal{M}_{2}^{\ddagger,n}\left(\boldsymbol{B}_{\cdot l},\boldsymbol{X}_{l},N_{l};v\right)-\mu_{\mathcal{M}^{\ddagger,n}}\left(v\right)\right|=O_{p}\left(\left(\frac{\mathrm{log}\left(L\right)}{Lh^{3+d}}\right)^{\nicefrac{1}{2}}\right).\label{eq:M_2_ddagger - miu_M_ddagger sup order bound}
\end{equation}

By the above result, Lemma \ref{Lemma 3}, (\ref{eq:phi_hat - phi rate})
and (\ref{eq:abc identity}), we have

\begin{equation}
\widehat{f}_{GPV}\left(v|\boldsymbol{x}\right)-f\left(v|\boldsymbol{x}\right)=\frac{1}{\varphi\left(\boldsymbol{x}\right)L}\sum_{l=1}^{L}\sum_{n\in\mathcal{N}}\left\{ \mathcal{M}_{2}^{n}\left(\boldsymbol{B}_{\cdot l},\boldsymbol{X}_{l},N_{l};v\right)-\mu_{\mathcal{M}^{n}}\left(v\right)\right\} +O_{p}\left(\left(\frac{\mathrm{log}\left(L\right)}{Lh^{1+d}}\right)^{\nicefrac{1}{2}}+\frac{\mathrm{log}\left(L\right)}{Lh^{3+d}}+h^{R}\right).\label{eq:f_hat - f bound 2}
\end{equation}

We showed that 
\[
\underset{v\in I\left(\boldsymbol{x}\right)}{\mathrm{sup}}\left|\mathrm{V}_{\mathcal{M}}\left(v|\boldsymbol{x},n\right)-\mathrm{Var}\left[h^{\nicefrac{\left(3+d\right)}{2}}\mathcal{M}_{2}^{\ddagger,n}\left(\boldsymbol{B}_{\cdot1},\boldsymbol{X}_{1},N_{1};v\right)\right]\right|=O\left(h^{3}\right)
\]
and 
\[
\underset{v\in I\left(\boldsymbol{x}\right)}{\mathrm{sup}}\left|\mathrm{Var}\left[\sum_{n\in\mathcal{N}}h^{\nicefrac{\left(3+d\right)}{2}}\mathcal{M}_{2}^{\ddagger,n}\left(\boldsymbol{B}_{\cdot1},\boldsymbol{X}_{1},N_{1};v\right)\right]-\sum_{n\in\mathcal{N}}\mathrm{Var}\left[h^{\nicefrac{\left(3+d\right)}{2}}\mathcal{M}_{2}^{\ddagger,n}\left(\boldsymbol{B}_{\cdot1},\boldsymbol{X}_{1},N_{1};v\right)\right]\right|=O\left(h^{3+d}\right)
\]
in the proof of Theorem 6.1. Therefore we have 
\begin{equation}
\underset{v\in I\left(\boldsymbol{x}\right)}{\mathrm{sup}}\left|\mathrm{Var}\left[\sum_{n\in\mathcal{N}}h^{\nicefrac{\left(3+d\right)}{2}}\mathcal{M}_{2}^{\ddagger,n}\left(\boldsymbol{B}_{\cdot1},\boldsymbol{X}_{1},N_{1};v\right)\right]-\sum_{n\in\mathcal{N}}\mathrm{V}_{\mathcal{M}}\left(v|\boldsymbol{x},n\right)\right|=O\left(h^{3}\right).\label{eq:Var=00005Bsum M_2_ddagger=00005D - sum V_M bound}
\end{equation}
Next, by (\ref{eq:abc identity}), we have
\begin{align*}
 & \frac{\widehat{f}_{GPV}\left(v|\boldsymbol{x}\right)-f\left(v|\boldsymbol{x}\right)}{\left(Lh^{3+d}\right)^{-\nicefrac{1}{2}}\left\{ \sum_{n\in\mathcal{N}}\varphi\left(\boldsymbol{x}\right)^{-2}\mathrm{V}_{\mathcal{M}}\left(v|\boldsymbol{x},n\right)\right\} ^{\nicefrac{1}{2}}}-\varGamma\left(v|\boldsymbol{x}\right)\\
= & \left(Lh^{3+d}\right)^{\nicefrac{1}{2}}\left\{ \frac{\varphi\left(\boldsymbol{x}\right)\left(\widehat{f}_{GPV}\left(v|\boldsymbol{x}\right)-f\left(v|\boldsymbol{x}\right)\right)}{\left(\sum_{n\in\mathcal{N}}\mathrm{V}_{\mathcal{M}}\left(v|\boldsymbol{x},n\right)\right)^{\nicefrac{1}{2}}}-\frac{\frac{1}{L}\sum_{l=1}^{L}\sum_{n\in\mathcal{N}}\left\{ \mathcal{M}_{2}^{\ddagger,n}\left(\boldsymbol{B}_{\cdot l},\boldsymbol{X}_{l},N_{l};v\right)-\mu_{\mathcal{M}^{\ddagger,n}}\left(v\right)\right\} }{\mathrm{Var}\left[\sum_{n\in\mathcal{N}}h^{\nicefrac{\left(3+d\right)}{2}}\mathcal{M}_{2}^{\ddagger,n}\left(\boldsymbol{B}_{\cdot1},\boldsymbol{X}_{1},N_{1}\right)\right]^{\nicefrac{1}{2}}}\right\} \\
= & \left(Lh^{3+d}\right)^{\nicefrac{1}{2}}\left\{ \frac{\varphi\left(\boldsymbol{x}\right)\left(\widehat{f}_{GPV}\left(v|\boldsymbol{x}\right)-f\left(v|\boldsymbol{x}\right)\right)-\frac{1}{L}\sum_{l=1}^{L}\sum_{n\in\mathcal{N}}\left\{ \mathcal{M}_{2}^{\ddagger,n}\left(\boldsymbol{B}_{\cdot l},\boldsymbol{X}_{l},N_{l};v\right)-\mu_{\mathcal{M}^{\ddagger,n}}\left(v\right)\right\} }{\left(\sum_{n\in\mathcal{N}}\mathrm{V}_{\mathcal{M}}\left(v|\boldsymbol{x},n\right)\right)^{\nicefrac{1}{2}}}\right.\\
 & +\frac{\frac{1}{L}\sum_{l=1}^{L}\sum_{n\in\mathcal{N}}\left(\mathcal{M}_{2}^{\ddagger,n}\left(\boldsymbol{B}_{\cdot l},\boldsymbol{X}_{l},N_{l};v\right)-\mu_{\mathcal{M}^{\ddagger,n}}\left(v\right)\right)}{\sum_{n\in\mathcal{N}}\mathrm{V}_{\mathcal{M}}\left(v|\boldsymbol{x},n\right)}\left(\mathrm{Var}\left[\sum_{n\in\mathcal{N}}h^{\nicefrac{\left(3+d\right)}{2}}\mathcal{M}_{2}^{\ddagger,n}\left(\boldsymbol{B}_{\cdot1},\boldsymbol{X}_{1},N_{1}\right)\right]^{\nicefrac{1}{2}}-\left(\sum_{n\in\mathcal{N}}\mathrm{V}_{\mathcal{M}}\left(v|\boldsymbol{x},n\right)\right)^{\nicefrac{1}{2}}\right)\\
 & +\frac{\frac{1}{L}\sum_{l=1}^{L}\sum_{n\in\mathcal{N}}\left(\mathcal{M}_{2}^{\ddagger,n}\left(\boldsymbol{B}_{\cdot l},\boldsymbol{X}_{l},N_{l};v\right)-\mu_{\mathcal{M}^{\ddagger,n}}\left(v\right)\right)}{\mathrm{Var}\left[\sum_{n\in\mathcal{N}}h^{\nicefrac{\left(3+d\right)}{2}}\mathcal{M}_{2}^{\ddagger,n}\left(\boldsymbol{B}_{\cdot1},\boldsymbol{X}_{1},N_{1}\right)\right]^{\nicefrac{1}{2}}\left(\sum_{n\in\mathcal{N}}\mathrm{V}_{\mathcal{M}}\left(v|\boldsymbol{x},n\right)\right)}\\
 & \left.\times\left(\mathrm{Var}\left[\sum_{n\in\mathcal{N}}h^{\nicefrac{\left(3+d\right)}{2}}\mathcal{M}_{2}^{\ddagger,n}\left(\boldsymbol{B}_{\cdot1},\boldsymbol{X}_{1},N_{1}\right)\right]^{\nicefrac{1}{2}}-\left(\sum_{n\in\mathcal{N}}\mathrm{V}_{\mathcal{M}}\left(v|\boldsymbol{x},n\right)\right)^{\nicefrac{1}{2}}\right)^{2}\right\} .
\end{align*}
Now 
\[
\underset{v\in I\left(\boldsymbol{x}\right)}{\mathrm{sup}}\left|\frac{\widehat{f}_{GPV}\left(v|\boldsymbol{x}\right)-f\left(v|\boldsymbol{x}\right)}{\left(Lh^{3+d}\right)^{-\nicefrac{1}{2}}\left\{ \sum_{n\in\mathcal{N}}\varphi\left(\boldsymbol{x}\right)^{-2}\mathrm{V}_{\mathcal{M}}\left(v|\boldsymbol{x},n\right)\right\} ^{\nicefrac{1}{2}}}-\varGamma\left(v|\boldsymbol{x}\right)\right|=O_{p}\left(\mathrm{log}\left(L\right)^{\nicefrac{1}{2}}h+\frac{\mathrm{log}\left(L\right)}{\left(Lh^{3+d}\right)^{\nicefrac{1}{2}}}+L^{\nicefrac{1}{2}}h^{\nicefrac{3+d}{2}+R}\right)
\]
follows from the above decomposition, (\ref{eq:M_2_ddagger - miu_M_ddagger sup order bound}),
(\ref{eq:f_hat - f bound 2}) and (\ref{eq:Var=00005Bsum M_2_ddagger=00005D - sum V_M bound}).

Next, we have
\begin{align*}
 & \frac{\widehat{f}_{GPV}\left(v|\boldsymbol{x}\right)-f\left(v|\boldsymbol{x}\right)}{\left(Lh^{3+d}\right)^{-\nicefrac{1}{2}}\widehat{\mathrm{V}}_{GPV}\left(v|\boldsymbol{x}\right)^{\nicefrac{1}{2}}}-\frac{\widehat{f}_{GPV}\left(v|\boldsymbol{x}\right)-f\left(v|\boldsymbol{x}\right)}{\left(Lh^{3+d}\right)^{-\nicefrac{1}{2}}\left\{ \sum_{n\in\mathcal{N}}\varphi\left(\boldsymbol{x}\right)^{-2}\mathrm{V}_{\mathcal{M}}\left(v|\boldsymbol{x},n\right)\right\} ^{\nicefrac{1}{2}}}\\
= & \left(Lh^{3+d}\right)^{\nicefrac{1}{2}}\left\{ \widehat{f}_{GPV}\left(v|\boldsymbol{x}\right)-f\left(v|\boldsymbol{x}\right)\right\} \left\{ \widehat{\mathrm{V}}_{GPV}\left(v|\boldsymbol{x}\right)^{-\nicefrac{1}{2}}-\left\{ \sum_{n\in\mathcal{N}}\varphi\left(\boldsymbol{x}\right)^{-2}\mathrm{V}_{\mathcal{M}}\left(v|\boldsymbol{x},n\right)\right\} ^{-\nicefrac{1}{2}}\right\} \\
= & O_{p}\left(\frac{\mathrm{log}\left(L\right)}{\left(Lh^{3+d}\right)^{\nicefrac{1}{2}}}+\mathrm{log}\left(L\right)^{\nicefrac{1}{2}}h^{R}\right),
\end{align*}
where the second equality follows from Theorem 6.2, (\ref{eq:M_2_ddagger - miu_M_ddagger sup order bound})
and Lemma \ref{Lemma 3}. The conclusion follows.\end{proof}
\begin{lem}
\label{Lemma 5}Suppose Assumptions 1 - 3 hold. Then 
\[
\underset{v\in I\left(\boldsymbol{x}\right)}{\mathrm{sup}}\left|\widetilde{f}^{*}\left(v,\boldsymbol{x},n\right)-\widetilde{f}\left(v,\boldsymbol{x},n\right)\right|=O_{p}^{*}\left(\left(\frac{\mathrm{log}\left(L\right)}{Lh^{1+d}}\right)^{\nicefrac{1}{2}}\right).
\]
\end{lem}
\begin{proof}[Proof of Lemma \ref{Lemma 5}]By the fact that the bids
in the bootstrap sample are conditionally i.i.d. and LIE, we have
\begin{eqnarray*}
\mathrm{E}^{*}\left[\widetilde{f}^{*}\left(v,\boldsymbol{x},n\right)\right] & = & \mathrm{E}^{*}\left[\mathbbm{1}\left(N_{1}^{*}=n\right)\frac{1}{N_{1}^{*}}\sum_{i=1}^{N_{1}^{*}}\frac{1}{h^{1+d}}K_{f}\left(\frac{V_{i1}^{*}-v}{h},\frac{\boldsymbol{X}_{1}^{*}-\boldsymbol{x}}{h}\right)\right]\\
 & = & \mathrm{E}^{*}\left[\mathbbm{1}\left(N_{1}^{*}=n\right)\mathrm{E}^{*}\left[\frac{1}{N_{1}^{*}}\sum_{i=1}^{N_{1}^{*}}\frac{1}{h^{1+d}}K_{f}\left(\frac{V_{i1}^{*}-v}{h},\frac{\boldsymbol{X}_{1}^{*}-\boldsymbol{x}}{h}\right)|\boldsymbol{X}_{1}^{*},N_{1}^{*}\right]\right]\\
 & = & \frac{1}{L}\sum_{l=1}^{L}\mathbbm{1}\left(N_{l}=n\right)\frac{1}{N_{l}}\sum_{i=1}^{N_{l}}\frac{1}{h^{1+d}}K_{f}\left(\frac{V_{il}-v}{h},\frac{\boldsymbol{X}_{l}-\boldsymbol{x}}{h}\right).
\end{eqnarray*}

By Jensen's inequality and LIE, we have 
\begin{eqnarray*}
\mathrm{E}^{*}\left[\left\{ \mathbbm{1}\left(N_{1}^{*}=n\right)\frac{1}{N_{1}^{*}}\sum_{i=1}^{N_{1}^{*}}\frac{1}{h^{1+d}}K_{f}\left(\frac{V_{i1}^{*}-v}{h},\frac{\boldsymbol{X}_{1}^{*}-\boldsymbol{x}}{h}\right)\right\} ^{2}\right] & \leq & \mathrm{E}^{*}\left[\mathbbm{1}\left(N_{1}^{*}=n\right)\frac{1}{N_{1}^{*}}\sum_{i=1}^{N_{1}^{*}}\frac{1}{h^{2\left(1+d\right)}}K_{f}\left(\frac{V_{i1}^{*}-v}{h},\frac{\boldsymbol{X}_{1}^{*}-\boldsymbol{x}}{h}\right)^{2}\right]\\
 & = & h^{-\left(1+d\right)}\left\{ \frac{1}{L}\sum_{l=1}^{L}\mathbbm{1}\left(N_{l}=n\right)\frac{1}{N_{l}}\sum_{i=1}^{N_{l}}\frac{1}{h^{1+d}}K_{f}\left(\frac{V_{il}-v}{h},\frac{\boldsymbol{X}_{l}-\boldsymbol{x}}{h}\right)^{2}\right\} .
\end{eqnarray*}
By applying the arguments used to show (\ref{eq:sup F average O_p(1)}),
we have 
\begin{equation}
\widehat{\sigma}_{V}^{2}\coloneqq\underset{v\in I\left(\boldsymbol{x}\right)}{\mathrm{sup}}\mathrm{E}^{*}\left[\left\{ \mathbbm{1}\left(N_{1}^{*}=n\right)\frac{1}{N_{1}^{*}}\sum_{i=1}^{N_{1}^{*}}\frac{1}{h^{1+d}}K_{f}\left(\frac{V_{i1}^{*}-v}{h},\frac{\boldsymbol{X}_{1}^{*}-\boldsymbol{x}}{h}\right)\right\} ^{2}\right]=O_{p}\left(h^{-\left(1+d\right)}\right).\label{eq:sigma_V_hat order}
\end{equation}

Let
\[
\mathcal{F}^{*,n}\left(\boldsymbol{u},\boldsymbol{z},m,;v\right)\coloneqq\mathbbm{1}\left(m=n\right)\frac{1}{m}\sum_{i=1}^{m}\frac{1}{h^{1+d}}K_{f}\left(\frac{u_{i}-v}{h},\frac{\boldsymbol{z}-\boldsymbol{x}}{h}\right).
\]
Standard arguments can be applied to verify that the class $\left\{ \mathcal{F}^{*,n}\left(\cdot;v\right):v\in I\left(\boldsymbol{x}\right)\right\} $
is uniformly VC-type with respect to the envelope 
\[
F_{\mathcal{F}^{*,n}}\left(\boldsymbol{z}\right)\coloneqq\left(\overline{C}_{D_{1}}+\overline{C}_{D_{2}}\right)\frac{1}{h^{1+d}}\left|K_{\boldsymbol{X}}^{0}\left(\frac{\boldsymbol{z}-\boldsymbol{x}}{h}\right)\right|.
\]
Now the CCK inequality yields 
\[
\mathrm{E}^{*}\left[\underset{v\in I\left(\boldsymbol{x}\right)}{\mathrm{sup}}\left|\widetilde{f}^{*}\left(v,\boldsymbol{x},n\right)-\widetilde{f}\left(v,\boldsymbol{x},n\right)\right|\right]\leq C_{1}\left\{ L^{-\nicefrac{1}{2}}\widehat{\sigma}_{V}\mathrm{log}\left(C_{2}L\right)^{\nicefrac{1}{2}}+L^{-1}\left\Vert F_{\mathcal{F}^{*}}\right\Vert _{\mathcal{X}}\mathrm{log}\left(C_{2}L\right)\right\} .
\]

The conclusion follows from this result, (\ref{eq:sigma_V_hat order})
and Lemma \ref{lem:auxiliary 3}. \end{proof}
\begin{lem}
\label{Lemma 6}Suppose Assumptions 1 - 3 hold. Then 
\[
\underset{n'\in\mathcal{N}}{\mathrm{max}}\underset{\mathbb{H}\left(\left(b',\boldsymbol{x}'\right),h\right)\subseteq\mathcal{S}_{B,\boldsymbol{X}}^{n'}}{\mathrm{sup}}\left|\widehat{G}^{*}\left(b',\boldsymbol{x}',n'\right)-\widehat{G}\left(b',\boldsymbol{x}',n'\right)\right|=O_{p}^{*}\left(\left(\frac{\mathrm{log}\left(L\right)}{Lh^{d}}\right)^{\nicefrac{1}{2}}\right)
\]
and 
\[
\underset{n'\in\mathcal{N}}{\mathrm{max}}\underset{\mathbb{H}\left(\left(b',\boldsymbol{x}'\right),h\right)\subseteq\mathcal{S}_{B,\boldsymbol{X}}^{n'}}{\mathrm{sup}}\left|\widehat{g}^{*}\left(b',\boldsymbol{x}',n'\right)-\widehat{g}\left(b',\boldsymbol{x}',n'\right)\right|=O_{p}^{*}\left(\left(\frac{\mathrm{log}\left(L\right)}{Lh^{1+d}}\right)^{\nicefrac{1}{2}}\right).
\]
\end{lem}
\begin{proof}[Proof of Lemma \ref{Lemma 6}]Fix $n'\in\mathcal{N}$.
Again by the fact that the bids in the bootstrap sample are conditionally
i.i.d. and LIE, we have 
\begin{eqnarray*}
\mathrm{E}^{*}\left[\widehat{G}^{*}\left(b',\boldsymbol{x}',n'\right)\right] & = & \mathrm{E}^{*}\left[\mathbbm{1}\left(N_{1}^{*}=n'\right)\frac{1}{N_{1}^{*}}\sum_{i=1}^{N_{1}^{*}}\mathbbm{1}\left(B_{i1}^{*}\leq b'\right)\frac{1}{h^{d}}K_{\boldsymbol{X}}\left(\frac{\boldsymbol{X}_{1}^{*}-\boldsymbol{x}'}{h}\right)\right]\\
 & = & \mathrm{E}^{*}\left[\mathbbm{1}\left(N_{1}^{*}=n'\right)\mathrm{E}^{*}\left[\frac{1}{N_{1}^{*}}\sum_{i=1}^{N_{1}^{*}}\mathbbm{1}\left(B_{i1}^{*}\leq b'\right)\frac{1}{h^{d}}K_{\boldsymbol{X}}\left(\frac{\boldsymbol{X}_{1}^{*}-\boldsymbol{x}'}{h}\right)|\boldsymbol{X}_{1}^{*},N_{1}^{*}\right]\right]\\
 & = & \frac{1}{L}\sum_{l=1}^{L}\mathbbm{1}\left(N_{l}=n'\right)\frac{1}{N_{l}}\sum_{i=1}^{N_{l}}\mathbbm{1}\left(B_{il}\leq b'\right)\frac{1}{h^{d}}K_{\boldsymbol{X}}\left(\frac{\boldsymbol{X}_{l}-\boldsymbol{x}'}{h}\right).
\end{eqnarray*}
By Jensen's inequality and LIE, we have 
\begin{eqnarray*}
\mathrm{E}^{*}\left[\left\{ \mathbbm{1}\left(N_{1}^{*}=n'\right)\frac{1}{N_{1}^{*}}\sum_{i=1}^{N_{1}^{*}}\mathbbm{1}\left(B_{i1}^{*}\leq b'\right)\frac{1}{h^{d}}K_{\boldsymbol{X}}\left(\frac{\boldsymbol{X}_{1}^{*}-\boldsymbol{x}'}{h}\right)\right\} ^{2}\right] & \leq & \mathrm{E}^{*}\left[\mathbbm{1}\left(N_{1}^{*}=n'\right)\frac{1}{N_{1}^{*}}\sum_{i=1}^{N_{1}^{*}}\mathbbm{1}\left(B_{i1}^{*}\leq b'\right)\frac{1}{h^{2d}}K_{\boldsymbol{X}}\left(\frac{\boldsymbol{X}_{1}^{*}-\boldsymbol{x}'}{h}\right)^{2}\right]\\
 & = & \frac{1}{L}\sum_{l=1}^{L}\mathbbm{1}\left(N_{l}=n'\right)\frac{1}{N_{l}}\sum_{i=1}^{N_{l}}\mathbbm{1}\left(B_{il}\leq b'\right)\frac{1}{h^{2d}}K_{\boldsymbol{X}}\left(\frac{\boldsymbol{X}_{l}-\boldsymbol{x}'}{h}\right)^{2}.
\end{eqnarray*}
By arguments used to show (\ref{eq:uniform convergence rate}), we
have 
\begin{align*}
 & \underset{\mathbb{H}\left(\left(b',\boldsymbol{x}'\right),h\right)\subseteq\mathcal{S}_{B,\boldsymbol{X}}^{n'}}{\mathrm{sup}}\left|\frac{1}{L}\sum_{l=1}^{L}\mathbbm{1}\left(N_{l}=n'\right)\frac{1}{N_{l}}\sum_{i=1}^{N_{l}}\mathbbm{1}\left(B_{il}\leq b'\right)\frac{1}{h^{d}}K_{\boldsymbol{X}}\left(\frac{\boldsymbol{X}_{l}-\boldsymbol{x}'}{h}\right)^{2}-\mathrm{E}\left[\mathbbm{1}\left(N_{1}=n'\right)\frac{1}{N_{1}}\sum_{i=1}^{N_{1}}\mathbbm{1}\left(B_{i1}\leq b'\right)\frac{1}{h^{d}}K_{\boldsymbol{X}}\left(\frac{\boldsymbol{X}_{1}-\boldsymbol{x}'}{h}\right)^{2}\right]\right|\\
= & O_{p}\left(\left(\frac{\mathrm{log}\left(L\right)}{Lh^{1+d}}\right)^{\nicefrac{1}{2}}\right).
\end{align*}
Now by the LIE and change of variables, we have
\[
\mathrm{E}\left[\mathbbm{1}\left(N_{1}=n'\right)\frac{1}{N_{1}}\sum_{i=1}^{N_{1}}\mathbbm{1}\left(B_{i1}\leq b'\right)\frac{1}{h^{d}}K_{\boldsymbol{X}}\left(\frac{\boldsymbol{X}_{1}-\boldsymbol{x}'}{h}\right)^{2}\right]=\int_{\mathcal{Y}}G\left(b',h\boldsymbol{y}+\boldsymbol{x}',n'\right)K_{\boldsymbol{X}}\left(\boldsymbol{y}\right)^{2}\mathrm{d}\boldsymbol{y}.
\]
Now it is clear that since $K_{1}$ is compactly supported on $\left[-1,1\right]$,
\[
\underset{\mathbb{H}\left(\left(b',\boldsymbol{x}'\right),h\right)\subseteq\mathcal{S}_{B,\boldsymbol{X}}^{n'}}{\mathrm{sup}}\mathrm{E}\left[\mathbbm{1}\left(N_{1}=n'\right)\frac{1}{N_{1}}\sum_{i=1}^{N_{1}}\mathbbm{1}\left(B_{i1}\leq b'\right)\frac{1}{h^{d}}K_{\boldsymbol{X}}\left(\frac{\boldsymbol{X}_{1}-\boldsymbol{x}'}{h}\right)^{2}\right]=O\left(1\right).
\]
Now it easily follows that 
\[
\widehat{\sigma}_{B}^{2}\coloneqq\underset{\mathbb{H}\left(\left(b',\boldsymbol{x}'\right),h\right)\subseteq\mathcal{S}_{B,\boldsymbol{X}}^{n'}}{\mathrm{sup}}\mathrm{E}^{*}\left[\left\{ \mathbbm{1}\left(N_{1}^{*}=n'\right)\frac{1}{N_{1}^{*}}\sum_{i=1}^{N_{1}^{*}}\mathbbm{1}\left(B_{i1}^{*}\leq b'\right)\frac{1}{h^{d}}K_{\boldsymbol{X}}\left(\frac{\boldsymbol{X}_{1}^{*}-\boldsymbol{x}'}{h}\right)\right\} ^{2}\right]=O_{p}\left(h^{-d}\right).
\]

Let
\[
\mathcal{G}^{*,n'}\left(\boldsymbol{u},\boldsymbol{z},m,;b',x'\right)\coloneqq\mathbbm{1}\left(m=n'\right)\frac{1}{m}\sum_{i=1}^{m}\frac{1}{h^{d}}\mathbbm{1}\left(u_{i}\leq b'\right)K_{\boldsymbol{X}}\left(\frac{\boldsymbol{z}-\boldsymbol{x}'}{h}\right).
\]
Standard arguments can be applied to verify that the class $\left\{ \mathcal{G}^{*,n'}\left(\cdot;b',x'\right):\mathbb{H}\left(\left(b',\boldsymbol{x}'\right),h\right)\subseteq\mathcal{S}_{B,\boldsymbol{X}}^{n'}\right\} $
is uniformly VC-type with respect to a constant envelope $h^{-d}\left(\overline{C}_{D_{1}}+\overline{C}_{D_{2}}\right)\underset{\boldsymbol{x}'\in\mathbb{R}^{d}}{\mathrm{sup}}\left|K_{\boldsymbol{X}}\left(\boldsymbol{x}'\right)\right|$.
Then the CCK inequality yields 
\begin{eqnarray*}
\mathrm{E}^{*}\left[\underset{\mathbb{H}\left(\left(b',\boldsymbol{x}'\right),h\right)\subseteq\mathcal{S}_{B,\boldsymbol{X}}^{n'}}{\mathrm{sup}}\left|\widehat{G}^{*}\left(b',\boldsymbol{x}',n'\right)-\widehat{G}\left(b',\boldsymbol{x}',n'\right)\right|\right] & \leq & C_{1}\left\{ L^{-\nicefrac{1}{2}}\widehat{\sigma}_{B}\mathrm{log}\left(C_{2}L\right)^{\nicefrac{1}{2}}+\left(Lh^{d}\right)^{-1}\mathrm{log}\left(C_{2}L\right)\right\} \\
 & = & O_{p}\left(\left(\frac{\mathrm{log}\left(L\right)}{Lh^{d}}\right)^{\nicefrac{1}{2}}\right).
\end{eqnarray*}

The first conclusion follows from Lemma \ref{lem:auxiliary 3} and
also the assumption that $\mathcal{N}$ is a bounded set of positive
integers. The second conclusion follows from similar arguments.\end{proof} 
\begin{lem}
\label{Lemma 7}Suppose that Assumptions 1 - 3 hold. Let $\boldsymbol{x}$
be an interior point of $\mathcal{X}$ and $n\in\mathcal{N}$ be fixed.
Let
\[
\widetilde{\mathbb{T}}_{il}^{*}\coloneqq\mathbbm{1}\left(\left(V_{il}^{*},\boldsymbol{X}_{l}^{*}\right)\in\mathbb{H}\left(\left(v,\boldsymbol{x}\right),\overline{\delta}\right)\right).
\]
Then 
\[
\widehat{f}_{GPV}^{*}\left(v,\boldsymbol{x},n\right)-\widetilde{f}^{*}\left(v,\boldsymbol{x},n\right)=\frac{1}{L}\sum_{l=1}^{L}\widetilde{\mathbb{T}}_{il}^{*}\frac{1}{h^{2+d}}K_{f}'\left(\frac{\widehat{V}_{il}^{*}-v}{h},\frac{\boldsymbol{X}_{l}^{*}-\boldsymbol{x}}{h}\right)\left(\widehat{V}_{il}^{*}-V_{il}^{*}\right)+O_{p}^{*}\left(\frac{\mathrm{log}\left(L\right)}{Lh^{3+d}}+h^{R}\right),
\]
where the remainder term is uniform in $v\in I\left(\boldsymbol{x}\right)$.
\end{lem}
\begin{proof}[Proof of Lemma \ref{Lemma 7}]It follows from Lemma
\ref{Lemma 6}, (\ref{eq:uniform convergence rate}) and \citet[Lemma S1]{Marmer_Shneyerov_Quantile_Auctions}
that 
\begin{gather}
\underset{n'\in\mathcal{N}}{\mathrm{max}}\underset{\mathbb{H}\left(\left(b',\boldsymbol{x}'\right),h\right)\subseteq\mathcal{S}_{B,\boldsymbol{X}}^{n'}}{\mathrm{sup}}\left|\widehat{G}^{*}\left(b',\boldsymbol{x}',n'\right)-G\left(b',\boldsymbol{x}',n'\right)\right|=O_{p}^{*}\left(\left(\frac{\mathrm{log}\left(L\right)}{Lh^{d}}\right)^{\nicefrac{1}{2}}+h^{1+R}\right)\nonumber \\
\underset{n'\in\mathcal{N}}{\mathrm{max}}\underset{\mathbb{H}\left(\left(b',\boldsymbol{x}'\right),h\right)\subseteq\mathcal{S}_{B,\boldsymbol{X}}^{n'}}{\mathrm{sup}}\left|\widehat{g}^{*}\left(b',\boldsymbol{x}',n'\right)-g\left(b',\boldsymbol{x}',n'\right)\right|=O_{p}^{*}\left(\left(\frac{\mathrm{log}\left(L\right)}{Lh^{1+d}}\right)^{\nicefrac{1}{2}}+h^{1+R}\right).\label{eq:uniform convergence rate bootstrap}
\end{gather}

We apply again the arguments used in the proof of Lemma \ref{lem:Lemma 1 Stochastic Expansion}.
First, we have
\begin{eqnarray}
\underset{i,l}{\mathrm{max}}\,\mathbb{T}_{il}^{*}\left|\widehat{V}_{il}^{*}-V_{il}^{*}\right| & \leq & \underset{i,l}{\mathrm{max}}\,\mathbb{T}_{il}^{*}\left|\frac{\widehat{G}^{*}\left(B_{il}^{*},\boldsymbol{X}_{l}^{*},N_{l}^{*}\right)-G\left(B_{il}^{*},\boldsymbol{X}_{l}^{*},N_{l}^{*}\right)}{g\left(B_{il}^{*},\boldsymbol{X}_{l}^{*},N_{l}^{*}\right)}\right|+\underset{i,l}{\mathrm{max}}\,\mathbb{T}_{il}^{*}\left|\frac{\widehat{G}^{*}\left(B_{il}^{*},\boldsymbol{X}_{l}^{*},N_{l}^{*}\right)\left(\widehat{g}^{*}\left(B_{il}^{*},\boldsymbol{X}_{l}^{*},N_{l}^{*}\right)-g\left(B_{il}^{*},\boldsymbol{X}_{l}^{*},N_{l}^{*}\right)\right)}{g\left(B_{il}^{*},\boldsymbol{X}_{l}^{*},N_{l}^{*}\right)^{2}}\right|\nonumber \\
 &  & +\underset{i,l}{\mathrm{max}}\,\mathbb{T}_{il}^{*}\left|\frac{\widehat{G}^{*}\left(B_{il}^{*},\boldsymbol{X}_{l}^{*},N_{l}^{*}\right)}{\widehat{g}^{*}\left(B_{il}^{*},\boldsymbol{X}_{l}^{*},N_{l}^{*}\right)}\frac{\left(\widehat{g}^{*}\left(B_{il}^{*},\boldsymbol{X}_{l}^{*},N_{l}^{*}\right)-g\left(B_{il}^{*},\boldsymbol{X}_{l}^{*},N_{l}^{*}\right)\right)^{2}}{g\left(B_{il}^{*},\boldsymbol{X}_{l}^{*},N_{l}^{*}\right)^{2}}\right|.\label{eq:max V_star - V bound 1}
\end{eqnarray}

Denote
\[
\overline{\mathbb{T}}_{il}^{*}\coloneqq\mathbbm{1}\left(\mathbb{H}\left(\left(B_{il}^{*},\boldsymbol{X}_{l}^{*}\right),h\right)\subseteq\mathcal{S}_{B,\boldsymbol{X}}^{N_{l}}\right).
\]
For the first term of the right hand side of (\ref{eq:max V_star - V bound 1}),
we have
\begin{align}
 & \underset{i,l}{\mathrm{max}}\,\mathbb{T}_{il}^{*}\left|\frac{\widehat{G}^{*}\left(B_{il}^{*},\boldsymbol{X}_{l}^{*},N_{l}^{*}\right)-G\left(B_{il}^{*},\boldsymbol{X}_{l}^{*},N_{l}^{*}\right)}{g\left(B_{il}^{*},\boldsymbol{X}_{l}^{*},N_{l}^{*}\right)}\right|\nonumber \\
= & \underset{i,l}{\mathrm{max}}\,\mathbb{T}_{il}^{*}\overline{\mathbb{T}}_{il}^{*}\left|\frac{\widehat{G}^{*}\left(B_{il}^{*},\boldsymbol{X}_{l}^{*},N_{l}^{*}\right)-G\left(B_{il}^{*},\boldsymbol{X}_{l}^{*},N_{l}^{*}\right)}{g\left(B_{il}^{*},\boldsymbol{X}_{l}^{*},N_{l}^{*}\right)}\right|+\underset{i,l}{\mathrm{max}}\,\mathbb{T}_{il}^{*}\left(1-\overline{\mathbb{T}}_{il}^{*}\right)\left|\frac{\widehat{G}^{*}\left(B_{il}^{*},\boldsymbol{X}_{l}^{*},N_{l}^{*}\right)-G\left(B_{il}^{*},\boldsymbol{X}_{l}^{*},N_{l}^{*}\right)}{g\left(B_{il}^{*},\boldsymbol{X}_{l}^{*},N_{l}^{*}\right)}\right|\label{eq:Lemma 1 order bound 1-1}
\end{align}
and by (\ref{eq:bids density bounded away from zero}) and (\ref{eq:uniform convergence rate bootstrap}),
\begin{eqnarray}
\underset{i,l}{\mathrm{max}}\,\mathbb{T}_{il}^{*}\overline{\mathbb{T}}_{il}^{*}\left|\frac{\widehat{G}^{*}\left(B_{il}^{*},\boldsymbol{X}_{l}^{*},N_{l}^{*}\right)-G\left(B_{il}^{*},\boldsymbol{X}_{l}^{*},N_{l}^{*}\right)}{g\left(B_{il}^{*},\boldsymbol{X}_{l}^{*},N_{l}^{*}\right)}\right| & \apprle & \underset{i,l}{\mathrm{max}}\,\overline{\mathbb{T}}_{il}^{*}\left|\widehat{G}^{*}\left(B_{il}^{*},\boldsymbol{X}_{l}^{*},N_{l}^{*}\right)-G\left(B_{il}^{*},\boldsymbol{X}_{l}^{*},N_{l}^{*}\right)\right|\nonumber \\
 & = & O_{p}^{*}\left(\left(\frac{\mathrm{log}\left(L\right)}{Lh^{d}}\right)^{\nicefrac{1}{2}}+h^{1+R}\right).\label{eq:Lemma 1 order bound 2}
\end{eqnarray}

It is argued in the proof of Lemma \ref{lem:Lemma 1 Stochastic Expansion}
that if $\underset{\left(\boldsymbol{x}',n'\right)\in\mathcal{X}\times\mathcal{N}}{\mathrm{sup}}\left|\widehat{\overline{b}}\left(\boldsymbol{x}',n'\right)-\overline{b}\left(\boldsymbol{x}',n'\right)\right|\leq h$,
$\underset{\boldsymbol{x}'\in\mathcal{X}}{\mathrm{sup}}\left|\widehat{\underline{b}}\left(\boldsymbol{x}'\right)-\underline{b}\left(\boldsymbol{x}'\right)\right|\leq h$
and $\mathbb{H}\left(\left(B_{il},\boldsymbol{X}_{l}\right),2h\right)\subseteq\mathcal{\widehat{S}}_{B,\boldsymbol{X}}^{N_{l}}$,
we must have $\mathbb{H}\left(\left(B_{il},\boldsymbol{X}_{l}\right),h\right)\subseteq\mathcal{S}_{B,X}^{N_{l}}$.
Now it is clear that when $h$ is sufficiently small,
\[
\mathrm{P}^{*}\left[\underset{i,l}{\mathrm{max}}\,\mathbb{T}_{il}^{*}\left(1-\overline{\mathbb{T}}_{il}^{*}\right)>0\right]\leq\mathbbm{1}\left(\underset{\left(x',n'\right)\in\mathcal{X}\times\mathcal{N}}{\mathrm{max}}\left|\widehat{\overline{b}}\left(\boldsymbol{x}',n'\right)-\overline{b}\left(\boldsymbol{x}',n'\right)\right|\leq h\right)+\mathbbm{1}\left(\underset{x'\in\mathcal{X}}{\mathrm{sup}}\left|\widehat{\underline{b}}\left(\boldsymbol{x}'\right)-\underline{b}\left(\boldsymbol{x}'\right)\right|\leq h\right)=o_{p}\left(1\right),
\]
where the equality follows from (\ref{eq:boundary estimator rate})
and Markov's inequality. Therefore we have 
\[
\underset{i,l}{\mathrm{max}}\,\mathbb{T}_{il}^{*}\left(1-\overline{\mathbb{T}}_{il}^{*}\right)\left|\frac{\widehat{G}^{*}\left(B_{il}^{*},\boldsymbol{X}_{l}^{*},N_{l}^{*}\right)-G\left(B_{il}^{*},\boldsymbol{X}_{l}^{*},N_{l}^{*}\right)}{g\left(B_{il}^{*},\boldsymbol{X}_{l}^{*},N_{l}^{*}\right)}\right|=o_{p}^{*}\left(\epsilon_{L}\right)
\]
for any null sequence $\epsilon_{L}\downarrow0$. By the above result,
(\ref{eq:Lemma 1 order bound 1-1}) and (\ref{eq:Lemma 1 order bound 2}),
we have
\begin{equation}
\underset{i,l}{\mathrm{max}}\,\mathbb{T}_{il}^{*}\left|\frac{\widehat{G}^{*}\left(B_{il}^{*},\boldsymbol{X}_{l}^{*},N_{l}^{*}\right)-G\left(B_{il}^{*},\boldsymbol{X}_{l}^{*},N_{l}^{*}\right)}{g\left(B_{il}^{*},\boldsymbol{X}_{l}^{*},N_{l}^{*}\right)}\right|=O_{p}^{*}\left(\left(\frac{\mathrm{log}\left(L\right)}{Lh^{d}}\right)^{\nicefrac{1}{2}}+h^{1+R}\right).\label{eq:sup V_hat - V first term order bound-1}
\end{equation}
Similarly, we have
\begin{equation}
\underset{i,l}{\mathrm{max}}\,\mathbb{T}_{il}^{*}\left|\frac{\widehat{G}^{*}\left(B_{il}^{*},\boldsymbol{X}_{l}^{*},N_{l}^{*}\right)\left(\widehat{g}^{*}\left(B_{il}^{*},\boldsymbol{X}_{l}^{*},N_{l}^{*}\right)-g\left(B_{il}^{*},\boldsymbol{X}_{l}^{*},N_{l}^{*}\right)\right)}{g\left(B_{il}^{*},\boldsymbol{X}_{l}^{*},N_{l}^{*}\right)^{2}}\right|=O_{p}^{*}\left(\left(\frac{\mathrm{log}\left(L\right)}{Lh^{1+d}}\right)^{\nicefrac{1}{2}}+h^{1+R}\right).\label{eq:Lemma 1 order bound 3}
\end{equation}

It follows from (\ref{eq:uniform convergence rate bootstrap}) that
\[
\underset{i,l}{\mathrm{max}}\,\overline{\mathbb{T}}_{il}^{*}\left|\widehat{g}^{*}\left(B_{il},\boldsymbol{X}_{l},N_{l}\right)-g\left(B_{il},\boldsymbol{X}_{l},N_{l}\right)\right|=o_{p}^{*}\left(1\right).
\]
It follows from the above result and (\ref{eq:bids density bounded away from zero})
that 
\[
\mathrm{P}^{*}\left[\underset{i,l}{\mathrm{max}}\,\overline{\mathbb{T}}_{il}^{*}\left|\widehat{g}^{*}\left(B_{il}^{*},\boldsymbol{X}_{l}^{*},N_{l}^{*}\right)^{-1}\right|\leq\left(\frac{\underline{C}_{g}}{2}\right)^{-1}\right]\rightarrow_{p}1,\textrm{ as \ensuremath{L\uparrow\infty}},
\]
which further implies $\underset{i,l}{\mathrm{max}}\,\overline{\mathbb{T}}_{il}^{*}\left|\widehat{g}^{*}\left(B_{il}^{*},\boldsymbol{X}_{l}^{*},N_{l}^{*}\right)^{-1}\right|=O_{p}^{*}\left(1\right)$.
Therefore we have
\[
\underset{i,l}{\mathrm{max}}\,\mathbb{T}_{il}^{*}\left|\frac{\widehat{G}^{*}\left(B_{il}^{*},\boldsymbol{X}_{l}^{*},N_{l}^{*}\right)}{\widehat{g}^{*}\left(B_{il}^{*},\boldsymbol{X}_{l}^{*},N_{l}^{*}\right)}\frac{\left(\widehat{g}^{*}\left(B_{il}^{*},\boldsymbol{X}_{l}^{*},N_{l}^{*}\right)-g\left(B_{il}^{*},\boldsymbol{X}_{l}^{*},N_{l}^{*}\right)\right)^{2}}{g\left(B_{il}^{*},\boldsymbol{X}_{l}^{*},N_{l}^{*}\right)^{2}}\right|=O_{p}^{*}\left(\frac{\mathrm{log}\left(L\right)}{Lh^{1+d}}+h^{2+2R}\right).
\]
It follows from the above result, (\ref{eq:max V_star - V bound 1}),
(\ref{eq:sup V_hat - V first term order bound-1}) and (\ref{eq:Lemma 1 order bound 3})
that
\begin{equation}
\underset{i,l}{\mathrm{max}}\,\mathbb{T}_{il}^{*}\left|\widehat{V}_{il}^{*}-V_{il}^{*}\right|=O_{p}^{*}\left(\left(\frac{\mathrm{log}\left(L\right)}{Lh^{1+d}}\right)^{\nicefrac{1}{2}}+h^{1+R}\right).\label{eq:V_hat - V Feasible trimming bootstrap}
\end{equation}

We showed in the proof of Lemma \ref{lem:Lemma 1 Stochastic Expansion}
that when $h$ is sufficiently small, for every $v\in I\left(\boldsymbol{x}\right)$,
$\left(V_{il}^{*},\boldsymbol{X}_{l}^{*}\right)\in\mathbb{H}\left(\left(v,\boldsymbol{x}\right),\overline{\delta}\right)$
implies $\mathbb{H}\left(\left(B_{il}^{*},\boldsymbol{X}_{l}^{*}\right),h\right)\subseteq\mathcal{S}_{B,\boldsymbol{X}}^{N_{l}}$.
Therefore we have
\begin{equation}
\underset{v\in I\left(\boldsymbol{x}\right)}{\mathrm{sup}}\underset{i,l}{\mathrm{max}}\,\widetilde{\mathbb{T}}_{il}^{*}\left|\widehat{V}_{il}^{*}-V_{il}^{*}\right|=O_{p}^{*}\left(\left(\frac{\mathrm{log}\left(L\right)}{Lh^{1+d}}\right)^{\nicefrac{1}{2}}+h^{1+R}\right).\label{eq:V_hat - V Infeasible Trimming bootstrap}
\end{equation}

Write
\[
\widehat{f}_{GPV}^{*}\left(v,\boldsymbol{x},n\right)=\frac{1}{L}\sum_{l=1}^{L}\mathbbm{1}\left(N_{l}^{*}=n\right)\frac{1}{N_{l}}\sum_{i=1}^{N_{l}}\widetilde{\mathbb{T}}_{il}^{*}\frac{1}{h^{1+d}}K_{f}\left(\frac{\widehat{V}_{il}^{*}-v}{h},\frac{\boldsymbol{X}_{l}^{*}-\boldsymbol{x}}{h}\right)+\kappa_{1}^{*}\left(v\right)+\kappa_{2}^{*}\left(v\right),
\]
where
\begin{gather*}
\kappa_{1}^{*}\left(v\right)\coloneqq\frac{1}{L}\sum_{l=1}^{L}\mathbbm{1}\left(N_{l}^{*}=n\right)\frac{1}{N_{l}^{*}}\sum_{i=1}^{N_{l}^{*}}\mathbb{T}_{il}^{*}\left(1-\widetilde{\mathbb{T}}_{il}^{*}\right)\frac{1}{h^{1+d}}K_{f}\left(\frac{\widehat{V}_{il}^{*}-v}{h},\frac{\boldsymbol{X}_{l}^{*}-\boldsymbol{x}}{h}\right)\\
\kappa_{2}^{*}\left(v\right)\coloneqq\frac{1}{L}\sum_{l=1}^{L}\mathbbm{1}\left(N_{l}^{*}=n\right)\frac{1}{N_{l}^{*}}\sum_{i=1}^{N_{l}^{*}}\widetilde{\mathbb{T}}_{il}^{*}\left(\mathbb{T}_{il}^{*}-1\right)\frac{1}{h^{1+d}}K_{f}\left(\frac{\widehat{V}_{il}^{*}-v}{h},\frac{\boldsymbol{X}_{l}^{*}-\boldsymbol{x}}{h}\right).
\end{gather*}

Since $K_{0}$ is supported on $\left[-1,1\right]$, $K_{0}\left(\nicefrac{\left(\widehat{V}_{il}^{*}-v\right)}{h}\right)$
is zero if $\widehat{V}_{il}^{*}$ is outside of a $h-$neighborhood
of $v$. Then by the triangle inequality, we have
\begin{eqnarray*}
\left|\kappa_{1}^{*}\left(v\right)\right| & \apprle & \frac{1}{L}\sum_{l=1}^{L}\mathbbm{1}\left(N_{l}^{*}=n\right)\frac{1}{N_{l}^{*}}\sum_{i=1}^{N_{l}^{*}}\mathbb{T}_{il}^{*}\left(1-\widetilde{\mathbb{T}}_{il}^{*}\right)h^{-\left(1+d\right)}\mathbbm{1}\left(\left|\widehat{V}_{il}^{*}-v\right|\leq h\right)\mathbbm{1}\left(\boldsymbol{X}_{l}^{*}\in\mathbb{H}\left(\boldsymbol{x},h\right)\right)\\
 & \apprle & \frac{1}{L}\sum_{l=1}^{L}\mathbbm{1}\left(N_{l}^{*}=n\right)\frac{1}{N_{l}^{*}}\sum_{i=1}^{N_{l}^{*}}\mathbb{T}_{il}^{*}\left(1-\widetilde{\mathbb{T}}_{il}^{*}\right)h^{-\left(1+d\right)}\mathbbm{1}\left(\left|V_{il}^{*}-v\right|\leq h+\underset{i,l}{\mathrm{max}}\,\mathbb{T}_{il}\left|\widehat{V}_{il}-V_{il}\right|\right)\mathbbm{1}\left(\boldsymbol{X}_{l}^{*}\in\mathbb{H}\left(\boldsymbol{x},h\right)\right).
\end{eqnarray*}
Therefore it is clear that
\[
\mathrm{P}^{*}\left[\underset{v\in I\left(\boldsymbol{x}\right)}{\mathrm{sup}}\left|\kappa_{1}^{*}\left(v\right)\right|>0\right]\leq\mathrm{P}^{*}\left[\underset{i,l}{\mathrm{max}}\,\mathbb{T}_{il}^{*}\left|\widehat{V}_{il}^{*}-V_{il}^{*}\right|\geq\frac{\overline{\delta}}{2}\right]=o_{p}\left(1\right),
\]
where the inequality holds when $h$ is sufficiently small. Therefore,
$\underset{v\in I\left(\boldsymbol{x}\right)}{\mathrm{sup}}\left|\kappa_{1}^{*}\left(v\right)\right|=o_{p}^{*}\left(\epsilon_{L}\right)$
for any null sequence $\epsilon_{L}\downarrow0$.

As argued in the proof of Lemma \ref{lem:Lemma 1 Stochastic Expansion},
when $h$ is sufficiently small, 
\begin{eqnarray*}
\underset{v\in I\left(\boldsymbol{x}\right)}{\mathrm{sup}}\left|\kappa_{2}^{*}\left(v\right)\right| & \apprle & h^{-\left(1+d\right)}\mathbbm{1}\left(\underset{\boldsymbol{x}'\in\mathbb{H}\left(\boldsymbol{x},3h\right)}{\mathrm{sup}}s\left(v_{u}\left(\boldsymbol{x}\right)+\overline{\delta},\boldsymbol{x}',n\right)+2h>\underset{\boldsymbol{x}'\in\mathbb{H}\left(\boldsymbol{x},3h\right)}{\mathrm{inf}}\overline{b}\left(\boldsymbol{x}',n\right)-\underset{\boldsymbol{x}'\in\mathcal{X}}{\mathrm{sup}}\left|\widehat{\overline{b}}\left(\boldsymbol{x}',n\right)-\overline{b}\left(\boldsymbol{x}',n\right)\right|\right)\\
 &  & +h^{-\left(1+d\right)}\mathbbm{1}\left(\underset{\boldsymbol{x}'\in\mathbb{H}\left(\boldsymbol{x},3h\right)}{\mathrm{inf}}s\left(v_{l}\left(\boldsymbol{x}\right)-\overline{\delta},\boldsymbol{x}',n\right)-2h<\underset{\boldsymbol{x}'\in\mathbb{H}\left(\boldsymbol{x},3h\right)}{\mathrm{sup}}\underline{b}\left(\boldsymbol{x}'\right)+\underset{\boldsymbol{x}'\in\mathcal{X}}{\mathrm{sup}}\left|\widehat{\underline{b}}\left(\boldsymbol{x}'\right)-\underline{b}\left(\boldsymbol{x}'\right)\right|\right).
\end{eqnarray*}
Now it follows that 
\begin{eqnarray*}
\mathrm{P}^{*}\left[\underset{v\in I\left(\boldsymbol{x}\right)}{\mathrm{sup}}\left|\kappa_{2}^{*}\left(v\right)\right|>0\right] & \leq & \mathbbm{1}\left(\underset{\boldsymbol{x}'\in\mathbb{H}\left(\boldsymbol{x},3h\right)}{\mathrm{sup}}s\left(v_{u}\left(\boldsymbol{x}\right)+\overline{\delta},\boldsymbol{x}',n\right)+2h>\underset{\boldsymbol{x}'\in\mathbb{H}\left(\boldsymbol{x},3h\right)}{\mathrm{inf}}\overline{b}\left(\boldsymbol{x}',n\right)-\underset{\boldsymbol{x}'\in\mathcal{X}}{\mathrm{sup}}\left|\widehat{\overline{b}}\left(\boldsymbol{x}',n\right)-\overline{b}\left(\boldsymbol{x}',n\right)\right|\right)\\
 &  & +\mathbbm{1}\left(\underset{\boldsymbol{x}'\in\mathbb{H}\left(\boldsymbol{x},3h\right)}{\mathrm{inf}}s\left(v_{l}\left(\boldsymbol{x}\right)-\overline{\delta},\boldsymbol{x}',n\right)-2h<\underset{\boldsymbol{x}'\in\mathbb{H}\left(\boldsymbol{x},3h\right)}{\mathrm{sup}}\underline{b}\left(\boldsymbol{x}'\right)+\underset{\boldsymbol{x}'\in\mathcal{X}}{\mathrm{sup}}\left|\widehat{\underline{b}}\left(\boldsymbol{x}'\right)-\underline{b}\left(\boldsymbol{x}'\right)\right|\right)\\
 & = & o_{p}\left(1\right),
\end{eqnarray*}
where the equality follows from (\ref{eq:boundary estimator rate})
and Markov's inequality. Therefore, $\underset{v\in I\left(\boldsymbol{x}\right)}{\mathrm{sup}}\left|\kappa_{2}^{*}\left(v\right)\right|=o_{p}^{*}\left(\epsilon_{L}\right)$
for any null sequence $\epsilon_{L}\downarrow0$.

A second-order Taylor expansion gives
\begin{eqnarray}
\widehat{f}_{GPV}^{*}\left(v,\boldsymbol{x},n\right)-\widetilde{f}^{*}\left(v,\boldsymbol{x},n\right) & = & \frac{1}{L}\sum_{l=1}^{L}\mathbbm{1}\left(N_{l}^{*}=n\right)\frac{1}{N_{l}^{*}}\sum_{i=1}^{N_{l}^{*}}\widetilde{\mathbb{T}}_{il}^{*}\frac{1}{h^{2+d}}K_{f}'\left(\frac{V_{il}^{*}-v}{h},\frac{\boldsymbol{X}_{l}^{*}-\boldsymbol{x}}{h}\right)\left(\widehat{V}_{il}^{*}-V_{il}^{*}\right)\nonumber \\
 &  & +\frac{1}{2}\frac{1}{L}\sum_{l=1}^{L}\mathbbm{1}\left(N_{l}^{*}=n\right)\frac{1}{N_{l}^{*}}\sum_{i=1}^{N_{l}^{*}}\widetilde{\mathbb{T}}_{il}^{*}\frac{1}{h^{3+d}}K_{f}''\left(\frac{\dot{V}_{il}^{*}-v}{h},\frac{\boldsymbol{X}_{l}^{*}-\boldsymbol{x}}{h}\right)\left(\widehat{V}_{il}-V_{il}\right)^{2},\label{eq:Lemma 1 approximation error 1-1}
\end{eqnarray}
for some mean value $\dot{V}_{il}^{*}$ that lies on the line joining
$\widehat{V}_{il}^{*}$ and $V_{il}^{*}$, with some remainder term
that is $o_{p}^{*}\left(\epsilon_{L}\right)$ for any null sequence
$\epsilon_{L}\downarrow0$. It follows from the triangle inequality
and the Lipschitz condition imposed on the kernel that
\begin{align}
 & \left|\frac{1}{L}\sum_{l=1}^{L}\mathbbm{1}\left(N_{l}^{*}=n\right)\frac{1}{N_{l}^{*}}\sum_{i=1}^{N_{l}^{*}}\widetilde{\mathbb{T}}_{il}^{*}\frac{1}{h^{3+d}}K_{f}''\left(\frac{\dot{V}_{il}^{*}-v}{h},\frac{\boldsymbol{X}_{l}^{*}-\boldsymbol{x}}{h}\right)\left(\hat{V}_{il}^{*}-V_{il}\right)^{2}\right|\nonumber \\
\apprle & \left(\frac{1}{L}\sum_{l=1}^{L}\mathbbm{1}\left(N_{l}^{*}=n\right)\frac{1}{N_{l}^{*}}\sum_{i=1}^{N_{l}^{*}}\widetilde{\mathbb{T}}_{il}^{*}\frac{1}{h^{3+d}}\left|K_{\boldsymbol{X}}^{0}\left(\frac{\boldsymbol{X}_{l}^{*}-\boldsymbol{x}}{h}\right)\right|\mathbbm{1}\left(\left|\dot{V}_{il}^{*}-v\right|\leq h\right)\right)\left(\underset{i,l}{\mathrm{max}}\,\widetilde{\mathbb{T}}_{il}^{*}\left(\widehat{V}_{il}^{*}-V_{il}^{*}\right)^{2}\right).\label{eq:Lemma 1 approximation error 2-1}
\end{align}

By the triangle inequality, we have
\begin{align}
 & \frac{1}{L}\sum_{l=1}^{L}\mathbbm{1}\left(N_{l}^{*}=n\right)\frac{1}{N_{l}^{*}}\sum_{i=1}^{N_{l}^{*}}\widetilde{\mathbb{T}}_{il}^{*}\frac{1}{h^{3+d}}\left|K_{\boldsymbol{X}}^{0}\left(\frac{\boldsymbol{X}_{l}^{*}-\boldsymbol{x}}{h}\right)\right|\mathbbm{1}\left(\left|\dot{V}_{il}^{*}-v\right|\leq h\right)\nonumber \\
\leq & \frac{1}{L}\sum_{l=1}^{L}\mathbbm{1}\left(N_{l}^{*}=n\right)\frac{1}{N_{l}^{*}}\sum_{i=1}^{N_{l}^{*}}\widetilde{\mathbb{T}}_{il}^{*}\left(1-\reallywidecheck{\mathbb{T}}_{il}^{*}+\reallywidecheck{\mathbb{T}}_{il}^{*}\right)\frac{1}{h^{3+d}}\left|K_{\boldsymbol{X}}^{0}\left(\frac{\boldsymbol{X}_{l}^{*}-\boldsymbol{x}}{h}\right)\right|\mathbbm{1}\left(\left|V_{il}^{*}-v\right|\leq h+\underset{j,k}{\mathrm{max}}\,\widetilde{\mathbb{T}}_{jk}^{*}\left|\dot{V}_{jk}^{*}-V_{jk}^{*}\right|\right)\nonumber \\
\leq & \frac{1}{L}\sum_{l=1}^{L}\mathbbm{1}\left(N_{l}^{*}=n\right)\frac{1}{N_{l}^{*}}\sum_{i=1}^{N_{l}^{*}}\reallywidecheck{\mathbb{T}}_{il}^{*}\frac{1}{h^{3+d}}\left|K_{\boldsymbol{X}}\left(\frac{\boldsymbol{X}_{l}^{*}-\boldsymbol{x}}{h}\right)\right|+\kappa_{3}^{*}\left(v\right),\label{eq:Lemma 1 approximation error indicator-1}
\end{align}
where
\[
\reallywidecheck{\mathbb{T}}_{il}^{*}\coloneqq\mathbbm{1}\left(\left|V_{il}^{*}-v\right|\leq2h\right)
\]
and
\[
\kappa_{3}^{*}\left(v\right)\coloneqq\frac{1}{L}\sum_{l=1}^{L}\mathbbm{1}\left(N_{l}^{*}=n\right)\frac{1}{N_{l}^{*}}\sum_{i=1}^{N_{l}^{*}}\widetilde{\mathbb{T}}_{il}^{*}\mathbbm{1}\left(\left|V_{il}^{*}-v\right|>2h\right)\frac{1}{h^{3+d}}\left|K_{\boldsymbol{X}}^{0}\left(\frac{\boldsymbol{X}_{l}^{*}-\boldsymbol{x}}{h}\right)\right|\mathbbm{1}\left(\left|V_{il}^{*}-v\right|\leq h+\underset{j,k}{\mathrm{max}}\,\widetilde{\mathbb{T}}_{jk}^{*}\left|\dot{V}_{jk}^{*}-V_{jk}^{*}\right|\right).
\]
Clearly we have 
\[
\mathrm{P}^{*}\left[\underset{v\in I\left(\boldsymbol{x}\right)}{\mathrm{sup}}\left|\kappa_{3}^{*}\left(v\right)\right|>0\right]\leq\mathrm{P}^{*}\left[\underset{v\in I\left(\boldsymbol{x}\right)}{\mathrm{sup}}\underset{j,k}{\mathrm{max}}\,\widetilde{\mathbb{T}}_{jk}^{*}\left|\dot{V}_{jk}^{*}-V_{jk}^{*}\right|>h\right]=o_{p}\left(1\right),
\]
where the equality follows from (\ref{eq:V_hat - V Infeasible Trimming bootstrap}).
Therefore, $\underset{v\in I\left(\boldsymbol{x}\right)}{\mathrm{sup}}\left|\kappa_{3}^{*}\left(v\right)\right|=o_{p}\left(\epsilon_{L}\right)$,
for any null sequence $\epsilon_{L}\downarrow0$.

By arguments used in the proof of Lemma \ref{Lemma 5}, we can easily
show
\begin{align*}
 & \underset{v\in I\left(\boldsymbol{x}\right)}{\mathrm{sup}}\left|\frac{1}{L}\sum_{l=1}^{L}\mathbbm{1}\left(N_{l}^{*}=n\right)\frac{1}{N_{l}^{*}}\sum_{i=1}^{N_{l}^{*}}\reallywidecheck{\mathbb{T}}_{il}^{*}\frac{1}{h^{1+d}}\left|K_{\boldsymbol{X}}\left(\frac{\boldsymbol{X}_{l}^{*}-\boldsymbol{x}}{h}\right)\right|-\mathrm{E}^{*}\left[\mathbbm{1}\left(N_{1}^{*}=n\right)\frac{1}{N_{1}^{*}}\sum_{i=1}^{N_{1}^{*}}\reallywidecheck{\mathbb{T}}_{i1}^{*}\frac{1}{h^{1+d}}\left|K_{\boldsymbol{X}}\left(\frac{\boldsymbol{X}_{1}^{*}-\boldsymbol{x}}{h}\right)\right|\right]\right|\\
= & O_{p}^{*}\left(\left(\frac{\mathrm{log}\left(L\right)}{Lh^{1+d}}\right)^{\nicefrac{1}{2}}\right).
\end{align*}
We also have
\[
\underset{v\in I\left(\boldsymbol{x}\right)}{\mathrm{sup}}\mathrm{E}^{*}\left[\mathbbm{1}\left(N_{1}^{*}=n\right)\frac{1}{N_{1}^{*}}\sum_{i=1}^{N_{1}^{*}}\reallywidecheck{\mathbb{T}}_{i1}^{*}\frac{1}{h^{1+d}}\left|K_{\boldsymbol{X}}\left(\frac{\boldsymbol{X}_{1}^{*}-\boldsymbol{x}}{h}\right)\right|\right]=\underset{v\in I\left(\boldsymbol{x}\right)}{\mathrm{sup}}\frac{1}{L}\sum_{l=1}^{L}\mathcal{F}^{n}\left(\boldsymbol{V}_{\cdot l},\boldsymbol{X}_{l},N_{l};v\right)=O_{p}\left(1\right),
\]
where the first equality follows from LIE and the fact that the bids
in the bootstrap sample are conditionally i.i.d. and the second equality
was shown in the proof of Lemma \ref{lem:Lemma 1 Stochastic Expansion}.
See (\ref{eq:sup F average O_p(1)}). Now it follows that 
\[
\frac{1}{L}\sum_{l=1}^{L}\mathbbm{1}\left(N_{l}^{*}=n\right)\frac{1}{N_{l}^{*}}\sum_{i=1}^{N_{l}^{*}}\reallywidecheck{\mathbb{T}}_{il}^{*}\frac{1}{h^{3+d}}\left|K_{\boldsymbol{X}}\left(\frac{\boldsymbol{X}_{l}^{*}-\boldsymbol{x}}{h}\right)\right|=O_{p}^{*}\left(h^{-2}\right),
\]
uniformly in $v\in I\left(\boldsymbol{x}\right)$. Then it follows
from the above result that
\[
\left(\frac{1}{L}\sum_{l=1}^{L}\mathbbm{1}\left(N_{l}^{*}=n\right)\frac{1}{N_{l}^{*}}\sum_{i=1}^{N_{l}^{*}}\widetilde{\mathbb{T}}_{il}^{*}\frac{1}{h^{3+d}}\left|K_{\boldsymbol{X}}^{0}\left(\frac{\boldsymbol{X}_{l}^{*}-\boldsymbol{x}}{h}\right)\right|\mathbbm{1}\left(\left|\dot{V}_{il}^{*}-v\right|\leq h\right)\right)\left(\underset{i,l}{\mathrm{max}}\,\widetilde{\mathbb{T}}_{il}^{*}\left(\widehat{V}_{il}^{*}-V_{il}^{*}\right)^{2}\right)=O_{p}^{*}\left(\frac{\mathrm{log}\left(L\right)}{Lh^{3+d}}+h^{2R}\right),
\]
uniformly in $v\in I\left(\boldsymbol{x}\right)$. The conclusion
follows.\end{proof}
\begin{lem}
\label{Lemma 8}Suppose that Assumptions 1 - 3 hold. Let $\boldsymbol{x}$
be an interior point of $\mathcal{X}$ and $n\in\mathcal{N}$ be fixed.
Then 
\[
\widehat{f}_{GPV}^{*}\left(v,\boldsymbol{x},n\right)-\widetilde{f}^{*}\left(v,\boldsymbol{x},n\right)=\frac{1}{L^{2}}\sum_{l=1}^{L}\sum_{k=1}^{L}\mathcal{M}^{n}\left(\left(\boldsymbol{B}_{\cdot l}^{*},\boldsymbol{X}_{l}^{*},N_{l}^{*}\right),\left(\boldsymbol{B}_{\cdot k}^{*},\boldsymbol{X}_{k}^{*},N_{k}^{*}\right);v\right)+O_{p}^{*}\left(\left(\frac{\mathrm{log}\left(L\right)}{Lh^{1+d}}\right)^{\nicefrac{1}{2}}+\frac{\mathrm{log}\left(L\right)}{Lh^{3+d}}+h^{R}\right),
\]
where the remainder term is uniform in $v\in I\left(\boldsymbol{x}\right)$.
\end{lem}
\begin{proof}[Proof of Lemma \ref{Lemma 8}]We have
\begin{align}
 & \frac{1}{L}\sum_{l=1}^{L}\mathbbm{1}\left(N_{l}^{*}=n\right)\frac{1}{N_{l}^{*}}\sum_{i=1}^{N_{l}^{*}}\widetilde{\mathbb{T}}_{il}^{*}\frac{1}{h^{2+d}}K_{f}'\left(\frac{V_{il}^{*}-v}{h},\frac{\boldsymbol{X}_{l}^{*}-\boldsymbol{x}}{h}\right)\left(\widehat{V}_{il}^{*}-V_{il}^{*}\right)\nonumber \\
= & -\frac{1}{L}\sum_{l=1}^{L}\mathbbm{1}\left(N_{l}^{*}=n\right)\frac{1}{N_{l}^{*}}\sum_{i=1}^{N_{l}^{*}}\widetilde{\mathbb{T}}_{il}^{*}\frac{1}{h^{2+d}}K_{f}'\left(\frac{V_{il}^{*}-v}{h},\frac{\boldsymbol{X}_{l}^{*}-\boldsymbol{x}}{h}\right)\frac{1}{N_{l}^{*}-1}\frac{G\left(B_{il}^{*},\boldsymbol{X}_{l}^{*},N_{l}^{*}\right)}{g\left(B_{il}^{*},\boldsymbol{X}_{l}^{*},N_{l}^{*}\right)^{2}}\left(\widehat{g}^{*}\left(B_{il}^{*},\boldsymbol{X}_{l}^{*},N_{l}^{*}\right)-g\left(B_{il}^{*},\boldsymbol{X}_{l}^{*},N_{l}^{*}\right)\right)\nonumber \\
 & +\varDelta_{1}^{*}\left(v\right)+\varDelta_{2}^{*}\left(v\right)+\varDelta_{3}^{*}\left(v\right),\label{eq:lemma 2 decomposition-1}
\end{align}
where $\varDelta_{1}^{*}\left(v\right)$, $\varDelta_{2}^{*}\left(v\right)$
and $\varDelta_{3}^{*}\left(v\right)$ are bootstrap versions of $\varDelta_{1}^{\ddagger}\left(v\right)$,
$\varDelta_{2}^{\ddagger}\left(v\right)$ and $\varDelta_{3}^{\ddagger}\left(v\right)$.
Now 
\[
\underset{v\in I\left(\boldsymbol{x}\right)}{\mathrm{sup}}\left|\varDelta_{2}^{*}\left(v\right)\right|=O_{p}^{*}\left(\frac{\mathrm{log}\left(L\right)}{Lh^{\nicefrac{3}{2}+d}}+h^{2R+1}\right)\textrm{ and }\underset{v\in I\left(\boldsymbol{x}\right)}{\mathrm{sup}}\left|\varDelta_{3}^{*}\left(v\right)\right|=O_{p}^{*}\left(\frac{\mathrm{log}\left(L\right)}{Lh^{2+d}}+h^{2R+1}\right)
\]
follow from 
\begin{equation}
\underset{v\in I\left(\boldsymbol{x}\right)}{\mathrm{sup}}\frac{1}{L}\sum_{l=1}^{L}\mathbbm{1}\left(N_{l}^{*}=n\right)\frac{1}{N_{l}^{*}}\sum_{i=1}^{N_{l}^{*}}\frac{1}{N_{l}^{*}-1}\frac{1}{h^{1+d}}\left|K_{f}'\left(\frac{V_{il}^{*}-v}{h},\frac{\boldsymbol{X}_{l}^{*}-\boldsymbol{x}}{h}\right)\right|\apprle\underset{v\in I\left(\boldsymbol{x}\right)}{\mathrm{sup}}\frac{1}{L}\sum_{l=1}^{L}\mathbbm{1}\left(N_{l}^{*}=n\right)\frac{1}{N_{l}^{*}}\sum_{i=1}^{N_{l}^{*}}\reallywidecheck{\mathbb{T}}_{il}^{*}\frac{1}{h^{1+d}}\left|K_{\boldsymbol{X}}\left(\frac{\boldsymbol{X}_{l}^{*}-\boldsymbol{x}}{h}\right)\right|,\label{eq:K_d K_prime bound-1}
\end{equation}
where the right hand side was shown to be $O_{p}^{*}\left(1\right)$
in the previous lemma and 
\[
\underset{v\in I\left(\boldsymbol{x}\right)}{\mathrm{sup}}\widetilde{\mathbb{T}}_{il}^{*}\left|\widehat{g}^{*}\left(B_{il}^{*},\boldsymbol{X}_{l}^{*},N_{l}^{*}\right)^{-1}\right|\leq\underset{i,l}{\mathrm{max}}\,\overline{\mathbb{T}}_{il}^{*}\left|\widehat{g}^{*}\left(B_{il}^{*},\boldsymbol{X}_{l}^{*},N_{l}^{*}\right)^{-1}\right|=O_{p}^{*}\left(1\right).
\]

Since $K_{0}'$ has a bounded support, the contribution of the trimmed
observations is asymptotically negligible. When $h$ is sufficiently
small, we have
\begin{equation}
\varDelta_{1}^{*}\left(v\right)=\frac{1}{L^{2}}\sum_{l=1}^{L}\sum_{k=1}^{L}\mathcal{G}^{n}\left(\left(\boldsymbol{B}_{\cdot l}^{*},\boldsymbol{X}_{l}^{*},N_{l}^{*}\right),\left(\boldsymbol{B}_{\cdot k}^{*},\boldsymbol{X}_{k}^{*},N_{k}^{*}\right);v\right),\textrm{ for all \ensuremath{v\in I\left(\boldsymbol{x}\right)}}.\label{eq:I_1 V_statistic representation-1}
\end{equation}
Let 
\[
\widehat{\mu}_{\mathcal{G}^{n}}\left(v\right)\coloneqq\mathrm{E}^{*}\left[\mathcal{G}^{n}\left(\left(\boldsymbol{B}_{\cdot1}^{*},\boldsymbol{X}_{1}^{*},N_{1}^{*}\right),\left(\boldsymbol{B}_{\cdot2}^{*},\boldsymbol{X}_{2}^{*},N_{2}^{*}\right);v\right)\right]=\varDelta_{1}^{\ddagger}\left(v\right)=O_{p}\left(h^{R}+\left(\frac{\mathrm{log}\left(L\right)}{Lh^{1+d}}\right)^{\nicefrac{1}{2}}+\frac{\mathrm{log}\left(L\right)}{Lh^{2+d}}\right),
\]
uniformly in $v\in I\left(\boldsymbol{x}\right)$, where the second
equality follows from LIE and the fact that the bids in the bootstrap
sample are conditionally i.i.d. and the third equality holds was shown
in the proof of Lemma \ref{Lemma 2}. Also denote
\[
\widehat{\mathcal{G}}_{1}^{n}\left(\boldsymbol{b}.,\boldsymbol{z},m;v\right)\coloneqq\mathrm{E}^{*}\left[\mathcal{G}^{n}\left(\left(\boldsymbol{b}.,\boldsymbol{z},m\right),\left(\boldsymbol{B}_{\cdot1}^{*},\boldsymbol{X}_{1}^{*},N_{1}^{*}\right);v\right)\right]\textrm{ and }\widehat{\mathcal{G}}_{2}^{n}\left(\boldsymbol{b}.,\boldsymbol{z},m;v\right)\coloneqq\mathrm{E}^{*}\left[\mathcal{G}^{n}\left(\left(\boldsymbol{B}_{\cdot1}^{*},\boldsymbol{X}_{1}^{*},N_{1}^{*}\right),\left(\boldsymbol{b}.,\boldsymbol{z},m\right);v\right)\right].
\]
The Hoeffding decomposition yields
\begin{align}
 & \frac{1}{L^{2}}\sum_{l=1}^{L}\sum_{k=1}^{L}\mathcal{G}^{n}\left(\left(\boldsymbol{B}_{\cdot l}^{*},\boldsymbol{X}_{l}^{*},N_{l}^{*}\right),\left(\boldsymbol{B}_{\cdot k}^{*},\boldsymbol{X}_{k}^{*},N_{k}^{*}\right);v\right)\nonumber \\
= & \widehat{\mu}_{\mathcal{G}^{n}}\left(v\right)+\left\{ \frac{1}{L}\sum_{l=1}^{L}\widehat{\mathcal{G}}_{1}^{n}\left(\boldsymbol{B}_{\cdot l}^{*},\boldsymbol{X}_{l}^{*},N_{l}^{*};v\right)-\widehat{\mu}_{\mathcal{G}^{n}}\left(v\right)\right\} +\left\{ \frac{1}{L}\sum_{l=1}^{L}\widehat{\mathcal{G}}_{2}^{n}\left(\boldsymbol{B}_{\cdot l}^{*},\boldsymbol{X}_{l}^{*},N_{l}^{*};v\right)-\widehat{\mu}_{\mathcal{G}^{n}}\left(v\right)\right\} \nonumber \\
 & +\frac{1}{L\left(L-1\right)}\sum_{l\neq k}\left\{ \mathcal{G}^{n}\left(\left(\boldsymbol{B}_{\cdot l}^{*},\boldsymbol{X}_{l}^{*},N_{l}^{*}\right),\left(\boldsymbol{B}_{\cdot k}^{*},\boldsymbol{X}_{k}^{*},N_{k}^{*}\right);v\right)-\widehat{\mathcal{G}}_{1}^{n}\left(\boldsymbol{B}_{\cdot l}^{*},\boldsymbol{X}_{l}^{*},N_{l}^{*};v\right)-\widehat{\mathcal{G}}_{2}^{n}\left(\boldsymbol{B}_{\cdot k}^{*},\boldsymbol{X}_{k}^{*},N_{k}^{*};v\right)+\widehat{\mu}_{\mathcal{G}^{n}}\left(v\right)\right\} \nonumber \\
 & +\frac{1}{L^{2}}\sum_{l=1}^{L}\mathcal{G}^{n}\left(\left(\boldsymbol{B}_{\cdot l}^{*},\boldsymbol{X}_{l}^{*},N_{l}^{*}\right),\left(\boldsymbol{B}_{\cdot l}^{*},\boldsymbol{X}_{l}^{*},N_{l}^{*}\right);v\right)-\frac{1}{L^{2}\left(L-1\right)}\sum_{l\neq k}\mathcal{G}^{n}\left(\left(\boldsymbol{B}_{\cdot l}^{*},\boldsymbol{X}_{l}^{*},N_{l}^{*}\right),\left(\boldsymbol{B}_{\cdot k}^{*},\boldsymbol{X}_{k}^{*},N_{k}^{*}\right);v\right).\label{eq:G_n_star decomposition}
\end{align}

Since $\left\{ \mathcal{G}^{n}\left(\cdot,\cdot;v\right):v\in I\left(\boldsymbol{x}\right)\right\} $
is uniformly VC-type with respect to the envelope (\ref{eq:F_G_n envelope}),
the CK inequality yields
\begin{align*}
 & \mathrm{E}^{*}\left[\underset{v\in I\left(\boldsymbol{x}\right)}{\mathrm{sup}}\left|\frac{1}{\left(L\right)_{2}}\sum_{\left(2\right)}\left\{ \mathcal{G}^{n}\left(\left(\boldsymbol{B}_{\cdot l}^{*},\boldsymbol{X}_{l}^{*},N_{l}^{*}\right),\left(\boldsymbol{B}_{\cdot k}^{*},\boldsymbol{X}_{k}^{*},N_{k}^{*}\right);v\right)-\widehat{\mathcal{G}}_{1}^{n}\left(\boldsymbol{B}_{\cdot l}^{*},\boldsymbol{X}_{l}^{*},N_{l}^{*};v\right)-\widehat{\mathcal{G}}_{2}^{n}\left(\boldsymbol{B}_{\cdot k}^{*},\boldsymbol{X}_{k}^{*},N_{k}^{*};v\right)+\widehat{\mu}_{\mathcal{G}^{n}}\left(v\right)\right\} \right|\right]\\
\apprle & L^{-1}\left(\mathrm{E}^{*}\left[F_{\mathcal{G}^{n}}\left(\boldsymbol{X}_{1}^{*},\boldsymbol{X}_{2}^{*}\right)^{2}\right]\right)^{\nicefrac{1}{2}}.
\end{align*}
By change of variables, we have
\begin{eqnarray*}
\mathrm{E}\left[\frac{1}{L^{2}}\sum_{l=1}^{L}\sum_{k=1}^{L}F_{\mathcal{G}^{n}}\left(\boldsymbol{X}_{l},\boldsymbol{X}_{k}\right)^{2}\right] & = & \frac{L-1}{L}\mathrm{E}\left[F_{\mathcal{G}^{n}}\left(\boldsymbol{X}_{1},\boldsymbol{X}_{2}\right)^{2}\right]+\frac{1}{L}\mathrm{E}\left[F_{\mathcal{G}^{n}}\left(\boldsymbol{X}_{1},\boldsymbol{X}_{1}\right)^{2}\right]\\
 & = & O\left(h^{-\left(4+2d\right)}\right).
\end{eqnarray*}
Then it follows that
\[
\mathrm{E}^{*}\left[F_{\mathcal{G}^{n}}\left(\boldsymbol{X}_{1}^{*},\boldsymbol{X}_{2}^{*}\right)^{2}\right]=\frac{1}{L^{2}}\sum_{l=1}^{L}\sum_{k=1}^{L}F_{\mathcal{G}^{n}}\left(\boldsymbol{X}_{l},\boldsymbol{X}_{k}\right)^{2}=O_{p}\left(h^{-\left(4+2d\right)}\right),
\]
where the second equality follows from the previous result and Markov's
inequality. Now it follows that 
\begin{align*}
 & \mathrm{E}^{*}\left[\underset{v\in I\left(\boldsymbol{x}\right)}{\mathrm{sup}}\left|\frac{1}{\left(L\right)_{2}}\sum_{\left(2\right)}\left\{ \mathcal{G}^{n}\left(\left(\boldsymbol{B}_{\cdot l}^{*},\boldsymbol{X}_{l}^{*},N_{l}^{*}\right),\left(\boldsymbol{B}_{\cdot k}^{*},\boldsymbol{X}_{k}^{*},N_{k}^{*}\right);v\right)-\widehat{\mathcal{G}}_{1}^{n}\left(\boldsymbol{B}_{\cdot l}^{*},\boldsymbol{X}_{l}^{*},N_{l}^{*};v\right)-\widehat{\mathcal{G}}_{2}^{n}\left(\boldsymbol{B}_{\cdot k}^{*},\boldsymbol{X}_{k}^{*},N_{k}^{*};v\right)+\widehat{\mu}_{\mathcal{G}^{n}}\left(v\right)\right\} \right|\right]\\
= & O_{p}\left(\left(Lh^{2+d}\right)^{-1}\right).
\end{align*}

Now by the LIE and the fact that the bids in the bootstrap sample
are conditionally i.i.d., we have
\[
\widehat{\mathcal{G}}_{1}^{n}\left(\boldsymbol{b}.,\boldsymbol{z},m;v\right)=-\mathbbm{1}\left(m=n\right)\frac{1}{m}\sum_{i=1}^{m}\frac{1}{h^{2+d}}K_{f}'\left(\frac{\xi\left(b_{i},\boldsymbol{z},m\right)-v}{h},\frac{\boldsymbol{z}-\boldsymbol{x}}{h}\right)\frac{\widehat{G}\left(b_{i},\boldsymbol{z},m\right)-G\left(b_{i},\boldsymbol{z},m\right)}{\left(m-1\right)g\left(b_{i},\boldsymbol{z},m\right)}.
\]
It follows from standard arguments that when $h$ is sufficiently
small, $\left\{ \widehat{\mathcal{G}}_{1}^{n}\left(\cdot;v\right):v\in I\left(\boldsymbol{x}\right)\right\} $
is VC-type with respect to the envelope
\[
F_{\widehat{\mathcal{G}}_{1}^{n}}\left(\boldsymbol{z}\right)\coloneqq\frac{\left(\overline{C}_{D_{1}}+\overline{C}_{D_{2}}\right)}{\left(n-1\right)\underline{C}_{g}h^{2+d}}\left|K_{\boldsymbol{X}}^{0}\left(\frac{\boldsymbol{z}-\boldsymbol{x}}{h}\right)\right|\left\{ \underset{\left(b',\boldsymbol{z}'\right)\in\mathcal{C}_{B,\boldsymbol{X}}^{n}}{\mathrm{sup}}\left|\widehat{G}\left(b',\boldsymbol{z}',n\right)-G\left(b',\boldsymbol{z}',n\right)\right|\right\} ,
\]
conditionally on the original sample. The VW inequality yields 
\begin{eqnarray*}
\mathrm{E}^{*}\left[\underset{v\in I\left(\boldsymbol{x}\right)}{\mathrm{sup}}\left|\frac{1}{L}\sum_{l=1}^{L}\widehat{\mathcal{G}}_{1}^{n}\left(\boldsymbol{B}_{\cdot l}^{*},\boldsymbol{X}_{l}^{*},N_{l}^{*};v\right)-\widehat{\mu}_{\mathcal{G}^{n}}\left(v\right)\right|\right] & \leq & L^{-\nicefrac{1}{2}}\left(\mathrm{E}^{*}\left[F_{\widehat{\mathcal{G}}_{1}^{n}}\left(\boldsymbol{X}_{1}^{*}\right)^{2}\right]\right)^{\nicefrac{1}{2}}\\
 & \apprle & L^{-\nicefrac{1}{2}}\left\{ \frac{1}{L}\sum_{l=1}^{L}\frac{1}{h^{4+2d}}K_{\boldsymbol{X}}^{0}\left(\frac{\boldsymbol{X}_{l}-\boldsymbol{x}}{h}\right)^{2}\right\} ^{\nicefrac{1}{2}}\left\{ \underset{\left(b',\boldsymbol{z}'\right)\in\mathcal{C}_{B,\boldsymbol{X}}^{n}}{\mathrm{sup}}\left|\widehat{G}\left(b',\boldsymbol{z}',n\right)-G\left(b',\boldsymbol{z}',n\right)\right|\right\} \\
 & = & O_{p}\left(\left(Lh^{4+d}\right)^{-\nicefrac{1}{2}}\right)O_{p}\left(\left(\frac{\mathrm{log}\left(L\right)}{Lh^{d}}\right)^{\nicefrac{1}{2}}+h^{1+R}\right),
\end{eqnarray*}
where the equality follows from change of variables, Markov's inequality
and (\ref{eq:uniform convergence rate}). 

Let 
\begin{eqnarray*}
\widehat{\mathcal{G}}_{2}^{\ddagger,n}\left(\boldsymbol{b}.,\boldsymbol{z},m;v\right) & \coloneqq & \mathrm{E}^{*}\left[\mathcal{G}^{\ddagger,n}\left(\left(\boldsymbol{B}_{\cdot1}^{*},\boldsymbol{X}_{1}^{*},N_{1}^{*}\right),\left(\boldsymbol{b}.,\boldsymbol{z},m\right);v\right)\right]\\
 & = & -\frac{1}{L}\sum_{k=1}^{L}\mathbbm{1}\left(N_{k}=n\right)\frac{1}{N_{k}}\sum_{j=1}^{N_{k}}\frac{1}{h^{2+d}}K_{f}'\left(\frac{\xi\left(B_{jk},\boldsymbol{X}_{k},N_{k}\right)-v}{h},\frac{\boldsymbol{X}_{k}-\boldsymbol{x}}{h}\right)\frac{1}{\left(N_{k}-1\right)g\left(B_{jk},\boldsymbol{X}_{k},N_{k}\right)}\\
 &  & \times\mathbbm{1}\left(m=N_{k}\right)\frac{1}{m}\sum_{i=1}^{m}\mathbbm{1}\left(b_{i}\leq B_{jk}\right)\frac{1}{h^{d}}K_{\boldsymbol{X}}\left(\frac{\boldsymbol{z}-\boldsymbol{X}_{k}}{h}\right),
\end{eqnarray*}
where the second equality follows from LIE and the fact that the bids
in the bootstrap sample are conditionally i.i.d.. It is straightforward
to verify that 
\[
\widehat{\mathcal{G}}_{2}^{n}\left(\boldsymbol{B}_{\cdot l}^{*},\boldsymbol{X}_{l}^{*},N_{l}^{*};v\right)-\widehat{\mu}_{\mathcal{G}^{n}}\left(v\right)=\widehat{\mathcal{G}}_{2}^{\ddagger,n}\left(\boldsymbol{B}_{\cdot l}^{*},\boldsymbol{X}_{l}^{*},N_{l}^{*};v\right)-\mathrm{E}^{*}\left[\widehat{\mathcal{G}}_{2}^{\ddagger,n}\left(\boldsymbol{B}_{\cdot1}^{*},\boldsymbol{X}_{1}^{*},N_{1}^{*};v\right)\right],\textrm{ for all \ensuremath{l=1,...,L}}
\]
and thus
\[
\frac{1}{L}\sum_{l=1}^{L}\widehat{\mathcal{G}}_{2}^{n}\left(\boldsymbol{B}_{\cdot l}^{*},\boldsymbol{X}_{l}^{*},N_{l}^{*};v\right)-\widehat{\mu}_{\mathcal{G}^{n}}\left(v\right)=\frac{1}{L}\sum_{l=1}^{L}\widehat{\mathcal{G}}_{2}^{\ddagger,n}\left(\boldsymbol{B}_{\cdot l}^{*},\boldsymbol{X}_{l}^{*},N_{l}^{*};v\right)-\mathrm{E}^{*}\left[\widehat{\mathcal{G}}_{2}^{\ddagger,n}\left(\boldsymbol{B}_{\cdot1}^{*},\boldsymbol{X}_{1}^{*},N_{1}^{*};v\right)\right].
\]

Since $\left\{ \mathcal{G}^{\ddagger,n}\left(\cdot,\cdot;v\right):v\in I\left(\boldsymbol{x}\right)\right\} $
is uniformly VC-type with respect to the envelope (\ref{eq:F_G_ddagger envelope})
(see the proof of Lemma \ref{Lemma 2}), it follows from \citet[Lemma 5.4]{Chen_Kato_U_Process}
that the class $\left\{ \widehat{\mathcal{G}}_{2}^{\ddagger,n}\left(\cdot;v\right):v\in I\left(\boldsymbol{x}\right)\right\} $
is uniformly VC-type with respect to the envelope 
\[
F_{\widehat{\mathcal{G}}_{2}^{\ddagger,n}}\left(\boldsymbol{z}\right)\coloneqq\frac{\left(\overline{C}_{D_{1}}+\overline{C}_{D_{2}}\right)}{\left(n-1\right)\underline{C}_{g}}\frac{1}{L}\sum_{k=1}^{L}\frac{1}{h^{2+2d}}\left|K_{\boldsymbol{X}}^{0}\left(\frac{\boldsymbol{X}_{k}-\boldsymbol{x}}{h}\right)\right|\left|K_{\boldsymbol{X}}\left(\frac{\boldsymbol{z}-\boldsymbol{X}_{k}}{h}\right)\right|,
\]
conditionally on the original sample. 

By Jensen's inequality, LIE and the fact that the bids in the bootstrap
sample are conditionally i.i.d., we have
\begin{align*}
 & \mathrm{E}^{*}\left[\widehat{\mathcal{G}}_{2}^{\ddagger,n}\left(\boldsymbol{B}_{\cdot1}^{*},\boldsymbol{X}_{1}^{*},N_{1}^{*};v\right)^{2}\right]\\
\leq & \mathrm{E}^{*}\left[\frac{1}{N_{1}^{*}}\sum_{i=1}^{N_{1}^{*}}\left\{ \frac{1}{L}\sum_{k=1}^{L}\mathbbm{1}\left(N_{k}=n\right)\frac{1}{N_{k}}\sum_{j=1}^{N_{k}}\frac{1}{h^{2+2d}}K_{f}'\left(\frac{\xi\left(B_{jk},\boldsymbol{X}_{k},N_{k}\right)-v}{h},\frac{\boldsymbol{X}_{k}-\boldsymbol{x}}{h}\right)\frac{\mathbbm{1}\left(N_{1}^{*}=N_{k}\right)\mathbbm{1}\left(B_{i1}^{*}\leq B_{jk}\right)}{\left(N_{k}-1\right)g\left(B_{jk},\boldsymbol{X}_{k},N_{k}\right)}K_{\boldsymbol{X}}\left(\frac{\boldsymbol{X}_{1}^{*}-\boldsymbol{X}_{k}}{h}\right)\right\} ^{2}\right]\\
= & \frac{1}{L}\sum_{l=1}^{L}\frac{1}{N_{l}}\sum_{i=1}^{N_{l}}\left\{ \frac{1}{L}\sum_{k=1}^{L}\mathbbm{1}\left(N_{k}=n\right)\frac{1}{N_{k}}\sum_{j=1}^{N_{k}}\frac{1}{h^{2+2d}}K_{f}'\left(\frac{\xi\left(B_{jk},\boldsymbol{X}_{k},N_{k}\right)-v}{h},\frac{\boldsymbol{X}_{k}-\boldsymbol{x}}{h}\right)\frac{\mathbbm{1}\left(N_{l}=N_{k}\right)\mathbbm{1}\left(B_{il}\leq B_{jk}\right)}{\left(N_{k}-1\right)g\left(B_{jk},\boldsymbol{X}_{k},N_{k}\right)}K_{\boldsymbol{X}}\left(\frac{\boldsymbol{X}_{l}-\boldsymbol{X}_{k}}{h}\right)\right\} ^{2}\\
= & \frac{1}{L^{3}}\sum_{l=1}^{L}\sum_{k=1}^{L}\sum_{k'=1}^{L}\mathcal{J}\left(\left(\boldsymbol{B}_{\cdot l},\boldsymbol{X}_{l},N_{l}\right),\left(\boldsymbol{B}_{\cdot k},\boldsymbol{X}_{k},N_{k}\right),\left(\boldsymbol{B}_{\cdot k'},\boldsymbol{X}_{k'},N_{k'}\right);v\right),
\end{align*}
where 
\begin{align*}
 & \mathcal{J}\left(\left(\boldsymbol{b}.,\boldsymbol{z},m\right),\left(\boldsymbol{b}.',\boldsymbol{z}',m'\right),\left(\boldsymbol{b}.'',\boldsymbol{z}'',m''\right);v\right)\\
\coloneqq & \frac{1}{h^{4\left(1+d\right)}}\frac{1}{m}\sum_{i=1}^{m}\mathbbm{1}\left(m'=n\right)\frac{1}{m'}\sum_{j=1}^{m'}K_{f}'\left(\frac{\xi\left(b_{j}',\boldsymbol{z}',m'\right)-v}{h},\frac{\boldsymbol{z}'-\boldsymbol{x}}{h}\right)\frac{\mathbbm{1}\left(m=m'\right)\mathbbm{1}\left(b_{i}\leq b_{j}'\right)}{\left(m'-1\right)g\left(b_{j}',\boldsymbol{z}',m'\right)}K_{\boldsymbol{X}}\left(\frac{\boldsymbol{z}-\boldsymbol{z}'}{h}\right)\\
 & \times\mathbbm{1}\left(m''=n\right)\frac{1}{m''}\sum_{j'=1}^{m''}K_{f}'\left(\frac{\xi\left(b_{j'}'',\boldsymbol{z}'',m''\right)-v}{h},\frac{\boldsymbol{z}''-\boldsymbol{x}}{h}\right)\frac{\mathbbm{1}\left(m=m''\right)\mathbbm{1}\left(b_{i}\leq b_{j'}''\right)}{\left(m''-1\right)g\left(b_{j'}'',\boldsymbol{z}'',m''\right)}K_{\boldsymbol{X}}\left(\frac{\boldsymbol{z}-\boldsymbol{z}''}{h}\right).
\end{align*}

Standard arguments can be applied to verify that $\left\{ \mathcal{J}\left(\cdot,\cdot,\cdot;v\right):v\in I\left(\boldsymbol{x}\right)\right\} $
is uniformly VC-type with respect to the envelope 
\begin{equation}
F_{\mathcal{J}}\left(\boldsymbol{z},\boldsymbol{z}',\boldsymbol{z}''\right)\coloneqq\frac{\left(\overline{C}_{D_{1}}+\overline{C}_{D_{2}}\right)^{2}}{\left(n-1\right)^{2}\underline{C}_{g}^{2}h^{4\left(1+d\right)}}\left|K_{\boldsymbol{X}}^{0}\left(\frac{\boldsymbol{z}'-\boldsymbol{x}}{h}\right)\right|\left|K_{\boldsymbol{X}}\left(\frac{\boldsymbol{z}-\boldsymbol{z}'}{h}\right)\right|\left|K_{\boldsymbol{X}}^{0}\left(\frac{\boldsymbol{z}''-\boldsymbol{x}}{h}\right)\right|\left|K_{\boldsymbol{X}}\left(\frac{\boldsymbol{z}-\boldsymbol{z}''}{h}\right)\right|.\label{eq:J envelope}
\end{equation}

By observing that $\mathcal{J}$ is symmetric with respect to the
second and the third arguments and the V-statistic decomposition argument
of \citet[5.7.3]{Serfling_Approximation_Theorems}, we have 
\begin{align*}
 & \frac{1}{L^{3}}\sum_{l=1}^{L}\sum_{k=1}^{L}\sum_{k'=1}^{L}\mathcal{J}\left(\left(\boldsymbol{B}_{\cdot l},\boldsymbol{X}_{l},N_{l}\right),\left(\boldsymbol{B}_{\cdot k},\boldsymbol{X}_{k},N_{k}\right),\left(\boldsymbol{B}_{\cdot k'},\boldsymbol{X}_{k'},N_{k'}\right);v\right)\\
= & \frac{1}{\left(L\right)_{3}}\sum_{\left(3\right)}\mathcal{J}\left(\left(\boldsymbol{B}_{\cdot l},\boldsymbol{X}_{l},N_{l}\right),\left(\boldsymbol{B}_{\cdot k},\boldsymbol{X}_{k},N_{k}\right),\left(\boldsymbol{B}_{\cdot k'},\boldsymbol{X}_{k'},N_{k'}\right);v\right)+\frac{O\left(L^{-1}\right)}{3L^{2}-2L}\left\{ \sum_{l=1}^{L}\mathcal{J}\left(\left(\boldsymbol{B}_{\cdot l},\boldsymbol{X}_{l},N_{l}\right),\left(\boldsymbol{B}_{\cdot l},\boldsymbol{X}_{l},N_{l}\right),\left(\boldsymbol{B}_{\cdot l},\boldsymbol{X}_{l},N_{l}\right);v\right)\right.\\
 & \left.+\sum_{l\neq k}\left(2\mathcal{J}\left(\left(\boldsymbol{B}_{\cdot l},\boldsymbol{X}_{l},N_{l}\right),\left(\boldsymbol{B}_{\cdot l},\boldsymbol{X}_{l},N_{l}\right),\left(\boldsymbol{B}_{\cdot k},\boldsymbol{X}_{k},N_{k}\right);v\right)+\mathcal{J}\left(\left(\boldsymbol{B}_{\cdot k},\boldsymbol{X}_{k},N_{k}\right),\left(\boldsymbol{B}_{\cdot l},\boldsymbol{X}_{l},N_{l}\right),\left(\boldsymbol{B}_{\cdot l},\boldsymbol{X}_{l},N_{l}\right);v\right)\right)\right\} .
\end{align*}
The Hoeffding decomposition yields 
\begin{align}
 & \frac{1}{\left(L\right)_{3}}\sum_{\left(3\right)}\mathcal{J}\left(\left(\boldsymbol{B}_{\cdot l},\boldsymbol{X}_{l},N_{l}\right),\left(\boldsymbol{B}_{\cdot k},\boldsymbol{X}_{k},N_{k}\right),\left(\boldsymbol{B}_{\cdot k'},\boldsymbol{X}_{k'},N_{k'}\right);v\right)\nonumber \\
= & \mu_{\mathcal{J}}\left(v\right)+\frac{1}{L}\sum_{l=1}^{L}\left(\mathcal{J}_{1}^{\left(1\right)}\left(\boldsymbol{B}_{\cdot l},\boldsymbol{X}_{l},N_{l};v\right)-\mu_{\mathcal{J}}\left(v\right)\right)+\frac{1}{L}\sum_{l=1}^{L}\left(\mathcal{J}_{2}^{\left(1\right)}\left(\boldsymbol{B}_{\cdot l},\boldsymbol{X}_{l},N_{l};v\right)-\mu_{\mathcal{J}}\left(v\right)\right)+\frac{1}{L}\sum_{l=1}^{L}\left(\mathcal{J}_{3}^{\left(1\right)}\left(\boldsymbol{B}_{\cdot l},\boldsymbol{X}_{l},N_{l};v\right)-\mu_{\mathcal{J}}\left(v\right)\right)\nonumber \\
 & +\Upsilon_{\mathcal{J}}^{1}\left(v\right)+\Upsilon_{\mathcal{J}}^{2}\left(v\right)+\Upsilon_{\mathcal{J}}^{3}\left(v\right)+\Psi_{\mathcal{J}}\left(v\right),\label{eq:J Hoeffding decomposition}
\end{align}
where the terms in the decomposition are defined by (\ref{eq:K Hoeffding term 1})
to (\ref{eq:K Hoeffding term 5}) with $\mathcal{K}$ replaced by
$\mathcal{J}$.

By the LIE, we have 
\begin{eqnarray*}
\mathcal{J}_{1}^{\left(1\right)}\left(\boldsymbol{b}.,\boldsymbol{z},m;v\right) & \coloneqq & \mathrm{E}\left[\mathcal{J}\left(\left(\boldsymbol{b}.,\boldsymbol{z},m\right),\left(\boldsymbol{B}_{\cdot1},\boldsymbol{X}_{1},N_{1}\right),\left(\boldsymbol{B}_{\cdot2},\boldsymbol{X}_{2},N_{2}\right);v\right)\right]\\
 & = & \frac{1}{m}\sum_{i=1}^{m}\frac{1}{h^{4\left(1+d\right)}}\left\{ \int_{\mathcal{X}}\int_{\underline{b}\left(\boldsymbol{z}'\right)}^{\overline{b}\left(\boldsymbol{z}',n\right)}K_{f}'\left(\frac{\xi\left(\boldsymbol{b}',\boldsymbol{z}',n\right)-v}{h},\frac{\boldsymbol{z}'-\boldsymbol{x}}{h}\right)\frac{\mathbbm{1}\left(m=n\right)}{n-1}\mathbbm{1}\left(b_{i}\leq b'\right)K_{\boldsymbol{X}}\left(\frac{\boldsymbol{z}-\boldsymbol{z}'}{h}\right)\mathrm{d}b'\mathrm{d}\boldsymbol{z}'\right\} ^{2}
\end{eqnarray*}
and 
\begin{align*}
 & \mathcal{J}_{2}^{\left(1\right)}\left(\boldsymbol{b}.,\boldsymbol{z},m;v\right)\\
\coloneqq & \mathrm{E}\left[\mathcal{J}\left(\left(\boldsymbol{B}_{\cdot1},\boldsymbol{X}_{1},N_{1}\right),\left(\boldsymbol{b}.,\boldsymbol{z},m\right),\left(\boldsymbol{B}_{\cdot2},\boldsymbol{X}_{2},N_{2}\right);v\right)\right]\\
= & \frac{1}{m}\sum_{i=1}^{m}\mathbbm{1}\left(m=n\right)K_{f}'\left(\frac{\xi\left(b_{i},\boldsymbol{z},m\right)-v}{h},\frac{\boldsymbol{z}-\boldsymbol{x}}{h}\right)\frac{1}{\left(m-1\right)g\left(b_{i},\boldsymbol{z},m\right)}\\
 & \times\int_{\mathcal{X}}\int_{\mathcal{X}}\int_{\underline{b}\left(\boldsymbol{z}''\right)}^{\overline{b}\left(\boldsymbol{z}'',n\right)}\frac{1}{h^{4\left(1+d\right)}}G\left(\mathrm{min}\left\{ b'',b_{i}\right\} ,\boldsymbol{z}',n\right)K_{\boldsymbol{X}}\left(\frac{\boldsymbol{z}'-\boldsymbol{z}}{h}\right)K_{f}'\left(\frac{\xi\left(b'',\boldsymbol{z}'',n\right)-v}{h},\frac{\boldsymbol{z}''-\boldsymbol{x}}{h}\right)\frac{1}{n-1}K_{\boldsymbol{X}}\left(\frac{\boldsymbol{z}'-\boldsymbol{z}''}{h}\right)\mathrm{d}b''\mathrm{d}\boldsymbol{z}''\mathrm{d}\boldsymbol{z}'.
\end{align*}

We observe that 
\[
\mu_{\mathcal{J}}\left(v\right)=\mathrm{E}\left[\mathcal{J}\left(\left(\boldsymbol{B}_{\cdot1},\boldsymbol{X}_{1},N_{1}\right),\left(\boldsymbol{B}_{\cdot2},\boldsymbol{X}_{2},N_{2}\right),\left(\boldsymbol{B}_{\cdot3},\boldsymbol{X}_{3},N_{3}\right);v\right)\right]=\mathrm{E}\left[\mathcal{G}_{2}^{\ddagger,n}\left(\boldsymbol{B}_{\cdot1},\boldsymbol{X}_{1},N_{1};v\right)^{2}\right]
\]
and thus
\begin{equation}
\underset{v\in I\left(\boldsymbol{x}\right)}{\mathrm{sup}}\left|\mu_{\mathcal{J}}\left(v\right)\right|=O\left(h^{-\left(1+d\right)}\right),\label{eq:sup miu_J bound}
\end{equation}
which was shown in the proof of Lemma \ref{Lemma 2}.

By Jensen's inequality, LIE and change of variables, we have
\begin{align*}
 & \mathrm{E}\left[\mathcal{J}_{1}^{\left(1\right)}\left(\boldsymbol{B}_{\cdot1},\boldsymbol{X}_{1},N_{1};v\right)^{2}\right]\\
\leq & \mathrm{E}\left[\frac{1}{N_{1}}\sum_{i=1}^{N_{1}}\frac{1}{h^{4\left(1+d\right)}}\frac{\mathbbm{1}\left(N_{1}=n\right)}{\left(n-1\right)^{4}}\left\{ \int_{\mathcal{X}}\int_{\underline{b}\left(\boldsymbol{z}'\right)}^{\overline{b}\left(\boldsymbol{z}',n\right)}\frac{1}{h^{1+d}}K_{f}'\left(\frac{\xi\left(\boldsymbol{b}',\boldsymbol{z}',n\right)-v}{h},\frac{\boldsymbol{z}'-\boldsymbol{x}}{h}\right)\mathbbm{1}\left(B_{i1}\leq b'\right)K_{\boldsymbol{X}}\left(\frac{\boldsymbol{X}_{1}-\boldsymbol{z}'}{h}\right)\mathrm{d}b'\mathrm{d}\boldsymbol{z}'\right\} ^{4}\right]\\
= & \frac{1}{\left(n-1\right)^{4}}\int_{\mathcal{X}}\int_{\underline{b}\left(\boldsymbol{z}\right)}^{\overline{b}\left(\boldsymbol{z},n\right)}\frac{1}{h^{4\left(1+d\right)}}\left\{ \int_{\mathcal{X}}\int_{\underline{b}\left(\boldsymbol{z}'\right)}^{\overline{b}\left(\boldsymbol{z}',n\right)}\frac{1}{h^{1+d}}K_{f}'\left(\frac{\xi\left(\boldsymbol{b}',\boldsymbol{z}',n\right)-v}{h},\frac{\boldsymbol{z}'-\boldsymbol{x}}{h}\right)\mathbbm{1}\left(b\leq b'\right)K_{\boldsymbol{X}}\left(\frac{\boldsymbol{z}-\boldsymbol{z}'}{h}\right)\mathrm{d}b'\mathrm{d}\boldsymbol{z}'\right\} ^{4}g\left(b,\boldsymbol{z},n\right)\mathrm{d}b\mathrm{d}\boldsymbol{z}\\
= & \frac{1}{\left(n-1\right)^{4}}\int_{\mathcal{Y}}\int_{\underline{b}\left(h\boldsymbol{y}+\boldsymbol{x}\right)}^{\overline{b}\left(h\boldsymbol{y}+\boldsymbol{x},n\right)}\frac{1}{h^{4+3d}}\left\{ \int_{\mathcal{Y}}\int_{\frac{\underline{v}\left(h\boldsymbol{y}'+\boldsymbol{x}\right)-v}{h}}^{\frac{\overline{v}\left(h\boldsymbol{y}'+\boldsymbol{x}\right)-v}{h}}K_{f}'\left(u,\boldsymbol{y}'\right)\mathbbm{1}\left(b\leq s\left(hu+v,h\boldsymbol{y}'+\boldsymbol{x},n\right)\right)K_{\boldsymbol{X}}\left(\boldsymbol{y}-\boldsymbol{y}'\right)s'\left(hu+v,h\boldsymbol{y}'+\boldsymbol{x},n\right)\mathrm{d}u\mathrm{d}\boldsymbol{y}'\right\} ^{4}\\
 & \times g\left(b,h\boldsymbol{y}+\boldsymbol{x},n\right)\mathrm{d}b\mathrm{d}\boldsymbol{y}
\end{align*}
and 
\begin{align*}
 & \mathrm{E}\left[\mathcal{J}_{2}^{\left(1\right)}\left(\boldsymbol{B}_{\cdot1},\boldsymbol{X}_{1},N_{1};v\right)^{2}\right]\\
\leq & \mathrm{E}\left[\frac{1}{N_{1}}\sum_{i=1}^{N_{1}}\mathbbm{1}\left(N_{1}=n\right)K_{f}'\left(\frac{\xi\left(B_{i1},\boldsymbol{X}_{1},N_{1}\right)-v}{h},\frac{\boldsymbol{X}_{1}-\boldsymbol{x}}{h}\right)^{2}\frac{1}{\left(N_{1}-1\right)^{2}g\left(B_{i1},\boldsymbol{X}_{1},N_{1}\right)^{2}}\frac{1}{h^{6+4d}}\right.\\
 & \left.\times\left\{ \int_{\mathcal{X}}\int_{\mathcal{X}}\int_{\underline{b}\left(\boldsymbol{z}''\right)}^{\overline{b}\left(\boldsymbol{z}'',n\right)}\frac{1}{h^{1+2d}}G\left(\mathrm{min}\left\{ b'',B_{i1}\right\} ,\boldsymbol{z}',n\right)K_{\boldsymbol{X}}\left(\frac{\boldsymbol{z}'-\boldsymbol{X}_{1}}{h}\right)K_{f}'\left(\frac{\xi\left(b'',\boldsymbol{z}'',n\right)-v}{h},\frac{\boldsymbol{z}''-\boldsymbol{x}}{h}\right)\frac{1}{n-1}K_{\boldsymbol{X}}\left(\frac{\boldsymbol{z}'-\boldsymbol{z}''}{h}\right)\mathrm{d}b''\mathrm{d}\boldsymbol{z}''\mathrm{d}\boldsymbol{z}'\right\} ^{2}\right]\\
= & \int_{\mathcal{X}}\int_{\underline{b}\left(\boldsymbol{z}\right)}^{\overline{b}\left(\boldsymbol{z},n\right)}\frac{1}{h^{6+4d}}K_{f}'\left(\frac{\xi\left(b,\boldsymbol{z},n\right)-v}{h},\frac{\boldsymbol{z}-\boldsymbol{x}}{h}\right)^{2}\frac{1}{\left(n-1\right)^{2}g\left(b,\boldsymbol{z},n\right)}\\
 & \times\left\{ \int_{\mathcal{X}}\int_{\mathcal{X}}\int_{\underline{b}\left(\boldsymbol{z}''\right)}^{\overline{b}\left(\boldsymbol{z}'',n\right)}\frac{1}{h^{1+2d}}G\left(\mathrm{min}\left\{ b'',b\right\} ,\boldsymbol{z}',n\right)K_{\boldsymbol{X}}\left(\frac{\boldsymbol{z}'-\boldsymbol{z}}{h}\right)K_{f}'\left(\frac{\xi\left(b'',\boldsymbol{z}'',n\right)-v}{h},\frac{\boldsymbol{z}''-\boldsymbol{x}}{h}\right)\frac{1}{n-1}K_{\boldsymbol{X}}\left(\frac{\boldsymbol{z}'-\boldsymbol{z}''}{h}\right)\mathrm{d}b''\mathrm{d}\boldsymbol{z}''\mathrm{d}\boldsymbol{z}'\right\} ^{2}\mathrm{d}b\mathrm{d}\boldsymbol{z}\\
= & \frac{1}{\left(n-1\right)^{4}}\frac{1}{h^{5+3d}}\int_{\mathcal{Y}}\int_{\frac{\underline{v}\left(h\boldsymbol{y}+\boldsymbol{x}\right)-v}{h}}^{\frac{\overline{v}\left(h\boldsymbol{y}+\boldsymbol{x}\right)-v}{h}}K_{f}'\left(u,\boldsymbol{y}\right)^{2}\frac{s'\left(hu+v,h\boldsymbol{y}+\boldsymbol{x},n\right)}{g\left(s\left(hu+v,h\boldsymbol{y}+\boldsymbol{x},n\right),h\boldsymbol{y}+\boldsymbol{x},n\right)}\left\{ \int_{\mathcal{Y}}\int_{\mathcal{Y}}\int_{\frac{\underline{v}\left(h\boldsymbol{y}''+\boldsymbol{x}\right)-v}{h}}^{\frac{\overline{v}\left(h\boldsymbol{y}''+\boldsymbol{x}\right)-v}{h}}K_{\boldsymbol{X}}\left(\boldsymbol{y}'-\boldsymbol{y}\right)K_{f}'\left(u'',\boldsymbol{y}''\right)\right.\\
 & \left.\times K_{\boldsymbol{X}}\left(\boldsymbol{y}'-\boldsymbol{y}''\right)G\left(\mathrm{min}\left\{ s\left(hu''+v,h\boldsymbol{y}''+\boldsymbol{x},n\right),s\left(hu+v,h\boldsymbol{y}+\boldsymbol{x},n\right)\right\} ,h\boldsymbol{y}'+\boldsymbol{x},n\right)s'\left(hu''+v,h\boldsymbol{y}''+\boldsymbol{x},n\right)\mathrm{d}u''\mathrm{d}\boldsymbol{y}''\mathrm{d}\boldsymbol{y}'\right\} ^{2}\mathrm{d}u\mathrm{d}\boldsymbol{y}.
\end{align*}

Now it is easy to observe
\begin{equation}
\sigma_{\mathcal{J}_{1}^{\left(1\right)}}^{2}\coloneqq\underset{v\in I\left(\boldsymbol{x}\right)}{\mathrm{sup}}\mathrm{E}\left[\mathcal{J}_{1}^{\left(1\right)}\left(\boldsymbol{B}_{\cdot1},\boldsymbol{X}_{1},N_{1};v\right)^{2}\right]=O\left(h^{-\left(4+3d\right)}\right)\label{eq:sigma_J_1 rate}
\end{equation}
and 
\[
\sigma_{\mathcal{J}_{2}^{\left(1\right)}}^{2}\coloneqq\underset{v\in I\left(\boldsymbol{x}\right)}{\mathrm{sup}}\mathrm{E}\left[\mathcal{J}_{2}^{\left(1\right)}\left(\boldsymbol{B}_{\cdot1},\boldsymbol{X}_{1},N_{1};v\right)^{2}\right]=O\left(h^{-\left(5+3d\right)}\right).
\]

Since $\left\{ \mathcal{J}\left(\cdot,\cdot,\cdot;v\right):v\in I\left(\boldsymbol{x}\right)\right\} $
is uniformly VC-type with respect to the envelope (\ref{eq:J envelope}),
it follows from \citet[Lemma 5.4]{Chen_Kato_U_Process} that $\left\{ \mathcal{J}_{1}^{\left(1\right)}\left(\cdot;v\right):v\in I\left(\boldsymbol{x}\right)\right\} $
is VC-type with respect to the envelope 
\[
F_{\mathcal{J}_{1}^{\left(1\right)}}\left(\boldsymbol{z}\right)\coloneqq\frac{\left(\overline{C}_{D_{1}}+\overline{C}_{D_{2}}\right)^{2}}{\left(n-1\right)^{2}\underline{C}_{g}^{2}}\int\int\frac{1}{h^{4\left(1+d\right)}}\left|K_{\boldsymbol{X}}^{0}\left(\frac{\boldsymbol{z}'-\boldsymbol{x}}{h}\right)\right|\left|K_{\boldsymbol{X}}\left(\frac{\boldsymbol{z}-\boldsymbol{z}'}{h}\right)\right|\left|K_{\boldsymbol{X}}^{0}\left(\frac{\boldsymbol{z}''-\boldsymbol{x}}{h}\right)\right|\left|K_{\boldsymbol{X}}\left(\frac{\boldsymbol{z}-\boldsymbol{z}''}{h}\right)\right|\varphi\left(\boldsymbol{z}'\right)\varphi\left(\boldsymbol{z}''\right)\mathrm{d}\boldsymbol{z}'\mathrm{d}\boldsymbol{z}''.
\]
Then the CCK inequality yields 
\begin{eqnarray*}
\mathrm{E}\left[\underset{v\in I\left(\boldsymbol{x}\right)}{\mathrm{sup}}\left|\frac{1}{L}\sum_{l=1}^{L}\mathcal{J}_{1}^{\left(1\right)}\left(\boldsymbol{B}_{\cdot l},\boldsymbol{X}_{l},N_{l};v\right)-\mu_{\mathcal{J}}\left(v\right)\right|\right] & \leq & C_{1}\left\{ L^{-\nicefrac{1}{2}}\sigma_{\mathcal{J}_{1}^{\left(1\right)}}\mathrm{log}\left(C_{2}L\right)^{\nicefrac{1}{2}}+L^{-1}\left\Vert F_{\mathcal{J}_{1}^{\left(1\right)}}\right\Vert _{\mathcal{X}}\mathrm{log}\left(C_{2}L\right)\right\} \\
 & = & O\left(\left(\frac{\mathrm{log}\left(L\right)}{Lh^{4+3d}}\right)^{\nicefrac{1}{2}}+\frac{\mathrm{log}\left(L\right)}{Lh^{4+2d}}\right),
\end{eqnarray*}
where the equality follows from (\ref{eq:sigma_J_1 rate}) and $\left\Vert F_{\mathcal{J}_{1}^{\left(1\right)}}\right\Vert _{\mathcal{X}}=O\left(h^{-\left(4+2d\right)}\right)$(which
follows from change of variables). 

Similarly, since $\left\{ \mathcal{J}_{2}^{\left(1\right)}\left(\cdot;v\right):v\in I\left(\boldsymbol{x}\right)\right\} $
is VC-type with respect to the envelope 
\[
F_{\mathcal{J}_{2}^{\left(1\right)}}\left(\boldsymbol{z}'\right)\coloneqq\frac{\left(\overline{C}_{D_{1}}+\overline{C}_{D_{2}}\right)^{2}}{\left(n-1\right)^{2}\underline{C}_{g}^{2}}\int\int\frac{1}{h^{4\left(1+d\right)}}\left|K_{\boldsymbol{X}}^{0}\left(\frac{\boldsymbol{z}'-\boldsymbol{x}}{h}\right)\right|\left|K_{\boldsymbol{X}}\left(\frac{\boldsymbol{z}-\boldsymbol{z}'}{h}\right)\right|\left|K_{\boldsymbol{X}}^{0}\left(\frac{\boldsymbol{z}''-\boldsymbol{x}}{h}\right)\right|\left|K_{\boldsymbol{X}}\left(\frac{\boldsymbol{z}-\boldsymbol{z}''}{h}\right)\right|\varphi\left(\boldsymbol{z}\right)\varphi\left(\boldsymbol{z}''\right)\mathrm{d}\boldsymbol{z}\mathrm{d}\boldsymbol{z}'',
\]
the CCK inequality yields 
\begin{eqnarray*}
\mathrm{E}\left[\underset{v\in I\left(\boldsymbol{x}\right)}{\mathrm{sup}}\left|\frac{1}{L}\sum_{l=1}^{L}\mathcal{J}_{2}^{\left(1\right)}\left(\boldsymbol{B}_{\cdot l},\boldsymbol{X}_{l},N_{l};v\right)-\mu_{\mathcal{J}}\left(v\right)\right|\right] & \leq & C_{1}\left\{ L^{-\nicefrac{1}{2}}\sigma_{\mathcal{J}_{2}^{\left(1\right)}}\mathrm{log}\left(C_{2}L\right)^{\nicefrac{1}{2}}+L^{-1}\left\Vert F_{\mathcal{J}_{2}^{\left(1\right)}}\right\Vert _{\mathcal{X}}\mathrm{log}\left(C_{2}L\right)\right\} \\
 & = & O\left(\left(\frac{\mathrm{log}\left(L\right)}{Lh^{5+3d}}\right)^{\nicefrac{1}{2}}+\frac{\mathrm{log}\left(L\right)}{Lh^{4+2d}}\right).
\end{eqnarray*}
Since $\mathcal{J}$ is symmetric with respect to the second and the
third arguments, we have 
\[
\mathrm{E}\left[\underset{v\in I\left(\boldsymbol{x}\right)}{\mathrm{sup}}\left|\frac{1}{L}\sum_{l=1}^{L}\mathcal{J}_{3}^{\left(1\right)}\left(\boldsymbol{B}_{\cdot l},\boldsymbol{X}_{l},N_{l};v\right)-\mu_{\mathcal{J}}\left(v\right)\right|\right]=O\left(\left(\frac{\mathrm{log}\left(L\right)}{Lh^{5+3d}}\right)^{\nicefrac{1}{2}}+\frac{\mathrm{log}\left(L\right)}{Lh^{4+2d}}\right).
\]

Let 
\[
F_{\mathcal{J}_{1}^{\left(2\right)}}\left(\boldsymbol{z},\boldsymbol{z}'\right)\coloneqq\int F_{\mathcal{J}}\left(\boldsymbol{z},\boldsymbol{z}',\boldsymbol{z}''\right)\varphi\left(\boldsymbol{z}''\right)\mathrm{d}\boldsymbol{z}''\textrm{, }F_{\mathcal{J}_{2}^{\left(2\right)}}\left(\boldsymbol{z},\boldsymbol{z}'\right)\coloneqq\int F_{\mathcal{J}}\left(\boldsymbol{z},\boldsymbol{z}'',\boldsymbol{z}'\right)\varphi\left(\boldsymbol{z}''\right)\mathrm{d}\boldsymbol{z}''\textrm{ and }F_{\mathcal{J}_{3}^{\left(2\right)}}\left(\boldsymbol{z},\boldsymbol{z}'\right)\coloneqq\int F_{\mathcal{J}}\left(\boldsymbol{z}'',\boldsymbol{z},\boldsymbol{z}'\right)\varphi\left(\boldsymbol{z}''\right)\mathrm{d}\boldsymbol{z}''.
\]
The CK inequality and change of variables yield
\begin{gather*}
\mathrm{E}\left[\underset{v\in I\left(\boldsymbol{x}\right)}{\mathrm{sup}}\left|\Upsilon_{\mathcal{J}}^{1}\left(v\right)\right|\right]\leq L^{-1}\left(\mathrm{E}\left[F_{\mathcal{J}_{1}^{\left(2\right)}}\left(\boldsymbol{X}_{1},\boldsymbol{X}_{2}\right)^{2}\right]\right)^{\nicefrac{1}{2}}=O\left(\left(Lh^{4+2d}\right)^{-1}\right),\\
\mathrm{E}\left[\underset{v\in I\left(\boldsymbol{x}\right)}{\mathrm{sup}}\left|\Upsilon_{\mathcal{J}}^{2}\left(v\right)\right|\right]\leq L^{-1}\left(\mathrm{E}\left[F_{\mathcal{J}_{2}^{\left(2\right)}}\left(\boldsymbol{X}_{1},\boldsymbol{X}_{2}\right)^{2}\right]\right)^{\nicefrac{1}{2}}=O\left(\left(Lh^{4+2d}\right)^{-1}\right),\\
\mathrm{E}\left[\underset{v\in I\left(\boldsymbol{x}\right)}{\mathrm{sup}}\left|\Upsilon_{\mathcal{J}}^{3}\left(v\right)\right|\right]\leq L^{-1}\left(\mathrm{E}\left[F_{\mathcal{J}_{3}^{\left(2\right)}}\left(\boldsymbol{X}_{1},\boldsymbol{X}_{2}\right)^{2}\right]\right)^{\nicefrac{1}{2}}=O\left(\left(Lh^{4+2d}\right)^{-1}\right)
\end{gather*}
and 
\[
\mathrm{E}\left[\underset{v\in I\left(\boldsymbol{x}\right)}{\mathrm{sup}}\left|\Psi_{\mathcal{J}}\left(v\right)\right|\right]\leq L^{-\nicefrac{3}{2}}\left(\mathrm{E}\left[F_{\mathcal{J}}\left(\boldsymbol{X}_{1},\boldsymbol{X}_{2},\boldsymbol{X}_{3}\right)^{2}\right]\right)^{\nicefrac{1}{2}}=O\left(L^{-\nicefrac{3}{2}}h^{4+\nicefrac{5d}{2}}\right).
\]

It follows from these bounds for expectations of suprema of empirical
and degenerate U-processes, (\ref{eq:J Hoeffding decomposition}),
(\ref{eq:sup miu_J bound}) and Markov's inequality that 
\[
\underset{v\in I\left(\boldsymbol{x}\right)}{\mathrm{sup}}\left|\frac{1}{\left(L\right)_{3}}\sum_{\left(3\right)}\mathcal{J}\left(\left(\boldsymbol{B}_{\cdot l},\boldsymbol{X}_{l},N_{l}\right),\left(\boldsymbol{B}_{\cdot k},\boldsymbol{X}_{k},N_{k}\right),\left(\boldsymbol{B}_{\cdot k'},\boldsymbol{X}_{k'},N_{k'}\right);v\right)\right|=O_{p}\left(h^{-\left(1+d\right)}+\left(\frac{\mathrm{log}\left(L\right)}{Lh^{5+3d}}\right)^{\nicefrac{1}{2}}+\frac{\mathrm{log}\left(L\right)}{Lh^{4+2d}}\right).
\]
It is also straightforward to check
\begin{eqnarray*}
\underset{v\in I\left(\boldsymbol{x}\right)}{\mathrm{sup}}\left|\frac{1}{L}\sum_{l=1}^{L}\mathcal{J}\left(\left(\boldsymbol{B}_{\cdot l},\boldsymbol{X}_{l},N_{l}\right),\left(\boldsymbol{B}_{\cdot l},\boldsymbol{X}_{l},N_{l}\right),\left(\boldsymbol{B}_{\cdot l},\boldsymbol{X}_{l},N_{l}\right);v\right)\right| & \apprle & \frac{1}{L}\sum_{l=1}^{L}\frac{1}{h^{4\left(1+d\right)}}K_{\boldsymbol{X}}^{0}\left(\frac{\boldsymbol{X}_{l}-\boldsymbol{x}}{h}\right)^{2}\\
 & = & O_{p}\left(h^{-\left(4+3d\right)}\right),
\end{eqnarray*}
\begin{align*}
 & \underset{v\in I\left(\boldsymbol{x}\right)}{\mathrm{sup}}\left|\frac{1}{L\left(L-1\right)}\sum_{l=1}^{L}\mathcal{J}\left(\left(\boldsymbol{B}_{\cdot l},\boldsymbol{X}_{l},N_{l}\right),\left(\boldsymbol{B}_{\cdot l},\boldsymbol{X}_{l},N_{l}\right),\left(\boldsymbol{B}_{\cdot k},\boldsymbol{X}_{k},N_{k}\right);v\right)\right|\\
\apprle & \frac{1}{L\left(L-1\right)}\sum_{l\neq k}\frac{1}{h^{4\left(1+d\right)}}\left|K_{\boldsymbol{X}}^{0}\left(\frac{\boldsymbol{X}_{l}-\boldsymbol{x}}{h}\right)\right|\left|K_{\boldsymbol{X}}^{0}\left(\frac{\boldsymbol{X}_{k}-\boldsymbol{x}}{h}\right)\right|\left|K_{\boldsymbol{X}}\left(\frac{\boldsymbol{X}_{l}-\boldsymbol{X}_{k}}{h}\right)\right|\\
= & O_{p}\left(h^{-\left(4+2d\right)}\right)
\end{align*}
and
\begin{eqnarray*}
\underset{v\in I\left(\boldsymbol{x}\right)}{\mathrm{sup}}\left|\frac{1}{L\left(L-1\right)}\sum_{l=1}^{L}\mathcal{J}\left(\left(\boldsymbol{B}_{\cdot k},\boldsymbol{X}_{k},N_{k}\right),\left(\boldsymbol{B}_{\cdot l},\boldsymbol{X}_{l},N_{l}\right),\left(\boldsymbol{B}_{\cdot l},\boldsymbol{X}_{l},N_{l}\right);v\right)\right| & \apprle & \frac{1}{L\left(L-1\right)}\sum_{l\neq k}\frac{1}{h^{4\left(1+d\right)}}K_{\boldsymbol{X}}^{0}\left(\frac{\boldsymbol{X}_{l}-\boldsymbol{x}}{h}\right)^{2}K_{\boldsymbol{X}}\left(\frac{\boldsymbol{X}_{l}-\boldsymbol{X}_{k}}{h}\right)^{2}\\
 & = & O_{p}\left(h^{-\left(4+2d\right)}\right),
\end{eqnarray*}
where the equalities follow from change of variables and Markov's
inequality. Then it follows that 
\begin{equation}
\sigma_{\widehat{\mathcal{G}}_{2}^{\ddagger,n}}^{2}\coloneqq\underset{v\in I\left(\boldsymbol{x}\right)}{\mathrm{sup}}\mathrm{E}^{*}\left[\widehat{\mathcal{G}}_{2}^{\ddagger,n}\left(\boldsymbol{B}_{\cdot1}^{*},\boldsymbol{X}_{1}^{*},N_{1}^{*};v\right)^{2}\right]=O_{p}\left(h^{-\left(1+d\right)}+\left(\frac{\mathrm{log}\left(L\right)}{Lh^{5+3d}}\right)^{\nicefrac{1}{2}}+\frac{\mathrm{log}\left(L\right)}{Lh^{4+2d}}\right)\label{eq:sigma G_ddagger_2_hat rate}
\end{equation}
and the CCK inequality yields
\begin{eqnarray*}
\mathrm{E}^{*}\left[\underset{v\in I\left(\boldsymbol{x}\right)}{\mathrm{sup}}\left|\frac{1}{L}\sum_{l=1}^{L}\widehat{\mathcal{G}}_{2}^{\ddagger,n}\left(\boldsymbol{B}_{\cdot l}^{*},\boldsymbol{X}_{l}^{*},N_{l}^{*};v\right)-\mathrm{E}^{*}\left[\widehat{\mathcal{G}}_{2}^{\ddagger,n}\left(\boldsymbol{B}_{\cdot1}^{*},\boldsymbol{X}_{1}^{*},N_{1}^{*};v\right)\right]\right|\right] & \leq & C_{1}\left\{ L^{-\nicefrac{1}{2}}\sigma_{\widehat{\mathcal{G}}_{2}^{\ddagger,n}}\mathrm{log}\left(C_{2}L\right)^{\nicefrac{1}{2}}+L^{-1}\left\Vert F_{\widehat{\mathcal{G}}_{2}^{\ddagger,n}}\right\Vert _{\mathcal{X}}\mathrm{log}\left(C_{2}L\right)\right\} \\
 & = & O_{p}\left(\left(\frac{\mathrm{log}\left(L\right)}{Lh^{1+d}}\right)^{\nicefrac{1}{2}}+\frac{\mathrm{log}\left(L\right)}{Lh^{2+d}}\right),
\end{eqnarray*}
where the equality follows from (\ref{eq:sigma G_ddagger_2_hat rate})
and $\left\Vert F_{\widehat{\mathcal{G}}_{2}^{\ddagger,n}}\right\Vert _{\mathcal{X}}=O_{p}\left(h^{-\left(2+d\right)}\right)$
(which follows from change of variables and Markov's inequality).

It is easy to check 
\begin{align*}
 & \underset{v\in I\left(\boldsymbol{x}\right)}{\mathrm{sup}}\left|\frac{1}{L^{2}\left(L-1\right)}\sum_{l\neq k}\mathcal{G}^{n}\left(\left(\boldsymbol{B}_{\cdot l}^{*},\boldsymbol{X}_{l}^{*},N_{l}^{*}\right),\left(\boldsymbol{B}_{\cdot k}^{*},\boldsymbol{X}_{k}^{*},N_{k}^{*}\right);v\right)\right|\\
\apprle & \frac{1}{L^{2}\left(L-1\right)}\sum_{l\neq k}\left\{ \frac{1}{h^{2+2d}}\left|K_{\boldsymbol{X}}^{0}\left(\frac{\boldsymbol{X}_{l}^{*}-\boldsymbol{x}}{h}\right)\right|\left|K_{\boldsymbol{X}}\left(\frac{\boldsymbol{X}_{l}^{*}-\boldsymbol{X}_{k}^{*}}{h}\right)\right|+\frac{1}{h^{2+d}}\left|K_{\boldsymbol{X}}^{0}\left(\frac{\boldsymbol{X}_{l}^{*}-\boldsymbol{x}}{h}\right)\right|\right\} \\
= & O_{p}^{*}\left(\left(Lh^{2}\right)^{-1}\right)
\end{align*}
and
\begin{eqnarray*}
\underset{v\in I\left(\boldsymbol{x}\right)}{\mathrm{sup}}\left|\frac{1}{L^{2}}\sum_{l=1}^{L}\mathcal{G}^{n}\left(\left(\boldsymbol{B}_{\cdot l}^{*},\boldsymbol{X}_{l}^{*},N_{l}^{*}\right),\left(\boldsymbol{B}_{\cdot l}^{*},\boldsymbol{X}_{l}^{*},N_{l}^{*}\right);v\right)\right| & \apprle & \frac{1}{L^{\text{2}}}\sum_{l=1}^{L}\left\{ \frac{1}{h^{2+2d}}\left|K_{\boldsymbol{X}}^{0}\left(\frac{\boldsymbol{X}_{l}^{*}-\boldsymbol{x}}{h}\right)\right|+\frac{1}{h^{2+d}}\left|K_{\boldsymbol{X}}^{0}\left(\frac{\boldsymbol{X}_{l}^{*}-\boldsymbol{x}}{h}\right)\right|\right\} \\
 & = & O_{p}^{*}\left(\left(Lh^{2+d}\right)^{-1}\right),
\end{eqnarray*}
where the equalities follow from change of variables, Markov's inequality
and Lemma \ref{lem:auxiliary 3}. Now 
\[
\underset{v\in I\left(\boldsymbol{x}\right)}{\mathrm{sup}}\left|\varDelta_{1}^{*}\left(v\right)\right|=O_{p}^{*}\left(\left(\frac{\mathrm{log}\left(L\right)}{Lh^{1+d}}\right)^{\nicefrac{1}{2}}+\frac{\mathrm{log}\left(L\right)}{Lh^{3+d}}+h^{R}\right)
\]
follows from order bounds derived for each of the terms in the decomposition
(\ref{eq:G_n_star decomposition}). Now the conclusion follows from
the definition of $\widehat{g}^{*}$.\end{proof}
\begin{lem}
\label{Lemma 9}Suppose that Assumptions 1 - 3 hold. Let $\boldsymbol{x}$
be an interior point of $\mathcal{X}$ and $n\in\mathcal{N}$ be fixed.
Then 
\begin{eqnarray*}
\widehat{f}_{GPV}^{*}\left(v,\boldsymbol{x},n\right)-\widetilde{f}^{*}\left(v,\boldsymbol{x},n\right) & = & \frac{1}{L^{2}}\sum_{l=1}^{L}\sum_{k=1}^{L}\mathcal{M}^{n}\left(\left(\boldsymbol{B}_{\cdot l},\boldsymbol{X}_{l},N_{l}\right),\left(\boldsymbol{B}_{\cdot k},\boldsymbol{X}_{k},N_{k}\right);v\right)\\
 &  & +\left\{ \frac{1}{L}\sum_{l=1}^{L}\mathcal{M}_{2}^{n}\left(\boldsymbol{B}_{\cdot l}^{*},\boldsymbol{X}_{l}^{*},N_{l}^{*};v\right)-\frac{1}{L}\sum_{l=1}^{L}\mathcal{M}_{2}^{n}\left(\boldsymbol{B}_{\cdot l},\boldsymbol{X}_{l},N_{l};v\right)\right\} \\
 &  & +O_{p}^{*}\left(\left(\frac{\mathrm{log}\left(L\right)}{Lh^{1+d}}\right)^{\nicefrac{1}{2}}+\frac{\mathrm{log}\left(L\right)}{Lh^{3+d}}+h^{R}\right),
\end{eqnarray*}
where the remainder term is uniform in $v\in I\left(\boldsymbol{x}\right)$.
\end{lem}
\begin{proof}[Proof of Lemma \ref{Lemma 9}]Let 
\[
\widehat{\mu}_{\mathcal{M}^{n}}\left(v\right)\coloneqq\mathrm{E}^{*}\left[\mathcal{M}^{n}\left(\left(\boldsymbol{B}_{\cdot1}^{*},\boldsymbol{X}_{1}^{*},N_{1}^{*}\right),\left(\boldsymbol{B}_{\cdot2}^{*},\boldsymbol{X}_{2}^{*},N_{2}^{*}\right);v\right)\right],
\]
\[
\widehat{\mathcal{M}}_{1}^{n}\left(\boldsymbol{b}.,\boldsymbol{z},m;v\right)\coloneqq\mathrm{E}^{*}\left[\mathcal{M}^{n}\left(\left(\boldsymbol{b}.,\boldsymbol{z},m\right),\left(\boldsymbol{B}_{\cdot1}^{*},\boldsymbol{X}_{1}^{*},N_{1}^{*}\right);v\right)\right]\textrm{ and }\widehat{\mathcal{M}}_{2}^{n}\left(\boldsymbol{b}.,\boldsymbol{z},m;v\right)\coloneqq\mathrm{E}^{*}\left[\mathcal{M}^{n}\left(\left(\boldsymbol{B}_{\cdot1}^{*},\boldsymbol{X}_{1}^{*},N_{1}^{*}\right),\left(\boldsymbol{b}.,\boldsymbol{z},m\right);v\right)\right].
\]
The Hoeffding decomposition yields
\begin{align}
 & \frac{1}{L^{2}}\sum_{l=1}^{L}\sum_{k=1}^{L}\mathcal{M}^{n}\left(\left(\boldsymbol{B}_{\cdot l}^{*},\boldsymbol{X}_{l}^{*},N_{l}^{*}\right),\left(\boldsymbol{B}_{\cdot k}^{*},\boldsymbol{X}_{k}^{*},N_{k}^{*}\right);v\right)\nonumber \\
= & \widehat{\mu}_{\mathcal{M}^{n}}\left(v\right)+\left\{ \frac{1}{L}\sum_{l=1}^{L}\widehat{\mathcal{M}}_{1}^{n}\left(\boldsymbol{B}_{\cdot l}^{*},\boldsymbol{X}_{l}^{*},N_{l}^{*};v\right)-\widehat{\mu}_{\mathcal{M}^{n}}\left(v\right)\right\} +\left\{ \frac{1}{L}\sum_{l=1}^{L}\widehat{\mathcal{M}}_{2}^{n}\left(\boldsymbol{B}_{\cdot l}^{*},\boldsymbol{X}_{l}^{*},N_{l}^{*};v\right)-\widehat{\mu}_{\mathcal{M}^{n}}\left(v\right)\right\} \nonumber \\
 & +\frac{1}{\left(L\right)_{2}}\sum_{\left(2\right)}\left\{ \mathcal{M}^{n}\left(\left(\boldsymbol{B}_{\cdot l}^{*},\boldsymbol{X}_{l}^{*},N_{l}^{*}\right),\left(\boldsymbol{B}_{\cdot k}^{*},\boldsymbol{X}_{k}^{*},N_{k}^{*}\right);v\right)-\widehat{\mathcal{M}}_{1}^{n}\left(\boldsymbol{B}_{\cdot l}^{*},\boldsymbol{X}_{l}^{*},N_{l}^{*};v\right)-\widehat{\mathcal{M}}_{2}^{n}\left(\boldsymbol{B}_{\cdot k}^{*},\boldsymbol{X}_{k}^{*},N_{k}^{*};v\right)+\widehat{\mu}_{\mathcal{M}^{n}}\left(v\right)\right\} \nonumber \\
 & +\frac{1}{L^{2}}\sum_{l=1}^{L}\mathcal{M}^{n}\left(\left(\boldsymbol{B}_{\cdot l}^{*},\boldsymbol{X}_{l}^{*},N_{l}^{*}\right),\left(\boldsymbol{B}_{\cdot l}^{*},\boldsymbol{X}_{l}^{*},N_{l}^{*}\right);v\right)-\frac{1}{L^{2}\left(L-1\right)}\sum_{l\neq k}\mathcal{M}^{n}\left(\left(\boldsymbol{B}_{\cdot l}^{*},\boldsymbol{X}_{l}^{*},N_{l}^{*}\right),\left(\boldsymbol{B}_{\cdot k}^{*},\boldsymbol{X}_{k}^{*},N_{k}^{*}\right);v\right).\label{eq:M_n_star decomposition}
\end{align}

By the LIE and the fact that the bids in the bootstrap sample are
conditionally i.i.d., we have 
\begin{equation}
\widehat{\mu}_{\mathcal{M}^{n}}\left(v\right)=\frac{1}{L^{2}}\sum_{l=1}^{L}\sum_{k=1}^{L}\mathcal{M}^{n}\left(\left(\boldsymbol{B}_{\cdot l},\boldsymbol{X}_{l},N_{l}\right),\left(\boldsymbol{B}_{\cdot k},\boldsymbol{X}_{k},N_{k}\right);v\right).\label{eq:eq:Lemma S9 bound  0}
\end{equation}

Since $\left\{ \mathcal{M}^{n}\left(\cdot,\cdot;v\right):v\in I\left(\boldsymbol{x}\right)\right\} $
is uniformly VC-type with respect to the envelope (\ref{eq:F_M_n envelope}),
the CK inequality yields 
\begin{align*}
 & \mathrm{E}^{*}\left[\underset{v\in I\left(\boldsymbol{x}\right)}{\mathrm{sup}}\left|\frac{1}{\left(L\right)_{2}}\sum_{\left(2\right)}\left\{ \mathcal{M}^{n}\left(\left(\boldsymbol{B}_{\cdot l}^{*},\boldsymbol{X}_{l}^{*},N_{l}^{*}\right),\left(\boldsymbol{B}_{\cdot k}^{*},\boldsymbol{X}_{k}^{*},N_{k}^{*}\right);v\right)-\widehat{\mathcal{M}}_{1}^{n}\left(\boldsymbol{B}_{\cdot l}^{*},\boldsymbol{X}_{l}^{*},N_{l}^{*};v\right)-\widehat{\mathcal{M}}_{2}^{n}\left(\boldsymbol{B}_{\cdot k}^{*},\boldsymbol{X}_{k}^{*},N_{k}^{*};v\right)+\widehat{\mu}_{\mathcal{M}^{n}}\left(v\right)\right\} \right|\right]\\
\apprle & L^{-1}\left(\mathrm{E}^{*}\left[F_{\mathcal{M}^{n}}\left(\boldsymbol{X}_{1}^{*},\boldsymbol{X}_{2}^{*}\right)^{2}\right]\right)^{\nicefrac{1}{2}}.
\end{align*}
Since
\begin{eqnarray*}
\mathrm{E}\left[\frac{1}{L^{2}}\sum_{l=1}^{L}\sum_{k=1}^{L}F_{\mathcal{M}^{n}}\left(\boldsymbol{X}_{l},\boldsymbol{X}_{k}\right)^{2}\right] & = & \frac{L-1}{L}\mathrm{E}\left[F_{\mathcal{M}^{n}}\left(\boldsymbol{X}_{1},\boldsymbol{X}_{2}\right)^{2}\right]+\frac{1}{L}\mathrm{E}\left[F_{\mathcal{M}^{n}}\left(\boldsymbol{X}_{1},\boldsymbol{X}_{1}\right)^{2}\right]\\
 & = & O\left(h^{-\left(6+2d\right)}\right),
\end{eqnarray*}
where the second equality follows from change of variables, we have
\[
\mathrm{E}^{*}\left[F_{\mathcal{M}^{n}}\left(\boldsymbol{X}_{1}^{*},\boldsymbol{X}_{2}^{*}\right)^{2}\right]=\frac{1}{L^{2}}\sum_{l=1}^{L}\sum_{k=1}^{L}F_{\mathcal{M}^{n}}\left(\boldsymbol{X}_{l},\boldsymbol{X}_{k}\right)^{2}=O_{p}\left(h^{-\left(6+2d\right)}\right),
\]
where the second equality follows from Markov's inequality. Therefore
we have
\begin{align}
 & \mathrm{E}^{*}\left[\underset{v\in I\left(\boldsymbol{x}\right)}{\mathrm{sup}}\left|\frac{1}{\left(L\right)_{2}}\sum_{\left(2\right)}\left\{ \mathcal{M}^{n}\left(\left(\boldsymbol{B}_{\cdot l}^{*},\boldsymbol{X}_{l}^{*},N_{l}^{*}\right),\left(\boldsymbol{B}_{\cdot k}^{*},\boldsymbol{X}_{k}^{*},N_{k}^{*}\right);v\right)-\widehat{\mathcal{M}}_{1}^{n}\left(\boldsymbol{B}_{\cdot l}^{*},\boldsymbol{X}_{l}^{*},N_{l}^{*};v\right)-\widehat{\mathcal{M}}_{2}^{n}\left(\boldsymbol{B}_{\cdot k}^{*},\boldsymbol{X}_{k}^{*},N_{k}^{*};v\right)+\widehat{\mu}_{\mathcal{M}^{n}}\left(v\right)\right\} \right|\right]\nonumber \\
= & O_{p}\left(\left(Lh^{3+d}\right)^{-1}\right).\label{eq:Lemma S9 bound  1}
\end{align}

By the LIE and the fact that the bids in the bootstrap sample are
conditionally i.i.d., we have 
\[
\widehat{\mathcal{M}}_{1}^{n}\left(\boldsymbol{b}.,\boldsymbol{z},m;v\right)=-\mathbbm{1}\left(m=n\right)\frac{1}{m}\sum_{i=1}^{m}\frac{1}{h^{2+d}}K_{f}'\left(\frac{\xi\left(b_{i},\boldsymbol{z},m\right)-v}{h},\frac{\boldsymbol{z}-\boldsymbol{x}}{h}\right)\frac{G\left(b_{i},\boldsymbol{z},m\right)}{\left(m-1\right)g\left(b_{i},\boldsymbol{z},m\right)}\left\{ \widehat{g}\left(b_{i},\boldsymbol{z},m\right)-g\left(b_{i},\boldsymbol{z},m\right)\right\} .
\]
It is clear from the definition that when $h$ is sufficiently small,
$\left\{ \widehat{\mathcal{M}}_{1}^{n}\left(\cdot;v\right):v\in I\left(\boldsymbol{x}\right)\right\} $
is uniformly VC-type with respect to the envelope
\[
F_{\widehat{\mathcal{M}}_{1}^{n}}\left(\boldsymbol{z}\right)\coloneqq\frac{\left(\overline{C}_{D_{1}}+\overline{C}_{D_{2}}\right)\overline{\varphi}}{\left(n-1\right)\underline{C}_{g}h^{2+d}}\left|K_{\boldsymbol{X}}^{0}\left(\frac{\boldsymbol{z}-\boldsymbol{x}}{h}\right)\right|\left\{ \underset{\left(b',\boldsymbol{z}'\right)\in\mathcal{C}_{B,\boldsymbol{X}}^{n}}{\mathrm{sup}}\left|\widehat{g}\left(b',\boldsymbol{z}',n\right)-g\left(b',\boldsymbol{z}',n\right)\right|\right\} .
\]
Then the VW inequality yields
\begin{align}
 & \mathrm{E}^{*}\left[\underset{v\in I\left(\boldsymbol{x}\right)}{\mathrm{sup}}\left|\frac{1}{L}\sum_{l=1}^{L}\widehat{\mathcal{M}}_{1}^{n}\left(\boldsymbol{B}_{\cdot l}^{*},\boldsymbol{X}_{l}^{*},N_{l}^{*};v\right)-\widehat{\mu}_{\mathcal{M}^{n}}\left(v\right)\right|\right]\nonumber \\
\leq & L^{-\nicefrac{1}{2}}\left(\mathrm{E}^{*}\left[F_{\widehat{\mathcal{M}}_{1}^{n}}\left(\boldsymbol{X}_{1}^{*}\right)^{2}\right]\right)^{\nicefrac{1}{2}}\nonumber \\
\apprle & L^{-\nicefrac{1}{2}}\left\{ \frac{1}{L}\sum_{l=1}^{L}\frac{1}{h^{4+2d}}K_{\boldsymbol{X}}^{0}\left(\frac{\boldsymbol{X}_{l}-\boldsymbol{x}}{h}\right)^{2}\right\} ^{\nicefrac{1}{2}}\left\{ \underset{\left(b',\boldsymbol{z}'\right)\in\mathcal{C}_{B,\boldsymbol{X}}^{n}}{\mathrm{sup}}\left|\widehat{g}\left(b',\boldsymbol{z}',n\right)-g\left(b',\boldsymbol{z}',n\right)\right|\right\} \nonumber \\
= & O_{p}\left(\left(Lh^{4+d}\right)^{-\nicefrac{1}{2}}\right)O_{p}\left(\left(\frac{\mathrm{log}\left(L\right)}{Lh^{1+d}}\right)^{\nicefrac{1}{2}}+h^{1+R}\right),\label{eq:Lemma S9 bound  2}
\end{align}
where the equality follows from change of variables, Markov's inequality
and (\ref{eq:uniform convergence rate}).

Let 
\[
\widehat{\mathcal{M}}_{2}^{\ddagger,n}\left(\boldsymbol{b}.,\boldsymbol{z},m;v\right)\coloneqq\mathrm{E}^{*}\left[\mathcal{M}^{\ddagger,n}\left(\left(\boldsymbol{B}_{\cdot1}^{*},\boldsymbol{X}_{1}^{*},N_{1}^{*}\right),\left(\boldsymbol{b}.,\boldsymbol{z},m\right);v\right)\right]
\]
and 
\[
\widehat{\mu}_{\mathcal{M}^{\ddagger,n}}\left(v\right)\coloneqq\mathrm{E}^{*}\left[\mathcal{M}^{\ddagger,n}\left(\left(\boldsymbol{B}_{\cdot1}^{*},\boldsymbol{X}_{1}^{*},N_{1}^{*}\right),\left(\boldsymbol{B}_{\cdot2}^{*},\boldsymbol{X}_{2}^{*},N_{2}^{*}\right);v\right)\right].
\]

Consider 
\[
\varDelta^{**}\left(v\right)\coloneqq\frac{1}{L}\sum_{l=1}^{L}\left\{ \left(\widehat{\mathcal{M}}_{2}^{n}\left(\boldsymbol{B}_{\cdot l}^{*},\boldsymbol{X}_{l}^{*},N_{l}^{*};v\right)-\widehat{\mu}_{\mathcal{M}^{n}}\left(v\right)\right)-\left(\mathcal{M}_{2}^{n}\left(\boldsymbol{B}_{\cdot l}^{*},\boldsymbol{X}_{l}^{*},N_{l}^{*};v\right)-\frac{1}{L}\sum_{k=1}^{L}\mathcal{M}_{2}^{n}\left(\boldsymbol{B}_{\cdot k},\boldsymbol{X}_{k},N_{k};v\right)\right)\right\} ,\textrm{ \ensuremath{v\in I\left(\boldsymbol{x}\right)}}.
\]
Since it is straightforward to check
\[
\widehat{\mathcal{M}}_{2}^{n}\left(\boldsymbol{B}_{\cdot l}^{*},\boldsymbol{X}_{l}^{*},N_{l}^{*};v\right)-\widehat{\mu}_{\mathcal{M}^{n}}\left(v\right)=\widehat{\mathcal{M}}_{2}^{\ddagger,n}\left(\boldsymbol{B}_{\cdot l}^{*},\boldsymbol{X}_{l}^{*},N_{l}^{*};v\right)-\widehat{\mu}_{\mathcal{M}^{\ddagger,n}}\left(v\right),\textrm{ for all \ensuremath{l=1,...,L}}
\]
and 
\[
\mathcal{M}_{2}^{n}\left(\boldsymbol{B}_{\cdot l}^{*},\boldsymbol{X}_{l}^{*},N_{l}^{*};v\right)-\frac{1}{L}\sum_{k=1}^{L}\mathcal{M}_{2}^{n}\left(\boldsymbol{B}_{\cdot k},\boldsymbol{X}_{k},N_{k};v\right)=\mathcal{M}_{2}^{\ddagger,n}\left(\boldsymbol{B}_{\cdot l}^{*},\boldsymbol{X}_{l}^{*},N_{l}^{*};v\right)-\frac{1}{L}\sum_{k=1}^{L}\mathcal{M}_{2}^{\ddagger,n}\left(\boldsymbol{B}_{\cdot k},\boldsymbol{X}_{k},N_{k};v\right),\textrm{ for all \ensuremath{l=1,...,L}},
\]
we have 
\[
\varDelta^{**}\left(v\right)=\frac{1}{L}\sum_{l=1}^{L}\left\{ \left(\widehat{\mathcal{M}}_{2}^{\ddagger,n}\left(\boldsymbol{B}_{\cdot l}^{*},\boldsymbol{X}_{l}^{*},N_{l}^{*};v\right)-\widehat{\mu}_{\mathcal{M}^{\ddagger,n}}\left(v\right)\right)-\left(\mathcal{M}_{2}^{\ddagger,n}\left(\boldsymbol{B}_{\cdot l}^{*},\boldsymbol{X}_{l}^{*},N_{l}^{*};v\right)-\frac{1}{L}\sum_{k=1}^{L}\mathcal{M}_{2}^{\ddagger,n}\left(\boldsymbol{B}_{\cdot k},\boldsymbol{X}_{k},N_{k};v\right)\right)\right\} ,\textrm{ \ensuremath{v\in I\left(\boldsymbol{x}\right)}}.
\]

Simple algebra yields 
\begin{align*}
 & \mathrm{E}^{*}\left[\left(\widehat{\mathcal{M}}_{2}^{\ddagger,n}\left(\boldsymbol{B}_{\cdot1}^{*},\boldsymbol{X}_{1}^{*},N_{1}^{*};v\right)-\mathcal{M}_{2}^{\ddagger,n}\left(\boldsymbol{B}_{\cdot1}^{*},\boldsymbol{X}_{1}^{*},N_{1}^{*};v\right)\right)^{2}\right]\\
= & \frac{1}{L^{\text{3}}}\sum_{l=1}^{L}\sum_{k=1}^{L}\sum_{k'=1}^{L}\mathcal{L}\left(\left(\boldsymbol{B}_{\cdot l},\boldsymbol{X}_{l},N_{l}\right),\left(\boldsymbol{B}_{\cdot k},\boldsymbol{X}_{k},N_{k}\right),\left(\boldsymbol{B}_{\cdot k'},\boldsymbol{X}_{k'},N_{k'}\right);v\right),
\end{align*}
where 
\begin{align*}
 & \mathcal{L}\left(\left(\boldsymbol{b}.,\boldsymbol{z},m\right),\left(\boldsymbol{b}.',\boldsymbol{z}',m'\right),\left(\boldsymbol{b}.'',\boldsymbol{z}'',m''\right);v\right)\\
\coloneqq & \left(\mathcal{M}^{\ddagger,n}\left(\left(\boldsymbol{b}.',\boldsymbol{z}',m'\right),\left(\boldsymbol{b}.,\boldsymbol{z},m\right);v\right)-\mathcal{M}_{2}^{\ddagger,n}\left(\boldsymbol{b}.,\boldsymbol{z},m;v\right)\right)\left(\mathcal{M}^{\ddagger,n}\left(\left(\boldsymbol{b}.'',\boldsymbol{z}'',m''\right),\left(\boldsymbol{b}.,\boldsymbol{z},m\right);v\right)-\mathcal{M}_{2}^{\ddagger,n}\left(\boldsymbol{b}.,\boldsymbol{z},m;v\right)\right).
\end{align*}
Standard arguments can be applied to verify that $\left\{ \mathcal{L}\left(\cdot,\cdot,\cdot;v\right):v\in I\left(\boldsymbol{x}\right)\right\} $
is VC-type with respect to the envelope 
\[
F_{\mathcal{L}}\left(\boldsymbol{z},\boldsymbol{z}',\boldsymbol{z}''\right)\coloneqq F_{\mathcal{M}^{\ddagger,n}}\left(\boldsymbol{z}',\boldsymbol{z}\right)F_{\mathcal{M}^{\ddagger,n}}\left(\boldsymbol{z}'',\boldsymbol{z}\right)+F_{\mathcal{M}_{2}^{\ddagger,n}}\left(\boldsymbol{z}\right)F_{\mathcal{M}^{\ddagger,n}}\left(\boldsymbol{z}'',\boldsymbol{z}\right)+F_{\mathcal{M}_{2}^{\ddagger,n}}\left(\boldsymbol{z}\right)F_{\mathcal{M}^{\ddagger,n}}\left(\boldsymbol{z}',\boldsymbol{z}\right)+F_{\mathcal{M}_{2}^{\ddagger,n}}\left(\boldsymbol{z}\right)^{2}.
\]

By observing that $\mathcal{J}$ is symmetric with respect to the
second and the third arguments and the V-statistic decomposition argument
of \citet[5.7.3]{Serfling_Approximation_Theorems}, we have 
\begin{align*}
 & \frac{1}{L^{3}}\sum_{l=1}^{L}\sum_{k=1}^{L}\sum_{k'=1}^{L}\mathcal{L}\left(\left(\boldsymbol{B}_{\cdot l},\boldsymbol{X}_{l},N_{l}\right),\left(\boldsymbol{B}_{\cdot k},\boldsymbol{X}_{k},N_{k}\right),\left(\boldsymbol{B}_{\cdot k'},\boldsymbol{X}_{k'},N_{k'}\right);v\right)\\
= & \frac{1}{\left(L\right)_{3}}\sum_{\left(3\right)}\mathcal{L}\left(\left(\boldsymbol{B}_{\cdot l},\boldsymbol{X}_{l},N_{l}\right),\left(\boldsymbol{B}_{\cdot k},\boldsymbol{X}_{k},N_{k}\right),\left(\boldsymbol{B}_{\cdot k'},\boldsymbol{X}_{k'},N_{k'}\right);v\right)+\frac{O\left(L^{-1}\right)}{3L^{2}-2L}\left\{ \sum_{l=1}^{L}\mathcal{L}\left(\left(\boldsymbol{B}_{\cdot l},\boldsymbol{X}_{l},N_{l}\right),\left(\boldsymbol{B}_{\cdot l},\boldsymbol{X}_{l},N_{l}\right),\left(\boldsymbol{B}_{\cdot l},\boldsymbol{X}_{l},N_{l}\right);v\right)\right.\\
 & \left.+\sum_{\left(2\right)}\left(2\mathcal{L}\left(\left(\boldsymbol{B}_{\cdot l},\boldsymbol{X}_{l},N_{l}\right),\left(\boldsymbol{B}_{\cdot l},\boldsymbol{X}_{l},N_{l}\right),\left(\boldsymbol{B}_{\cdot k},\boldsymbol{X}_{k},N_{k}\right);v\right)+\mathcal{L}\left(\left(\boldsymbol{B}_{\cdot k},\boldsymbol{X}_{k},N_{k}\right),\left(\boldsymbol{B}_{\cdot l},\boldsymbol{X}_{l},N_{l}\right),\left(\boldsymbol{B}_{\cdot l},\boldsymbol{X}_{l},N_{l}\right);v\right)\right)\right\} .
\end{align*}
The Hoeffding decomposition yields
\begin{align*}
 & \frac{1}{\left(L\right)_{3}}\sum_{\left(3\right)}\mathcal{L}\left(\left(\boldsymbol{B}_{\cdot l},\boldsymbol{X}_{l},N_{l}\right),\left(\boldsymbol{B}_{\cdot k},\boldsymbol{X}_{k},N_{k}\right),\left(\boldsymbol{B}_{\cdot k'},\boldsymbol{X}_{k'},N_{k'}\right);v\right)\\
= & \frac{1}{\left(L\right)_{2}}\sum_{\left(2\right)}\mathcal{L}_{3}^{\left(2\right)}\left(\left(\boldsymbol{B}_{\cdot l},\boldsymbol{X}_{l},N_{l}\right),\left(\boldsymbol{B}_{\cdot k},\boldsymbol{X}_{k},N_{k}\right);v\right)\\
 & +\frac{1}{\left(L\right)_{3}}\sum_{\left(3\right)}\left\{ \mathcal{L}\left(\left(\boldsymbol{B}_{\cdot l},\boldsymbol{X}_{l},N_{l}\right),\left(\boldsymbol{B}_{\cdot k},\boldsymbol{X}_{k},N_{k}\right),\left(\boldsymbol{B}_{\cdot k'},\boldsymbol{X}_{k'},N_{k'}\right);v\right)-\mathcal{L}_{3}^{\left(2\right)}\left(\left(\boldsymbol{B}_{\cdot k},\boldsymbol{X}_{k},N_{k}\right),\left(\boldsymbol{B}_{\cdot k'},\boldsymbol{X}_{k'},N_{k'}\right);v\right)\right\} ,
\end{align*}
where the terms in the decomposition are defined by (\ref{eq:K Hoeffding term 1})
to (\ref{eq:K bound 7}) with $\mathcal{K}$ replaced by $\mathcal{L}$.
Note that it is easy to check that all other terms in the Hoeffding
decomposition vanish.

Denote
\[
F_{\mathcal{L}_{3}^{\left(2\right)}}\left(\boldsymbol{z},\boldsymbol{z}'\right)\coloneqq\int F_{\mathcal{L}}\left(\boldsymbol{z}'',\boldsymbol{z},\boldsymbol{z}'\right)\varphi\left(\boldsymbol{z}''\right)\mathrm{d}\boldsymbol{z}''.
\]
The CK inequality yields 
\[
\mathrm{E}\left[\underset{v\in I\left(\boldsymbol{x}\right)}{\mathrm{sup}}\left|\frac{1}{\left(L\right)_{2}}\sum_{\left(2\right)}\mathcal{L}_{3}^{\left(2\right)}\left(\left(\boldsymbol{B}_{\cdot l},\boldsymbol{X}_{l},N_{l}\right),\left(\boldsymbol{B}_{\cdot k},\boldsymbol{X}_{k},N_{k}\right);v\right)\right|\right]\leq L^{-1}\left(\mathrm{E}\left[F_{\mathcal{L}_{3}^{\left(2\right)}}\left(\boldsymbol{X}_{1},\boldsymbol{X}_{2}\right)^{2}\right]\right)^{\nicefrac{1}{2}}=O\left(\left(Lh^{6+2d}\right)^{-1}\right)
\]
and 
\begin{align*}
 & \mathrm{E}\left[\underset{v\in I\left(\boldsymbol{x}\right)}{\mathrm{sup}}\left|\frac{1}{\left(L\right)_{3}}\sum_{\left(3\right)}\left\{ \mathcal{L}\left(\left(\boldsymbol{B}_{\cdot l},\boldsymbol{X}_{l},N_{l}\right),\left(\boldsymbol{B}_{\cdot k},\boldsymbol{X}_{k},N_{k}\right),\left(\boldsymbol{B}_{\cdot k'},\boldsymbol{X}_{k'},N_{k'}\right);v\right)-\mathcal{L}_{3}^{\left(2\right)}\left(\left(\boldsymbol{B}_{\cdot k},\boldsymbol{X}_{k},N_{k}\right),\left(\boldsymbol{B}_{\cdot k'},\boldsymbol{X}_{k'},N_{k'}\right);v\right)\right\} \right|\right]\\
\leq & L^{-\nicefrac{3}{2}}\left(\mathrm{E}\left[F_{\mathcal{L}}\left(\boldsymbol{X}_{1},\boldsymbol{X}_{2},\boldsymbol{X}_{3}\right)^{2}\right]\right)^{\nicefrac{1}{2}}\\
= & L^{-\nicefrac{3}{2}}h^{-\left(6+\nicefrac{5d}{2}\right)},
\end{align*}
where the equalities follow from change of variables.

It is also straightforward to check that
\begin{align*}
 & \underset{v\in I\left(\boldsymbol{x}\right)}{\mathrm{sup}}\left|\frac{1}{L}\sum_{l=1}^{L}\mathcal{L}\left(\left(\boldsymbol{B}_{\cdot l},\boldsymbol{X}_{l},N_{l}\right),\left(\boldsymbol{B}_{\cdot l},\boldsymbol{X}_{l},N_{l}\right),\left(\boldsymbol{B}_{\cdot l},\boldsymbol{X}_{l},N_{l}\right);v\right)\right|\\
\leq & \frac{1}{L}\sum_{l=1}^{L}F_{\mathcal{M}^{\ddagger,n}}\left(\boldsymbol{X}_{l},\boldsymbol{X}_{l}\right)^{2}+\frac{2}{L}\sum_{l=1}^{L}F_{\mathcal{M}_{2}^{\ddagger,n}}\left(\boldsymbol{X}_{l}\right)F_{\mathcal{M}^{\ddagger,n}}\left(\boldsymbol{X}_{l},\boldsymbol{X}_{l}\right)+\frac{1}{L}\sum_{l=1}^{L}F_{\mathcal{M}_{2}^{\ddagger,n}}\left(\boldsymbol{X}_{l}\right)^{2}\\
= & O_{p}\left(h^{-\left(6+3d\right)}\right),
\end{align*}
\begin{align*}
 & \underset{v\in I\left(\boldsymbol{x}\right)}{\mathrm{sup}}\left|\frac{1}{\left(L\right)_{2}}\sum_{\left(2\right)}\mathcal{L}\left(\left(\boldsymbol{B}_{\cdot l},\boldsymbol{X}_{l},N_{l}\right),\left(\boldsymbol{B}_{\cdot l},\boldsymbol{X}_{l},N_{l}\right),\left(\boldsymbol{B}_{\cdot k},\boldsymbol{X}_{k},N_{k}\right);v\right)\right|\\
\leq & \frac{1}{\left(L\right)_{2}}\sum_{\left(2\right)}F_{\mathcal{M}^{\ddagger,n}}\left(\boldsymbol{X}_{k},\boldsymbol{X}_{l}\right)F_{\mathcal{M}^{\ddagger,n}}\left(\boldsymbol{X}_{l},\boldsymbol{X}_{l}\right)+\frac{1}{\left(L\right)_{2}}\sum_{\left(2\right)}F_{\mathcal{M}_{2}^{\ddagger,n}}\left(\boldsymbol{X}_{l}\right)F_{\mathcal{M}^{\ddagger,n}}\left(\boldsymbol{X}_{k},\boldsymbol{X}_{l}\right)\\
 & +\frac{1}{\left(L\right)_{2}}\sum_{\left(2\right)}F_{\mathcal{M}_{2}^{\ddagger,n}}\left(\boldsymbol{X}_{l}\right)F_{\mathcal{M}^{\ddagger,n}}\left(\boldsymbol{X}_{l},\boldsymbol{X}_{l}\right)+\frac{1}{L}\sum_{l=1}^{L}F_{\mathcal{M}_{2}^{\ddagger,n}}\left(\boldsymbol{X}_{l}\right)^{2}\\
= & O_{p}\left(h^{-\left(6+2d\right)}\right)
\end{align*}
and 
\begin{align*}
 & \underset{v\in I\left(\boldsymbol{x}\right)}{\mathrm{sup}}\left|\frac{1}{\left(L\right)_{2}}\sum_{\left(2\right)}\mathcal{L}\left(\left(\boldsymbol{B}_{\cdot k},\boldsymbol{X}_{k},N_{k}\right),\left(\boldsymbol{B}_{\cdot l},\boldsymbol{X}_{l},N_{l}\right),\left(\boldsymbol{B}_{\cdot l},\boldsymbol{X}_{l},N_{l}\right);v\right)\right|\\
\leq & \frac{1}{\left(L\right)_{2}}\sum_{\left(2\right)}F_{\mathcal{M}^{\ddagger,n}}\left(\boldsymbol{X}_{l},\boldsymbol{X}_{k}\right)^{2}+\frac{2}{\left(L\right)_{2}}\sum_{\left(2\right)}F_{\mathcal{M}_{2}^{\ddagger,n}}\left(\boldsymbol{X}_{k}\right)F_{\mathcal{M}^{\ddagger,n}}\left(\boldsymbol{X}_{l},\boldsymbol{X}_{k}\right)+\frac{1}{L}\sum_{l=1}^{L}F_{\mathcal{M}_{2}^{\ddagger,n}}\left(\boldsymbol{X}_{l}\right)^{2}\\
= & O_{p}\left(h^{-\left(6+2d\right)}\right),
\end{align*}
where the equalities follow from change of variables and Markov's
inequality. 

Now 
\begin{eqnarray}
\widehat{\sigma}_{\varDelta}^{2} & \coloneqq & \underset{v\in I\left(\boldsymbol{x}\right)}{\mathrm{sup}}\,\mathrm{E}^{*}\left[\left\{ \widehat{\mathcal{M}}_{2}^{\ddagger,n}\left(\boldsymbol{B}_{\cdot1}^{*},\boldsymbol{X}_{1}^{*},N_{1}^{*};v\right)-\mathcal{M}_{2}^{\ddagger,n}\left(\boldsymbol{B}_{\cdot1}^{*},\boldsymbol{X}_{1}^{*},N_{1}^{*};v\right)\right\} ^{2}\right]\nonumber \\
 & = & O_{p}\left(\left(Lh^{6+2d}\right)^{-1}\right)\label{eq:sigma_DELTA rate}
\end{eqnarray}
follows from these bounds and Markov's inequality.

Since $\left\{ \mathcal{M}^{\ddagger,n}\left(\cdot,\cdot;v\right):v\in I\left(\boldsymbol{x}\right)\right\} $
is uniformly VC-type with respect to the envelope (\ref{eq:F_M_ddagger}),
it follows from \citet[Lemma 5.4]{Chen_Kato_U_Process} that $\left\{ \widehat{\mathcal{M}}_{2}^{\ddagger,n}\left(\cdot;v\right):v\in I\left(\boldsymbol{x}\right)\right\} $
is uniformly VC-type with respect to the envelope
\[
F_{\widehat{\mathcal{M}}_{2}^{\ddagger,n}}\left(\boldsymbol{z}\right)\coloneqq\frac{\overline{\varphi}\left(\overline{C}_{D_{1}}+\overline{C}_{D_{2}}\right)\overline{C}_{K_{g}}}{\left(n-1\right)\underline{C}_{g}}\frac{1}{L}\sum_{k=1}^{L}\frac{1}{h^{3+2d}}\left|K_{\boldsymbol{X}}^{0}\left(\frac{\boldsymbol{X}_{k}-\boldsymbol{x}}{h}\right)\right|\left|K_{\boldsymbol{X}}\left(\frac{\boldsymbol{z}-\boldsymbol{X}_{k}}{h}\right)\right|,
\]
conditionally on the original sample. Now $\left\Vert F_{\widehat{\mathcal{M}}_{2}^{\ddagger,n}}\right\Vert _{\mathcal{X}}=O_{p}\left(h^{-\left(3+d\right)}\right)$
follows from change of variables and Markov's inequality. It follows
from \citet[Lemma 16]{nolan1987u} that $\left\{ \widehat{\mathcal{M}}_{2}^{\ddagger,n}\left(\cdot;v\right)-\mathcal{M}_{2}^{\ddagger,n}\left(\cdot;v\right):v\in I\left(\boldsymbol{x}\right)\right\} $
is uniformly VC-type with respect to $F_{\widehat{\mathcal{M}}_{2}^{\ddagger,n}}+F_{\mathcal{M}_{2}^{\ddagger,n}}$,
conditionally on the original sample. The CCK inequality yields 
\begin{eqnarray}
\mathrm{E}^{*}\left[\underset{v\in I\left(\boldsymbol{x}\right)}{\mathrm{sup}}\left|\varDelta^{**}\left(v\right)\right|\right] & \leq & C_{1}\left\{ L^{-\nicefrac{1}{2}}\widehat{\sigma}_{\varDelta}\mathrm{log}\left(C_{2}L\right)^{\nicefrac{1}{2}}+L^{-1}\left(\left\Vert F_{\widehat{\mathcal{M}}_{2}^{\ddagger,n}}\right\Vert _{\mathcal{X}}+\left\Vert F_{\mathcal{M}_{2}^{\ddagger,n}}\right\Vert _{\mathcal{X}}\right)\mathrm{log}\left(C_{2}L\right)\right\} \nonumber \\
 & = & O_{p}\left(\frac{\mathrm{log}\left(L\right)}{Lh^{3+d}}\right),\label{eq:Lemma S9 bound  3}
\end{eqnarray}
where the equality follows from (\ref{eq:sigma_DELTA rate}), $\left\Vert F_{\widehat{\mathcal{M}}_{2}^{\ddagger,n}}\right\Vert _{\mathcal{X}}=O_{p}\left(h^{-\left(3+d\right)}\right)$
and $\left\Vert F_{\mathcal{M}_{2}^{\ddagger,n}}\right\Vert _{\mathcal{X}}=O\left(h^{-\left(3+d\right)}\right)$. 

It is also straightforward to check that 
\begin{eqnarray*}
\underset{v\in I\left(\boldsymbol{x}\right)}{\mathrm{sup}}\left|\frac{1}{L^{2}}\sum_{l=1}^{L}\mathcal{M}^{n}\left(\left(\boldsymbol{B}_{\cdot l}^{*},\boldsymbol{X}_{l}^{*},N_{l}^{*}\right),\left(\boldsymbol{B}_{\cdot l}^{*},\boldsymbol{X}_{l}^{*},N_{l}^{*}\right);v\right)\right| & \apprle & \frac{1}{L^{2}}\sum_{l=1}^{L}\frac{1}{h^{3+2d}}\left|K_{\boldsymbol{X}}^{0}\left(\frac{\boldsymbol{X}_{l}^{*}-\boldsymbol{x}}{h}\right)\right|+\frac{1}{L^{2}}\sum_{l=1}^{L}\frac{1}{h^{2+d}}\left|K_{\boldsymbol{X}}^{0}\left(\frac{\boldsymbol{X}_{l}^{*}-\boldsymbol{x}}{h}\right)\right|\\
 & = & O_{p}^{*}\left(\left(Lh^{3+d}\right)^{-1}\right)
\end{eqnarray*}
and
\begin{align*}
 & \underset{v\in I\left(\boldsymbol{x}\right)}{\mathrm{sup}}\left|\frac{1}{L^{2}\left(L-1\right)}\sum_{l\neq k}\mathcal{M}^{n}\left(\left(\boldsymbol{B}_{\cdot l}^{*},\boldsymbol{X}_{l}^{*},N_{l}^{*}\right),\left(\boldsymbol{B}_{\cdot k}^{*},\boldsymbol{X}_{k}^{*},N_{k}^{*}\right);v\right)\right|\\
\apprle & \frac{1}{L^{2}\left(L-1\right)}\sum_{l\neq k}\frac{1}{h^{3+2d}}\left|K_{\boldsymbol{X}}^{0}\left(\frac{\boldsymbol{X}_{l}^{*}-\boldsymbol{x}}{h}\right)\right|\left|K_{\boldsymbol{X}}\left(\frac{\boldsymbol{X}_{k}^{*}-\boldsymbol{X}_{l}^{*}}{h}\right)\right|+\frac{1}{L^{2}}\sum_{l=1}^{L}\frac{1}{h^{2+d}}\left|K_{\boldsymbol{X}}^{0}\left(\frac{\boldsymbol{X}_{l}^{*}-\boldsymbol{x}}{h}\right)\right|\\
= & O_{p}^{*}\left(\left(Lh^{3}\right)^{-1}\right),
\end{align*}
where the equalities follow from change of variables, Markov's inequality
and Lemma \ref{lem:auxiliary 3}. 

The conclusion follows from these results, (\ref{eq:M_n_star decomposition}),
(\ref{eq:eq:Lemma S9 bound  0}), (\ref{eq:Lemma S9 bound  1}), (\ref{eq:Lemma S9 bound  2})
and (\ref{eq:Lemma S9 bound  3}). \end{proof}
\begin{lem}
\label{Lemma 10}Suppose Assumptions 1 - 3 hold. Then 
\[
\underset{v\in I\left(\boldsymbol{x}\right)}{\mathrm{sup}}\left|Z^{*}\left(v|\boldsymbol{x}\right)-\varGamma^{*}\left(v|\boldsymbol{x}\right)\right|=O_{p}\left(\mathrm{log}\left(L\right)^{\nicefrac{1}{2}}h+\frac{\mathrm{log}\left(L\right)}{\left(Lh^{3+d}\right)^{\nicefrac{1}{2}}}+L^{\nicefrac{1}{2}}h^{\nicefrac{\left(3+d\right)}{2}+R}\right).
\]
\end{lem}
\begin{proof}[Proof of Lemma \ref{Lemma 10}]It is straightforward
to verify that 
\begin{equation}
\mathcal{M}_{2}^{n}\left(B_{\cdot l}^{*},\boldsymbol{X}_{l}^{*},N_{l}^{*};v\right)-\widehat{\mu}_{\mathcal{M}_{2}^{n}}\left(v\right)=\mathcal{M}_{2}^{\ddagger,n}\left(B_{\cdot l}^{*},\boldsymbol{X}_{l}^{*},N_{l}^{*};v\right)-\widehat{\mu}_{\mathcal{M}_{2}^{\ddagger,n}}\left(v\right),\textrm{ for all \ensuremath{l=1,...,L}}.\label{eq:M_2_n - miu_hat_M_2_n =00003D M_2_ddagger_n - miu_M_2_ddagger_n}
\end{equation}
It follows from (\ref{eq:M_2_n - miu_hat_M_2_n =00003D M_2_ddagger_n - miu_M_2_ddagger_n})
that
\[
\frac{1}{L}\sum_{l=1}^{L}\mathcal{M}_{2}^{n}\left(\boldsymbol{B}_{\cdot l}^{*},\boldsymbol{X}_{l}^{*},N_{l}^{*};v\right)-\frac{1}{L}\sum_{l=1}^{L}\mathcal{M}_{2}^{n}\left(\boldsymbol{B}_{\cdot l},\boldsymbol{X}_{l},N_{l};v\right)=\frac{1}{L}\sum_{l=1}^{L}\mathcal{M}_{2}^{\ddagger,n}\left(\boldsymbol{B}_{\cdot l}^{*},\boldsymbol{X}_{l}^{*},N_{l}^{*};v\right)-\frac{1}{L}\sum_{l=1}^{L}\mathcal{M}_{2}^{\ddagger,n}\left(\boldsymbol{B}_{\cdot l},\boldsymbol{X}_{l},N_{l};v\right).
\]
Note that by the LIE and the fact that the bids in the bootstrap sample
are conditionally i.i.d., we have
\[
\mathrm{E}^{*}\left[\mathcal{M}_{2}^{\ddagger,n}\left(\boldsymbol{B}_{\cdot1}^{*},\boldsymbol{X}_{1}^{*},N_{1}^{*};v\right)\right]=\frac{1}{L}\sum_{l=1}^{L}\mathcal{M}_{2}^{\ddagger,n}\left(\boldsymbol{B}_{\cdot l},\boldsymbol{X}_{l},N_{l};v\right).
\]

Let 
\begin{eqnarray*}
\widehat{\sigma}_{\mathcal{M}_{2}^{\ddagger,n}}^{2} & \coloneqq & \underset{v\in I\left(\boldsymbol{x}\right)}{\mathrm{sup}}\mathrm{E}^{*}\left[\mathcal{M}_{2}^{\ddagger,n}\left(\boldsymbol{B}_{\cdot1}^{*},\boldsymbol{X}_{1}^{*},N_{1}^{*};v\right)^{2}\right]\\
 & \leq & h^{-\left(3+d\right)}\left\{ \underset{v\in I\left(\boldsymbol{x}\right)}{\mathrm{sup}}\mathrm{E}\left[h^{3+d}\mathcal{M}_{2}^{\ddagger,n}\left(\boldsymbol{B}_{\cdot1},\boldsymbol{X}_{1},N_{1};v\right)^{2}\right]\right.\\
 &  & \left.+\underset{v\in I\left(\boldsymbol{x}\right)}{\mathrm{sup}}\left|\frac{1}{L}\sum_{l=1}^{L}h^{3+d}\mathcal{M}_{2}^{\ddagger,n}\left(\boldsymbol{B}_{\cdot l},\boldsymbol{X}_{l},N_{l};v\right)^{2}-\mathrm{E}\left[h^{3+d}\mathcal{M}_{2}^{\ddagger,n}\left(\boldsymbol{B}_{\cdot1},\boldsymbol{X}_{1},N_{1};v\right)^{2}\right]\right|\right\} .
\end{eqnarray*}
We have
\[
\underset{v\in I\left(\boldsymbol{x}\right)}{\mathrm{sup}}\mathrm{E}\left[h^{3+d}\mathcal{M}_{2}^{\ddagger,n}\left(\boldsymbol{B}_{\cdot1},\boldsymbol{X}_{1},N_{1};v\right)^{2}\right]=O\left(1\right)
\]
since it was shown in the proof of Theorem 6.1 that 
\[
\underset{v\in I\left(\boldsymbol{x}\right)}{\mathrm{sup}}\left|\mathrm{E}\left[h^{3+d}\mathcal{M}_{2}^{\ddagger,n}\left(\boldsymbol{B}_{\cdot1},\boldsymbol{X}_{1},N_{1};v\right)^{2}\right]-\mathrm{V}_{GPV}\left(v|\boldsymbol{x},n\right)\right|\rightarrow0,\textrm{ as \ensuremath{L\uparrow\infty}}.
\]
It follows from \citet[Corollary A.1]{chernozhukov2014gaussian} that
$\left\{ h^{3+d}\mathcal{M}_{2}^{\ddagger,n}\left(\cdot;v\right)^{2}:v\in I\left(\boldsymbol{x}\right)\right\} $
is uniformly VC-type with respect to the envelope 
\[
\widetilde{F}_{\mathcal{M}_{2}^{\ddagger,n}}\left(\boldsymbol{z}\right)\coloneqq\left(\frac{\overline{\varphi}\left(\overline{C}_{D_{1}}+\overline{C}_{D_{2}}\right)\overline{C}_{K_{g}}}{\left(n-1\right)\underline{C}_{g}}\right)^{2}\frac{1}{h^{3+3d}}\left(\int\left|K_{\boldsymbol{X}}^{0}\left(\frac{\boldsymbol{z}'-\boldsymbol{x}}{h}\right)\right|\left|K_{\boldsymbol{X}}\left(\frac{\boldsymbol{z}'-\boldsymbol{z}}{h}\right)\right|\varphi\left(\boldsymbol{z}'\right)\mathrm{d}\boldsymbol{z}'\right)^{2}.
\]
It follows from change of variables that $\left\Vert \widetilde{F}_{\mathcal{M}_{2}^{\ddagger,n}}\right\Vert _{\mathcal{X}}=O\left(h^{-\left(3+d\right)}\right)$.
Let 
\[
\widetilde{\sigma}_{\mathcal{M}_{2}^{\ddagger,n}}^{2}\coloneqq\underset{v\in I\left(\boldsymbol{x}\right)}{\mathrm{sup}}\mathrm{E}\left[h^{2\left(3+d\right)}\mathcal{M}_{2}^{\ddagger,n}\left(\boldsymbol{B}_{\cdot1},\boldsymbol{X}_{1},N_{1};v\right)^{4}\right]=O\left(h^{-\left(1+d\right)}\right),
\]
where the second equality is shown in the proof of Theorem 6.4. Now
the CCK inequality yields
\begin{align*}
 & \mathrm{E}\left[\underset{v\in I\left(\boldsymbol{x}\right)}{\mathrm{sup}}\left|\frac{1}{L}\sum_{l=1}^{L}h^{3+d}\mathcal{M}_{2}^{\ddagger,n}\left(\boldsymbol{B}_{\cdot l},\boldsymbol{X}_{l},N_{l};v\right)^{2}-\mathrm{E}\left[h^{3+d}\mathcal{M}_{2}^{\ddagger,n}\left(\boldsymbol{B}_{\cdot1},\boldsymbol{X}_{1},N_{1};v\right)^{2}\right]\right|\right]\\
\leq & C_{1}\left\{ L^{-\nicefrac{1}{2}}\widetilde{\sigma}_{\mathcal{M}_{2}^{\ddagger,n}}\mathrm{log}\left(C_{2}L\right)^{\nicefrac{1}{2}}+L^{-1}\left\Vert \widetilde{F}_{\mathcal{M}_{2}^{\ddagger,n}}\right\Vert _{\mathcal{X}}\mathrm{log}\left(C_{2}L\right)\right\} \\
= & O\left(\left(\frac{\mathrm{log}\left(L\right)}{Lh^{1+d}}\right)^{\nicefrac{1}{2}}+\frac{\mathrm{log}\left(L\right)}{Lh^{3+d}}\right).
\end{align*}

Now it follows that $\widehat{\sigma}_{\mathcal{M}_{2}^{\ddagger,n}}^{2}=O_{p}\left(h^{-\left(3+d\right)}\right)$.
Since $\left\Vert F_{\mathcal{M}_{2}^{\ddagger,n}}\right\Vert _{\mathcal{X}}=O\left(h^{-\left(3+d\right)}\right)$,
the CCK inequality yields
\begin{align*}
 & \mathrm{E}^{*}\left[\underset{v\in I\left(\boldsymbol{x}\right)}{\mathrm{sup}}\left|\frac{1}{L}\sum_{l=1}^{L}\mathcal{M}_{2}^{\ddagger,n}\left(\boldsymbol{B}_{\cdot l}^{*},\boldsymbol{X}_{l}^{*},N_{l}^{*};v\right)-\frac{1}{L}\sum_{l=1}^{L}\mathcal{M}_{2}^{\ddagger,n}\left(\boldsymbol{B}_{\cdot l},\boldsymbol{X}_{l},N_{l};v\right)\right|\right]\\
\leq & C_{1}\left\{ L^{-\nicefrac{1}{2}}\widehat{\sigma}_{\mathcal{M}_{2}^{\ddagger,n}}\mathrm{log}\left(C_{2}L\right)^{\nicefrac{1}{2}}+L^{-1}\left\Vert F_{\mathcal{M}_{2}^{\ddagger,n}}\right\Vert _{\mathcal{X}}\mathrm{log}\left(C_{2}L\right)\right\} \\
= & O_{p}\left(\left(\frac{\mathrm{log}\left(L\right)}{Lh^{3+d}}\right)^{\nicefrac{1}{2}}\right).
\end{align*}

The conclusion follows from the above result, Lemmas \ref{Lemma 2},
\ref{Lemma 5}, \ref{Lemma 9}, Theorem 6.2, Lemma \ref{lem:auxiliary 3},
the decomposition (\ref{eq:f_GPV_hat_star - f_GPV_hat}) and \citet[Lemma S.1]{Marmer_Shneyerov_Quantile_Auctions}.\end{proof}

\section{Justification of Remark 3.4 and Remark 6.1}

To prove the result stated in Remark 3.4, note that
\begin{eqnarray*}
\frac{\mathrm{V}_{GPV}\left(v\right)}{\mathrm{V}_{QB}\left(v\right)}=\frac{\int\left\{ \int K'\left(u\right)K\left(w-s'\left(v\right)u\right)\mathrm{d}u\right\} ^{2}\mathrm{d}w}{s'\left(v\right)^{2}\int K'\left(u\right)^{2}\mathrm{d}u} & = & \frac{\int\left\{ \int K\left(u\right)K'\left(w-s'\left(v\right)u\right)\mathrm{d}u\right\} ^{2}\mathrm{d}w}{\int K'\left(u\right)^{2}\mathrm{d}u}\\
 & \leq & \frac{\int\int K\left(u\right)K'\left(w-s'\left(v\right)u\right)^{2}\mathrm{d}u\mathrm{d}w}{\int K'\left(u\right)^{2}\mathrm{d}u},
\end{eqnarray*}
where the second equality follows from integration by parts and the
inequality follows from Jensen's inequality. It follows easily from
the Fubini-Tonelli theorem and the fact $\int K\left(u\right)\mathrm{d}u=1$
that 
\[
\int\int K\left(u\right)K'\left(w-s'\left(v\right)u\right)^{2}\mathrm{d}u\mathrm{d}w=\int K\left(u\right)\left\{ \int K'\left(w-s'\left(v\right)u\right)^{2}\mathrm{d}w\right\} \mathrm{d}u=\int K'\left(w\right)^{2}\mathrm{d}w.
\]
Now it follows that $\mathrm{V}_{GPV}\left(v\right)\leq\mathrm{V}_{QB}\left(v\right)$. 

Next, we proceed to the general case with a random number of bidders
and auction-specific heterogeneity. We provide details about the proof
of the result stated in Remark 6.1. \citet*{Marmer_Shneyerov_Quantile_Auctions}
showed that the quantile-based estimator $\widehat{f}_{QB}\left(v|\boldsymbol{x},n\right)$
is asymptotically normal:
\[
\left(Lh^{3+d}\right)^{\nicefrac{1}{2}}\left(\widehat{f}_{QB}\left(v|\boldsymbol{x},n\right)-f\left(v|\boldsymbol{x}\right)\right)\rightarrow_{d}\mathrm{N}\left(0,\mathrm{V}_{QB}\left(v|\boldsymbol{x},n\right)\right),
\]
where
\begin{align*}
\mathrm{V}_{QB}\left(v|\boldsymbol{x},n\right)\coloneqq & \frac{1}{n\left(n-1\right)^{2}}\frac{F\left(v|\boldsymbol{x}\right)^{2}f\left(v|\boldsymbol{x}\right)^{4}}{\pi\left(n|\boldsymbol{x}\right)\varphi\left(\boldsymbol{x}\right)g\left(s\left(v,\boldsymbol{x},n\right)|\boldsymbol{x},n\right)^{5}}\left(\int K\left(u\right)^{2}\mathrm{d}u\right)^{d}\int K'\left(u\right)^{2}\mathrm{d}u
\end{align*}
and $K$ is a kernel function. See \citet*[Theorem 2]{Marmer_Shneyerov_Quantile_Auctions}.

Now instead of Assumption 2(b), we assume $K_{0}=K_{1}=K$, for some
common second-order kernel function $K$ that is supported on $\left[-1,1\right]$
and used in both steps. Denote 
\[
K_{d}\left(\boldsymbol{x}\right)\coloneqq\prod_{k=1}^{d}K\left(x_{k}\right),\textrm{ for }\boldsymbol{x}=\left(x_{1},...,x_{d}\right)\in\mathbb{R}^{d}.
\]
Now the ratio between $\mathrm{V}_{QB}\left(v|\boldsymbol{x},n\right)$
and $\mathrm{V}_{GPV}\left(v|\boldsymbol{x},n\right)$ is given by
\begin{eqnarray}
\frac{\mathrm{V}_{QB}\left(v|\boldsymbol{x},n\right)}{\mathrm{V}_{GPV}\left(v|\boldsymbol{x},n\right)} & = & \frac{f\left(v|\boldsymbol{x}\right)^{2}}{g\left(s\left(v,\boldsymbol{x},n\right)|\boldsymbol{x},n\right)^{2}}\frac{\left(\int K\left(u\right)^{2}\mathrm{d}u\right)^{d}\int K'\left(u\right)^{2}\mathrm{d}u}{\int\int\left\{ \int\int K_{d}\left(\boldsymbol{y}\right)K'\left(w\right)K_{d}\left(\boldsymbol{y}-\boldsymbol{z}\right)K\left(u-s_{v}w-s_{\boldsymbol{x}}^{\mathrm{T}}\boldsymbol{y}\right)\mathrm{d}w\mathrm{d}\boldsymbol{y}\right\} ^{2}\mathrm{d}u\mathrm{d}\boldsymbol{z}}\nonumber \\
 & = & \frac{s_{v}^{2}\left(\int K\left(u\right)^{2}\mathrm{d}u\right)^{d}\int K'\left(u\right)^{2}\mathrm{d}u}{\int\int\left\{ \int\int K_{d}\left(\boldsymbol{y}\right)K'\left(w\right)K_{d}\left(\boldsymbol{y}-\boldsymbol{z}\right)K\left(u-s_{v}w-s_{\boldsymbol{x}}^{\mathrm{T}}\boldsymbol{y}\right)\mathrm{d}w\mathrm{d}\boldsymbol{y}\right\} ^{2}\mathrm{d}u\mathrm{d}\boldsymbol{z}}.\label{eq:variance ratio}
\end{eqnarray}

Since the kernel is supported on $\left[-1,1\right]$. By integration
by parts, we have 
\[
\int_{-1}^{1}K'\left(w\right)K\left(u-s_{v}w-s_{\boldsymbol{x}}^{\mathrm{T}}\boldsymbol{y}\right)\mathrm{d}w+\int_{-1}^{1}K\left(w\right)K'\left(u-s_{v}w-s_{\boldsymbol{x}}^{\mathrm{T}}\boldsymbol{y}\right)\left(-s_{v}\right)\mathrm{d}w=0
\]
for any $w\in\mathbb{R}$. We now have 
\begin{align}
 & \int\int\left\{ \int\int K_{d}\left(\boldsymbol{y}\right)K'\left(w\right)K_{d}\left(\boldsymbol{y}-\boldsymbol{z}\right)K\left(u-s_{v}w-s_{\boldsymbol{x}}^{\mathrm{T}}\boldsymbol{y}\right)\mathrm{d}w\mathrm{d}\boldsymbol{y}\right\} ^{2}\mathrm{d}u\mathrm{d}\boldsymbol{z}\nonumber \\
= & s_{v}^{2}\int\int\left\{ \int\int K_{d}\left(\boldsymbol{y}\right)K\left(w\right)K'\left(u-s_{v}w-s_{\boldsymbol{x}}^{\mathrm{T}}\boldsymbol{y}\right)K_{d}\left(\boldsymbol{y}-\boldsymbol{z}\right)\mathrm{d}w\boldsymbol{\mathrm{d}y}\right\} ^{2}\mathrm{d}u\mathrm{d}\boldsymbol{z}\nonumber \\
\leq & s_{v}^{2}\int\int\int\int K_{d}\left(\boldsymbol{y}\right)K\left(w\right)K'\left(u-s_{v}w-s_{\boldsymbol{x}}^{\mathrm{T}}\boldsymbol{y}\right)^{2}K_{d}\left(\boldsymbol{y}-\boldsymbol{z}\right)^{2}\mathrm{d}w\mathrm{d}\boldsymbol{y}\mathrm{d}u\mathrm{d}\boldsymbol{z},\label{eq:variance ratio inequality}
\end{align}
where the inequality follows from Jensen's inequality under the assumption
that $K$ is second-order (i.e., $K$ is a probability density function).
It is clear that the integrand of the multiple integral on the third
line of (\ref{eq:variance ratio inequality}) is supported on a compact
set. Since $K$ and $K'$ are continuous on $\mathbb{R}$, the Fubini-Tonelli
Theorem applies and it gives that
\begin{align}
 & \int\int\int\int K_{d}\left(\boldsymbol{y}\right)K\left(w\right)K'\left(u-s_{v}w-s_{\boldsymbol{x}}^{\mathrm{T}}\boldsymbol{y}\right)^{2}K_{d}\left(\boldsymbol{y}-\boldsymbol{z}\right)^{2}\mathrm{d}w\mathrm{d}\boldsymbol{y}\mathrm{d}u\mathrm{d}\boldsymbol{z}\nonumber \\
= & \int\int K_{d}\left(\boldsymbol{y}\right)K\left(w\right)\left(\int K'\left(u-s_{v}w-s_{\boldsymbol{x}}^{\mathrm{T}}\boldsymbol{y}\right)^{2}\mathrm{d}u\right)\left(\int K_{d}\left(\boldsymbol{y}-\boldsymbol{z}\right)^{2}\mathrm{d}\boldsymbol{z}\right)\mathrm{d}w\mathrm{d}\boldsymbol{y}.\label{eq:variance ratio fubini}
\end{align}

For any $\left|w\right|\leq1$, $\boldsymbol{y}\in\mathbb{H}\left(\boldsymbol{0},1\right)$,
by a simple change of variable argument, we have
\begin{equation}
\int K'\left(u-s_{v}w-s_{\boldsymbol{x}}^{\mathrm{T}}\boldsymbol{y}\right)^{2}\mathrm{d}u=\int K'\left(u\right)^{2}\mathrm{d}u\label{eq:variance ratio change of variable 1}
\end{equation}
and similarly, for each $\boldsymbol{y}\in\mathbb{H}\left(\boldsymbol{0},1\right)$,
\begin{equation}
\int K_{d}\left(\boldsymbol{y}-\boldsymbol{z}\right)^{2}\mathrm{d}\boldsymbol{z}=\left(\int K\left(u\right)^{2}\mathrm{d}u\right)^{d}.\label{eq:variance ratio change of variable 2}
\end{equation}

It follows from (\ref{eq:variance ratio inequality}) to (\ref{eq:variance ratio change of variable 2})
that
\[
\int\int\left\{ \int\int K_{d}\left(y\right)K'\left(w\right)K_{d}\left(\boldsymbol{y}-\boldsymbol{z}\right)K\left(u-s_{v}w-s_{\boldsymbol{x}}^{\mathrm{T}}\boldsymbol{y}\right)\mathrm{d}w\mathrm{d}\boldsymbol{y}\right\} ^{2}\mathrm{d}u\mathrm{d}\boldsymbol{z}\leq s_{v}^{2}\left(\int K'\left(u\right)^{2}\mathrm{d}u\right)\left(\int K\left(u\right)^{2}\mathrm{d}u\right)^{d}.
\]
It now follows from the above result and (\ref{eq:variance ratio})
that
\[
\frac{\mathrm{V}_{GPV}\left(v|\boldsymbol{x},n\right)}{\mathrm{V}_{QB}\left(v|\boldsymbol{x},n\right)}\leq1.
\]

\section{Additional Monte Carlo Simulations\label{sec:Monte-Carlo-Simulations supp}}

In this section, we report the simulation results for the case with
auction-specific heterogeneity captured by a single covariate. To
allow for covariates, we extend the design of \citet{Marmer_Shneyerov_Quantile_Auctions}.
We consider a covariate $X$, which is a truncated normal random variable
truncated between $0$ and $2$, with the mean equal to $1$, and
some variance $\sigma^{2}$. The PDF of $X$ is given by:
\[
\varphi\left(x\right)=\frac{\phi\left(\frac{x-1}{\sigma}\right)}{\sigma\left\{ \Phi\left(\frac{2-1}{\sigma}\right)-\Phi\left(\frac{0-1}{\sigma}\right)\right\} }\mathbbm{1}\left(0\leq x\leq2\right),
\]
where $\phi$ is the standard normal PDF, and $\Phi$ is the standard
normal CDF. 

Given $X$, the valuations are drawn from the CDF $F(\cdot|X)$:
\[
F(v|X)\coloneqq\begin{cases}
0, & v<0\\
v^{X}, & 0\leq v\leq1\\
1, & v\geq1.
\end{cases}
\]
In an auction with $N$ bidders, the Bayesian Nash equilibrium bidding
strategy is now given by
\[
s\left(u|X\right)=\left(1-\frac{1}{X(N-1)+1}\right)u.
\]

We choose the triweight kernel when implementing the two-step estimator.
We use the second-order triweight kernel in the second step and used
the fourth-order triweight kernel in the first step. 

We again follow Silverman's rule-of-thumb approach when selecting
the bandwidths. We use $h_{g}=3.72\cdot\widehat{\sigma}_{b}\cdot\left(N\cdot L\right)^{-\nicefrac{1}{6}}$
and $h_{X,1}=3.72\cdot\widehat{\sigma}_{X}\cdot\left(N\cdot L\right)^{-\nicefrac{1}{6}}$
as the first-step bandwidths, where $\widehat{\sigma}_{X}$ is the
estimated standard deviation of the observed covariates. We use $h_{f}=3.15\cdot\widehat{\sigma}_{v}\cdot\left(\left(N\cdot L\right)_{\mathbb{T}}\right)^{-\nicefrac{1}{5}}$
and $h_{X,2}=3.15\cdot\widehat{\sigma}_{X}\cdot\left(N\cdot L\right)^{-\nicefrac{1}{6}}$
as the second-step bandwidths. The rate is changed to $-\nicefrac{1}{6}$
because now there is also smoothing over the values of the covariate.
. For Nadaraya-Watson estimation of $\varphi\left(x\right)$ in the
second step, we use the fourth-order triweight kernel and a rule-of-thumb
bandwidth selection rule: $h_{X,3}=3.72\cdot\widehat{\sigma}_{X}\cdot\left(N\cdot L\right)^{-\nicefrac{1}{5}}$. 

The results are reported in Table \ref{tab:Coverage-probabilities-covariates}.
Under auction-specific heterogeneity, the coverage accuracy of the
IGA uniform confidence bands is less precise than that in the case
with no covariates, especially for the wider range of values $v\in[0.2,0.8]$.
We believe this is due to smaller effective sample sizes, which arise
because of smoothing over the covariate values. Our conjecture is
that increasing the number of auctions $L$ would restore the accuracy.

\begin{table}
\caption{Coverage probabilities of the Intermediate Gaussian Approximation
uniform confidence band $CB^{*}$ in presence of auction-specific
heterogeneity, for the number of bidders $N=3,5,7$, the distribution
parameter $\sigma=0.8,1,1.2$, different ranges of values $v$, and
the nominal coverage probability $=0.90,0.95,0.99$. The number of
auctions $L$ is determined by $N\cdot L=2100$\label{tab:Coverage-probabilities-covariates}}

$\,$
\centering{}%
\begin{tabular}{cc>{\centering}p{2cm}>{\centering}p{2cm}>{\centering}p{2cm}}
\toprule 
Range of $v$ & $N$ & 0.90 & 0.95 & 0.99\tabularnewline
\midrule
 &  & \multicolumn{3}{c}{$\sigma=0.8$}\tabularnewline
$\left[0.3,0.7\right]$ & $3$ & 0.818 & 0.888 & 0.958\tabularnewline
 & $5$ & 0.804 & 0.892 & 0.962\tabularnewline
 & $7$ & 0.816 & 0.892 & 0.964\tabularnewline
$\left[0.2,0.8\right]$ & $3$ & 0.642 & 0.762 & 0.924\tabularnewline
 & $5$ & 0.768 & 0.864 & 0.930\tabularnewline
 & $7$ & 0.798 & 0.862 & 0.952\tabularnewline
\midrule
 &  & \multicolumn{3}{c}{$\sigma=1$}\tabularnewline
$\left[0.3,0.7\right]$ & $3$ & 0.798 & 0.892 & 0.966\tabularnewline
 & $5$ & 0.810 & 0.880 & 0.948\tabularnewline
 & $7$ & 0.818 & 0.898 & 0.952\tabularnewline
$\left[0.2,0.8\right]$ & $3$ & 0.694 & 0.806 & 0.928\tabularnewline
 & $5$ & 0.764 & 0.860 & 0.932\tabularnewline
 & $7$ & 0.832 & 0.880 & 0.922\tabularnewline
\midrule
 &  & \multicolumn{3}{c}{$\sigma=1.2$}\tabularnewline
$\left[0.3,0.7\right]$ & $3$ & 0.862 & 0.916 & 0.968\tabularnewline
 & $5$ & 0.826 & 0.896 & 0.942\tabularnewline
 & $7$ & 0.834 & 0.888 & 0.952\tabularnewline
$\left[0.2,0.8\right]$ & $3$ & 0.700 & 0.818 & 0.936\tabularnewline
 & $5$ & 0.804 & 0.878 & 0.934\tabularnewline
 & 7 & 0.812 & 0.886 & 0.942\tabularnewline
\bottomrule
\end{tabular}
\end{table}


%% file: manuscript_with_supplement_for_arXiv.bbl
\begin{thebibliography}{63}
\expandafter\ifx\csname natexlab\endcsname\relax\def\natexlab#1{#1}\fi
\expandafter\ifx\csname url\endcsname\relax
  \def\url#1{\texttt{#1}}\fi
\expandafter\ifx\csname urlprefix\endcsname\relax\def\urlprefix{URL }\fi

\bibitem[{Armstrong and Koles{\'a}r(2016)}]{armstrong2016simple}
Armstrong, T.~B., Koles{\'a}r, M., 2016. Simple and honest confidence intervals
  in nonparametric regression, working paper, Yale University.

\bibitem[{Athey and Haile(2002)}]{athey2002identification}
Athey, S., Haile, P.~A., 2002. Identification of standard auction models.
  Econometrica 70~(6), 2107--2140.

\bibitem[{Athey and Haile(2007)}]{athey2007nonparametric}
Athey, S., Haile, P.~A., 2007. Nonparametric approaches to auctions. Handbook
  of Econometrics 6, 3847--3965.

\bibitem[{Augenblick(2015)}]{augenblick2015sunk}
Augenblick, N., 2015. The sunk-cost fallacy in penny auctions. Review of
  Economic Studies 83~(1), 58--86.

\bibitem[{Bickel and Rosenblatt(1973)}]{bickel1973some}
Bickel, P.~J., Rosenblatt, M., 1973. On some global measures of the deviations
  of density function estimates. Annals of Statistics 1~(6), 1071--1095.

\bibitem[{Calonico et~al.(2014)Calonico, Cattaneo, and
  Titiunik}]{calonico2014robust}
Calonico, S., Cattaneo, M.~D., Titiunik, R., 2014. Robust nonparametric
  confidence intervals for regression-discontinuity designs. Econometrica
  82~(6), 2295--2326.

\bibitem[{Chen and Kato(2017)}]{Chen_Kato_U_Process}
Chen, X., Kato, K., 2017. Jackknife multiplier bootstrap: Finite sample
  approximations to the {U}-process supremum with applications, working paper,
  University of Illinois at Urbana-Champaign.

\bibitem[{Chernozhukov et~al.(2014{\natexlab{a}})Chernozhukov, Chetverikov, and
  Kato}]{chernozhukov2014anti}
Chernozhukov, V., Chetverikov, D., Kato, K., 2014{\natexlab{a}}.
  Anti-concentration and honest, adaptive confidence bands. Annals of
  Statistics 42~(5), 1787--1818.

\bibitem[{Chernozhukov et~al.(2014{\natexlab{b}})Chernozhukov, Chetverikov, and
  Kato}]{chernozhukov2014gaussian}
Chernozhukov, V., Chetverikov, D., Kato, K., 2014{\natexlab{b}}. Gaussian
  approximation of suprema of empirical processes. Annals of Statistics 42~(4),
  1564--1597.

\bibitem[{Chernozhukov et~al.(2016)Chernozhukov, Chetverikov, and
  Kato}]{chernozhukov2016empirical}
Chernozhukov, V., Chetverikov, D., Kato, K., 2016. Empirical and multiplier
  bootstraps for suprema of empirical processes of increasing complexity, and
  related {G}aussian couplings. Stochastic Processes and their Applications
  126~(12), 3632--3651.

\bibitem[{de~Castro and Paarsch(2010)}]{de2010testing}
de~Castro, L.~I., Paarsch, H.~J., 2010. Testing affiliation in private-values
  models of first-price auctions using grid distributions. Annals of Applied
  Statistics, 2073--2098.

\bibitem[{Dudley(2002)}]{dudley2002real}
Dudley, R.~M., 2002. Real Analysis and Probability. Cambridge University Press.

\bibitem[{Fang and Tang(2014)}]{fang2014inference}
Fang, H., Tang, X., 2014. Inference of bidders' risk attitudes in ascending
  auctions with endogenous entry. Journal of Econometrics 180~(2), 198--216.

\bibitem[{Flambard and Perrigne(2006)}]{flambard2006asymmetry}
Flambard, V., Perrigne, I., 2006. Asymmetry in procurement auctions: evidence
  from snow removal contracts. Economic Journal 116~(514), 1014--1036.

\bibitem[{Folland(1999)}]{Folland_text_book}
Folland, G.~B., 1999. Real Analysis: Modern Techniques and Their Applications,
  2nd Edition. John Wiley \& Sons.

\bibitem[{Gentry and Li(2014)}]{gentry2014identification}
Gentry, M., Li, T., 2014. Identification in auctions with selective entry.
  Econometrica 82~(1), 315--344.

\bibitem[{Ghosal et~al.(2000)Ghosal, Sen, and van~der
  Vaart}]{ghosal2000testing}
Ghosal, S., Sen, A., van~der Vaart, A.~W., 2000. Testing monotonicity of
  regression. Annals of statistics 28~(4), 1054--1082.

\bibitem[{Gimenes(2017)}]{gimenes2017econometrics}
Gimenes, N., 2017. Econometrics of ascending auctions by quantile regression.
  Review of Economics and Statistics 99~(5), 944--953.

\bibitem[{Gimenes and Guerre(2016)}]{gimenes2016quantile}
Gimenes, N., Guerre, E., 2016. Quantile methods for first-price auction: A
  signal approach, working paper, PUC-Rio.

\bibitem[{Gin{\'e} and Nickl(2015)}]{gine2015mathematical}
Gin{\'e}, E., Nickl, R., 2015. Mathematical Foundations of Infinite-dimensional
  Statistical Models. Cambridge University Press.

\bibitem[{Guerre et~al.(1995)Guerre, Perrigne, and Vuong}]{GPV1995}
Guerre, E., Perrigne, I., Vuong, Q., 1995. Nonparametric estimation of
  first-price auctions, working paper.

\bibitem[{Guerre et~al.(2000)Guerre, Perrigne, and
  Vuong}]{Guerre_Perrigne_Vuong_Auction_Econometrica_2000}
Guerre, E., Perrigne, I., Vuong, Q., 2000. Optimal nonparametric estimation of
  first-price auctions. Econometrica 68~(3), 525--574.

\bibitem[{Guerre et~al.(2009)Guerre, Perrigne, and
  Vuong}]{guerre2009nonparametric}
Guerre, E., Perrigne, I., Vuong, Q., 2009. Nonparametric identification of risk
  aversion in first-price auctions under exclusion restrictions. Econometrica
  77~(4), 1193--1227.

\bibitem[{Haile et~al.(2003)Haile, Hong, and Shum}]{haile2003nonparametric}
Haile, P.~A., Hong, H., Shum, M., 2003. Nonparametric tests for common values
  at first-price sealed-bid auctions, {NBER} Working paper No.10105.

\bibitem[{Haile and Tamer(2003)}]{haile2003iim}
Haile, P.~A., Tamer, E., 2003. Inference with an incomplete model of english
  auctions. Journal of Political Economy 111~(1), 1--51.

\bibitem[{Henderson et~al.(2012)Henderson, List, Millimet, Parmeter, and
  Price}]{henderson_2012_JOE}
Henderson, D.~J., List, J.~A., Millimet, D.~L., Parmeter, C.~F., Price, M.~K.,
  2012. Empirical implementation of nonparametric first-price auction models.
  Journal of Econometrics 168~(1), 17--28.

\bibitem[{Hendricks and Porter(2007)}]{hendricks2007empirical}
Hendricks, K., Porter, R.~H., 2007. An empirical perspective on auctions.
  Handbook of Industrial Organization 3, 2073--2143.

\bibitem[{Hickman and Hubbard(2014)}]{hickman_hubbard_2014_JAE}
Hickman, B.~R., Hubbard, T.~P., 2014. Replacing sample trimming with boundary
  correction in nonparametric estimation of first-price auctions. Journal of
  Applied Econometrics 30, 739--762.

\bibitem[{Hickman et~al.(2012)Hickman, Hubbard, and
  Sa{\u{g}}lam}]{hickman2012structural}
Hickman, B.~R., Hubbard, T.~P., Sa{\u{g}}lam, Y., 2012. Structural econometric
  methods in auctions: A guide to the literature. Journal of Econometric
  Methods 1~(1), 67--106.

\bibitem[{Hill and Shneyerov(2013)}]{hill2013there}
Hill, J.~B., Shneyerov, A., 2013. Are there common values in first-price
  auctions? a tail-index nonparametric test. Journal of Econometrics 174~(2),
  144--164.

\bibitem[{Hu et~al.(2013)Hu, McAdams, and Shum}]{hu2013identification}
Hu, Y., McAdams, D., Shum, M., 2013. Identification of first-price auctions
  with non-separable unobserved heterogeneity. Journal of Econometrics 174~(2),
  186--193.

\bibitem[{Hubbard et~al.(2012)Hubbard, Li, and
  Paarsch}]{hubbard2012semiparametric}
Hubbard, T.~P., Li, T., Paarsch, H.~J., 2012. Semiparametric estimation in
  models of first-price, sealed-bid auctions with affiliation. Journal of
  Econometrics 168~(1), 4--16.

\bibitem[{Jun et~al.(2010)Jun, Pinkse, and Wan}]{jun2010consistent}
Jun, S.~J., Pinkse, J., Wan, Y., 2010. A consistent nonparametric test of
  affiliation in auction models. Journal of Econometrics 159~(1), 46--54.

\bibitem[{Kato and Sasaki(2017)}]{Kato_Sasaki_2}
Kato, K., Sasaki, Y., 2017. Uniform confidence bands for nonparametric
  errors-in-variables regression, working paper, University of Tokyo.

\bibitem[{Kato and Sasaki(2018)}]{Kato_Sasaki_1}
Kato, K., Sasaki, Y., 2018. Uniform confidence bands in deconvolution with
  unknown error distribution. Journal of Econometrics 207~(1), 129--169.

\bibitem[{Kawai and Nakabayashi(2015)}]{kawai2014detecting}
Kawai, K., Nakabayashi, J., 2015. Detecting large-scale collusion in
  procurement auctions, working paper, University of California, Berkeley.

\bibitem[{Kosorok(2007)}]{kosorok2007introduction}
Kosorok, M.~R., 2007. Introduction to Empirical Processes and Semiparametric
  Inference. Springer.

\bibitem[{Krasnokutskaya(2011)}]{krasnokutskaya2011identification}
Krasnokutskaya, E., 2011. Identification and estimation of auction models with
  unobserved heterogeneity. Review of Economic Studies 78~(1), 293--327.

\bibitem[{Krasnokutskaya and Seim(2011)}]{krasnokutskaya2011bid}
Krasnokutskaya, E., Seim, K., 2011. Bid preference programs and participation
  in highway procurement auctions. American Economic Review 101~(6),
  2653--2686.

\bibitem[{Lahiri(2013)}]{Lahiri_Dependent_Bootstrap}
Lahiri, S.~N., 2013. Resampling Methods for Dependent Data. Springer.

\bibitem[{Li and Racine(2007)}]{li2007net}
Li, Q., Racine, J.~S., 2007. Nonparametric Econometrics: Theory and Practice.
  Princeton University Press, Princeton, New Jersey.

\bibitem[{Li et~al.(2002)Li, Perrigne, and Vuong}]{li2002structural}
Li, T., Perrigne, I., Vuong, Q., 2002. Structural estimation of the affiliated
  private value auction model. RAND Journal of Economics, 171--193.

\bibitem[{Li et~al.(2003)Li, Perrigne, and Vuong}]{li2003semiparametric}
Li, T., Perrigne, I., Vuong, Q., 2003. Semiparametric estimation of the optimal
  reserve price in first-price auctions. Journal of Business \& Economic
  Statistics 21~(1), 53--64.

\bibitem[{Li and Zhang(2010)}]{li2010testing}
Li, T., Zhang, B., 2010. Testing for affiliation in first-price auctions using
  entry behavior. International Economic Review 51~(3), 837--850.

\bibitem[{Li and Zheng(2009)}]{li2009entry}
Li, T., Zheng, X., 2009. Entry and competition effects in first-price auctions:
  theory and evidence from procurement auctions. Review of Economic Studies
  76~(4), 1397--1429.

\bibitem[{Liu and Luo(2017)}]{liu2017nonparametric}
Liu, N., Luo, Y., 2017. A nonparametric test for comparing valuation
  distributions in first-price auctions. International Economic Review 58~(3),
  857--888.

\bibitem[{Liu and Vuong(2013)}]{liuvuong2013}
Liu, N., Vuong, Q., 2013. Nonparametric test of monotonicity of bidding
  strategy in first price auctions, working paper, Shanghai University of
  Finance and Economics.

\bibitem[{Luo and Wan(2018)}]{luo2017integrated}
Luo, Y., Wan, Y., 2018. Integrated-quantile-based estimation for first-price
  auction models. Journal of Business \& Economic Statistics 36~(1), 173--180.

\bibitem[{Mammen(1992)}]{Mammen_Bootstrap_and_Normality_1992}
Mammen, E., 1992. Bootstrap, wild bootstrap, and asymptotic normality.
  Probability Theory and Related Fields 93~(4), 439--455.

\bibitem[{Mammen et~al.(2012)Mammen, Rothe, Schienle,
  et~al.}]{mammen2012nonparametric}
Mammen, E., Rothe, C., Schienle, M., et~al., 2012. Nonparametric regression
  with nonparametrically generated covariates. Annals of Statistics 40~(2),
  1132--1170.

\bibitem[{Marmer and Shneyerov(2012)}]{Marmer_Shneyerov_Quantile_Auctions}
Marmer, V., Shneyerov, A., 2012. Quantile-based nonparametric inference for
  first-price auctions. Journal of Econometrics 167~(2), 345--357.

\bibitem[{Marmer et~al.(2013)Marmer, Shneyerov, and Xu}]{marmer2013model}
Marmer, V., Shneyerov, A., Xu, P., 2013. What model for entry in first-price
  auctions? {A} nonparametric approach. Journal of Econometrics 176~(1),
  46--58.

\bibitem[{Newey(1994)}]{Newey_Kernel_ET_1994}
Newey, W.~K., 1994. Kernel estimation of partial means and a general variance
  estimator. Econometric Theory 10~(2), 1--21.

\bibitem[{Nolan and Pollard(1987)}]{nolan1987u}
Nolan, D., Pollard, D., 1987. U-processes: rates of convergence. Annals of
  Statistics 15~(2), 780--799.

\bibitem[{Paarsch(1997)}]{paarsch1997deriving}
Paarsch, H.~J., 1997. Deriving an estimate of the optimal reserve price: an
  application to {B}ritish {C}olumbian timber sales. Journal of Econometrics
  78~(2), 333--357.

\bibitem[{Pinkse(2001)}]{pinkse2001nonparametric}
Pinkse, J., 2001. Nonparametric regression estimation using weak separability.
  Unpublished manuscript.

\bibitem[{Rilstone(1996)}]{rilstone1996nonparametric}
Rilstone, P., 1996. Nonparametric estimation of models with generated
  regressors. International Economic Review 37~(2), 299--313.

\bibitem[{Roberts(2013)}]{roberts2013unobserved}
Roberts, J.~W., 2013. Unobserved heterogeneity and reserve prices in auctions.
  RAND Journal of Economics 44~(4), 712--732.

\bibitem[{Roberts and Sweeting(2013)}]{roberts2013should}
Roberts, J.~W., Sweeting, A., 2013. When should sellers use auctions? American
  Economic Review 103~(5), 1830--1861.

\bibitem[{Serfling(2009)}]{Serfling_Approximation_Theorems}
Serfling, R.~J., 2009. Approximation Theorems of Mathematical Statistics. John
  Wiley \& Sons.

\bibitem[{van~der Vaart(2000)}]{VanDerVaart_Asymptotic_Statistics_Book}
van~der Vaart, A.~W., 2000. Asymptotic Statistics. Cambridge University Press.

\bibitem[{van~der Vaart and Wellner(1996)}]{van1996weak}
van~der Vaart, A.~W., Wellner, J.~A., 1996. Weak Convergence and Empirical
  Processes with Applications to Statistics. Springer.

\bibitem[{Zincenko(2018)}]{zincenko2018nonparametric}
Zincenko, F., 2018. Nonparametric estimation of first-price auctions with
  risk-averse bidders. Journal of Econometrics 205~(2), 303--335.

\end{thebibliography}
